\documentclass[hyperref,12pt]{cernyrep}
\usepackage{hyperref}
\usepackage{ifpdf}
\usepackage{url}
\usepackage{subfigure}
\usepackage{rotating}
\usepackage{units}
\usepackage[english]{babel}
\usepackage{feynmp,xcolor}
\usepackage{graphicx}
\usepackage{cite}
\usepackage{color}
\usepackage[utf8]{inputenc}
\usepackage{enumitem}
\usepackage{xspace}
\usepackage[compat=1.1.0]{tikz-feynman}

\newcommand{\be}{\begin{equation}}
\newcommand{\ee}{\end{equation}}
\newcommand{\bea}{\begin{eqnarray}}
\newcommand{\eea}{\end{eqnarray}}
\newcommand{\mtlep}{\ensuremath{m_{\mathrm T}^{\ell}}}
\def\etmiss{\ensuremath{E_{\mathrm{T}}^{\mathrm{miss}}}\xspace}

\def\ptmiss{\ensuremath{\vec p^{\mathrm{\ miss}}_\mathrm{T}\xspace}}
\newcommand{\mttwo}{\ensuremath{m_{\mathrm{T2}}}\xspace}
\newcommand{\mttwoblj}{\ensuremath{m_{\mathrm{T2}_{blj}}}\xspace}
\newcommand{\amttwo}{\ensuremath{am_{\mathrm{T2}}}\xspace}
\def\ttbar{\ensuremath{t\bar{t}}}

\def\TeV{\ifmmode {\mathrm{\ Te\kern -0.1em V}}\else
                   \textrm{Te\kern -0.1em V}\fi}%
\def\GeV{\ifmmode {\mathrm{\ Ge\kern -0.1em V}}\else
                   \textrm{Ge\kern -0.1em V}\fi}%
\newcommand{\pt}{\ensuremath{p_{\mathrm{T}}}\xspace}

\hypersetup{
  colorlinks,
  citecolor=blue,
  linkcolor=red,
  urlcolor=blue}

\usepackage{tikz}
\usetikzlibrary{trees}
\usetikzlibrary{decorations.pathmorphing}
\usetikzlibrary{decorations.markings}

  \definecolor{jblue}  {RGB}{20,50,100}
  \definecolor{npurple}  {RGB} {153, 51, 204}
  \definecolor{wred}   {RGB}{217,0,56}
  \definecolor{white}   {RGB}{255,255,255}
  
  \definecolor{korange}   {RGB}{235, 80,  43}
  \definecolor{korange2}   {RGB}{245, 100,  63}
  \definecolor{kyelloworange}   {RGB}{255, 210,  110}
  \definecolor{kyelloworange2}   {RGB}{240, 170,  90}
  \definecolor{kred}   {RGB}{204,  102, 153}
  \definecolor{kpurple}   {RGB}{153,  61, 190}
  \definecolor{kpurplelight}   {RGB}{213,  161, 230}

	\tikzset{
	  photon/.style={decorate, decoration={snake}, draw=npurple,very thick},
	  boson/.style={decorate, decoration={snake}, draw=npurple,very thick},
	  electron/.style={draw=jblue,very thick, postaction={decorate},
	           decoration={markings,mark=at position .55 with {\arrow[draw=jblue]{>}}}
	  },
	  electron2/.style={draw=jblue,very thick, postaction={decorate},
	           decoration={markings,mark=at position .55 with {\arrow[draw=jblue]{<}}}
	  },
	  fermion/.style={draw=jblue,very thick, postaction={decorate},
	            decoration={markings,mark=at position .55 with {\arrow[draw=jblue]{}}}
	  },
	  gluon/.style={decorate, draw=korange,very thick, 
	    decoration={coil,amplitude=4pt, segment length=6pt}},
	  higgs/.style={draw=wred,very thick, postaction={decorate},
	           decoration={markings,mark=at position .55 with {\arrow[draw=wred]{>}}}
	  },
	  graviton/.style={draw=wred,very thick, postaction={decorate},
	           decoration={snake}
	  },
	  nothing/.style={draw=white,very thick}
	}

\addto\captionsenglish{}

\makeatletter
\def\myaddcontentsline@toc#1#2#3{%
   \addtocontents{#1}{\protect\thispagestyle{empty}}%
   \addtocontents{#1}{\protect \vspace{-.5cm}}
   \addtocontents{#1}{\protect\contentsline{#2}{#3}{}{}}
   \addtocontents{#1}{\protect \vspace{-.3cm}}}
\def\myaddcontentsline#1#2#3{%
  \expandafter\@ifundefined{addcontentsline@#1}%
  {\addtocontents{#1}{\protect \vspace{-.5cm}}
  \addtocontents{#1}{\protect\contentsline{#2}{#3}{}{}}
  \addtocontents{#1}{\protect \vspace{-.3cm}}}
  {\csname addcontentsline@#1\endcsname{#1}{#2}{#3}}{}}
\makeatother

\makeatletter
\def\mysuperaddcontentsline@toc#1#2#3{%
   \addtocontents{#1}{\protect\thispagestyle{empty}}%
   \addtocontents{#1}{\protect \vspace{.1cm}}
   \addtocontents{#1}{\protect\contentsline{#2}{#3}{\thepage}{}}
   \addtocontents{#1}{\protect \vspace{-.1cm}}}
\def\mysuperaddcontentsline#1#2#3{%
  \expandafter\@ifundefined{addcontentsline@#1}%
  {\addtocontents{#1}{\protect \vspace{.1cm}}
  \addtocontents{#1}{\protect\contentsline{#2}{#3}{\thepage}{}}
  \addtocontents{#1}{\protect \vspace{-.1cm}}}
  {\csname addcontentsline@#1\endcsname{#1}{#2}{#3}}{}}
\makeatother

\newcommand{\superpart}[1]{\clearpage \thispagestyle{empty}  \vphantom{blabla}\vspace{5cm} \centerline{\huge #1} 
\mysuperaddcontentsline{toc}{part}{\protect\numberline{} \hspace{-2cm}#1}%
\clearpage}
   
\newcommand{\AddToContent}[1]{\myaddcontentsline{toc}{chapter}{\protect  \textit{\hspace{1.5cm} \small#1}}}

\renewcommand{\thefigure}{\arabic{figure}}
\renewcommand{\thetable}{\arabic{table}}
\renewcommand{\theequation}{\arabic{equation}}

\def\etmiss{E_T^{miss}}

\catcode`\@=11
\font\manfnt=manfnt
\def\Watchout{\@ifnextchar [{\W@tchout}{\W@tchout[1]}}
\def\W@tchout[#1]{{\manfnt\@tempcnta#1\relax%
  \@whilenum\@tempcnta>\z@\do{%
    \char"7F\hskip 0.3em\advance\@tempcnta\m@ne}}}
\let\foo\W@tchout
\def\dubious{\@ifnextchar[{\@dubious}{\@dubious[1]}}

\def\@dubious[#1]{%
  \setbox\@tempboxa\hbox{\@W@tchout#1}
  \@tempdima\wd\@tempboxa
  \list{}{\leftmargin\@tempdima}\item[\hbox to 0pt{\hss\@W@tchout#1}]}
\def\@W@tchout#1{\W@tchout[#1]}
\catcode`\@=12

\newcommand{\eq}[1]{Eq.~(\ref{#1})}

\newcommand{\leer}[1]{}

\newcommand{\mt}[1]{\ensuremath{m_{t'}}}
\newcommand{\mb}[1]{\ensuremath{m_{b'}}}
\newcommand{\me}[1]{\ensuremath{m_{\tau'}}}
\newcommand{\mv}[1]{\ensuremath{m_{\nu'}}}

\begin{document}

\setcounter{tocdepth}{0}
\thispagestyle{empty}

\vspace{1cm}

\begin{center}
{\Large {\bf LES HOUCHES 2017: PHYSICS AT TEV COLLIDERS \\[4mm]}}
{\Large {\bf NEW PHYSICS WORKING GROUP REPORT}}
\end{center}

\vspace{0.1cm}
\begin{center}
\textbf{G.~Brooijmans}$^{1}$, 
\textbf{M.~Dolan}$^{2}$, %
\textbf{S.~Gori}$^{3}$, 
\textbf{F.~Maltoni}$^{4}$, 
\textbf{M.~McCullough}$^{5}$, %
\textbf{P.~Musella}$^{6}$, %
\textbf{L.~Perrozzi}$^{6}$, %
\textbf{P.~Richardson}$^{5,7}$ 
\textbf{and}
\textbf{F.~Riva}$^{5,8}$, 
\textbf{(convenors)}\\
A.~Angelescu$^{9}$, 
S.~Banerjee$^{7,10}$, 
D.~Barducci$^{11}$, 
G.~B\'elanger$^{10}$, 
B.~Bhattacherjee$^{12}$, 
M.~Borsato$^{13}$, 
A.~Buckley$^{14}$, 
J.~M.~Butterworth$^{15}$, 
G.~Cacciapaglia$^{16}$, 
H.~Cai$^{16}$, 
A.~Carvalho$^{17}$, 
A.~Chakraborty$^{18}$, 
G.~Cottin$^{19}$, 
A.~Deandrea$^{16,20}$, 
J.~de~Blas$^{21,22}$, 
N.~Desai$^{23}$, 
M.~Endo$^{18}$, 
N.~Ezroura$^{24}$, 
G.~Facini$^{15}$, 
S.~Fichet$^{25}$, 
L.~Finco$^{16}$, 
T.~Flacke $^{26}$, 
B.~Fuks$^{20,27}$, 
P.~Gardner$^{24}$, 
S.~Gascon-Shotkin$^{16}$, 
A.~Goudelis$^{27,28}$, 
P.~Gras$^{29}$, 
D.~Grellscheid$^{7}$, 
R.~Gr\"ober$^{7}$, 
D.~Guadagnoli$^{10}$, 
U.~Haisch$^{5,30}$, 
J.~Harz$^{27,28}$, 
J.~Heisig$^{31}$, 
B.~Herrmann$^{10}$, 
J.~Hewett$^{32}$, 
T.~Hryn'ova$^{33}$, 
J.~F.~Kamenik$^{34,35}$, 
S.~Kraml$^{36}$, 
U.~Laa$^{37,38}$, 
K.~Lane$^{39}$, 
A.~Lessa$^{40}$, 
S.~Liebler$^{41}$, 
K.~Lohwasser$^{42}$, 
D.~M.~Lombardo$^{8}$, 
D.~Majumder$^{43}$, 
A.~Malinauskas$^{30}$, 
O.~Mattelaer$^{4}$, 
K.~Mimasu$^{4}$, 
G.~Moreau$^{44}$, 
M.~M\"uhlleitner$^{41}$, 
A.~E.~Nelson$^{45}$, 
J.~M.~No$^{46}$, 
M.~M.~Nojiri$^{18,47}$, 
P.~Pani$^{48}$, 
L.~Panizzi$^{49,50}$, 
M.~Park$^{45}$, 
G.~Polesello$^{51}$, 
W.~Porod$^{52}$, 
L.~Pritchett$^{39}$, 
H.~B.~Prosper$^{53}$, 
A.~Pukhov$^{54}$, 
J.~Quevillon$^{36}$, 
T.~Rizzo$^{32}$, 
P.~Roloff$^{48}$, 
H.~Rzehak$^{55}$, 
S.~Sekmen$^{56}$, 
D.~Sengupta$^{57}$, 
M.~Spira$^{58}$, 
C.~Vernieri$^{59}$, 
D.~G.~E.~Walker$^{24}$, 
D.~Yallup$^{15}$, 
B.~Zaldivar$^{10}$, 
S.~Zhang$^{16,60}$, 
J.~Zurita$^{61,62}$ 
\end{center}

 \vspace{1cm}
\begin{center}
{\large {\bf Abstract}}\\[.2cm]
\end{center}
We present the activities of the `New Physics' working group for the `Physics at
TeV Colliders' workshop (Les Houches, France, 5--23 June, 2017). Our report
includes new physics studies connected with the Higgs boson and its properties,
direct search strategies, reinterpretation of the LHC results in the building
of viable models and new computational tool developments. 
\vspace{1cm}
\begin{center}
{\bf Acknowledgements}\\[.2cm]
\end{center}
We would like to heartily thank all funding bodies, the organisers
(N.~Berger, F.~Boudjema, C.~Delaunay, M.~Delmastro, S.~Gascon, P.~Gras,
J.P.~Guillet, B.~Herrmann, S.~Kraml, G.~Moreau, E.~Re,
P.~Slavich and D.~Zerwas), the staff and all participants of the Les
Houches workshop for providing a stimulating and lively atmosphere in which to
work.

\newpage

\vspace{1cm}

\thispagestyle{empty}
\setcounter{page}{2}

\noindent
{\footnotesize
$^1$ Physics Department, Columbia University, New York, NY 10027, USA\\ 
$^{2}$ ARC Centre of Excellence for Particle Physics at the Terascale, School of Physics, The University of Melbourne, Victoria 3010, Australia\\ 
$^{3}$  Department of Physics, University of Cincinnati, Cincinnati, Ohio 45221, USA\\
$^{4}$ Centre for Cosmology, Particle Physics and Phenomenology (CP3), Universit\'e catholique de Louvain, Chemin du Cyclotron, 2, B-1348 Louvain-la-Neuve, Belgium \\ 
$^{5}$ Theoretical Physics Department, CERN, Geneva, Switzerland\\
$^{6}$ ETH Zurich - Institute for Particle Physics and Astrophysics (IPA), Zurich, Switzerland \\ 
$^{7}$ Institute for Particle Physics Phenomenology, Department of Physics, Durham University, DH1 3LE, Durham, UK \\ 
$^{8}$ D\'epartement  de  Physique  Th\'eorique, Universit\'e  de  Gen\`eve,  24  quai  E.  Ansermet,  CH-1211  Geneva,  Switzerland \\ 
$^{9}$  Department of Physics and Astronomy, University of Nebraska-Lincoln, Lincoln, NE, 68588, USA\\
$^{10}$ Univ. Grenoble Alpes, CNRS, USMB, LAPTh, 74000 Annecy, France \\ 
$^{11}$ SISSA and INFN, Sezione di Trieste, via Bonomea 265, 34136 Trieste, Italy\\ 
$^{12}$ Centre for High Energy Physics, Indian Institute of Science, Bangalore 560012, India\\ 
$^{13}$ Universidade de Santiago de Compostela, Santiago de Compostela, Spain\\ 
$^{14}$ SUPA, School of Physics and Astronomy, University of Glasgow, Glasgow G12 8QQ, UK\\ 
$^{15}$ Department of Physics \& Astronomy, UCL, London, UK\\ 
$^{16}$ Univ. Lyon, Universit\'e Claude Bernard Lyon 1, CNRS/IN2P3, IPNL, 69622 Villeurbanne, France\\ 
$^{17}$ National Institute for Chemical Physics and Biophysics, Akadeemia tee, 10143 Tallinn, Estonia\\ 
$^{18}$ Theory Center, Institute of Particle and Nuclear Studies, KEK, 1-1 Oho, Tsukuba, Ibaraki 305-0801, Japan\\ 
$^{19}$ Department of Physics, National Taiwan University, Taipei 10617, Taiwan \\ 
$^{20}$ Institut Universitaire de France, 103 boulevard Saint-Michel, 75005 Paris, France\\
$^{21}$ Dipartimento di Fisica e Astronomia ``Galileo Galilei'', Universit\`a di Padova, I-35131 Padova, Italy \\
$^{22}$ INFN, Sezione di Padova, I-35131 Padova, Italy\\
$^{23}$ Laboratoire Univers et Particules de Montpellier, CNRS-Universit\'e de Montpellier, 34095 Montpellier, France\\
$^{24}$ Department of Physics and Astronomy, Dartmouth College, Hanover, NH 03755, USA\\
$^{25}$ ICTP-SAIFR \& IFT-UNESP, R. Dr. Bento Teobaldo Ferraz 271, S\~ao Paulo, Brazil\\
$^{26}$ Center for Theoretical Physics of the Universe, Institute for Basic Science (IBS), Daejeon, 34126, Korea\\
$^{27}$ Sorbonne Universit\'e, CNRS, Laboratoire de Physique Th\'eorique et Hautes \'Energies, LPTHE, F-75005 Paris, France\\
$^{28}$ Sorbonne Universit\'e, Institut Lagrange de Paris (ILP), 75014 Paris, France\\
$^{29}$ IRFU, CEA, Universit\'e Paris-Saclay, Gif-sur-Yvette, France\\
$^{30}$ Rudolf Peierls Centre for Theoretical Physics, University of Oxford, OX1 3PN Oxford, United Kingdom\\
$^{31}$ Institute for Theoretical Particle Physics and Cosmology, RWTH Aachen University, 52056 Aachen, Germany\\
$^{32}$ SLAC National Accelerator Laboratory, 2575 Sand Hill Rd., Menlo Park, CA, 94025, USA\\
$^{33}$ Univ. Grenoble Alpes, Univ. Savoie Mont Blanc, CNRS, LAPP, 74000 Annecy, France\\
$^{34}$ Jo\v zef Stefan Institute, Jamova 39, 1000 Ljubljana, Slovenia\\
$^{35}$ Faculty of Mathematics and Physics, University of Ljubljana, Jadranska 19, 1000 Ljubljana, Slovenia\\
$^{36}$ Univ. Grenoble Alpes, CNRS, Grenoble INP, LPSC-IN2P3, 38000 Grenoble, France\\
$^{37}$ School of Physics and Astronomy, Monash University, Clayton, VIC 3800, Australia\\
$^{38}$ Department of Econometrics and Business Statistics, Monash University, Clayton, VIC 3800, Australia\\
$^{39}$ Department of Physics, Boston University, Boston, MA 02215, USA\\
$^{40}$ Centro de Ci\^encias Naturais e Humanas, Universidade Federal do ABC, Santo Andr\'e, 09210-580 SP, Brazil\\
$^{41}$ Institute for Theoretical Physics, Karlsruhe Institute of Technology, 76131 Karlsruhe, Germany\\
$^{42}$ Department of Physics \& Astronomy, Sheffield University, UK\\
$^{43}$ The University of Kansas, Lawrence, KS, USA\\
$^{44}$ Laboratoire de Physique Th\'eorique, B\^at. 210, CNRS, Univ. Paris Sud, Universit\'e Paris-Saclay, F-91405 Orsay Cedex, France\\
$^{45}$ Department of Physics, Box 1560, University of Washington, Seattle, WA 98195-1560 USA\\
$^{46}$ Department of Physics, King's College London, Strand, WC2R 2LS London, UK\\
$^{47}$ Kavli IPMU (WPI), University of Tokyo, Kashiwa, Chiba 277-8583, Japan\\
$^{48}$ CERN, Experimental Physics Department, CH-1211 Geneva 23, Switzerland\\
$^{49}$ Dipartimento di Fisica, Universit\`a di Genova and INFN, Sezione di Genova, 16146 Genova, Italy\\
$^{50}$ School of Physics and Astronomy, University of Southampton, Highfield, Southampton SO17 1BJ, UK\\
$^{51}$ INFN, Sezione di Pavia, Via Bassi 6, 27100 Pavia, Italy\\
$^{52}$ Institut f\"ur Theoretische Physik und Astrophysik, Uni.\ W\"urzburg, 97070 W\"urzburg, Germany \\
$^{53}$ Department of Physics, Florida State University, Tallahassee, Florida 32306, USA\\
$^{54}$ Skobeltsyn Institute of Nuclear Physics, Moscow State University, Moscow 119991, Russia\\
$^{55}$ CP3-Origins, University of Southern Denmark, Odense, Denmark \\
$^{56}$ Kyungpook National University, Physics Dept., Daegu, South Korea\\
$^{57}$ Department of Physics and Astronomy, Michigan State University, East Lansing, Michigan-48824, USA\\
$^{58}$ Paul Scherrer Institut, CH--5232 Villigen PSI, Switzerland\\
$^{59}$ Fermi National Accelerator Laboratory, Batavia, IL, 60510, USA\\
$^{60}$ Institute of High Energy Physics/University of the Chinese Academy of Sciences, Shijingshan District, Beijing, China 100049\\
$^{61}$ Institute for Theoretical Particle Physics (TTP), Karlsruhe Institute of Technology, 76128 Karlsruhe, Germany \\
$^{62}$ Institute for Nuclear Physics (IKP), KIT Karlsruhe Institute of Technology, D-76344 Eggenstein-Leopoldshafen, Germany
}


\newpage

\tableofcontentscern

\newpage

\noindent {\Large {\bf Introduction}}
\vspace{.5cm}

{\it G.~Brooijmans, M.~Dolan, S.~Gori, F.~Maltoni, M.~McCullough, P.~Musella, L.~Perrozzi, P.~Richardson, F.~Riva}\\[.4cm]

This document is the report of the New Physics session of the 2017 Les Houches
Workshop `Physics at TeV Colliders'. The workshop brought together theorists
and experimenters who discussed a significant number of novel ideas related to Higgs and
beyond the Standard Model physics.  New computational methods
and techniques were considered, with the aim of improving the technology
available for theoretical, phenomenological and experimental new physics studies.

More precisely, one set of studies undertaken during
the workshop
concerns investigations associated with specific new physics models either
constructed from a top-down approach or built following a bottom-up path. A
second set of studies is connected to the Higgs boson discovered a few years
ago. Its properties are now measured with increasing accuracy at the
LHC, constraining the construction of any realistic
new physics theory correspondingly. Finally,
recasting techniques are the subject of a third
series of contributions, including suggestions on the way experimental
information could be presented.

In the first section searches for new physics beyond the Standard Model are presented, covering diverse frontiers in the hunt for new states, from new two-body resonances to new particles with macroscopic lifetimes.  A first contribution reviews two-body resonance searches at the LHC, highlighting a few cases that are not currently covered.  Many models contain vector-like quarks, for which single production can significantly enhance the search range.  In direct single production, next-to-leading order effects, studied in a second contribution, can have significant effects on distributions used to separate the signal from the Standard Model background.  The third contribution explores another way to produce single vector-like quarks: the production of new heavy spin-0 or spin-1 bosons that decay to a vector-like quark and a Standard Model fermion, which again leads to changes in distributions used to discriminate against the backgrounds.
Three contributions tackle macroscopic lifetimes.  The first reveals the connection between searches for long-lived particles (LLPs) and a compelling paradigm for dark matter production, known as ``freeze-in''.  This connection relates the dark matter abundance in freeze-in models to the lifetime of the LLP produced at the LHC, making the lifetime determination a key target for such models.  A complementary study determines the accuracy with which one could hope to answer this question, revealing how detector effects or analysis cuts could influence the accuracy with which the LLP lifetime could be determined.  A related study in the tools sections exposes how attempts to recast current LHC LLP searches may be hampered by the format in which analysis details are presented.  Consequentially, this contribution makes recommendations on the presentation of analysis details to maximise the impact of LHC searches in recasting for alternative scenarios.

In the top sector, two studies have been performed.  One explores the ``mono-top'' signature of dark matter production, and a second flavour-violating top squark decays, showing the limitations of simplied models in quantifying LHC sensitivity to new physics.  This is followed by an examination of the sensitivity of the LHC to intermediate mass (pseudo-)scalars produced in association with heavy flavor quark partners, indicating that with some optimization the LHC could cover the relevant areas in parameter space.

Another area in which the face of new physics may be already partially revealed is flavour physics, particularly concerning b-quarks, where a number of mild anomalies in individual measurements of different b-meson decay final states may be consistently pointing towards evidence of violation of lepton-flavour universality.  The status of these anomalies and potential theories of new physics that may coherently explain the measurements through the existence of new particles is reviewed.

The second section includes contributions related to the physics of the Higgs boson and the electroweak symmetry breaking sector. Some contributions focus on effective field theories (EFTs) for Higgs physics, some on models with new light Higgs bosons, and some on models with additional heavy Higgs bosons. 
In particular, in the context of EFTs, one study compares different ways of experimentally accessing Beyond the Standard Model (BSM) physics via EFTs: either via a direct search dedicated to these effects, or via Simplified Template Cross Sections (STXS). Another study discusses, instead, the EFT reach at linear colliders (using CLIC as benchmark) in diboson processes, and designs a dedicated search to maximise it. Another study compares different EFT bases numerically and identifies the higher order effects that differ between them. Finally, the last study investigates the potential of the measured Higgs fiducial cross sections for deriving constraints on BSM Higgs production.
In the context of models with new light Higgs bosons, one contribution investigates the bounds on models containing new scalars addressing the $Z\to b\bar b$ ALEPH anomaly, using Contur. A study of collider bounds on light pseudoscalars with a mass below 50 GeV is also presented, focusing on the mass regions [3, 5] GeV and [9, 11] GeV, where the mixing of the pseudoscalar with QCD bound states has to be included. A last project analyses the LHC prospects to discover a light scalar with mass below 65 GeV produced in association with a $Z$ boson.
Finally, in the context of models with new heavy Higgs bosons, a study presents the full NLO corrections to Higgs gluon fusion in SUSY QCD, within the framework of the MSSM including the full mass dependence of the particles running in the loop. We also report the first sensitivity study of the channel: $H_1 \to H_2 H_2 \to b \bar b b \bar b$, with both $H_1$ and $H_2$ states beyond the Standard Model, and we discuss the importance of interference effects in the search for heavy Higgs bosons decaying into $hh$ and $Zh$.

Finally, the third section presents progress specific to software tools and methods that are crucial for any new physics investigation.
Four contributions are included.
The first focuses on the sensitivity of SM LHC measurements to new particles simulated through simplified models,
using pairs of photons in the final state.
The exercise shows that the generic light scalar models considered imply significant contributions to differential
cross sections involving weak bosons and/or isolated photons which have already been measured at the LHC and shown to be consistent with the Standard Model,
posing stringent constraints on the new physics parameter space. 
The second contribution proposes a first benchmark comparison assessing the performance of different public recasting tools in reproducing 
ATLAS and CMS searches with Monte Carlo simulation. 
The analyses considered show good agreement between the different frameworks and detector simulation techniques.
The proposed method can be further applied to assess the reliability of the recasting methods in, 
e.g. extreme regions of phase space and/or for very different signal hypotheses than the one the analyses have been designed for.
The third deals with the recasting of Long-Lived Particles Searches.
In fact, extrapolating LHC search limits to other scenarios often proves to be a difficult task outside the experimental collaborations.
The study proves that without detailed object reconstruction and selection efficiencies 
a satisfactory recasting can not be performed, and provides recommendations to
the experimental collaborations to include cut-flow tables and limits for at least two models or topologies as a sanity check.
The last contribution investigates the usage of an analysis description language for LHC result reinterpretations,
to be employed to describe in an unambiguous and concise manner a data analysis including all the details needed for recasting.

The meeting in Les Houches has fostered a large number of discussions between
theorists and experimenters. In-depth studies could however only be completed
for several of the generated ideas on the required timescale. It is clear that
even those that could not converge to a written contribution have paid off
through the breadth of searches conducted by experimenters and the understanding 
of the challenges placed on an experiment by the ever-changing theoretical landscape.
We expect that
many more future results will  benefit from the discussions held at the
workshop.

\addtocontents{toc}{\protect\contentsline{part}{\protect\numberline{} \hspace{-2cm}Introduction}{6}{}}
\AddToContent{G.~Brooijmans, M.~Dolan, S.~Gori, F.~Maltoni, M.~McCullough, P.~Musella, L.~Perrozzi, P.~Richardson, F.~Riva}

\setcounter{figure}{0}
\setcounter{table}{0}
\setcounter{section}{0}
\setcounter{equation}{0}
\setcounter{footnote}{0}
\clearpage


\superpart{ New physics }


\chapter{Overview of two-body resonant searches at LHC}

{\it G.~Brooijmans, G.~Facini, J.~Hewett, T.~Hryn'ova, T.~Rizzo}





This contribution presents a novel approach to categorizing the existing two-body resonant searches at LHC and is aimed to 
identify some potential search channels that have been so far overlooked. This work is inspired by Ref.~\cite{Craig:2016rqv}.

A typical resonance search looks for a bump on top of a smooth background. Some searches are intrinsically 
model independent (e.g. the inclusive, opposite-sign dilepton searches~\cite{Aaboud:2017buh}). Other searches are more fine-tuned 
to a specific new physics model, such as the  same-sign dilepton search which looks in particular for the pair-production of the doubly charged 
Higgs bosons~\cite{Aaboud:2017qph} so that an additional invariant mass requirement is added to the more general search criteria. 
Here we will specifically highlight these general searches in our categorization in order to make it as model-independent at possible:
\begin{itemize}
\item Existing two body resonant searches are summarized in Table~\ref{tab:bsmsearch-1}. 
\item Existing two body resonant searches which have additional particle 
or double-production requirements are summarized in Table~\ref{tab:bsmsearch-2}. 
\item If no search is performed in either of the above categories, the corresponding channel is marked with "0", the channels covered only in one of two tables are marked by ``*''.  
\item The neutrino ($\nu$) category in this table experimentally corresponds to a missing transverse momentum requirement. Typical searches (except in the $l\nu$ channel) assume not a Standard 
Model neutrino as the MET source, but some yet to be discovered stable or long-living neutral particle. 
\end{itemize}
The tables show the most recently available results from the ATLAS or CMS collaborations with highest available integrated luminosity for the largest possible center-of-mass energy. 

Some single production searches, e.g. $l\nu$ (W'), will not cover the pair-production of similar resonances ($l\nu l\nu$), because this search is performed employing the transverse mass 
variable and assumes that all of the missing transverse energy comes from a single particle. It also does not cover associated production (e.g. $ll\nu$) because it has a second lepton 
veto~\cite{Aaboud:2017efa}. We note the cases where the final state {\it is} covered, but the  search was not done using the corresponding mass distribution, which might reduce its sensitivity 
to find a resonance. For example, in Ref.~\cite{Aaboud:2017qpr} ($Zt$ channel) a single bin analysis is performed for the pair production search for vector-like-quarks in the one lepton, jets, plus 
missing transverse momentum channel. In Ref.~\cite{Aad:2015mba} ($Wt$ channel) both a cut-based analysis and a boosted-decision-tree approach are used. The latter might not be 
easily reinterpretable for other models.

The following channels are identified as completely uncovered by present searches, although they would be 
interesting to pursue in the context of certain models as 
indicated below: 
\begin{itemize}
\item $be$, $te$, $b\mu$, $t\mu$ - leptoquarks
\item $\tau\gamma$, $b\gamma$, $t\gamma$ - excited leptons and quarks
\item $tj$ - vector-like top quark
\item $hj$ - vector-like light flavour quark
\item $h\gamma$ - Kaluza-Klein excitation of Higgs
\end{itemize}

Some of the channels not covered by the dedicated searches above ($be$, $b\mu$, $b\gamma$) are instead covered in the so-called general search~\cite{ATLAS-CONF-2017-001}, 
in which various combinations and multiplicities of electrons, muons, photons, jets, b-jets and missing transverse momentum are scanned for deviations from the Standard Model Monte
Carlo prediction in the distributions of the effective mass and the total visible invariant mass of the reconstructed objetcs.

We believe it might be interesting for the ATLAS and CMS collaborations to adopt this approach in the presentation of summaries of the two-body resonant search results in 
addition to their current summary tables.  Furthermore, this analysis should be extended to three-body resonance searches, as in 
many new physics models the largest production cross-sections are not for the lightest new particles, and heavier particles 
are naturally strongly coupled to lighter ones.  

\begin{table} 
\begin{center}
\caption{Existing two-body exclusive final state resonance searches performed by ATLAS or CMS. Only highest luminosity and largest center-of-mass energy publication is shown, the default being the full 2015+2016
dataset. References in {\it italic} employ only the 2015 or partial 2015-2016 datasets.
\underline{Underlined references} use Run 1 data. Note that $e/\mu+MET$ had extra lepton veto applied. 
}\label{tab:bsmsearch-1}
\hspace*{-2cm}
\begin{tabular}{c|c|c|c|c|c|c|c|c|c|c|c}\hline\hline &      $e$ & $\mu$ & $\tau$ & $\nu$ & $j$ & $b$ & $t$ & $\gamma$ &  $Z$ & $W$ & h \\\hline
$e$ &  $\pm\mp$\cite{Aaboud:2017buh},$\pm\pm$\cite{Aaboud:2017qph} &  $\pm\pm$\cite{Aaboud:2017qph}    & {\it \cite{Aaboud:2016hmk}} & \cite{Aaboud:2017efa} & \underline{\cite{Aad:2013gma}}&  0   & 0    &  *   & \underline{\cite{Aad:2015dha}} &  *   & *  \\
$\mu$ & - &  $\pm\mp$\cite{Aaboud:2017buh},$\pm\pm$\cite{Aaboud:2017qph}  &  {\it \cite{Aaboud:2016hmk}}  &   \cite{Aaboud:2017efa} & \underline{\cite{Aad:2013gma}} &  0   &  0   &  *   &  \underline{\cite{Aad:2015dha}}   &  *   &  * \\
$\tau$ & - & - &  \cite{Aaboud:2017sjh} & \underline{\cite{Khachatryan:2015pua},\cite{CMS-PAS-EXO-16-006}} &   *   & *    & *  &  0   &  *   &  *   & *  \\
$\nu$ & - & - &  - &  * (?)     & *    &  *   &  *   &  \cite{Sirunyan:2017nyt}   &  *   &  *   & *  \\
$j$ & - & - & - & - & \cite{Aaboud:2017yvp} & {\it \cite{Aaboud:2016nbq}} & 0 & \cite{Aaboud:2017nak} & \cite{Sirunyan:2017acf} & \cite{Sirunyan:2017acf} & 0\\
$b$ & - & - & - & - & - & {\it \cite{Aaboud:2016nbq}} & \cite{Sirunyan:2017vkm} & 0 & * & * & * \\
$t$ & - & - & - & - & - & - & {\it \cite{ATLAS-CONF-2016-014,Sirunyan:2017uhk}} & 0 & \cite{Sirunyan:2017ynj} & * & * \\
$\gamma$ & - & - & - & - & - & - & - & \cite{Aaboud:2017yyg} & {\it \cite{Aaboud:2016trl}} & \underline{\cite{Aad:2014fha}} & 0 \\
$Z$ & - & - & - & - & - & - & - & - & \cite{Sirunyan:2017acf} & \cite{Sirunyan:2017acf} & \cite{Sirunyan:2017wto} \\
$W$ & - & - & - & - & - & - & - & - & - & \cite{Sirunyan:2017acf} & \cite{Sirunyan:2017wto} \\
$h$ & - & - & - & - & - & - & - & - & - & - & \cite{Sirunyan:2017isc,Sirunyan:2017guj} \\
\end{tabular}
\end{center}
\end{table}

\begin{sidewaystable}
\begin{center}
\caption{Existing two-body exclusive final state resonance searches 
(with extra particle or double production requirement) by ATLAS or CMS.
Only highest luminosity and largest center-of-mass energy publication is shown, the default is the full 2015+2016 dataset. References in {\it italic} use only the 2015 or partial 2015-2016 datasets.
\underline{Underlined references} use Run 1 data.
* means that this state is covered exclusively in Table~\ref{tab:bsmsearch-1}.
x denotes double production requirement. The channels shown in red are those in which 
analyses might be not optimal for a resonant search or reinterpretation tricky because of the 
use of a multi-variate technique.
}\label{tab:bsmsearch-2}
\hspace*{-1cm}
\begin{tabular}{c|c|c|c|c|c|c|c|c|c|c|c}\hline\hline
&      $e$ & $\mu$ & $\tau$ & $\nu$ & $j$ & $b$ & $t$ & $\gamma$ &  $Z$ & $W$ & h \\\hline
$e$ & x~{$\pm\pm$\cite{Aaboud:2017qph}} & x~{$\pm\pm$\cite{Aaboud:2017qph}} & * & * & \underline{e~\cite{Khachatryan:2015qda}}, x~{\it \cite{Aaboud:2016qeg}}\underline{\cite{Khachatryan:2015vaa}}&  0   &  0   &  $e$\underline{\cite{Khachatryan:2015scf}}   &  $e$\underline{\cite{Khachatryan:2015scf}},$\nu/l V/h$\cite{Sirunyan:2017qkz}   &  $\nu/l V/h$\cite{Sirunyan:2017qkz}   & $\nu/l V/h$\cite{Sirunyan:2017qkz}  \\
$\mu$ & - & x~{$\pm\pm$\cite{Aaboud:2017qph}}  & * & * & \underline{$\mu$~\cite{Khachatryan:2015qda}}, x~{\it \cite{Aaboud:2016qeg}}\underline{\cite{Khachatryan:2015vaa}} &  0   &  0   &  $\mu$\underline{\cite{Khachatryan:2015scf}}   & $\mu$\underline{\cite{Khachatryan:2015scf}},$\nu/l V/h$\cite{Sirunyan:2017qkz}    &  $\nu/l V/h$\cite{Sirunyan:2017qkz}   & $\nu/l V/h$\cite{Sirunyan:2017qkz}  \\
$\tau$ & - & - & * &  * & x {\it \cite{Sirunyan:2017yrk}} & x\underline{\cite{Khachatryan:2014ura}}   & x\underline{\cite{Khachatryan:2015bsa}}    &  0   &  $\nu/l V/h$\cite{Sirunyan:2017qkz}   &  $\nu/l V/h$\cite{Sirunyan:2017qkz}   &  $\nu/l V/h$\cite{Sirunyan:2017qkz} \\
$\nu$ & - & - &  - &  j\cite{Aaboud:2017phn,Sirunyan:2017jix}, $\gamma$\cite{Aaboud:2017dor}, H\cite{Aaboud:2017yqz} & $ej,~\mu j$\underline{\cite{Khachatryan:2015vaa}} & x\underline{\cite{Chatrchyan:2012st},\cite{Aad:2015caa}} &   x\underline{\cite{Aad:2015caa}}  &  *   &  $\nu/l V/h$\cite{Sirunyan:2017qkz}   &  $\nu/l V/h$\cite{Sirunyan:2017qkz}   & $\nu/l V/h$\cite{Sirunyan:2017qkz}\\
 &  &  &   &  W\cite{Sirunyan:2017jix}, Z\cite{Aaboud:2017bja,Sirunyan:2017qfc,Sirunyan:2017jix} & & &  &   &   &    & \\
$j$ & - & - & - & - & j\cite{Sirunyan:2017nvi} & * & 0 & * & * & * & 0 \\
$b$ & - & - & - & - & - & * & * & 0 & \underline{bj;V/H$+$b\cite{Aad:2014efa}} & x\cite{Sirunyan:2017pks,Aaboud:2017zfn} & \underline{b\cite{Aad:2016shx};V/H$+$b/t\cite{Aad:2015kqa}} \\
$t$ & - & - & - & - & - & - & * & 0 & \color{red}{ x \cite{Aaboud:2017qpr}} & \color{red}{\underline{V/H$+$b/t\cite{Aad:2015mba}}} & \color{red}{\it V/H$+$b/t\cite{ATLAS-CONF-2016-032}}\\
$\gamma$ & - & - & - & - & - & - & - & \underline{$\gamma$/x\cite{Aad:2015bua}} & * & * & 0 \\
$Z$ & - & - & - & - & - & - & - & - & * & * & * \\
$W$ & - & - & - & - & - & - & - & - & - & * & * \\
$h$ & - & - & - & - & - & - & - & - & - & - & * \\
\end{tabular}
\end{center}
\end{sidewaystable}






 
\AddToContent{G.~Brooijmans, G.~Facini, J.~Hewett, T.~Hryn'ova, T.~Rizzo}
\renewcommand{\thesection}{\arabic{section}}

\newcommand{\DB}[1]{{\bf{{\color{red}[DB: #1]}}}}
\newcommand{\redcomment}[1]{{\bf{{\color{red}[#1]}}}}
\newcommand{\cyan}{\color{cyan}}

\newcommand{\TR}[1]{{\color{blue} \bf TR~\![#1]}} 
\newcommand{\RG}[1]{{\color{magenta} \bf RG~\![#1]}} 
\newcommand{\HC}[1]{{\color{blue} \bf HC~\![#1]}} 
\newcommand{\WP}[1]{{\cyan WP~\![#1]}}
\newcommand{\TF}[1]{{\color{green} \bf TF~\![#1]}} 
\newcommand{\gmu}{\gamma^\mu}

\newcommand{\qn}[1]{\textbf #1} 
\def\bsp#1\esp{\begin{split}#1\end{split}}


\chapter{Exotic decays of heavy boson into SM quarks and vector-like quarks}

{\it D.~Barducci, H.~Cai, T.~Flacke, B.~Fuks, R.~Gr\"ober, W.~Porod, T.~Rizzo}



\begin{abstract}
We identify the quantum numbers of heavy scalar or vector resonances that can be singly produced via proton-proton scattering at the LHC. We then classify the quantum numbers of heavy vector-like quarks into which the heavy bosons can decay in association with a Standard Model quark. We subsequently briefly discuss the phenomenology of these non-standard signatures at the LHC.
\end{abstract}

 \section{INTRODUCTION}

Searches for the on-shell production of heavy resonances are among the
priorities of the Large Hadron Collider (LHC) physics program and a powerful tool to probe various new physics (NP) scenarios.  In fact many beyond the Standard
Model (BSM) realizations formulated to address the shortcomings of the Standard Model (SM) predict the presence of unstable spin-0 or spin-1 states which can promptly decay into a pair of SM particles. The latter can generally be reconstructed within the LHC detectors and consequently provide sensitivity to the possible presence of such BSM states.
The most simple examples of this program are the searches for peaks in, \emph{e.g.}, the $\gamma\gamma$, $jj$ and $\ell^+\ell^-$ invariant-mass distributions. Final state consisting of a pair of unstable SM states, such as $t\bar t$, $ZZ$
and $W^+ W^-$, can also be exploited, thanks to the generally good reconstruction efficiency for such objects.

On general grounds, in order to have a significant number of signal events, the heavy resonance should decay copiously into the chosen SM final state with the event rate (within the narrow width approximation) being determined by  the product of the on-shell resonance production cross section and the corresponding branching fraction into the specific final state of interest.
However, in many NP models there exist additional decay channels for such heavy states that are open and so decays into non-SM final states can become the dominant ones.
This happens, for example, in Composite Higgs Models (CHMs), where new spin-1 resonances can have a sizeable branching fractions into a pair of vector-like quarks (VLQs) or a VLQ and a SM quark~\cite{Barducci:2012kk,Greco:2014aza,Deandrea:2017rqp} or of supersymmetric models with extended gauge symmetries, where the heavy $Z^\prime$ and $W^\prime$ can directly decay predominantly into non-SM states~\cite{Basso:2015pka,Araz:2017qcs,Araz:2017wbp}.
The ``depletion`` of the heavy resonances branching ratios into SM states can reduce the reach of the NP searches performed at the LHC and relax the constraints that can be enforced on the masses of such objects.\footnote{Note however that by reinterpreting non-dedicated analyses these bounds can be recovered~\cite{Barducci:2015vyf}.} In order to be sensitive to the maximum number of NP configurations possible, recently the experimental collaborations have started to pursue analyses targeting possible non-minimal decays of heavy resonances. This 
is, for example, the case of the CMS search of Ref.~\cite{Sirunyan:2017bfa}, wherein a heavy spin-1 resonance is looked for in a final state containing a top quark and a VLQ with an electric charge equal to 2/3.

Motivated by this analysis, we categorize in this note the possible final state configurations arising from the decay of a heavy spin-0 or spin-1 particle that can be resonantly produced in the $s$-channel via proton-proton collisions, and that can decay into a SM quark and a VLQ. We identify the SM quantum numbers of such bosonic resonances as well as the quantum numbers of the VLQs that can be present in their decays. We then discuss the associated phenomenology highlighting which channels could be experimentally covered by the reinterpretation of existing experimental analyses and which ones require a new dedicated search strategy.

\section{HEAVY RESONANCES PRODUCTION}
In order for a bosonic resonances to be produced via the $s$-channel in
proton-proton collisions, they should couple to SM quarks and/or gluons
whose quantum number under $\mathcal G_{SM}=SU(3)_c \times SU(2)_L \times U(1)_Y$ are provided for the sake of clarity in the 
following Table~\footnote{We adopt the convention $Q_{\rm em}=T^3_L+Y$.},

\renewcommand{\arraystretch}{1.3}
\begin{center}
\begin{tabular}{c||c|c|c|c}
Field       & $q_L$  & $u_R$ & $d_R$ & $g$ \\
\hline
$\mathcal G_{SM}$ &  (\textbf{3}, \textbf{2}, 1/6) & (\textbf{3}, \textbf{1}, 2/3) & (\textbf{3}, \textbf{1}, -1/3) & (\textbf{8}, \textbf{1}, 0)  \\
\end{tabular}
\end{center}
In the case of a vector resonance, the interaction structure with the SM quarks is of the form $\bar{q}_L^c \gmu u_R$, $\bar{q}_L^c \gmu d_R$, $\bar{q}_L
\gmu q_L$, $\bar{u}_R \gmu u_R$, $\bar{d}_R \gmu d_R$,  $\bar{u}_R\gmu d_R$,  while it is of the form $\bar{q}_L
u_R$, $\bar{q}_L d_R$, $\bar{q}_L q_L^c$, $\bar{u}_R^c u_R$, $\bar{u}^c_R d_R$ or $\bar{d}_R^c d_R$ in the case of a scalar
resonance. Here we have defined $\psi_L^c=(\psi_L)^c=C\gamma_0 \psi_L^*$, with
$\psi_L^c$ transforming like a right-handed field and $\psi_R^c$ like a left-handed field and with $C=i\gamma^2 \gamma^0$ in Dirac notation.
We then provide in Tab.~\ref{VSBrep} the quantum number under
$\mathcal G_{SM}=SU(3)_c \times SU(2)_L \times U(1)_Y$ of the new resonances that can be singly produced via proton-proton
collisions, together with their electric charges and the schematic form of their interaction with the proton constituents; more information is also available
in Refs.~\cite{Biggio:2016wyy,DelNobile:2009st}.
We however do not account in for the case where an interaction between a new scalar and the SM fermions arises due to mixing, as this would be for example the case for a $(\qn{1},\qn{1},0)$ scalar that acquires a vacuum expectation value 
and its interactions to the SM fermions would then stem from a mixing with the SM Higgs boson.

\begin{table}[!h]
\centering
\hspace*{-0.9cm}\begin{tabular}{|c|c|c||c|c|c|}
\hline
\multicolumn{3}{|c||}{{\bf Vectors}} & \multicolumn{3}{|c|}{{\bf Scalars}} \\
\hline $\cal G_{SM}$                           & Q$_{\rm em}$ &
Interaction                                    & $\cal G_{SM}$   &
Q$_{\rm em}$ & Interaction\\  \hline \hline $(\qn{1}\oplus\qn{8},\qn{1},0)$ & $0$ &
$\bar u u, \bar d d, \bar q q, (gg)$    & $(\qn{1},\qn{2},1/2)$ & $0,1$ & $\bar q
u, \bar q d $ \\
 \hline ($\qn{1}\oplus\qn{8}$,\qn{1},1) & 1 &
$u \bar d $ & $(\qn{8},\qn{2},1/2)$ & $0,1$ & $\bar q u, \bar q d $\\
\hline (\qn{1},\qn{3},0) & 1,0,-1           & $\bar q q$ &
$(\qn{3}\oplus \bar{\qn{6}},\qn{1},-4/3)$ & -4/3& $\bar u^c
u$\\ \hline
$(\qn{3}\oplus \bar{\qn{6}}, \qn{2}, 1/6)$
& 2/3,-1/3        & $\bar q^c d$ &
$(\qn{3}\oplus \bar{\qn{6}},\qn{1},-1/3)$ &-1/3& $\bar
q^c q, \bar d^c u$
\\ \hline
$(\qn{3}\oplus \bar{\qn{6}},\qn{2}, -5/6)$
&-1/3,-4/3         & $\bar q^c u$ &
$(\qn{3}\oplus \bar{\qn{6}},\qn{1},2/3)$ & $2/3$& $\bar d^c d$\\
\hline$(\qn{8},\qn{3},0)$ & 1,0,-1 & $\bar q q$
&$ (\qn{3}\oplus \bar{\qn{6}},\qn{3},-1/3)$ &2/3,-1/3,-4/3& $\bar q^c q$\\
\hline
\end{tabular}
\caption{The new bosons quantum numbers under the SM gauge group together with their electric charge and schematic interaction
  structure with the SM quarks and gluons, for both cases of vector (left) and
  scalar (right) resonances. The chirality of the SM quarks is implicit from the interaction
structure. In the vector case we moreover do not explicitly write the $\gmu$ factor. \label{VSBrep}}
\end{table}

From a model building point of view new vectors usually arise either in strongly-interacting theories (similar to the $\rho$ meson in QCD) or in weakly-interacting theories as part of extended gauge groups. This usually implies that there is a whole new plethora of particles which 
may participate in the vector decays. While in the former case the theory can be regarded as an effective field theory valid up to some cut-off scale and hence it can be described by a non-renormalizable theory, in the latter case, on which we focus on here,  it would be desirable 
to restore renormalizablity.  This in turns implies that certain constraints on the possible interactions between the various states must be fulfilled if the model is to be ultraviolet (UV) complete, {\emph{e.g.}} that the interactions are built up from gauge-covariant quantities. Conversely, for 
spin-0 resonances the situation is more straightforward and the SM can be simply extended by a new scalar multiplet.\footnote{If BSM scalar fields participate in electroweak symmetry breaking they can mix with the SM Higgs, and the mass mixing affects production and decay of 
the new scalar resonances (as well as of the Higgs boson). In this analysis we focus on heavy new states for which we expect scalar mass mixing effects and the effects from electroweak symmetry breaking in interactions to be suppressed by $\mathcal{O} (v/M_\phi)$, with $M_\phi$ being the scalar mass.}
Moreover, since the purpose of this study is to categorize the possible SM quantum numbers of the resonances that can be produced on-shell at the LHC and that can decay into a SM quark and a VLQ, we do not consider any other possible interaction among the heavy vectors 
and scalars with the SM fields except the ones responsible for these production and decay mechanisms. Restricting then to just gauge invariant and renormalizable interactions, the generic Lagrangians for the production of the resonances of Tab.~\ref{VSBrep} are given in Eq.~\eqref{eq:vector_prod1} and Eq.~\eqref{eq:vector_prod2} for the vector case and in Eq.~\eqref{eq:scalar_prod1} and Eq.~\eqref{eq:scalar_prod2} for the scalar case, where $\sigma^a$ are the Pauli matrices
acting in the $SU(2)_L$ space, $T^A$ are the $SU(3)_c$ generators in the fundamental representation with 
$A$=1,...,8, $\lambda, \kappa$ are generic coupling parameters and the subscripts on $V^\mu$ vector resonances and $S$ scalar resonances indicate their quantum numbers under $\mathcal G_{SM}$.\footnote{In Tab.~\ref{VSBrep}, we indicate a possible production from 
gluon fusion of a color octet vector which is allowed by virtue of the conservation of all quantum numbers, but does not follow from Eq.~\eqref{eq:scalar_prod1}. Such an interaction is absent at tree-level but is  not forbidden by the Landau Yang theorem \cite{Cacciari:2015ela} and could be 
induced at higher order.}

\begin{equation}
\bsp
  \mathcal{L}_{1\oplus8}=  &\  \kappa_{1R} \overline{u_{R}}  \gamma_{\mu}  d_{R} V_{\qn{1},\qn{1},1}^{\mu} +\kappa_{8R} \overline{u_{R}}  \gamma_{\mu} T^A d_{R}
V_{\qn{8},\qn{1},1}^{\mu,
 A} \\ & \
 +\left( \kappa_{qL}  \overline{q_{L}}  \gamma_{\mu} T^A q_{L}
+ \kappa_{uR} \overline{u_{R}}  \gamma_{\mu} T^A u_{R} +
\kappa_{dR} \overline{d_{R}}  \gamma_{\mu} T^A d_{R} \right)
V_{\qn{8},\qn{1},0}^{\mu,
 A} \\
& \  + \left( \kappa_{qL}^{\prime}\overline{q_{L}}  \gamma_{\mu}
q_{L} + \kappa_{uR}^\prime \overline{u_{R}}  \gamma_{\mu}  u_{R} +
\kappa_{dR}^\prime \overline{d_{R}}  \gamma_{\mu}  d_{R} \right)
 V_{\qn{1},\qn{1},0}^{\mu} \\
 & \ + \kappa_{3L} \overline{q_{L}} \sigma^a \gamma_{\mu} T^A q_{L}
V_{\qn{8},\qn{3},0}^{\mu,
 A, a}+ \kappa_{3L}^{\prime}\overline{q_{L}} \sigma^a \gamma_{\mu} q_{L}
 V_{\qn{1},\qn{3},0}^{\mu,a} + {\rm h.c.} \,.
\esp \label{eq:vector_prod1}
\end{equation}
\be\bsp
 \mathcal{L}_{3 \oplus \bar{6}}= &\ \kappa_{2}\overline{q_{L}^{c}} i \sigma_2 \gamma_{\mu}d_{R}  V_{\qn{3} \oplus \bar{\qn{6}},\qn{2},1/6}^{\mu}
 +\kappa_{2}^{\prime}\overline{q_{L}^{c}} i \sigma_2 \gamma_{\mu}u_{R} V_{\qn{3} \oplus \bar{\qn{6}},\qn{2},-5/6}^{\mu}+{\rm h.c}.\,,
 \label{eq:vector_prod2}
\esp\ee
\be
\mathcal{L}_{1\oplus8}=
 \lambda_{u}\overline{q_{L}} T^A i \sigma_2 \, u_{R} S_{\qn{8},\qn{2},1/2}^{A *} + \lambda_{u}^\prime \overline{q_{L}}  i \sigma_2 \, u_{R} S_{\qn{1},\qn{2},1/2}^*
 + \lambda_{d}\overline{q_{L}} T^A  d_{R} S_{\qn{8},\qn{2},1/2}^A +
\lambda_{d}^\prime \overline{q_{L}}   d_{R} S_{\qn{1},\qn{2},1/2}
\label{eq:scalar_prod1}
\ee
\be\bsp \mathcal{L}_{3 \oplus \bar{6}}= &\  \left(\lambda_{qL} \overline{q_{L}^{c}}i\sigma_{2}q_{L}+\lambda_{R}^{1/3}\overline{u_{R}^{c}}d_{R}\right)S_{\qn{3} \oplus \bar{\qn{6}},\qn{1},-1/3}+ \lambda_{R}^{2/3}\overline{d_{R}^{c}}d_{R}S_{\qn{3} \oplus \bar{\qn{6}},\qn{1},2/3} \\
& \    +  \,\, \lambda_{R}^{4/3}\overline{u_{R}^{c}}u_{R} S_{\qn{3}
\oplus \bar{\qn{6}},\qn{1},-4/3} +
  \lambda_{3L}\overline{q_{L}^{c}}i\sigma_{2} \, \sigma^a \, q_{L}\,  S_{\qn{3} \oplus \bar{\qn{6}},\qn{3},-1/3}^a + {\rm h.c.}
\esp \label{eq:scalar_prod2}
\ee

 \section{HEAVY RESONANCE DECAY}

Having classified the possible resonances that can be  singly produced on-shell
at the LHC, we identify in this Section the quantum numbers of the VLQs,
{\emph i.e.} the fermions lying in the fundamental representation ($\bf 3$) of
$SU(3)_c$, into which the heavy resonance can decay in association with with a SM quark, thus assuming this process to be kinematically allowed. We discuss separately the cases of vector and scalar resonances, and categorize the VLQs according to the representation of $SU(2)_L$ in which the corresponding field lies. By matching the VLQ hypercharge in order to have gauge invariant interactions we can identify the electric charge of the various components of the VLQ multiplets.

\subsection{Vectors}
\label{sec:vec-decay}

If the new vectors lie in the singlet or octet representation of $SU(3)_c$ then gauge invariant and renormalizable interactions with a VLQ and a SM quark can be written only for VLQ with weak isospin up to 3/2, while the maximum allowed weak isospin is 1 in the case where 
the new vectors lie in the triplet or anti-sextet representation of $SU(3)_c$.

 \subsubsection{Case of $1\oplus8$}
\begin{table}
\centering
 \begin{tabular}{|c|c|c||c||c|c|c||c|}
 \hline
  \multicolumn{4}{|c||}{$\psi_{\rm VLQ}=(\qn{3},\qn{1},Y)$} &   \multicolumn{4}{|c|}{$\psi_{\rm VLQ}=(\qn{3},\qn{2},Y)$} \\
  \hline
 $\psi_{\rm SM}$ &  vector                                  & $\psi_{\rm VLQ}$  & $Q_{\rm VLQ}$  & $\psi_{\rm SM}$ &    vector                                  & $\psi_{\rm VLQ}$  & $Q_{\rm VLQ}$      \\
 \hline
 $u_R$               &  (\qn{1} $\oplus$ \qn{8},\qn{1},0)                     & (\qn{3},\qn{1},2/3)             & 2/3                        & $q_L$                 &  (\qn{1}$\oplus$ \qn{8},\qn{1},0)                     & (\qn{3},\qn{2},1/6)                       &  (2/3,-1/3)                                         \\ \hline
                             &   (\qn{1}$\oplus$ \qn{8},\qn{1}, $\pm$1)& (\qn{3},\qn{1},-1/3)                & -1/3    &  &  (\qn{1}$\oplus$ \qn{8},\qn{1},$\pm$1)               & (\qn{3},\qn{2},7/6)                       &  (5/3,2/3)                                                               \\
                             &                                              & (\qn{3},\qn{1},5/3)             & 5/3     & &                                   & (\qn{3},\qn{2},-5/6)                          &  (-1/3,-4/3)                                            \\ \hline
 $d_R$               &  (\qn{1}$\oplus$ \qn{8},\qn{1},0)                    & (\qn{3},\qn{1},-1/3)                & -1/3     & &  (\qn{1}$\oplus$ \qn{8},\qn{3},0)                     & (\qn{3},\qn{2},1/6)                       &  (2/3,-1/3)                                                          \\ \hline
                             &   (\qn{1}$\oplus$ \qn{8},\qn{1}, $\pm$1)   & (\qn{3},\qn{1},2/3)             & 2/3                        \\
                             &                                              & (\qn{3},\qn{1},-4/3)                & -4/3                           \\
                              \cline{1-4}
 \end{tabular}

 \vspace{.5cm}

 \begin{tabular}{|c|c|c||c||c|c|c||c|}
 \hline
  \multicolumn{4}{|c||}{$\psi_{\rm VLQ}=(\qn{3},\qn{3},Y)$} &   \multicolumn{4}{|c|}{$\psi_{\rm VLQ}=(\qn{3},\qn{4},Y)$} \\
 \hline
 $\psi_{\rm SM}$ &  vector                                  & $\psi_{\rm VLQ}$                                        & $Q_{\rm VLQ}$      &  $\psi_{\rm SM}$ & vector                                  & $\psi_{\rm VLQ}$                                        & $Q_{\rm VLQ}$               \\ \hline \hline
 $u_R$               &  (\qn{1}$\oplus$ \qn{8},\qn{3},0)                                 & (\qn{3},\qn{3},2/3)                                                   & (5/3,2/3,-1/3)  &  $q_L$                 &  (\qn{1}$\oplus$ \qn{8},\qn{3},0)                                 & (\qn{3},\qn{3},1/6)                                                   & (5/3,2/3,-1/3,-4/3)                  \\ \hline
                                     $d_R$               &  (\qn{1}$\oplus$ \qn{8},\qn{3},0)                                 & (\qn{3},\qn{3},-1/3)                                                      & (2/3,-1/3,-4/3)                  \\
 \cline{1-4}
 \end{tabular}
 \caption{Quantum numbers of the VLQs into which a color singlet or octet vector resonance can decay  together with the indicated SM quark for the case of a VLQ lying in the singlet (upper left), double (upper right), triplet (lower left)
  and fourplet (lower right) representations of $SU(2)_L$.}
  \label{vector_18}
 \end{table}

The categorization of the allowed VLQ quantum numbers in cases where the vector
resonance lies in the singlet or octet representation of $SU(3)_c$ is given
in Tab.~\ref{vector_18}, assuming decays into either
the SM quark weak doublet $q_L$ or the weak singlets $u_R$ and $d_R$.
We observe that most of the VLQs lying in the singlet, doublet and triplet
representation of $SU(2)_L$ are  ``standard'' VLQ representations that generally arise in CHM, {\emph{i.e.}} representations for which it is possible to write gauge invariant, renormalizable Yukawa type interactions that mix the SM quarks and the VLQs once electroweak symmetry is broken. 
Through this mass mixing these VLQ can decay into a SM boson ($W$, $Z$ or $h$) and a SM quark, and these decay channels have been largely explored at the LHC in conventional VLQ searches (see {\emph{e.g.}} in Refs.~\cite{Aaboud:2017zfn, Aaboud:2017qpr,Sirunyan:2017usq,Sirunyan:2017pks}), albeit with the assumption that the branching fractions for these three final states sum to unity.
However, for the special assignments $(\qn{3},\qn{1},5/3)$ and $(\qn{3},\qn{1},-4/3)$  these types of interactions are not possible. The same is true for the $SU(2)_L$ quadruplet, for which a dimension-4 Yukawa type interaction with the SM Higgs is forbidden, (see {\emph e.g} in Ref.~\cite{delAguila:2000rc}). Consequently these VLQs will decay back into a SM quark and the new resonance through which they were produced which will, however, be off-shell and will itself subsequently decay into a pair of SM quarks or gluons. This process gives thus rise to a $qqqq$ or $ggqq$ final state, where $q$ could be a light quark, a $b$ quark or a top quark.
While these ``backward decays'' are possible also in the case where Yukawa type interactions are allowed, they will generically be suppressed with respect to the  $VLQ\to SM\;SM$ decay pattern, being the former a three- instead than a two-body decay proceeding through an off-shell state. We thus expect that these decay patterns do not affect the reach of conventional VLQ experimental searches, as long as the couplings to the SM states are not strongly suppressed compared to the ones to the new vector.


 \subsubsection{Case of $3\oplus \bar 6$}

In the case of vectors lying in the triplet or the anti-sextet representation of $SU(3)_c$, the allowed quantum numbers for the VLQ are given in Tab.~\ref{vector_36bis}.  Only standard quantum numbers for the VLQs, {\emph{i.e.}} quantum numbers that allow for Yukawa type interactions with the SM fields are found in these cases.

\begin{table}
\centering
 \begin{tabular}{|c|c|c||c||c|c|c||c|}
 \hline
  \multicolumn{4}{|c||}{$\psi_{\rm VLQ}=(\qn{3},\qn{1},Y)$} &   \multicolumn{4}{|c|}{$\psi_{\rm VLQ}=(\qn{3},\qn{2},Y)$}\\
 \hline
 $\psi_{\rm SM}$ &  vector                    & $\psi_{\rm VLQ}$             & $Q_{\rm VLQ}$  &  $\psi_{\rm SM}$ & vector                                                          & $\psi_{\rm VLQ}$                                       & $Q_{\rm VLQ}$   \\ \hline \hline
 $q_L$               &  ($\qn{3}\oplus \bar{\qn{6}}$, \qn{2},-5/6)     &  (\qn{3},\qn{1},2/3)        & 2/3     & $u_R$               &  ($\qn{3}\oplus \bar{\qn{6}}$, \qn{2},-5/6)          & (\qn{3},\qn{2},1/6)                       &  (2/3,-1/3)                         \\ \hline
            &  ($\qn{3}\oplus \bar{\qn{6}}$, \qn{2}, 1/6)                                  & (\qn{3},\qn{1},-1/3)                                                         & -1/3      & &  ($\qn{3}\oplus \bar{\qn{6}}$ \qn{2}, 1/6)          &  (\qn{3},\qn{2},-5/6)                           &  (-1/3,-4/3)          \\
                         \cline{1-8}
\multicolumn{4}{c||}{} & $d_R$               &  ($\qn{3}\oplus \bar{\qn{6}}$, \qn{2},- 5/6)          & (\qn{3},\qn{2},7/6)                       &  (5/3,2/3) \\
\multicolumn{4}{c||}{} &                         &  ($\qn{3}\oplus
\bar{ \qn{6}}$, \qn{2}, 1/6)          & (\qn{3},\qn{2},1/6) &
(2/3,-1/3)
\\  \cline{5-8}
 \end{tabular}

\vspace{0.5cm}

\begin{tabular}{|c|c|c||c|}
 \hline
  \multicolumn{4}{|c|}{$\psi_{\rm VLQ}=({\bf 3}, {\bf 3},Y)$} \\
 \hline
 $\psi_{\rm SM}$ &  vector                             & $\psi_{\rm VLQ}$        & $Q_{\rm VLQ}$     \\ \hline \hline
 $q_L$               &  $(\qn{3}\oplus \bar{\qn{6}}$, \qn{2},-5/6)     & (\qn{3},\qn{3},2/3)                      & (5/3,2/3,-1/3)        \\ \hline
 &  ($\qn{3}\oplus \bar{\qn{6}}$, \qn{2},1/6)                           & (\qn{3},\qn{3},-1/3)       & (2/3,-1/3,-4/3)                             \\ \hline
 \end{tabular}
 \caption{ Quantum numbers of the VLQs into which a color triplet or anti-sextet vector resonance can decay itogether with the indicated SM quark for the case of a VLQ lying in the singlet (upper left), doublet (upper right) or triplet
 (lower) representations of $SU(2)_L$.}
 \label{vector_36bis}
 \end{table}

\subsection{Scalars}

In this Section we perform the same classification as in  Sec.~\ref{sec:vec-decay} above for the case of the scalar resonances reported in Tab.~\ref{VSBrep}.
Again, the new scalar is still assumed to decay into a SM-fermion plus a new vector-like quark as before although with a different chirality structure.

 \subsubsection{Case of $1\oplus8$}

The quantum numbers and charges of the possible vector-like quarks $\psi_{VLQ}$
in cases where the scalar resonance lies in the trivial or adjoint representation of $SU(3)_c$ can be found in Tab.~\ref{scalar_18}. Also in this case we see that only standard quantum numbers for the VLQs, {\emph{i.e.}} are found. 

\begin{table}
\centering
 \begin{tabular}{|c|c|c||c||c|c|c||c|}
 \hline
  \multicolumn{4}{|c||}{$\psi_{\rm VLQ}=(\qn{3},\qn{1},Y)$} &   \multicolumn{4}{|c|}{$\psi_{\rm VLQ}=(\qn{3},\qn{2},Y)$} \\
 \hline
 $\psi_{\rm SM}$ &  scalar                                  & $\psi_{\rm VLQ}$                                        & $Q_{\rm VLQ}$      &  $\psi_{\rm SM}$ & scalar                                  & $\psi_{\rm VLQ}$                                        & $Q_{\rm VLQ}$               \\ \hline \hline
 $q_L$               &  (\qn{1}$\oplus$ \qn{8},\qn{2},1/2)                                    & (\qn{3},\qn{1},2/3)                                                   & 2/3 &  $t_R$                 &  (\qn{1}$\oplus$ \qn{8},2,1/2)                                        & (\qn{3},\qn{2},1/6)                                               & (2/3,-1/3)                   \\
                                         &                          & (\qn{3},\qn{1},-1/3)                                                      & -1/3      & &       & (\qn{3},\qn{2},7/6) & (5/3,4/3) \\ \hline
                                     \multicolumn{4}{c||}{}  & $b_R$ & (\qn{1}$\oplus$ \qn{8},\qn{2},1/2) & (\qn{3},\qn{2},1/6) & (2/3,-1/3) \\
                                      \multicolumn{4}{c||}{}  & & & (\qn{3},\qn{2},-5/6) & ( -2/3,-4/3) \\ \cline{5-8}

 \end{tabular}

\vspace{0.5cm}

 \begin{tabular}{|c|c|c||c|}
 \hline
  \multicolumn{4}{|c|}{$\psi_{\rm VLQ}=(\qn{3},\qn{3},Y)$} \\
 \hline
 $\psi_{\rm SM}$ &  scalar                                                          & $\psi_{\rm VLQ}$                                        & $Q_{\rm VLQ}$     \\ \hline \hline
 $q_L$               &  (\qn{1}$\oplus$ \qn{8},\qn{2},1/2)                                    &(\qn{3},\qn{3},2/3)                                                    &(5/3, 2/3, -1/3)                         \\  & &     (\qn{3},\qn{3},-1/3)                                  &(2/3,-1/3,-4/3)     \\ \hline
 \end{tabular}
 \caption{Quantum numbers of the VLQs into which a color singlet or octet scalar resonance can decay together with the indicated SM quark for the case of a VLQ lying in the singlet (upper left), doublet (upper right) and triplet (lower)
representations of $SU(2)_L$.
 }
 \label{scalar_18}
 \end{table}

 \subsubsection{Case of $3\oplus \bar{6}$}

The same classification can be made for a scalar resonance lying in the triplet
or sextet ($3\oplus \bar{6}$) representation of the strong gauge group. The
results are reported in Tab.~\ref{scalar_36}. As in the vector $1\oplus 8$ cases above we now observe the appearance of non-standard VLQ quantum numbers.

 \begin{table}
\centering
 \begin{tabular}{|c|c|c||c||c|c|c||c|}
 \hline
  \multicolumn{4}{|c||}{$\psi_{\rm VLQ}=(\qn{3},\qn{1},Y)$} &   \multicolumn{4}{|c|}{$\psi_{\rm VLQ}=(\qn{3},\qn{2},Y)$} \\
 \hline
 $\psi_{\rm SM}$ &  scalar                                  & $\psi_{\rm VLQ}$                                        & $Q_{\rm VLQ}$      &  $\psi_{\rm SM}$ & scalar                                  & $\psi_{\rm VLQ}$                                        & $Q_{\rm VLQ}$               \\ \hline \hline
 $u_R$               &  ($\qn{3} \oplus \bar{\qn{6}}$,\qn{1},-4/3)         & (\qn{3},\qn{1},2/3)                                                    & 2/3  &  $q_L$                 &  ($\qn{3} \oplus \bar{ \qn{6}}$,\qn{1},-4/3)                                       & (\qn{3},\qn{2}, 7/6)                                                  & (2/3,5/3)                  \\     \hline &
 ($\qn{3}\oplus \bar{\qn{6}}$,\qn{1},2/3)                    & (\qn{3},\qn{1},-4/3)                                                      & -4/3      & &    ($\qn{3} \oplus \bar{\qn{6}}$,\qn{1},2/3)  & (\qn{3},\qn{2},-5/6) & (-1/3,-4/3) \\ \hline
  & ($\qn{3} \oplus \bar{\qn{6}}$,\qn{1},-1/3) & (\qn{3},\qn{1},-1/3) & -1/3 &  & ($\qn{3} \oplus \bar{\qn{6}}$,\qn{1},-1/3) & (\qn{3},\qn{2},1/6) & (2/3,-1/3) \\ \hline
  $d_R$ & ($\qn{3} \oplus \bar{\qn{6}}$,\qn{1},-4/3)   & (\qn{3},\qn{1},5/3) & 5/3 &   \multicolumn{4}{c}{}  \\ \cline{1-4}
   & ($\qn{3} \oplus \bar{\qn{6}}$,\qn{1},2/3) & (\qn{3},\qn{1},-1/3) & -1/3 & \multicolumn{4}{c}{}  \\ \cline{1-4}
   & ($\qn{3} \oplus \bar{\qn{6}}$,\qn{1},-1/3)  & (\qn{3},\qn{1},2/3) & 2/3 &\multicolumn{4}{c}{}  \\ \cline{1-4}
 \end{tabular}

\vspace{0.5cm}

 \begin{tabular}{|c|c|c||c||c|c|c|c|}
 \hline
   \multicolumn{4}{|c||}{$\psi_{\rm VLQ}=(\qn{3},\qn{3},Y)$} &   \multicolumn{4}{|c|}{$\psi_{\rm VLQ}=(\qn{3},\qn{4},Y)$}\\
 \hline
  $\psi_{\rm SM}$ & scalar   & $\psi_{\rm VLQ}$   & $Q_{\rm VLQ}$  &    $\psi_{\rm SM}$ &  scalar                     & $\psi_{\rm VLQ}$                                        & $Q_{\rm VLQ}$             \\
   \hline \hline
$u_R$& ($\qn{3} \oplus \bar{\qn{6}}$,\qn{3},-1/3) &
(\qn{3},\qn{3},-1/3)& (2/3,-1/3,-4/3) &  $q_L$   &  ($\qn{3}
\oplus
\bar{\qn{6}}$,\qn{3},-1/3)    & (\qn{3},\qn{4},1/6) &(5/3,2/3,-1/3,-4/3) \\
\hline $d_R$ & ($\qn{3} \oplus \bar{\qn{6}}$,\qn{3},-1/3) &
(\qn{3},\qn{3},2/3)& (5/3,2/3,-1/3)\\ \cline{1-4}
 \end{tabular}
 \caption{Quantum numbers of the VLQs into which a color triplet or anti-sextet scalar resonance can decay together with the indicated SM quark for the case of a VLQ lying in the singlet (upper left), doublet (upper right), triplet (lower
left) and fourplet (lower right) representations of $SU(2)_L$.}
 \label{scalar_36}
 \end{table}

\section{PHENOMENOLOGY}

Having identified all the possible quantum numbers of VLQs arising from the decay of an on-shell vector or scalar resonance singly produced at the LHC and decaying into a SM quark and a VLQ, we now give an overview of the phenomenology which is expected from these production and decay
patterns\footnote{Some of the scalars and vectors could be leptoquarks, e.g.\ couple to
leptons and quarks. We assume here
that the corresponding couplings are zero as these are heavily constraint by the non-observation 
of proton decay \cite{Dorsner:2016wpm}.}.

\begin{itemize}
\item {\bf{Prompt decay into a VLQ and a SM quark:}} If the new VLQs have the same color and electric charge quantum numbers as do the SM fermions we can write down a mixing term generated by a coupling to the SM Higgs boson. This would hence lead to decays of the new vector-like quarks into either a Higgs boson and a
SM fermion, a $Z$-boson and a SM fermion or a $W$-boson and a SM fermion, see e.g.~\cite{Aguilar-Saavedra:2013qpa}.
 Searches for such modes have been recently performed by the CMS collaboration~\cite{Sirunyan:2017bfa}.
\item {\bf{Displaced vertices:}} Conversely, if the new fermion has exotic quantum numbers, such mixing terms are not allowed. In these cases the VLQ will decay back into an off-shell heavy boson which will then decay back to the SM. Depending on the lifetime of the VLQ, both prompt multijet final states, with the potential presence of top quarks, or signatures exhibiting a displaced vertex will be possible.
\item {\bf{R hadron:}} For long enough lifetimes the VLQ will hadronize before decaying, allowing in this way for the formation of new exotic and heavy bounds states~\cite{Buchkremer:2012dn, DiLuzio:2015oha}.
\item {\bf{Associated production of new scalars:}}
Producing the new scalars in $q\bar q$ annihilation together with quarks of the first and/or second generation could be strongly constrained by flavour observables. However, one can produce the scalar in association with a $b\bar b$ or a
$t \bar t$ pair. The allowed quantum numbers for the scalar remain the same of Tab.~\ref{VSBrep}.
\item {\bf{Reversed mass hierarchy:}}
While throughout our discussion we have assumed a mass hierarchy such as the new
boson can decay into a VLQ and a SM quark, the opposite hierarchy can also give
rise to an interesting phenomenology. In that case the VLQ will undergo a decay into a heavy boson and a SM quark. This configuration has recently received some attention, also due to the possibility that bounds on the mass of the VLQs could be relaxed~\cite{Serra:2015xfa,Anandakrishnan:2015yfa,Brooijmans:2016vro,Banerjee:2016wls}.\footnote{Note that exclusion bounds for VLQs with decays into new particles can also be strengthened, for instance in the case where they decay into a stable scalar, such that stop searches can be reinterpreted \cite{Chala:2017xgc}. This is typically the case in non-minimal Composite Higgs Models \cite{Chala:2018qdf}.}
\end{itemize}

\section{CONCLUSIONS}
New bosons of spin-0 and/or spin-1 are common in many extensions of the Standard Model, in particular in composite Higgs models or extra-dimensional models. We have classified all spin-0 and spin-1 states that can be produced at the LHC through initial state quarks by an $s$-channel exchange. Several existing LHC searches set bounds on such resonances when they decay into SM states. Less explored is the possibility that the $s$-channel resonance decays into non-SM states, as predicted in many new physics models. In such a case dedicated searches should be performed. We have concentrated here on the case where the new resonance decays into a VLQ and a SM fermion and identified all the possible quantum numbers of the VLQ; our list contains VLQs with charges of $5/3$, $2/3$, $-1/3$ and $-4/3$. We have also commented on the phenomenology of these new states.
While for some specific quantum numbers the VLQs can mix with the SM quarks through a Yukawa type interaction and hence decay into SM states, for some of the cases we have observed that this was not possible. The VLQs can then decay only via ``backwards'' decay, meaning via the (off-shell) 
heavy boson through which they were produced. Depending on the coupling strength between these states, the VLQ might be long-lived giving rise to a peculiar phenomenology which deserves a deeper investigation.

\section*{ACKNOWLEDGEMENTS}
The authors would like to thank the organizers of the Les Houches workshop where this work
was initiated.
RG is supported by a European Union COFUND/Durham Junior Research Fellowship under the EU grant number 609412. TF is supported by IBS under the
project code IBS-R018-D1. TR was supported by the Department of Energy, Contract DE-AC02-76SF00515, and BF has been partly supported by French state funds managed by the Agence Nationale de la Recherche (ANR), in the context of the LABEX ILP (ANR-11-IDEX-0004-02, ANR-10-LABX-63).
WP is supported by the DFG, project nr.\ PO 1337-7/1.



\AddToContent{D.~Barducci, H.~Cai, T.~Flacke, B.~Fuks, R.~Gr\"ober, W.~Porod, T.~Rizzo}
\renewcommand{\thesection}{\arabic{section}}

\graphicspath{{VLQ_NLO/}}


\chapter{Precision predictions for the single production of third generation vector-like quarks}

{\it G.~Cacciapaglia, A.~Carvalho, A.~Deandrea, T.~Flacke, B.~Fuks, D.~Majumder, L.~Panizzi}



\begin{abstract}
We study the effects of next-to-leading-order corrections in QCD on the single
production of third generation vector-like quarks, assuming standard couplings of the extra
quarks to the weak gauge and Higgs bosons so that they could decay into one of
these bosons and a Standard Model quark.
\end{abstract}

\section{INTRODUCTION}

Vector-like quarks (VLQs), {\it i.e.} coloured heavy fermions that have non-chiral couplings to the 
Standard Model (SM) gauge interactions, are a common ingredient of many models
of new physics.
In particular, when they couple to the third generation of SM quarks, they
often play a role in the fine-tuning problematics of the
Higgs-boson mass. In addition, they also appear in models with extra space dimensions. 
These reasons, together with the fact that they can be copiously
produced at hadron colliders, make them an ideal target to be searched for at the LHC and at 
future hadron colliders. 

Many searches targeting VLQs coupled to third generation quarks are performed by
both the ATLAS and CMS collaborations~\cite{Sirunyan:2017ynj,Sirunyan:2018fjh,Sirunyan:2016ipo,%
Sirunyan:2017tfc,Biedermann:2016jph}, although specific searches complementarily
focus on VLQs coupling to light quarks~\cite{Aad:2015tba,Sirunyan:2017lzl}. In
the third generation case, the considered final state contains one third generation quark (top or bottom) and one SM weak or Higgs boson ($W$-boson, $Z$-boson
or a Higgs boson). The corresponding branching ratios in each channel depend on
the details of the model, and in particular on the dimension of the SU(2)$_L$
multiplet the VLQ belongs to and on the electroweak symmetry breaking pattern.
In a previous Les Houches workshop~\cite{Brooijmans:2014eja}, some of the authors of this contribution worked out a model-independent
strategy to study the most general decay pattern relying on a parameterisation
of the couplings in terms of the physical branching
ratios~\cite{Buchkremer:2013bha,Cacciapaglia:2010vn,Cacciapaglia:2011fx}.
However, non-standard decay modes may still be allowed in specific models, for
which a full classification can be found in Ref.~\cite{Brooijmans:2016vro}
and dedicated analyses in Refs~\cite{Serra:2015xfa,Anandakrishnan:2015yfa,Fan:2015sza,Dolan:2016eki,Dobrescu:2016pda,Aguilar-Saavedra:2017giu}.

In this project, we focus on the standard channels, but we aim at studying in detail effects
that arise from next-to-leading order (NLO) corrections in QCD. While such effects are well studied for the QCD production of a pair of VLQs,
which is analogous to top-antitop pair production, no such studies exist for VLQ
single production.
In Ref.~\cite{Fuks:2016ftf}, some of us published a {\sc FeynRules}~\cite{Alloul:2013bka} implementation of a general VLQ model that
includes full NLO effects in QCD. This model was developed as part of a previous Les Houches
project~\cite{Brooijmans:2016vro}, and applied first to the study of di-Higgs
final states originating from the decays of
on-shell VLQs that couple to first generation quarks~\cite{Cacciapaglia:2017gzh}.
We now use this implementation to study the kinematic distributions of jets
produced in association with a single VLQ of third generation, the most
well-known examples of such a new physics state being top partners,
{\it i.e.} VLQs with the same SM quantum numbers as the top quark.

\section{THE MODEL AT NLO}

There are four types of VLQs that can decay directly into a SM quark plus a boson, and they are distinguished by 
their electromagnetic charge: two have the same charge $e_Q$ as the top and bottom
quarks respectively, and we call them $T$ ($e_T = 2/3$) and $B$ ($e_B = -1/3$), 
and two exhibit exotic charges that differ by one unit from the standard ones,
$X$ ($e_X=5/3$) and $Y$ ($e_Y = -4/3$).
The leading order Lagrangian that we have implemented reads~\cite{Fuks:2016ftf}
\begin{eqnarray}
& \mathcal{L}_{\rm LO} = i \bar{Q} \slashed{D} Q - m_Q \bar{Q} Q - h \left[ \bar{B} \left( \hat{\kappa}_L^B P_L + \hat{\kappa}^B_R P_R \right) B + \bar{T} \left( \hat{\kappa}_L^T P_L + \hat{\kappa}_R^T P_R \right) T + \mathrm{h.c.} \right] & \nonumber \\
& + \frac{g}{2 c_W} \left[ \bar{B} \slashed{Z} \left( \tilde{\kappa}_L^B P_L + \tilde{\kappa}^B_R P_R \right) b +  \bar{T} \slashed{Z} \left( \tilde{\kappa}_L^T P_L + \tilde{\kappa}^T_R P_R \right) t + \mathrm{h.c.} \right] & \nonumber \\
& + \frac{g}{\sqrt{2}} \left[ \bar{B} \slashed{W}^- \left( \kappa_L^B P_L + \kappa_R^B P_R \right) t + \bar{T} \slashed{W}^+ \left( \kappa_L^T P_L + \kappa_R^T P_R \right) b + \mathrm{h.c.} \right] & \nonumber \\
& + \frac{g}{\sqrt{2}} \left[ \bar{X} \slashed{W}^+ \left( \kappa_L^X P_L + \kappa_R^X P_R \right) t + \bar{Y} \slashed{W}^- \left( \kappa_L^Y P_L + \kappa_R^Y P_R \right) b + \mathrm{h.c.} \right]\,, &
\end{eqnarray}
where $Q = X, T, B, Y$. The covariant derivative only contains gauge
interactions from QCD and QED, the couplings of a pair of VLQs to $W$-bosons and
$Z$-bosons being omitted as they are very model dependent and give minor
contributions to the production cross sections~\cite{Buchkremer:2013bha}.
This model differs slightly from the parameterisation proposed in Ref.~\cite{Buchkremer:2013bha} in the mass dependence of the couplings that has been
removed. The reason behind this choice is to render the NLO implementation easier, as the couplings can be renormalised independently of the
masses. There is however a qualitative (and quantitative) difference between the
VLQ coupling to the Higgs-boson $h$ and that to the gauge bosons.
The former corresponds to a Yukawa coupling while the latter to a mixing angle, and the relation between the two is a factor $v/m_Q$ where $v$ is the
vacuum expectation value of the SM Higgs field~\cite{Cacciapaglia:2011fx}.

Our simulations for the LHC make use of an NLO model file encoded in the UFO
format~\cite{Degrande:2011ua} that has been generated with the
{\sc FeynRules}~\cite{Alloul:2013bka} and NLOCT~\cite{Degrande:2014vpa}
packages. The resulting UFO library contains tree-level vertices as well as
ingredients necessary for the evaluation of one-loop diagrams in {\sc MadGraph5}\_aMC@NLO~\cite{Alwall:2014hca}. For more details on the validation of our implementation we refer to Ref.~\cite{Fuks:2016ftf}.
We then use the {\sc MadGraph5}\_aMC@NLO platform for generating events both at
the LO and NLO accuracy in QCD. We have used the NNPDF2.3 (LO QCD + LO QED)~\cite{Ball:2014uwa} for the LO processes, while for the NLO processes we have used the NNPDF3.0 (NLO) set. The simulation of the QCD environment (parton
showering and hadronisation) has been achieved with
{\sc Pythia}~8.2~\cite{Sjostrand:2014zea}, while the jet reconstruction has been made by using using the anti-k$_T$ algorithm~\cite{Cacciari:2008gp} with radius 0.4 and b-jet tagging with distance $\Delta$R=0.5, implemented in {\sc FastJet}~3.2.1~\cite{Cacciari:2011ma}.

\subsection{Simulation results}
\begin{figure}
\centering
\begin{minipage}{.2\textwidth}
\includegraphics[width=\textwidth]{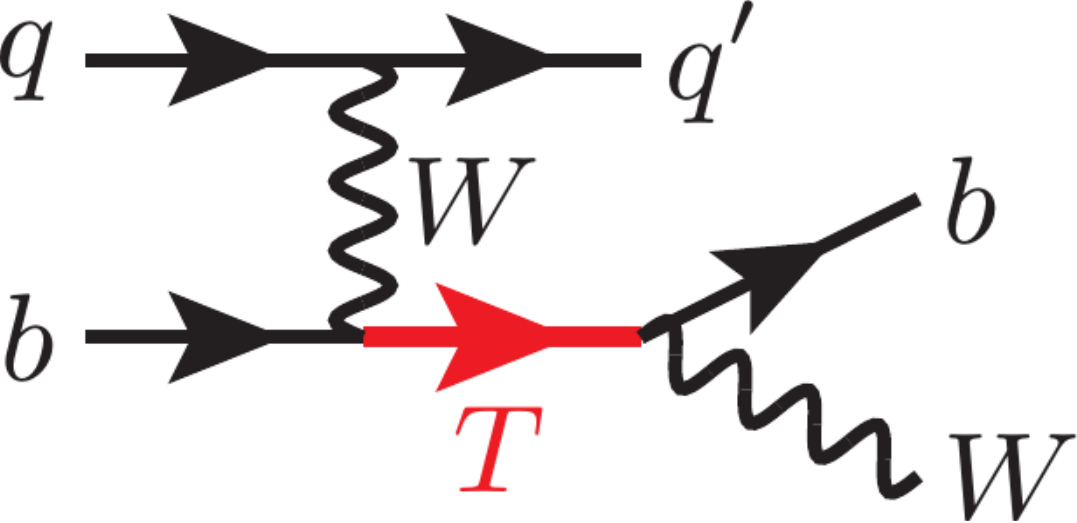}
\end{minipage}\hskip 30pt
\begin{minipage}{.25\textwidth}
\includegraphics[width=\textwidth]{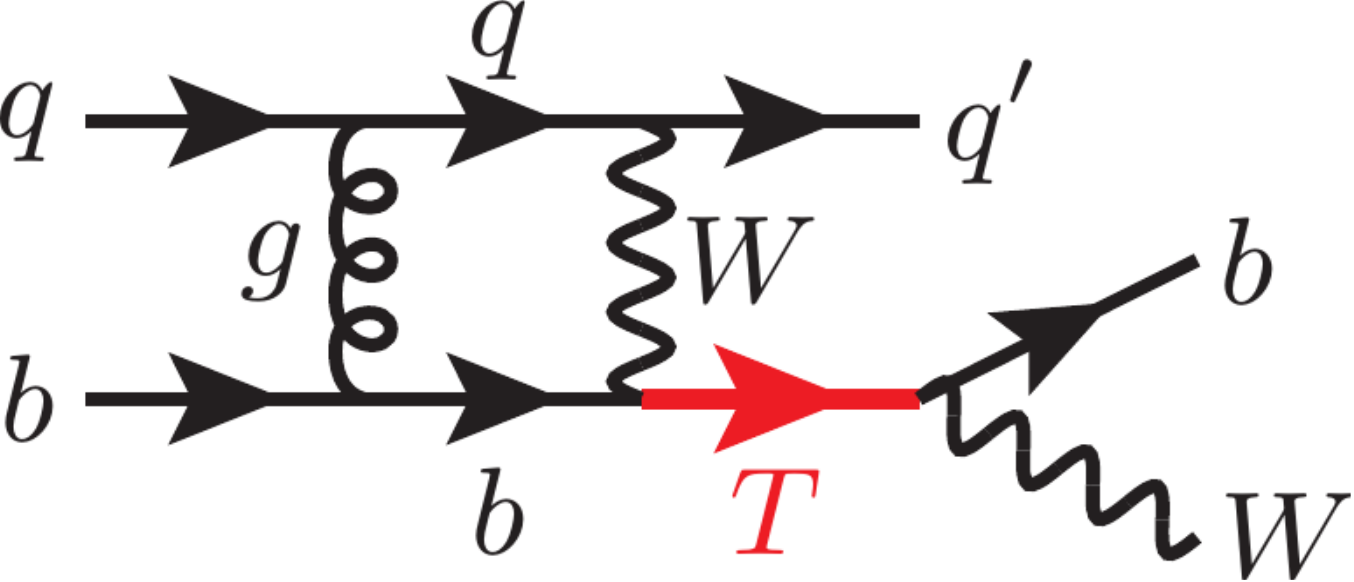}
\end{minipage}\hskip 30pt
\begin{minipage}{.25\textwidth}
\includegraphics[width=\textwidth]{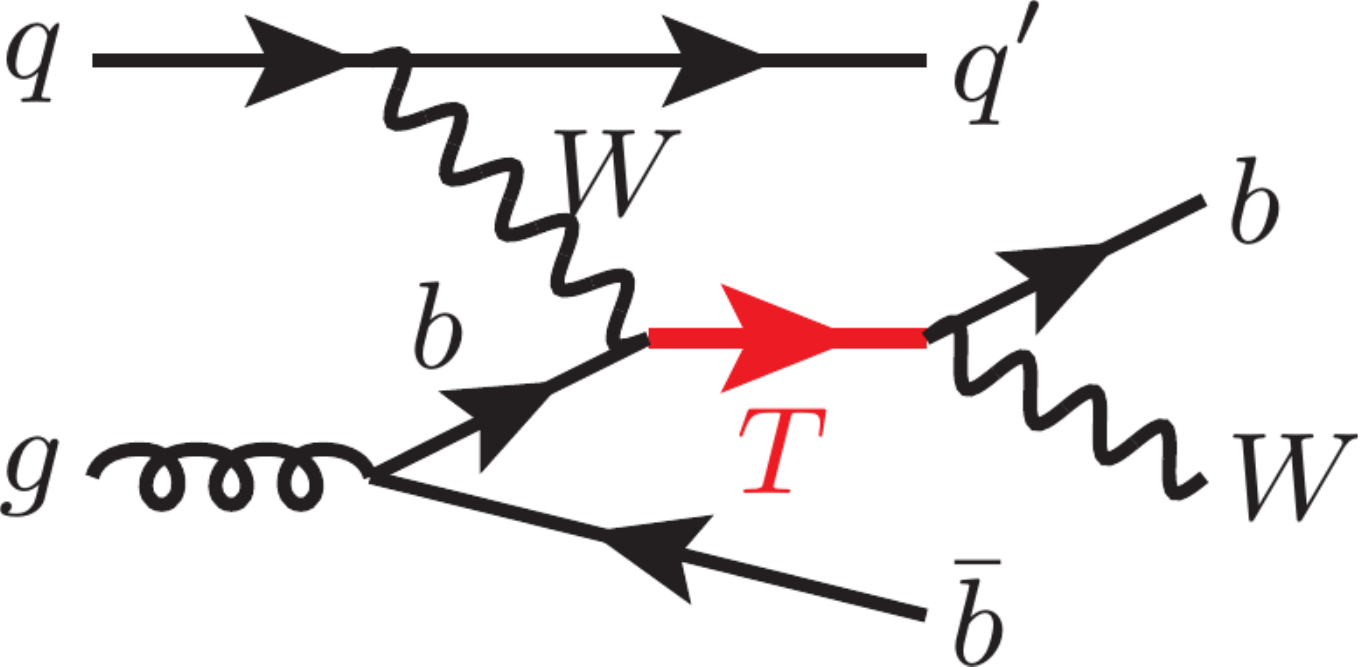}
\end{minipage}
\caption{\label{fig:topologies}Subset of topologies for single $T$ production at the LHC. Left panel: $Tj$ production at LO in the 5FNS; central panel: NLO QCD contribution to the $Tj$ process in the 5FNS; right panel: real emission for the $Tj$ process in the 5FNS and LO topology for the $Tjb$ process in the 4FNS.}
\end{figure}

\begin{figure}
\begin{center}
\includegraphics[width=0.45\textwidth]{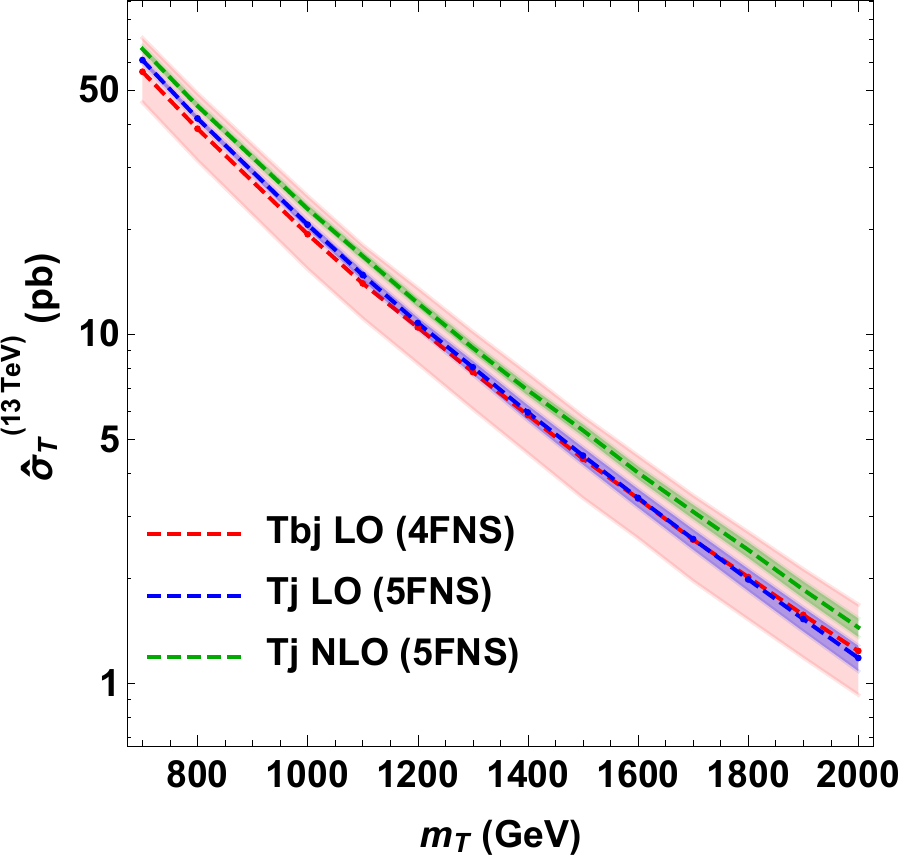}
\includegraphics[width=0.45\textwidth]{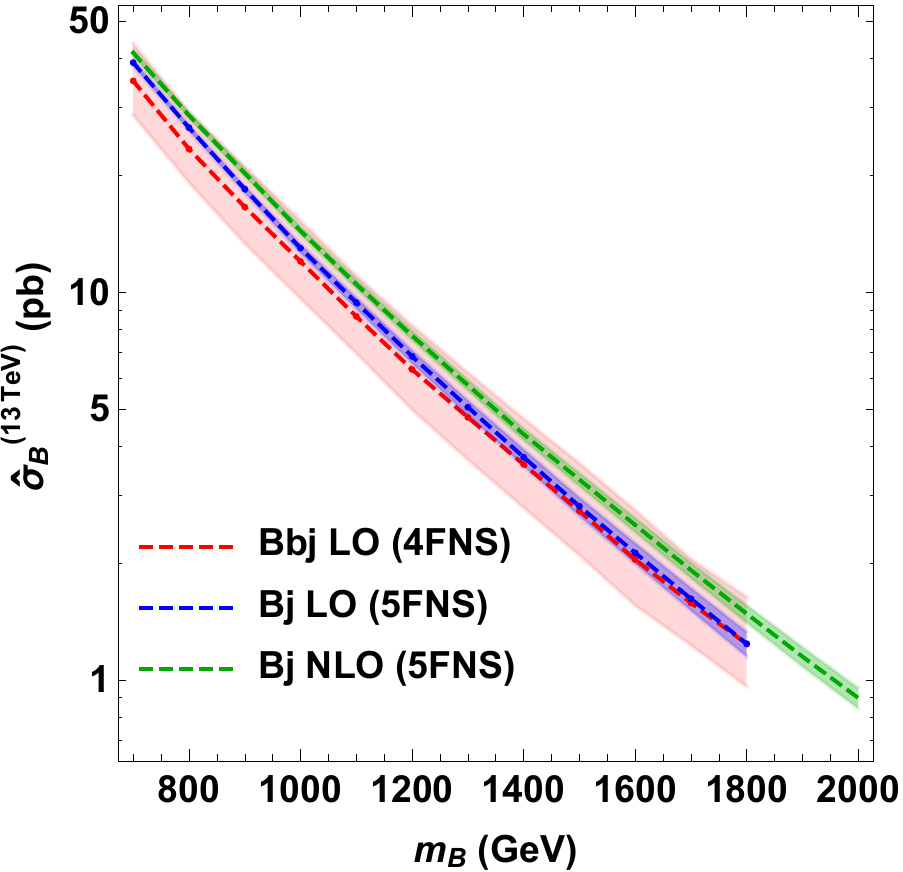}
 \caption{Production cross section of a single VLQ in association with jets for
  LHC collisions at a centre-of-mass energy of 13~TeV. We present results in the
  case of a $T$ (left) and $B$ (right) quark, and show their dependence on the
  VLQ mass. The cross sections are normalised to $\kappa^T=\tilde{\kappa}^B=1$.}
\label{fig:prodXnlo}
\end{center}
\end{figure}

One of the main impacts of the QCD corrections to the single production of a VLQ is
to modify the corresponding production cross-section. We focus in this work on
single VLQ production in association with a jet, as illustrated by the
representative Feynman diagrams of Fig.~\ref{fig:topologies} for single $T$
production. At tree-level, such a process occurs through VLQ
couplings to the $W$-boson ({\it i.e.} the $\kappa_{L/R}^T$ interactions), while
for single $B$ production it occurs through the VLQ coupling to the
$Z$-boson  ({\it i.e.} the $\tilde{\kappa}_{L/R}^B$ couplings).
Total rate results given as a function of the mass of the $T$ and $B$ quark, and
for LHC collisions at a centre-of-mass energy of 13~TeV, are shown in
Fig.~\ref{fig:prodXnlo}. We compare predictions in the 5-flavour-number
scheme (5FNS) with predictions in the 4-flavour-number scheme (4FNS). In the 5FNS,
bottom quark contributions to the parton densities of the proton are included,
while in 4FNS, initial bottom quarks originate from gluon splitting. The results
are normalised to $\kappa$ parameter values equals to 1, and include
contributions from both VLQ and anti-VLQ production. While calculations in the
5FNS are easier and hence allows to include higher-order corrections, 4FNS
results are known to better describe the shapes of the kinematic distributions.

The global effect of the NLO corrections is to increase the cross-section value
and to generally reduce the scale uncertainties, although new subprocesses may
appear at NLO and contribute significantly enough to spoil the reduction of the
uncertainties. Results in the 4FNS and 5FNS agree, after accounting for the
uncertainties. Significant NLO effects are however expected when considering
more exclusive observables like those related to the final-state jet properties.
For instance, the kinematics of the $b$-jet produced in association with the VLQ
is crucial. While such a jet already appears at tree-level in the 4FNS, as this
consists in a $2\to3$ process (see Fig.~\ref{fig:topologies}), NLO corrections
are required in the 5FNS as $b$-jets arise at the lowest order through
radiative contributions.
To ascertain which strategy better characterises the kinematic properties of the event, we compare below distributions obtained by the three calculations,
{\it i.e.} in the 4FNS (at LO, $2\to3$ process) and in the 5FNS (at LO and NLO).
NLO corrections to the 4FNS results are left to future work.

\begin{figure}
\begin{center}
\includegraphics[width=0.35\textwidth]{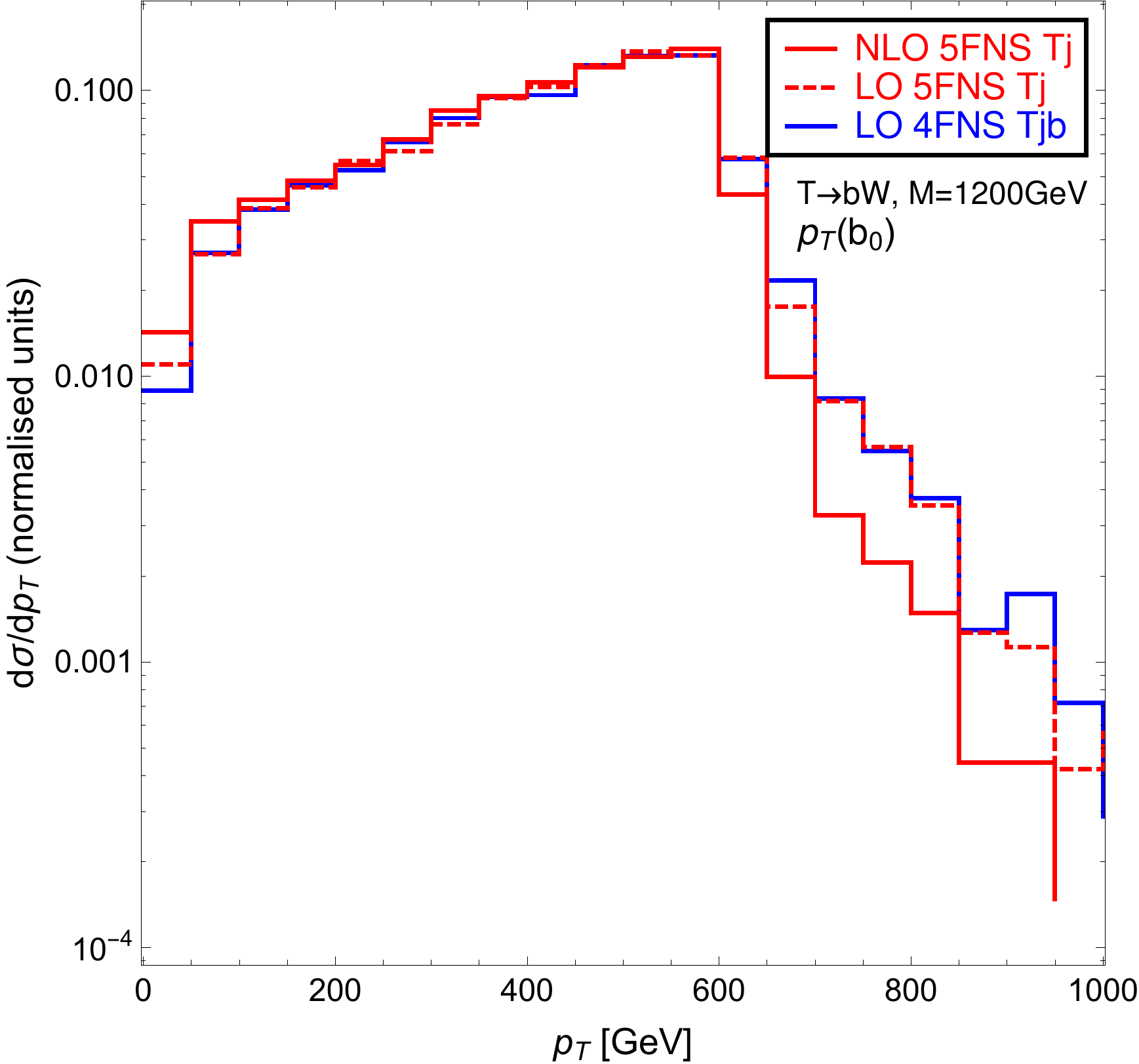} \hskip 20pt
\includegraphics[width=0.322\textwidth]{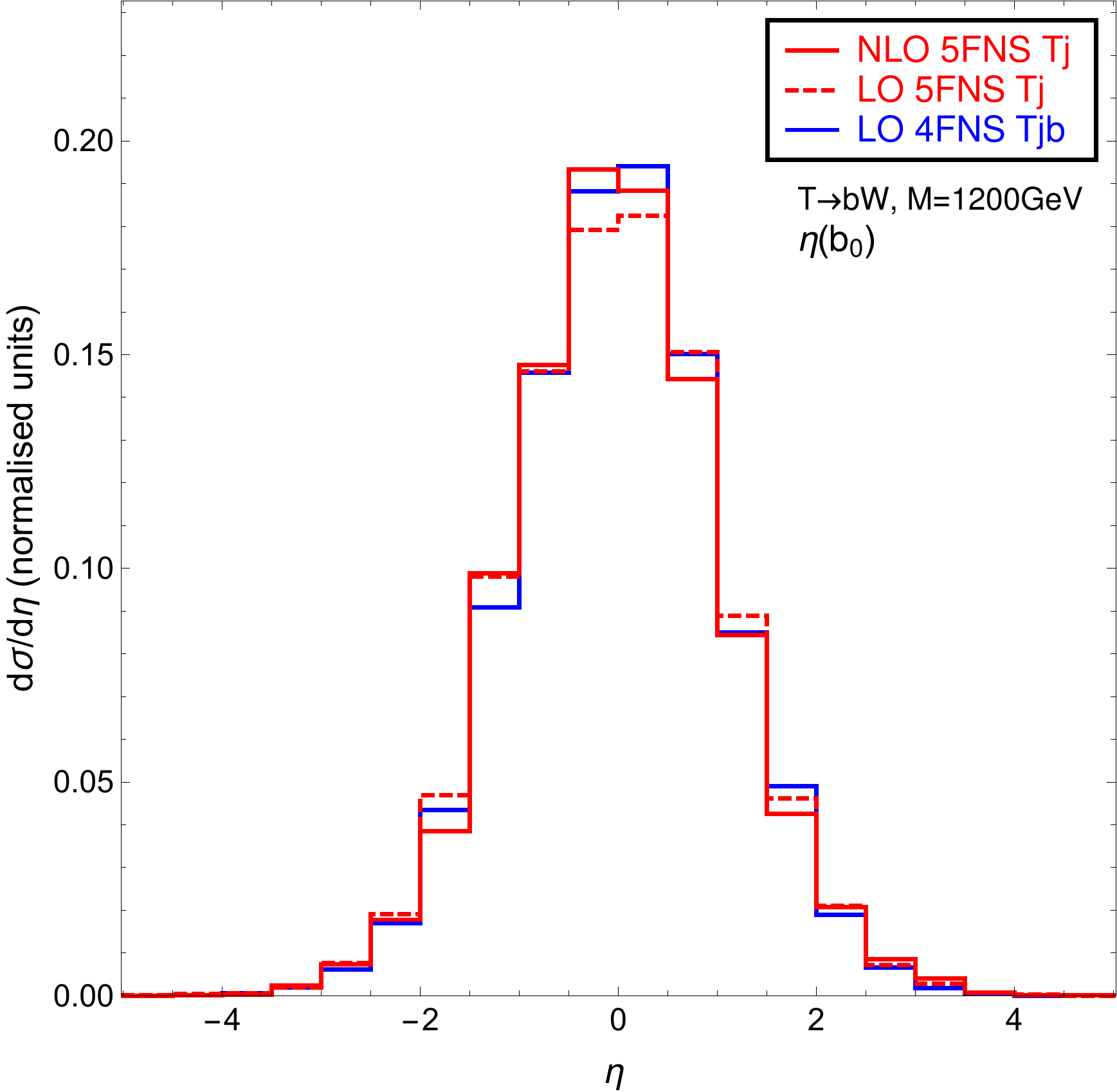}\\
\includegraphics[width=0.35\textwidth]{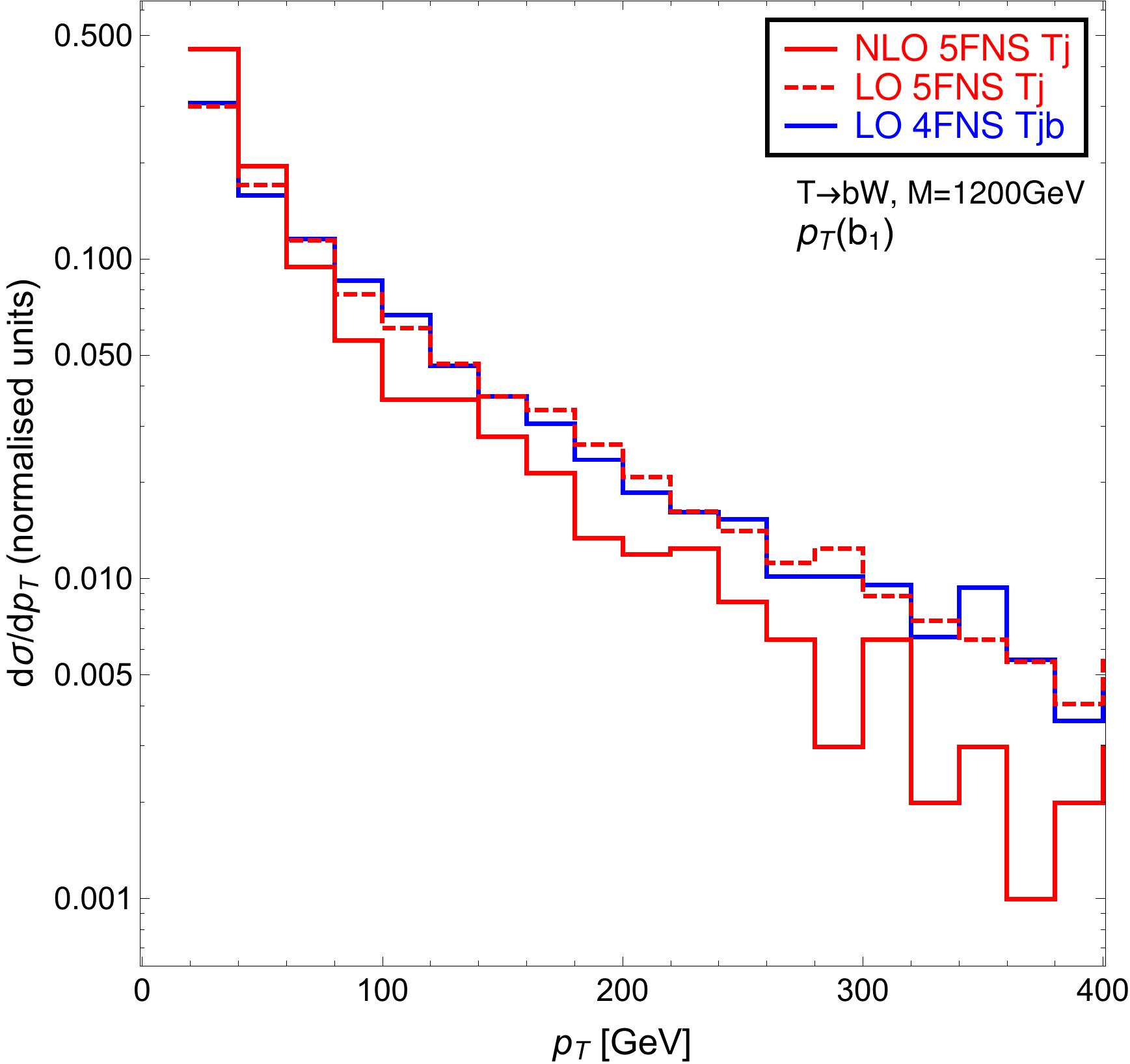} \hskip 20pt
\includegraphics[width=0.322\textwidth]{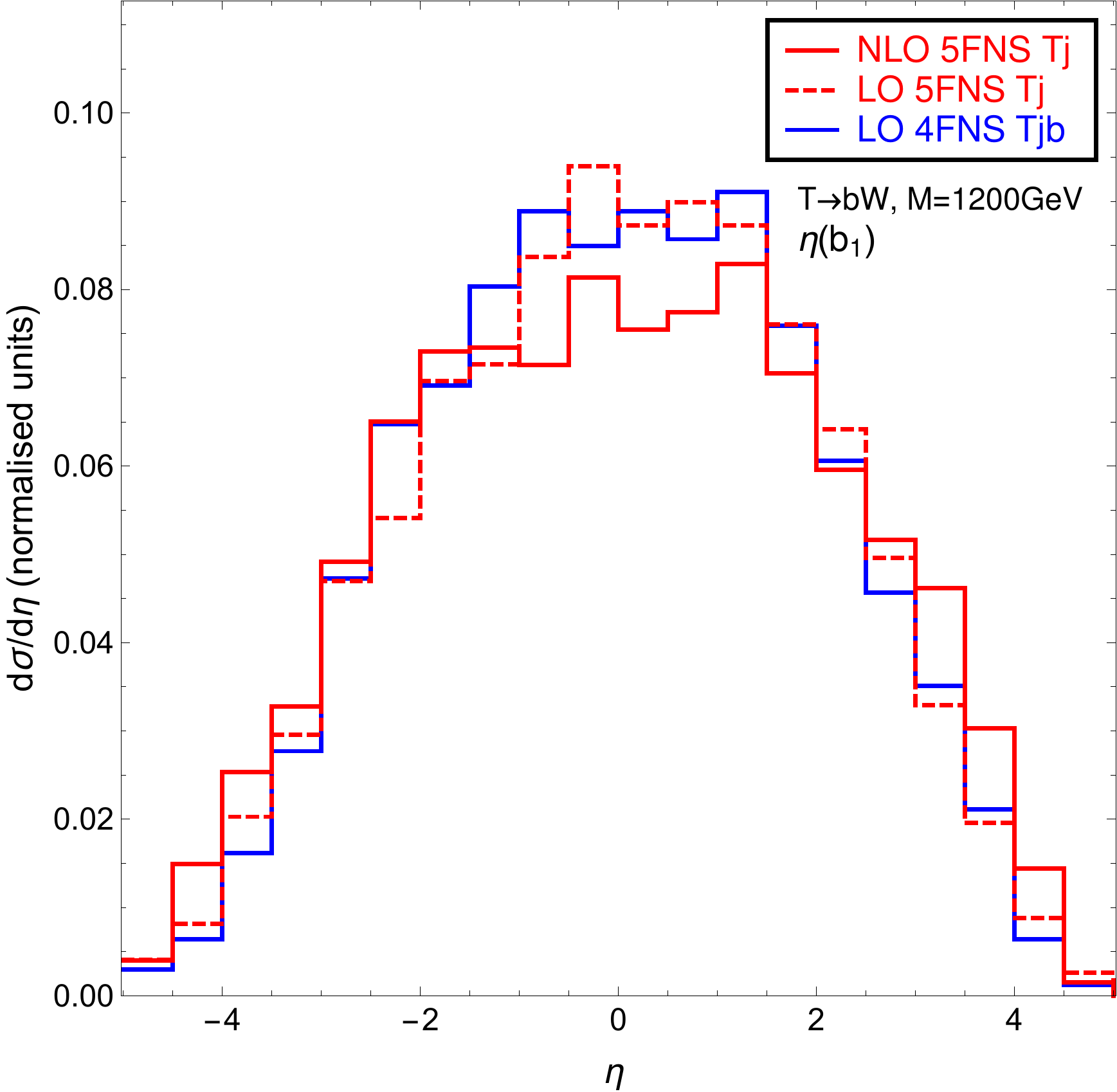}
\caption{\label{fig:distributions} Normalised distributions of the transverse
   momentum and pseudorapidity of the leading (top row) and sub-leading (bottom
   row) reconstructed $b$-jets for single $T$ production, {\it i.e.}
   $pp\to T j$ in the 5FNS at LO and NLO and $pp\to Tbj$ in the 4FNS at LO after
   including a subsequent $T\to bW$ decay for $M_T$=1200 GeV.}
\includegraphics[width=0.35\textwidth]{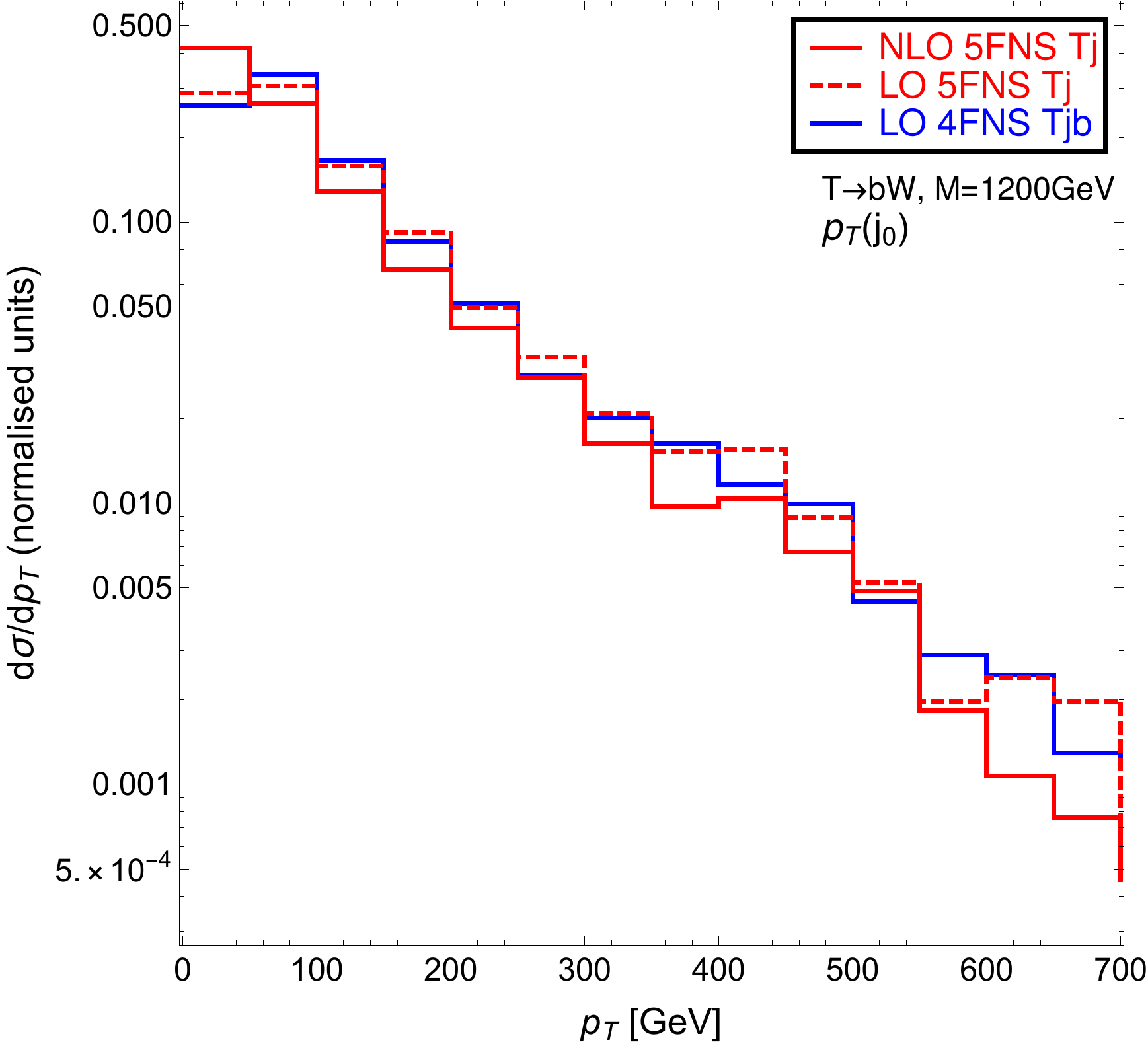} \hskip 20pt
\includegraphics[width=0.322\textwidth]{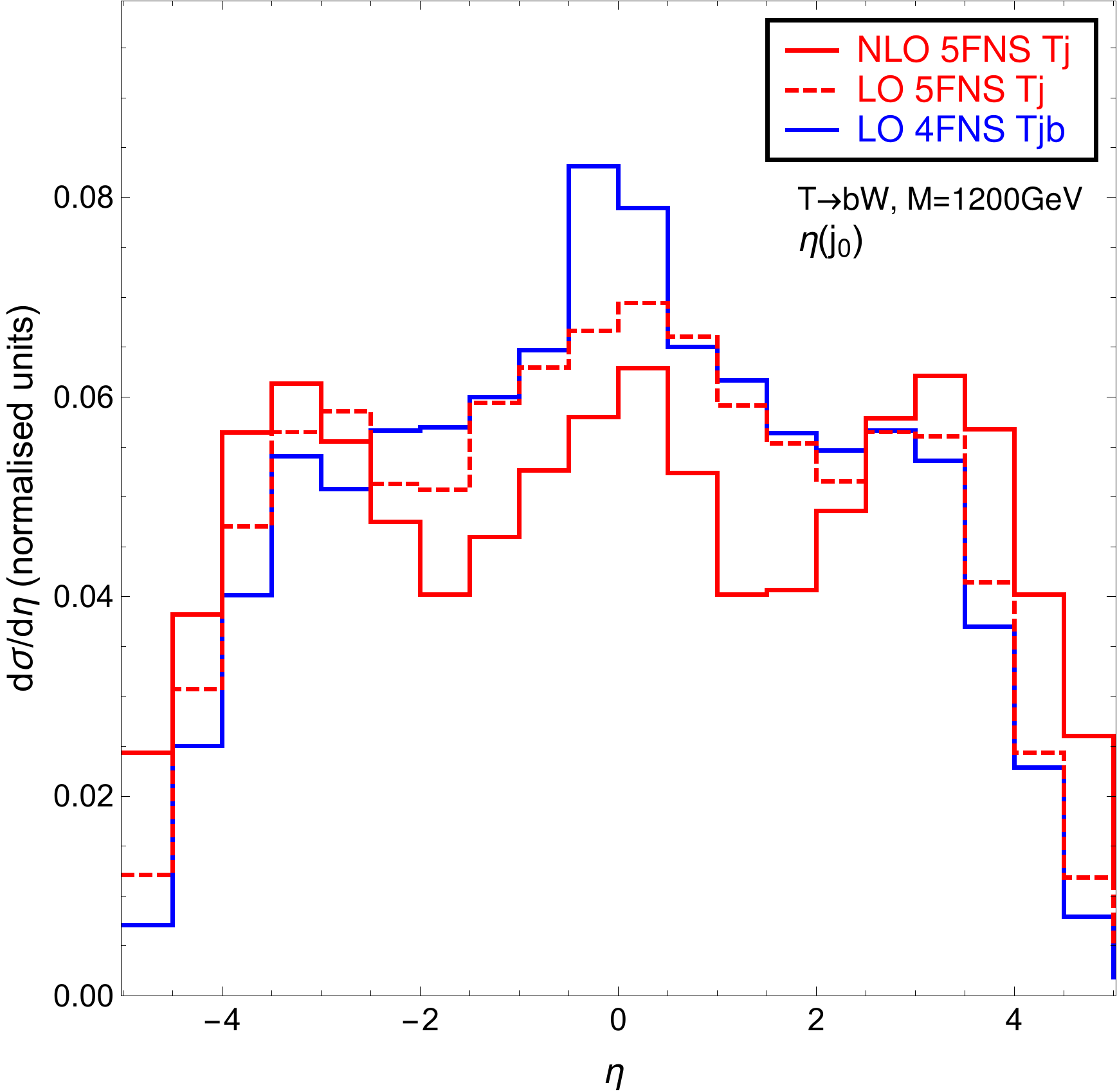}
\caption{\label{fig:distributions2} Normalised distributions of the transverse momentum and pseudorapidity of the leading jets (non-b) for the processes $PP\to T j$ (5FNS at LO and NLO) and $PP\to Tbj$ (4FNS at LO) with subsequent $T\to bW$ decay for $M_T$=1200 GeV. We select leptonic decays for the $W$.}
\end{center}
\end{figure}

To this end, in the remaining of this contribution we consider  single-$T$ quark
production where the extra quark decays with a 100\% branching fraction into a
$Wb$ system. In Fig.~\ref{fig:distributions}, we compare the (normalised)
distributions of the transverse momentum ($p_T$) and pseudorapidity ($\eta$) of
the leading (top row) and sub-leading (bottom row) reconstructed $b$-jets, for a
$T$ mass set to $M_T = 1200$~GeV, and after applying cuts on the b-jet $p_T>20$ GeV and
$|\eta|<5$.
The markedly central pseudorapidity distribution of the leading $b$-jet and the
steep fall of the $p_T$ distribution for $p_T$ values larger than
$M_T/2 = 600$~GeV show that this jet can clearly be associated with the bottom
quark originating from the $T$-quark decay, independently on the scheme in which
the calculation has been made. Furthermore, as the decay of the $T$-quark is
computed at the LO accuracy in all cases, the leading $b$-jet distributions do
not exhibit clear differences between the LO and NLO results, except for a
slight tendency towards softer $p_T$ values at NLO as shown by the reduction of
events with $p_T$ larger than $M_T/2 = 600$~GeV.
The sub-leading $b$-jet distributions shows as well a slightly
different behaviour. The soft $p_T$ spectrum is in agreement with the fact
that this jet originates from radiation, as it is the case for all the events
in the 5FNS-LO. The pseudorapidity shows a slight difference 
between the two LO results and the NLO one, the latter exhibiting a tendency to more forward distribution of $b$-jets. 
A more thorough investigation of these results is needed to draw definite conclusions.
Our preliminary results show, nevertheless, that properly accounted NLO effects
are crucial for an accurate description of the kinematic distributions of the additional
jets accompanying the singly produced VLQ.

For completeness, in Fig.~\ref{fig:distributions2} we show the transverse momentum and pseudorapidity distributions for the leading (non-$b$) jet.
In this case, we observe that the NLO results showcase more forward jets, together with a population of jets which are more markedly central. Note that we selected only events featuring a leptonic decay of the $W$-boson, so that the jet distributions do not include the ones from hadronic $W$ decays (which tend to have higher transverse momentum and be central). Further investigation of the features of these results, together with NLO distributions in the 4FNS, are under way.

\section*{CONCLUSIONS}
We have studied the single production of third generation VLQs, which can decay into a SM quark
plus a SM boson, at NLO in QCD. While the total single production cross-section
is only slightly affected by the corrections (as the main production diagrams
are of electroweak origin), we have shown that the detailed distributions are in
contrast significantly impacted.

\section*{ACKNOWLEDGEMENTS}
We would like to thank the organisers of the 2017 Les Houches workshop on TeV
colliders for the hospitality and the nice atmosphere offered during which some
of the work contained herein was performed. TF is supported by IBS under the
project code IBS-R018-D1, and BF is partly supported by French state funds managed by the Agence Nationale de la Recherche (ANR), in the context of the LABEX ILP (ANR-11-IDEX-0004-02, ANR-10-LABX-63).
GC and AD acknowledge partial support from the Labex-LIO (Lyon Institute of Origins) under grant ANR-10-LABX-66 and FRAMA (FR3127, F\'ed\'eration de Recherche ``Andr\'e Marie Amp\`ere''). 


 
\AddToContent{G.~Cacciapaglia, A.~Carvalho, A.~Deandrea, T.~Flacke, B.~Fuks, D.~Majumder, L.~Panizzi}
\renewcommand{\thesection}{\arabic{section}}


\graphicspath{{LLP_DM/}}

\newcommand{\df}{\dfrac}

\chapter{Long-lived particles at the LHC and freeze-in dark matter}

{\it G.~B\'elanger, H.~Cai, N.~Desai, A.~Goudelis, J.~Harz, A.~Lessa, J.~M.~No, A.~Pukhov, S.~Sekmen, D.~Sengupta, B.~Zaldivar and J.~Zurita}



\begin{abstract}
Long-lived particles appearing in models in which dark matter is produced via the freeze-in mechanism can be probed at the LHC. This is illustrated for the case of a long-lived charged fermion which decays into dark matter and a lepton (electron or muon), using a search for heavy stable charged particles and a displaced lepton search by the CMS collaboration. 

\end{abstract}

\section{INTRODUCTION}
The search for long-lived particles (LLPs) at the Large Hadron Collider (LHC) has recently gained momentum in the high-energy physics community. One obvious reason for this tendency is the lack of evidence for physics Beyond the Standard Model (BSM) in more traditional searches involving, for example, several jets along with missing transverse momentum. Thus, other --more exotic-- analyses should be invoked in order to make the most of the present experimental capabilities.

From an experimental standpoint, an LLP is a BSM state with a macroscopic lifetime, typically longer than a few hundreds of ps. From the theory side, there are essentially two ways in which a particle produced at the LHC can decay slowly enough to be considered an LLP: 1) the decay is kinematically suppressed because the particle is part of a new sector characterised by a compressed enough mass spectrum, or 2) the decay is suppressed due to small (effective) couplings to the ``daughter'' particles. The latter can arise in several ways (mass suppression, breaking of symmetries, fine-tuning, etc) , and we are ultimately agnostic about its origin. The case of kinematic suppression has so far been the most studied one, for example in the context of Supersymmetry \cite{Aad:2014nra,CMS-PAS-SUS-13-009,Khachatryan:2014doa,Aad:2014mha,ATLAS:2012tna,Aad:2014qaa,Khachatryan:2014rra}. In this work we will instead focus on the case in which long particle lifetimes are due to coupling suppression.

At the same time, the search for dark matter (DM)\footnote{As usual, note that the LHC itself cannot determine whether a particle escaping the detectors is (at least) part of the observed DM in the universe, since it requires complementary information from other DM-related experiments.} at the LHC is currently one of the most active topics of research. Typically, DM searches are interpreted in the framework of WIMPs (Weakly Interacting Massive Particles). These DM candidates are characterised by couplings to the SM of the order of the electroweak interactions, and thus, they could be copiously produced at the LHC. However, as mentioned before, no evidence of WIMPs has appeared so far, which motivates the consideration of other types of DM candidates. For example, DM could be made up of particles whose interactions with the SM are extremely suppressed such that, contrary to WIMPs, their production in the early universe would be out of thermal equilibrium with the SM (or, more generally, the visible) sector. Such types of DM candidates have been dubbed FIMPs (Feebly Interacting Massive Particles), and they can be produced through the so-called freeze-in mechanism \cite{McDonald:2001vt,Hall:2009bx}. The purpose of this work is to establish a link between the search for LLPs at the LHC and the freeze-in production of FIMP dark matter.

Indeed, the process through which a particle produced at the LHC decays into DM far away from the collision point could also be the one responsible for the DM production in the early universe. To the best of our knowledge, there are just a few examples in the literature which have studied this connection see, \textit{e.g.} \cite{Co:2015pka,Hessler:2016kwm,Ghosh:2017vhe}. Here, we study the connection between LLPs and FIMPs in one of the simplest freeze-in DM models that could give rise to observable signals at the LHC. We consider the case in which an electrically charged mother particle decays into a neutral one (a DM candidate) along with a lepton. The lifetime of the mother particle is such that the corresponding signature consists of a Heavy Stable Charged Particle (HSCP) producing a heavily ionised track or a displaced vertex.

In this model, there is a one-loop contribution to $\mu\to e,\gamma$. However, for the values of the couplings under consideration, this constribution is much below the current sensitivity~\cite{TheMEG:2016wtm}.

\section{THE MODEL}
\label{sec:model}
We consider an extension of the Standard Model by an additional real scalar field $s$ that transforms trivially under $SU(3)_c \times SU(2)_L \times U(1)_Y$ as well as an additional vector-like charged lepton $E$ transforming as $\left( \mathbf{1}, \mathbf{1}, \mathbf{-1} \right)$
\footnote{The vector-like nature of $E$ ensures that the model is anomaly-free.}
. Both particles are taken to be odd under a discrete ${\cal{Z}}_2$ symmetry, whereas all Standard Model fields are taken to be even. Under these assumptions, the Lagrangian of the model reads
\begin{align}\label{eq:lag}
{\cal{L}} & = {\cal{L}}_{\rm SM} + \left(\partial_\mu s\right)\left(\partial^\mu s\right) - \frac{\mu_s^2}{2} s^2 - \frac{\lambda_s}{4} s^4 - \lambda_{sh} s^2 \left(H^\dagger H\right) \\ \nonumber
& + i \left( \bar{E}_L \slashed{D}~ E_L  + \bar{E}_R \slashed{D}~ E_R \right) - \left( m_{E} \bar{E}_L E_R + y_{e} s \bar{E}_L e_R + y_{\mu} s \bar{E}_L \mu_R + {\rm{h.c.}} \right),
\end{align}
where $E_{L,R}$ and $e_R$, $\mu_R$ are the left- and right-handed components of the heavy lepton and the right-handed component of the Standard Model electron and muon, respectively. For simplicity we have neglected couplings to the third generation leptons. The model is described by six free parameters, namely
\begin{equation}
\mu_s, \ \lambda_s, \ \lambda_{sh}, \ m_E, \ y_{e}, \ y_{\mu}
\end{equation}
out of which $\lambda_s$ is irrelevant for our purposes whereas $\mu_s$ can be traded for the physical mass of $s$ through
$\mu_s^2 = m_s^2 + \lambda_{sh} v^2$,
where $v$ is the Higgs vacuum expectation value. For simplicity, we will also take the coupling $\lambda_{sh}$ to be identically zero. These choices leave us with only four free parameters
\begin{equation}
m_s, \ m_E, \ y_{e}, \ y_{\mu}.
\end{equation}
Due to its electric charge, the heavy lepton $E$ is kept in thermal equilibrium with the SM thermal bath in the early Universe. For $m_s < m_E$, the scalar $s$ becomes stable and can play the role of a dark matter candidate.

Note that the model described by Lagrangian (\ref{eq:lag}) can, for light enough values of $m_E$, lead to substantial contributions to the SM $Z$ boson decay width. Throughout the following, we will always place ourselves in the situation $m_E > m_Z/2$.

\section{FREEZE-IN PRODUCTION}
\label{sec:decay}
The dominant processes contributing to DM production are $E\to e s$ and $E\to \mu s$. Additional contributions can come from scattering processes, which we have found to be subleading for our choices of parameter values. The 	Boltzmann equation for DM can be written as:
\bea  
\dot n_s + 3 H n_s
 &=&\sum_i \int \df{d^3p_E}{(2\pi)^3 2E_E} \df{d^3p_i}{(2\pi)^3 2E_i} \df{d^3p_s}{(2\pi)^3 2E_s} (2\pi)^4\delta^{(4)}(P_E-P_i-P_s)|{\cal M}_i|^2 \nonumber \\ 
&\times& [f_E (1-f_i)(1+f_s) - f_i f_s (1- f_E)]~,
\label{Boltz1}
\eea
where the sum runs over the two processes with $i=e,\mu$. Besides, $n_s$ is the DM number density, $H$ the Hubble parameter, $P_k = (E_k, p_k)$ the four-momentum of particle $k$ with distribution function $f_k$. $\cal M$ denotes the amplitude of the process.
\newline\newline\noindent
{\it Simplifying assumptions}. The standard freeze-in computation relies on the following assumptions: 1) the initial density of DM particles is zero such that (for small enough couplings) the annihilation term can be neglected, 2) DM production occurs during the radiation dominated era, and 3) Maxwell-Boltzmann distribution functions are assumed for all bath particles (i.e. no difference between bosons and fermions)\footnote{See \cite{Belanger:2018ccd} and \cite{Blennow:2013jba} for a more detailed discussion on the distribution functions of the bath particles. }.

By adopting these simplifications the comoving DM number density (or yield $Y_s$) is given by the following expression:
\be 
Y_s \approx \df{45\xi M_{\rm Pl}}{8\pi^4\cdot 1.66} \df{g_E}{m_E^2}\Gamma \int_{m_E/T_R}^{m_E/T_0} dx~x^3 \df{K_1(x)}{g^s_*(x)\sqrt{g_*(x)}},
\label{eq:DMyield}
\ee
where $\xi=2$ since the decaying particle $E$ is not self-conjugate (otherwise $\xi=1$), $g_E$ are the internal degrees of freedom of $E$ and $\Gamma$ the sum of all partial decay widths into DM: $\Gamma = \Gamma_e + \Gamma_\mu$. $M_{\rm Pl}=1.2\times 10^{19}$ GeV is the Planck mass, $T_R$ the reheating temperature of the universe (an input for freeze-in calculations), $T_0$ is the temperature today, $K_1(x)$ is the modified Bessel function of the second kind of degree 1, and $g_*,g^s_*$ the effective degrees of freedom for the energy and entropy densities, respectively. The relation between today's relic abundance and yield of DM is\cite{Hall:2009bx}:
\be 
\Omega_s h^2\approx \df{m_s Y_s}{3.6\times 10^{-9} {\rm GeV}}~.
\ee

Most of the details of the model in Eq.~(\ref{eq:DMyield}) are encoded in the expression for the decay width $\Gamma$, which leads to the lifetime
\be 
c\tau\sim 10^3~{\rm cm}\left(\frac{10^{-9}}{y^2_e+y^2_\mu}\right)\left(\frac{{\rm TeV}}{m_E}\right)~.
\ee
 Consequently, by assuming that freeze-in via decay of the LLP is the dominant mechanism responsible for DM abundance, we can make a fairly model-independent connection between the lifetime of the LLP and the LLP and DM masses by requiring the correct DM abundance via freeze-in:
\be 
c\tau \approx 4.5~{\rm m}~\xi g_E \left(\df{0.12}{\Omega_s h^2}\right)
\left(\df{m_s}{100{\rm keV}}\right)\left(\df{200{\rm GeV}}{m_E}\right)^2
\left[\df{\int_{m_E/T_R}^{m_E/T_0} dx~x^3 K_1(x)}{3\pi/2}\right]~,
\label{eq:ctau}
\ee

where we have evaluated $g_*(x),g_*^s(x)$ on $x=3$ \footnote{This turns out to be a good approximation since for this model, most of the production occurs around the freeze-in temperature $T\approx m_E/3$. }. Note the large hierarchy of masses between the DM and the mother particle, needed in order to obtain the observed relic abundance while having a sufficiently long lifetime for the LLP, provided $m_E\ll T_R$.\footnote{For $T_0\ll m_E\ll T_R$, the ratio in squared brackets in Eq.~(\ref{eq:ctau}) will approach to 1.} Alternatively, a freeze-in solution for DM masses $m_s\gg$ MeV could be obtained by decreasing the reheating temperature, such that $m_E\gtrsim T_R$. This means essentially that the DM production history is shorter, relying only on the Boltzmann tail of the mother particle. In this work we will adopt the former regime.

We have solved numerically the Boltzmann equation (\ref{Boltz1}) with the {\tt
micrOMEGAs 5.0} code \cite{Belanger:2018ccd} under the assumptions discussed
above. The results are in 
good agreement with the
analytical approximation (\ref{eq:ctau}) shown in Fig.\ref{fig:ctau}.

\section{LHC CONSTRAINTS}
\label{sec:lhc}

\begin{figure}
\centering
\begin{tikzpicture}
\begin{feynman}
  \node [blob] (a);
  \vertex [above left=of a] (i1) {$p$};
  \vertex [below left=of a] (i2) {$p$};
  \vertex [right=of a] (b);
  \node [above right=of b,circle,fill,inner sep=1pt] (E1);
  \node [below right=of b,circle,fill,inner sep=1pt] (E2);
  \vertex [above right=of E1] (l1) {$l$};
  \vertex [below=of l1] (s1) {$s$};
  \vertex [below right=of E2] (l2) {$l$};
  \vertex [above=of l2] (s2) {$s$};

\diagram*{
   (i1) -- [double] (a) -- [double] (i2),
   (E1) -- [fermion,edge label'={$E$}] (b) -- [fermion,edge label'={$E$}] (E2),
   (s1) -- [scalar] (E1) -- [anti fermion] (l1), 
   (s2) -- [scalar] (E2) -- [fermion] (l2),
   (a) -- [photon,edge label'={$\gamma,Z$}] (b),
};
\end{feynman}
\end{tikzpicture}
\caption{\it Diagram for the main production and decay process of $E$
at the LHC.}
\label{fig:process}
\end{figure}
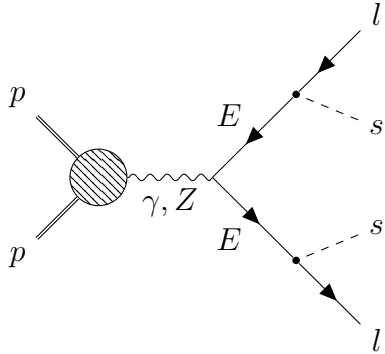

As illustrated in Fig.~\ref{fig:ctau}, in order for the 
scenario described in Sec.~\ref{sec:model} to produce the observed
dark matter relic abundance, the vector-like lepton ($E$) lifetime 
has to be larger than $\simeq 0.01$ ns (corresponding to $c\tau \sim 0.1$ m).
Consequently, if $E$ is produced at the LHC with moderate to high
velocities, it will cross a macroscopic distance in the detector.
Searches for long lived particles (LLPs)
can then be used to constrain this scenario.
In our model, $E$ will
always be pair-produced via a Drell-Yan process at the LHC
and will then decay via the $s-E-e/\mu$ coupling,
as shown in Fig.~\ref{fig:process}.
The LLP signature associated with the production of $E$'s
strongly depends on their lifetime. For $\tau \lesssim 10$~ns,
the decay occurs mostly inside the tracker, leading to a displaced
lepton or tracks with kinks. If $E$ is sufficiently long lived
to decay outside the detector ($\tau \gtrsim 100$~ns) or
outside the tracker ($\tau \gtrsim 10$~ns), it will
appear as a heavy stable charged particle (HSCP).
Below we will discuss how the LHC searches for displaced leptons
and HSCPs constrain the parameter space of the model.

\subsection{HSCP Searches}
\label{sec:hscp}

If the vector-like lepton ($E$) has a $c\tau$ of the order of a few meters, it
will deposit a considerable fraction of its energy in the tracker.
Due to its large mass, the long-lived particle will be produced
in the non-ultrarelativistic regime, thus leading to a  highly ionized
track, which can be distinguished from the ultrarelativistic tracks
produced by long-lived Standard Model particles.
Furthermore, if the LLP traverses the muon chambers, it will
produce a non-ultrarelativistic track with
a larger (or anomalous) {\it time-of-flight} (TOF) than the one
expected from muons, since $\beta_{\text{HSCP}} < 1$.
This anomalous TOF can also be used when searching for HSCPs.
Both ATLAS and CMS have performed HSCP searches at 7, 8 and
13~TeV~\cite{Chatrchyan:2012sp,Aad:2012pra,Chatrchyan:2013oca,ATLAS:2014fka,Aaboud:2016dgf,CMS-PAS-EXO-16-036}.
Because of their small background, these searches are highly sensitive to charged
LLPs and model independent, since no special veto (except for some basic
HSCP isolation) is imposed on the signal.
Here we will consider the results obtained by the CMS 8~TeV
search~\cite{Chatrchyan:2013oca}, as it provides all the detailed
information required for re-interpretation of the HSCP limits. 
The 8~TeV search presents limits for
charged LLPs using only tracker data as well as tracker plus muon chamber
(or time-of-flight) data. As mentioned above, the former is more sensitive
to lifetimes satisfying $3~\text{m} \lesssim c \tau \lesssim 10$~m, while the
latter is more sensitive for $c \tau \gtrsim 10$~m.
Since the $E$ lifetime can vary in a wide
range of values (\textit{cf Sec.~\ref{sec:model}}), we will consider both the tracker-only and the tracker plus
time-of-flight  constraints.
In order to compute the constraints on our model, we use
{\sc MadGraph5\_aMC@NLO}~\cite{Alwall:2014hca} and \textsc{Pythia}~8.2~\cite{Sjostrand:2014zea}
to simulate events for the pair production of $E$'s.

\begin{figure}
\centering
\includegraphics[scale=0.5]{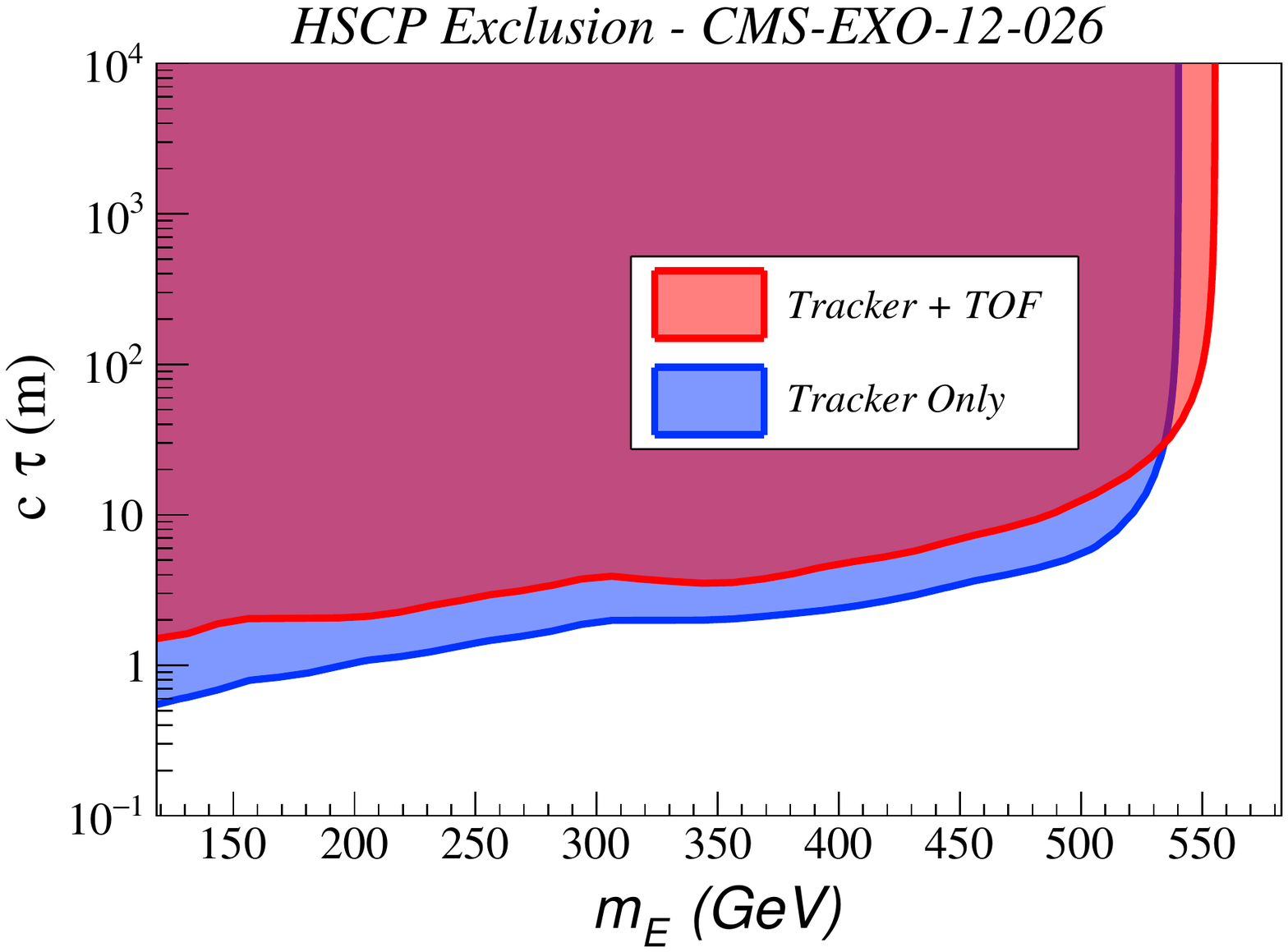}
\caption{\it LHC contraints on the LLP vector-like lepton model. The red
area corresponds to the region of the FIMP parameter space
excluded by the 8 TeV CMS search for HSCPs~\cite{Khachatryan:2015lla} using the
tracker plus time-of-flight data.
The red region shows the corresponding bounds using the tracker-only data. See
text for more details.}
\label{fig:hscpLimits}
\end{figure}

\begin{figure}
  \centering
  \includegraphics[width=0.75\textwidth]{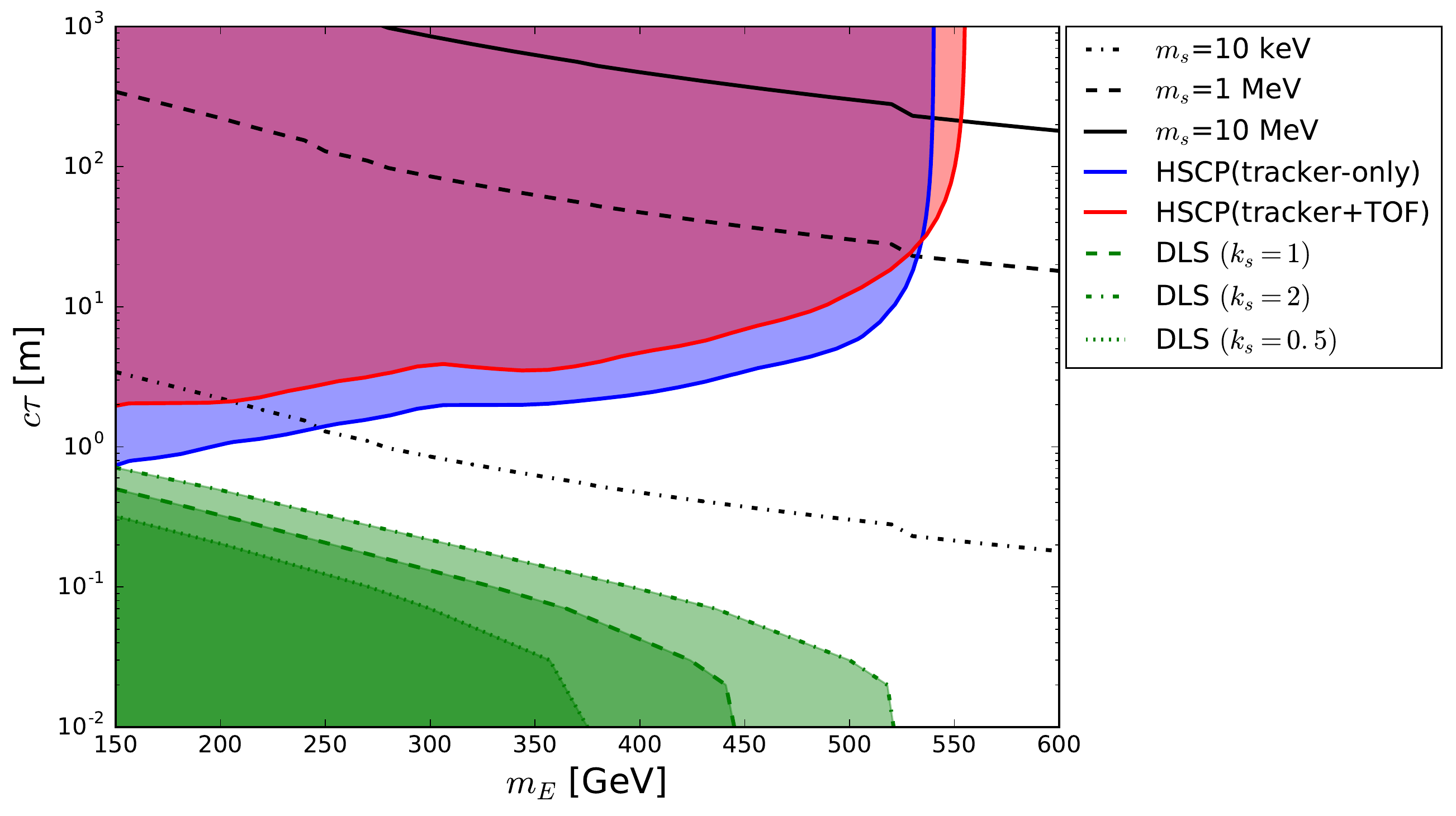} 
  \caption{\it Summary plot of this work. Iso-contours of the DM relic abundance $\Omega_sh^2 = 0.12$ are shown for three different masses: $m_s=$10 MeV (black solid), 1 MeV (black dashed) and 10 keV (black dot dashed), according to the freeze-in calculation, where the reheating temperature is taken as $T_R=10^{10}$ GeV. Red and blue regions are excluded by the HSCP searches, wheres green regions are excluded by the displaced lepton searches. See text for more details.}
\label{fig:ctau}
\end{figure}

Using the full recasting of the tracker plus
time-of-flight analysis discussed in Appendix \ref{hscp_recast}, we computed the
expected number of signal events for the FIMP scenario described
in Sec.~\ref{sec:model}. The signal yield, along with
the number of observed and expected background events provided
by CMS~\cite{Khachatryan:2015lla}, allows us to
constrain the model parameter space.
The red region in Fig.~\ref{fig:hscpLimits} shows the
region in the $c\tau$ vs $m_{E}$ plane excluded at 95\% C.L.
by the tracker plus TOF data.
For very large lifetimes we obtain a constraint $m_{E} > 550$~GeV.
It is important to point out that due to its vector-like
and fermionic nature, the cross-section for $E$ pair production is
significantly higher than the corresponding cross-section for charged
scalars. For this reason the constraints on $m_E$ are stronger than the
ones obtained by CMS for pair production of staus~\cite{Chatrchyan:2013oca}.
Once the vector-like lepton is no longer stable at detector scales, the
limits on its mass become increasingly weaker. In particular,
for $c \tau \simeq 2$~m, we have $m_{E} \gtrsim 200$~GeV.

Although the limits obtained using the tracker plus TOF data become
weaker once $c \tau \lesssim 5$~m, if the LLP decay occurs outside
the tracker volume, it is still possible to constrain our scenario using
the tracker-only data. In Ref.~\cite{Chatrchyan:2013oca} CMS has
provided cross-section upper limits (as a function of the LLP mass) for the tracker-only
analysis. However, the corresponding trigger and selection efficiencies
are not publicly available, hence a full recasting of the tracker only
search is not feasible.   
Nonetheless, since the Drell-Yan process for production of the vector-like
leptons $E$ shown in Fig.~\ref{fig:process} is kinematically similar to pair
production of staus, it is still possible to re-interpret the CMS limits for long-lived staus and use
them to constrain the FIMP scenario. 
The CMS limits from Ref.~\cite{Chatrchyan:2013oca}
are given for the total production cross-section of staus as a
function of its mass under the assumption that the staus
are stable at detector scales ($c \tau \gg 10$~m).
Therefore we can not directly apply the limits to the $E$ production
cross-section, $\sigma(EE)$, if the vector-like lepton has a finite lifetime.
In order to account for the finite lifetime -- induced -- suppression of the limits, we
compute an effective production cross section using:
\begin{equation}
\sigma_{eff}(EE) = \sigma(EE) \times f_{L}
\end{equation}
where $f_L$ represents the effective fraction of HSCPs which
have decayed at a distance $L$ from the primary vertex.
This fraction depends on the LLP lifetime and the size of
the tracker, which we assume to be $3$~m, since it approximately
corresponds to the maximum tracker radius in CMS.
For the specific details on the calculation of $f_{L}$, see 
Appendix \ref{hscp_recast}.

Once the effective cross-section is computed for each value of $m_E$
and $\tau$, we can directly compare it to the corresponding cross-section
upper limit ($\sigma_{UL}$) presented by CMS in Ref.~\cite{Chatrchyan:2013oca}.
If $\sigma_{eff}(EE) > \sigma_{UL}$ we consider the point in parameter space to
be excluded by the CMS 8~TeV search.
The results are shown by the blue region in Fig.~\ref{fig:hscpLimits}.
As expected, for small lifetimes ($c \tau \lesssim 30$~m) the constraints are
more severe than the ones obtained previously, while for large lifetimes, the tracker-only 
limits are slightly weaker than the tracker plus TOF ones, resulting in $m_E >
540$~GeV instead.

\subsection{Displaced Lepton Searches: $e\mu$}
\label{sec:dlep}

For LLP decay lengths in the range $c \tau \sim 0.1 - 100$ mm, the LLP can decay at a sizable distance from the interaction point.  If the decay products are charged, this leads to tracks with a non-zero impact parameter (which can further be used to reconstruct displaced vertices). The simplest of such searches is the CMS search for events with oppositely charged, displaced electrons and muons, conducted at both 8 TeV~\cite{CMS:emu8TeV} and 13 TeV~\cite{CMS:2016isf} runs.  This search is potentially sensitive to our model if $E$ decays with similar branching fractions to electrons and muons ($y_e \simeq y_{\mu}$). 

Here we analyze the bounds from the 13 TeV CMS search with 
$2.6$ fb$^{-1}$~\cite{CMS:2016isf}. The discriminating variable is the transverse impact parameter $d_0$, defined as the 
closest distance between the beam axis and the track in the transverse plane.  For our study, we use  generator-level information to calculate the transverse impact parameter of the lepton as:
\begin{equation}
d^{\ell}_0 = \frac{\left| p^{\ell}_x L_y - p^{\ell}_y L_x  \right|}{p^{\ell}_T}
\end{equation}
where $L_{x,y}$ the distance in $x$, $y$ travelled by the LLP before decaying, $p^{\ell}_T$ the 
transverse momentum of the lepton and $p^{\ell}_{x,y}$ the $x$, $y$ components 
of the lepton and LLP 3-momenta. 

The displaced-$e\mu$ CMS search selects events with exactly one electron and one muon with 
$\left|\eta_{\ell} \right| < 2.4$ and $p^{\ell}_T > 42$ ($40$) GeV for electrons (muons), in addition to 
$\Delta R_{e\mu} > 0.5$ and further isolation criteria for both 
leptons\footnote{A lepton is considered ``isolated" if the sum of the $p_T$ of *all other* particles in a 
cone of radius $R$ around it, normalized to its own $p_T$, is below a cut-off value $\epsilon$. For electrons in the barrel, 
electrons in the endcap and muons we have $(R,\epsilon) = (0.3, 3.5 \%), (0.3, 6.5 \%)$ and $(0.4, 15 \%)$ respectively.}.
The search then defines three non-overlapping signal regions (SR): 

\begin{itemize}
 
 \item SR III: Both $d^e_0$ and $d^{\mu}_0 \in [1, 100]$ mm. 
 
 \item SR II: Both $d^e_0$ and $d^{\mu}_0 > 0.5$ mm but one or both leptons fail SR III.
 
 \item SR I: Both $d^e_0$ and $d^{\mu}_0 > 0.2$ mm but one or both leptons fail SR III and SR II.
 
\end{itemize}

We use {\sc MadGraph5\_aMC@NLO}~\cite{Alwall:2014hca} to simulate $E^+ E^-$ production from Drell-Yan, with the 
vector-like leptons $E$s decaying to electrons and muons with equal branching fractions. We include a flat 
NLO $\kappa$-factor for Drell-Yan $E^+ E^-$ production, $\kappa \sim 1.2$~\cite{Beenakker:1999xh}. We use identification efficiencies  
for electrons and muons as a function of $p^{\ell}_T$ and $d^{\ell}_0$
publicly provided in the displaced-$e\mu$ 8 TeV CMS analysis~\cite{CMS:emu8TeV}.  Since it has been shown that using this 
parametrisation leads to a factor of two mismatch when extrapolating to 13 TeV\footnote{For validation of 
the displaced-$e\mu$ 13 TeV CMS analysis using these efficiencies, see Cottin et al. in the same proceedings volume.}, 
we show all limits with a rescaling of the signal $s_i$ ($i =$ I, II, III) by $\kappa_s = 2,\,1,\,1/2$.

Once we obtain the number of expected signal events in SR I, II, III, $s_i (c \tau, m_{\mathrm{LLP}})$, we perform 
a likelihood analysis to obtain the 95 \% C.L. exclusion limit from the 13 TeV displaced-$e\mu$ CMS search in 
the ($c \tau, m_{\mathrm{LLP}}$) plane. Our likelihood function is
built as a product of Poisson probabilities for the three SR
\begin{eqnarray}
L^{\ell}(s_i,\kappa_s) = \prod_{i=\mathrm{I,II,III}} \,e^{-(\kappa_s s_i +\, b_i)}\, \frac{(\kappa_s s_i + b_i)^{n_i}}{n_i !} \,,   
\label{likelihood_NS}
\end{eqnarray}
with $b_i$ the number of predicted background events in SR I, II, III, being respectively $ < 3.2$, $< 0.50$, $< 0.020$ (for the present analysis, 
we assume these inequalities are saturated), and $n_i$ the number of observed events in SR I, II, III, being respectively $1$, $0$, $0$.
The significance is estimated via the test statistic $Q^{\ell}_{\kappa_s}$, 
\begin{eqnarray}
Q^{\ell}_{\kappa_s} \equiv -2\, \mathrm{Log} \left[\frac{L^{\ell}(s_i,\kappa_s)}{L^{\ell}(0)} \right]\,,
\label{likelihood_1}
\end{eqnarray}
and the 95 \% C.L. exclusion limit is given by $Q^{\ell}_{\kappa_s}(s_i (c \tau, m_{\mathrm{LLP}})) = 3.84$.
We show the respective limits for $\kappa_s = 2,\,1,\,1/2$ in Fig.~\ref{fig:ctau}. These limits are highly complementary to those from 
HSCP searches discussed in section~\ref{sec:hscp}, ruling out $c \tau < 30$ cm for $m_E = 200$ GeV (for $\kappa_s = 1$). For 
LLP decays $c \tau \sim 1$ cm, the bounds go as far 
as imposing $m_E > 440$ GeV. The effect of a possible higher efficiency in moving from 8 TeV to 13 TeV, $\kappa_s = 2$, is also apparent from 
Fig.~\ref{fig:ctau}, extending the limits to rule out $m_E < 520$ GeV for $c \tau = 1$ cm and $c \tau < 50$ cm for $m_E = 200$ GeV. 
In such a case the combination of the 13 TeV displaced $e- \mu$ CMS search and HSCP searches can rule 
out $m_E < 150$ GeV through the entire freeze-in parameter space of the model.

In any case, we need to stress again that the bounds from the 13 TeV displaced-$e\mu$ CMS search shown in Fig.~\ref{fig:ctau} apply 
in the limit $y_e = y_{\mu}$, but as soon as one departs from this scenario and one of the $E \to \ell s$ ($\ell = e$, $\mu$) branching 
fraction dominates over the other, the bounds are correspondingly weaker as the signal is proportional to the product of the two branching ratios.

\section*{CONCLUSIONS}
Dark matter is undoubtedly a driver for the construction of Standard
Model extensions. Here we have studied the connection between the
freeze-in mechanism and the LHC searches for long-lived particles. An
alternative scenario to the vanilla thermal paradigm, freeze-in drives
out-of-equilibrium dark matter production via tiny ${\cal{O}}(10^{-10})$
couplings which guarantee macroscopic lifetimes irrespectively of the
mass splittings between the new states. Employing a simplified model
featuring a scalar dark matter particle and a long-lived charged
fermion, we have used two out of the ample LHC LLP searches, namely the
Heavy Stable Charged Particles (HSCP) and the high-impact displaced
lepton ones, to show that LLP masses up to about 550 GeV can be probed
by the current LHC.

The case of a very light dark matter (below the MeV scale) which is
associated with a shorter lifetime of the new heavy charged lepton is the most
challenging to probe. Nevertheless, a combination of
these two searches can rule out masses of the new charged lepton up to 
150 GeV for the full parameter space compatible with the freeze-in scenario
within the assumption that the heavy lepton decays with the same
branching fraction into electrons and muons. We further expect the
constraints from HSCP searches to improve with the use of 13 TeV data.
early results show that the gain in sensitivity in the case of stable
staus reaches about 200GeV with early 13 TeV
data~\cite{Khachatryan:2016sfv}. Cosmologically viable scenarios
involving heavier dark matter and shorter heavy lepton lifetimes are
likely to require a modification of the thermal history of the Universe
and will be studied in future work.

\section*{ACKNOWLEDGEMENTS}
The work of G.B., A.P. and B.Z. was supported in part by the French ANR, Project DMAstro-LHC ANR-12-BS05-0006, by the {\it Investissements d'avenir}, Labex ENIGMASS,
and by the Research Executive Agency (REA) of the European Union under the Grant Agreement PITN-GA2012-316704 (``HiggsTools").
A.P. also acknowledges support from  MESRI, France (progamme ACCES). 
J.M.N. is supported by the European Research Council under the European Union's Horizon 2020 program (ERC Grant Agreement no.648680 DARKHORIZONS).
The work of A.L. was supported by the S\~ao Paulo Research Foundation (FAPESP), projects
2015/20570-1 and 2016/50338-6.
N.D. acknowledges the support of the OCEVU Labex (ANR-11-LABX-0060) and the A*MIDEX project (ANR-11-IDEX-0001-02) funded by the "Investissements d'Avenir" French government program managed by the ANR.
J. H. and A.G. were supported by the Labex ILP (reference ANR-10-LABX-63) part of the Idex SUPER, and received financial state aid managed by the Agence Nationale de la Recherche, as part of the programme Investissements davenir under the reference ANR-11-IDEX-0004-02.

\appendix
\section{Details of the CMS HSCP analysis recast}
\label{hscp_recast}
Here we briefly elucidate the procedure employed to recast the CMS HSCP analysis for  constraints from the tracker plus
TOF scenario. In Ref.~\cite{Khachatryan:2015lla}, CMS has provided efficiencies
for the trigger and event selection of HSCPs for the
tracker plus TOF search.\footnote{For more details on 
the recasting procedure see Ref.~\cite{Heisig:2015yla}.} 
The CMS efficiencies are given as a function of the HSCP truth level
kinematics, $\vec{k} = \left(p_T,\eta,\beta\right)$ such that no detector simulator is
required.
Since the signal selection requires at least one HSCP in each event, 
the total trigger (or selection) efficiency for an event containing two
isolated HSCPs (such as pair production of vector-like leptons) is given by:
\begin{equation}
\epsilon_{T}^{a} =  \epsilon_1^a
\times\left(1-\epsilon_2^a \right) + \epsilon_2^a
 \times \left(1-\epsilon_1^a \right) 
+  \epsilon_1^a \times \epsilon_2^a \;, 
\label{eq:efftwo}
\end{equation}
where $a =$ trigger or selection, $\epsilon^a_i$ represents the
efficiency for the $i$-th HSCP and $\epsilon_T^a$ the combined efficiency.
The first two terms in Eq.~(\ref{eq:efftwo}) correspond to the probability of at
least one HSCP passing the trigger or selection, while the last term corresponds to the probability of
both particles being selected.
With the above definitions, the final event efficiency is simply given by: 
\begin{equation}
\epsilon_{\text{event}} = \epsilon_T^{trigger} \times \epsilon_T^{selection} .
\label{eq:hscpEff}
\end{equation}
Finally, to compute the total signal efficiency, one must sum over the efficiencies
of all events:
\begin{equation}
\epsilon_{signal} = \frac{1}{N} \sum_{events} \epsilon_{event} \;,
\label{eq:effSignal} 
\end{equation}  
where $N$ is the total number of events generated.

All the above efficiencies correpond to (detector) stable LLPs.
However, if the long lived particle has a finite lifetime, the event efficiency
must be rescaled by the fraction of LLPs which cross the detector without
decaying. Eq.~(\ref{eq:hscpEff}) must then be modified if
$\tau$ is finite:
\begin{equation}
\epsilon_{event} = \epsilon_T^{trigger} \times \epsilon_T^{selection} \times f_L \;,
\label{eq:eventTau}
\end{equation}
where $f_L$ is the
effective fraction of HSCPs which cross a distance $L$ of the detector without
decaying.
For events with a single LLP the rescaling is trivial and $f_L$
is simply given by
\begin{equation}
f_L = F \equiv e^{- m L/\left(c \tau |\vec{p}|\right)} \label{eq:Flong}
\end{equation}
In the above expression $L$ is the detector radius, $m$ is the LLP mass, $\tau$
its proper lifetime and $\vec{p}$ its 3-momentum in the event. For the full CMS
detector we take $L = 9$~m, 10~m or 11~m for a pseudo-rapity $|\eta| < 0.8$, 1.1
or 2.5.
However, for events with two LLPs, the effective fraction of LLPs is
given by:
\begin{eqnarray}
f_L = F_1 \times F_2 &+& F_1 \left(1 - F_2\right) \times
\left(
\epsilon_1^{trigger}\epsilon_1^{selection}\right)/\left(\epsilon_T^{trigger}
\times \epsilon_T^{selection}\right) \nonumber\\
&+& F_2 \left(1 - F_1\right) \times
\left( \epsilon_2^{trigger}\epsilon_2^{selection} \right)/
\left(\epsilon_T^{trigger} \times \epsilon_T^{selection}\right) \;,
\label{eq:effLong}
\end{eqnarray}
where $F_i$ is the fraction in Eq.~(\ref{eq:Flong}) computed for the $i$-th LLP in
the event. The first term corresponds
to both LLPs decaying outside the detector, while
the last two terms correspond to only one LLP
crossing the detector without decaying.
Eqs.~(\ref{eq:effLong}), (\ref{eq:eventTau}) and (\ref{eq:effSignal}) can then
be used to compute the total signal efficiency for a given model with finite lifetime.
Finally, we point out that we expect the signal efficiencies of the
tracker-only and the tracker plus TOF analyses to rescale equally with lifetime.
Therefore, although the values of $f_L$  are computed explicitly using the 
trigger and selection efficiencies for the tracker plus TOF analysis,
we use the same values when rescaling the results for the tracker only 
analysis. The only difference is that for the tracker only analysis
we take $L$ to be the size of the CMS tracker ($L \simeq 3$~m).


 
\AddToContent{G.~B\'elanger, H.~Cai, N.~Desai, A.~Goudelis, J.~Harz, A.~Lessa, J.~M.~No, A.~Pukhov, S.~Sekmen, D.~Sengupta, B.~Zaldivar and J.~Zurita}
\renewcommand{\thesection}{\arabic{section}}
\renewcommand{\thesubsection}{\thesection.\arabic{subsection}}
\renewcommand{\thesubsubsection}{\thesubsection.\arabic{subsubsection}}
\renewcommand{\thefigure}{\arabic{figure}}
\renewcommand{\theequation}{\arabic{equation}}
\renewcommand{\thetable}{\arabic{table}}
\renewcommand{\thefootnote}{\arabic{footnote}}

\graphicspath{{LLPlifetime/}}
\newcommand{\bit}{\begin{itemize}}
\newcommand{\eit}{\end{itemize}}
\newcommand{\ii}{\item}
\newcommand{\cw}[1][{}]{\ensuremath{\cos^{#1} \theta_{W}}}
\newcommand{\sw}[1][{}]{\ensuremath{\sin^{#1} \theta_{W}}}
\newcommand{\tw}[1][{}]{\ensuremath{\tan^{#1} \theta_{W}}}
\newcommand{\ctw}[1][{}]{\ensuremath{\cot^{#1}\theta_{W}}}
\newcommand{\mink}[2]{\ensuremath{\,#1\! \cdot \! #2 \,}}
\newcommand{\cv}[1][{}]{\ensuremath{\cos^{#1} \beta}}
\newcommand{\sv}[1][{}]{\ensuremath{\sin^{#1} \beta}}
\newcommand{\tv}[1][{}]{\ensuremath{\tan^{#1} \beta}}
\newcommand{\M}{\ensuremath{\mathcal{M}}}
\newcommand{\ssl}[1]{\ensuremath{\slashed{#1}}}

\def\gsim{\lower0.5ex\hbox{$\:\buildrel >\over\sim\:$}}
\def\lsim{\lower0.5ex\hbox{$\:\buildrel <\over\sim\:$}}

\def\xutchi{BR($\tilde{u_{1}} \to t \chi^{0}_{1}$)}
\def\xucchi{BR($\tilde{u_{1}} \to c \chi^{0}_{1}$)}

\def\met{\rm E{\!\!\!/}_T}

\def\tcb{\textcolor{blue}}
\def\tcr{\textcolor{red}}



\chapter{Towards determining the lifetime of long-lived particles at the LHC}

{\it S.~Banerjee, D.~Barducci, B.~Bhattacherjee, A.~Goudelis, B.~Herrmann, D.~Sengupta}


%
%
%
%
%
%


\begin{abstract}
We address the question of measuring the lifetime of a long-lived particle (LLP), assuming 
evidence for a displaced vertex at the Large Hadron Collider. In particular, we 
analyse to which precision it will be possible to access the lifetime experimentally. 
Based on a simplified framework, we investigate the dependence of the lifetime estimation 
on several factors, \textit{viz.}, the mass of the LLP, its momentum distribution, 
the experimental cuts imposed and the final statistics. We also discuss the potential 
impact of smearing effects, on the lifetime estimation.  
\end{abstract}


\section{Introduction}
The Large Hadron Collider (LHC) is currently operating in its Run-2 phase and pursues the quest for New Physics. Up to now, no direct or indirect signal of new particles has been observed in the existing search channels. Consequently, it is important to also consider alternative possibilities going beyond the standard assumptions adopted in traditional searches for physics beyond the Standard Model (SM). One such possibility is that some of the produced particles are long-lived, {\emph{i.e.}}\ that the secondary vertices through which they decay are macroscopically displaced with respect to the primary interaction point. While some relevant studies are already being pursued at at the LHC, see \textit{e.g.}\ \cite{Aaboud:2017mpt, Aaboud:2017iio, Khachatryan:2016sfv, Sirunyan:2017jdo, Sirunyan:2017sbs}, most new physics searches are targeting scenarios where beyond the SM states undergo a prompt decay.

Typically, long-lived particles (LLPs) are states with a proper lifetime $\tau$ greater than $\sim 100$ ps. Such lifetimes can be induced either by very small couplings or in specific kinematic configurations involving small mass splittings and/or large propagator masses. They appear in a large variety of New Physics frameworks such as supersymmetry \cite{Giudice:1998bp, Meade:2010ji, Arvanitaki:2012ps, Cerdeno:2013oya, Burdman:2006tz}, Twin Higgs models \cite{Chacko:2005pe}, dark matter \cite{Hall:2009bx, TuckerSmith:2001hy, Co:2015pka, Hessler:2016kwm}, Hidden Valley models \cite{Strassler:2006im, Strassler:2006ri, Strassler:2006qa} or baryogenesis \cite{Cui:2014twa} and they can be either neutral or charged, see also \cite{Banerjee:2017hmw} and references therein.  Charged long-lived particles typically lead to disappearing or kinked tracks, while neutral ones to displaced vertices.

Regardless of the underlying model, LLPs introduce an additional complication for experimental searches, related to the particle's lifetime. For instance, from a theorist's standpoint, an electron is a universally-defined entity. However, experimentally, an electron that appears within the tracker  is a completely different object than an electron which appears, \textit{e.g.}, in the electromagnetic calorimeter. This implies that an experimental search for neutral particles decaying into a pair of visible objects at different parts of the LHC detectors could necessitate radically different analyses which can be more or less challenging. In this work we will focus on the case in which a -- larger or smaller -- fraction of LLPs decay into pairs of charged particles within the tracker detector. The associated displaced vertex signature consists a pair of ``emergent'' tracks accompanied by signals in (some of the) other parts of the detector.

Our goal is not to study the discovery potential of the LHC for such scenarios, but to place ourselves in the situation in which a signal is observed and investigate the capacity of the LHC to reconstruct the lifetime of the decaying particle. To the best of our knowledge, only a handful of such studies have been performed in the literature, focusing on different LLP decay channels \cite{Asai:2009ka, Ambrosanio:2000zu}. Although our study will be performed within a toy framework, the method is fairly generic, leaving more concrete realisations of this scenario in terms of models for future work. We will explore various models elucidating the efficacy of this method in a future work.

\section{Long-lived particle lifetime reconstruction}

Measured in the laboratory frame, the decay length of a particle is given by
\begin{equation}
	d ~=~ \beta \gamma c\tau  \,,
	\label{LLPlifetime_eq1}
\end{equation}
where $\tau$ is the proper decay time of the decaying particle, {\emph{i.e.}}\ the time interval until the particle decays as measured in its own rest frame, $\gamma=E/m=(1-\beta^2)^{-1/2}$ is the relativistic factor with $\beta=v/c=|\vec{p}|/E$, $v$ is the velocity of the decaying particle and $c$ denotes the speed of light. In the LHC setting, if we consider the production of a number $N_0$ of such unstable particles with proper decay times $\tau_i$ and mean (proper) lifetime $\tau$, the expected number of decay events as a function of $\tau_i$ is given by the usual exponentially decreasing distribution
\begin{equation}
	N_i ~=~ N_0 ~ e^{-\tau_i/\tau} \,.
	\label{LLPlifetime_eq2}
\end{equation}
By measuring the decay length $d_i$ of each event, together with the corresponding kinematical factor $\beta_i$, we can compute the proper decay time associated to the event. Ideally then, it is possible to infer the values of $N_0$ and $\tau$ by performing an exponential fit of the sample data, provided that enough statistics is available. 

Here we will consider the production of a neutral long-lived particle, hereafter denoted by $X$, decaying into a pair of leptons. As already mentioned in the Introduction, the part of the LHC detectors in which a decay event occurs drastically alters the amount of information that can be extracted for this event. Here we study the case in which the particle $X$ decays inside the tracker. As a first approximation, let us assume that the momenta of the decay products as well as the position of the secondary vertex can be reconstructed with infinite precision. These are, of course, simplifying assumptions, which will not hold in a real experimental analysis. In the following we will relax the former condition, whereas the impact of vertex position measurement uncertainties is a heavily experiment-dependent issue which is difficult to address without a fair amount of knowledge on technical aspects of the LHC detectors. Note also that we neglect the effects of initial or final state radiation, which can provide additional information on the reconstructed events.

In order to study this ideal situation we have have used \texttt{Pythia 6} \cite{Sjostrand:2006za} to generate data samples of 10000 events each, assuming different lifetimes and masses for the particle $X$. For the ideal case where we assume the four momenta of the long-lived particles to be precisely reconstructed it is sufficient to generate parton-level events, $q q \to X X$. However, we must mention here that the $\beta \gamma$ spectra of the LLPs will vary depending on their production mode.

\begin{figure}
	\begin{center}
 		\includegraphics[width=.48\textwidth]{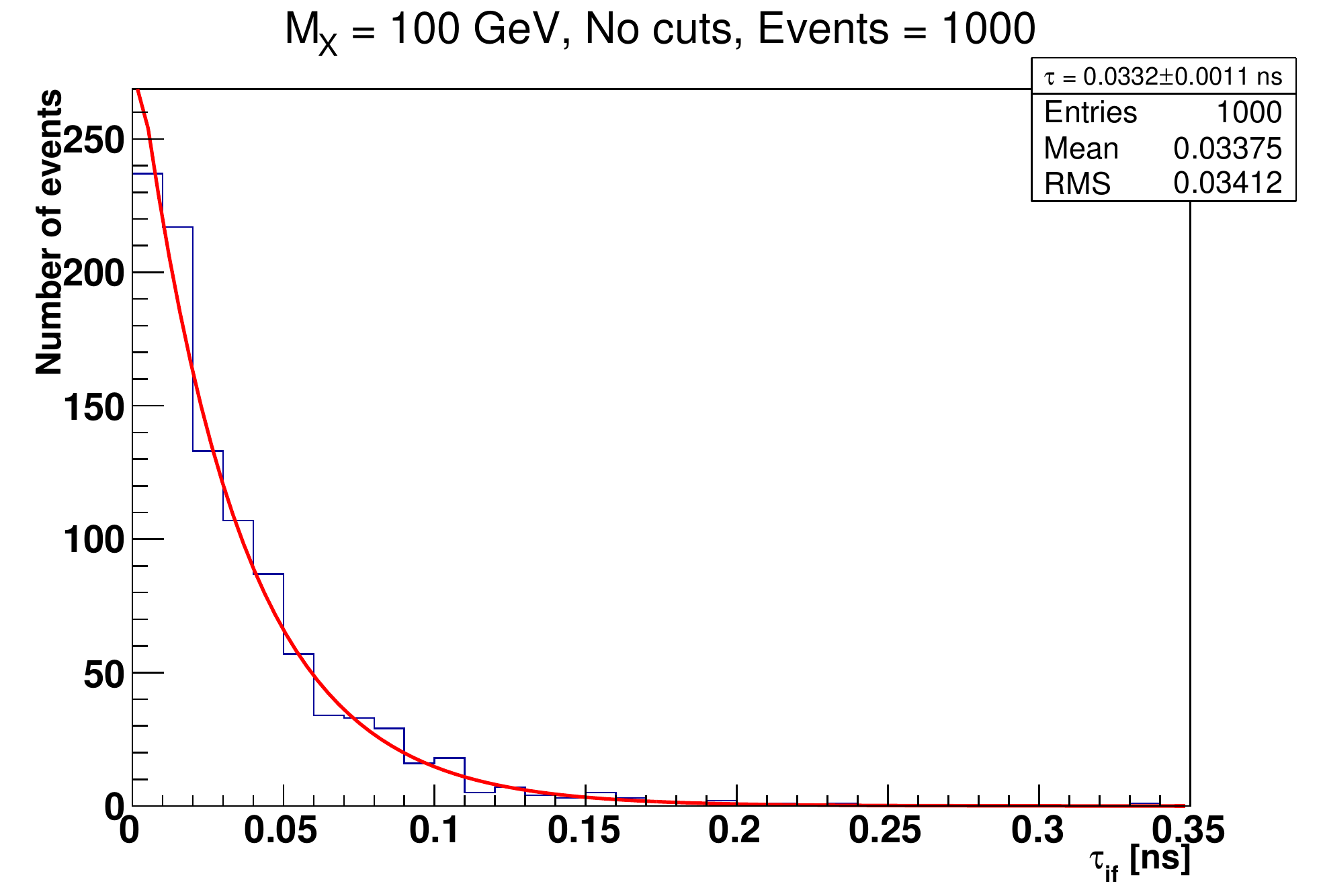}
 		\includegraphics[width=.48\textwidth]{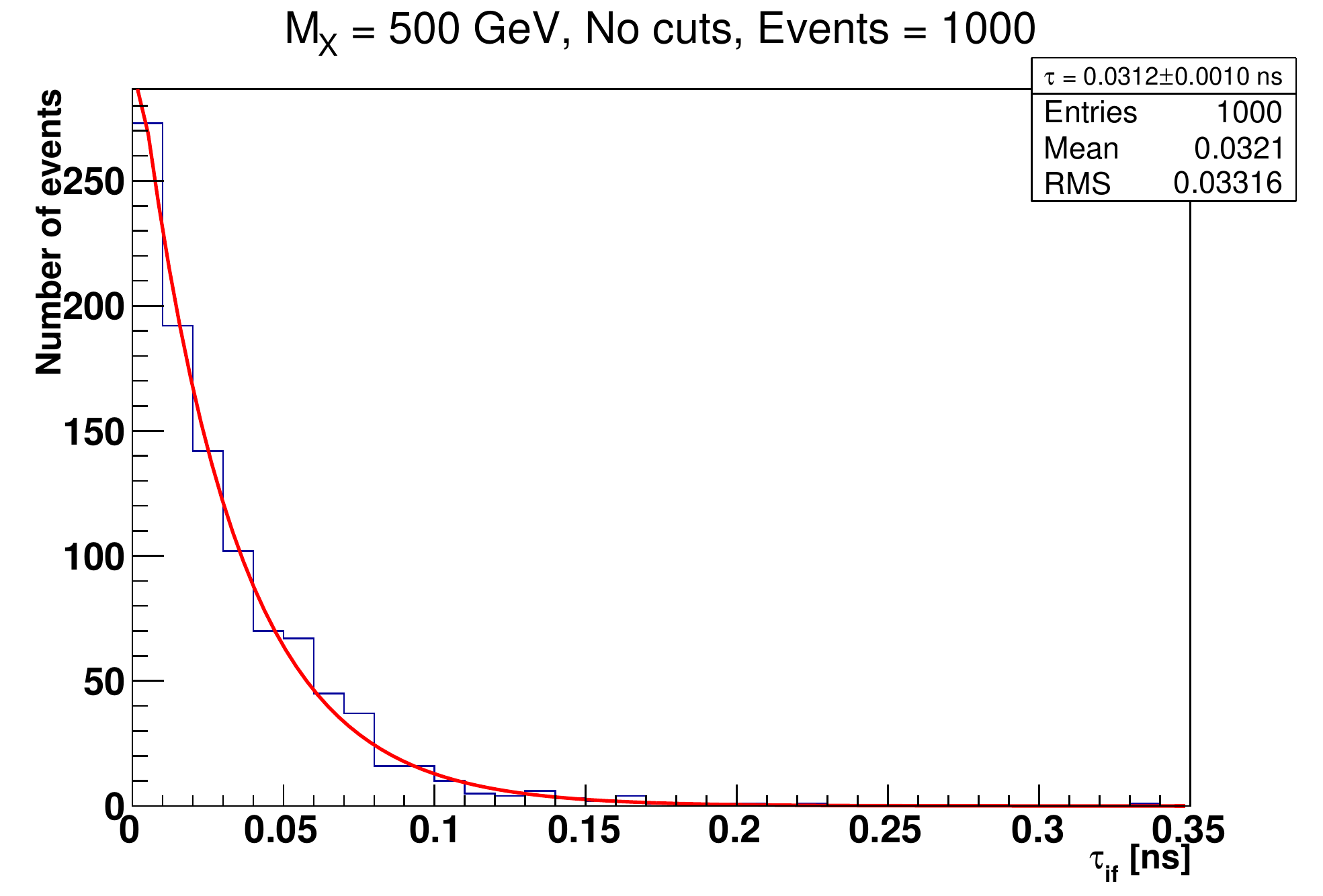}\\
 		\includegraphics[width=.48\textwidth]{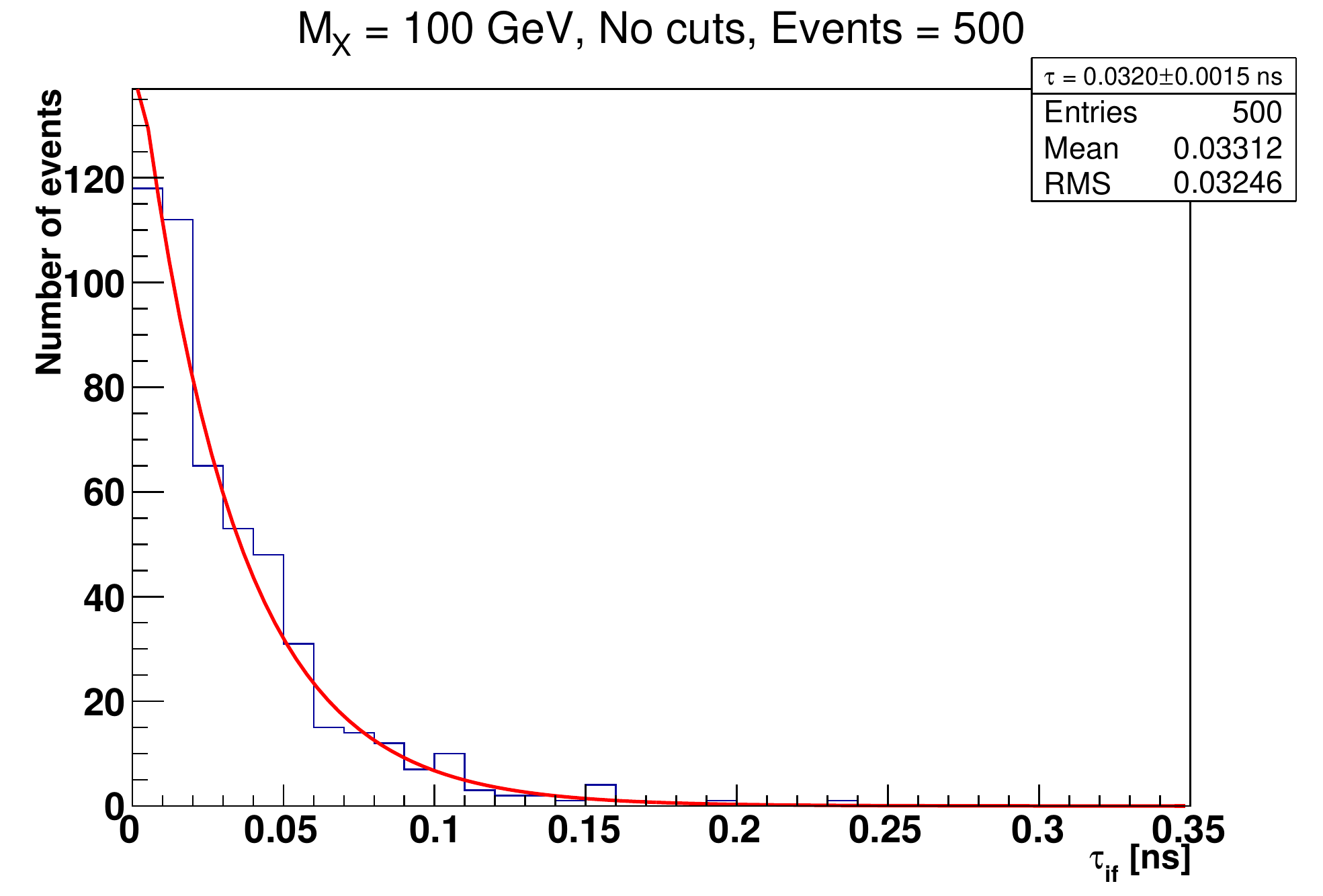}
 		\includegraphics[width=.48\textwidth]{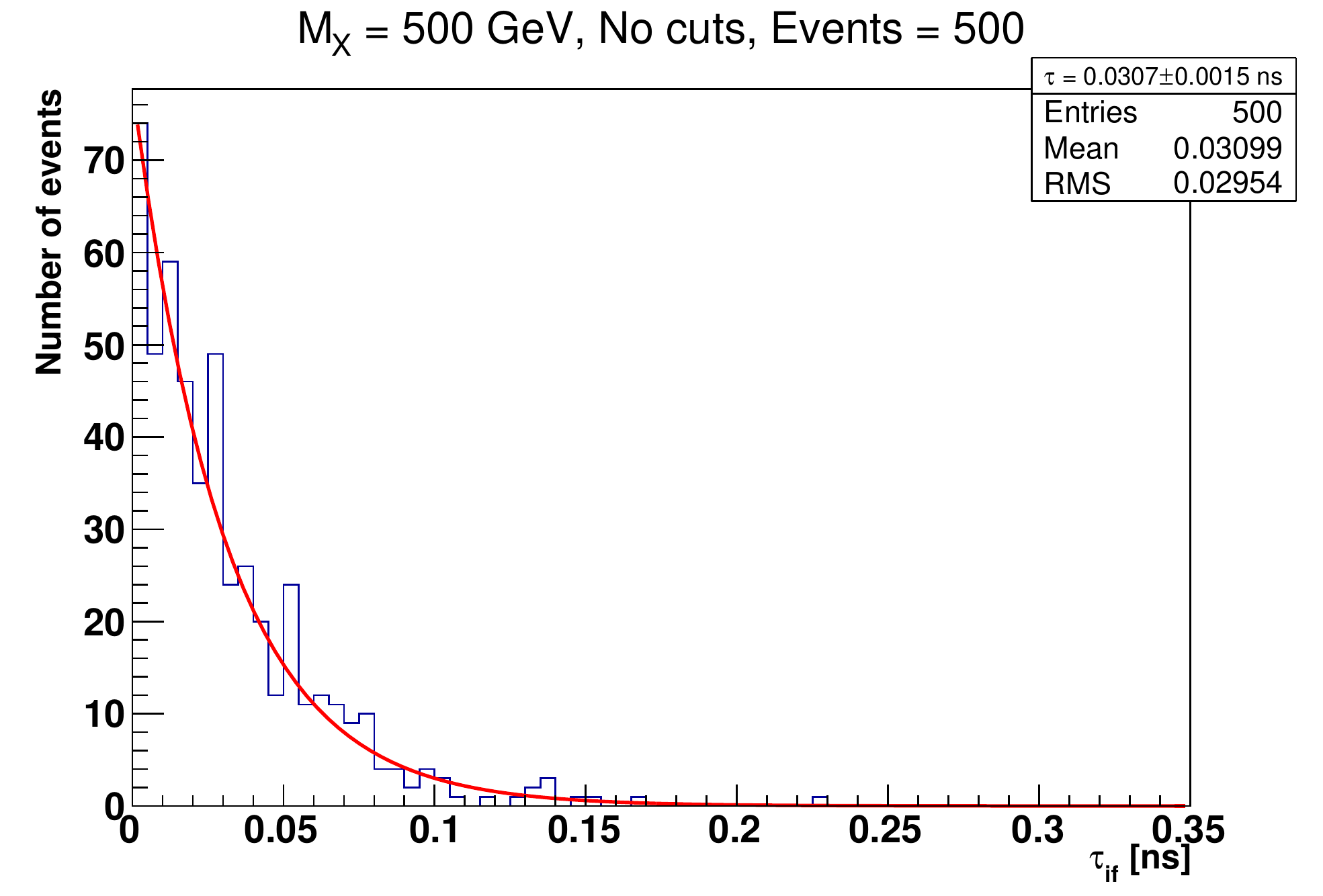}
	\end{center}
	\vspace*{-5mm}
	\caption{Lifetime determination using an exponential fit for two LLP masses (100 and 500 GeV, left and right columns) for samples containing 1000 and 500 events (top and bottom rows respectively).}
	\label{LLPlifetime_fig1}
\end{figure}

In Fig.\ \ref{LLPlifetime_fig1} we show the results of a simple exponential fit as performed using the \texttt{TF1} class integrated in the \texttt{ROOT} \cite{Brun:1997pa} environment, to a randomly selected subset of our samples for two LLP masses (100 and 500 GeV, left and right columns respectively) assuming two different numbers of observed events (1000 and 500, top and bottom rows) and a lifetime of $0.033$ ns, without imposing any kinematic cuts on the events contained in the samples. Naturally, we see that in this idealised case the lifetime of the parent particle can be reconstructed with an excellent precision of $\sim 3\%$ assuming 1000 events and of $5\%$ assuming 500 events for $m_X = 100$ GeV.

For the sake of a realistic study, experimental cuts must be imposed. A basic set of requirements consists of demanding a pseudorapidity $\left| \eta \right| < 2.4$ and a transverse momentum $p_T > 5$ GeV. Samples satisfying only these two conditions will be hereafter referred to as BCA (Basic Cuts Applied). In addition to the BCA cuts, given the increasing difficulty in reconstructing the secondary vertex as the latter approaches the outer surface of the tracker, we will study event samples satisfying an additional condition on the displacement of the secondary vertex with respect to the interaction point, $\beta_i \gamma_i c \tau_i$. Concretely, we consider three such extra cuts (EC), $\beta_i \gamma_i c \tau_i < 20$ mm (EC1), $50$ mm (EC2) and $100$ mm (EC3)~\footnote{Note that current ATLAS and CMS searches are sensitive to displaced vertices between 10 mm to $\sim$ 300 mm with a peak sensitivity at around 100 mm. \cite{Aad:2015rba,Sirunyan:2017jdo}}. The strongest constraint on $\beta_i \gamma_i c \tau_i$ is inspired from the 8 TeV CMS study pertaining to the reconstruction of leptons as functions of the transverse impact parameter ($d_0$), in supersymmetric scenarios involving displaced vertices \cite{twiki}.

The imposition of the EC cuts, however, introduces an important subtlety: by rejecting large values of $\beta_i \gamma_i c \tau_i$ we are inevitably biasing our event samples in favour of events characterised by smaller proper decay times. This implies that the expected exponential distribution is skewed and our estimate for the reconstructed lifetime will also be biased towards smaller values. In other words, provided enough statistics is available, every lifetime reconstruction based on samples with EC cuts tends to underestimate the true LLP lifetime. This effect is exemplified in Tab.\ \ref{LLPlifetime_tab:misreconstructed}, where we present the fitting procedure results for three different combinations of LLP masses and lifetimes assuming the four sets of cuts, BCA and BCA$+$EC.

\begin{table}[h]
\begin{tabular}{ccccccc} 
\hline
Events & $M_{X}$ & $l$ (mm) & $\tau_{\rm BCA}$ (ns)   & $\tau_{\rm EC1}$ (ns)   & $\tau_{\rm EC2}$ (ns)   & $\tau_{\rm EC3}$ (ns) \\\hline
1000   & 200     &  10      & $0.0338 \pm 11\times 10^{-4}$     & $0.0196 \pm 8\times 10^{-4}$     & $0.0258 \pm 9\times 10^{-4}$     & $0.0309 \pm 10\times 10^{-4}$     \\
1000   & 200     &  5       & $0.0170 \pm 6\times 10^{-4}$     & $0.0196 \pm 8\times 10^{-4}$     & $0.0127 \pm 5\times 10^{-4}$     & $0.0164 \pm 5\times 10^{-4}$     \\
1000   & 2500    &  10      & $0.0326 \pm 9\times 10^{-4}$     & $0.0304 \pm 9\times 10^{-4}$     & $0.0325 \pm 9\times 10^{-4}$     & $0.0326 \pm 9\times 10^{-4}$     \\\hline
\end{tabular}
\caption{(mis-)Reconstructed lifetimes. $M_X$ is given in GeV. In all cases, we start with 1000 Monte Carlo events which are reduced upon imposing each cut.}
\label{LLPlifetime_tab:misreconstructed}
\end{table}

We can, indeed, clearly see that when cutting harder (EC1) on $\beta_i \gamma_i c \tau_i$, we obtain an estimate of the LLP lifetime which, although seemingly accurate, can be false. The effect is more pronounced for larger lifetimes, as can be seen by comparing the first with the second row of Tab.\ \ref{LLPlifetime_tab:misreconstructed} since, all other quantities kept constant, a smaller lifetime implies overall smaller values of $c \tau_i$ and, hence, that a greater number of events is concentrated in a smaller area of $\beta_i \gamma_i c \tau_i$. Similarly, by comparing the first and the third rows of Tab.\ \ref{LLPlifetime_tab:misreconstructed}, we can see that the bias induced by the EC cuts is larger for smaller LLP masses. This is due to the fact that heavy LLPs are globally characterised by smaller $\beta_i \gamma_i$ values, hence, the impact of the EC cuts on the selected proper decay times is milder. 

\begin{figure}
	\begin{center}
 		\includegraphics[width=.48\textwidth]{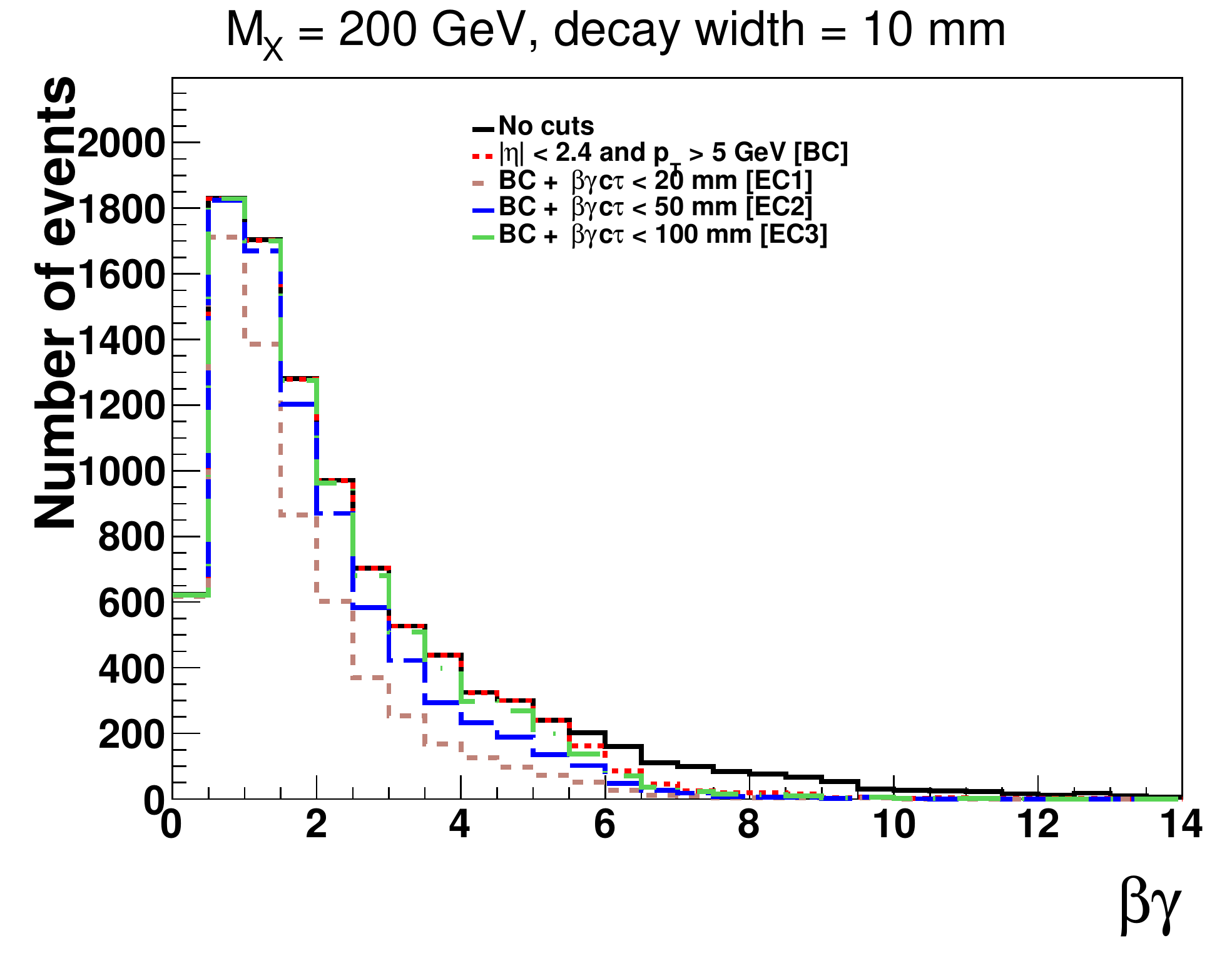}
 		\includegraphics[width=.48\textwidth]{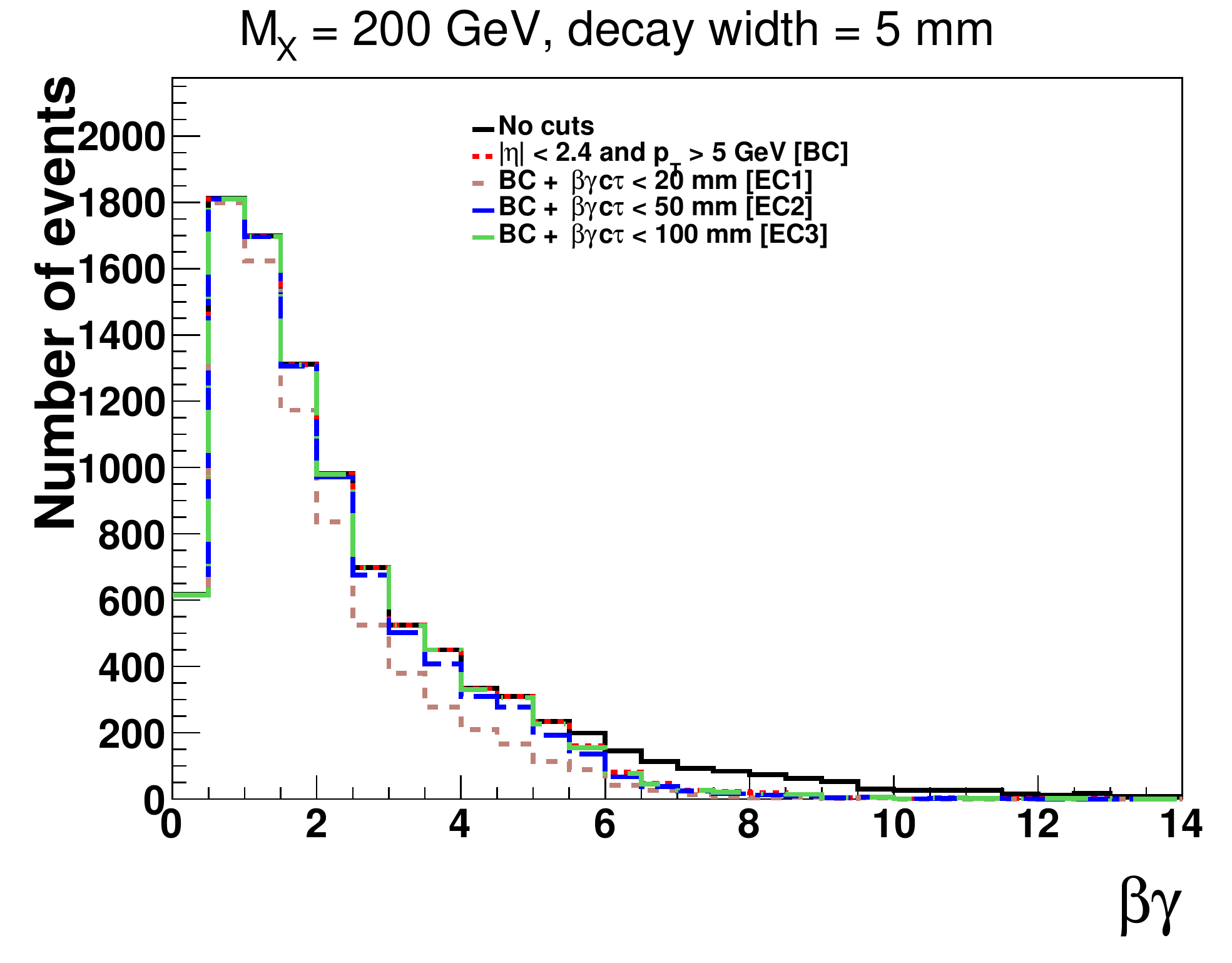}\\
 		\includegraphics[width=.48\textwidth]{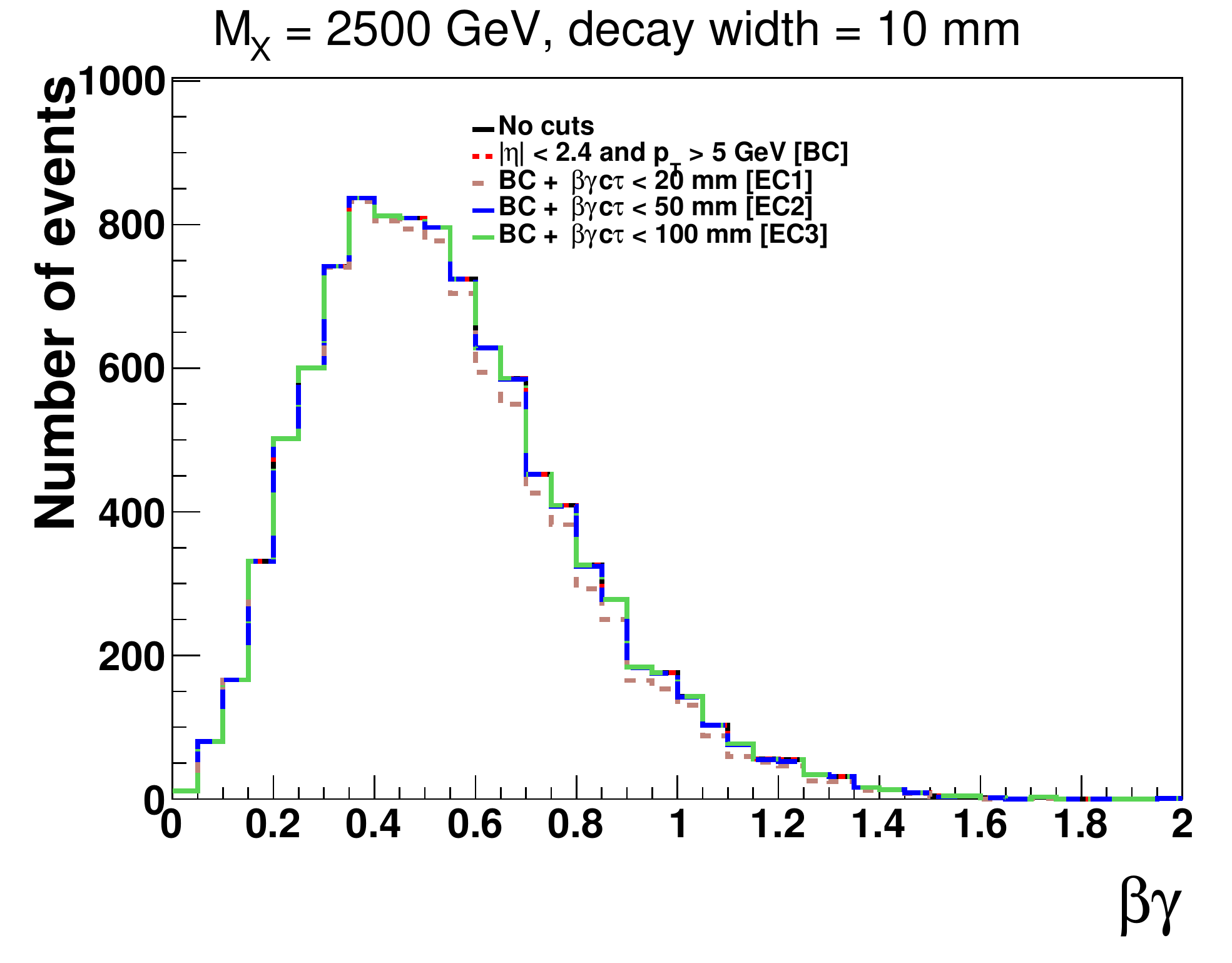}
	\end{center}
	\vspace*{-5mm}
	\caption{Number of events as a function of $\beta\gamma$ for the three mass and lifetime combinations of Tab.~\ref{LLPlifetime_tab:misreconstructed}.}
	\label{LLPlifetime_fig2}
\end{figure}

These comments are further illustrated in Fig.\ \ref{LLPlifetime_fig2}, where we show the number of events as a function of $\beta\gamma$ for the three benchmarks of Tab.\ \ref{LLPlifetime_tab:misreconstructed} after the successive imposition of the BCA and EC cuts.

In light of these observations, we need to devise a method in order to reconstruct the LLP lifetime without biasing the signal. The crucial observation is that, as long as a sufficient number of events with large $\tau_i$ values is kept in the sample, the estimate tends asymptotically towards the true LLP lifetime from below as the considered $\tau_i$ region becomes larger. Then, given an EC cut, we can successively compute lifetime estimates based on an increasing number of events with large proper decay times which can, in turn, be achieved simply by restricting the $\beta_i \gamma_i$ region to smaller and smaller values. If, below a certain value of $\beta_i \gamma_i$, the estimate saturates, this means that we have included a sufficient number of events with large $\tau_i$ values and the estimate for the true LLP lifetime, along with its associated uncertainty, can be trusted. If, on the other hand, the estimate does not saturate, it can only be viewed as a lower bound on the true LLP lifetime. 

\begin{figure}
	\begin{center}
 		\includegraphics[width=.48\textwidth]{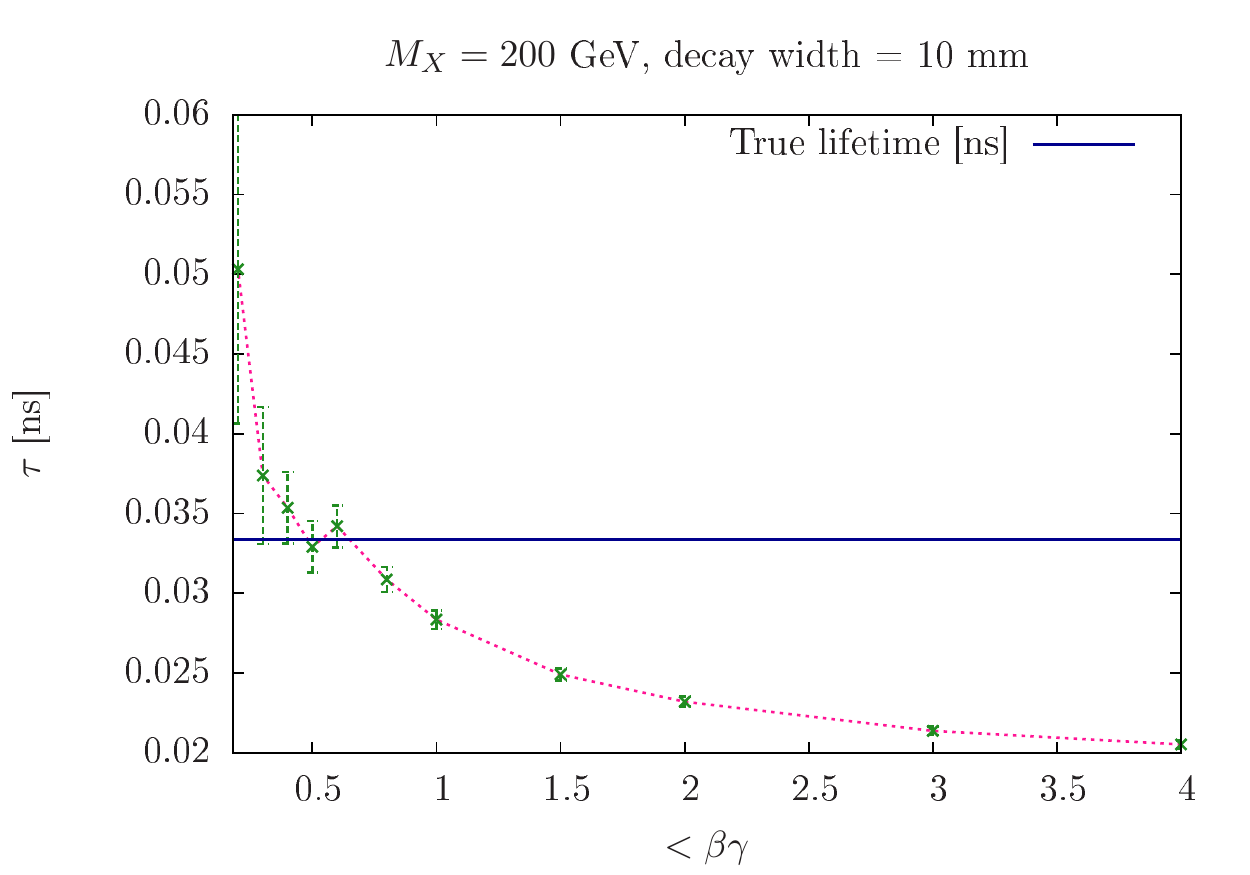}
 		\includegraphics[width=.48\textwidth]{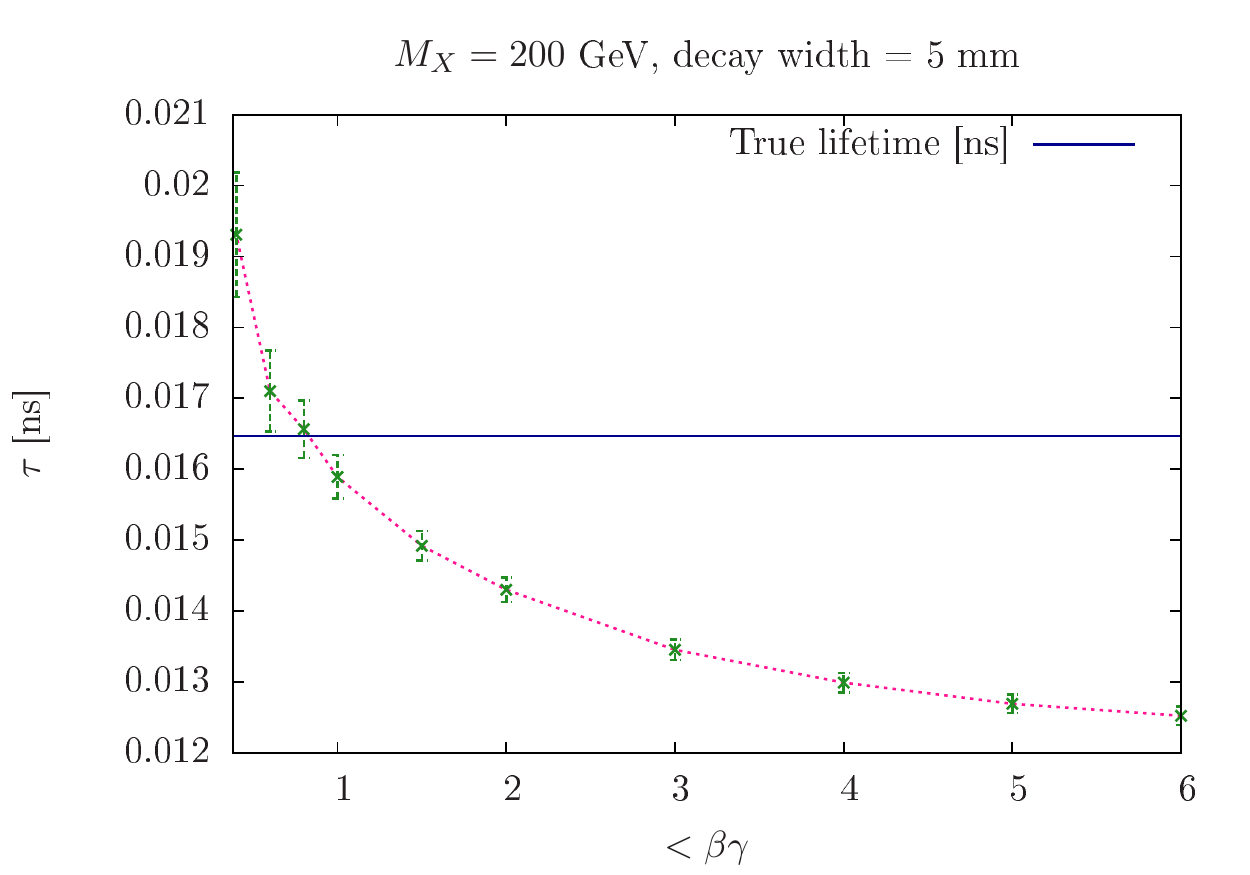}
	\end{center}
	\vspace*{-5mm}
	\caption{Reconstructed LLP lifetime along with its associated uncertainty as a function of the highest $\beta\gamma$ considered, for $M_X = 200$ GeV and two different true lifetime values.}
	\label{LLPlifetime_fig3}
\end{figure}

The results obtained with this method are shown in Fig.~\ref{LLPlifetime_fig3} for two examples assuming an LLP mass $M_X = 200$ GeV and two different decay lengths (lifetimes) of $10$ mm (left) and $5$ mm (right). In the former case, we can clearly observe the saturation of the estimated LLP lifetime once an upper bound of about $0.6$ is imposed on $\beta_i \gamma_i$. In the latter case, although perhaps less visible graphically, a similar situation occurs around $\beta_i \gamma_i < 1$ whereas, as expected, the estimated lifetime globally varies in a much narrower range. Besides, in both cases, we observe that restricting $\beta_i \gamma_i$ to extremely small values leads to an overestimate of the LLP lifetime. This simply reflects the fact that below a certain threshold for $\beta_i \gamma_i$ there is not enough statistics to obtain an accurate estimate. Still, the true value of the lifetime lies within $2\sigma$ to $3\sigma$ from the estimated one.

So far, we have made the assumption that the lepton momenta can be reconstructed with infinite precision. In order to study the impact of the associated uncertainties, we consider a more generic process where the LLP decays into three very light daughter particles~\footnote{The motivation for studying three- rather than two-body decays lies with the fact that the greater the number of daughter particles, the more important the smearing effect will be on the reconstructed mother LLP.}. Using the standard smearing algorithm of \texttt{Pythia 6}, we parametrise the shift in the transverse momentum ($p_T$) measurement of each daughter particle as $\Delta p_T = a \sqrt{p_T}$, where $a$ denotes the smearing parameter and is assumed to be the same for all the daughter particles. Hence, we obtain $\frac{\Delta p_T}{p_T} = \frac{a}{\sqrt{p_T}}$. We generate random Gaussian numbers for each daughter particle, smear their $p_T$ and, using standard kinematic relations, obtain their smeared four-momenta. For this study, we focus on a simple scenario where a massive LLP decays to very light final state particles. These daughter particles, in turn, will be highly boosted, rendering $\frac{\Delta p_T}{p_T}$ negligible. Upon choosing two benchmark values of $a = 1$ and $1.5$ (which correspond to 10\% and 15\% smearing on each particle), we do not obtain any observable effect on the lifetime fit. However, we must stress that a scenario involving one or more heavy daughter particles or the parent LLP being much lighter, will amount to the daughter particles being less boosted and, consequently, the smearing effects may play an important role in the lifetime fit. Lastly, we must mention here that in actual experiments, the uncertainties on measuring the secondary vertex may be dependent on the $p_T$ measurements of the tracks used to reconstruct them, which in turn will depend on the particle smearings. In this study, we test the effect of smearing on the benchmark $M_X = 200$ GeV, decay length 10 mm, decaying to very light daughter particles.

\section{Conclusions and perspectives}

In this note, we examined the capacity of the LHC to reconstruct the mean lifetime of long-lived particles. At first, we investigated an ideal situation in which an LLP decays inside the tracker into an all-visible final state and its four momenta, as well as the position of the secondary vertex, can be precisely measured. We highlighted the fact that the experimental cuts on the displacement of the secondary vertex introduce a bias in the reconstructed LLP lifetime and proposed a method to overcome it. We saw that, under these idealised conditions, following this method it is indeed possible to measure, or at least to put a lower limit on, the lifetime of the LLP to a good degree. 

As a second step, we studied a scenario involving decays of an LLP into an all-visible final state and introduced $p_T$ smearing on the daughter particles. We found that for highly boosted final state particles, the effects of smearing are negligible on the lifetime measurement. However, for less boosted final states, momentum smearing will have a more severe impact on the measurement.

There are several layers of complication that can, and \textit{should}, be added to this first study. In particular, having computed $\tau_i$ from the ratio between $\beta_i \gamma_i c \tau_i$ and $\beta_i \gamma_i c$, it is important to properly consider the error propagation corresponding to this ratio. Moreover, in several New Physics scenarios involving long-lived particles, at least one of the LLPs can have invisible decay products along with --at least -- two visible ones (the minimum number of visible decay products required to reconstruct the secondary vertex is two). In such cases, the LHC experiments can reconstruct the secondary vertex (our $\beta_i \gamma_i c \tau_i$) for each event, but not the exact four-momenta ($\beta_i \gamma_i$) of the LLPs. In such scenarios, some observables like $M_{T2}$ may provide handles on the mass of the LLP, making it possible to extract its four-momenta as well. 
Besides, here we have not considered effects related to the initial state particles. The shape of the $\beta \gamma$ spectra will be different depending on whether the process is gluon-gluon, quark-anti-quark or quark-gluon initiated. Such effects may affect the lifetime measurement in a substantial manner.

\subsection*{Acknowledgements}

We would like to thank the organizers of the 2017 ``Les Houches – Physics at TeV colliders'' workshop where this work was initiated.
S.B.\ thanks Shilpi Jain for helpful discussions regarding fitting in \texttt{ROOT}. The work of B.H.\ and S.B.\ are partially supported by Investissements d'avenir, Labex ENIGMASS, contrat ANR-11-LABX-0012. The works of S.B.\ and B.B.\ are partially supported by the CNRS LIA-THEP (Theoretical High Energy Physics) and the INFRE-HEPNET (IndoFrench 
Network on High Energy Physics) of CEFIPRA/IFCPAR (Indo-French Centre for the Promotion of Advanced 
Research). The work of S.B.\ is also partially supported by a Durham Junior Research Fellowship COFUNDed
between Durham University and the European Union under grant agreement number 609412. The work of B.B.\ is 
also supported by the Department of Science and Technology, Government of India, under the Grant 
Agreement number IFA13-PH-75 (INSPIRE Faculty Award). A.G.\ was supported by the Labex ILP (reference ANR-10-LABX-63) part of the Idex SUPER, and received financial state aid managed by the Agence Nationale de la Recherche, as part of the programme Investissements d’avenir under the reference ANR-11-IDEX-0004-02. The work of D.S.\ is supported by the National Science Foundation under Grant No. 1519045.



\AddToContent{S.~Banerjee, D.~Barducci, B.~Bhattacherjee, A.~Goudelis, B.~Herrmann, D.~Sengupta}
\renewcommand{\thesection}{\arabic{section}}

\graphicspath{{DMST/}}

\newcommand{\bjet}{\ensuremath{b}-jet}
\newcommand{\btagged}{\ensuremath{b}-tagged}

\chapter{Associated production of a single top with dark matter in a
  two-Higgs-doublet plus pseudoscalar mediator model}

{\it G.~B\'elanger, B.~Fuks, F.~Maltoni, J.~M.~No, P.~Pani, G.~Polesello, B.~Zaldivar}


\begin{abstract}
This proceeding aims to extend the studies already available in literature on
the associated production of a single top quark and dark matter in a model with
two Higgs doublets and a pseudoscalar mediator. We put the model into a broader
context of astrophysical quantities and measurements and we extend the
interpretation in terms of selections and parameter space.
\end{abstract}


\section{Introduction}

The sensitivity of the LHC experiments to the associated production of dark
matter (DM) with a single top has been recently studied \cite{Pani:2017qyd} in the
framework of an extension of the standard model featuring two Higgs doublets
and an additional pseudo-scalar
mediator \cite{Bauer:2017fsw,Bauer:2017ota} (2HDM+$a$). The study in
Ref.~\cite{Pani:2017qyd} and the one of this contribution extend the results previously
available
in the literature \cite{Pinna:2017tay,Arcadi:2017wqi,Plehn:2017bys}, that
show the importance of final states involving a single top quark
and DM (DM$t$) by means of a simplified model.
Like single top production within the Standard Model (SM), the DM$t$ signature in the
2HDM+$a$ model receives three different types of contributions at leading order (LO) in QCD. These are $t$-channel production, $s$-channel production and
associated production together with a $W$ boson ($tW$). The presence of an extended Higgs sector, contrary to the case of a singlet mediator, 
ensures perturbative unitarity of the $pp \rightarrow t\chi \bar \chi
+X$ class of processes where $\chi$ denotes the DM particle. Moreover, the fact
that $tW\chi\chi$ final states can be produced with sizeable cross sections
in diagrams involving the on-shell production of intermediate charged Higgs
bosons renders the phenomenology of this model particularly interesting. In
this contribution to the Les Houches proceedings, we aim to extend the study of
Ref.~\cite{Pani:2017qyd} by broadening the interpretation of the model in terms of
collider searches. Furthermore, we include the estimation of the sensitivity of
direct dark matter experiments on the parameter space and we propose new selection strategies for
analyses targeting events with a single lepton in the final states and assess the parameter-space
coverage of these new selections assuming an integrated luminosity of
300~fb$^{-1}$ of proton-proton collisions at a centre-of-mass energy of
$14\,\TeV$.

\section{Cosmological constraints} 
The model in question contains a fermionic DM candidate $\chi$ that is
assumed to be a Weakly Interacting Massive Particle (WIMP). The DM particle
communicates with the SM via exchanges with a pseudo-scalar mediator that mixes with
the Higgs sector and it thus couples to all SM particles. Then,
depending on the spectrum and the mixing angles, the relic abundance
of DM will be dominated by DM annihilation into pairs of $b$-quarks, top quark
pairs or pairs of mediators $a$, as well as in final states comprised of a $ah$
or $aZ$ system, $h$ and $Z$ respectively denoting the SM Higgs boson and
$Z$-boson. The latter
channels represent the most important difference between this model
and the corresponding simplified-model implementation, which only
couples the mediator to the fermionic sector of the SM. Consequently,
the prediction of the relic abundance will be very different from one
model to another, in the relevant region of the parameter space where
those new channels dominate. This region is defined by the condition
$2m_\chi\gtrsim (m_a+m_{h,Z})$. For more details on the relic density
predictions of the model, we refer the reader to Ref.~\cite{Goncalves:2016iyg}.

On the other hand DM indirect detection constraints are potentially
very important. For a pseudo-scalar mediator, DM annihilation into a fermion pair
consists in an $s$-wave process and there is thus no velocity suppression of the
cross section at present time. On the other hand, the $\chi\bar\chi\to aa$
annihilation channel is $p$-wave suppressed, but the $\chi\bar\chi\to
ah$ and $aZ$ ones are not and the relative importance of the latter on the
cosmological consequences for the model have not been addressed.
For the region of parameter space where the fermionic channels dominate,
predictions for the DM annihilation cross section at present times coincide with
those estimated in the simplified model context. We therefore refer the reader
to Ref.~\cite{Banerjee:2017wxi} for a detailed analysis of the indirect
detection constraints in the 2HDM<+$a$ framework. 

On different lines, DM direct detection searches are not relevant since the
strongest constraints from that front come from the presence of spin-independent
interactions in the theory that our model does not feature at tree level. As in
contrast, the featured DM spin-dependent interactions lead to a velocity
suppression allowing to evade any direct detection bound.

\section{Dedicated LHC analysis for 300~fb$^{-1}$} 
A detailed analysis explicitly optimised for the signature depicted above, and
targeting the full projected statistics of the LHC Run 3 of 300~fb$^{-1}$
is described in Ref.~\cite{Pani:2017qyd}.
We  briefly summarise in the following the methods
and the results of this analysis, and we extend it by projecting
the results on additional slices of the parameter
space. We moreover include signatures that are relevant and that have been
neglected before.

\subsection{Summary of the available LHC projection}
We base the simulation of the DM signal and the SM background following the
strategy introduced in Ref.~\cite{Pani:2017qyd}. We simulate a full set of SM processes
leading to the presence of one or two final-state leptons ($e$, $\mu$)
originating from the decay of a $W$-boson, a $Z$-boson or a $\tau$ lepton.
Events are generated within either the {\tt POWHEG~BOX}
framework~\cite{Alioli:2010xd} or the {\tt MadGraph5\_aMC@NLO}
platform~\cite{Alwall:2014hca}, and the simulation of the QCD environment
(parton showering and hadronisation) has been performed with
{\tt PYTHIA 8}~\cite{Sjostrand:2014zea}.
Hard-scattering signal events have been produced with {\tt MadGraph5\_aMC@NLO}
on the basis of the UFO model~\cite{Degrande:2011ua} provided together with
Ref.~\cite{Bauer:2017ota}, and parton showering and hadronisation have been
again simulated with {\tt PYTHIA 8}.
We finally include detector effects by smearing the properties of
the final-state physics objects ({\it i.e.} electrons, muons, jets and
$\etmiss$) in a way reproducing the measured performance of the
ATLAS detector.

We specifically focus on a final state comprised of two $W$-bosons, a jet
issued from the fragmentation of a $b$-quark ({\it i.e.} a $b$-jet), and 
missing transverse momentum associated with the presence of the two undetected
DM particles. 
Two analyses have been developed, requiring either
one or two leptons in the final state, corresponding to the  the cases
where only one or both of the two $W$ bosons decays into leptons respectively.
The selection criteria are based on constraints imposed on a set of dimensionful
transverse variables, such as the transverse mass $\mtlep$ \cite{Patrignani:2016xqp}, $\mttwo$
\cite{Lester:1999tx,Barr:2003rg}, and $\amttwo$ \cite{Konar:2009qr,Lester:2014yga}.
All those variables exhibit a kinematic endpoint in the SM context, when all the
missing energy arises from the neutrinos originating from $W$-boson decays. In
the signal case, the presence of additional $\etmiss$ induced by the
DM particles violates these bounds, and provides a handle 
for the separation of  the signal from the background.
Prior studies
(see Fig.~8 of Ref.~\cite{Pani:2017qyd}) have determined the regions of the parameter
space (presented in the $(m(H^{\pm}), \tan\beta)$ plane) that can be excluded at
the 95\% confidence level (CL), for an integrated luminosity of
$300 \, {\rm fb}^{-1}$ of proton-proton collisions at a centre-of-mass energy of
$14 \, {\rm TeV}$ LHC. The results are given for a fixed mediator mass of $m(a)=150$~GeV. The
two-lepton signature, which has lower signal statistics but allows for a softer
kinematic selection, dominates for lower $m(H^{\pm})$ masses, whereas the
single-lepton signature allows to get sensitivity to charged-Higgs masses above
$m(H^{\pm})=600$~GeV.

\subsection{Reinterpretation of the LHC reach}
The analysis in Ref.~\cite{Pani:2017qyd} concentrated on assessing the dependence of the
LHC reach on the main model parameters $m(H^{\pm})$, $\tan\beta$ and $m(a)$.
It is however interesting to compare the parameter space which can
be covered by different analyses, as it can help prioritising
the search strategy at the LHC. A rather complete survey
is provided in Ref.~\cite{Bauer:2017ota}, in which the coverage expected from
the analysis of different signatures is presented in the $(m(a)$, $\tan\beta)$
plane for four different model configurations. Whilst for maximal mixing
of the singlet and doublet pseudo-scalars ({\it i.e.} $\sin\theta\sim0.7$) the
dominant model signature consists of the production of a single Higgs boson
recoiling against a pair of DM particles, mono-$Z$-boson probes become
predominant for cases in which the $a$ boson is mostly doublet-like
($\sin\theta\sim0.35$).
\begin{figure}
\begin{center}
\includegraphics[width=0.48\textwidth]{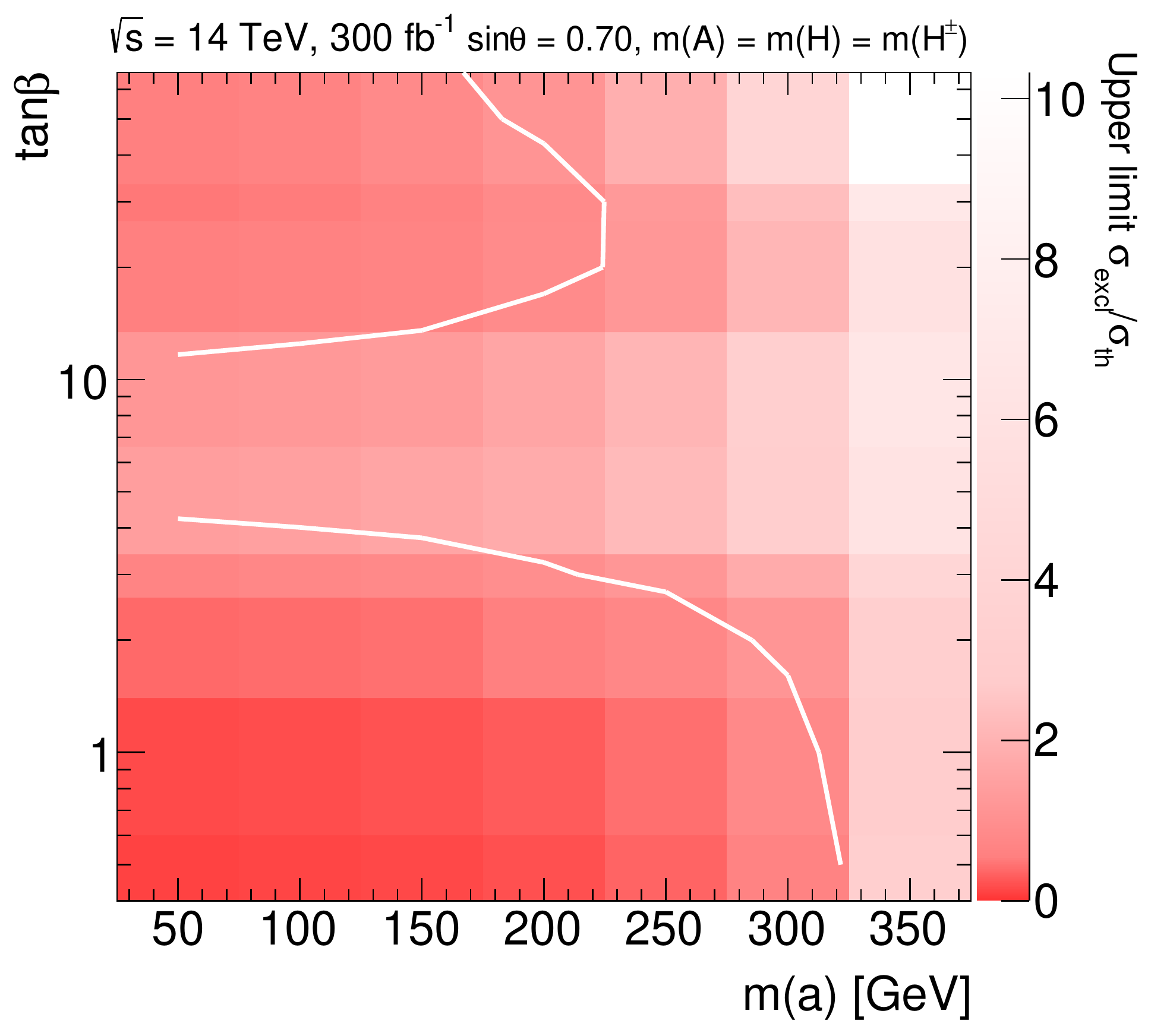}
\includegraphics[width=0.48\textwidth]{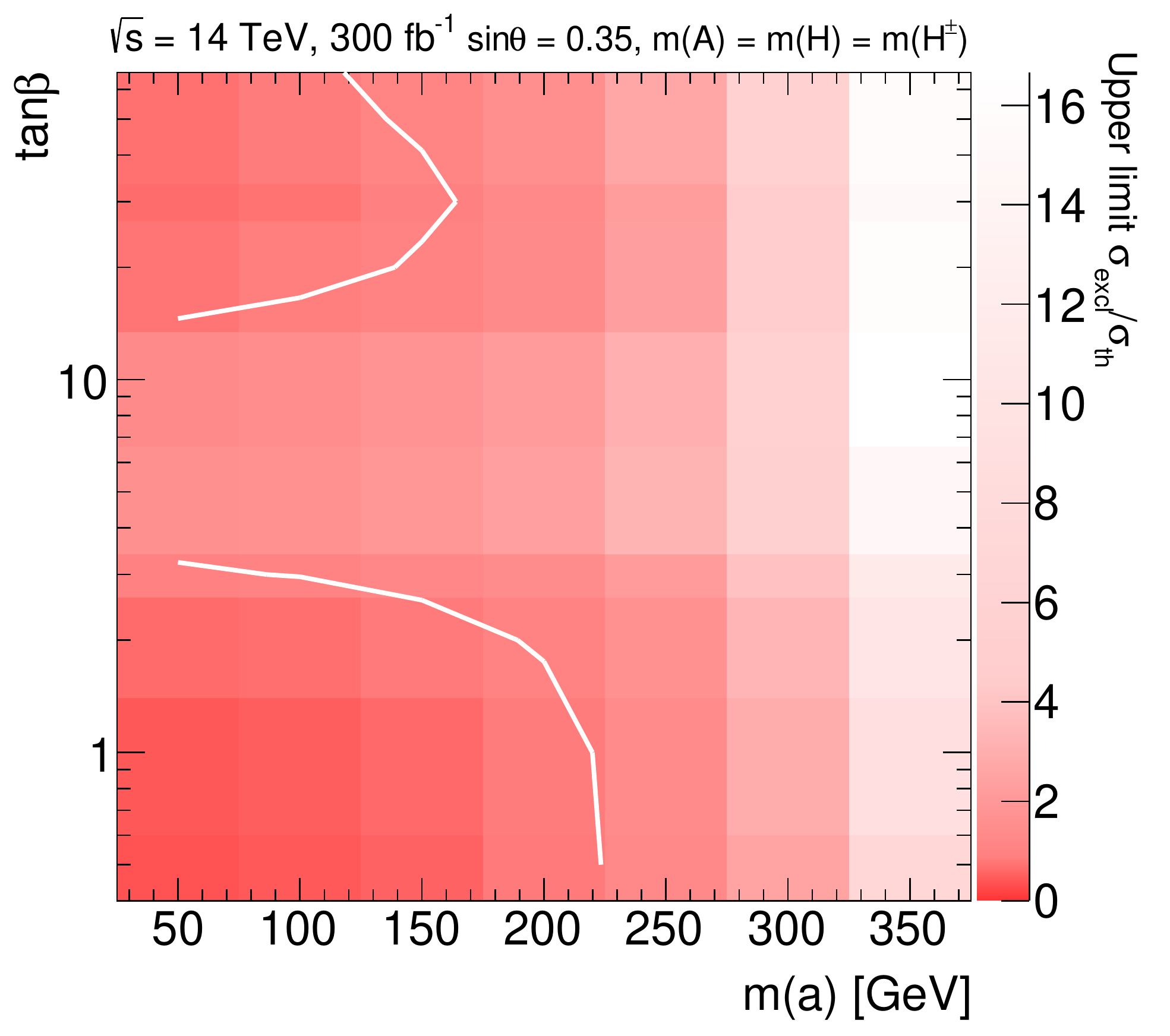}
\caption{Regions in the $(m(a), \tan\beta)$ planes which can be excluded at
  the 95\%~CL through single-lepton and di-lepton searches. In the left panel,
  the singlet-doublet pseudo-scalar mixing is maximal ($\sin\theta\sim0.7$)
  whilst in the right panel it is smaller ($\sin\theta\sim0.35$).
  The $z$-axis palette shows the ratio between the excluded cross section and the theoretical one. 
  For both cases, we have fixed $m(A)=m(H^{\pm})=m(H)=500$~GeV and the results
  assume $300 \, {\rm fb}^{-1}$ of $14 \, {\rm TeV}$ LHC data and a systematic
  uncertainty of 20\% on the SM background and of 5\% on the signal.}
\label{2HDMST:95clmatb}
\end{center}
\end{figure}
We recast in Figure~\ref{2HDMST:95clmatb} the reach of the di-lepton analysis
described in Ref.~\cite{Pani:2017qyd} and present it in the $(m(a), \tan\beta)$ plane,
for $m(A)=m(H^{\pm})=m(H)=500$~GeV and for the two mixing scenarios
discussed above. By comparing these results with the sensitivity projections
for the same integrated luminosity shown in Ref.~\cite{Bauer:2017ota}, we find
out that the associated production of a single top and a charged Higgs boson
$H^{\pm}$ covers a region of parameter space comparable to the one for which the
mono-$Z$ and mono-$h$ probes are sensitive to.

\subsection{Additional handles on the model}

\begin{figure}
\includegraphics[width=.48\textwidth]{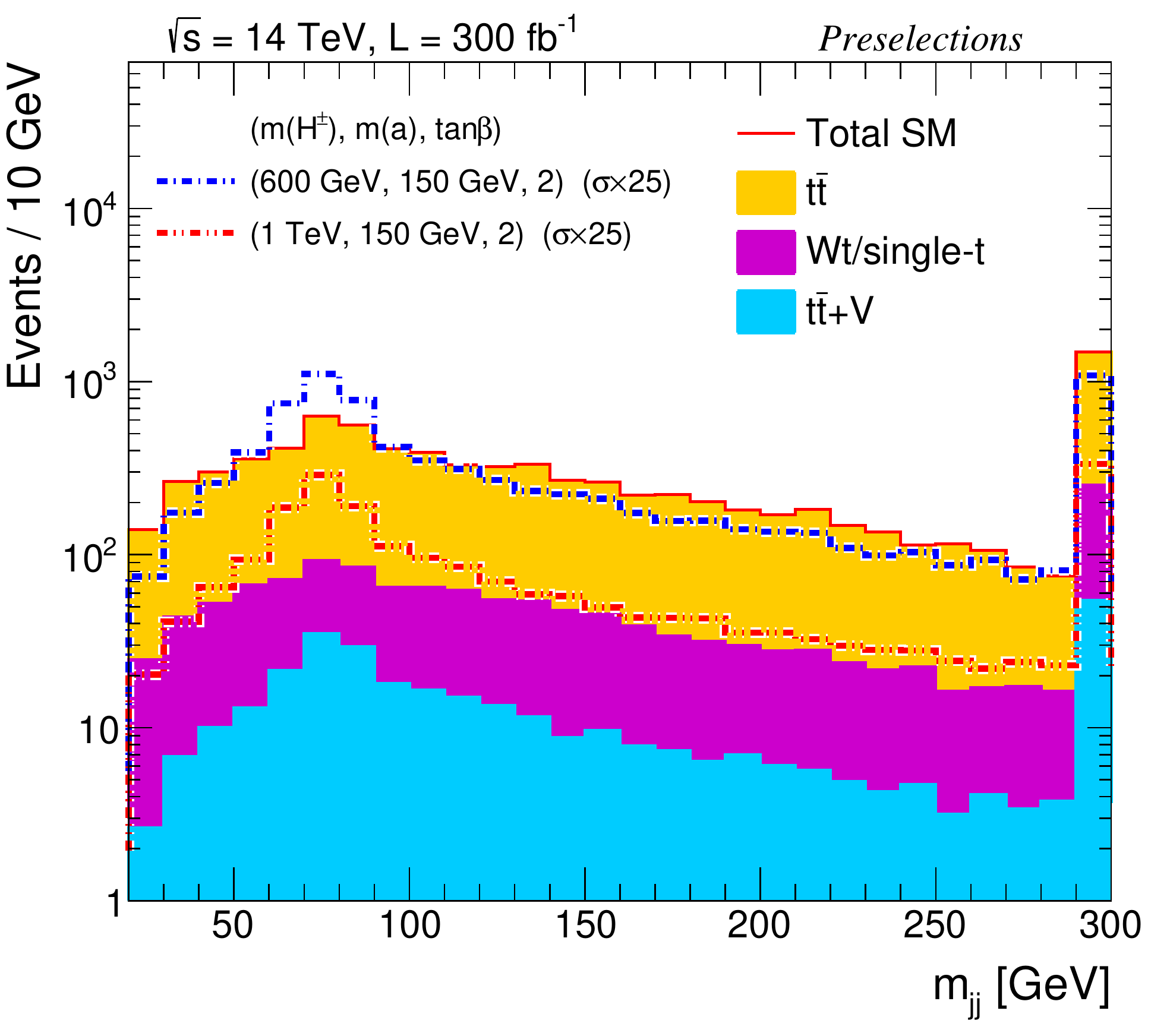}
\includegraphics[width=.48\textwidth]{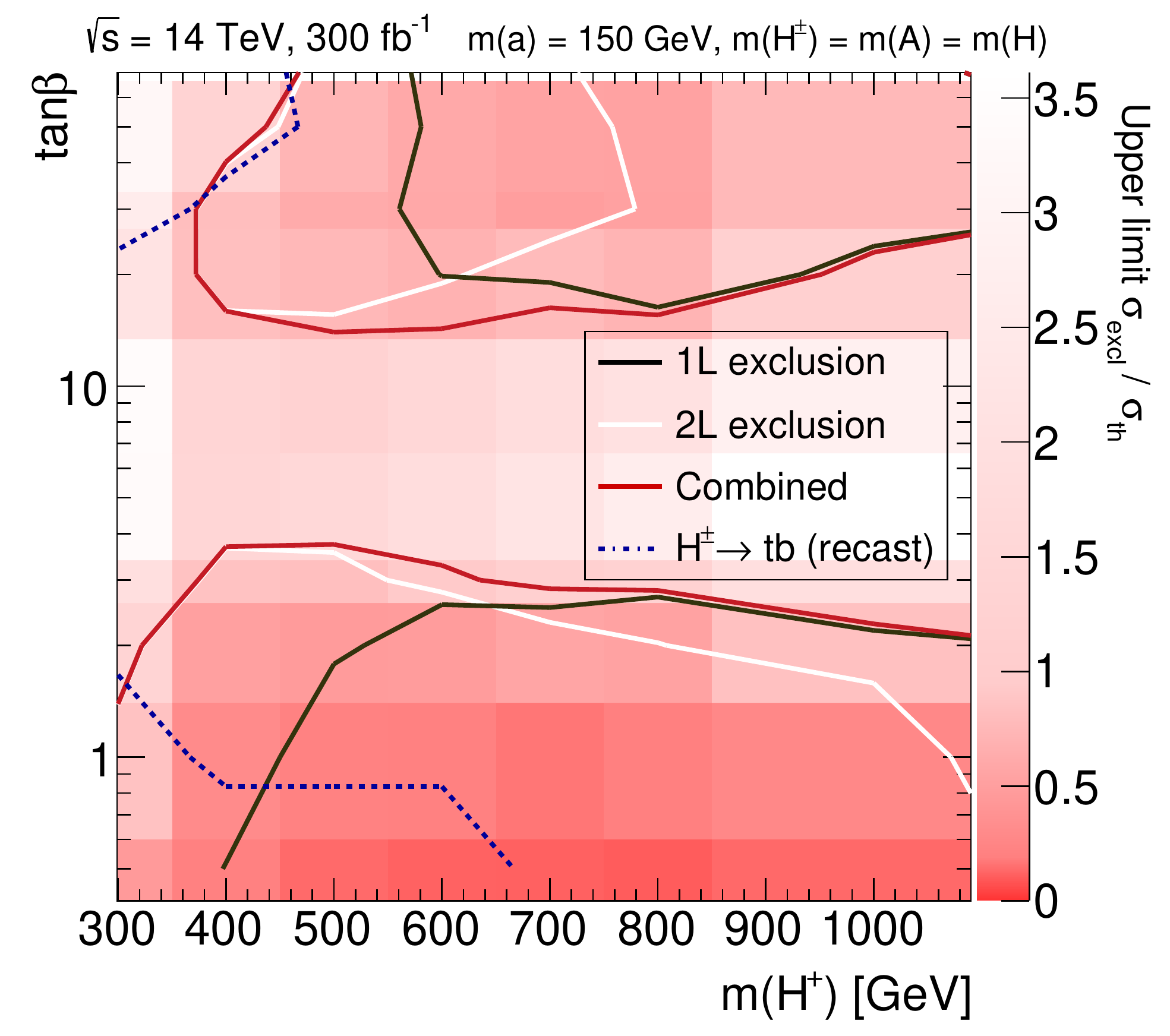}
\caption{(Left) Distribution of the hadronic
$W$-boson mass for a relaxed version of the analysis
selections. (Right) 95\%~CL exclusion limits for the combination
of the two one-lepton regions described in the 
text and the two-lepton selections of Ref.~\cite{Pani:2017qyd}.
  We assumes $300 \, {\rm fb}^{-1}$ of $14 \, \TeV$ LHC data and a systematic  uncertainty of 20\% (5\%) on the SM background (signal).}
\label{2HDMST;mWkinematic}
\end{figure}

A limitation of the analysis described in Ref.~\cite{Pani:2017qyd} lies in the fact 
that the selections for the single-lepton analysis are focused 
on a situation where the lepton is produced in the decay 
chain of the charged Higgs boson $H^{\pm}$, which implies a
hadronically-decaying top quark recoiling against a lepton-$\etmiss$ system that
is very hard. The selection cuts explicitly exploit this situation, killing in
this way both the topology where the lepton originates from the top quark decay
and the $t$-channel contribution, which becomes 
relevant at high $H^{\pm}$ masses.
In order to increase the LHC sensitivity to the model, we develop a new selection
strategy, focusing on the situation where the lepton arises from the decay of
the top quark. 
Two non-overlapping signal selections are defined, respectively targeting the
$H^\pm \rightarrow \ell\nu\chi\chi$ and $H^\pm \rightarrow jj\chi\chi$ decays.
They will be referred to in the following as leptonic-$H^\pm$ and hadronic-$H^\pm$ selections, respectively. 
In the leptonic-$H^\pm$ selection, we require that the system comprised of the lepton and the
leading  \bjet\ has a large invariant mass, $m(b1,\ell) > 150$~GeV, and that the
system comprised of the leading light jet and leading \bjet\ has a small
invariant mass, $m(b1,j1) < 150$~GeV. In the hadronic-$H^\pm$ selections, we in contrast
require that $m(b1,\ell) < 150$ GeV and $m(b1,j1) > 130$ GeV,
 where the former cut ensure the orthogonality between the regions. We moreover
improve the background rejection by requesting that the invariant mass of the
system comprised to the two leading light jets lies between $60$ and $105$~GeV.
This requirement, applied in both selections, is extremely efficient in
separating the signal from the background, as shown in the left panel of
Fig.~\ref{2HDMST;mWkinematic}.

We impose that all reconstructed jets with a transverse momentum
$p_\mathrm{T}^j >25 \, \GeV$ and a pseudorapidity satisfying $|\eta_{j}|<2.5$
are well separated from the missing momentum, requiring that the jet that is the
closest to the $\ptmiss$ in azimuth satisfies $|\Delta\phi_{\rm min}|>1.0$.
At least one jet is required to be \btagged\ and events with a second
\btagged\ jet with $\pt > 50\,\GeV$ are rejected.
The asymmetric
stransverse mass,  $\amttwo$, is required to be at least of 230 (200) GeV
in the leptonic-$H^\pm$ (hadronic-$H^\pm$) selection. 
The rest of the selections is designed to
exploit in the signal the fact that the boost of the charged Higgs is
either propagated to the leptonic or to the hadronic part of the event.

In the leptonic-$H^\pm$ selection, we additionally
require three jets of relatively low transverse momenta ($\pt > 50,50,20$ GeV)
and exactly one isolated lepton ($e$ or $\mu$) with 
$\pt > 120$ GeV, $|\eta_{\ell}|<2.5$. The missing energy is imposed to be larger than
250~GeV and the lepton-$\etmiss$ system transverse mass has to satisfy
$\mtlep > 300$ GeV. In addition, we impose that
$\etmiss+0.4\cdot\mtlep > 375$ GeV. 

For the hadronic-$H^\pm$ selection
we instead require the presence of three relatively hard jets
($\pt > 100,50,40$ GeV)
and exactly one isolated lepton with low transverse momentum $\pt >
25$ GeV. The requirement on $\mtlep$ is kept at the minimum value
needed to suppress the background contributions of semi-leptonic \ttbar\
decays, $\mtlep > 120$ GeV. 
Finally the events are selected if $\etmiss > 400$ GeV. As in this
topology the $\etmiss$ and the hadronically decaying $W$-boson
come from the decay of the same particle, it is convenient to define
a proxy for the $H^\pm$ transverse mass as the
invariant mass on the transverse plane of the hadronic $W$-boson,
constructed from the two leading light jets, and the $\etmiss$. This
variable is required to be of at least $680$ GeV, in order to enhance the
sensitivity of this selection to charged Higgs bosons of about 1~TeV.

The total background in the leptonic-$H^\pm$ selection amounts for approximately 4
events, dominantly arising from $tt+V$ and $tZ$ production.
For charged Higgs masses ranging from 
$500 \, \GeV $ to $1\, \TeV$ the signal acceptance is of $[0.12, 0.35]\%$
($[0.12, 0.31]\%$) for $m(a) = 150\, \GeV$ and $\tan\beta = 1 \, (20)$.
In the hadronic-$H^\pm$ selection, the total background is expected to be approximately 2
events, also dominantly arising from $tt+V$ and $tZ$ production.
For charged-Higgs masses ranging again from
$500 \, \GeV $ to $1\, \TeV$ the signal acceptance
is found to be $[9\cdot 10^{-5}, 0.17]\%$ ($[9\cdot 10^{-5}, 0.16]\%$) for $m(a) = 150\, \GeV$ and 
$\tan\beta = 1 \, (20)$.
The corresponding exclusion limits are presented in the
right panel of Fig.~\ref{2HDMST;mWkinematic} and compared with the di-lepton
selection from Ref.~\cite{Pani:2017qyd}. We observe that the new selection extends
the reach of the analysis towards high $H^\pm$ masses. However, the
improvement is tamed by the strong
kinematic similarity of the hadronic-$H^\pm$ signal to the
background, which can only be suppressed by very aggressive selection
requirements that are characterised by a low signal efficiency.

\section{Conclusions}

The prospects of future LHC runs for probing
interactions between a DM particle and top quarks via the $t + \etmiss$
signature have been studied. We observed that the sensitivity of the single top
signal is complementary to the one of the mono-Higgs and mono-$Z$-boson probes,
once we include searches targetting a single-leptonic and di-leptonic final
state. In particular, thanks to a dedicated optimisation of the
analysis focusing on the single-lepton final state, we have shown that the LHC
is in principle sensitive to charged Higgs bosons of about $1$ TeV, for a
large range of $\tan\beta$ values and assuming an integrated luminosity of
$300$fb$^{-1}$.




\AddToContent{G.~B\'elanger, B.~Fuks, F.~Maltoni, J.~M.~No, P.~Pani, G.~Polesello, B.~Zaldivar}
\renewcommand{\thesection}{\arabic{section}}

\graphicspath{{nmfv/}}


\def\xutchi{BR($\tilde{u}_1 \to t \tilde{\chi}^0_1$)}
\def\xucchi{BR($\tilde{u}_1 \to c \tilde{\chi}^0_1$)}

\def\met{\rm E{\!\!\!/}_T}
\def\PET{\rm p{\!\!\!/}_T}

\def\tcb{\textcolor{blue}}
\def\tcr{\textcolor{red}}


\def\ptl{\ensuremath{\vec p^{\mathrm{\ \ell}}_\mathrm{T}\xspace}}
\def\dphimin{\ensuremath{\Delta\phi_{\mathrm{min}}}}
\newcommand{\pbll}{\ensuremath{p_{\mathrm{b}}^{ll}}\xspace}
\newcommand{\ptll}{\ensuremath{p_{\mathrm{T}}^{ll}}\xspace}
\newcommand{\ttbarW}{\ensuremath{t\bar{t}W}\xspace}
\newcommand{\ttbarZ}{\ensuremath{t\bar{t}Z}\xspace}
\newcommand{\mTlep}{\ensuremath{m_\mathrm{T}^{lep}}\xspace}
\newcommand{\Wjets}{\ensuremath{W}+jets\xspace}
\newcommand{\Zjets}{\ensuremath{Z}+jets\xspace}
\newcommand{\W}{\ensuremath{W}}




\chapter{Probing flavour-violating decays of squarks at the LHC}

{\it A.~Chakraborty, M.~Endo, B.~Fuks, B.~Herrmann, M.~M.~Nojiri, P.~Pani and G.~Polesello}


%
%


\begin{abstract}
We study squark decays beyond minimally flavour-violating supersymmetry at the
LHC. Considering second and third generation squark mixings, we consider a
simplified model with two active squark flavours and evaluate the sensitivity
of current squarks searches at the Run-1 and Run-2 of the LHC. We moreover
investigate the gain in sensitivity of a dedicated search strategy involving
leptons, jets and missing transverse energy at the high luminosity run of LHC.
\end{abstract}


\section{Introduction}
\label{lhcnmfv_sec:intro}

Despite the absence of any experimental evidence at the Large Hadron Collider (LHC), from a theoretical and phenomenological point of view supersymmetry (SUSY) remains an attractive extension of the Standard Model (SM) of particle physics.
Whereas current experimental searches are unfruitful, supersymmetry can still be
viable either after introducing heavy superpartners, or by considering
non-minimal realisations of the theory. While the former explanation is rather unattractive from the phenomenological point of view, the latter one calls for studies of supersymmetric frameworks beyond the ``usual'' Minimal Supersymmetric Standard Model (MSSM).

Without adding extra fields or extending the gauge symmetry groups, going beyond
the ``standard'' MSSM can be achieved by considering additional mixing between the states of the model. More precisely, in addition to the helicity mixing between ``left-handed'' and ``right-handed'' states, the model symmetries allow for
inter-generational mixing of the scalar partners beyond the commonly assumed Cabibbo-Kobayashi-Maskawa (CKM)- and Pontecorvo-Maki-Nakagawa-Sakata (PMNS)-induced terms. In the following, we consider this non-minimally flavour-violating (NMFV) framework and assume additional mixing between the second and third generation of squarks. The latter is indeed allowed to be sizeable, in contrast to any
mixing involving the first generation that is largely constrained by flavour
data. In particular, at that stage of our study, we focus on mixing between charm and top flavours in the squark sector. 

From the model-building point of view, non-minimal flavour violation can be motivated from Grand Unified Theories, possibly in combination with certain flavour symmetries at the high scale (see, {\it e.g.} Refs.~\cite{Dimou:2015yng, Dimou:2015cmw}). In the present study, however, we introduce the corresponding terms directly at the TeV scale. More precisely, we consider the generation-mixing entries of the sfermion mass matrices as free parameters, in the same way as the diagonal entries in such a phenomenological TeV-scale setup. Although there are stringent constraints on the flavour-violating neutral currents induced by these additional generation-mixing entries, a considerable part of the resulting parameter space is in agreement with theoretical constraints and current experimental measurements \cite{DeCausmaecker:2015yca}. Moreover, the additional charm-stop mixing leads to characteristical signatures at the LHC \cite{Hurth:2009ke, Bruhnke:2010rh, Bartl:2010du, Bartl:2011wq, Bartl:2012tx,Backovic:2015rwa,Crivellin:2016rdu}. A particular feature is that, if the lightest up-type squark is a mixture of charm and top flavour, its decays into charm and top quarks, together with a neutralino, may be simultaneously open \cite{Bartl:2010du}. 

Experimental ATLAS and CMS searches have led to stringent limits on the
production cross sections of charmed and top squarks~\cite{Aad:2015gna,%
Aaboud:2017aeu, Sirunyan:2017xse, Aaboud:2017phn, Sirunyan:2017pjw,%
Sirunyan:2017leh, Sirunyan:2017kiw, Aaboud:2017ayj}. However, these searches are
mainly based on simplified models assuming no generation-mixing entries in the squark mass matrices. As adding such entries may altern the decay pattern of the squarks, it is an interesting question to evaluate their impact on the present mass limits and evaluate the sensitivity of new dedicated searches at the LHC. 
This is the goal of the present work. Within an {\it ad-hoc} simplified model containing two squark flavours, we recast mass limits from previous ATLAS and CMS
analyses. We then estimate the sensitivity of dedicated searches at the LHC
with a centre-of-mass energy of $\sqrt{s} = 14$ TeV and assuming 300 fb$^{-1}$ of integrated luminosity.

\section{Model setup and existing LHC limits}
\label{lhcnmfv_sec:model}

In this section, we present the general setup of our study. We first introduce
the simplified model which we base our analysis on. We then discuss the adopted search strategy and recast recent squark searches at the LHC.

\subsection{A simplified model for squark flavour violation}
\label{lhcnmfv_SecModel}

While in the general MSSM, each squark eigenstate is an admixture of the six
flavours eigenstates (see Ref.\ \cite{DeCausmaecker:2015yca}), the present
analysis is based on a simplified model capturing the essential features of
non-minimal flavour violation in the squark sector once existing bounds are
accounted for. In our setup, the squark sector consists of two active flavours,
namely a right-handed stop and a right-handed scharm, and their mixing leads to
two physical eigenstates $\tilde{u}_1$ and $\tilde{u}_2$ defined by
\begin{equation}
	\begin{pmatrix} \tilde{u}_1 \\ \tilde{u}_2 \end{pmatrix} =
	\begin{pmatrix} ~~\cos\theta_{tc} & \sin\theta_{tc} \\ -\sin\theta_{tc} & \cos\theta_{tc} \end{pmatrix}
	\begin{pmatrix} \tilde{c}_R \\ \tilde{t_R} \end{pmatrix} \ .
\end{equation}
Here, $\tilde{u}_1$ is assumed to be the lighter of the two squark mass
eigenstates. In addition, we include one neutralino $\tilde{\chi}^0_1$ whose
mass is fixed to $m_{\chi^0_1} = 50$~GeV and composition\
taken bino-like\footnote{These assumptions do not have a significant impact on the analysis, the main ingredient being the considered branching fractions of the squarks into quarks and the neutralino.}. The rest of the spectrum is assumed to be decoupled and ignored in the following. Our simplified setup is thus governed by three parameters: the masses $m_{\tilde{u}_1}$ and $m_{\tilde{u}_2}$ of the squarks together with the squark mixing angle $\theta_{tc}$. This choice of a
mixing between the ``right-handed'' squark flavours allows for more flexibility,
as less affected by constraints originating from $B$-physics, contrary to a
mixing between ``left-handed'' states.

Being an admixture of $\tilde{t}_R$ and $\tilde{c}_R$ eigenstates, each physical
squark may decay either into a top or a charm quark,
\begin{equation}
	\tilde{u}_i \to t \tilde{\chi}^0_1 \,, \qquad \tilde{u}_i \to c \tilde{\chi}^0_1 \qquad\text{with}\ \ i=1,2\ .
\end{equation}
Once pair-produced at the LHC, the pair-produced squarks hence give rise to two
final-state neutralinos manifesting themselves as missing transverse energy
($\etmiss$). Assuming squark-antisquark production at the LHC, typical existing
search strategies focus on the processes
\begin{equation}
  pp \to t\bar{t} + \etmiss \qquad\text{and}\qquad
  pp \to c\bar{c} + \etmiss \,.
\end{equation}
However, after considering additional flavour mixing as introduced above, another process may be relevant,
\begin{equation}
	pp \to tc + \etmiss \to \ell b c + \etmiss \,,
\end{equation}
where one squark decays into a top and the other into a charm
quark~\cite{Bartl:2010du} and the final step assumes leptonic top decay. We hence propose to
target a final state comprised of one isolated lepton, one $b$-tagged jet, one
$c$-tagged jet, and a large amount of missing transverse energy.

\begin{figure}[!htb]
	\begin{center}
	\includegraphics[width=.45\textwidth]{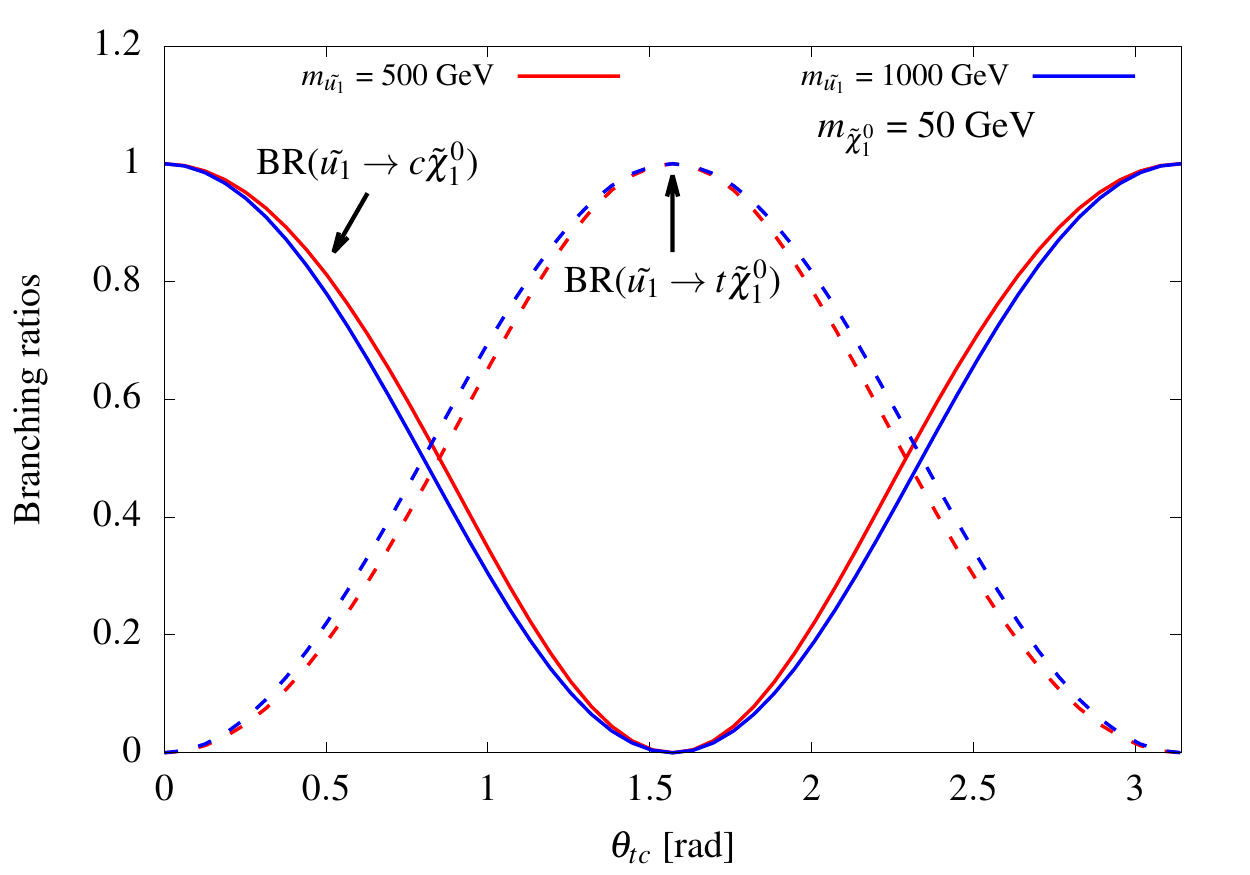}
	\includegraphics[width=.45\textwidth]{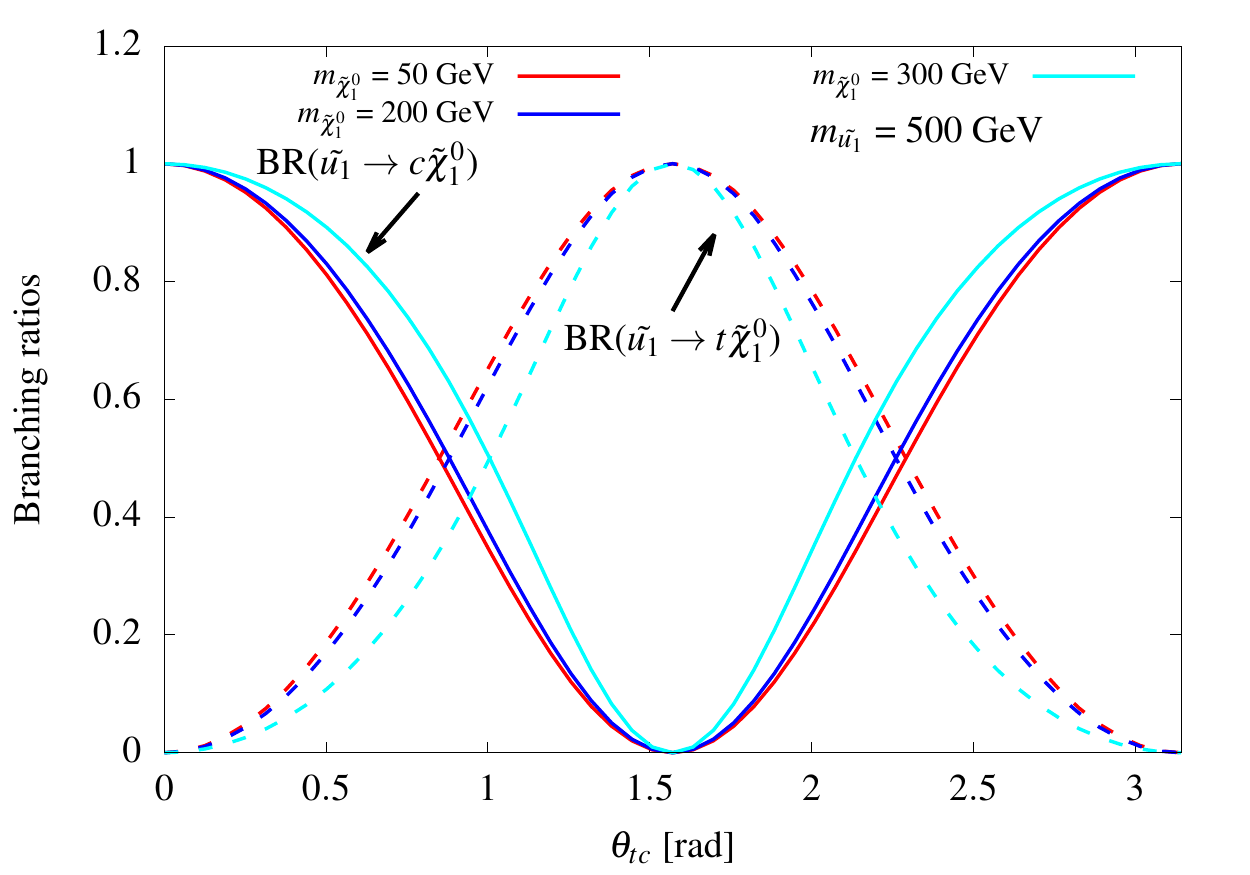}
	\end{center}
	\vspace*{-5mm}
	\caption{Branching ratios of the $\tilde{u}_1 \to t \tilde{\chi}^0_1 $
  (dashed line) and $\tilde{u}_1 \to c \tilde{\chi}^0_1$ (solid line) decays,
  shown as functions of the mixing angle $\theta_{tc}$. In the Left figure, the red and blue curves correspond to squark masses of $m_{\tilde{u}_1} = 500$ GeV and $m_{\tilde{u}_1} = 1000$ GeV, respectively with a fixed $m_{\chi^0_1} = 50$~GeV. In the right plot, we fix $m_{\tilde{u}_1}$ to 500 GeV and vary $m_{\chi^0_1}$ for three different values 50 GeV (red), 200 GeV (blue) and 300 GeV (cyan).}
	\label{lhcnmfv_fig:bratio}
\end{figure}

We present the two relevant branching ratios as function of the generation
mixing in Fig.~\ref{lhcnmfv_fig:bratio}. The branching fractions are found to
vary significantly as a function of the value of the mixing angle $\theta_{tc}$. In
particular, we can obtain a situation where both decay modes have comparable
rates. There exists another key parameter which controls these branching ratios which is 
the mass difference between the squark and the neutralino. In the right panel of Fig.~\ref{lhcnmfv_fig:bratio} 
we show the dependence of the branching ratios on the mass of the neutralino by setting $m_{\chi^0_1}$ to three different values 50 GeV (red), 
200 GeV (blue) and 300 GeV (cyan) for a fixed $m_{\tilde{u}_1}$ = 500 GeV. We observe that, even though the 
impact on the squark mass itself is marginal, the branching ratios can vary significantly with the the mixing angle as soon as the mass difference between the squark and neutralino becomes close to the top mass threshold. In the rest of our analysis, the neutralino mass is held fixed at 50 GeV. 
%
\subsection{Recasting LHC limits}

Dedicated searches for scalar top quarks decaying into a $t\bar t ~+~\etmiss$
system or scalar charm quarks decaying into a $c\bar c ~+~\etmiss$ system have
been performed by both the ATLAS and CMS collaborations at the LHC
\cite{Aad:2015gna, Aaboud:2017aeu, Sirunyan:2017xse, Aaboud:2017phn, Sirunyan:2017pjw, Sirunyan:2017leh, Sirunyan:2017kiw, Aaboud:2017ayj}. The observed exclusion limits in case of light neutralino masses are of the order of the TeV for top squarks and of about $500$ GeV for charmed squarks\footnote{The charm squark analysis is published only for a centre-of-mass energy of 8~TeV.}.
As a comparison, the present limits for flavour-agnostic squark searches, summed over  
left and right squarks and over four flavours are around $1.5 \TeV$~\cite{Aaboud:2017vwy,Sirunyan:2017cwe}

\begin{figure}[htb!]
	\begin{center}
	\includegraphics[width=.49\textwidth]{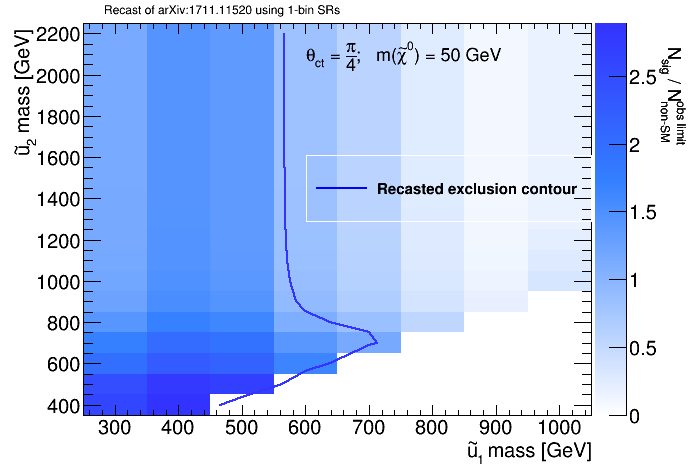}
	\includegraphics[width=.49\textwidth]{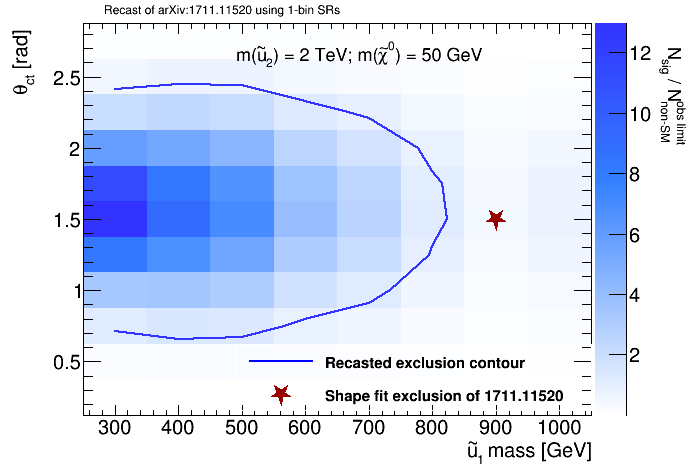}
	\end{center}
	\vspace*{-5mm}
	\caption{Reinterpretation of the ATLAS search for top squarks in the single lepton final state \cite{Aaboud:2017aeu} in the
  $(m_{\tilde{u}_1}$, $m_{\tilde{u}_2})$ (left) and $(m_{\tilde{u}_1}$,
  $\theta_{tc})$ (right) planes.}
	\label{lhcnmfv_fig:recast}
\end{figure}

Using the available information provided by the ATLAS collaboration for the search for top squarks in the single lepton 
final state \cite{Aaboud:2017aeu}, we recast their observed limits in terms of the three-dimensional parameter space of our model ($m_{\tilde{u}_1}$, $m_{\tilde{u}_2}$ and $\theta_{tc}$), as shown in Fig.~\ref{lhcnmfv_fig:recast}. 
The acceptances and efficiencies for each of the ``discovery tN\_med" and ``discovery tN\_high" regions have been used to estimate the signal yield ($N_{\rm sig}$) in these regions, considering the appropriate decay branching ratios for each parameter choice. These two signal regions are optimized specifically looking at moderate and higher stop masses respectively.

The signal yields were then compared to the model-independent upper limit
($N^{\rm obs~limit}_{\rm non-SM}$) 
provided in Ref.\ \cite{Aaboud:2017aeu} for each of the regions. When the ratio of these two yields exceeds one, the signal point is considered to be excluded. For the final exclusion limit estimation, the ATLAS analysis does not consider the ``discovery regions" as they are, but exploits a multi-bin fit in the most sensitive distribution. As the reinterpretation performed in this paper cannot exploit the full multi-bin information, due to the lack of details in the reference, the exclusion contour presented in Fig.\ \ref{lhcnmfv_fig:recast} can be considered as a conservative limit. The official observed limit of Ref.\ \cite{Aaboud:2017aeu} is therefore shown as a star on the right panel of Fig.\ \ref{lhcnmfv_fig:recast} in order to provide a comparative assessment of the multi-bin effects. It is interesting to mention that the star corresponds to the 
case in which the lightest squark dominantly (almost 100\%) decays into the top and neutralino system, while the charm mode is negligible. Consequently, a
non-negligible mixing angle yields a significant production of charm quarks in
the decays (see Fig.\ \ref{lhcnmfv_fig:bratio}), that importantly impacts the
current exclusion limits.

\section{Collider projections for the reach of the $tc$ channel} 
\label{lhcnmfv_sec:collider}
In this section, we describe a dedicated analysis aimed at evaluating
the reach for the signature under consideration for the full LHC
statistics and for the HL-LHC. We first
provide a brief outline of the Monte-Carlo (MC) simulations
used to generate signal and background events. The reconstruction of the
final-state leptons, jets and missing transverse momentum (\ptmiss) along with
some handy kinematic variables is also discussed. In order to optimise the
signal selection together with the rejection of the SM backgrounds, we impose
several kinematic requirements on the final state, that we provide below
together with our findings.

\subsection{Monte Carlo simulation}

Signal events originating from squark pair-production were produced using 
{\tt MadGraph5\_aMC@NLO}~\cite{Alwall:2014hca}, the hard-scattering matrix
elements being convoluted with the {\tt NNPDF3.0} parton 
distribution functions~\cite{Ball:2014uwa}. We 
use {\tt Feynrules 2.0} \cite{Alloul:2013bka} to obtain the model files 
incorporating the mixing between the charm and top squarks.
The mass of the squark with a dominant stop component (denoted stop
hereafter) was scanned in steps of 100 \GeV~in the [600 GeV, 1.4 TeV] mass
window. The lightest neutralino mass was kept fixed to 50~\GeV, as discussed in
Sec.~\ref{lhcnmfv_SecModel}, and the squark mixing angle was set to
$\theta_{tc} = \pi/4$. The generated events were passed to
{\tt PYTHIA 8.2}~\cite{Sjostrand:2014zea} for parton showering and hadronisation
and the events have been reweighted to a production cross-section at the
NLO+NLL accuracy~\cite{Borschensky:2014cia}. In the results presented below, we
consider LHC proton-proton collisions at a centre-of-mass energy of 14~TeV.

For the backgrounds, we focus on SM processes with one or two final-state
leptons originating from the decay of either a vector boson, or of a tau lepton,
and that precisely consists in \ttbar, $Wt$, $t$-channel single top, \ttbarW,
\ttbarZ, $tWZ$, $tZ$, \Wjets, \Zjets,  $WW$, $WZ$ and $ZZ$ production.
\ttbar\ and single top events are simulated at the next-to-leading (NLO) order 
accuracy within the {\tt POWHEG BOX} framework~\cite{Alioli:2010xd}, while the
\Wjets, \Zjets and $tWZ$ samples are generated at leading order (LO) with 
{\tt MadGraph5\_aMC@NLO}, merging samples containing up to four additional jets
at the level of the matrix element matched with the CKKW prescription
as implemented in PYTHIA8 \cite{Lonnblad:2011xx}. \ttbarW and \ttbarZ production has
been
achieved at the LO accuracy, matrix elements containing up to two extra jets
having been merged this time, with the $W$-boson being forced to decay
leptonically and the $Z$-boson invisibly. All these background samples have been
normalised to the next-to-next-to-leading order (NNLO) cross-sections, if
available, or to the NLO cross section otherwise. For details on the 
background simulation and the normalization technique, we refer the reader to Ref.\ \cite{Pani:2017qyd}.  

In order to perform a realistic analysis, we need to consider the detector effects on 
the various reconstructed objects, namely leptons ($e$ and $\mu$), jets and $\etmiss$.
Jets are reconstructed based on the MC truth particles using
{\tt FASTJET}~\cite{Cacciari:2011ma}, relying on the anti-$k_T$
algorithm~\cite{Cacciari:2008gp} with a radius parameter $R = 0.4$, whereas the
missing transverse energy is defined as the vector sum of the transverse momenta
of all the invisible particles.
We then make use of smearing functions tuned to mimic the
performance of the ATLAS detector~\cite{Aad:2008zzm,Aad:2009wy}, as
described in Ref.~\cite{Pani:2017qyd}. We have
validated our results by comparing with a reduced statistical sample in which the
simulation of the detector is performed within the publicly available code
{\tt Delphes}~\cite{deFavereau:2013fsa}.

\subsection{Variable definition and event selection}

Several variables exploiting the kinematic difference between the 
signal and the backgrounds are used in our analysis. The construction of these 
discriminating variables is based on the assumption that the events are selected with exactly 
one lepton (electron or muon) and one and only one \btagged\ jet. The final
state is further allowed to contain extra jets, and missing transverse energy. More
precisely, we preselect events containing exactly one electron or one muon with a
transverse momentum and pseudorapidity fulfilling $\pt>25\, \GeV$ and
$|\eta|<2.5$, and exactly one \btagged\ jet with $\pt>30\, \GeV$
within $|\eta|<2.5$. The chosen
$b$-tagging working point corresponds to an average tagging efficiency of 77\%
and is based on the parameterisations given in
Ref.~\cite{ATL-PHYS-PUB-2015-022}. We moreover demand the presence of at least
one additional jet with a transverse momentum $\pt>100\, \GeV$
and $|\eta|<2.5$ failing the $b$-tagging algorithm.

\begin{figure}
	\begin{center}
	\includegraphics[width=0.49\textwidth]{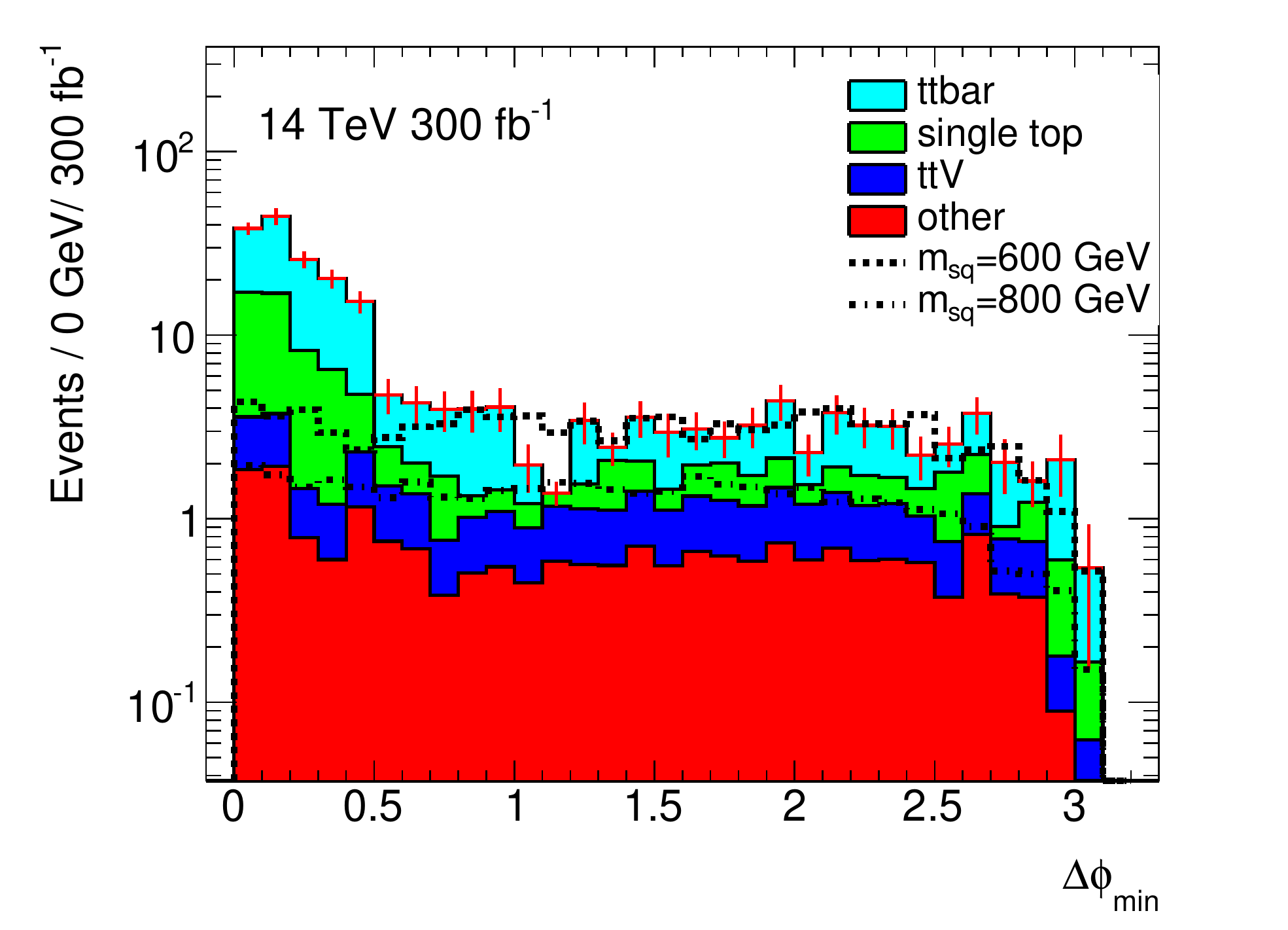}
	\includegraphics[width=0.49\textwidth]{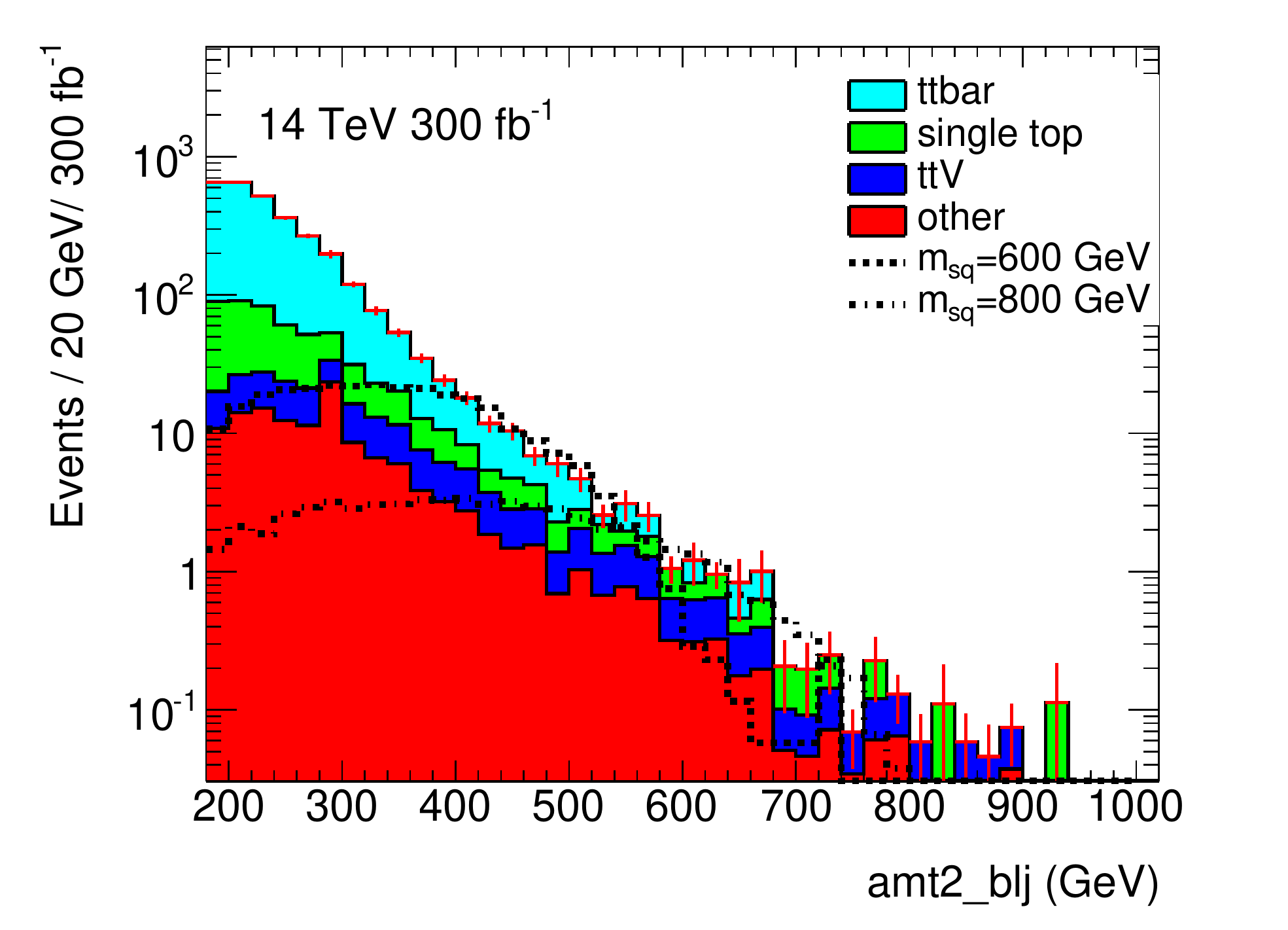}
	\end{center}
	\vspace*{-8mm}
	\caption{Distributions in the $|\Delta\phi_{\rm min}|$ (left) and
  \mttwoblj (right) variables after imposing all cuts of our analysis, excepted
  the one on the represented variable. We present results for the different
  background contributions and for two representative signal scenarios. The
  $|\Delta\phi_{\rm min}|$ results include a selection on the \mttwoblj variable
  that is imposed to be larger than 400 GeV.}
	\label{lhcnmfv_fig:nm1}
\end{figure}

The dominant background for the present analysis arises from \ttbar\ production,
and the main handles for reducing this background consist of a set of
kinematic variables in the transverse plane with respect to the LHC beams that
are bounded from above by the fact that the invisible particles are always
neutrinos issued from $W$-boson decays in the SM.
\begin{itemize}
\item
\underline{Transverse mass of lepton and $\etmiss$ (\mTlep)}: This variable,
defined as
\begin{equation}
	\mTlep \equiv\sqrt{2\,|\ptl|\,|\ptmiss|\,(1-\cos\Delta\phi_{\ptl\ptmiss})} \ ,
\end{equation}
is built from the lepton transverse momentum ($\ptl$) and the vector sum of the
transverse momenta of the invisible particles ($\ptmiss$).
For all processes where the lepton and the missing transverse energy are issued
from the decay of a single $W$-boson, this variable has a kinematic end-point at
80~GeV, whilst it is allowed to reach much larger values for the signal.
Imposing a lower bound on this variable therefore strongly suppresses the \ttbar\
background, killing events where only one leg decays leptonically, as well as
the $W$+jets background. We impose $\mTlep>160\, \GeV$.

\item
\underline{Asymmetric \mttwo (\amttwo)}:
After the \mTlep\ requirement, the background is still
dominated by $t\bar{t}$ events that decay into two leptons,
but with one of the two leptons that is not identified in the detector.
The variable $\amttwo$ \cite{Konar:2009qr,Lester:2014yga} was developed to
tame this kind of background, and consists in
a generalisation of the \mttwo variable~\cite{Lester:1999tx,Barr:2003rg}.
It is built by considering the production of two particles that each decay into
visible objects and missing energy, forming hence two legs. For the first leg,
the visible momentum is defined as the vector sum of the momenta of the
\btagged\ jet and the lepton, whilst a vanishing test mass is introduced to
kinematically derive the missing momentum. For the second leg, the visible
momentum is fixed to that of the additional jet, that is most likely to be a
\bjet, and the invisible transverse vector is derived from a test mass set to
80 GeV. In real experimental conditions, the jet employed for the second
leg is the second hardest \btagged\ jet, or, if there isn't any,
the light jet with the highest weight returned by the $b$-tagging algorithm.
Since this information not available within our parametric detector simulation,
we use the information at truth level. 
We identify as the second leg either the hardest
non-\btagged\ $b$-jet or the hardest $c$-jet or the hardest
light jet. For top quark pairs decaying into two leptons where one lepton is lost, this
variable has an end-point at around 150~\GeV, which provides a handle to
suppress background contamination. We require $\amttwo>200\, \GeV$.

\item
\underline{$m_{b\ell}$}: The invariant mass of the system made of the \btagged\ 
jet and lepton provides a good discriminant against backgrounds
not including a top quark, since if the \bjet\ and the lepton
originate from the decay of a single top quark, the corresponding distribution
exhibits an endpoint at $\sim160\, \GeV$; we impose
$m_{bl}<160 \, \GeV$.

\item
\underline{$\Delta\phi_{\rm min}$}:
The minimum azimuthal angle between any jet and the
missing transverse momentum \ptmiss\ can be used to increase the background
rejection. In the extreme configurations stemming from the selections on the
various \mttwo-like variables, the backgrounds tend to exhibit missing transverse energy
aligned with a hard jet, whereas there is no correlation for the signal. We
therefore require that $|\Delta\phi_{\rm min}|>0.6$.

\item
\underline{$\Delta R_{b\ell}$}: The distance in the transverse plane between the
lepton and the leading \btagged\ jet is used as an additional discriminant, and we
demand that $\Delta R_{b\ell}<1.75$.

\item
\underline{\mttwoblj}: We compute the \mttwo variable as usual by using two legs
with a visible and invisible component. The first leg is defined as above,
whilst the second leg takes as visible momentum the momentum of the hardest
non-$b$-tagged jet. Both test masses are put to zero. This variable
is  dependent on the mass difference of the squark and 
the lightest neutralino, as it tries, from the definition of
the second leg, to approximate the Jacobian of the squark into charm and
$\tilde\chi^0_1$ decay. This is therefore the final discriminant used in the analysis, 
and different selections are imposed optimising the sensitivity for each
targeted squark mass. As can be seen in the right panel of
Figure~\ref{lhcnmfv_fig:nm1}, the minimal cut to reduce the \ttbar\ background
to the same level as the signal for the minimum squark mass considered
in the analysis is approximately $400 \GeV$, and a cut at $\sim550 \GeV$
mostly removes the \ttbar\ background.
\end{itemize}

In Fig.\ \ref{lhcnmfv_fig:nm1} we present the distributions in
$|\Delta\phi_{\rm min}|$ and \mttwoblj, obtained after imposing all selection
cuts but the one that is shown, for the different background contributions and
two representative signal benchmarks. The $|\Delta\phi_{\rm min}|$ results
include a \mttwoblj $>$~400~\GeV~selection.

\subsection{Results}

\begin{table}
\begin{center}
\begin{tabular}{|cccccc|}
\hline
$m_{{\tilde u}_1}$  & $m_{{\tilde \chi}^{0}_1}$ &\mttwoblj cut &   $N_s$  & $N_b$ &  $\sigma^{\mathrm{excl}}/\sigma^{\mathrm{SUSY}}$ \\
(GeV) & (GeV)    & (GeV)        &          &      &          \\
\hline
600 & 50 &  400 &  124.9 &  63.0  &  0.23\\
700  & 50  & 450  & 55.6 &  31.5 &   0.30\\
800  & 50  & 500  & 25.8 &  15.3  &   0.41\\
900  & 50  & 500  & 17.0 &  15.3 &  0.63\\
1000  & 50  & 550  & 8.3 &   7.6 &   0.89\\
1100  & 50  & 600  & 4.0 &   4.4  &   1.41\\
\hline
\end{tabular}
\caption{Number of background ($N_b$) and signal ($N_s$) events surviving our
  selection, for different benchmark models and optimised selections on the
  \mttwoblj variable. We present the results under the form of upper limits, at
  the 95\% CL, on the ratio of the signal yield to the corresponding benchmark
  predictions. We assume an integrated luminosity of $300 \, {\rm fb}^{-1}$ of
  proton-proton collisions at $14 \, \TeV$, and systematic uncertainties of 20\%
  (5\%) on the SM background (signal).}
\label{lhcnmfv_tab:res300fb}

\vspace{.5cm}

\begin{tabular}{|cccccc|}
\hline
$m_{{\tilde u}_1}$  & $m_{{\tilde \chi}^{0}_1}$ &\mttwoblj cut &   $N_s$  & $N_b$ &  $\sigma^{\mathrm{excl}}/\sigma^{\mathrm{SUSY}}$ \\
(GeV) & (GeV)    & (GeV)        &          &      &          \\
\hline
600 & 50 &  400 &  1249 &  630  &  0.18\\
700  & 50  & 500  & 321.5 &  152.6 &   0.19\\
800  & 50  & 550  & 161.1 &  76.4  &   0.21\\
900  & 50  & 550  & 117.8 &  76.4 &  0.28\\
1000  & 50  & 700  & 26.7 &   11.1 &   0.33\\
1100  & 50  & 700  & 22.5 &   11.1  &   0.44\\
1200  & 50  & 700  & 15.2 &   11.1 &   0.59\\
1300  & 50  & 700  & 10.2 &   11.1 &   0.87\\
1400  & 50  & 750  & 5.1 &   6.8 &   1.38\\
\hline
\end{tabular}
\caption{Same as Table~\ref{lhcnmfv_tab:res300fb} but for a luminosity of
  $3000 \, {\rm fb}^{-1}$.}
\label{lhcnmfv_tab:res3000fb}
\end{center}
\end{table}

We now estimate the sensitivity of our analysis to the presence of mixed stop
states and present our results in Tables~\ref{lhcnmfv_tab:res300fb} and
\ref{lhcnmfv_tab:res3000fb} for integrated luminosities of 300 and 3000~fb$^{-1}$ 
respectively (without modifying the detector performance). We
consider multiple signal scenarios for which we
introduce different selections on the \mttwoblj variable, the threshold being
obtained by maximising the sensitivity with a scan in steps of 50 GeV.

A profile likelihood test statistic is then used to evaluate the
upper limit on the ratio of the signal yield to the one predicted in the context
of the considered simplified model, and we use the CLs method~\cite{Read:2002hq}
to derive exclusion limits at the 95\% confidence level (CL).
The statistical analysis has been performed by employing the {\tt RooStat}
toolkit \cite{Moneta:2010pm} and the discovery/exclusion reaches assume
systematic uncertainties of 20\% (5\%) for the SM background (signal), as well
as a 3\% uncertainty on the luminosity. We observe that mixed stop scenarios
with a squark mass up  to about $1~\TeV$ would yield a $2\sigma$ excess already with an
integrated luminosity of 300~${\rm fb^{-1}}$, this range being extended to
$1.3~\TeV$ for the high-luminosity LHC run. It is thus crucial to extend the
current LHC search program and include analyses dedicated to the pair-production
of top partners decaying into a single top quark and a lighter jet.

\section{Outlook}
\label{lhcnmfv_sec:outlook}
In the analysis presented above, we have targeted a supersymmetric scenario
where right-handed charm and top squarks mix and where squark pair-production
could yield a signature made of charm jets, top quarks or both. In principle,
mixing between the first and third generation could be allowed, although the
situation is severely constrained by flavour data. It might be nevertheless
interesting to design a strategy that may effectively discriminate these two
possibilities at the LHC. This requires the ability to tag jets originating from
the fragmentation of $c$-quarks, as opposed to jets arising from the
fragmentation of light quarks or $b$-quarks. Charm tagging is currently being
addressed by both LHC collaborations, as described {\it e.g.}~by
ATLAS~\cite{ATL-PHYS-PUB-2015-001}. We have however ignored charm tagging
in our analysis, relying only on $b$-tagging, as there is currently no public
information on the correlations between the $b$-tagging and $c$-tagging
algorithms. The latter are indeed necessary for a meaningful study.
An alternative interesting possibility to discriminate the actual squark mixing
could be to study the dependence of the results on the $b$-tagging working
point, as the amount of light, $c$-jet (mis)identification varies
significantly. Comparisons could indeed yield indications on the fraction of
$c$-jets that is selected. An excellent control of the variations of the
background composition with the $b$-tagging working point should however be
necessary for this exercise, and robust predictions therefore require
dedicated investigations.

\section*{Acknowledgements}

The authors would like to thank Michihisa Takeuchi for many useful discussions
and comments. This work has been partially supported by French state funds
managed by the Agence Nationale de la Recherche (ANR) in the context of the
Investissements d'avenir Labex ENIGMASS (ANR-11-LABX-0012) and Labex ILP
(ANR-11-IDEX-0004-02, ANR-10-LABX-63), and by the Grant-in-Aid for
Scientific Research on Scientific Research B (No.16H03991) and Innovative Areas
(16H06492).



\AddToContent{A.~Chakraborty, M.~Endo, B.~Fuks, B.~Herrmann, M.~M.~Nojiri, P.~Pani and G.~Polesello}
\renewcommand{\thesection}{\arabic{section}}

\graphicspath{{interm-higgs/}}

\chapter{Phenomenological Aspects of Intermediate Higgs (or Natural Composite Higgs) Models}

{\it N.~Ezroura, P.~Gardner, A.~E.~Nelson, M.~Park and D.~G.~E.~Walker}



\begin{abstract}
Intermediate Higgs Models~\cite{Katz:2005au} (IH), rechristened in the literature as natural composite Higgs models, have a large region of parameter space yet to be constrained by the Large Hadron Collider (LHC).  We explore the parameter space of these models by exploring the possibility of resolving low-mass boosted objects produced in association with colored partners of the top and bottom quark.  
For a simple model with 
an approximate $SU(4)/Sp(4)$ coset space, we consider unconstrained benchmark points that are consistent with precision electroweak data.  
We find that extending current searches for boosted objects to very low invariant masses ($\sim 25$ GeV) provides an improved sensitivity to this theory relative to standard searches for top and bottom partners.  
Although we focus on a simple IH model, our analysis is useful in any scenario with new colored partners with approximate global symmetries, as are common for natural solutions to the hierarchy problem.  The full analysis appears in~\cite{Park:2018}.
\end{abstract}

\section{INTRODUCTION}

In this summary, we explore some phenomenological consequences of Intermediate Higgs model~\cite{Katz:2005au} at the Large Hadron Collider (LHC).  IH models, rechristened in the literature as natural composite Higgs models, allow for a light Higgs boson where all of the new physics appears at a new scale which is naturally larger than the Higgs mass by a loop factor.  This scenario is largely consistent with current measurements at the LHC.  Given the lack of new physics beyond the Standard Model (SM), this loop factor has become insufficient.  If naturalness arguments are relevant for the weak scale, new physics associated with compositeness must be at a much larger scale.  IH scenarios allow for this by adding new top partners which cancel to corrections to the Higgs mass from the top loop.  %
%
%
%
Here we study a simple IH model based on an $SU(4)/Sp(4)$ coset space.  This $SU(4)/Sp(4)$ extension yields five new pNGB degrees of freedom, one of which is a new pseudo-scalar, with couplings to SM quarks through a new Yukawa sector (heavy top and bottom partners). The pseudo-scalar and top partners feature exotic LHC signatures at hadron collider scales.

\section{THE MODEL}
 We use a nonlinear sigma model with a  antisymmetric unitary matrix $\Sigma$ that transforms under \textit{SU(4)} as:
 \begin{equation}
\Sigma \rightarrow V\Sigma V^\dagger 
\end{equation}
where $V$ are $SU(4)$ representations. $\Sigma$ is defined as:
\begin{equation}
\Sigma(x)=e^{2i\Pi/f}\Sigma_0 
\end{equation}
where $f$ is a decay constant, $\Sigma_0$ is a background field invariant under $\textit{Sp(4)}$, and  $\Pi$ represents the fluctuations of the Nambu-Goldstone bosons about $\Sigma_0$ in the direction of the broken generators:
\begin{equation}
\Sigma_0  = \begin{pmatrix}
    i\sigma_2 & \\
              & i\sigma_2 
\end{pmatrix}  \qquad \qquad
\Pi = \pi^a X^a
\end{equation}
The Higgs bosons are in the coset space $\Pi$ which is defined as:
\begin{equation}
\Pi = \frac{1}{2\sqrt{2}}\begin{pmatrix}
A & H \\
H^\dagger & -A
\end{pmatrix}
\end{equation}
Here $A$ and $H$ are matrices such that:
\begin{equation}
A = \begin{pmatrix}
a & \\
 & a
\end{pmatrix} \qquad \qquad H = \begin{pmatrix}
h^0 + ih_3 & ih^2+h_2 \\
ih_1 - h_2 & h^0 - ih_3
\end{pmatrix}\,.
\end{equation}
Here $a$ is the electroweak singlet. $H$ must satisfy $\sigma_2 H - H^* \sigma_2 =0$ and $h_i$ are real.

The new physics for a low energy effective Higgs theory with an $SU(4) / Sp(4)$ coset space is the existence of a single additional pseudo-scalar particle $a$. Such a particle would be produced with electroweak couplings and decay hadronically. This particle would also be generically lighter than the Higgs boson, making it one the worst case scenarios at a hadron collider like the LHC, as a leptophobic timid pseudo-scalar~\cite{Cacciapaglia:2017iws}.

Addressing the hierarchy problem requires the addition of a colored multiplet of vector-like fermionic partner states $\Psi = (Q_T, \, Q_B, \, T, \, B )$ and $\overline{\Psi}$ that mix with the Standard Model top and bottom quarks $q = (q_T, \, q_B )$, $\bar{t}_3$ and $\bar{b}_3$. The extended theory introduces a new vector-like doublet $(Q^\prime, \overline{Q}^\prime)$, and has the following gauge invariant Yukawa sector.
\begin{eqnarray}
  \mathcal{L}_{\rm yukawa} &\supset& -i \lambda_1\,f\,\Psi \,\Sigma\,\overline{\Psi} + \lambda_2 f \, q \,\overline{Q} + \lambda_3\,f\,T\,\bar{t}_3\, + \lambda_4\,f\,B\,\bar{b}_3\, + {\rm h.c.} \\
  &+&  M^\prime \,Q^\prime\,\overline{Q}^\prime\, + \tilde{\lambda}_1 \,f\,  \overline{Q}^\prime \,\left( q_3\, + Q \right) \, + \tilde{\lambda}_2 \,f\,Q^\prime\,\overline{Q}\, + {\rm h.c.}
\end{eqnarray}
The addition of the extra doublet alleviates tension with precision electroweak data by allowing for small mixings between the SM doublet and the lightest partner states, while SM top and bottom mass ratio is still determined largely by the ratio $m_{\rm top}/ m_{\rm bottom} \sim \lambda_3 / \lambda_4$. We choose two benchmark points from this extension that are within the bounds of precision electroweak data. We then develop search strategies for resolving the pseudo-scalar in its low mass regime through its production in association with the top and bottom partners via the decays such as $Q \rightarrow a \, b$.

\subsection{Simulation results}
For our study of the process at the LHC ($pp$ collisions at the $\sqrt s = 13$ TeV), we have chosen two benchmark points to demonstrate the analysis. The minimal Yukawa sector couplings are set to a common value $\lambda_1 = 2, \, \lambda_2 = 0.2, \, \lambda_3 = 2.9, \, \lambda_4 = 0.04$, the mass of the additional doublets is set to $M^\prime = 1000$ GeV, and the pseudo-scalar mass is fixed at $m_a = 25$ GeV. With these values there is a parameter subspace that interpolates between regions with low pseudo-scalar production rates due to large mass splittings of the partner states ($\tilde{\lambda}_1 > \tilde{\lambda}_2$) and one where their masses are fairly degenerate ($\tilde{\lambda}_1 < \tilde{\lambda}_2$). The chiral symmetry breaking scale $f$ raises the partner masses uniformly and is chosen to fix the lightest partner (which is always the bottom quark) at the mass of its current LHC limit $m ~\sim 750$ GeV ~\cite{Khachatryan:2015gza}.
\vspace{0.1cm}
\begin{itemize}
\item Signal Benchmark A: $\tilde{\lambda}_1 = 2.9$, $\tilde{\lambda}_2 = 0.81$ $f=500$~GeV
\item Signal Benchmark B: $\tilde{\lambda}_1 = 0.79$, $\tilde{\lambda}_2 = 2.8$, $f=650$~GeV
\end{itemize}
\vspace{0.1cm}
Event generation for all signal and background processes was performed using {\tt MadGraph v2.6.0}, with parton showering by {\tt Pythia8} and detector simulations using {\tt Delphes3}. Signal benchmarks A and B were simulated using model files created in {\tt UFO} format with {\tt FeynCalc} with the full spectrum of tree level decays. Our study closely resembles the lepton + jets analysis performed recently by CMS~\cite{Sirunyan:2017usq}. The requirements of a high-$p_T$ isolated lepton ($>50$ GeV) and sizable $\slashed{E}_T$ in this channel are used to minimize the contribution from QCD multijets, while the dominant remaining backgrounds come from $t \bar{t}$+jets, $W/Z$+jets, diboson, and single-top production in association with $W$ bosons and $b$ jets. Hadronically decaying boosted objects are resolved by defining ``fat jets'' with large radii $\Delta R = 0.8$ which are divided into exclusive categories as either ``H-tagged'' or ``V-tagged'', based on the number of b-tagged subjects and their invariant mass which is required to be in the range $\left[ 60, 160 \right]$. H-tagged fat jet are then divided into two categories $H_1$ and $H_2$ based on the number of subjets that pass the b-tag requirement. \\
\vspace{0.1cm}
\begin{figure}[ht]
  \begin{center}
    \includegraphics[scale=0.32]{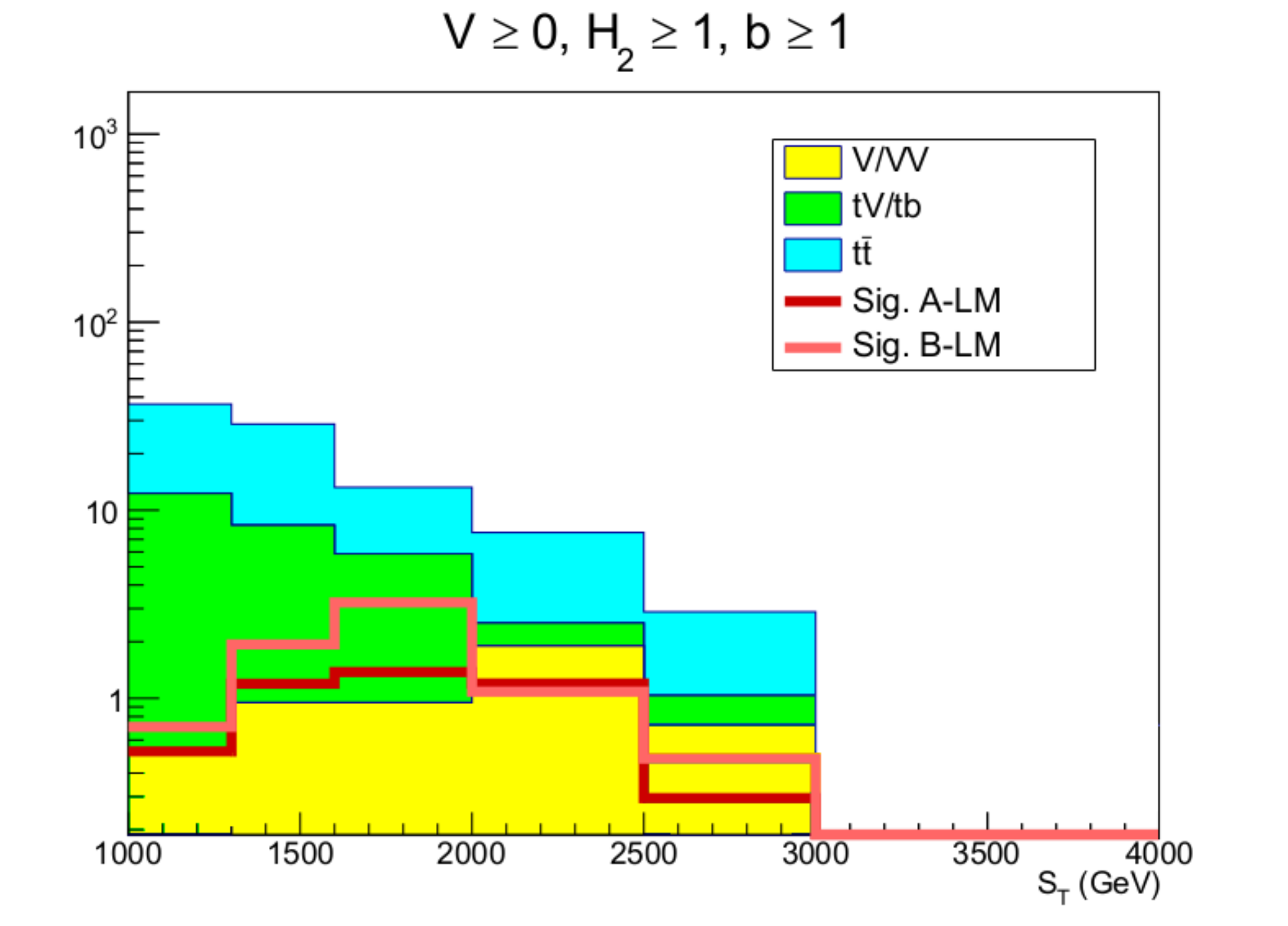} \hspace{0.1cm}
    \includegraphics[scale=0.32]{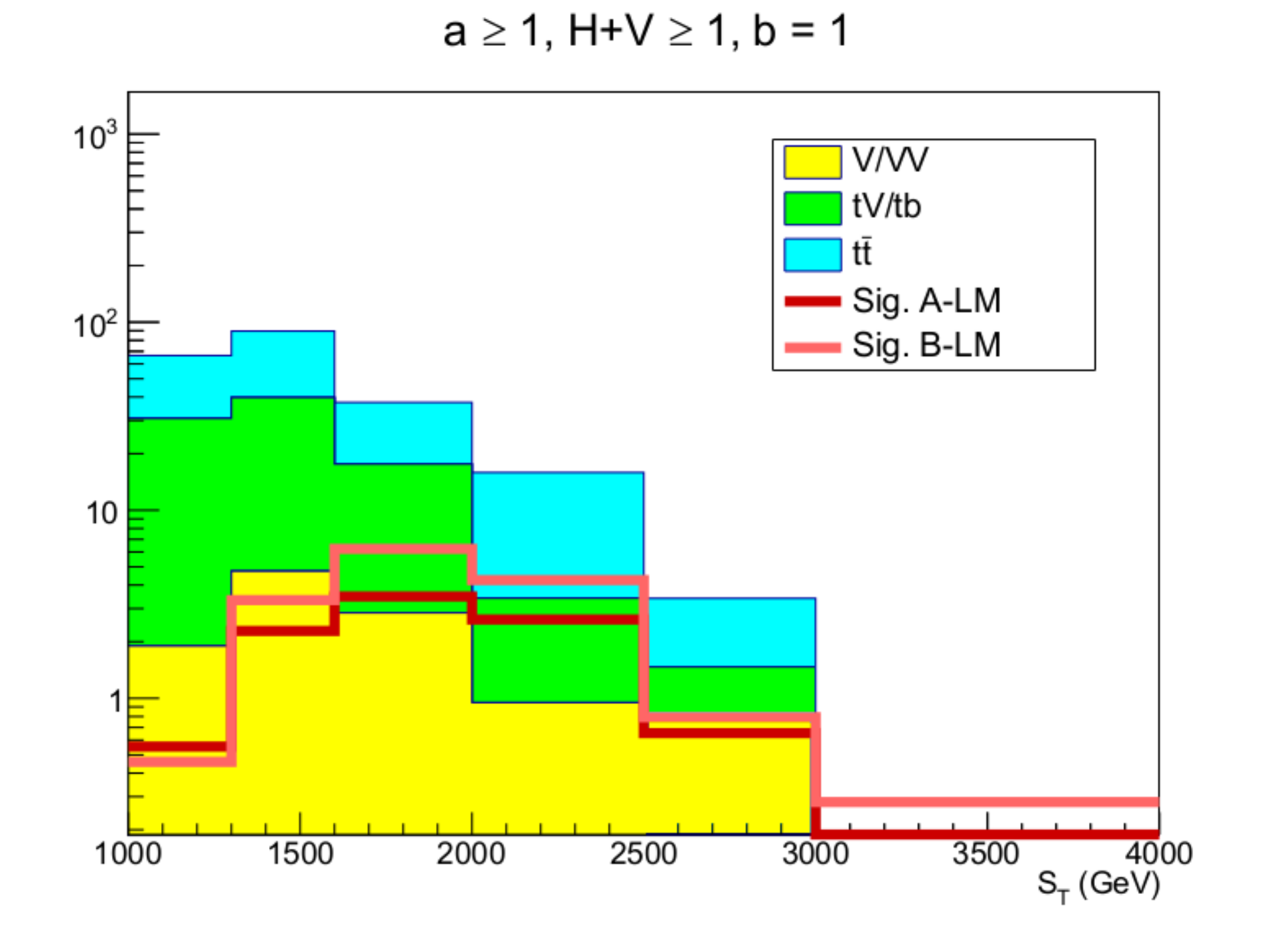} \vspace{-0.75cm}
  \end{center}
  \caption{$S_T$ distributions for two of the proposed search categories with the highest signal efficiency.}\label{fig:kinematic}
\end{figure}\vspace{-0.5cm}
\begin{table}[ht]
  \centering
  \begin{tabular}{| c | c | c |}
    \hline
    \hline
    \multicolumn{3}{| c |}{\bf Signal Region Selection Efficiencies} \\ [0.5ex] 
    \hline
    \hline
    Final State & $H$-tag Channel Signal A/B & $a$-tag Channel Signal A/B \\
    \hline
    $V \geq 1$, $H \geq 1$, $a = 0$ & 2.6\% / 3.0\%  & 2.6\% / 3.4\%  \\ [0.5ex] 
    \hline
    $V \geq 1$, $H = 0$, $a \geq 1$ & 4.1\% / 2.9\% & 5.3\% / 5.0\% \\ 
    \hline
    $V = 0$, $H \geq 1$, $a \geq 1$ & 0.06\% / 0.14\% & 1.1\% / 0.3\% \\
    \hline
    $V \geq 1$, $H \geq 1$, $a \geq 1$ & 5.3\% / 5.2\% & 5.3\% / 5.4\% \\
    \hline
  \end{tabular}
 \caption{Signal efficiencies in the proposed search categories. These efficiencies are comparable to those used to set the current strongest limits.~\cite{Sirunyan:2017usq}} \label{tab:benchmarks}
\end{table}
\vspace{0.1cm}
\noindent
With these cuts, events are divided into exclusive categories based on the number of tagged boosted $H_1$, $H_2$ and $V$ bosons, and binned by the total momentum scale of the event $S_T$. Additional categories focused on selecting boosted objects with low invariant masses, in the range of $10$ - $40$ GeV are then studied. This new category targets a low mass ($25$ GeV) pseudo-scalar $a$ decaying predominantly to $b \bar{b}$ pairs, aimed at probing theory spaces beyond the simplified limit. We thus define an object to be ``a-tagged'' if it lies in the invariant mass window $[10, 40]$ GeV, has $p_T > 70$ GeV, and massdrop variable $\mu < 0.8$, and has at least one $b$-tagged subjet. We find that binning events in these new categories by their $S_T$ provides signal-to-background ratios comparable to the signal categories in the CMS search~\cite{Sirunyan:2017usq}. The selection efficiencies for the signal region by various final states is given in Table [\ref{tab:benchmarks}], and $S_T$ distributions for the two best categories are Figure [\ref{fig:kinematic}].

\section*{CONCLUSIONS}
The existence of a low mass pseudo-scalar particle, that is produced with electroweak cross sections but decays with dominantly hadronic branching fractions, presents a surmountable challenge for the Large Hadron Collider. In a forthcoming publication we demonstrate the effectiveness of boosted object taggers at low invariant masses in identifying these objects, if they are produced in association with other heavy states.




\AddToContent{N.~Ezroura, P.~Gardner, A.~E.~Nelson, M.~Park and D.~G.~E.~Walker}
\renewcommand{\thesection}{\arabic{section}}

\newcommand{\Ell}{\mathscr{L}}
\newcommand{\HNP}{\mathscr{H}_{\rm NP}}
\newcommand{\Op}{\mathcal{O}}
\newcommand{\mc}{\mathcal}
\renewcommand{\th}{{\tilde h}}
\newcommand{\tH}{{\tilde H}}
\newcommand{\tphi}{{\tilde \phi}}
\newcommand{\<}{\langle}
\renewcommand{\>}{\rangle}
\newcommand{\Bsmumu}{\overline B_s \rightarrow \mu^+ \mu^-}
\newcommand{\re}{{\rm Re}}
\newcommand{\im}{{\rm Im}}
\newcommand{\ra}{\rightarrow}
\newcommand{\alem}{\alpha_{\rm em}}
\newcommand{\sh}{\hat s}
\renewcommand{\th}{\hat t}
\newcommand{\uh}{\hat u}
\newcommand{\mh}{\hat m}
\newcommand{\Mh}{\hat M}
\newcommand{\mumu}{\mu^+ \mu^-}

\newcommand{\nn}{\nonumber}

\newcommand{\tmpbf}[1]{{\bf \boldmath [#1]}}
\newcommand{\tmpbfdg}[1]{{\bf \boldmath \textcolor{red}{[DG: #1]}}}
\newcommand{\tmpcite}{{\bf \textcolor{red}{[cite?]}}}
\newcommand{\tmpred}[1]{\textcolor{red}{#1}}
\newcommand{\tmpblue}[1]{\textcolor{blue}{#1}}


\chapter{Review of flavour anomalies}

{\it M.~Borsato, D.~Guadagnoli}

\noindent {\bf Introduction --} A whole range of $b \to s$ data involving a $\mumu$ pair display a consistent pattern, with experimental data below the respective Standard-Model (SM) prediction, for di-lepton invariant masses below the charmonium threshold. This is true for the $B^{0} \to K^{0} \mumu$, the $B^{+} \to K^{+} \mumu$ and the $B^{+} \to K^{*+} \mumu$ decays \cite{Aaij:2014pli}, for the $B^0_s \to \phi \mumu$ decay \cite{Aaij:2015esa} and, very recently, even in hyperon channels for the $\Lambda_b \to \Lambda \mumu$ decay \cite{Aaij:2015xza,Detmold:2016pkz}. With these data alone, however, it is presently impossible to establish beyond-SM effects, as branching-ratio measurements suffer in general from sizable theoretical uncertainties due to hadronic form factors. On the other hand, such problems are basically absent if one considers suitable ratios of branching ratios to different lepton channels. Dedicated measurements exist on such ratios, and actually constitute the most alluring set of anomalies. In the $b \to s$ case these measurements are \cite{Aaij:2014ora,Aaij:2017vbb}
\bea
\label{eq:RK}
&R_K([1, 6] {\textrm{GeV}}^2) ~\equiv~ \frac{\mc B (B^+ \to K^+ \mu^+\mu^-)}{\mc B (B^+ \to K^+ e^+e^-)}|_{q^2 \in [1, 6] {\textrm{GeV}}^2} ~=~
0.745^{+0.090}_{-0.074}\,{\textrm{(stat)}}\pm 0.036\,{\textrm{(syst)}}~,&\nn\\
&R_{K^{*0}}([0.045, 1.1] \, {\textrm{GeV}}^2) ~=~ 0.660 ^{+0.110}_{-0.070} \pm 0.024~,&\\
&R_{K^{*0}}([1.1, 6] \, {\textrm{GeV}}^2) ~=~ 0.685 ^{+0.113}_{-0.069} \pm 0.047~,&\nn
\eea
where we have omitted the definition of $R_{K^{*0}}$, analogous to the $R_K$ one, and where $q^2$ denotes the invariant mass squared of the di-lepton pair. All of the above measurements are predicted to be unity (first and third of them) and respectively close to it (second one) within the SM, with a few-percent accuracy \cite{Bordone:2016gaq} (see also \cite{Bobeth:2007dw,Bouchard:2013mia,Hiller:2003js}). Therefore, the $R_{K}$ and $R_{K^{*0}}$ measurements each imply a discrepancy between 2 and $2.6\sigma$ \cite{Aaij:2014ora,Aaij:2017vbb}, at face value signalling lepton-universality violation (LUV) beyond the SM. 

The electron-channel measurement would be an obvious culprit for the discrepancies, because of bremsstrahlung and lower statistics. However, the measurement agrees with the SM. It is instead in the muon channel that experiment tends to depart from the SM \cite{Aaij:2014pli,Aaij:2012vr,Aaij:2016flj}, with data {\em below} predictions. Note that such a pattern is coherent with $R_K$ being below unity. Besides, muons are among the most reliable objects within LHCb, and this would disfavour (although of course not exclude) systematic effects as an explanation. These arguments provide circumstantial support to the overall coherence of the experimental picture, that suggests a violation of lepton universality, with effects in muons, and not in electrons.

Importantly, the emerging picture can be established from ratios alone, but it is supported by the other measurements, whose theory error is more debated. Some comments are deserved in particular for the $B \rightarrow K^{*} \mu \mu$ angular analysis \cite{Aaij:2013qta,Aaij:2015oid,Abdesselam:2016llu,Wehle:2016yoi,Atlas,CMS}, exhibiting a discrepancy in one combination of the angular-expansion coefficients, known as $P_{5}'$. The theory error on this observable is a matter of debate because, while it is designed to minimize f.f. dependence \cite{Descotes-Genon:2013vna}, what cancels is the dependence on the infinite-$m_b$ form factors. The crucial issue is how important departures from the infinite-$m_b$ limit are as the di-lepton invariant mass squared $q^2$ approaches the charmonium threshold $4 m_c^2$ -- in particular, departures due to $c \bar c$-loop contributions, that at present are still incalculable. Such contributions are formally power suppressed in a $1/m_b$ expansion, but also come with a factor of $1 / (q^2 - 4 m_c^2)$ \cite{Khodjamirian:2010vf}, that becomes larger and larger as $q^2$ approaches the charmonium threshold. On the different approaches towards assigning a significance to the $P_{5}'$ discrepancy, see in particular \cite{Descotes-Genon:2013wba,Lyon:2014hpa,Jager:2014rwa,Ciuchini:2015qxb}. Nonetheless, this observable again supports, even quantitatively, the picture emerging from ratios alone.

Equally interesting results come from measurements of the ratios $R(D^{(*)}) \equiv \mc B (B \to D^{(*)} \tau \nu) / \mc B (B \to D^{(*)} \ell \nu)$ \cite{Lees:2012xj,Aaij:2015yra,Huschle:2015rga,Belle-semilep}. Here the theory error is less intuitive, as the $\tau$ mass is not negligible with respect to $m_B$, and f.f. (form factor) dependence does not quite cancel in these ratios. However, they can be constrained thanks to accurate LQCD determinations (in the case of $R_D$) \cite{Na:2015kha,Lattice:2015rga} or to other experimental measurements ($R_{D^*}$) \cite{Fajfer:2012vx,Kamenik:2008tj}. Accordingly, a simultaneous fit to all these $R(D)$ and $R(D^*)$ measurements yields a discrepancy with respect to the SM predictions with a significance of about 4$\sigma$ \cite{Amhis:2016xyh}, comparable to the global significance of $b \to s$ anomalies. Note also that $b \to c$ anomalies come jointly from several experiments: $B$ factories and LHCb.

\bigskip

\noindent {\bf Theory considerations --} The above-mentioned discrepancies are intriguing for at least the following reasons: {\em (a)} they concern two sets of observables ($b \to s$ and $b \to c$) related by the SM $SU(2)_L$ symmetry \cite{Bhattacharya:2014wla}; {\em (b)} both $b \to s$ and $b \to c$ measurements suggest dynamics that distinguishes between the different species of leptons, i.e. beyond-SM Lepton-Universality Violation'' (LUV); {\em (c)} discrepancies appear to obey a pattern, i.e. data are always on a given side with respect to the SM predictions; {\em (d)} this pattern finds a straightforward interpretation within what is called an effective-field-theory (EFT) framework, to be expanded upon next.

Let us consider the following Hamiltonian, which is part of the full $\bar b \to \bar s \ell \ell$ one
\bea
\label{eq:HSMNP}
\hspace{-0.3cm}\mc H_{\rm SM+NP}(\bar b \to \bar s\ell^+\ell^-) = -\frac{4G_F}{\sqrt{2}}V_{tb}^* V_{ts}\frac{\alpha_{em}(m_b)}{4\pi} \times \nn \\
\left[\bar b_L \gamma^\lambda s_L \, \bar \ell \left(C_9^{(\ell)} \gamma_\lambda 
 + C_{10}^{(\ell)} \gamma_\lambda \gamma_5\right)\ell \right]  + {\rm H.c.}\,,
\eea
where the index ${(\ell)}$ on the Wilson coefficients $C_{9,10}$ denotes that the corresponding new-physics shift distinguishes between lepton flavours, whereas the SM contribution doesn't, as well known. The SM contributions are such that $C_9 \simeq - C_{10}$ at the $m_b$ scale, yielding (accidentally) an approximate $(V-A) \times (V-A)$ structure. Advocating likewise $C_{9, \rm NP}^{(\mu)} = - C_{10, \rm NP}^{(\mu)}$ for the new-physics shifts (note, in the $\mu$-channel only) turns out to account at one stroke for all $b \to s$ discrepancies \cite{Hiller:2014yaa,Ghosh:2014awa}. Further global fits by different groups consistently show that the most favourite solutions are either a negative new-physics (NP) contribution to $C_9$, with $C_{9, \rm NP}^{(\mu)} \sim - 30\% \, C_{9,\rm SM}$, or NP in the mentioned SU(2)$_L$-invariant direction $C_{9, \rm NP}^{(\mu)} = - C_{10, \rm NP}^{(\mu)} \simeq - 12\% \, |C_{9,\rm SM}|$. Note that such a solution is approximately RGE-stable.

The latter solution is especially interesting from a UV point of view, because it amounts to a $(V-A)_{\rm quark} \times (V-A)_{\rm lepton}$ operator (see eq. (\ref{eq:HSMNP})), that can in turn be promoted to an $SU(2)_L$-invariant, which is what one would expect of interactions arising above the EWSB scale. Let us then focus on this solution: $C_9^{(\ell)} \approx - C_{10}^{(\ell)}$ and $|C_{9,\rm NP}^{(\mu)}| \gg |C_{9,\rm NP}^{(e)}|$. Such a pattern, with effects much larger for muons than for electrons, can be generated from a purely third-generation interaction \cite{Glashow:2014iga} 
\be
\label{eq:HNP}
\mc H_{\rm NP} = G \, (\bar b_L' \gamma^\lambda b_L') \, (\bar \tau_L' \gamma_\lambda \tau_L')~,
\ee
with $G = 1 / \Lambda_{\rm NP}^2$ a new Fermi-like coupling, corresponding to a NP scale $\Lambda_{\rm NP}$ in the TeV ballpark. The interaction in eq. (\ref{eq:HNP}) is expected, e.g., in partial-compositeness frameworks \cite{Gripaios:2014tna}. The prime on the fields indicates that they are in the ``gauge'' basis, i.e. that below the EWSB scale they need to be rotated to the mass eigenbasis by usual chiral unitary transformations of the form
\bea
\label{eq:U}
b_L' \equiv (d_L')_3 = (U_L^d)_{3i} (d_L)_i~, ~~~~~~~
\tau_L' \equiv (\ell_L')_3 = (U_L^\ell)_{3i} (\ell_L)_i~,
\eea
whereby the r.h.s. fields represent the mass eigenbasis. In the absence of further assumptions on the structure of the rotation matrices, they will in general induce LUV {\em and} Lepton-Flavor Violation (LFV) effects alike \cite{Glashow:2014iga}. This is also a general expectation in a `top-bottom' approach: consider a new, LUV interaction introduced to explain $R_K$, and defined above the electroweak symmetry breaking (EWSB) scale. Such interaction may be of the kind $\bar \ell Z^\prime \ell$, with $Z^\prime$ a new vector boson, or $\bar \ell \phi q$, with $\phi$ a leptoquark. The question arises, in what basis are quarks and leptons in the above interaction. Generically, it is not the mass eigenbasis -- this basis does not yet even exist, as we are above the EWSB scale. Rotation to the mass eigenbasis generates LFV effects, although the initial interaction was introduced to produce only LUV ones.

With the above ingredients we can straightforwardly explain $b \to s$ data. In particular, neglecting the negligible phase-space difference between the electron and the muon channels, one finds
\be
\label{eq:RKmodel}
R_K \approx \frac{|C_9^{(\mu)}|^2 + |C_{10}^{(\mu)}|^2}{|C_9^{(e)}|^2 + |C_{10}^{(e)}|^2} \simeq \frac{2 |C_{10, \rm SM} + C_{10, \rm NP}^{(\mu)}|^2}{2 |C_{10, \rm SM}|^2}~,
\ee
where the factors of 2 on the r.h.s. are due to the contributions from $|C_9|$ and $|C_{10}|$ being equal by assumption. Note as well that
\be
\label{eq:0.77}
0.77 \pm 0.20 = \frac{\mc B(B_s \to \mu \mu)_{\rm exp}}{\mc B(B_s \to \mu \mu)_{\rm SM}} = \frac{\mc B(B_s \to \mu \mu)_{\rm SM + NP}}{\mc B(B_s \to \mu \mu)_{\rm SM}} = \frac{|C_{10, \rm SM} + C_{10, \rm NP}^{(\mu)}|^2}{|C_{10, \rm SM}|^2}~,
\ee
implying, within the model in ref. \cite{Glashow:2014iga}, the correlations (see also \cite{Hiller:2014yaa})
\be
\frac{\mc B(B_s \to \mu \mu)_{\rm exp}}{\mc B(B_s \to \mu \mu)_{\rm SM}} \simeq R_K \simeq \frac{\mc B(B^+ \to K^+ \mu \mu)_{\rm exp}}{\mc B(B^+ \to K^+ \mu \mu)_{\rm SM}}~.
\ee
This relation states that the measurement-over-SM ratio for $\mc B(B_s \to \mu \mu)$ provides a proxy for $R_K$. This is one more good reason to pursue accuracy in the $\mc B(B_s \to \mu \mu)$ measurement. To the extent that the central value on the l.h.s. of eq. (\ref{eq:0.77}) remains low, this test will be a sensitive one already by the end of Run 2, because the $\mc B(B_s \to \mu \mu)$ total error (dominated by the experimental component) is anticipated to be around 10\% \cite{Guadagnoli:2013mru,Bediaga:2012py}.

Concerning the expected LFV, the crucial question of course is whether it is experimentally accessible. The interaction (\ref{eq:HNP}), plus the measured amount of LUV pointed to by $R_K$, provide a general argument \cite{Guadagnoli:2017jcl,Glashow:2014iga} on the LFV rates to expect. In fact, $R_K$ yields the ratio
\be
\rho_{\rm NP} = -0.159^{+0.060}_{-0.070}
\ee
between the NP and the SM+NP contribution to $C_9^{(\mu)}$. Then, for {\em any} decay of the kind $B \to K \ell_i^\pm \ell_j^\mp$, where $i \ne j$ amounts to LFV, one can write
\be
\label{eq:BKll}
\frac{\mc B (B \to K \ell_i^\pm \ell_j^\mp)}{\mc B (B^+ \to K^+ \mu^+ \mu^-)} ~\simeq~ 
2 \rho_{\rm NP}^2 
\frac{| (U^\ell_L)_{3i} |^2 | (U^\ell_L)_{3j} |^2}{|(U^\ell_L)_{32}|^4}~,~~
\ee
where we used the $U$-matrix transformations (\ref{eq:U}), and normalised to the well-measured $\mc B (B^+ \to K^+ \mu^+ \mu^-) \simeq 4.3 \times 10^{-7}$ \cite{Aaij:2014pli}. One thereby obtains
\bea
\label{eq:BKlilj}
\mc B (B \to K \ell_i^\pm \ell_j^\mp) \simeq 5\% \, \cdot \, \mc B (B^+ \to K^+ \mu^+ \mu^-) \, \cdot \, \frac{| (U^\ell_L)_{3i} |^2 | (U^\ell_L)_{3j} |^2}{|(U^\ell_L)_{32}|^4} \simeq &&\nn \\
[0.1cm]
\simeq 2.2 \times 10^{-8} \, \cdot \, \frac{| (U^\ell_L)_{3i} |^2 | (U^\ell_L)_{3j} |^2}{|(U^\ell_L)_{32}|^4}~.\hspace{2.4cm}&&
\eea
Eq. (\ref{eq:BKlilj}) neglects all terms proportional to lepton masses, which in the case of $\tau$ leptons produce corrections of several tens of percent. Such effects are unimportant in the context of the present argument, whose aim is to produce order-of-magnitude estimates. Eq. (\ref{eq:BKlilj}) tells us that LFV $B \to K$ decays are expected to be of order $10^{-8}$ times an {\em unknown factor} involving $U_L^\ell$ matrix entries. Then, the argument about the expected size of LFV effects boils down to an argument about such factor.

In the $\ell_i \ell_j = e \mu$ case, this ratio reads $|(U_L^\ell)_{31} / (U_L^\ell)_{32}| \lesssim 3.7$ \cite{Glashow:2014iga}, implying that the $B \to K \mu e$ rate may be around $10^{-8}$, or much less if $|(U_L^\ell)_{31} / (U_L^\ell)_{32}| \ll 1$. The latter possibility would suggest $U_L^\ell$ entries that decrease in magnitude with the distance from the diagonal. But then one may expect the ratio $|(U_L^\ell)_{33} / (U_L^\ell)_{32}| > 1$, implying a $B \to K \mu \tau$ rate of O($10^{-8}$) or above! In short, assuming the interaction (\ref{eq:HNP}), one can expect just because of the overall unitarity of the $U^\ell_L$ matrix that at least one LFV $B \to K$ decay rate be in the ballpark of $10^{-8}$ \cite{Glashow:2014iga}, which happens to be within reach at LHCb's Run 2.\footnote{For more quantitative studies, see e.g. \cite{Guadagnoli:2015nra,Boucenna:2015raa,Guadagnoli:2016erb}.} An entirely analogous reasoning applies for the purely leptonic modes $B_s \to \ell_i^\pm \ell_j^\mp$.\footnote{%
We should keep in mind that at Run 2 the LHCb is expected \cite{Bediaga:2012py} to provide a first measurement of $\mc B (B_d \to \mu^+ \mu^-)$, which in the SM is as small as $1 \cdot 10^{-10}$.}

Being defined above the EWSB scale, the operator in eq. (\ref{eq:HNP}) can, and should be promoted to a full $SU(2)_L$ invariant \cite{Alonso:2014csa}, as mentioned above. This operation yields interactions of the kind $(\bar Q_L^{\prime i} \gamma^\lambda Q_L^{\prime i})\,(\bar L_L^{\prime j} \gamma_\lambda L_L^{\prime j})$ and $(\bar Q_L^{\prime i} \gamma^\lambda Q_L^{\prime j})\,(\bar L_L^{\prime j} \gamma_\lambda L_L^{\prime i})$, with $i, j$ SU(2)$_L$ indices and $Q_L^\prime$, $L_L^\prime$ the SM quark and lepton doublets in the gauge basis. The second interaction yields in turn charged currents like $(\bar t^\prime_L \gamma^\lambda b_L^\prime)(\bar \tau^\prime_L \gamma_\lambda \nu_{\tau L}^\prime)$. After rotation to the mass eigenbasis, the last structure contributes to $\Gamma(b \to c \tau \nu_\tau)$ \cite{Bhattacharya:2014wla}, thereby allowing to explain the LHCb and $B$-factories deviations on $R(D^{(*)})$.

While the above scenario is very attractive, it has to withstand non-trivial constraints, in particular from $B \to K \bar \nu \nu$ (see also \cite{Calibbi:2015kma}), from LEP-measured $Z \to \ell \ell$ couplings, and, most importantly, from LUV effects in $\tau \to \ell \nu \nu$ decays \cite{Feruglio:2016gvd}. The latter constraints are especially dangerous, as they are tested to per mil accuracy, and they turn out to ``strongly disfavor an explanation of the $R(D^{(*)})$ anomaly  model-independently'' \cite{Feruglio:2016gvd}. The same argument shows that also LFV decays of leptons are generated, and that they provide probes well competitive with the ones pointed out above, in particular $\mc B(\tau \to 3\mu)$, $\mc B(\tau \to \mu \rho) \sim 5 \times 10^{-8}$ \cite{Feruglio:2016gvd}.

\bigskip

\noindent {\bf Model-building considerations --} Up to now we have restricted ourselves to EFT considerations. Needless to say, we would expect the EFT picture to be the result of some new UV dynamics. Progress in this respect has to face a few challenging obstacles. The first one is the fact that, in the SM, $B \to D^{(*)} \tau \nu$ and $B \to K^{(*)} \ell \ell$ decays arise respectively at tree and loop level, whereas the NP corrections hinted at by data are in either case of O(15-25\%). This issue is relevant if we seek a common explanation of $b \to c \tau \nu$ and $b \to s \ell \ell$ discrepancies.

A second obstacle is inherent in the fact that the needed NP is of the kind $J_{\rm q} \times J_{\rm \ell}$, i.e. the product of a quark and a lepton current. In most UV setups, such operators are typically accompanied by $J_{\rm q} \times J_{\rm q}$ and $J_{\rm \ell} \times J_{\rm \ell}$ structures, that are severely constrained by data, respectively from $B_s$-mixing observables, and from purely leptonic LFV or LUV decays.

Finally, a third obstacle emerges from the observation that most model-building attempts advocate new charged, and possibly colored, states, with masses not larger than O(tens of TeV) and with significant couplings to 3$^{\rm rd}$-generation SM fermions. These conditions make constraints from direct searches, in particular of resonances decaying to $\tau \tau$ pairs, especially relevant, see \cite{Greljo:2015mma,Faroughy:2016osc}.

With the above qualifications, many proposals of plausible UV models for the anomalies have been made, typically involving a new Lorentz scalar or vector, with any\footnote{Compatibly with gauge invariance.} of the following transformation properties under the SM gauge group: a singlet or a triplet under SU(3)$_c$ (the latter case is known as leptoquark); a singlet or a doublet or a triplet under SU(2)$_L$.

It follows a very short review of the models proposed to account simultaneously for the $b \to s$ and the $b \to c$ anomalies. This review is far from exhaustive, in particular it limits itself to work published at the time of this workshop (summer 2017). Our aim is primarily to expose the non-trivial challenges that such simultaneous explanation poses in the face of all the existing constraints. 

\bigskip

\noindent {\bf Explicit Models --} A first natural possibility among those mentioned before is that of a color-singlet, weak-triplet vector field, i.e. a heavier replica of the $W^\pm, Z^0$ bosons \cite{Greljo:2015mma,Boucenna:2016wpr}. Ref. \cite{Greljo:2015mma} discusses in detail the bounds imposed by $\tau \to \ell \nu \nu$, $B_s$-mixing, and direct searches. The minimal model turns out to be ruled out by searches of resonance-produced $\tau$ pairs. Non-minimal versions of the model can avoid this constraint at the price of a larger parameter space. Ref. \cite{Boucenna:2016wpr} aims at a general discussion of gauge extensions with LUV, and concludes that within these models universality violation is most likely a signature of Yukawa couplings between the SM fermions and new vector-like fermions. The other color-neutral possibilities (weak triplet of scalars, and weak doublets or singlets) either amount to extended Higgs sectors, that in general have a tree-level FCNC problem, or else fail to fulfil gauge invariance.

Let us now turn to color-triplet scalars or vectors. Color triplets are usually referred to as leptoquarks (LQ) \cite{Davidson:1993qk}, i.e. states coupled to a quark and a lepton. By definition they thus avoid tree-level contributions to meson mixings and purely leptonic LUV/LFV decays. Besides, vector LQs are ubiquitous in grand-unified theory (GUT) scenarios. Ref. \cite{Bauer:2015knc} proposes an $SU(2)_L$-singlet Lorentz-scalar LQ able to explain both $R_{K^{(*)}}$ and $R(D^{(*)})$ with loop corrections to the former and tree corrections to the latter. While such dynamics is rather appealing, in that it reproduces the same suppression pattern as the SM, the model produces too large corrections to the measured ratio $\mc B (B \to D \mu \nu) / \mc B (B \to D e \nu)$ \cite{Becirevic:2016oho}. Ref. \cite{Fajfer:2015ycq} proposes a weak-triplet Lorentz-vector with completely general flavour couplings $g_{ij}$ to a $\bar Q^i_L L^j_L$ bilinear, $i,j$ denoting flavour indices. This scenario generalises \cite{Calibbi:2015kma}, where the LQ is assumed to couple only to the third-generation fermions in the weak basis. From a fit to data, one obtains $g_{b \mu}^* g_{s \mu} \sim 10^{-3} \cdot (M_U/\mbox{TeV})^2$ and $|g_{b \tau}| \gtrsim 2$ from $b \to s$ and respectively $b \to c$ anomalies, with $M_U$ the mass of the LQ. One may argue that this hierarchy introduces another flavour problem. One further problem, actually common to all cases involving {\em vector} LQs, will be emphasized below.

Ref. \cite{Barbieri:2015yvd} interprets the patterns of corrections required by the anomalies as the result of a broken flavour symmetry. Specifically, the authors note that $b \to c \tau \nu$ anomalies involve the $3^{\rm rd}$ generation of leptons, whereas $b \to s \ell \ell$ ones concern light generations of leptons only. They then invoke a flavour group $G_F$ and a tree-level LQ exchange such that, in the limit of exact $G_F$, the LQ couples only to the $3^{\rm rd}$ generation of SM fermions, and the needed NP effects arise from the $G_F$ breaking. This implies singly suppressed corrections to $b \to c \tau \nu$ and three times suppressed effects in $b \to s \ell \ell$. Ref. \cite{Barbieri:2015yvd} encompasses the cases of a weak-singlet vector, and of a weak-singlet scalar or vector LQ. While the mechanism is very plausible, the generated EFT operators do not escape, in general, the argument in ref. \cite{Feruglio:2016gvd}, because the only $G_F$-invariant SM fermions are the left-handed doublets. Among the merits of ref. \cite{Barbieri:2015yvd} is also the fact that it exposes a problem that is common to all models with {\em massive vector} LQs, and already signalled at the end of the previous paragraph. It is the problem of {\em power-like} sensitivity to the UV cutoff. This sensitivity manifests itself in the power-like divergence of 2-, 3-point functions, and box diagrams. This issue, on the one hand {\em de facto} reintroduces at one loop the constraints from, e.g., $B$-meson mixings, that for LQs is, as mentioned above, absent at tree level; on the other hand, it prevents a reliable assessment of such constraints. This is problematic in the same way it was to allow for $M_{W,Z} \neq 0$ before the introduction of the Higgs mechanism and the discovery of its renormalizability. In short, such cutoff sensitivity in the words of ref. \cite{Barbieri:2015yvd} ``cries out for an explicit UV completion'', if only for the sake of calculability. An attempt (as of the present writing) in this direction is ref. \cite{Barbieri:2016las}, although a separate challenge is represented by the detailed verification that this UV completion does withstand all the existing constraints.

Finally, ref. \cite{Becirevic:2016yqi} advocates a weak-doublet scalar LQ coupled to the bilinears $\bar d_R L_L$ and $\bar Q_L \nu_R$ through $Y_L$ and $Y_R$ Yukawa couplings, whereby the right-handed $\nu$ field is required to have negligibly small mass. By virtue of the $(V + A)_{\rm quark} \times (V-A)_{\rm lepton}$ current invoked, this setup is not affected by the constraint in ref. \cite{Feruglio:2016gvd}, but predicts $R_{K^*} > 1$, in tension with the recent measurement \cite{Aaij:2017vbb}.

\bigskip

\noindent {\bf Further tests --} The above discussion highlights that the UV aspects of the dynamics responsible for the anomalies are not quite as established as the EFT picture. Of course, if anomalies are here to stay, the correct UV picture will eventually emerge, guided from data. Further tests that can help consolidating the picture may be classified into three broad categories:
\begin{itemize}
\item Measurements of additional LUV ratios;
\item Extraction of long-distance effects from {\em data};
\item Definition and measurement of new observables sensitive to $C_9$ and $C_{10}$.
\end{itemize}
The first direction is rather evident. Other planned measurements include $R_\phi$, $R_{K_0(1430)}$, $R_{f_0}$, and the inclusive $R_{X_s}$. An interesting test \cite{Hiller:2014ula} is to define the double ratios
\be
X_H \equiv \frac{R_H}{R_K}~,
\ee
with $H = K^*,$ $\phi$, $K_0(1430)$, $f_0$ or $X_s$. Deviations from unity in $X_H$ can only come from right-handed quark currents.

As concerns the second item, it should be recalled that, especially in $b \to s \ell \ell$ modes an important obstacle towards a robust comparison of data with theory is the presence of long-distance (LD) effects due to $c \bar c$ loops. (I reiterate, however, that while this may be an issue for branching ratios, it is not for ratios such as $R_K$.) Encouraging is the fact that, in many cases, this matter seems amenable to be sorted out experimentally, by measuring the $m_{\mumu}$ spectrum, including the $c \bar c$ resonances, and fitting it to suitable parameterizations such as \cite{Lyon:2014hpa,Kruger:1996cv,Guadagnoli:2017quo}. An application of this approach to the $B^+ \to K^+ \mumu$ case was recently  presented in ref. \cite{Aaij:2016cbx}. Interestingly, the measurement yields a result compatible with previous measurements \cite{Aaij:2014pli} and, again, {\em below} the SM prediction \cite{Bailey:2015dka}.

Finally, the third item above suggests to pursue measurements of new observables, independently sensitive to $C_9$ and $C_{10}$. One example is the $B_s \to \mumu \gamma$ decay \cite{Dincer:2001hu,Melikhov:2004mk}, whose spectrum is, for low $q^2$, sensitive to electromagnetic-dipole operators, and, in the whole $q^2$ range, to interactions in eq. (\ref{eq:HSMNP}) as well as to their right-handed counterparts. Besides, its total branching ratio is one order of magnitude above the $B_s \to \mumu$ one, since the chiral suppression in the latter decay is replaced by an $\alpha_{\rm em} / \pi$ factor \cite{Dincer:2001hu}. (For this very reason, the radiative decay is promising even for lepton-flavor-violating searches \cite{Guadagnoli:2016erb,Guadagnoli:2017jcl}.) However, a measurement of the $B_s \to \mumu \gamma$ decay by direct detection of the photon poses a major challenge at hadron colliders, because photons are also the typical signature of the ubiquitous $\pi^0$, and because photons leave only (if at all) calorimetric information. Ref. \cite{Dettori:2016zff} points out that, for large $q^2$, $B_s \to \mumu \gamma$ events may actually be searched for in the very same event sample selected for the $\mc B(B_s \to \mumu)$ measurement, by enlarging the $q^2$ signal window beneath the peak region $q^2 \simeq m_{B_s}^2$. Hence this method combines the advantage of a large and ever increasing $B_s \to \mumu$ event sample with the advantage of $B_s \to \mumu \gamma$, that probes the interactions hinted at by the anomalies more thoroughly than $B_s \to \mumu$. From an experimental point of view, this method may be the only one practicable at hadron colliders, for the reasons already stated at the beginning of this paragraph. We also note that this method would provide the first experimental determination (ever) of $\mc B(B_s \to \mumu \gamma)$, because the PDG does not even quote a bound on this mode.


\AddToContent{M.~Borsato, D.~Guadagnoli}
\renewcommand{\thesection}{\arabic{section}}


\superpart{ The Higgs Boson }

\graphicspath{{STSXvsBDT/}}

 \newcommand{\sss}{\scriptscriptstyle}
 \renewcommand{\phi}{\varphi}


\def\lsim{\mathrel{\raise.3ex\hbox{$<$\kern-.75em\lower1ex\hbox{$\sim$}}}}
\def\gsim{\mathrel{\raise.3ex\hbox{$>$\kern-.75em\lower1ex\hbox{$\sim$}}}}
\newcommand{\red}[1]{\textcolor{red}{#1}}

%
\chapter{Simplified Template Cross Sections: sensitivity to dimension-6 interactions at the LHC}

{\it J.~de~Blas, K.~Lohwasser, P.~Musella, and K.~Mimasu}


\begin{abstract}
We perform a sensitivity study of the simplified template cross section (STXS) measurements to dimension-6 interactions within the Standard Model Effective Field Theory framework. We focus on energy dependent effects in Higgs production in association with a $Z$-boson, $p p \to Z H \to \ell^+\ell^- b\bar{b}$. Several benchmark points are considered, with different values of a representative Wilson coefficient, alongside the Standard Model prediction as well as the dominant $Zb\bar{b}$ background. We contrast the expected sensitivity obtained by the STXS to an analysis exploiting multivariate techniques via a boosted decision tree classifier. The aim of this exercise is to estimate the amount information retained in the STXS binning, and therefore the power of the framework for model-independent hypothesis testing in Higgs physics. We observe that the final performance of the BDT analysis does not differ significantly from the differential information in $Z$-boson $p_T$ offered by the STXS, with one notable exception. This would suggest that, once the sensitivity of the STXS measurements is saturated, moving towards optimised multivariate methods remains well-motivated. 
\end{abstract}



\section{Introduction}
\label{stsxsec:intro}
The Standard Model Effective Field Theory (SMEFT) is, by now, a well established framework for parametrising new physics effects in the interactions of Standard Model (SM) particles in a model independent way. It has been and continues to be a key part of the LHC programme, complementary to direct searches for new physics. The framework employs an operator expansion in canonical dimension suppressed by a generic cutoff scale, $\Lambda$, assumed to be much larger than the electroweak (EW) scale. The leading new physics contributions supplement the SM Lagrangian with dimension 6 operators\footnote{There is also one dimension-5 operator, the Weinberg operator, which generates neutrino masses.},
\vspace{-0.05cm}
\begin{align}
    \mathcal{L}=\mathcal{L}_{SM}+\sum_i \frac{C_i}{\Lambda^2}\mathcal{O}^i_{D=6}+\cdots.
\end{align} 
\vspace{-0.05cm}
New physics effects are then always suppressed by $q^2/\Lambda^2$, where $q<\Lambda$ is a given mass scale, e.g. $q=v$ the Higgs vacuum expectation value, or $q=E$ the typical energy scale of a physical process.

One of the main strengths of the LHC in this respect is its ability to probe the high energy regime, 
in which it is expected that the sensitivity to the $E^2/\Lambda^2$ effects will be maximised. Furthermore, the discovery of the Higgs boson in 2012~\cite{Aad:2012tfa,Chatrchyan:2012xdj} has opened a brand new avenue in constraining the SMEFT parameter space consisting of the various operators involving Higgs fields. Measurements of Higgs production and decay modes have already provided new constraints on many operators and have also helped to constrain some blind directions in existing fits to low-energy data such as precision electroweak measurements at LEP. 

In the first run of the LHC, a very successful programme of signal strength measurements took place, in which information from many searches was combined into a global fit to overall coupling modifiers between the Higgs and the rest of the SM particles~\cite{Khachatryan:2016vau}. The natural evolution of these measurements for Run 2 is to subdivide the phase space and work towards differential observables in Higgs production and decay. To this end, a staged approach termed Simplified Template Cross Sections (STXS) is being developed~\cite{deFlorian:2016spz}, consisting of an increasingly fine-grained binning of kinematic observables, separated by production and decay mode. The aim is to provide measurements in mutually exclusive regions of phase space, performed in simplified fiducial volumes and unfolded to remove detector and acceptance effects.

Being one of the main elements of LHC searches for non-SM physics, it is of great interest to evaluate the sensitivity of the STXS measurements to SMEFT effects in Higgs boson interactions, particularly since they will be able to access these high energy tails of kinematic distributions. In particular, one would like to know how the information provided by a generic framework such as the STXS would compare to an optimised, dedicated search for SMEFT effects. Naively, one may expect some loss of information given, \emph{e.g.}, the finite binning of the distributions. In this study, we aim to quantify this difference by comparing and contrasting the ability to constrain SMEFT effects in Higgs production between the STXS measurements and an optimised analysis making use of multivariate methods to extract the maximum classification power of the SMEFT signals. We consider the concrete scenario of the (ZH) production of a Higgs boson decaying into a pair of $b$-quarks in association with a $Z$-boson  decaying to a pair of leptons, in the presence of a single EFT operator. We simulate several benchmark values for the operator Wilson coefficient consistent with current constraints, along with the dominant reducible SM background, and evaluate the statistical discriminating power of a hypothesis test using the STXS measurements versus a multivariate Boosted Decision Tree (BDT) classifier.

The paper is organised as follows. We first outline the Monte Carlo event generation procedure for the SM and EFT benchmarks in Section~\ref{stsxsec:gen}, then in Section~\ref{stsxsec:tools} we describe the fiducial selection employed, the training and analysis implemented using the BDT classifier and the STXS binning used for ZH. In Section~\ref{sec:test}, we summarise the results of the selections and binning, and perform a statistical hypothesis test to quantify the relative strengths of the two methods. We conclude by laying out the avenues for further investigation in Section~\ref{stsxsec:conclusions}.

\section{Generated Models} 
\label{stsxsec:gen}
%
The production of a Higgs boson in association with an EW gauge boson can be considered one of the canonical LHC processes sensitive to SMEFT effects. Evidence for this process involving the $b\bar{b}$ decay mode of the Higgs and leptonic vector boson decays was finally observed in 2017~\cite{Aaboud:2017xsd,Sirunyan:2017elk}. Some of the operators which modify the Higgs coupling to these gauge bosons introduce $E^2/\Lambda^2$ effects in the production rate, enhancing it at high energies.
The associated production process can naturally access this region of phase space since the Higgs is produced recoiling against the associated vector, meaning that the $p_T$ of the Higgs or vector boson are a faithful proxy for the energy flowing through the EFT vertex. Of the many dimension-6 operators that can contribute to this process, we consider 
\begin{align}
    \mathcal{O}_{\sss HW} &= 
    \frac{ig}{2\Lambda^2} \big[D^\mu \phi^\dag \sigma_{k} D^\nu \phi\big] 
    W^{k}_{\mu \nu},
\end{align}
an operator from the so called strongly interacting light Higgs (SILH) basis~\cite{Giudice:2007fh,Contino:2013kra}. Here, $\sigma_{k}$ refers to the Pauli matrices and the covariant derivative, $D_\mu$, for the Higgs field is defined as
\begin{align}
    D_\mu\phi &= \partial_\mu \phi -  i g \frac{\sigma_{k}}{2} W_\mu^k \phi - \frac12 i g' B_\mu \phi,
\end{align}
with $g$ and $g^\prime$ the weak and hypercharge gauge couplings respectively.

A global fit~\cite{Ellis:2014jta} combining information from precision measurements at LEP and LHC Run 1 data constrains the Wilson coefficient, $C_{\sss HW}$, to lie in the range 
\begin{align}
    \frac{m_{\sss W}^2}{\Lambda^2}C_{\sss HW} \in [-0.07,\,0.03]~\mbox{ at 95\% C.L.}
\end{align}
A sensitivity estimate for LHC Run 2 was also performed in Ref.~\cite{Degrande:2016dqg} by projecting an 8 TeV ZH analysis~\cite{TheATLAScollaboration:2013lia} to 13 TeV. The results of the study indicated that the previous bound could be improved by an order of magnitude. This indicates that the STXS measurements are likely to provide even greater sensitivity to this parameter.

Motivated by the limits from the global fit, we select the following benchmark values for $c_{\sss HW}\equiv C_{\sss HW}~\! m_{\sss W}^2/\Lambda^2$:
\begin{align}
    c_{\sss HW} = \pm 0.03 \text{ and } \pm0.01.
\end{align}
The first value roughly saturates the positive end of the limit, and has been shown to yield drastic effects in the kinematic tails of distributions~\cite{Degrande:2016dqg}. The second corresponds to a smaller, yet potentially accessible value of the parameter that may better test the relative discriminating power of relatively small effects between the two methods we investigate. We simulate our Monte Carlo samples at leading order using {\sc MadGraph5\_aMC@NLO}~\cite{Alwall:2014hca} with the public {\sc HELatNLO}~\cite{Degrande:2016dqg,Alloul:2013naa} {\sc FeynRules}~\cite{Alloul:2013bka} model, exploiting reweighting methods to simulate multiple parameter space points (including the SM) simultaneously. We also include the dominant irreducible background contribution from $Z b\bar{b}$ production with the $Z$-boson decaying leptonically. Showering and hadronisation, as well as the Higgs boson decay to $b\bar{b}$ are performed with {\sc PYTHIA8}~\cite{Sjostrand:2014zea} and the events are reconstructed from hadron level with {\sc MadAnalysis5}~\cite{Conte:2012fm} which makes use of {\sc FastJet}~\cite{Cacciari:2011ma}.

\section{Analysis}
\label{stsxsec:tools}
\subsection{Fiducial selection}
\label{sec:fiducial}
We first perform a simple fiducial selection on the event samples, to emulate a typical LHC selection that would be performed for the ZH process. To this end, we also implement a $p_T$ and $|\eta|$ dependent smearing function on the $b$-jet momenta to approximate finite detector resolution effects following the parametrised functions determined by the CMS particle-flow performance analysis~\cite{Sirunyan:2017ulk}.

Jets are clustered with the anti-$k_T$ algorithm with a radius parameter of 0.4 and required to have $p_T > 20$ GeV. Events are required to have two leptons satisfying $p_T > 25$ GeV and $|\eta|< 2.5$. Exactly two $b$-jets, as identified using truth-level information by {\sc MadAnalysis5}, are required satisfying $p_T > 20$ GeV and $|\eta|< 2.5$. We assume a flat $b$-tagging efficiency of $70\%$, corresponding to the DeepCSV medium working point defined in Ref.~\cite{Sirunyan:2017ezt}. Additionally, $Z$- and Higgs-boson mass windows are imposed on the invariant masses of the lepton and $b$-jet pairs such that $75 < M_{\ell\ell} < 105$ GeV and  $60 < M_{bb} < 140$ GeV. This defines our fiducial volume on which both the BDT training and STXS binning are performed. Table~\ref{tab:FiducialXS} summarises the cross sections obtained after the fiducial selection for the Monte Carlo samples generated. The $H\to b\bar{b}$ branching fraction is computed in the SM and the SMEFT benchmark points using the e{\sc HDECAY}~\cite{Contino:2014aaa} interface of {\sc Rosetta}~\cite{Falkowski:2015wza} and folded into the cross section results.
\begin{table}[h!]
    \centering
    \begin{tabular}{|l|l|}
        \hline
        $pp\to b\,\bar{b}\,\ell^+\,\ell^-$& $\sigma_{\text{fid.}} [fb]$\tabularnewline
        \hline
        $ZH$ SM&2.72\tabularnewline
        $ZH$ $c_{\sss HW} = 0.03$&3.64\tabularnewline
        $ZH$ $c_{\sss HW} = -0.03$&2.21\tabularnewline
        $ZH$ $c_{\sss HW} = 0.01$&3.38\tabularnewline
        $ZH$ $c_{\sss HW} = -0.01$&2.50\tabularnewline
        $Z b\bar{b}$ SM&291.3\tabularnewline
        \hline
    \end{tabular}
    \caption{\label{tab:FiducialXS} Cross sections obtained at LO after imposing the fiducial selection cuts described in Section~\ref{stsxsec:tools}.}
\end{table}
Clearly, the $Z b\bar{b}$ background is overwhelmingly large even after the Higgs mass
window selection. A realistic analysis will employ multivariate analyses techniques to
reduce this background. As explained in the next section, we mimic this aspect of the
experimental analyses by training a BDT discriminant to optimally reject this background
in favour of the SM $ZH$ process.

\subsection{Kinematic discriminants}
\label{sec:training}

We built a set of gradient BDT classifiers to efficiently discriminate
between the different classes of event hypotheses involved in the analysis, namely $Z b\bar{b}$,
SM $Z H$, and BSM $Z H$ production. 
The classifiers receive kinematic variables related to the event as inputs and approximate
the likelihood for each event to belong to any of the three classes. The likelihoods are parameterised as
\begin{equation}
p_{i}(\vec x) = \frac{ e^{\beta_{i}(\vec x) } }{ \sum_j e^{\beta_{j}(\vec x) } },
\end{equation}
where $\vec x$ represents all the input variables to the discriminant, that are shown in
Table~\ref{tab:bdt_features}, and $\beta_{i}(\vec x)$ are non-linear functions of these
variables.

These kind of algorithms are regularly used by the experimental analyses and often provide a comparable performance to those of more sophisticated techniques such as matrix element methods. 
\begin{table}[h!]
\centering
\begin{tabular}{|c|c|}
\hline
Variable name & Variable definition \\
\hline
$p_T(l_1)$ & leading lepton $p_T$ \\
$p_T(l_2)$ & sub-leading lepton $p_T$\\
$p_T(b_1)$ & leading b-jet $p_T$ \\
$p_T(b_2)$ & sub-leading b-jet $p_T$\\
$\eta(l_1)$ & leading lepton pseudo-rapidity \\
$\eta(l_2)$ & sub-leading lepton pseudo-rapidity\\
$\eta(b_1)$ & leading b-jet pseudo-rapidity \\
$\eta(b_2)$ & sub-leading b-jet  pseudo-rapidity\\
$y(Z)$ & Z-candidate rapidity  \\
$y(H)$ & H-candidate rapidity \\
$p_T(Z)$ & Z-candidate $p_T$ \\
$m_{bb}$ & H-candidate invariant mass \\
$m_{ZH}$ & HZ invariant mass  \\
\hline
\end{tabular}
\caption{
\label{tab:bdt_features}
Input variables used by the kinematic BDT discriminants.
}
\end{table}

Five sets of discriminants were trained:
\begin{enumerate}
\item A binary discriminant to separate $Z b\bar{b}$ production from the SM $Z H$ production.
\item Four sets of three-class discriminants (one for each of the $c_{\sss HW}$ benchmarks) to
    discriminate between $Z b\bar{b}$, SM $Z H$ and BSM $Z H$ production.
\end{enumerate}
The first discriminant was used to reduce the $Z b\bar{b}$ background in the template
cross section analysis, in such a way to mimic the experimental analyses.
The last four discriminants were used to first reduce the $Z b\bar{b}$ background and then to further classify the selected events to discriminate between the SM and BSM hypotheses.

The BDTs were trained using the scikit-learn~\cite{scikit-learn} and xgboost packages~\cite{xgboost}. To this end, the events were
split into two statistically independent samples, with a ratio of 3:1, used respectively for training and
application of the discriminants. 
We used the categorical cross-entropy loss function, and the algorithm hyperparameters were
optimised on the training sample using stochastic grid search and the mean
k-fold cross-validation loss as figure of merit, with $k = 5$.
The optimized hyperparameters are shown in Table~\ref{tab:bdt_hpars}.

\begin{table}
\centering
\begin{tabular}{|c|c|}
\hline
Parameter & Value \\
\hline
number of trees & 600 \\
maximum depth & 5 \\
bagging fraction & 0.8 \\
learning rate & 0.05 \\
$L_2$ regularisation strength & 1 \\
\hline
\end{tabular}
\caption{
\label{tab:bdt_hpars}
Optimized BDT training hyperparameters.
}
\end{table}

After training, the $p(Z b\bar{b})$ ($\equiv p_{Z b\bar{b}}(\vec x)$) variables were used to define selection criteria
for the events to be considered for analysis. The maximum allowed value for $p(Z b\bar{b})$
was determined in order to maximise the quantity $\varepsilon^2(SM Z H) / \varepsilon(Z
b\bar{b})$, where $\varepsilon$ is the selection efficiency of a given sample. Such a
figure of merit is, in the Gaussian limit, proportional to the squared median expected
discovery significance to observe the standard model production of of $Z H$.
The choice was made again to mimic the experimental analyses, for which the observation of
the SM Higgs signal will be the primary goal.

\begin{figure}
\centering
\includegraphics[width=\textwidth]{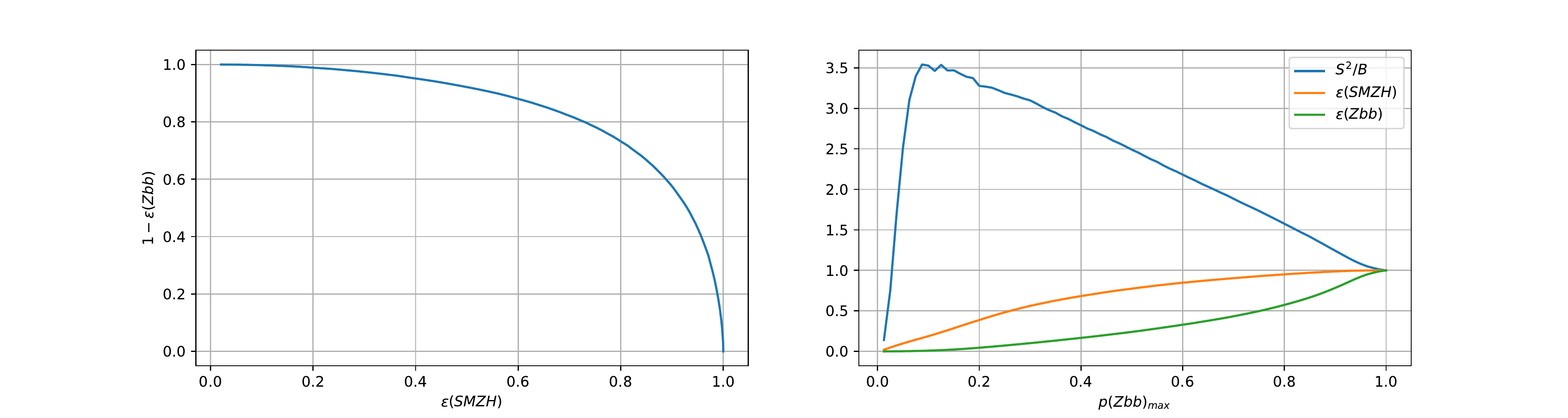}
\caption{
\label{fig:s2_over_b}
Optimisation analysis to determine the maximum value of $p(Z b\bar{b})$ for events
considered in the analysis. (left) {\it ROC} curve for $Z b\bar{b}$ vs SM $Z H$
separation. (right) Regularised $\varepsilon^2(SM Z H) /
\varepsilon(Zb\bar{b})$ as a function of the $p(Z b\bar{b})_{max}$.
}
\end{figure}

In order to ensure a smooth behaviour, the figure $\varepsilon^2(SM Z H) /
\varepsilon(Zb\bar{b})$ was regularised by replacing $\varepsilon(Zb\bar{b})$ with
$\varepsilon(Zb\bar{b}) \oplus \varepsilon_0$ where $\oplus$ denotes the sum in quadrature
and $\varepsilon_0 = 0.03$.
Figure~\ref{fig:s2_over_b} shows, as an example, the result of the optimisation scan for
the discriminant trained to separate $Z b\bar{b}$ production from the SM $Z H$
production. The analysis was repeated separately for each of the 5 discriminants and
similar results were obtained in all cases.

\begin{figure}[h!]
\centering
\includegraphics[width=0.45\linewidth]{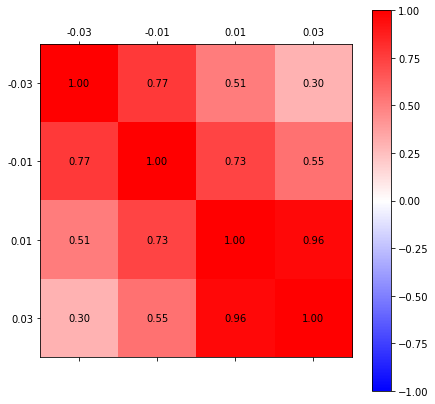}
\caption{\label{fig:bdt_corr}
  Linear correlation coefficient between $p(BSM)$ for
  each of the four scenarios, evaluated on SM $Z H$ production events.
  Rows and columns are labelled by the value of $c_{\sss HW}$.
}
\end{figure}

For this exploratory study, we only considered four BSM benchmarks, varying $c_{\sss HW}$ and testing the
sensitivity to each of these benchmarks using a dedicated set of kinematic
discriminants. While the design of an optimal discriminant that continuously depends on
the BSM parameters is beyond the scope of this work, we investigated the degree of
correlation between the BSM discriminants for each of the scenarios.
Figure~\ref{fig:bdt_corr} shows the linear correlation coefficient between $p(BSM)$ for
each of the four scenarios, evaluated on SM $Z H$ production events. The $p(BSM)$
BDT outputs estimate the likelihood for an event to come from each of the considered BSM
scenarios.
As can be seen,
the linear correlation varies between 0.3 and 0.96, and it increases as the distance
between the benchmark points decreases. This suggests that the information used to
discriminate the different benchmarks is similar, but that optimal results are obtained
when a specific benchmark is targeted.

\subsection{STXS binning}
\label{sec:stxs}
\begin{figure}[h!]
\centering
\includegraphics[width=0.8\textwidth]{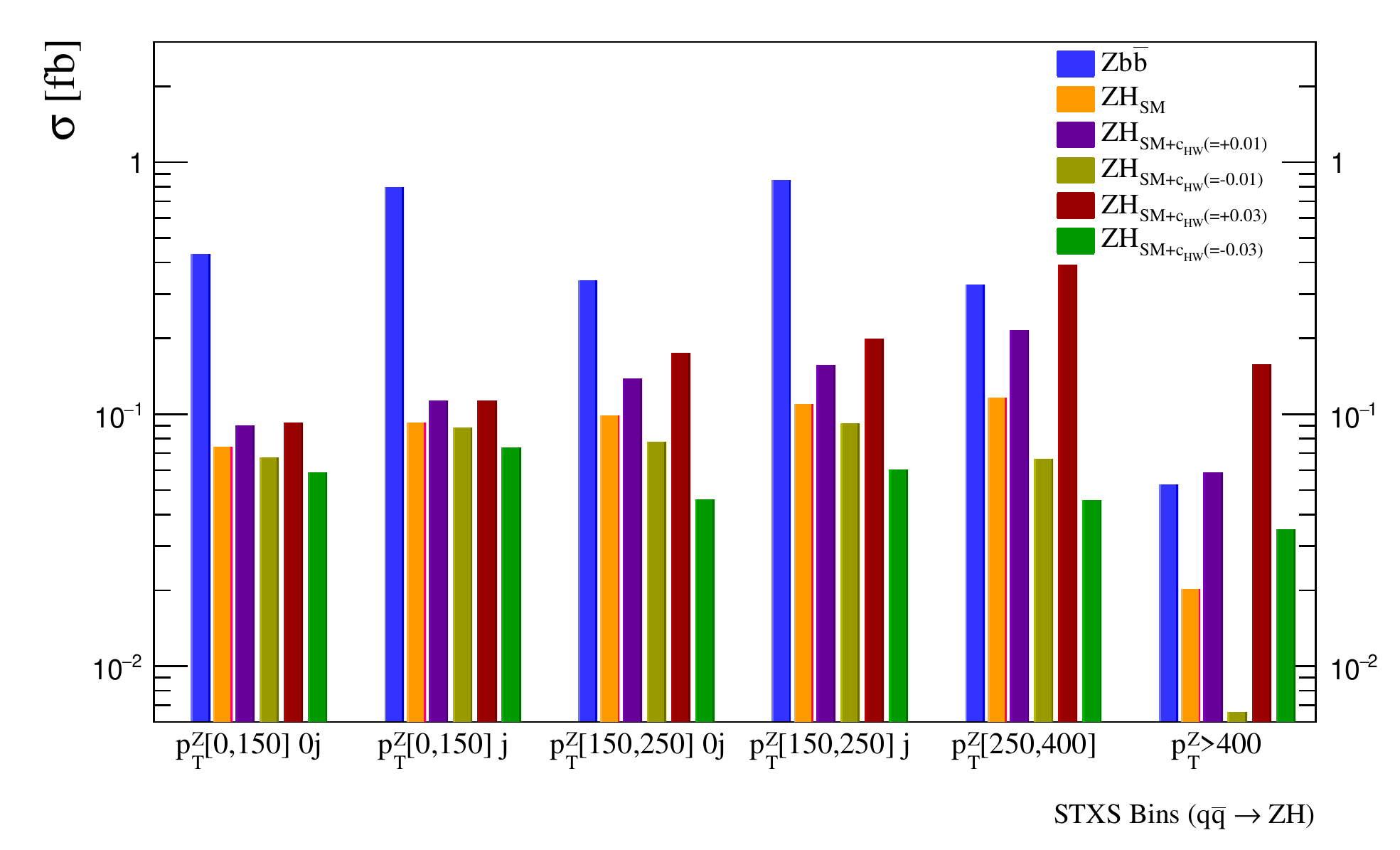}
\caption{
\label{fig:stxs_crosssec}
Predicted cross sections in the stage 2, $ZH$ STXS bins for SM $ZH$ production as well as our four EFT benchmarks and the $Zb\bar{b}$ background. The cross sections correspond to events passing the basic fiducial selection of Section~\ref{sec:fiducial}, accounting for $b$-tagging efficiencies and after applying a cut on the $Zb\bar{b}$ BDT discriminant as described in the text.
    }
\end{figure}

For the STXS sensitivity analysis, the generated samples for all signal benchmarks as well as the backgrounds (SM $pp \to ZH, H\to b\bar{b}$ and $pp\to Zb\bar{b}$) are categorized according to the STXS proposal for the $VH$ channel in Ref.~\cite{deFlorian:2016spz}. Within this framework, different regions of the phase space -- referred to as ``bins'' for simplicity -- are defined, with the purpose of optimizing the sensitivity of the measurements while at the same time minimizing their dependence on theory assumptions. The different STXS bins are defined specifically for each Higgs production mode. For our process of interest ($pp \to ZH \to \ell^+ \ell^- b\bar{b}$) the different stages of the categorization and the resulting bins can be summarized as follows (see \cite{deFlorian:2016spz} for details):
\vspace{0.5cm}
\begin{itemize}
{\item  {\bf Stage 0:} Events with $\left|y_H\right| <2.5$ are selected.}
{\item {\bf Stage 1:} $ZH$ production is split into $q\bar{q}$ and $gg$ initial states (our process sample was generated at LO and therefore only contains $q\bar{q}\to ZH$ events). Events are subsequently classified according to the value of $p_T^Z$ and number of extra jets in the event as follows:
\vspace{0.25cm}
\begin{center}
${\bf \underline{q\bar{q}\to ZH}}$
\end{center}
\begin{eqnarray}
p_T^Z \in& [0,150]~\mathrm{GeV},& \nonumber\\
p_T^Z \in& [150,250]~\mathrm{GeV}&(0\mbox{-}\mathrm{j}),\\
p_T^Z \in& [150,250]~\mathrm{GeV}&(\geq 1\mbox{-}\mathrm{j}),\nonumber\\
p_T^Z >&250~\mathrm{GeV}.\nonumber&
\end{eqnarray}
}
{\item {\bf Stage 2:} In this last stage the low $p_T^Z$ bins are further separated according to the number of extra jets, while the high-$p_T^Z$ region is split at 400 GeV. The final set of STXS bins that apply in our case are the following six:
\vspace{0.25cm}
\begin{center}
${\bf \underline{q\bar{q}\to ZH}}$
\end{center}
\begin{eqnarray}
p_T^Z \in& [0,150]~\mathrm{GeV}&(0\mbox{-}\mathrm{j}),\nonumber\\
p_T^Z \in& [0,150]~\mathrm{GeV}&(\geq 1\mbox{-}\mathrm{j}),\nonumber\nonumber\\
p_T^Z \in& [150,250]~\mathrm{GeV}&(0\mbox{-}\mathrm{j}),\\
p_T^Z \in& [150,250]~\mathrm{GeV}&(\geq 1\mbox{-}\mathrm{j}),\nonumber\\
p_T^Z \in& [250,400]~\mathrm{GeV},&\nonumber\\
p_T^Z >&400~\mathrm{GeV}.\nonumber&
\end{eqnarray}
}
\end{itemize}
In order to profit from the maximum amount of available information, we use the stage 2 categorisation to compare with the multivariate analysis. Additionally, we use the BDT discriminant, $p(Z b\bar{b})$, trained to reject the $Zb\bar{b}$ background in favour of SM $ZH$, described in Section~\ref{sec:training} to purify our event sample. Our STXS yields are computed after cutting on this discriminant with 18.6\% efficiency for SM $ZH$  and $> 99\%$ rejection for $Zb\bar{b}$. No extra information or discriminant to enhance sensitivity to new physics is used, consistently with the STXS hypotheses.  Figure~\ref{fig:stxs_crosssec} shows the predicted cross sections for the various samples in the STXS bins. The $Zb\bar{b}$ contribution has clearly been brought under control by the BDT discriminant. In the bins that have been split by jet multiplicity, this contribution appears to have a larger relative increase when going from zero to one or more jets, probably owing to the dominant $gg$-initiated contribution to this process increasing the radiation probability. We see that the EFT contributions diverge from the SM prediction with increasing $p_T$, as expected, and the $Zb\bar{b}$ background also becomes less and less important. 

\section{Statistical Hypothesis testing}
\label{sec:test}
A statistical analysis is carried out to estimate the sensitivity to SMEFT effects in Higgs boson interactions using the STXS measurements of Section~\ref{sec:stxs}.
We compare that sensitivity to the one obtained from a dedicated analysis using the multivariate classifier described in Section~~\ref{sec:training}. To this effect, a simple significance analysis is used based on the {\sc RooStats} framework~\cite{Moneta:2010pm} which determines the expected significance using an asymptotic calculator with nominal Asimov data sets and a one-sided profile likelihood. 

A few approximations have been made. No systematic uncertainties have been considered, even if the generated events have been smeared to reflect the limited resolution and corrected for the finite efficiencies of $b$-tagging. The measurement in the STXS bins requires an extrapolation from the measured phase space, which includes those selections applied specifically to reject backgrounds, in this case from the $Z b\bar{b}$ process. In the present analysis this is achieved using a BDT specifically trained to select $H\to b\bar{b}$ over $Z b\bar{b}$ events. For the STXS analysis, a BDT cut is applied, retaining 18.6\% of all SM Higgs events but less than 1\% of the $Z b\bar{b}$ background. For the BDT analysis targeting EFT operators, two settings are explored: using the same BDT requirement as in the STXS analysis and additionally using a selection requirement of the BDT that leads to similar acceptances as in the STXS case. 

In general, the sensitivity to any
 non-SM contribution to a measured cross section can only come from the events recorded
 in the fiducial volume, prior to any model-dependent extrapolations.
 Therefore, we do not use such extrapolations in this analysis. We perform our study 
 directly in the region of the phase space selected by the BDT cuts.
 In a real-life analysis, however, some degree of extrapolation is always performed. 
 Different acceptances of SM and BSM can therefore play a role in BSM searches, if in the latter the events selected at detector level are extrapolated assuming SM acceptances alone. Nevertheless, if one models the acceptances of the BDT-selection properly for both cases, this kind of effects can be accounted for.
The information on acceptances is, however, not always reported in the experimental analyses.
 
 The acceptances of our BDT-selection for both SM and BSM events with $c_{\sss HW} \neq 0$ are summarized in Table~\ref{tab:bkg_acceptance}. 
The acceptance for the SM Higgs can (depending on the BDT cut) be very similar for both STXS BDT (i.e.~the BDT used to reject $Zb\bar{b}$) and EFT optimized-BDT (around 19\%). The acceptance for events with a Wilson coefficient $c_{\sss HW}=0.03$ is larger than that by about a factor of 1.5 whilst it is smaller by the 25\% for $c_{\sss HW}=-0.03$. For the samples produced with a smaller Wilson coefficient, $c_{\sss HW} = \pm0.01$, the acceptances are slightly closer to the SM Higgs scenario, which is expected since for $c_{\sss HW} \to 0$ the SM is restored. The smaller the Wilson coefficients, the smaller the issues from acceptance effects.

\begin{table}[!h]
\begin{center}
{\scriptsize
\begin{tabular}{|l|c|c|c|}
\hline  
Sample		&  STXS: Acceptance (BDT$_{\mathrm{SM}}$) [\%] &  BDT: Acceptance (BDT$_{\mathrm{SM}}$, &  BDT: Acceptance (BDT$_{\mathrm{SM}}$, 	\\ 
		&  &  same as STXS SM-acceptance) [\%] &  same cut as STXS-BDT) [\%] 	\\ \hline

Higgs (BDT for $c_{\sss HW} = +0.03$) & 	18.6	&  18.5	& 29.6	\\
Higgs (BDT for $c_{\sss HW} = -0.03$)& 	18.6	&  18.3	& 33.1	\\
Higgs (BDT for $c_{\sss HW} = +0.01$)& 	18.6	&  18.6	& 33.4	\\ 
Higgs (BDT for $c_{\sss HW} = -0.01$)& 	18.6	&  19.8	& 34.2	\\ \hline

$c_{\sss HW} = 0.03$ &	31.1 & 31.8& 42.7\\
$c_{\sss HW} = -0.03$ &	14.4 &13.4& 28.2\\

$c_{\sss HW} = 0.01$ &	22.9	&23.0& 37.9\\
$c_{\sss HW} = -0.01$ &	15.9	&17.2& 31.7\\ \hline

$Z b\bar{b}$ &	$<1.0$	& $\sim$1.0 & $\sim$3.0\\ \hline

\end{tabular}
}
\vskip0.5truecm
\caption{Acceptances (in \%) of the first BDT selection, meant to separate $H\to b\bar{b}$ from $Z b\bar{b}$ production.}
\label{tab:bkg_acceptance}
\end{center}
\end{table} 

\begin{figure}[htb]
\centering
      \includegraphics[width=0.45\textwidth]{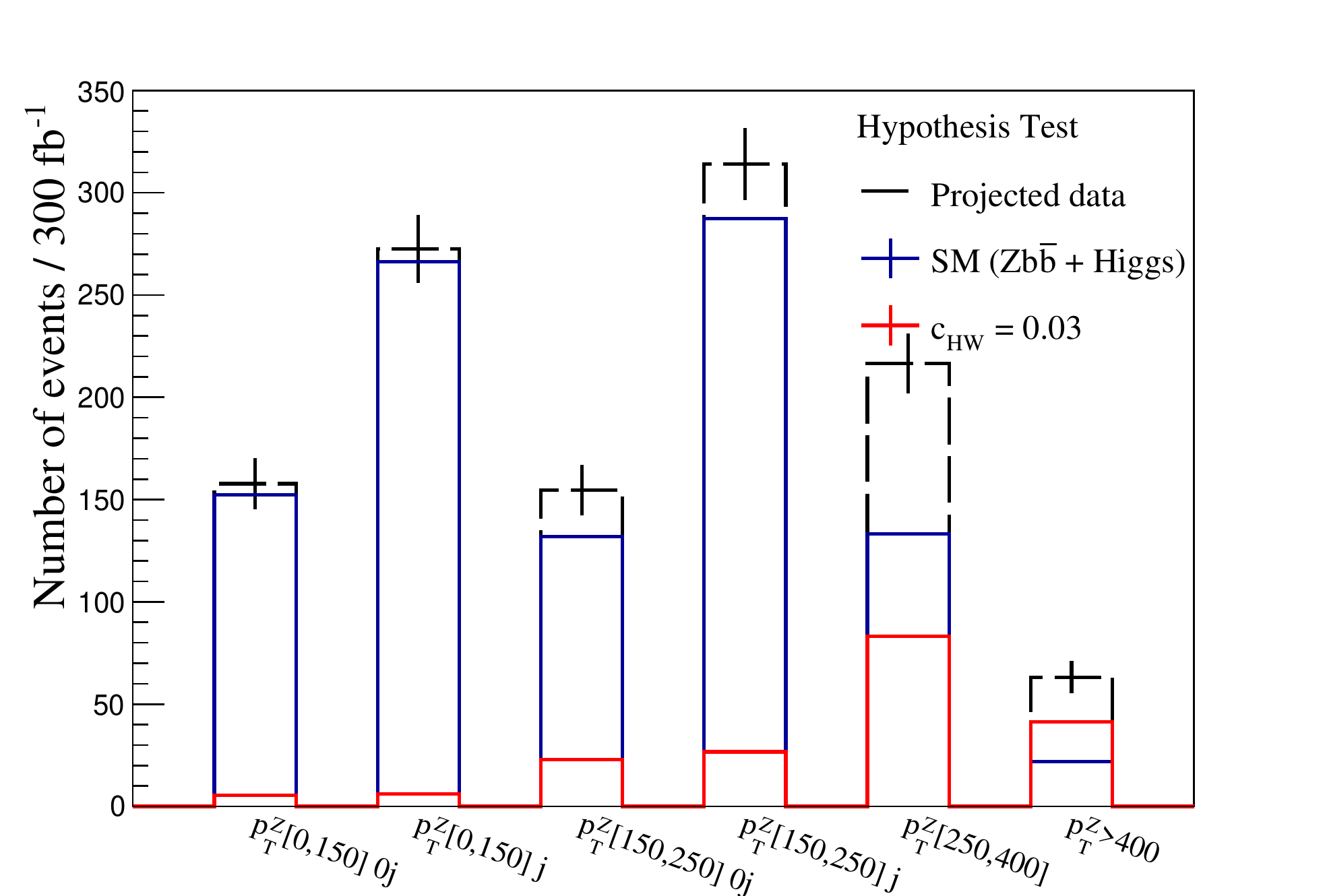}
      \includegraphics[width=0.45\textwidth]{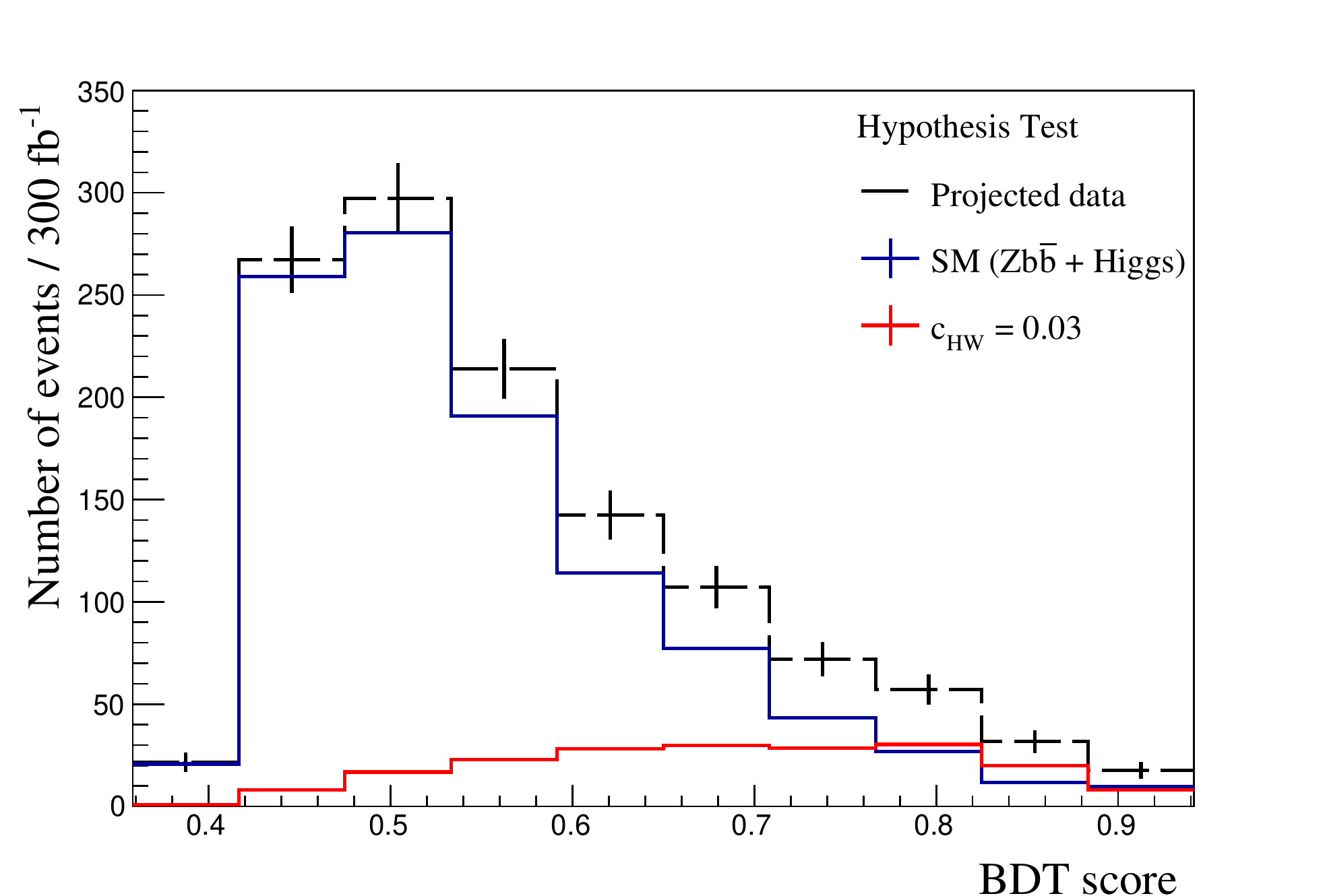}
      \caption{Predicted number of events selected for 300 fb$^{-1}$. Since the acceptance extrapolation into the STXS phase space does not change the available data statistics, it it not applied here, i.e.~the distributions are shown after the first cut on the BDT classifier to reject the $Z b\bar{b}$ background.}
\label{fig:hypotest}
\end{figure}

After application of the BDT requirements to reject $Z b\bar{b}$, either the STXS binning or the distribution of the other BDT classifiers are used to estimate the sensitivity to new physics $c_{\sss HW} \neq 0$. Three different luminosity scenarios are investigated: the full LHC Run-2 results, corresponding to 150 fb$^{-1}$, the integrated luminosity projected for LHC Run-3 (300 fb$^{-1}$) and the expected data collected at the High-Luminosity LHC (3000 fb$^{-1}$). The significance is determined simultaneously in the 6 STXS bins (see Section~\ref{sec:stxs}) and in the BDT discriminant distributed in 10 equal bins. (We checked that a finner binning did not significantly change the significance.) For both, STXS and BDT discriminant, no uncertainties on the shapes of these distributions are assumed. Figure~\ref{fig:hypotest} depicts the distributions used as inputs in the STXS (left) and the BDT (right) case for 300 fb$^{-1}$. The SM hypothesis ($Z b\bar{b}$ + $ZH$) is shown as blue line, whereas the BSM signal with $c_{\sss HW} = 0.03$ is shown as red line. They are added in the signal+background hypothesis which is depicted as dashed black line. A simple significance test for the signal+background hypothesis is carried out using the {\sc RooStats} framework~\cite{Moneta:2010pm} for these binned distributions.
 
To get a feeling of how realistic the scenario considered here is, the significance of a Higgs discovery in the STXS scenario is also investigated, in addition to the BSM sensitivity. The expected significances for the 2-lepton channel we studied here are 1.9 for ATLAS~\cite{Aaboud:2017xsd} and 1.8 for CMS~\cite{Sirunyan:2017elk} for $\sim$36 fb$^{-1}$. This is about what is expected from the simplified studies herein, which do not account for systematic uncertainties (which make up to half the total uncertainty in the measurements) but are not optimized for a Higgs observation. The expected significance of a Higgs signal compared to a background-only (i.e.~$Z b\bar{b}$) sample is shown in Table~\ref{tab:significances}.

Table~\ref{tab:significances} also summarizes the significances found for the three luminosity scenarios for the hypothesis tests for the STXS and the BDT approach for Wilson coefficients of $c_{\sss HW} = \pm0.03$ and $c_{\sss HW} = \pm0.01$. In the case of the BDT approach and $c_{\sss HW} = \pm0.03$, three alternatives were tested. For these, either the first BDT selection with the same STXS SM-acceptance or the same BDT cut as STXS are investigated. In addition, an alternative non-optimal BDT discriminant is used. This was trained not on the targeted Wilson coefficient (e.g. $c_{\sss HW}=+0.03$), but on the one with opposite sign (-0.03). 

\begin{table}[!h]
\begin{center}
{\scriptsize
\begin{tabular}{|l|c|c|c|c|c|c|c|c|c|}
\hline  
Hypothesis test		&  Full Run-2 (150 fb$^{-1}$) 	& LHC Run-3 (300 fb$^{-1}$) 	& HL-LHC  (3000 fb$^{-1}$) \\ \hline

STXS: Higgs discovery	& 		3.01 &  	3.70
			& 8.06 \\ \hline

STXS: $c_{\sss HW} = 0.03$ &		6.44 	&	8.82	& 26.46\\
STXS: $c_{\sss HW} = -0.03$ &		1.66&	2.24 &  6.44\\
BDT: $c_{\sss HW} = 0.03$ (STXS SM-acceptance)&		 6.29&	8.58 & 25.61\\
BDT: $c_{\sss HW} = -0.03$ (STXS SM-acceptance)&		1.80&	2.44& 7.24\\ \hline
BDT: $c_{\sss HW} = 0.03$ (same BDT cut as STXS)&	6.17&	8.64 & 27.03\\
BDT: $c_{\sss HW} = -0.03$ (same BDT cut as STXS)&	1.74 &	2.08 & 7.50\\ \hline
BDT: $c_{\sss HW} = 0.03$ (alt BDT cut)		&4.40	&	6.15	&19.18\\
BDT: $c_{\sss HW} = -0.03$ (alt BDT cut)	&1.44	&	2.41	& 6.69\\ \hline
STXS: $c_{\sss HW} = 0.01$ &		2.26		&	3.04			& 8.78\\
STXS: $c_{\sss HW} = -0.01$ &		1.08		&	1.46			& 4.30\\ 
BDT: $c_{\sss HW} = 0.01$ &		2.62		&	3.07			& 8.90\\
BDT: $c_{\sss HW} = -0.01$ &		1.44		&	1.99			& 6.10\\ \hline
\end{tabular}
}
\vskip0.5truecm
\caption{Expected significances for the different scenarios.}
\label{tab:significances}
\end{center}
\end{table}

\section{Conclusions}
\label{stsxsec:conclusions}
We have performed an exploratory study comparing the sensitivity to higher dimensional operators of the proposed STXS measurements in $ZH$ production to an optimised analysis exploiting multivariate methods. We considered four benchmark scenarios in which the $\mathcal{O}_{\sss HW}$ operator coefficient is set to values $c_{HW}=\pm 0.03$ and $c_{HW}=\pm 0.01$. The former case corresponds to saturating existing limits from a global fit to LHC Run 1 data and precision electroweak measurements, while the latter case intends to showcase a scenario with smaller deviations from SM expectations. The sensitivity was quantified by the expected statistical significance against the SM hypothesis obtainable after a collected integrated luminosity of 150, 300 and 3000 fb$^{-1}$, taking into account the presence of the dominant SM background of $Zb\bar{b}$ production. The contribution of this background is efficiently mitigated by training a BDT classifier to distinguish this process from SM $ZH$ production and first cutting on this discriminant before performing the two alternative SM vs EFT significance analyses. The discriminant was able to effectively reduce this background contribution down by two orders of magnitude.

Overall, very large significances can be expected for the benchmarks saturating the current limits, while the benchmarks for the smaller Wilson coefficients are not likely to be identified beyond 3$\sigma$ until the High-Luminosity LHC run. We observe that the final performance of the BDT analysis does not differ significantly from the differential information in $Z$-boson $p_T$ offered by the STXS, with the exception of the $c_{\sss HW}=-0.01$ case, which predicts the smallest deviation from the SM case. Here, the discriminating power of the BDT output over the differential $p_T$ distributions becomes apparent, suggesting that once the sensitivity of the STXS measurements is saturated, moving towards optimised multivariate methods remains well-motivated. The exercise was performed in a simplified situation, largely ignoring detector effects besides a parametrised $b$-jet smearing implementation and $b$-tagging efficiency corrections as well as all other potential sources of systematic uncertainty. We leave a more thorough investigation, including these effects as well as the possibility of including other significant backgrounds to a follow-up study. 

By comparing the optimised BDT discriminants for the different EFT benchmarks, we conclude that there is significant information overlap between them but that some parameter dependence remains. This means that one would benefit from a parametrised learning approach, in which the new physics parameter is also fed in as an input to the discriminant training. This can be understood from the presence of both an interference and squared contribution of the EFT $ZH$ amplitude in the new physics signal. The shape of EFT squared contribution has the benefit of being independent of the value of the Wilson coefficient, while the relative impact of the interference term depends very much on this value. In the `large' $c_{\sss HW}$ benchmarks, the contribution from the quadratic term in the Wilson coefficient is clearly dominant at high energies, as evidenced by the positive relative contribution over the SM prediction for both $\pm0.03$ in the STXS overflow bin, see Figure~\ref{fig:stxs_crosssec}. On the other hand the $-0.01$ case consistently predicts a deficit with respect to the SM. The fact that the BDT outperforms for this benchmarks may imply that one can obtain better sensitivity using these methods for EFT signals in which the interference with the SM amplitude is significant, which may also be considered more `well-behaved' concerning the EFT expansion. However, this effect may also be caused by the greater loss in acceptance post $Zb\bar{b}$ BDT cut suffered by the negative $c_{\sss HW}$ benchmarks and should be further investigated.

One should bear in mind that in this first study, a rather kinematically simple process has been chosen.
Indeed, in $ZH$ production the $p_T$ $Z$-boson is strongly correlated to the energy flow through the production vertex, in which the EFT effects occur. It is therefore not surprising that we do not observe a huge difference in significance in this case. Further investigations concerning a comparison between BDT and STXS for more complicated kinematic environments would be interesting, \emph{e.g.}, for other $2\to3$ production modes such as vector boson fusion or $t\bar{t}H$ associated production.  Furthermore, it should be noted that although our BDT analysis is touted as an `optimised' discriminating method, the fully potential of the BDT information was not exploited in this analysis. In short, by cutting on the SM vs $Zb\bar{b}$ variable and fitting on the resulting one-dimensional discriminant, some amount of exclusion power was sacrificed for the sake of simplicity. In the ideal case, a two-dimensional fit on the initial BDT classifier would be performed, an exercise which we leave for the follow-up study.

\section*{Acknowledgments}

We thank the organizers and conveners of the Les Houches workshop, ``Physics
at TeV Colliders'', for a stimulating meeting. K.L. is supported by the European Union's
Horizon 2020 research and innovation programme under ERC grant agreement
No. 715871. K. M. is supported in part by the Belgian Federal Science Policy Office
through the Interuniversity Attraction Pole P7/37 and by the European Union's Horizon 2020
research and innovation programme under the Marie Sk\l{}odowska-Curie grant agreement
No. 707983. 
P. M. was supported in part by the Swiss National Science Foundation under the project
200020\_162665.



\AddToContent{J.~de~Blas, K.~Lohwasser, P.~Musella, and K.~Mimasu}
\renewcommand{\thesection}{\arabic{section}}


\renewcommand{\eq}[1]{Eq.~(\ref{#1})}

%
%
%



\chapter{Improved BSM Sensitivity in Diboson Processes at Linear Colliders}

{\it D.~M.~Lombardo, F.~Riva, P.~Roloff}


\begin{abstract}
We study diboson processes at CLIC and asses the reach of linear colliders to new physics effects in the form of effective field theory (EFT) operators. Given that inclusive measurements in diboson process suffer from certain SM-BSM non-interference rules, we perform non-inclusive analyses that include azimuthal differential information.
\end{abstract}

\section{Motivation}
Standard Model (SM) precision tests are an increasingly important tool in new physics searches. New dynamics at a  mass scale $M$ can leave an imprint in observables at energies $E\ll M$, whose size is proportional to some power of $E/M$. These effects can be captured generically and systematically through an Effective Field Theory (EFT), organised as operators of increasing dimension, in addition to the SM. In most scenarios, the leading effects arise at the dimension-6 level (see however Refs.~\cite{Liu:2016idz,Bellazzini:2017bkb} for exceptions).


Given the upcoming European Strategy for particle physics, it is important to find benchmarks and scenarios that can be readily accessed by different types of experiments; these scenarios allow for a comparison of the reach of radically different machines. In this note we study $W^+W^-$ production at future linear $e^+e^-$ colliders (see also \cite{Battaglia:2004mw,Ellis:2017kfi,Rahaman:2017aab,GutierrezRodriguez:2008nk,Falkowski:2014tna}).  
Diboson processes have a rich variety of physical information, incapsulated at high-energy in the lognitudinal and transverse polarizations: from a BSM perspective these can be considered genuinely different processes testing genuinely different physics.
Beside their interest for future planning, linear colliders also offer an interesting playground for complex studies: they allow to isolate and understand in detail analyses that can then be brought over to the framework of hadron machines.
  $W^+W^-$ processes are a perfect example of this. Indeed, for what concerns new physics in the transverse polarizations, SM and BSM exhibit different helicity structures (see Ref.~\cite{Azatov:2016sqh} for a recent discussion), so that the two amplitudes do not interfere in inclusive measurements: an important drawback of traditional analyses in the context of a precision program. Non-inclusive differential distributions in the azimuthal angles of the $W$ boson decay planes do bear the interference information (see \cite{Panico:2017frx}), in the form of a modulating signal that vanishes once integrated over (to reproduce the non-interference results)\footnote{Interference information is also present in the off-shell region \cite{Helset:2017mlf}.}. At linear colliders this modulating signal is very visible, as we will show in this note. This will allow us to discuss it in isolation and quantify the impact of interference.
  
Even when new physics is in the longitudinal polarizations, BSM searches as precision tests are challenging. In the SM, the unpolarized cross section is dominated by the transverse-transverse components, and the longitudinals are small. Therefore, while the SM and BSM do interfere here, the dominant SM contribution acts as an irreducible background that reduces the sensitivity of the experiment. In this note we point out that beam polarization can play a crucial r\^ole in this context, as it can substantially reduce the transverse component, which receives the largest contribution from a t-channel involving left-handed electrons.

\section{BSM in Transverse Polarizations}

Diboson processes are often presented as measurements of anomalous trilinear gauge couplings (TGCs), associated with the parameters $\lambda_\gamma$, $g_1^Z$ and $\kappa_\gamma$ of Ref. \cite{Hagiwara:1986vm}. These are in correspondence with dimension-6 operators.

For our scope, these effects  can be divided into two classes: couplings that contribute to the transverse and longitudinal amplitude. In the former class, which we discuss in this section, we have the CP-even operator 
\begin{equation}\label{opo3w}
{\cal O}_{3W}=\epsilon^{ijk}W_\mu^{i \nu}W_\nu^{j\rho}W_\rho^{k \mu}\,.
\end{equation}
BSM effects in the transverse amplitudes are difficult to test, as we now explain.
The problem of the BSM ${\cal O}_{3W}$ operator is that it produces, at tree-level and at high-energy, dominantly $++$ or $--$  helicities in the final states, with amplitudes ${\cal A}^{++}_{BSM}={\cal A}^{--}_{BSM}$. This does not interfere, in inclusive $2\to 2$ scattering, with the SM amplitude ${\cal A}_{SM}$ \cite{Azatov:2016sqh}. 
SM processes have, in the high-energy and classical limits, dominantly $+-$, $-+$ or $00$ helicity. The latter is however smaller and has little impact on this part of the analysis.

Nevertheless, the amplitude for $e^+e^-\to 4f$ decays into fermions can in principle interfere. This interference  is proportional to a function of the azimuthal angles  of the decay planes of the fermion/anti-fermion originating from the $W^+$ and $W^-$ respectively. 
\begin{figure}
\centering
\includegraphics[width=0.45\textwidth]{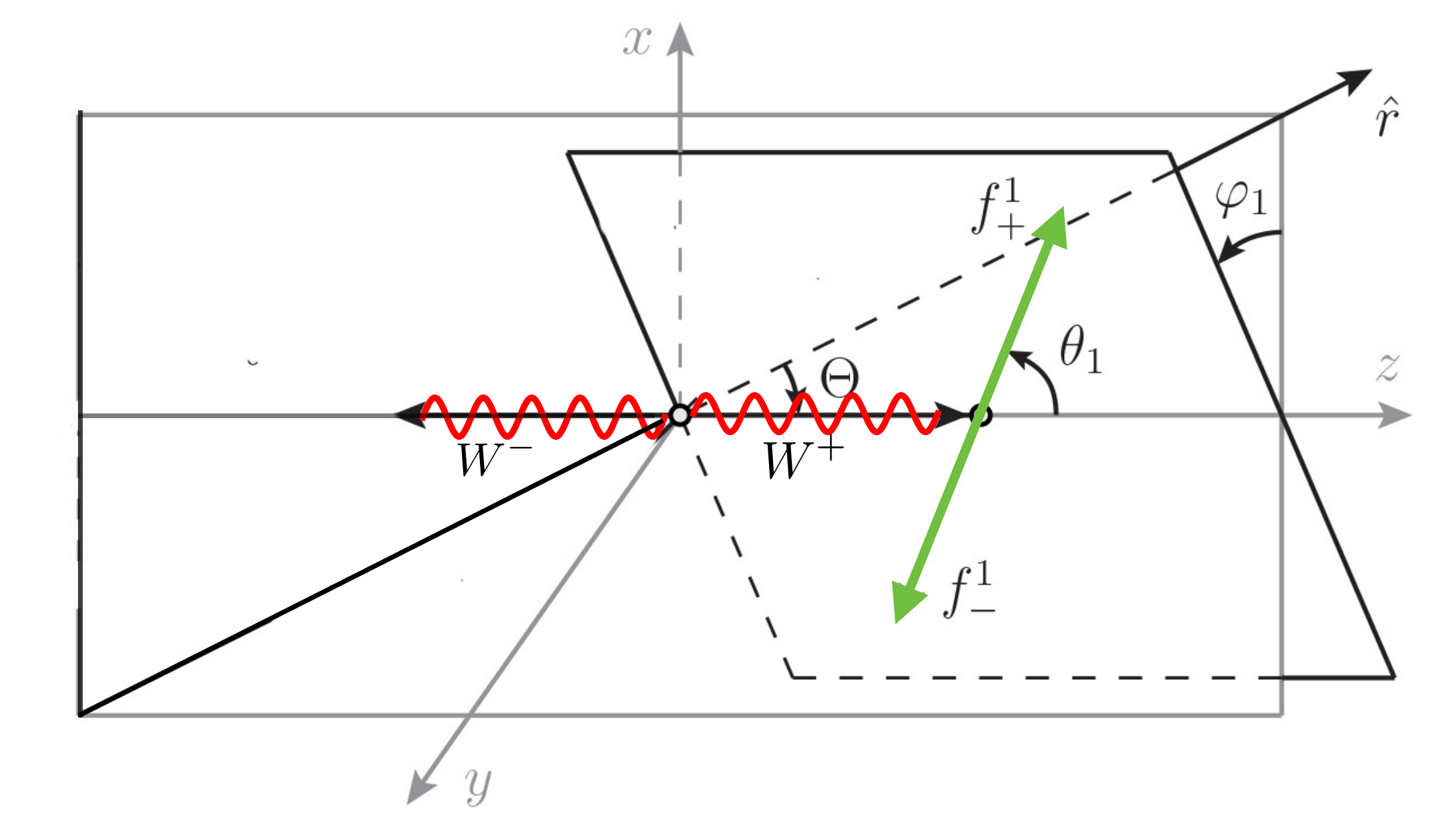}
\caption{Definitions of the polar angle $\Theta$ and azimuthal angle $\varphi$.}
\label{fig:plane}
\end{figure}

In this note we focus on a single-differential distribution and study the azimuthal distribution of the decay products of one of the two $W$s only, as illustrated in Fig.~\ref{fig:plane}.
 We remain inclusive about the other $W$, which can then  be thought as a state of well defined helicity. The interference term, between the transverse-transverse amplitudes, reads~\cite{Jacob:1959at}
\begin{eqnarray}\label{intWg}
&&I^{WW}\hspace{-2pt}\propto {\mathcal{A}}_{++}^{\textrm{\tiny{BSM}}}\hspace{-2pt}
\left[\hspace{-1pt}{\mathcal{A}}_{-+}^{\textrm{\tiny{SM}}}\hspace{-4pt}+\hspace{-2pt}{\mathcal{A}}_{+-}^{\textrm{\tiny{SM}}}\hspace{-2pt}
\right]\hspace{-2pt}\cos{2\varphi}\,,
\end{eqnarray}
see also \cite{Panico:2017frx} for more details.

Interference vanishes when integrated over, reproducing the above non-interference result.
The question we want to address here is how this measurement performs in lepton colliders: how much an azimuthal differential distribution improves upon an inclusive measurement, in terms of BSM reach?

\subsection{Ambiguity and Channel Selection}

The angle $\varphi$  is defined in Fig.~\ref{fig:plane} making reference to the outgoing fermion of positive helicity ($f_+$ in the figure). When the $W$ is decaying hadronically, this information is unaccessible\footnote{In this context it would be interesting to study decays including charm quarks; we leave this for the future.}, implying an ambiguity
\begin{equation}
\{\Theta,\varphi\}\leftrightarrow \{\pi-\Theta,\varphi+\pi\}\,.
\end{equation}
Such ambiguity doesn't prevent us, however, from observing distributions of the form \eq{intWg}.

For leptonically decaying $W$-bosons the situation is different: here the charged lepton has always well-defined helicity (depending of whether it's a $W^+$ or $W^-$), but the plane is defined only if the neutrino momentum is also known. This can be in principle reconstructed from the kinematics if only one neutrino is present in the event.

Here we focus on the semileptonic channel $\nu l^+\bar q q $. This channel has a large branching ratio, because of the hadronic contribution, and at the same time allows to (almost) fully reconstruct the neutrino, for which the transverse components of momentum can be identified as missing energy.
By requiring that the invariant mass of charged lepton and neutrino exactly reconstruct the $W$-mass, one finds an equation with two solutions for the longitudinal momentum of the neutrino. At hadron machines it is impossible to single out which of these two solutions corresponds to the real one, and this introduces a further ambiguity in the angle reconstruction that forbids, for instance, observation of CP-odd effects. At linear lepton colliders, however, the total center-of-mass energy is known, up to initial state radiation (ISR) and beam-strahlung. Now, the two solutions have  different neutrino longitudinal momentum and therefore different amounts of ISR, hence different energy. So, it is possible, by appropriately cutting on the energy of the hadronically decaying $W$, to avoid the ambiguity completely. We assume here that this is the case, and leave a more thorough study of this possibility for future work. In this work we therefore assume that the decay plane of the leptonic $W$ is fully reconstructible and study its distribution, while ignore the hadronic $W$ distribution.

\subsection{Amplitudes and Choice of Cuts}

\begin{figure}[ht]
\centering
\includegraphics[width=0.45\textwidth]{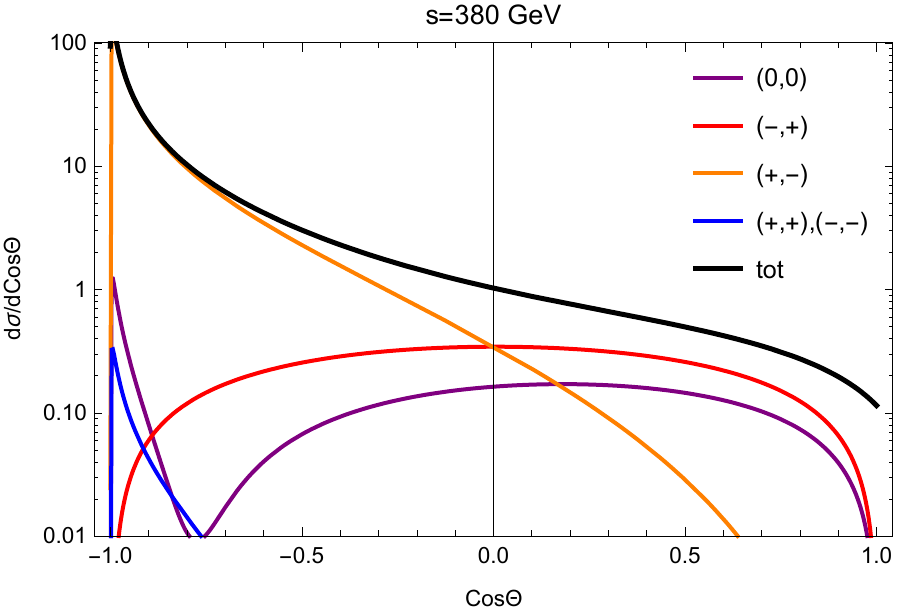}
\caption{Polar angle distribution of the different helicity amplitudes in the SM.}
\label{fig:dist}
\end{figure}

Knowledge of the SM amplitude can guide us through the most appropriate choice of cuts and binning.
In the high-energy limit, the SM amplitudes for inclusive dibosons read
\begin{equation}\label{ampssm}
{\cal A}^{-+}_{\textrm{\tiny{SM}}}=-g^2 \sin\Theta\quad\quad
{\cal A}^{+-}_{\textrm{\tiny{SM}}}=2g^2  \sin ^4\frac{\Theta}{2}  \csc \Theta\quad \quad
{\cal A}^{00}_{\textrm{\tiny{SM}}}=\frac{1}{2}  (g^2+{g^\prime}^2) \sin\Theta
\end{equation}
where $\Theta$ is the polar angle, corresponding to the angle between the incoming electron and the outgoing $W^+$. 
The SM amplitudes are illustrated in Fig.~\ref{fig:dist} for $\sqrt{s}=380$ GeV.
The BSM amplitude is instead
\begin{equation}
{\mathcal{A}}_{+\,+}^{\textrm{\tiny{BSM}}}={\mathcal{A}}_{-\,-}^{\textrm{\tiny{BSM}}}\approx C_{3W} 6e\sqrt{2}M^2_{W\gamma} \sin\Theta \,
\end{equation}
where $C_{3W}$ is the (dimensionfull) coefficient of the ${\cal O}_{3W}$ operator as it appears in the Lagrangian.

The important lessons here are:
\begin{itemize}
\item In the forward region $\cos\Theta\approx 1$ both SM and BSM vanish, so that this region is not favorable
\item In the backward region $\cos\Theta\approx -1$ the BSM vanishes and the SM explodes because of the $t$-channel neutrino pole; the interference term is in fact finite. The signal over sqrt-background vanishes in the backward point, but increases rapidly ($\sim\Theta^{3/2}$) as we approach the central region, so that even this backward region can have interesting information. 
\item In the central region  $\cos\Theta\approx 1$ the BSM amplitude has its maximum, and the SM switches from being dominated by the $+-$ to being dominated by $-+$. Most importantly, since the latter SM amplitudes have opposite sign (see \eq{ampssm}): the overall SM amplitude changes sign!
\end{itemize}

In light of these, we understand that the most important region for our analysis will be $\cos\Theta\sim 0$. Moreover, it is important to separate the analysis (or implement an asymmetry) for 
\begin{equation}
\cos\Theta<0\quad \textrm{and}\quad \cos\Theta> 0;
\end{equation}
because of the opposite SM amplitude sign, the sum of the interference terms from these distinct regions tends to cancel (see Fig.~\ref{fig:plotssmsbm}). So, in what follows, we consider 4 bins in polar angle
\begin{equation}\label{polarbin}
\cos\Theta\in [-1,-0.5,0,0.5,1]
\end{equation}
with particular hopes on the central bins.

\section{Analysis}

As benchmarks for future colliders we consider CLIC at 380 GeV, with 500 fb$^{-1}$ of integrated luminosity, and CLIC at 3 TeV with 3 ab$^{-1}$ of luminosity~\cite{Battaglia:2004mw}. We leave the study of a richer variety of scenarios for the future.

For this preliminary study we use {\sc MadGraph}~\cite{Alwall:2014hca} and simulate the process $e^+e^-\to W^+ W^-$ where we take the $W^-$ to decay hadronically and the $W^+$ to decay into $e^+ + \nu$; we then multiply the crossection by a factor of 4 to account for decays into muons, and for the charge inverse process; we include an acceptance of 50\% to be conservatives. 

\begin{figure}[ht]
\centering
\includegraphics[width=0.3\textwidth]{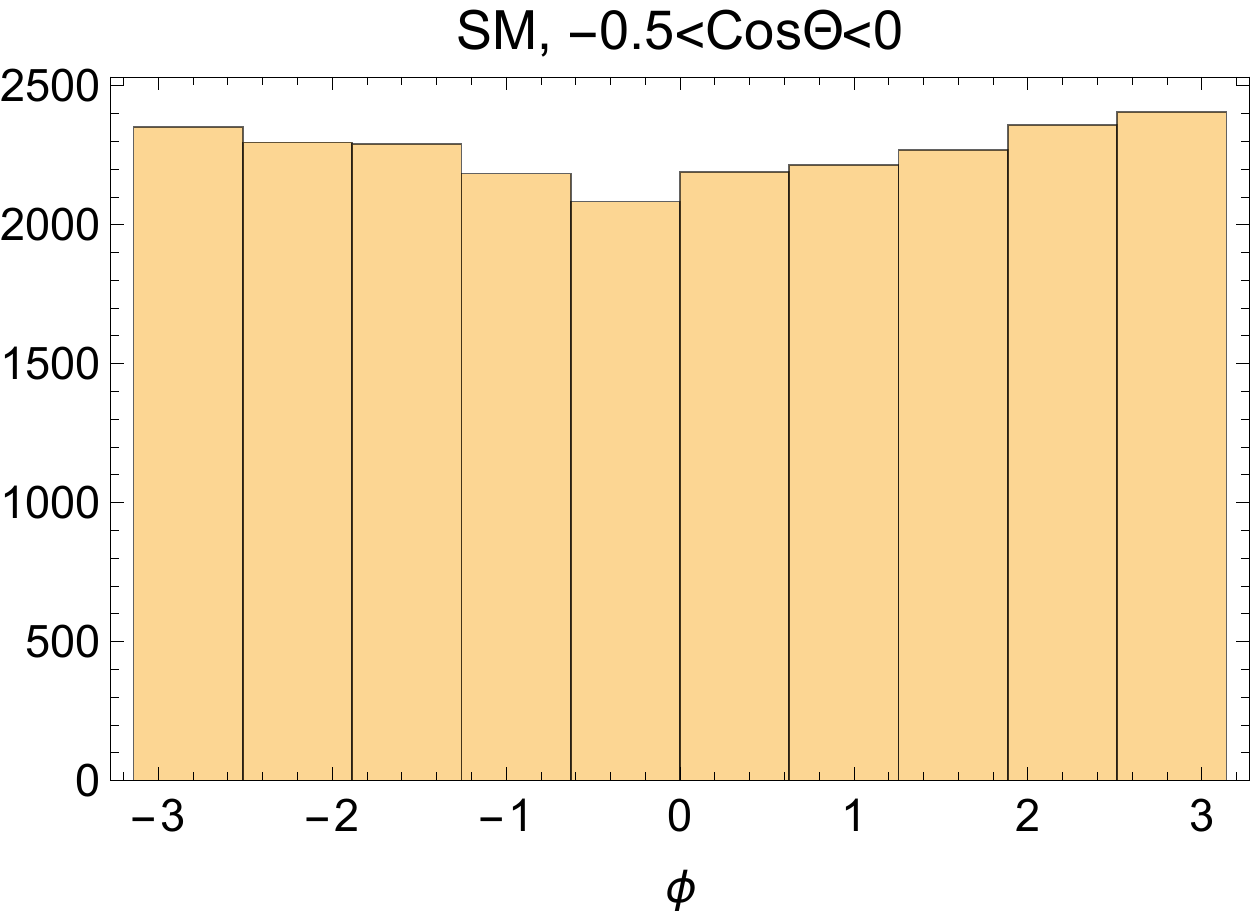}
\includegraphics[width=0.3\textwidth]{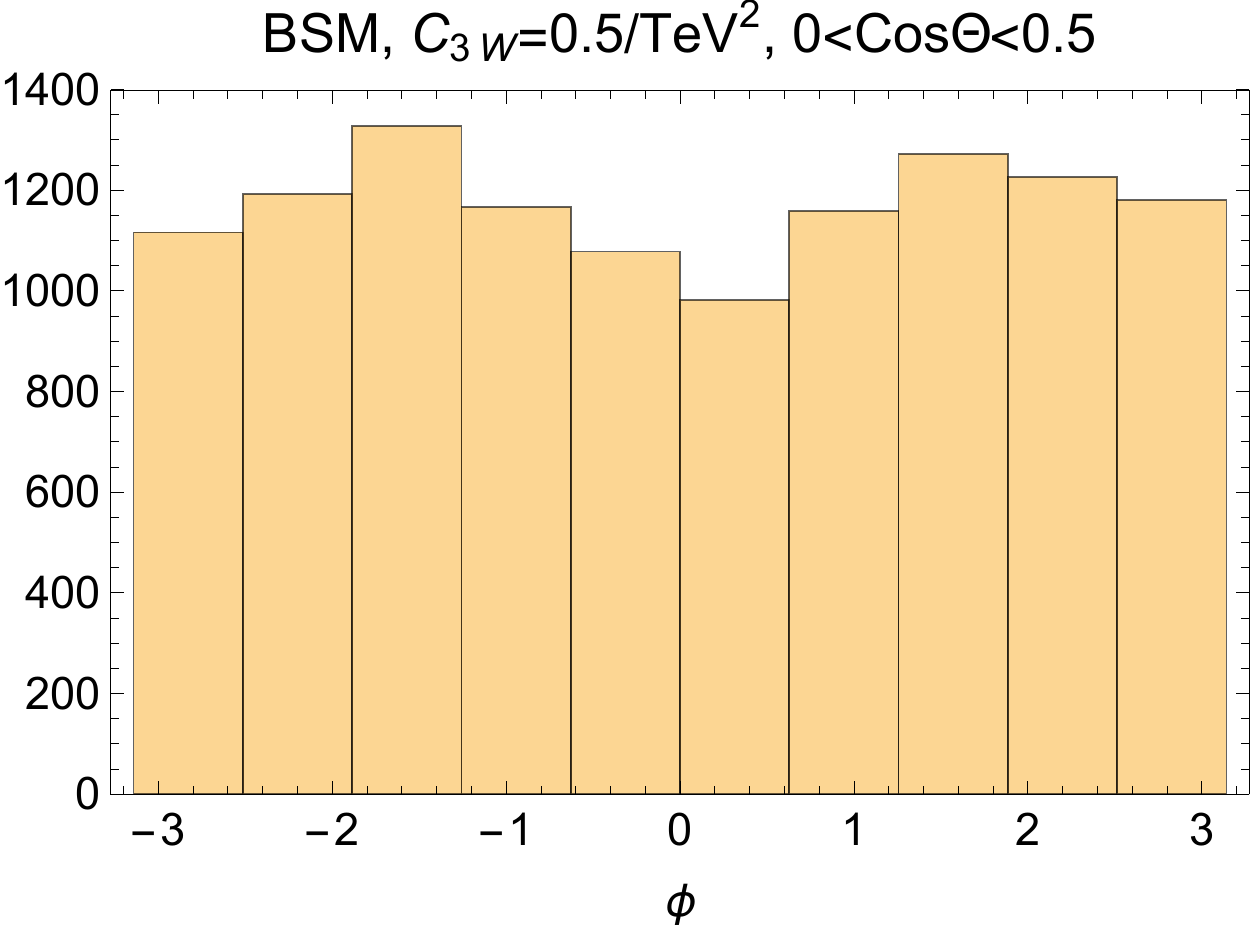}
\includegraphics[width=0.3\textwidth]{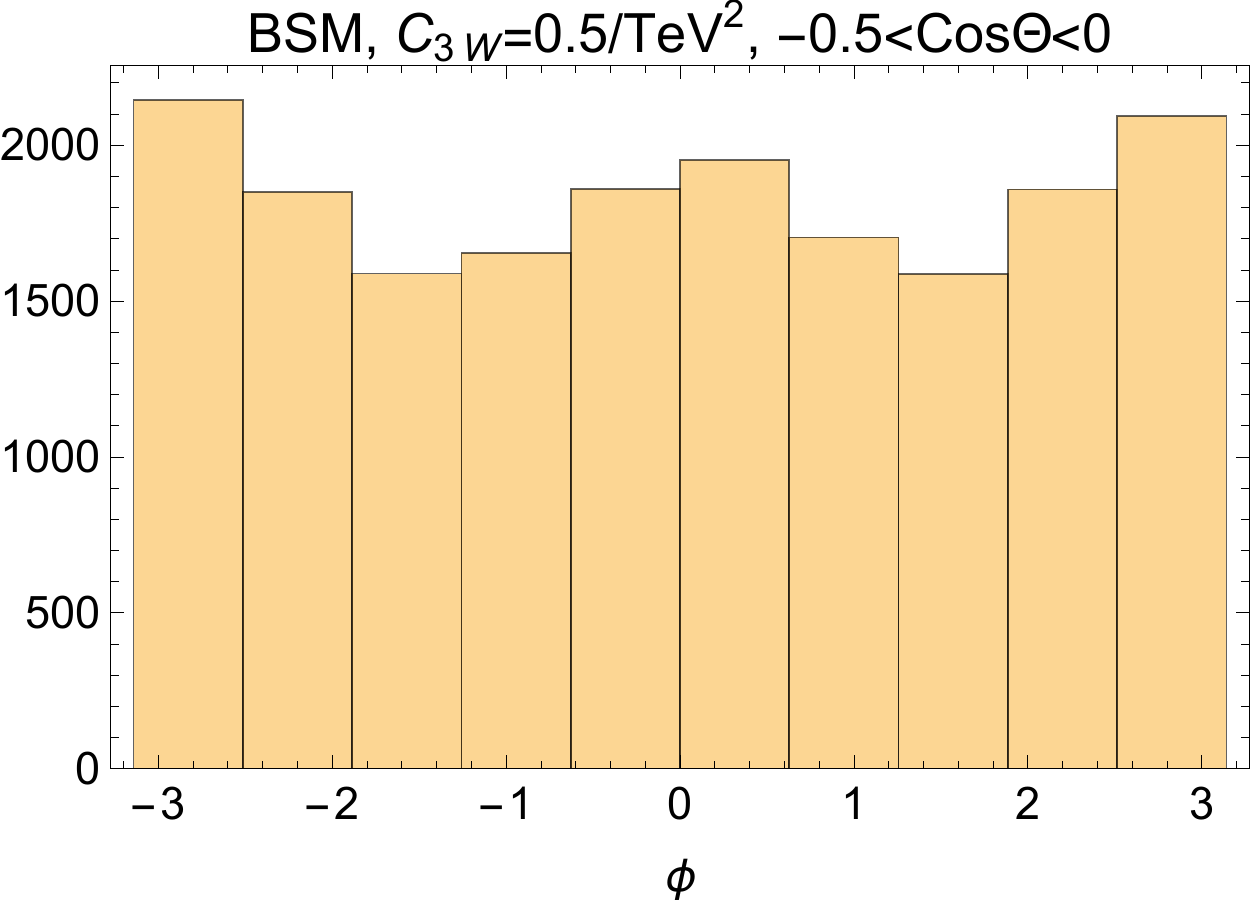}
\caption{Un-normalized histograms for the azimuthal differential distribution for different values of the BSM parameters.}
\label{fig:plotssmsbm}
\end{figure}

An example of the azimuthal distribution that we are interested in, is shown in Fig.~\ref{fig:plotssmsbm}. There it is also visible the fact that in the regions $\cos\Theta>0$ and $\cos\Theta<0$ the SM-BSM interference changes sign. The SM distribution is flat.

To access the azimuthal distribution, we bin $\varphi$ in 10 parts (in addition to the 4 bins in polar angle \eq{polarbin}) and perform a binned $\chi^2$-analysis in two different ways: differential in every $\varphi$ bin and inclusive. We study the reach in two different CLIC configurations: a low-energy one at 380 GeV, and a high-energy one at 3 TeV.
We include a 50\% acceptance, but extend the analysis to both elctrons and muons and to leptonic decays of either $W^+$ or $W^-$.\\

\noindent
{\bf Low Energy Run (380 GeV).} We assume a luminosity 500 fb$^{-1}$ and compare two scenarios with 1\% and 10\% systematic uncertainty $\delta_{syst}$ in all bins. Without azimuthal binning we find that the CLIC reach on the (dimensionful) Wilson coefficient of the operator \eq{opo3w}, is 
\begin{align}
\textrm{Without }\phi\textrm{ distribution:    } & c_{3W}\in[-1.4,1.1]\, \textrm{TeV}^{-2} \,\, (\delta_{syst}=1\%)\\  & c_{3W}\in[-15,6]\, \textrm{TeV}^{-2}\,\,\, (\delta_{syst}=10\%)
\end{align}
while including azimuthal data we obtain
\begin{align}
\textrm{With }\phi\textrm{ distribution:    } & c_{3W}\in[-0.6,0.6]\, \textrm{TeV}^{-2} \,\, (\delta_{syst}=1\%)\\  & c_{3W}\in[-2.5,2]\, \textrm{TeV}^{-2}\,\,\, (\delta_{syst}=10\%)
\end{align}
 We also find that the bounds are dominated by the central bins, as expected.\\
 
\noindent
{\bf High Energy Run (3000 GeV).} In this case we take 3 ab$^{-1}$ of luminosity and find
\begin{align}
\textrm{Without }\phi\textrm{ distribution:    } & c_{3W}\in[-0.12,0.12]\, \textrm{TeV}^{-2} \,\, (\delta_{syst}=1\%)\\  & c_{3W}\in[-0.13,0.13]\, \textrm{TeV}^{-2}\,\,\, (\delta_{syst}=10\%)
\end{align}
\begin{align}
\textrm{With }\phi\textrm{ distribution:    } & c_{3W}\in[-0.10,0.10]\, \textrm{TeV}^{-2} \,\, (\delta_{syst}=1\%)\\  & c_{3W}\in[-0.11,0.11]\, \textrm{TeV}^{-2}\,\,\, (\delta_{syst}=10\%)
\end{align}

\section{Outlook}

In this short note we have initiated an improved study of the BSM search prospects in diboson processes at future linear colliders. In particular, to overcome the fact that SM and BSM amplitudes do not interfere in inclusive measurements, we have studied the possibility of accessing interference via differential azimuthal measurements. Focussing on the angle spanned by a leptonically decaying $W$ in semileptonic processes, we have compared the reach with and without the azimuthal measurement, and found that the former is indeed  better.
In the near future we plan to extend the analysis in a number of ways that we mention in what follows.

First, complete double differential distribution (in both azimuthal angles) can be advantageous. This should be studied together with the possibility of looking at the fully hadronic amplitude (and investigate whether the ambiguity of not knowing which one is the positively charged $W$ has an impact on the BSM reach).
This also opens the door to accessing independently the CP even and CP odd distributions.
Finally, a more refined simulation setup, including WHIZARD \cite{Kilian:2007gr}, fast and full simulations, should be performed.

On a different front, it is interesting to study  BSM physics in the longitudinal polarizations (e.g. the operator ${\cal O}_W$). This certainly interferes with the SM, but suffers here from another problem: the longitudinal amplitude is  small in the SM compared to the transverse one, and acts in measurements of the inclusive (longitudinal+transverse) crossection as a background. In \cite{Franceschini:2017xkh}, in the context of LHC, a possible way out of this is described: selecting the central region, where the longitudinals are  suppressed.
At linear lepton colliders it is possible to polarize the beam. Taking the electron to be right-handed, implies that (in the high-E limit and assuming massless electrons) diboson processes can occur only through an s-channel diagram that allows only for longitudinal final states. Thus, beam polarization  kills the transverse channel that dominates the crossection (which is what makes the analysis of this channel poor). Interestingly, because of the equivalence theorem, new physics that modifies  $WW$ processes, also affects $Zh$ processes (see e.g. \cite{Craig:2015wwr,Cohen:2016bsd,Beneke:2014sba}): it would be nice to compare the reach of these different probes (along the lines of \cite{Durieux:2017rsg}).

We leave all these open question for a future detailed study.



\AddToContent{D.~M.~Lombardo, F.~Riva, P.~Roloff}
\renewcommand{\thesection}{\arabic{section}}



\chapter{Comparing effective field theory operator bases numerically}

{\it R.~Gr\"ober, O.~Mattelaer, K.~Mimasu}


\begin{abstract}
We compare numerically different processes computed with different operator bases in Standard Model effective field theory.
We show that while as expected at the dimension six level, they lead to the same numerical results for zero width of the propagating particles in the Feynman diagrams, once a non-zero width is introduced, different bases can lead to different results at the dimension six level. As in the SM this is related to the breaking of gauge invariance once particle widths are introduced. We show how the width can be consistently included in Standard Model effective field theory and provide first steps towards a consistent inclusion into the package {\tt MadGraph5\_aMC@NLO}.
\end{abstract}

\section{INTRODUCTION}
So far no signs of new physics has been detected  at the LHC. It is hence sensible to assume that there is a gap between the electroweak scale and the new physics scale. In this paradigm, effective field theories (EFT) are a reasonable, model independent tools to interpret searches for Beyond the Standard Model (BSM) interactions. They supplement the Standad Model (SM) Lagrangian with higher dimensional operators that parametrise deviations in interactions between SM fields. The way in which these are organised/classified depends on the set of assumptions one makes with respect to the possible global and gauge symmetries imposed at the Lagrangian level. Since all current Higgs boson measurements point to a SM like, CP-even Higgs boson that is part of an $SU(2)$ doublet, one can for instance assume that the $SU(2)\times U(1)$ symmetry is linearly realised.
Then the SM effective Lagrangian can be organised such that operators with higher dimensionality are suppressed by $(v/\Lambda)^{d-4}$, where $d$ denotes the dimensionality of the operator, $v$ the electroweak scale and $\Lambda$ the new physics scale. 
\\
Given the SM field content and assuming flavour universality as well as lepton and baryon number conservation, the leading BSM effects can be parametrised in terms of 59 dimension six operators~\cite{Grzadkowski:2010es}. In an earlier construction of an EFT basis \cite{Buchmuller:1985jz} 80 operators were found, however the differences were shown to be redundant operators that could be removed by equations of motions. So by using equations of motions or, alternatively, by redefinition of the SM fields different sets of operators can be obtained. Along this line, different operator bases have been proposed in the literature, such as the Warsaw basis \cite{Grzadkowski:2010es}, the BSM primaries basis \cite{Gupta:2014rxa, Ellis:2014jta, Falkowski:2014tna, Efrati:2015eaa} and the SILH basis \cite{Giudice:2007fh, Contino:2013kra}. The different bases are equivalent up to suppressed higher dimensional effects of $d >6$. A tool to translate between the different bases is {\tt Rosetta} \cite{Falkowski:2015wza}.
\\
Our goal in this study is to verify that different operators connected by a field redefinition really lead to numerically equivalent results. It is expected that, up to dimension-6, one should obtain identical results for on-shell scattering amplitudes for two equivalent, dimension-6 operator sets. In order to test this, we pick a single operator that we add to the Standard Model and then compare the results for several processes to the ones obtained from the operator set in which the original operator has been removed by performing a field redefinition. 
\\
Computations with unstable particles require the inclusion of the width of the particle. We will test how the inclusion of the width affects our comparison and will comment on how to consistently include the width in SMEFT. 
\\
This contribution is structured as follows: in section \ref{sec:operators} we give more details on the operators we chose and how we redefine the fields to obtain a new basis and in section \ref{sec:numanalysis} we show our numerical results and comment on the inclusion of the width for unstable particles. 
\section{OPERATORS \label{sec:operators}}
We exemplify the effect of the change in the basis by using the SM and a single new operator
\begin{equation}
\text{Model 1:} \hspace*{0.5cm}\mathcal{L}=\mathcal{L}_{SM}+ \frac{c}{\Lambda^2} i(H^{\dagger}\sigma^i \overleftrightarrow{D}_{\mu}H)( \bar{q}_{3}\gamma_{\mu}\sigma^i q_3) \label{eq:model1}
\end{equation}
with $\sigma^i$ ($i=1,2,3$) the SU(2) generators and $\{\sigma_i, \sigma_j\}=\delta_{ij}/2$, and $q_3$ are the third generation fermion SU(2) doublets. The scale $\Lambda$ is assumed to be much larger than the electroweak scale. The Wilson coefficient $c$ is generic.
We define 
\begin{equation}
(H^{\dagger}\sigma^i \overleftrightarrow{D}_{\mu}H)= H^{\dagger} \sigma^i (D_{\mu}H)- (D_{\mu}H)^{\dagger} \sigma^i H
\end{equation}
and $H=1/\sqrt{2}\,(0,v+h)^T$ in the unitary gauge. This operator leads to a modification of the $Wtb$, $Ztt$ and $Zbb$ vertices, as well as associated contact interactions with one and two Higgs bosons.
We now eliminate the dimension-6 operator in eq.~\eqref{eq:model1} by a field redefinition.
\begin{equation}
W_a^{\mu}\to W_a^{\mu}+ \frac{\tilde{c}}{\Lambda^2} \bar{q}_3\gamma^{\mu} \sigma_a q_3\,.
\end{equation}
The field redefinition adds a contribution to the original Lagrangian at order $1/\Lambda^2$ that is proportional to the $W$ equation of motion. The operator is removed by setting $\tilde{c}=-c$. Note that an alternative field redefinition could also be used to remove the operator in eq.~\eqref{eq:model1}, namely  $W_a^{\mu}\to W_a^{\mu}+ i\frac{\tilde{c}}{\Lambda^2} (H^{\dagger}\sigma^i \overleftrightarrow{D}_{\mu}H)$. Such a redefinition leads to an operator contribution to muon decay, which would change the definition of the electroweak vacuum expectation value as a function of the Fermi constant. This leads to a proliferation of `hidden' EFT effects which, in numerical computations such as those we will be performing in this study, unavoidably lead to higher order terms in the EFT expansion parameter via. \emph{e.g.}, the squaring of EW parameters in matrix element evaluations.
Such effects are both known and distinct from the comparison study we would like to perform and would complicate the task by having to disentangle two effects. We therefore refrain from discussing it here any further. 
\\
We can now define the second model, 
\begin{equation}
\text{Model 2:} \hspace*{0.5cm}\mathcal{L}=\mathcal{L}_{SM}- \frac{c}{\Lambda^2} \left[ \sum_{q}(\bar{q} \gamma_{\mu} \sigma^i q) (\bar{q}_3 \gamma^{\mu}  \sigma^i q_3)+ \sum_{l} ( \bar{l} \gamma_{\mu} \sigma^i l) (\bar{q}_3 \gamma^{\mu}  \sigma^i q_3)+\frac{1}{g}(\bar{q}_3 \gamma^{\mu}  \sigma^i q_3) D_{\nu}W^i_{\nu\mu}\right], \label{eq:model2}
\end{equation}
with $l$ and $q$ denoting the lepton and quark doublets and $g$ the SU(2) coupling. The sums run over the three generations of fermion doublets. We have also defined
\begin{equation}
D_{\rho}W^i_{\mu\nu}=\partial_{\rho} W^{i}_{\mu\nu} +  g \epsilon^{ijk} W_{\rho}^j W_{\mu\nu}^k,
\end{equation}
with $W_{\mu\nu}$ the usual SU(2) field strength. We see that this operator can be traded for some 4-fermion operators involving lepton and quark doublets, as well as a higher-derivative gauge boson interaction with the fermion current.
\section{NUMERICAL ANALYSIS \label{sec:numanalysis}}

The model described in the previous section have been implemented into {\tt FeynRules} \cite{Alloul:2013bka, Christensen:2008py} to produce Universal FeynRules Output (UFO)~\cite{Degrande:2011ua} models.\footnote{We have cross-checked our model files by two independent implementations.} For an implementation of the Higgs effective Lagrangian see \cite{Alloul:2013naa}. Our numerical results are obtained using the {\tt MadGraph5\_aMC@NLO} package \cite{Alwall:2014hca} (dubbed {\tt MG5aMC}). To show the level of agreement between the two models, we start by evaluating the interference term between the SM and the EFT contributions at a single phase-space point for a series of key processes. Those phase-space points are generated with RAMBO \cite{Kleiss:1985gy} at a centre-of-mass energy of $1 \text{ TeV}$. 

In table~\ref{tab:agreement}, we have set $\frac{c}{\Lambda^2} = 1 \text{ TeV}^{-2}$ and all of the particle widths to zero. While in this proceeding we restrict ourselves to only five different processes for brevity, we emphasise that more than one hundred different processes were checked in the complete study.
\begin{table}
\renewcommand{\arraystretch}{1.3}
\begin{center}
\begin{tabular}{cccc}
\hline
process (interference) & Model 1 & Model 2 & relative difference \\ \hline
$b \bar{b} \to w^+ w^- z$& $-2.560957\cdot10^{-06}$ & $-2.560957\cdot10^{-06}$ & $1.95\cdot10^{-13}$ \\
$\bar{d} u \to \bar{b} t z$& $8.006993\cdot10^{-09}$ & $8.006993\cdot10^{-09}$ & $1.24\cdot10^{-15}$ \\
$b w^+ \to b w^+$& $2.708079\cdot10^{-01}$ & $2.708079\cdot10^{-01}$ & $3.69\cdot10^{-15}$ \\
$h w^+ \to t \bar{t} w^+$& $-1.767258\cdot10^{-05}$ & $-1.767258\cdot10^{-05}$ & $1.96\cdot10^{-14}$ \\
$b \bar{b} \to t \bar{t}$& $3.380570\cdot10^{-03}$ & $3.380570\cdot10^{-03}$ & $8.79\cdot10^{-15}$ \\
\end{tabular}
\end{center}
\caption{Comparison between Model 1 and 2 setting the width in the propagators to zero.\label{tab:agreement}}
\end{table}
As can be inferred from table~\ref{tab:agreement} we observe a perfect agreement of both models at the pure dimension-6 level (the dimension-8 contribution originating from the dimension-6 contribution squared will obviously also be identical in this case).
In order to study the impact of the width in the propagators, we present table~\ref{tab:SMwidth} were we have kept all particle widths to their SM value.
\begin{table}
\renewcommand{\arraystretch}{1.3}
\begin{center}
\begin{tabular}{cccc}
\hline
process (interference) & Model 1 & Model 2 & relative difference \\ \hline
$b \bar{b} \to w^+ w^- z$& $-2.562000\cdot10^{-06}$ & $-2.571304\cdot10^{-06}$ & $1.81\cdot10^{-03}$ \\
$\bar{d} u \to \bar{b} t z$& $8.005881\cdot10^{-09}$ & $8.003428\cdot10^{-09}$ & $1.53\cdot10^{-04}$ \\
$b w^+ \to b w^+$& $2.708134\cdot10^{-01}$ & $2.706435\cdot10^{-01}$ & $3.14\cdot10^{-04}$ \\
$h w^+ \to t \bar{t} w^+$& $-1.760324\cdot10^{-05}$ & $-1.754204\cdot10^{-05}$ & $1.74\cdot10^{-03}$ \\
$b \bar{b} \to t \bar{t}$& $3.380570\cdot10^{-03}$ & $3.380510\cdot10^{-03}$ & $8.79\cdot10^{-06}$ \\
\end{tabular}
\end{center}
\caption{Comparison between Model 1 and 2 setting the width in the propagators to their SM values.\label{tab:SMwidth}}
\end{table}
We now see that the two models do not agree anymore. This disagreement is related to the fact that the width is formally a higher order effect and its inclusion amounts to a mixing of perturbative orders. This also breaks gauge invariance once a decay width is introduced into the gauge boson propagators. This is a well known problem that can be for instance solved by employing the complex mass scheme \cite{Denner:1999gp}. The level of disagreement worsens with the number of possible propagators with modified widths appearing in the Feynman diagrams for each process. We stress that the observed difference is not an artefact of introducing the wrong width in the propagator (i.e.  not including the effects of dimension-6 operators consistently). In order to show this explicitly we perform the same computation but also including the dimension-6 contribution to the width, see table~\ref{tab:dim6width}.

For the sake of the example let's focus of the width of the top (the same can be done for the $Z$ propagator). We dubbed  $\Gamma^t_{SM}$ the standard model contribution and  $\Gamma^t_6$ the dimension-6 contribution to the width (the interference term):
\begin{equation}
\Gamma^t_{SM}= \frac{(M_t^2 - M_w^2)e^2}{64\pi M_t^3s_w^2} (M_t^2 - 2M_w^2+\frac{M_t^4}{M_w^2}),
\end{equation}
\begin{equation}
\Gamma^t_{6}= \frac{c}{\Lambda^2} \frac{v^2}{2} \Gamma^t_{SM}. 
\end{equation}
Computing Feynman rules with $\Gamma^t_{SM} + \Gamma^t_{6}$ is actually not consistent for an EFT point of view since due to the presence of the width in the propagator, this is equivalent to add higher order term of the EFT 
inside the computation. To avoid such higher order term we use a Taylor expansion of the propagator:
\begin{equation}
\frac{1}{q^2-m^2+im(\Gamma^t_{SM}+\Gamma^t_{6})} \approx  \frac{1}{q^2-m^2+im\Gamma^t_{SM}}-\frac{im\Gamma^t_6}{\left(q^2-m^2+im\Gamma^t_{SM}\right)^2}  \label{eq:prop}
\end{equation}
Such expansion is converging for all phase-space points as long as $\Gamma^t_{6} < \Gamma^t_{SM}$, i.e.~as long as $\frac{c}{\Lambda^2} \frac{v^2}{2} < 1$.
We can then see the second term of the Taylor expansion as generating some additional effective vertex and compute their interference with the SM amplitude.
Interestingly, this then extends the impact of the EFT operators to processes where they do not contribute otherwise, as e.g. to $e^+e^-\to \mu^+\mu^-$ using only the operators of eq.~\eqref{eq:model1} or \eqref{eq:model2}. 

The computation of Eq.~\ref{eq:prop} cannot be performed using the {\tt MG5aMC} package `out of the box'. To allow the numerical computation using this method, we then created an extension of the usermod of  {\tt MG5aMC}
in order to generate a new dedicated UFO package. The idea is to add two new particles ($\tilde t$, $\tilde Z$) associated to a custom propagator \cite{deAquino:2011ub,Christensen:2013aua} that provides the contribution from the second term in eq.~\eqref{eq:prop}. Then for each interaction with either a top or $Z$ boson, we add to the model one copy of that interaction with (at most one) top/$Z$ replaced by it's equivalent $\tilde t$, $\tilde Z$.
The coupling of such interactions is also tagged in a way to allow to compute interference terms in {\tt MG5aMC}. Additionally the parameter $\Gamma_6$ is added to the model as an internal parameter associated to the analytical formula automatically extracted from the {\it decays.py} file \cite{Alwall:2014bza} from the original model. A tool fully automating such feature is in preparation.\footnote{The semi-automatic version can be downloaded with the following command:
{\tt bzr branch lp:~maddevelopers/mg5amcnlo/eft\_width\_expansion}.}

With this consistent inclusion of the width at dimension-6 level we obtain the results presented in table~\ref{tab:dim6width}, in which it can be seen that the level of agreement is not improved.
\begin{table}
\renewcommand{\arraystretch}{1.3}
\begin{center}
\begin{tabular}{cccc}
\hline
process (interference) & Model 1 & Model 2 & relative difference \\ \hline
$b \bar{b} \to w^+ w^- z$& $-2.561587\cdot10^{-06}$ & $-2.570890\cdot10^{-06}$ & $1.81\cdot10^{-03}$ \\
$\bar{d} u \to \bar{b} t z$& $8.005881\cdot10^{-09}$ & $8.003428\cdot10^{-09}$ & $1.53\cdot10^{-04}$ \\
$b w^+ \to b w^+$& $2.708370\cdot10^{-01}$ & $2.706671\cdot10^{-01}$ & $3.14\cdot10^{-04}$ \\
$h w^+ \to t \bar{t} w^+$& $-1.760429\cdot10^{-05}$ & $-1.754309\cdot10^{-05}$ & $1.74\cdot10^{-03}$ \\
$b \bar{b} \to t \bar{t}$& $3.380538\cdot10^{-03}$ & $3.380479\cdot10^{-03}$ & $8.79\cdot10^{-06}$ \\
\end{tabular}
\end{center}
\caption{Comparison between Model 1 and 2 using the implementation of the width as given in eq.~\eqref{eq:prop}.\label{tab:dim6width}}
\end{table}
This confirms our previous statement, that the problem is deeper that just consistently including the dimension-6 contributions to the width and points to an issue regarding the presence of the decay width itself. 
Comparing table~\ref{tab:SMwidth} and \ref{tab:dim6width} one can also see the impact of including the dimension-6 terms to the width. The differences displayed here are quite small but one should keep in mind that such statement is highly phase-space dependent and  
should be maximal for onshell decay where it is of the order of $\Gamma_6/\Gamma_{SM}$ and therefore proportional to the EFT expansion parameter $\frac{c}{\Lambda^2}$ 

\begin{table}
\renewcommand{\arraystretch}{1.3}
\begin{center}
\begin{tabular}{cccc}
\hline
process (interference) & Model 1 & Model 2 & relative difference \\ \hline
$e^+ e^- \to b b \bar{b} \bar{b}$& $1.434685\cdot10^{-12}$ & $1.434685\cdot10^{-12}$ & $2.96\cdot10^{-15}$ \\
$b \bar{b} \to e^+ \nu_e e^- \bar{\nu}_e$& $7.661298\cdot10^{-14}$ & $7.661298\cdot10^{-14}$ & $1.07\cdot10^{-14}$ \\
$g g \to b \bar{b} e^+ \nu_e \mu^- \bar{\nu}_{\mu}$& $5.186028\cdot10^{-20}$ & $5.186028\cdot10^{-20}$ & $2.44\cdot10^{-15}$ \\
$u \bar{u} \to b \bar{b} \mu^+ \mu^-$& $-4.947679\cdot10^{-15}$ & $-4.947679\cdot10^{-15}$ & $3.12\cdot10^{-14}$ \\
$b \bar{b} \to e^+ \nu_e \mu^- \bar{\nu}_{\mu}$& $2.045529\cdot10^{-14}$ & $2.045529\cdot10^{-14}$ & $7.71\cdot10^{-15}$ \\
\end{tabular}
\end{center}
\caption{Comparison between Model 1 and 2 using the standard model width using the complex mass scheme. The width is set to his SM value. \label{tab:dim6cms}}
\end{table}
As stated above, in order to include the width effect in a fully consistent way inside the SM to insure gauge invariance, one needs to use a dedicated method like for instance the complex mass scheme \cite{Denner:1999gp} or the overall-factor scheme \cite{Baur:1991pp, Baur:1995aa} (for an application to BSM see \cite{Grober:2015fia}).
It therefore makes sense to do the same in our two models. However, in order to have consistent result in the complex mass scheme one has to decay all particles. Therefore we can not present results for the same set of processes as shown before.\footnote{We also test close to one hundred different processes in this case.}.
As a first step we neglect the effect of $\Gamma_6$ and use only the SM width (see table~\ref{tab:dim6cms}). This result shows that using the complex mass scheme is actually crucial in the context of EFT in order to have results independent of the basis. 

The latest result is however not fully satisfactory since it does not include correctly the modification by the EFT operator to the width. On the one hand we have to include the dimension-6 contribution in the propagators as we have outlined before. On the other hand,
for a fully consistent result at the dimension-6 level, contributions stemming from the replacement of the mass by 
\begin{equation}
m^2 \to m^2 + i m\, (\Gamma_{SM}+\Gamma_{6})
\end{equation}
or the replacement of the weak mixing angle by
\begin{equation}
\begin{split}
\cos^2\theta_W=\frac{m_W^2}{m_Z^2} \to &\frac{m_W^2 + i m_W (\Gamma_{SM}^W +\Gamma_6^W)}{m_Z^2 + i m_Z (\Gamma_{SM}^Z +\Gamma_6^Z)} \\ \approx& \frac{m_W^2 + i m_W \Gamma_{SM}^W}{m_Z^2 + i m_Z \Gamma_{SM}^Z}+ \frac{i m_W \Gamma_{6}^W }{m_Z^2 + i m_Z (\Gamma_{SM}^Z)}-\frac{i m_Z \Gamma_{6}^Z (m_W^2 + i m_W \Gamma_{SM}^W)}{(m_Z^2 + i m_Z \Gamma_{SM}^Z)^2}
\end{split}
\end{equation}
with $\Gamma^{Z/W}_{SM,6}$ denoting the SM (dimension-6) contributions to the width of the $Z$/$W$ boson,
in the complex mass scheme in the SM matrix elements is necessary. Note that for our operators $\Gamma_{6}^W=0$. The correct solution would be to do a Taylor expansion not only on the propagator like in Eq.~\ref{eq:prop} but on all parts of the matrix elements which now depends on the width, which again includes dimension-6 contributions, due to the complex mass scheme approach. This is in principle similar to the shifts in the SM dependent parameters in the presence of dimension-6 operators \cite{Brivio:2017vri}, as we discussed briefly in the previous section. These effects will be the focus of future work.
As a first approximation we present in table \ref{tab:dim6cms+eft} results using 
Eq.~\ref{eq:prop} on top of using the complex mass scheme for the SM width. The two models are in perfect agreement but we emphasize again, that this does not mean that all the dimension-6 effects are correctly included.
\begin{table}
\renewcommand{\arraystretch}{1.3}
\begin{center}
\begin{tabular}{cccc}
\hline
process (interference) & Model 1 & Model 2 & relative difference \\ \hline
$e^+ e^- \to b b \bar{b} \bar{b}$& $1.442082\cdot10^{-12}$ & $1.442082\cdot10^{-12}$ & $1.61\cdot10^{-15}$ \\
$b \bar{b} \to e^+ \nu_e e^- \bar{\nu}_e$& $7.664749\cdot10^{-14}$ & $7.664749\cdot10^{-14}$ & $1.07\cdot10^{-14}$ \\
$g g \to b \bar{b} e^+ \nu_e \mu^- \bar{\nu}_{\mu}$& $5.185849\cdot10^{-20}$ & $5.185849\cdot10^{-20}$ & $3.54\cdot10^{-15}$ \\
$u \bar{u} \to b \bar{b} \mu^+ \mu^-$& $-4.927938\cdot10^{-15}$ & $-4.927938\cdot10^{-15}$ & $3.47\cdot10^{-14}$ \\
$b \bar{b} \to e^+ \nu_e \mu^- \bar{\nu}_{\mu}$& $2.045606\cdot10^{-14}$ & $2.045606\cdot10^{-14}$ & $1.28\cdot10^{-14}$ \\
\end{tabular}
\end{center}
\caption{Comparison between Model 1 and 2 using the standard model width using the complex mass scheme for the SM width and including the effect due to the modification of the width only via the propagator as given in eq.~\eqref{eq:prop}. \label{tab:dim6cms+eft}}
\end{table}

\section*{CONCLUSIONS}
We have shown by example that, while different EFT operator bases are equivalent under field redefinitions if the width of the propagating particles in the numerical evaluation is set to zero, this is not the case if a non-zero width is used. As in the SM, this is related to the fact that the width is already a higher order effect in the perturbative expansion and formally breaks gauge invariance. In analogy to the SM the problem can be addressed by a complex mass scheme, where all the masses in propagators and couplings are replaced by a complex parameter including the physical width of the particle in the imaginary component. We showed that indeed this resolves the issue. 
The inclusion of the width however gives additional contributions at the dimension-6 level: 1.) the width in the particle propagators gets a dimension-6 piece. This part has to be Taylor expanded up to first order to keep it strictly at the dimension-6 level. We have provided a model for that in {\tt MG5aMC}. 2.) Employing a complex mass scheme redefines also the masses of the unstable particles in the couplings, that compared to the SM complex mass scheme, now also obtain a contribution at the dimension-6 level from the particle's width. For consistency this contribution also needs to be included. We leave the numerical impact of this contribution to future work. 
\section*{ACKNOWLEDGEMENTS}
We thank the organisers of the Les Houches Workshop Series ``Physics at TeV Colliders'' 2017 for the nice and fruitful atmosphere during the workshop. We thank C\'eline Degrande for useful discussions.
RG is supported by a European Union COFUND/Durham Junior Research Fellowship under the EU grant number 609412. KM is supported in part by the Belgian Federal Science Policy Office through the Interuniversity Attraction 
Pole P7/37 and by the European Union’s Horizon 2020 research and innovation programme under the Marie Skłodowska-Curie grant agreement No. 707983.



\AddToContent{R.~Gr\"ober, O.~Mattelaer, K.~Mimasu}
\renewcommand{\thesection}{\arabic{section}}

\graphicspath{{FiducialDistributions/}}

\def\dd{\ensuremath{\rm d}}
\def\pp{\ensuremath{pp}}
\def\W{\ensuremath{W}}

\def\mh{\ensuremath{m_H}}
\def\pt{\ensuremath{p_{\rm T}^{}}}

\def\mgg{\ensuremath{m_{\gamma\gamma}}}
\def\ptgg{\ensuremath{p^{\gamma\gamma}_{\rm T}}}
\def\pttgg{\ensuremath{p^{\gamma\gamma}_{\rm Tt}}}

\def\dr{\ensuremath{\Delta R}}
\def\met{\ensuremath{E_{T}^{\rm miss}}}
\def\drjl{\ensuremath{\dr_{j,\ell}}}
\def\ptjl{\ensuremath{p_{T}^{ j_1}}}
\def\ptjsl{\ensuremath{p_{T}^{ j_2}}}
\def\ptjssl{\ensuremath{p_{T}^{ j_3}}}
\def\htj{\ensuremath{H_{T}}}
\def\yjl{\ensuremath{|y_{j_1}|}}
\def\yjsl{\ensuremath{|y_{j_2}|}}
\def\ptggjj{\ensuremath{p^{}_{\rm T, \gamma\gamma{}jj}}}
\def\ygg{\ensuremath{|y_{\gamma\gamma}|}}
\def\mjj{\ensuremath{m_{jj}}}
\def\deltayjj{\ensuremath{|\Delta{y_{jj}|}}}
\def\deltaygg{\ensuremath{|\Delta{y_{\gamma\gamma}|}}}
\def\dphiggjj{\ensuremath{|\Delta\phi_{\gamma\gamma,jj}|}}
\def\dphijj{\ensuremath{|\Delta\phi_{jj}|}} 
\def\njet{\ensuremath{N_{\rm jets}}}
\def\costhetastar{\ensuremath{\rm | cos \, \theta^{*}|}}
\def\taujet{\ensuremath{\tau_{1}}}
\def\sumtaujet{\ensuremath{\sum_i \tau_{i}}}

\chapter{On the use of Higgs fiducial cross sections for constraining new physics}

{\it S.~Kraml, U.~Laa, K.~Lohwasser}



\begin{abstract}
  We discuss the potential of the measured Higgs fiducial cross sections for deriving constraints 
  on anomalous Higgs production from BSM processes. 
  Using the examples of three SUSY processes, we show that these constraints can be complementary to those 
  from dedicated searches for new physics.
\end{abstract}

\section{INTRODUCTION}

The increasingly precise data on the 125~GeV Higgs boson from Run~1 and Run~2 of the LHC provide severe constraints on new physics beyond the Standard Model (BSM). A well-established approach to assess the compatibility of a BSM Higgs boson with the experimental results 
is the use of signal strengths, which compare---preferably in a detailed breakdown of production$\times$decay modes---the observed state to Standard Model (SM) expectations; see \cite{LHCHiggsCrossSectionWorkingGroup:2012nn,Heinemeyer:2013tqa,Boudjema:2013qla} for detailed discussions. 
There has indeed been a boom of phenomenological studies making use of the signal strength measurements 
to work out the implications of the 125~GeV Higgs boson for non-standard Higgs sectors by simple scaling of production cross sections and decay branching ratios relative to the SM. 
In-depth studies concerned two-Higgs-doublet models, supersymmetric (SUSY) models, Randall-Sundrum (with Higgs-radion mixing) models, technicolor, little Higgs, composite Higgs models, and so on. The underlying assumption is that the signal selection efficiencies are to good approximation 
the same in the new model and the SM.  

In situations in which the kinematic distribution of the signal depends on model parameters, simple scaling of production cross sections and decay branching ratios relative to the SM is, however, not appropriate --- one must account for the change in the signal selection efficiencies. 
Such situations can arise from the presence of new tensor structures (anomalous couplings, higher-dimensional operators), 
as well as from the presence of new Higgs production modes through decays of heavier new states.
To address these cases, \cite{Boudjema:2013qla} advocated the measurement of fiducial cross sections, i.e.\ cross sections, whether total or differential, for specific final states within the phase space defined by the experimental selection and acceptance cuts. 
Fiducial cross sections can be interpreted in the context of whatever theoretical model, 
provided it is possible to compute its predictions for the given fiducial volume (typically by means of a Monte Carlo event simulation) 
and that no significant extrapolation has been applied to define the fiducial cross section (e.g., to correct for an event-level selection based on machine learning that has a different acceptance for SM and BSM events). 
Fiducial cross sections also have the advantage of largely separating experimental and theoretical uncertainties, such that a re-evaluation of constraints is possible when more precise theoretical prediction become available without re-analysing the data.  

Higgs fiducial cross sections were the subject of a dedicated task force for the Yellow \mbox{Report 4~\cite{deFlorian:2016spz}} 
of the LHC Higgs cross section working group. 
Differential measurements are particularly interesting in this context, as the shapes of distributions may provide more sensitive tests than integrated event rates, for instance in the presence of interference effects.\footnote{See also the contribution by A.~Carvalho, R.~Gr\"ober, S.~Liebler and J.~Quevillon in these proceedings.}

Both ATLAS and CMS provide total and differential fiducial cross section measurements for specific Higgs decay modes, concretely 
$H\to\gamma\gamma$~\cite{Aad:2014lwa,Khachatryan:2015rxa,CMS-PAS-HIG-17-015,Aaboud:2018xdt}, 
$H\to ZZ^*\to 4\,$leptons~\cite{Aad:2014tca,Khachatryan:2015yvw,Aaboud:2017oem} and 
$H\to WW^*\to 2l2\nu$~\cite{Aad:2016lvc}. 
These measurements are agnostic of the Higgs production mode and thus potentially sensitive for constraining additional Higgs 
production from BSM (cascade) decays.\footnote{In contrast, the so-called ``simplified template cross sections'' (STXS)~\cite{Badger:2016bpw,deFlorian:2016spz} are cross sections per production mode, split into mutually exclusive kinematic bins for each of the main production
modes. They are determined from the experimental categories by a global fit that combines all decay channels, with the SM serving as kinematic template.} 
The facts that the definition of the fiducial volume is based on simple cuts and the results are unfolded to the particle level 
are big advantages for interpretation studies. The downside is that precise SM predictions to compare to require highly sophisticated computations. 

In this contribution, we investigate the use of fiducial cross section measurements to constrain new Higgs boson production modes in BSM models. 
We concentrate on the $H\to\gamma\gamma$~\cite{Aad:2014lwa} and $H\to ZZ^*\to 4\ell$~\cite{Aad:2014tca} measurements from ATLAS 
at $\sqrt{s}=8$~TeV, for which detailed HEPData entries~\cite{hepdata:diphoton,hepdata:fourlepton} and validated Rivet~\cite{Buckley:2010ar} routines~\cite{ATLAS:2014:I1306615,ATLAS:2014:I1310835} are available. 
(A Rivet routine is also available for the combination of the $H\to\gamma\gamma$ and $H\to ZZ^*\to 4\ell$ analyses of ATLAS \cite{Aad:2015lha}, but this effectively assumes stable Higgs bosons; CMS provides the results for $H\to\gamma\gamma$ at 8 TeV on HEPData, but no Rivet routine.)

\section{BACK-OF-THE-ENVELOP CONSIDERATIONS}

In~\cite{Aad:2014lwa}, ATLAS reported the $pp\to H\to\gamma\gamma$  fiducial cross section at 8~TeV as  
$43.2 \pm 9.4$(stat.) $^{+3.2}_{-2.9}$(syst.) $\pm 1.2$(lumi)~fb for a Higgs boson of mass 125.4~GeV decaying to two isolated photons 
with pseudorapidity $|\eta|<2.37$ and $p_T^{}/m_{\gamma\gamma}>0.35$ (0.25) for the leading (subleading) photon. 
The SM prediction is $30.5 \pm 3.3$~\cite{Heinemeyer:2013tqa} following the recommendatons of the HXSWG, but ranges from 
$27^{+3.6}_{-3.2}$~\cite{Grazzini:2013mca} to $34.1^{+3.6}_{-3.5}$~\cite{Stewart:2013faa} using other calculations. 
An overview of the experimental and theoretical values for the seven fiducial regions  considered in~\cite{Aad:2014lwa} is given in Table~\ref{tab:fxs-atlas-gamgam}. We see that that data agree quite well with SM expectations, but uncertainties are sizable, and there is still room for 
contributions from new physics.  

\begin{table}[h!]
\centering
  \begin{tabular}{| l | c | c |}
  \hline
  Fiducial region & Measured cross section (fb) & SM predictions (fb) \\
  \hline
Baseline & $43.2 \pm  10.0$ & [24.0,\,37.7] \\        
$N_{\rm jets} \geq 1$ & $21.5 \pm 5.8$ & [8.1,\,15.5] \\
$N_{\rm jets} \geq 2$& $9.2 \pm  3.1$  & [3.4,\,6.52] \\
$N_{\rm jets} \geq 3$& $4.0 \pm  1.5$  & $0.94\pm 0.15$ \\
VBF-enhanced  & $1.68 \pm  0.63$  & $0.87\pm 0.08$   \\
  \hline
$N_{\rm leptons} \geq 1$ & $< 0.80$   & $0.27\pm 0.02$   \\
$\met > 80$~GeV  & $< 0.74 $    & $0.14\pm 0.01$ \\
\hline
  \end{tabular}
  \caption{Measured $pp\to H\to\gamma\gamma$  fiducial cross section at $\sqrt{s}=8$~TeV from ATLAS~\cite{Aad:2014lwa} in the baseline, $N_{\rm jets} \geq 1$, $N_{\rm jets} \geq 2$, $N_{\rm jets} \geq 3$ and VBF-enhanced fiducial regions, and cross-section limits at 95\% confidence level in the single-lepton and high-\met\ fiducial regions. The ranges given for the SM prediction correspond to the envelope of the theory predictions, incl.\ $1\sigma$ uncertainties, quoted in Table~4 of \cite{Aad:2014lwa}. 
  }
  \label{tab:fxs-atlas-gamgam}
\end{table}

With these numbers, we can make some back-of-the envelop estimates. The total SM Higgs production cross section at 8~TeV is 
$19.15$~pb for $m_h=125.4$~GeV~\cite{Heinemeyer:2013tqa}. With BR$(H\to\gamma\gamma)=0.228\%$, this gives a total cross section 
in the diphoton channel of $43.66$~fb, which means the diphoton baseline fiducial volume contains about 70\% of the total production. 
Approximating the SM prediction for the baseline fiducial region as $30.85\pm 6.85$ and summing experimental and theoretical uncertainties in quadrature, an additional BSM contribution of up to about 36~fb would be allowed at $2\sigma$ (not accounting for additional BSM uncertainties). Assuming Higgs branching ratios and a fiducial acceptance like in the SM, this means up to about 23~pb of Higgs production from BSM processes is allowed.  
If, however, the BSM Higgs production always includes 2, 3, or more hard jets, e.g., because of cascade decays, this room shrinks considerably, 
see  Table~\ref{tab:fxs-atlas-gamgam}. For Higgs + 3~jets, a total BSM cross section of the order of 4~pb can be excluded, still assuming 
BR$(H\to\gamma\gamma)=0.228\%$ and an acceptance of around 0.7. 
With the same reasoning, one could expect to exclude BSM Higgs + $W$ associated production above about 2~pb due to the  $N_{\rm leptons} \geq 1$ 
fiducial limit.\footnote{In reality this is somewhat too optimistic, partly because of the $p_T > 15$ GeV and $|\eta|< 2.47$ requirements for letopns.}
Likewise, if BSM Higgs production always leads to large $\met>80$~GeV, its total cross section should be below about 0.4~pb. 
Of course, these are very rough estimates which may easily be off by a factor of a few, especially because the acceptance does not stay constant. 
(Besides also the Higgs branching ratios can vary in BSM models.) 
Nonetheless they may serve as guidelines for the order of magnitude of possible constraints on new physics. 
We also note that stronger constraints can come from the differential distributions presented in \cite{Aad:2014lwa}, 
limiting for instance highly boosted Higgs production.

Turning to the $H\to ZZ^*\to 4\ell$ channel, Ref.~\cite{Aad:2014tca} reports a total fiducial cross section of 
$\sigma_{\rm fid}^{\rm tot}=2.11^{+0.53}_{-0.47}$\,(stat.)$\pm 0.08$\,(syst.)~fb, to be compared to the theoretical prediction in the SM 
of $1.30\pm 0.13$~fb~\cite{Heinemeyer:2013tqa} for a Higgs boson mass of $125.4$ GeV. 
The paper moreover shows the differential fiducial cross sections as a function of
$p_{T,H}$, $y_H$, $m_{34}$, $|\cos\theta_*|$, $N({\rm jets})$, and $p_{T,\rm jet}$. 
All these results are available on HEPData. Moreover, upon request, also the SM predictions used 
in \cite{Aad:2014tca} were made available on HEPData, which is extremely useful for the 
purpose of constraining additional BSM contributions.\footnote{Unfortunately, the same was not done for the 
$pp\to H\to\gamma\gamma$ differential distributions.}
From the numbers above, a BSM contribution to $\sigma_{\rm fid}^{\rm tot}$ of 1.9~fb or larger can be excluded. 
Performing the same exercise as above, that is taking BR$(H\to ZZ^*\to 4\ell)=1.286\times 10^{-4}$ and assuming an acceptance 
similar to the one in the SM, this would correspond to a total inclusive BSM production cross section of 28~pb, i.e.\ comparable 
but a bit larger than for $H\to\gamma\gamma$. 
More sensitive constraints come again from the differential distributions, in particular when the BSM production leads to 
high $p_{T,H}$ or high jet multiplicity.

\section{HIGGS PRODUCTION FROM SUSY CASCADES}

Anomalous Higgs production was considered by ATLAS and CMS in the context of SUSY searches. 
This provides a welcome possibility to compare constraints from Higgs fiducial cross sections to constraints from dedicated BSM searches. 

\begin{figure}[t!]\centering
\includegraphics[width=0.25\textwidth]{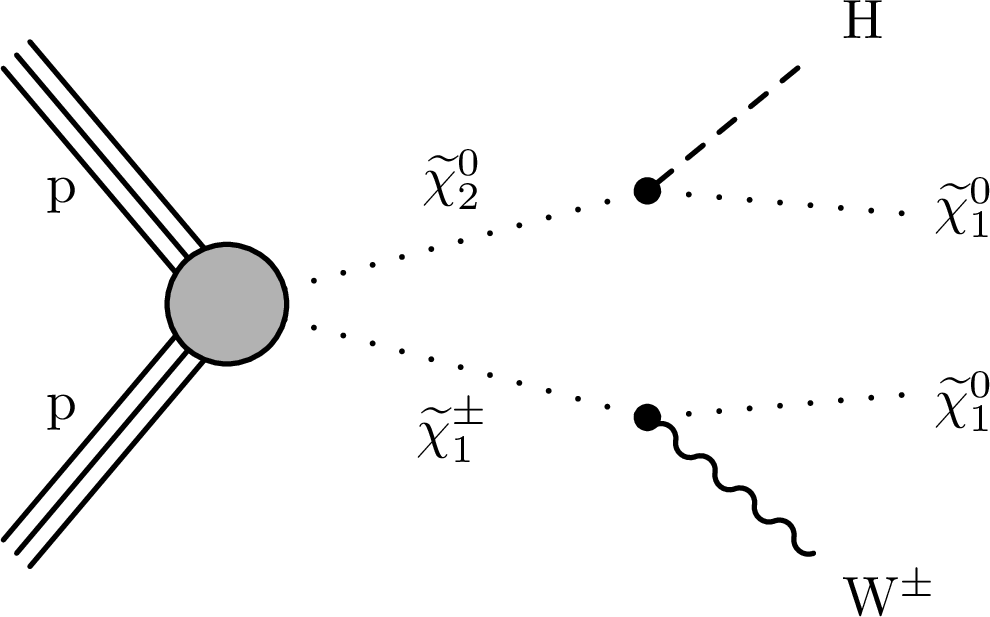}\qquad\includegraphics[width=0.25\textwidth]{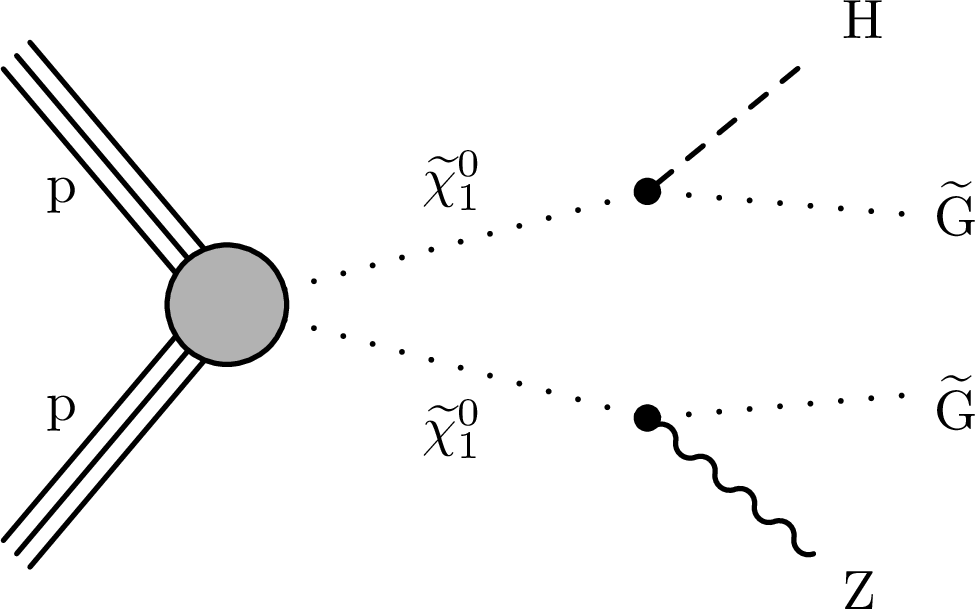}\qquad\includegraphics[width=0.2\textwidth]{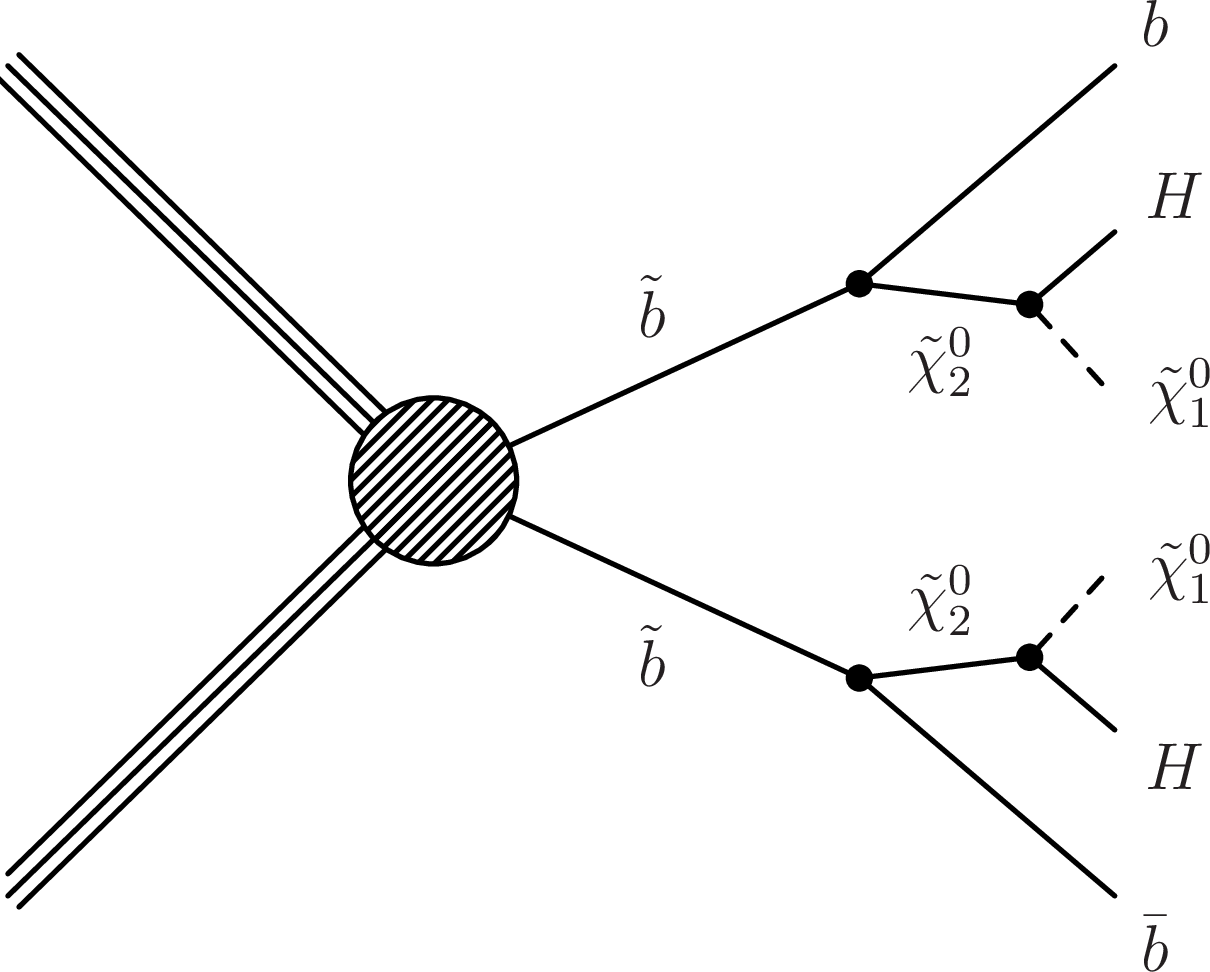}
\caption{Examples for Higgs production in SUSY cascade decays considered in ATLAS and/or CMS SUSY searches.}
  \label{fig:fxs-diagrams}
\end{figure}

\begin{figure}[t!]\centering
\includegraphics[width=0.55\textwidth]{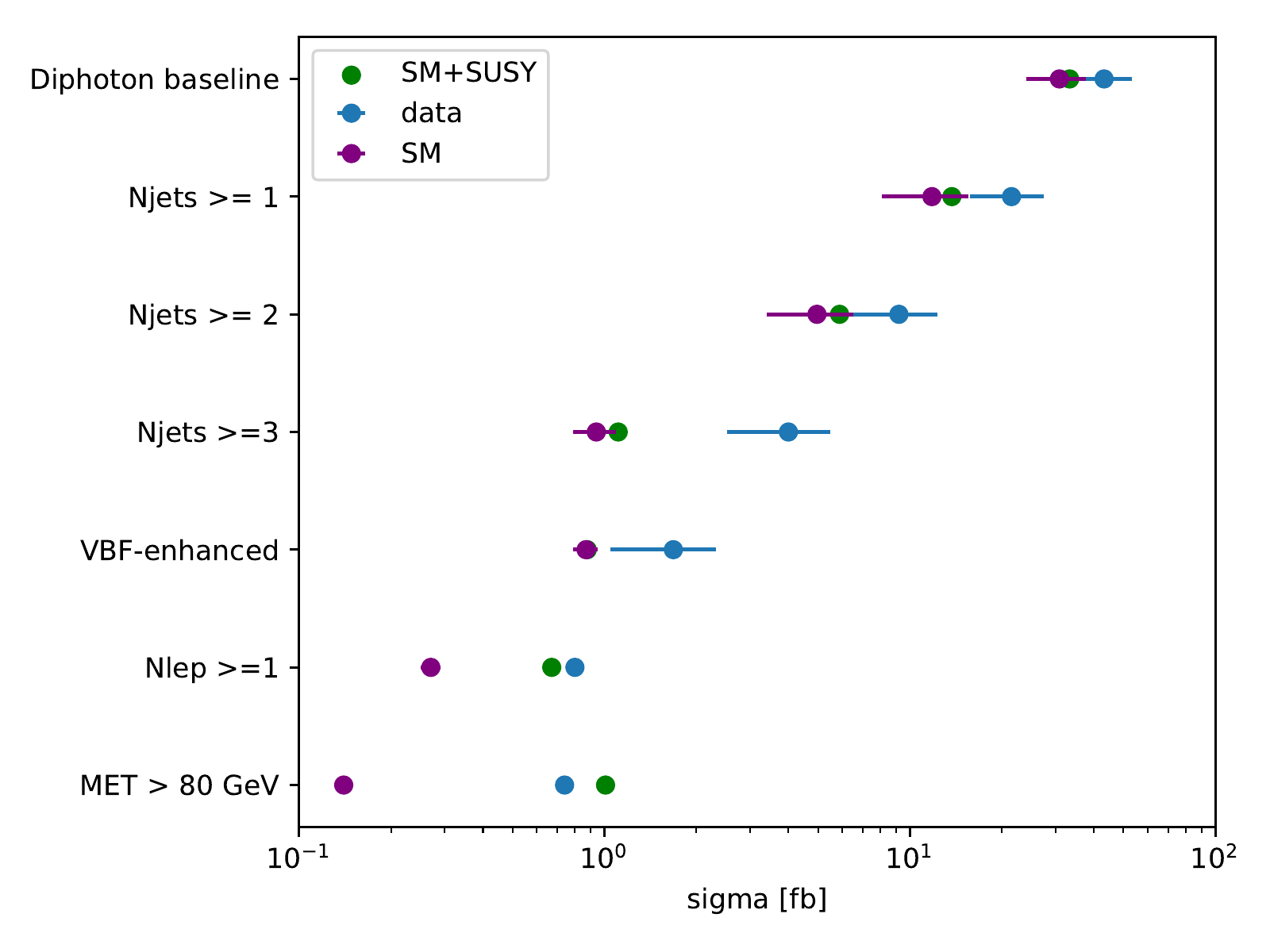}
\caption{Results for the $H\to \gamma\gamma$ fiducial regions from ATLAS~\cite{Aad:2014lwa}. The SUSY signal considered 
is $pp\to\tilde\chi^\pm_1\tilde\chi^0_2$ followed by $\tilde\chi^\pm_1\to W^\pm \tilde\chi^0_1$ and $\tilde\chi^0_2\to H \tilde\chi^0_1$ 
for $m_{\tilde\chi^\pm_1,\tilde\chi^0_2}=150$~GeV and $m_{\tilde\chi^0_1}=20$~GeV and $\sigma_{\rm tot}(\tilde\chi^\pm_1\tilde\chi^0_2)=2.41$~pb.}
  \label{fig:fxs-ex1aa}
\end{figure}

\begin{figure}[t!]\centering
\includegraphics[width=0.5\textwidth]{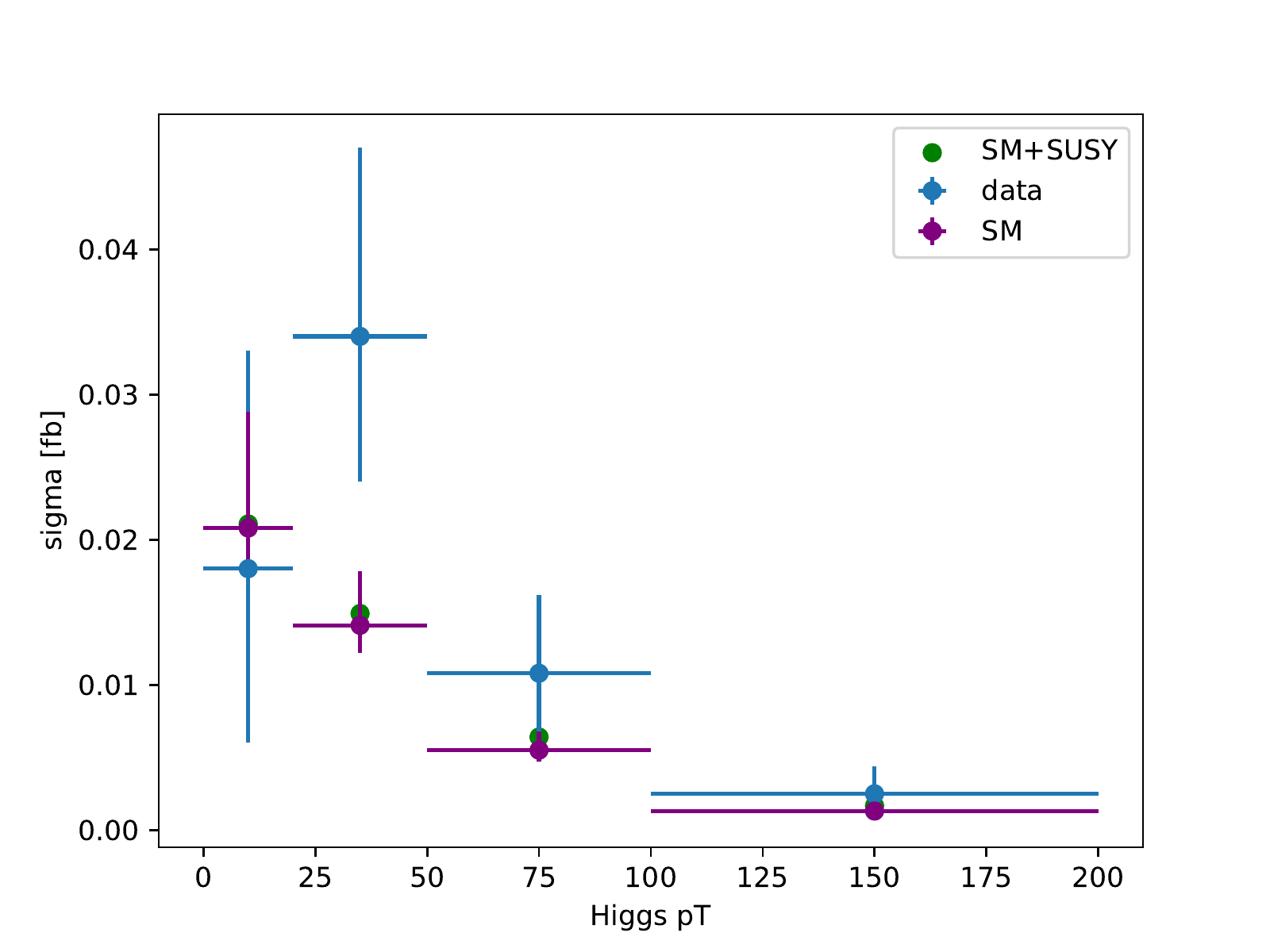}\includegraphics[width=0.5\textwidth]{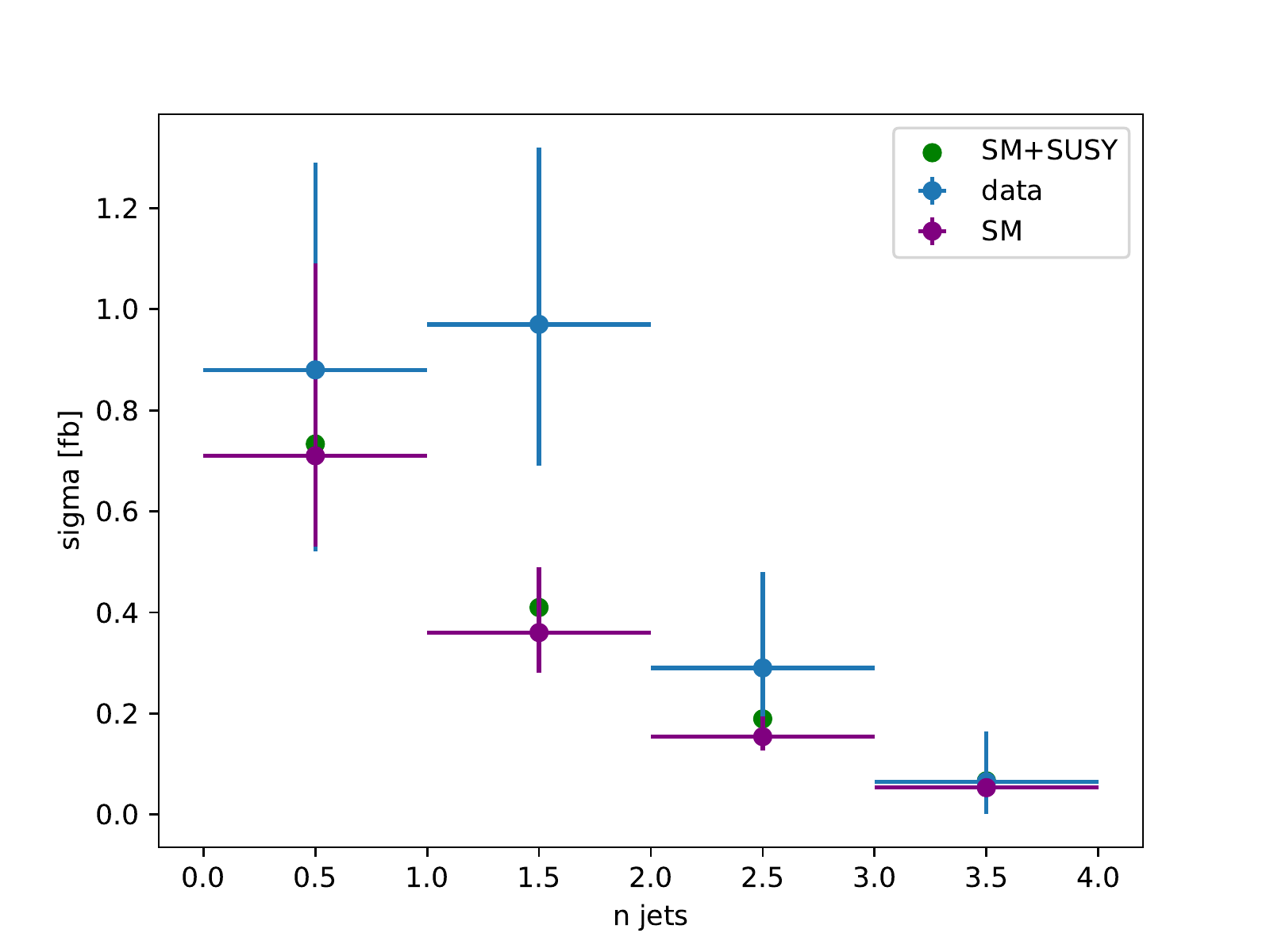}
\caption{Differential distributions of $p_{T,H}$ (left) and N(jets) for $H\to ZZ^*\to 4\ell$ of ATLAS~\cite{Aad:2014tca}. 
The SUSY signal considered 
is $pp\to\tilde\chi^\pm_1\tilde\chi^0_2$ followed by $\tilde\chi^\pm_1\to W^\pm \tilde\chi^0_1$ and $\tilde\chi^0_2\to H \tilde\chi^0_1$ 
for $m_{\tilde\chi^\pm_1,\tilde\chi^0_2}=150$~GeV and $m_{\tilde\chi^0_1}=20$~GeV and $\sigma_{\rm tot}(\tilde\chi^\pm_1\tilde\chi^0_2)=2.41$~pb.}
  \label{fig:fxs-ex1ZZ}
\end{figure}

As first example, we consider chargino-neutralino production in the MSSM, $pp\to\tilde\chi^\pm_1\tilde\chi^0_2$ followed 
by $\tilde\chi^\pm_1\to W^\pm \tilde\chi^0_1$ and $\tilde\chi^0_2\to H \tilde\chi^0_1$, 
see the left diagram in Fig.~\ref{fig:fxs-diagrams}.\footnote{For simplicity, we keep using upper case $H$ for the SM-like Higgs boson also in the SUSY case.} 
The 8~TeV ATLAS analysis \cite{Aad:2015jqa} puts a limit of $m_{\tilde\chi^\pm_1,\tilde\chi^0_2}\gtrsim 170$~GeV 
for $m_{\tilde\chi^0_1}\lesssim 40$~GeV in the $\ell\gamma\gamma+\met$ channel ($W\to\ell\nu$, $H\to\gamma\gamma$). 
The median expected limit is $m_{\tilde\chi^\pm_1,\tilde\chi^0_2}\gtrsim 154$--160~GeV for $m_{\tilde\chi^0_1}\lesssim 30$~GeV with a 
very large uncertainty. 
The 8~TeV CMS analysis \cite{Khachatryan:2014mma} includes a search in the $H\to\gamma\gamma$ channel with $W\to\ell\nu$ or $W\to 2$\,jets. 
Combining the result with that of \cite{Khachatryan:2014qwa} the CMS limit reaches $m_{\tilde\chi^\pm_1,\tilde\chi^0_2}\gtrsim 210$~GeV for (very) 
small $\tilde\chi^0_1$ mass; at $m_{\tilde\chi^\pm_1,\tilde\chi^0_2}=150$~GeV, the reach in $m_{\tilde\chi^0_1}$ is about $20$~GeV. 
We therefore take $m_{\tilde\chi^\pm_1,\tilde\chi^0_2}=150$~GeV and $m_{\tilde\chi^0_1}=20$~GeV as our test point. 
The production cross section at NLO+NLL accuracy is 2.41~pb \cite{susyxs} for wino-like $\tilde\chi^\pm_1$ and $\tilde\chi^0_2$. 
We assume ${\rm BR}(\tilde\chi^\pm_1\to W^\pm \tilde\chi^0_1) = {\rm BR}(\tilde\chi^0_2\to H \tilde\chi^0_1)=
1$, and a perfectly SM-like Higgs boson $H$ 
with mass of 125~GeV. The SUSY signal is simulated with Pythia~8.2~\cite{Sjostrand:2014zea} and then passed 
to the Rivet analysis routines~\cite{ATLAS:2014:I1306615,ATLAS:2014:I1310835}. 
The result for the $H\to\gamma\gamma$ fiducial regions from Table~\ref{tab:fxs-atlas-gamgam} is shown in Fig.~\ref{fig:fxs-ex1aa}, 
examples for two differential distributions for the $H\to ZZ^*\to 4\ell$ selection in Fig.~\ref{fig:fxs-ex1ZZ}. 
While the $H\to ZZ^*\to 4\ell$ distributions are not very sensitive to this signal, the limit in the $\met > 80$~GeV  fiducial region of 
the $H\to\gamma\gamma$ measurements excludes the benchmark point. The signal in the 
$N_{\rm leptons} \geq 1$ fiducial region is also close to the 95\% CL limit, consistent with the naive expectations 
from the previous section.

\begin{figure}[t!]\centering
\includegraphics[width=0.55\textwidth]{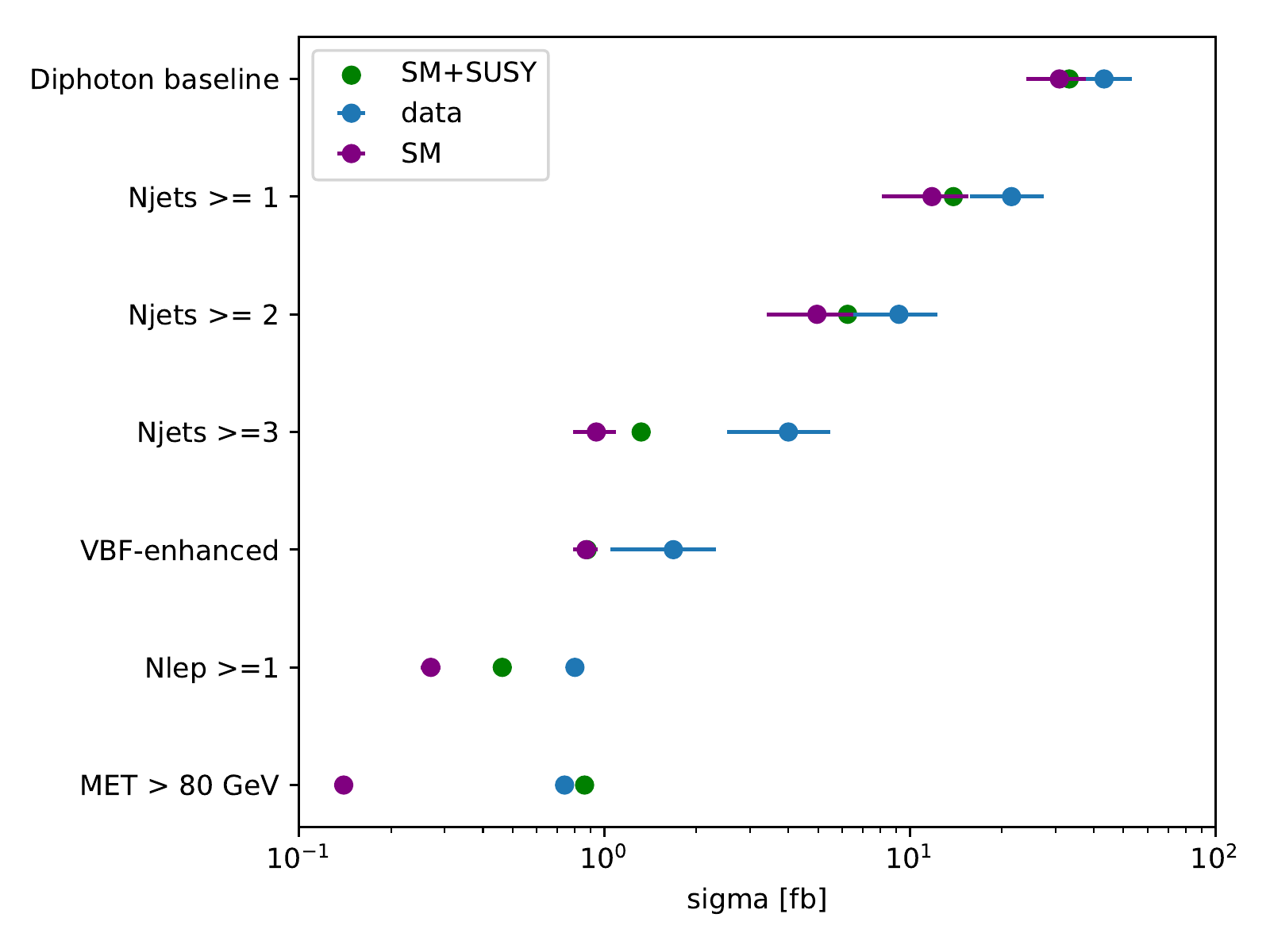}
\caption{Results for the $H\to \gamma\gamma$ fiducial regions from ATLAS~\cite{Aad:2014lwa}. The SUSY signal considered
is higgsino production with ${\rm BR}(\tilde\chi^0_1\to Z \tilde G)={\rm BR}(\tilde\chi^0_1\to H \tilde G) = 0.5$
for $m_{\tilde\chi^0_1}=150$~GeV and $m_{\tilde G}=1$~GeV and $\sigma_{\rm tot}(higgsino)=2.14$~pb.}
  \label{fig:fxs-ex2aa}
\end{figure}

Our second example is a related topology giving $HZ+\met$ or $HH+\met$ final states. 
This was considered by CMS in \cite{Khachatryan:2014mma,Sirunyan:2017eie} in the context of higgsino-like neutralino production 
with $\tilde\chi^0_1$ decaying into Higgs or $Z$ and a gravitino $\tilde G$, shown as the middle diagram in Fig.~\ref{fig:fxs-diagrams}. 
Although in the MSSM the $\tilde\chi^0_1\to Z \tilde G$ always dominates over $\tilde\chi^0_1\to H \tilde G$, the topology is interesting 
per se. Adopting one of the simplified models of the CMS study, we assume 
${\rm BR}(\tilde\chi^0_1\to Z \tilde G)={\rm BR}(\tilde\chi^0_1\to H \tilde G) = 0.5$. 
Moreover, to have a concrete benchmark point, we fix $m_{\tilde\chi^0_1}=150$~GeV and $m_{\tilde G}=1$~GeV. 
The expected limit of the 8 TeV CMS search \cite{Khachatryan:2014mma} in the $\ell\gamma\gamma+\met$ channel is about 15~pb  
for this point, an order of magnitude larger than the total higgsino production cross section, see Fig.~19 of \cite{Khachatryan:2014mma}. 
(It has to be added that a much stronger limit is obtained in the $3\ell + \met$ final state and the analysis actually excludes higgsino masses 
below about 300~GeV by combining all channels.)
The result from using the $H\to\gamma\gamma$ fiducial cross section is shown in Fig.~\ref{fig:fxs-ex2aa}. Following CMS, the cross section 
obtained from Pythia has been rescaled to the total higgsino production cross section of 2.14~pb \cite{susyxs}, assuming the decays of 
$\tilde\chi^\pm_1,\tilde\chi^0_2$ to the $\tilde\chi^0_1$ plus soft pions are effectively invisible.
As in the example for SUSY $WH$ production above, the scenario is excluded by the limit in the $\met > 80$~GeV  fiducial region, 
although by a smaller margin. 

\begin{figure}[t!]
\hspace*{-2mm}\includegraphics[width=0.54\textwidth]{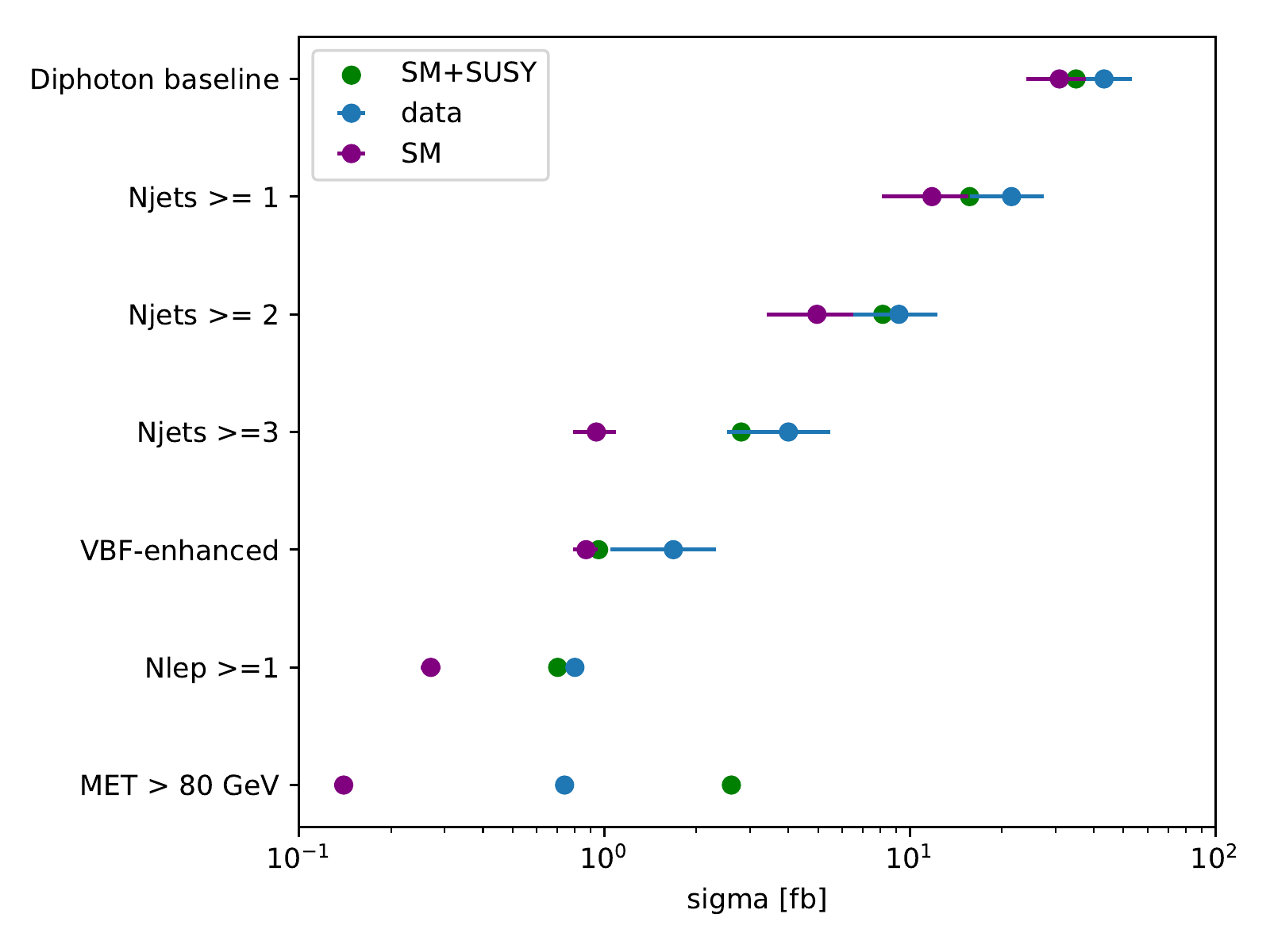}\includegraphics[width=0.52\textwidth]{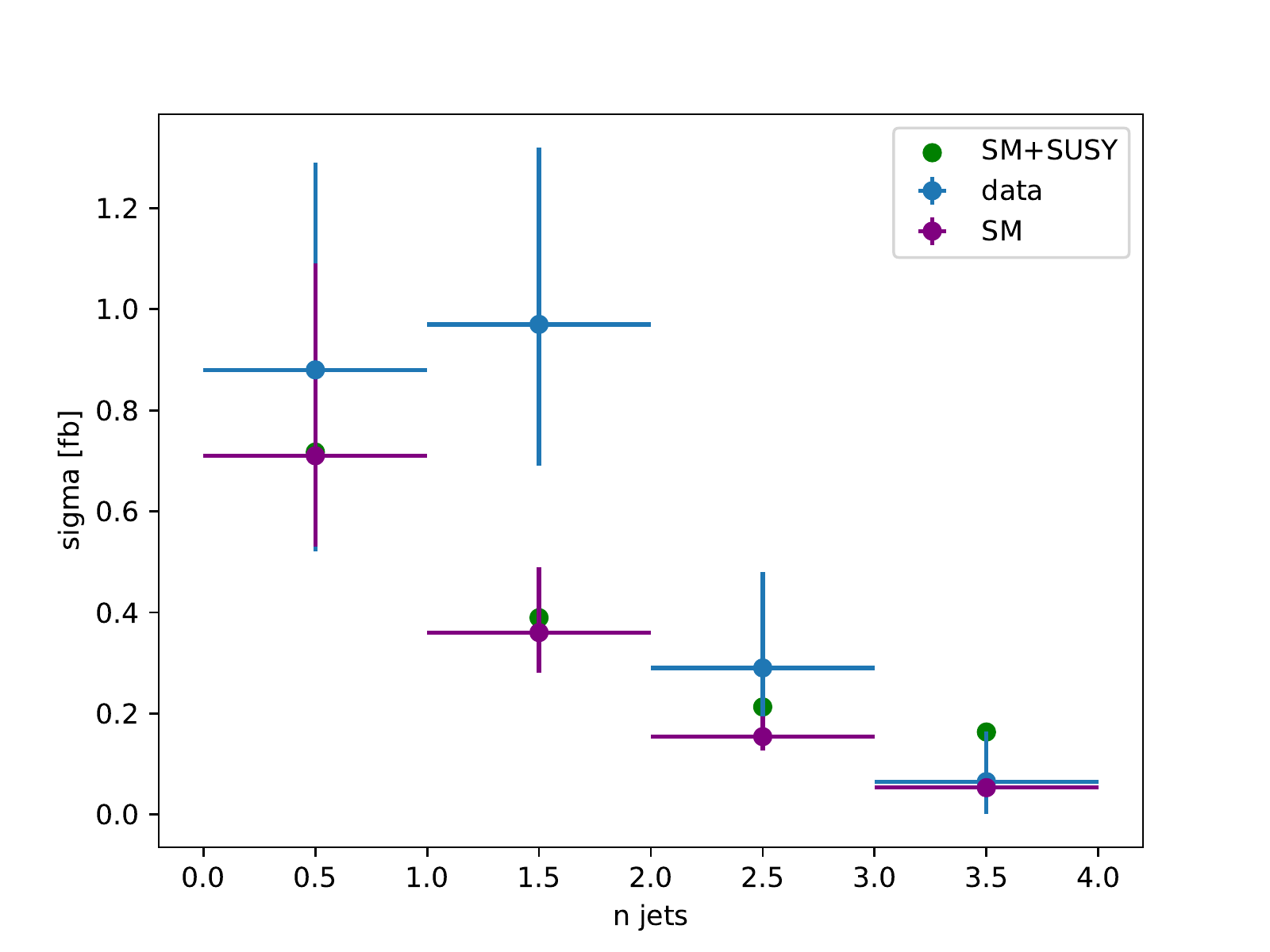}
\caption{On the left, results for the $H\to \gamma\gamma$ fiducial regions from ATLAS~\cite{Aad:2014lwa}. 
On the right, the N(jets) distribution for $H\to ZZ^*\to 4\ell$ of ATLAS~\cite{Aad:2014tca}.
The SUSY signal considered 
is $pp\to\tilde b_1\tilde b_1$ followed by  $\tilde b_1\to b\tilde\chi^0_2$ and $\tilde\chi^0_2\to H \tilde\chi^0_1$ 
for $m_{\tilde b_1} = 300$, $m_{\tilde\chi^0_2} = 280$ and $m_{\tilde\chi^0_1} = 150$~GeV with a total inclusive cross section 
of about 2~pb.}
  \label{fig:fxs-ex3}
\end{figure}

Finally, we consider sbottom-pair production with both sbottoms decaying into $\tilde b_1\to b\tilde\chi^0_2$ followed by 
$\tilde\chi^0_2\to H\tilde\chi^0_1$ (right diagram in Fig.~\ref{fig:fxs-diagrams}). 
This was proposed in \cite{Duarte:2017bbq} to explain a local $2.9\sigma$ excess in 
the 8 TeV search for electroweak SUSY partners in $H\to\gamma\gamma$~+~1~jet events~\cite{CMS-PAS-SUS-14-017}, 
and followed up by a dedicated interpretation in the 13~TeV CMS analysis~\cite{Sirunyan:2017eie}.
In this contribution, we choose a benchmark point with $m_{\tilde b_1} = 300$, $m_{\tilde\chi^0_2} = 280$ and $m_{\tilde\chi^0_1} = 150$~GeV, 
which lies just outside the CMS SUSY exclusion at 13~TeV (because the spectrum is rather compressed). 
The sbottom-pair production cross section for 300 GeV at 8 TeV is 1.996~pb \cite{susyxs}. 
The simulation is again done with Pythia~8.2 and the events fed to Rivet. The result for the diphoton fiducial regions from ATLAS 
is shown in Fig.~\ref{fig:fxs-ex3}\,(left).  
Note the important SUSY contribution to the $N_{\rm jets} \geq 2$ and $N_{\rm jets} \geq 3$ fiducial regions. 
Moreover, because the signal contains two Higgs bosons, $H\to\gamma\gamma$ and $H\to WW^*, ZZ^*$ combinations 
can give $H\to\gamma\gamma$ events with additional leptons, getting us close to the 95\% CL exclusion in the 
$N_{\rm leptons} \geq 1$ fiducial region. Finally, the $\met > 80$~GeV fiducial region excludes the benchmark point. 
The presence of the additional jets in the SUSY cascade process also leads to a small effect in the N(jets) distribution for 
$H\to ZZ^*\to 4\ell$, see Fig.~\ref{fig:fxs-ex3}\,(right). 
 
\section*{CONCLUSIONS}

We showed that fiducial Higgs cross section measurements can provide interesting constraints on anomalous Higgs production from BSM processes. 
These are complementary to and sometimes extend the constraints from dedicated BSM searches in final states with Higgs bosons. 
This is particularly useful for interpretation studies when the scenario of interest has not been considered by the ATLAS and CMS collaborations in any of their searches and/or 
when the relevant BSM searches cannot easily be reproduced outside the experimental collaboration, e.g., because they use a signal selection based on machine learning. In turn, it can be instructive to consider the existing constraints from fiducial measurements when constructing a search/interpretation for a new BSM signal. 

Here we focussed on SUSY scenarios, for which dedicated searches exist. Analogous studies will be interesting for non-SUSY models 
featuring new heavy scalars, new vector bosons, etc., which can decay into the SM-like Higgs boson. 
We leave this for future work, noting that fiducial measurements for 36~fb$^{-1}$ of data at 13~TeV are already available 
for $H\to\gamma\gamma$~\cite{Aaboud:2018xdt} and 
$H\to ZZ^*\to 4\,\ell$~\cite{Aaboud:2017oem} from ATLAS.

To make the fiducial Higgs cross section measurements maximally useful, we kindly ask the ATLAS and CMS collaborations to provide the SM 
predictions used in the plots of differential distributions available in HEPData, together with the measured data. Moreover, it is highly appreciated when Rivet routines are provided, as they greatly facilitate the re-use of these important data. 
The ATLAS $H\to ZZ^*\to 4\ell$ fiducial measurements from Run~1 are an example of good practise and we hope that other analyses will follow this example.

\section*{ACKNOWLEDGEMENTS}
We thank the ATLAS Higgs group for providing the SM predictions used in \cite{Aad:2015lha} on HEPData, and 
Christian Gutschow and Jonathan Stahlman for providing a Rivet routine for this analysis in time for being used in these proceedings. 

K.L. is supported by the European Union's Horizon 2020 research and innovation programme under ERC grant agreement No.\ 715871.



\AddToContent{S.~Kraml, U.~Laa, K.~Lohwasser}
\renewcommand{\thesection}{\arabic{section}}

\graphicspath{{LH2HDM/}}

\newcommand\ltap{\
  \raise.3ex\hbox{$<$\kern-.75em\lower1ex\hbox{$\sim$}}\ }
\newcommand\gtap{\
  \raise.3ex\hbox{$>$\kern-.75em\lower1ex\hbox{$\sim$}}\ }

\newcommand\simge{\mathrel{%
   \rlap{\raise 0.511ex \hbox{$>$}}{\lower 0.511ex \hbox{$\sim$}}}}
\newcommand\simle{\mathrel{
   \rlap{\raise 0.511ex \hbox{$<$}}{\lower 0.511ex \hbox{$\sim$}}}}

\newcommand{\slashchar}[1]%
        {\kern .25em\raise.18ex\hbox{$/$}\kern-.75em #1}
\def\lsim{\mathrel{\raise.3ex\hbox{$<$\kern-.75em\lower1ex\hbox{$\sim$}}}}
\def\gsim{\mathrel{\raise.3ex\hbox{$>$\kern-.75em\lower1ex\hbox{$\sim$}}}}
\newcommand\CA{{\cal A}}\newcommand\CCA{$\CA$}
\newcommand\CB{{\cal B}}\newcommand\CCB{$\CB$}
\newcommand\CC{{\cal C}}\newcommand\CCC{$\CC$}
\newcommand\CD{{\cal D}}\newcommand\CCD{$\CD$} 
\newcommand\CE{{\cal E}}\newcommand\CCE{$\CE$}
\newcommand\CF{{\cal F}}\newcommand\CCF{$\CF$}
\newcommand\CG{{\cal G}}\newcommand\CCG{$\CG$}
\newcommand\CH{{\cal H}}\newcommand\CCH{$\CH$}
\newcommand\CI{{\cal I}}\newcommand\CCI{$\CI$}
\newcommand\CJ{{\cal J}}\newcommand\CCJ{$\CJ$}
\newcommand\CK{{\cal K}}\newcommand\CCK{$\CK$}
\newcommand\CL{{\cal L}}\newcommand\CCL{$\CL$}
\newcommand\CM{{\cal M}}\newcommand\CCM{$\CM$}
\newcommand\CN{{\cal N}}\newcommand\CCN{$\CN$}
\newcommand\CO{{\cal O}}\newcommand\CCO{$\CO$}
\newcommand\CP{{\cal P}}\newcommand\CCP{$\CP$}
\newcommand\CQ{{\cal Q}}\newcommand\CCQ{$\CQ$}
\newcommand\CR{{\cal R}}\newcommand\CCR{$\CR$}
\newcommand\CS{{\cal S}}\newcommand\CCS{$\CS$}
\newcommand\CT{{\cal T}}\newcommand\CCT{$\CT$}
\newcommand\CU{{\cal U}}\newcommand\CCU{$\CU$}
\newcommand\CV{{\cal V}}\newcommand\CCV{$\CV$}
\newcommand\CW{{\cal W}}\newcommand\CCW{$\CW$}
\newcommand\CX{{\cal X}}\newcommand\CCX{$\CX$}
\newcommand\CY{{\cal Y}}\newcommand\CCY{$\CY$}
\newcommand\CZ{{\cal Z}}\newcommand\CCZ{$\CZ$}
\newcommand\ba{\begin{array}}
\newcommand\ea{\end{array}}
\newcommand\tx{\textstyle}
\newcommand\whW{\widehat W}
\newcommand\wtW{\widetilde W}
\newcommand\whD{\widehat D}
\newcommand\wtD{\widetilde D}
\newcommand\homega{\widehat\omega_{Da}}
\newcommand\whomega{\widehat\omega_8}
\newcommand\whomegaz{\widehat\omega_{8,0}}
\newcommand\dtwoE{d^2E(W)/dq\Lambda^2}
\newcommand{\bket}{\ensuremath{|B^0 \rangle}}
\newcommand{\bbra}{\ensuremath{\langle B^0|}}
\newcommand{\bbarket}{\ensuremath{|\bar{B}^0\rangle}}
\newcommand{\bbarbra}{\ensuremath{\langle \bar{B}^0|}}
\newcommand{\lcp}{\ensuremath{\lambda_{CP}}}
\newcommand{\dmu}{\ensuremath{\partial_{\mu}}}
\newcommand{\dmup}{\ensuremath{\partial^{\mu}}}
\newcommand{\dnu}{\ensuremath{\partial_{\nu}}}
\newcommand{\sigbar}{\ensuremath{\bar{\sigma}^{\mu}\cdot}}
\newcommand{\sig}{\ensuremath{\sigma^{\mu}\cdot}}
\newcommand{\VA}{\ensuremath{\gamma^{\mu}(1 - \gamma_5)}}
\newcommand{\gm}{\ensuremath{\gamma^{\mu}}}
\newcommand{\gd}{\ensuremath{\gamma_{\mu}}}
\newcommand{\gammat}{\ensuremath{\left( \begin{array}{cc} 0 & \sigma^{\mu} \\
        \bar{\sigma}^{\mu} & 0 \\ \end{array} \right)}}
\newcommand{\gndbra}{\ensuremath{\langle \Omega |}}
\newcommand{\gndket}{\ensuremath{|\Omega \rangle}}
\newcommand{\half}{\ensuremath{\frac{1}{2}}}
\newcommand{\third}{\ensuremath{\frac{1}{3}}}
\newcommand{\fourth}{\ensuremath{\frac{1}{4}}}
\newcommand{\thalf}{\ensuremath{\textstyle{\frac{1}{2}}}}
\newcommand{\tthird}{\ensuremath{\textstyle{\frac{1}{3}}}}
\newcommand{\tfourth}{\ensuremath{\textstyle{\frac{1}{4}}}}
\newcommand{\thhalf}{\ensuremath{\frac{3}{2}}}
\newcommand{\fourthirds}{\ensuremath{\frac{4}{3}}}
\newcommand{\hsig}{\ensuremath{\frac{\sigma_a}{2}}}
\newcommand{\BD}{\ensuremath{B_d}}
\newcommand{\BDbar}{\ensuremath{\bar B_d}}
\newcommand{\stwobeta}{\ensuremath{\sin{2\beta}}}
\newcommand{\thw}{\ensuremath{\theta_W}}
\newcommand\thc{\theta_C}
\newcommand\thy{\theta_Y}
\newcommand\dagg{\dagger}
\newcommand\ts{\thinspace}
\newcommand\Ra{\Rightarrow}
\newcommand\Lra{\Longrightarrow}
\newcommand\longra{\longrightarrow}
\newcommand\leftra{\leftrightarrow}
\newcommand\llra{\longleftrightarrow}
\newcommand\olra{\overleftrightarrow}
\newcommand\ol{\bar}
\newcommand\mev{{\rm MeV}}
\newcommand\gev{{\rm GeV}}
\newcommand\tev{{\rm TeV}}
\newcommand\MeV{{\rm MeV}}
\newcommand\nb{{\rm nb}}
\newcommand\pb{{\rm pb}}
\newcommand\ipb{{\rm pb}^{-1}}
\newcommand\fb{{\rm fb}}
\newcommand\ifb{{\rm fb}^{-1}}
\newcommand\ecm{\sqrt{s}}
\newcommand\rshat{\sqrt{\shat}}
\newcommand\shat{\hat s}
\newcommand\nin{\noindent}
\newcommand\lvac{\langle \Omega \vert}
\newcommand\rvac{\vert \Omega \rangle}
\newcommand\condt{\langle \bar T T\rangle}
\newcommand\condtt{\langle \bar T^t_L T^t_R\rangle}
\newcommand\condtb{\langle \bar T^b_L T^b_R\rangle}
\newcommand\condtl{\langle \bar T^l_L T^l_R\rangle}
\newcommand\condaa{\langle \bar T^1_L T^1_R\rangle}
\newcommand\condbb{\langle \bar T^2_L T^2_R\rangle}
\newcommand\condab{\langle \bar T^1_L T^2_R\rangle}
\newcommand\condba{\langle \bar T^2_L T^1_R\rangle}
\newcommand\condij{\langle \bar T^i_L T^j_R\rangle}
\newcommand\conduij{\langle \bar U_{iL} U_{jR}\rangle}
\newcommand\conddij{\langle \bar D_{iL} D_{jR}\rangle}
\newcommand\condtbt{\langle \bar t t\rangle}
\newcommand\condbbb{\langle \bar b b\rangle}
\newcommand\et{E_T}
\newcommand\emiss{\slashchar{E}}
\newcommand\cstar{\cos \theta^*}
\newcommand\sscy{10 \ts \fb^{-1}}
\newcommand\hl{10^{33} \ts {\rm cm}^{-2} \ts {\rm s}^{-1}}
\newcommand\hly{10^{40} \ts {\rm cm}^{-2}}
\newcommand\uhl{10^{34} \ts {\rm cm}^{-2} \ts {\rm s}^{-1}}
\newcommand\uhly{10^{41} \ts {\rm cm}^{-2}}
\newcommand\sscd{100 \ts \fb^{-1}}
\newcommand\yr{{\rm yr}}
\newcommand\cmsec{{\rm cm^{-2}sec^{-1}}}
\newcommand\cm{{\rm cm}}
\newcommand\ellm{\ell^-}
\newcommand\ellpm{\ell^\pm}
\newcommand\ellp{\ell^+}
\newcommand\epm{e^\pm}
\newcommand\ep{e^+}
\newcommand\h{H^0}
\newcommand\mee{M_{e^+ e^-}}
\newcommand\mgaga{M_{\gamma \gamma}}
\newcommand\mll{M_{\ell^+ \ell^-}}
\newcommand\mmm{M_{\mu^+ \mu^-}}
\newcommand\mm{{\rm mm}}
\newcommand\mum{\mu^-}
\newcommand\mupm{\mu^\pm}
\newcommand\mup{\mu^+}
\newcommand\qqbar{q \ol q}
\newcommand\cond{\langle \ol T T\rangle}
\newcommand\getc{g_{ETC}}
\newcommand\Gtc{G_{TC}}
\newcommand\Gew{SU(2)\otimes U(1)}
\newcommand\Getc{G_{ETC}}
\newcommand\su{SU(3)}
\newcommand\sufive{SU(5)}
\newcommand\sofive{SO(5)}
\newcommand\suc{SU(3)_C}
\newcommand\Ntc{N_{TC}}
\newcommand\sutc{SU(N_{TC})}
\newcommand\uone{U(1)_1}
\newcommand\utwo{U(1)_2}
\newcommand\uy{U(1)_Y}
\newcommand\sutwow{SU(2)_W}
\newcommand\sutwop{SU(2)'}
\newcommand\sutwoc{SU(2)_C}
\newcommand\suone{SU(3)_1}
\newcommand\sutwo{SU(3)_2}
\newcommand\suthree{SU(3)_3}
\newcommand\aqcd{\alpha_{S}}
\newcommand\atc{\alpha_{TC}}
\newcommand\aetc{\alpha_{ETC}}
\newcommand\Metc{M_{ETC}}
\newcommand\Letc{\Lambda_{ETC}}
\newcommand\Ltc{\Lambda_{TC}}
\newcommand\Leff{{\cal L}_{\rm eff}}
\newcommand\Lsig{{\cal L}_{\Sigma}}
\newcommand\LFF{{\cal L}_{\rm gauge}}
\newcommand\LWZW{{\cal L}_{\rm WZW}}
\newcommand\Lff{{\cal L}_{\bar f f}}
\newcommand\Lpifbf{{\cal L}_{\tpi \bar f f}}
\newcommand\grpp{g_{\rho_T\pi_T\pi_T}}
\newcommand\condtc{{\langle \ol T T \rangle}_{TC}}
\newcommand\condetc{{\langle \ol T T \rangle}_{ETC}}
\newcommand\vev{\langle \phi \rangle}
\newcommand\tom{\omega_{T}}
\newcommand\tro{\rho_{T}}
\newcommand\atro{\alpha_{\rho_T}}
\newcommand\aat{\alpha_{a_T}}
\newcommand\trho{\rho_T}
\newcommand\ta{a_T}
\newcommand\at{a_T}
\newcommand\taz{a_T^0}
\newcommand\tapm{a_T^\pm}
\newcommand\tros{\rho_{T8}^{0}}
\newcommand\troct{\rho_{T8}}
\newcommand\tropm{\rho_{T}^\pm}
\newcommand\trop{\rho_{T}^+}
\newcommand\trom{\rho_{T}^-}
\newcommand\troz{\rho_{T}^0}
\newcommand\toppi{\pi_t}
\newcommand\tpi{\pi_T}
\newcommand\tpipm{\pi_T^\pm}
\newcommand\tpimp{\pi_T^\mp}
\newcommand\tpip{\pi_T^+}
\newcommand\tpim{\pi_T^-}
\newcommand\tpiz{\pi_T^0}
\newcommand\tpipr{\pi_T^{0 \prime}}
\newcommand\tpilq{\pi_{L\ol Q}}
\newcommand\tpiql{\pi_{Q\ol L}}
\newcommand\tpioct{\pi_{T8}}
\newcommand\etat{\eta_{_{T}}}
\newcommand\trou{\rho_{_{\ol U U}}}
\newcommand\trod{\rho_{_{\ol D D}}}
\newcommand\rhol{\rho_{_{\ol L L}}}
\newcommand\rhoq{\rho_{_{\ol Q Q}}}
\newcommand\rhon{\rho_{_{\ol N N}}}
\newcommand\rhoe{\rho_{_{\ol E E}}}
\newcommand\piqq{\pi_{_{\ol Q Q}}}
\newcommand\piql{\pi_{_{\ol Q L}}}
\newcommand\piuu{\pi_{_{\ol U U}}}
\newcommand\pidd{\pi_{_{\ol D D}}}
\newcommand\piud{\pi_{_{\ol U D}}}
\newcommand\pidu{\pi_{_{\ol D U}}}
\newcommand\pinu{\pi_{_{\ol N U}}}
\newcommand\piun{\pi_{_{\ol U N}}}
\newcommand\pieu{\pi_{_{\ol E U}}}
\newcommand\piue{\pi_{_{\ol U E}}}
\newcommand\pind{\pi_{_{\ol N D}}}
\newcommand\pidn{\pi_{_{\ol D N}}}
\newcommand\pied{\pi_{_{\ol E D}}}
\newcommand\pide{\pi_{_{\ol D E}}}
\newcommand\chipr{\chi^{\ts \prime}}
\newcommand\afb{A_{FB}}
\newcommand\pbp{\ol p p}
\newcommand\Mv{M_{V_8}}
\newcommand\Mtt{\CM_{\ol t t}}
\newcommand\MMtt{\langle \Mtt \rangle}
\newcommand\MSMtt{\langle \Mtt^2 \rangle}
\newcommand\RMStt{{\langle \Mtt^2 \rangle^{1/2}}}
\newcommand\Deltt{\Delta \Mtt}
\newcommand\jet{{\rm jet}}
\newcommand\jets{{\rm jets}}
\newcommand\stt{\sigma(\tbt)}
\newcommand\QbQ{\ol Q Q}
\newenvironment{changemargin}[2]{\begin{list}{}{
        \setlength{\topsep}{0pt}\setlength{\leftmargin}{0pt}
        \setlength{\rightmargin}{0pt}
        \setlength{\listparindent}{\parindent}
        \setlength{\itemindent}{\parindent}
        \setlength{\parsep}{0pt plus 1pt}
        \addtolength{\leftmargin}{#1}\addtolength{\rightmargin}{#2}
        }\item }{\end{list}}
%
\chapter{Death and the Model: A Contur Case Study}

{\it J.~M.~Butterworth, D.~Grellscheid,
K.~Lane, K.~Lohwasser and L.~Pritchett}



\begin{abstract}
  This report summarizes the use of Contur to constrain a two-Higgs-doublet
  model explanation of an apparent $30\,\gev$ dimuon resonance observed in a reanalysis
  of $Z \to \bar bb$ events in ALEPH data taken in 1992--95 and reported in
  arXiv:1610.06536. The model was proposed by two of us in
  arXiv:1701.07376. Contur is used to limit the mass of the model's charged
  Higgs boson, $h^\pm$, which is produced in pairs or in association with
  the extra neutral CP-even and odd scalars $h$ and $\eta_A$. The limit
  obtained excludes $h^\pm$ and the 2HDM model for the ALEPH dimuon excess.

\end{abstract}


\section{Introduction}
\label{sec2hdm:intro}
In addition to the extensive programme of searches, the growing `library' of
measurements from the LHC experiments is placing ever-more-stringent
requirements on proposed extensions of the Standard Model (SM).  Measurements
defined in terms of final-state particles, in fiducial regions reflecting the
acceptance of the detectors, are rather model-independent and are thus
particularly suited for confrontation with the predictions of new models.
Many of these measurements are made differentially in key kinematic
variables.  In this contribution, we take an extension of the SM which has
been proposed~\cite{Lane:2017gry} to address a feature in ALEPH
data~\cite{Heister:2016stz}, and confront it with such data from ATLAS and
CMS.  The model and its motivation are outlined in Sec.~\ref{sec2hdm:2hdmtext}.
The analysis tools we use are described in Sec.~\ref{sec2hdm:tools}, and the
analysis and results are presented in Sec.~\ref{sec2hdm:analysis}, before we
present our conclusions.

\section{Two-Higgs-doublet model of the $30\,\gev$ dimuon} 
\label{sec2hdm:2hdmtext}
\begin{figure}[!ht]
 \begin{center}
\includegraphics[width=2.60in, height=2.60in]{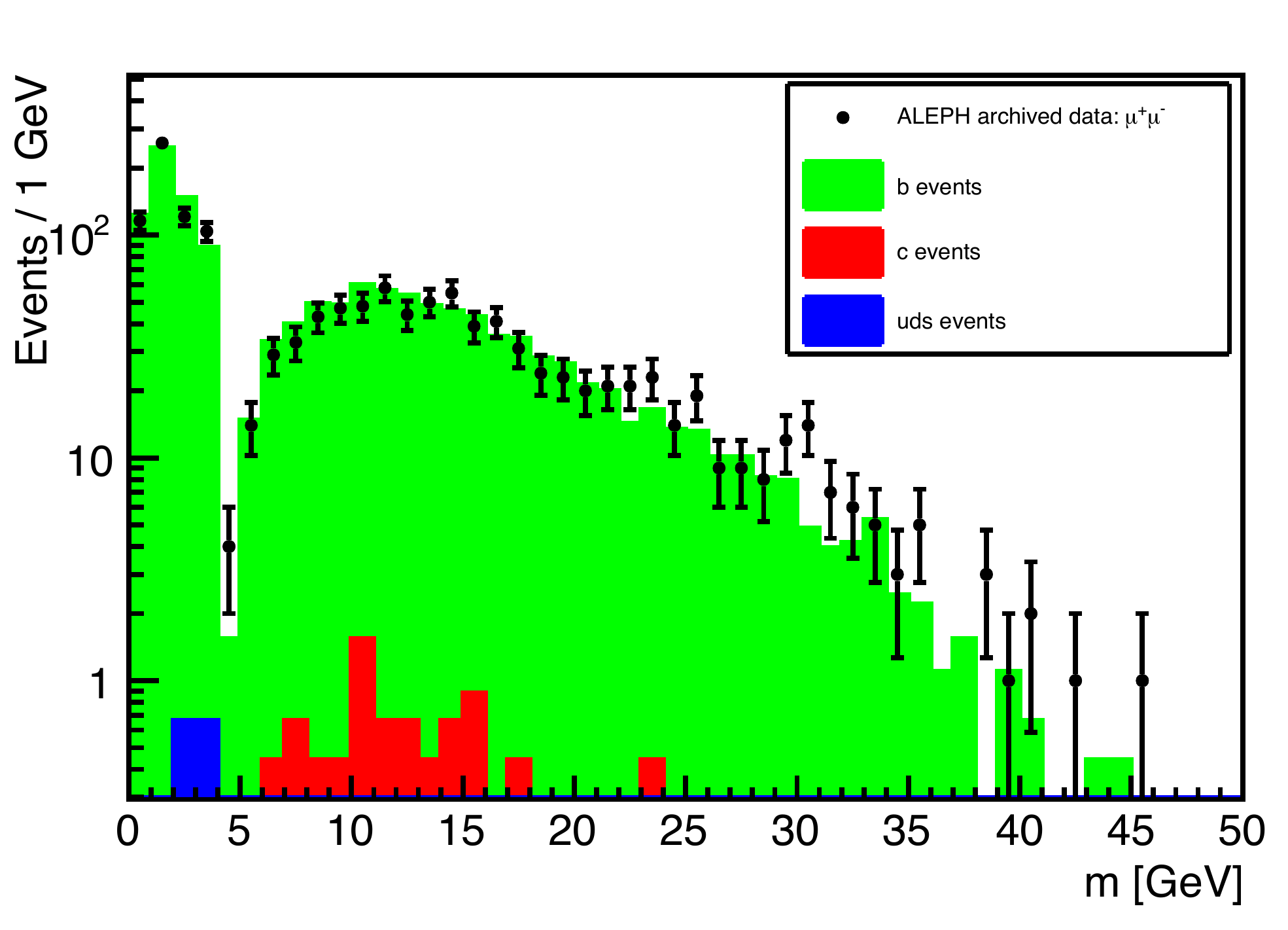}
\includegraphics[width=2.60in, height=2.60in]{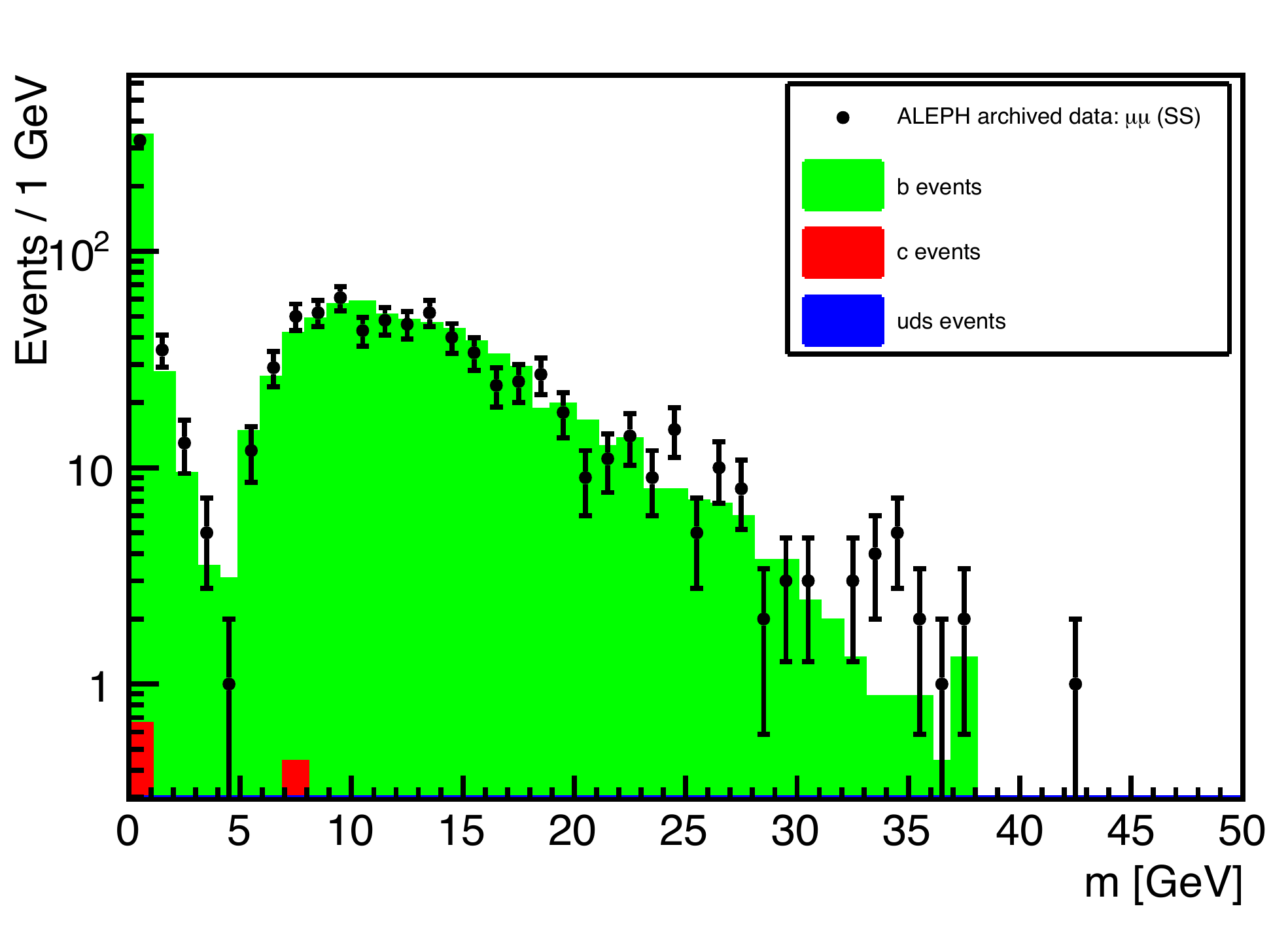}
\caption{The opposite-sign~(left) and same-sign~(right) dimuon mass spectra
  in $Z \to \bar bb \mu\mu$ data taken by the ALEPH Collaboration; from
  Ref.~\cite{Heister:2016stz}.}
  \label{fig:Fig1ab}
 \end{center}
 \end{figure}

 In 2016 Heister presented an analysis of archived data of the ALEPH
 experiment at LEP and found evidence for a narrow dimuon ($\mu^+ \mu^-$)
 resonance at $30\,\gev$~\cite{Heister:2016stz}. The data, taken in 1992-95,
 involve 1.9~million hadronic decays of $Z$-bosons produced at rest in
 $e^+e^-$ annihilation. This excess appears in $Z \to \bar bb \mu^+\mu^-$
 events. The opposite-sign dimuon spectrum data is shown in
 Fig.~\ref{fig:Fig1ab} (left) along with the expected background. The
 same-sign dimuon spectrum in Fig.~\ref{fig:Fig1ab} (right) has no significant
 excesses. The data have the following characteristics:
\begin{itemize}

\item[1.)] Two benchmark methods were used to estimate the significance of
  the excess. One gave a local significance of about $2.6\,\sigma$, the other
  $5.4\,\sigma$. The second method requires using the look-elsewhere effect;
  it reduces its significance by~1.4--$1.6\,\sigma$. See
  Ref.~\cite{Heister:2016stz} for details.

\item[2.)] There is an excess of $32\pm 11$ events in the resonant peak of
  Fig.~\ref{fig:Fig2} corresponding to a mass of $30.40\,\gev$ with a
  Breit-Wigner width of $1.78\,\gev$ (Gaussian width of $0.74\,\gev$),
  consistent with the expected ALEPH dimuon mass reconstruction performance
  at $30\,\gev$. Using the ALEPH $b$-tag and single-muon-ID efficiencies of
  $38\%$ and $86\%$~\cite{Heister:2016stz} and
  $B(Z \to \bar bb)/B(Z \to {\rm hadrons}) = 0.216$ yields the branching
  ratio
\be\label{eq:BZbbmm}
B(Z \to \bar b b\, X (\to \mu^+ \mu^-)) = (2.77 \pm 0.95)\times 10^{-4}.
\ee
If the dimuon excess is due to the decay of a new particle $X$, it is not
known whether it is emitted from the $Z$, as in $Z \to Z^* X$ with
$Z^* \to \bar b b$ and $X \to \mu^+\mu^-$, or from one of the $b$-quarks, as
in $Z \to \bar bb \to \bar bb + X$, or from two new particles, $Z \to XY$,
with $X \to \mu^+ \mu^-$ and $Y \to \bar bb$.

\begin{figure}[!t]
 \begin{center}
\includegraphics[width=3.15in, height=3.15in]{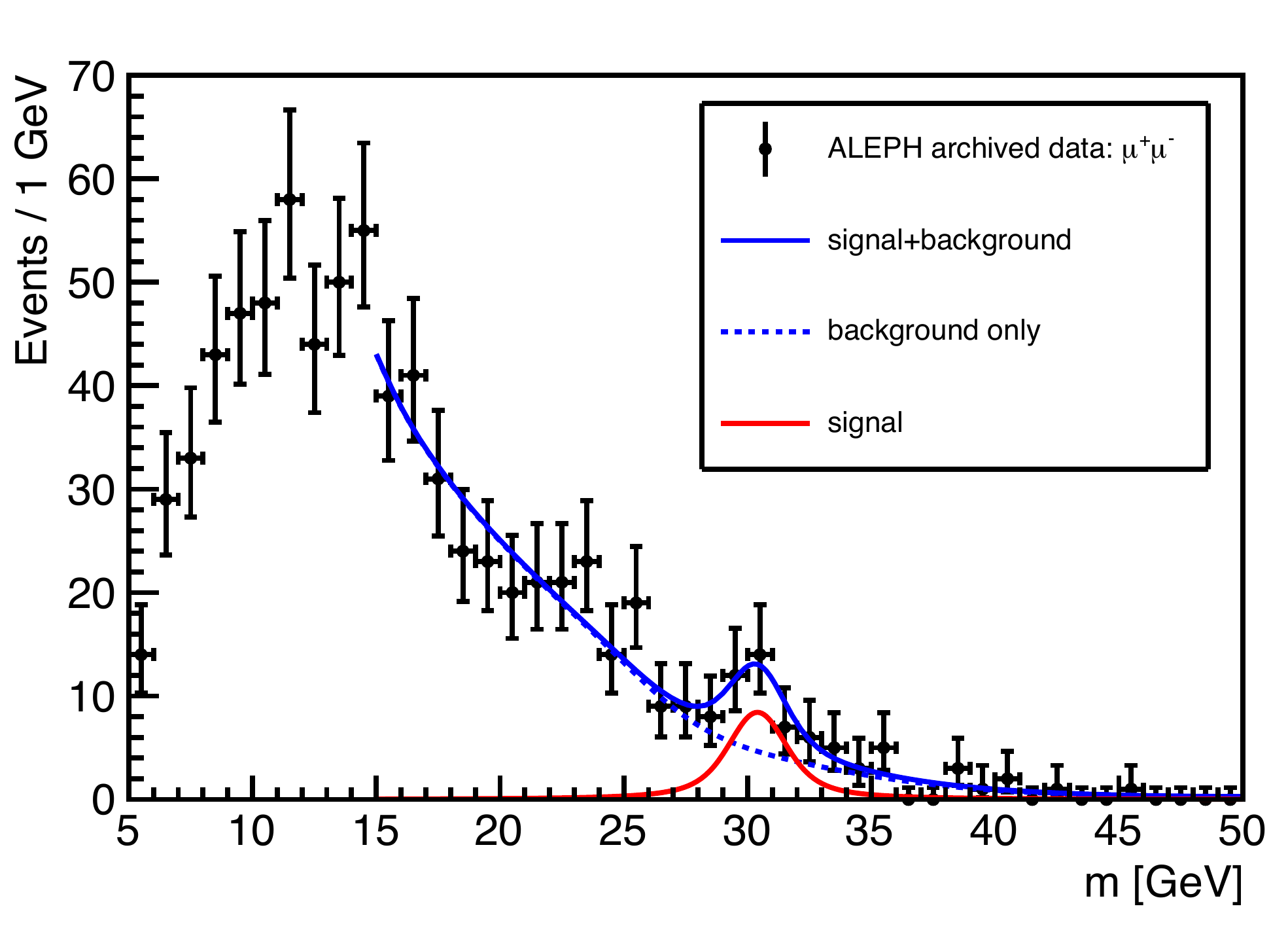}
\caption{ALEPH $Z \to \bar bb \mu^+\mu^-$ data with signal$+$background model
  used to extract the 30~GeV signal parameters in
  Ref.~\cite{Heister:2016stz}.}
  \label{fig:Fig2}
 \end{center}
 \end{figure}

\item[3.)] There is a small excess of $8.0 \pm 4.5$ events near $M_{e^+ e^-}
  = 30\,\gev$ in the $Z \to \bar bb e^+ e^-$ data. This is not considered in
  the model described below.

\item[4.)] There is no evidence for the 30~GeV dimuon excess in events for
  which the $b$-tag has been inverted; see Fig.~\ref{fig:Fig3}, from
  Ref.~\cite{Heister:2016stz}. 

\end{itemize}

\begin{figure}[!t]
 \begin{center}
\includegraphics[width=3.15in, height=3.15in]{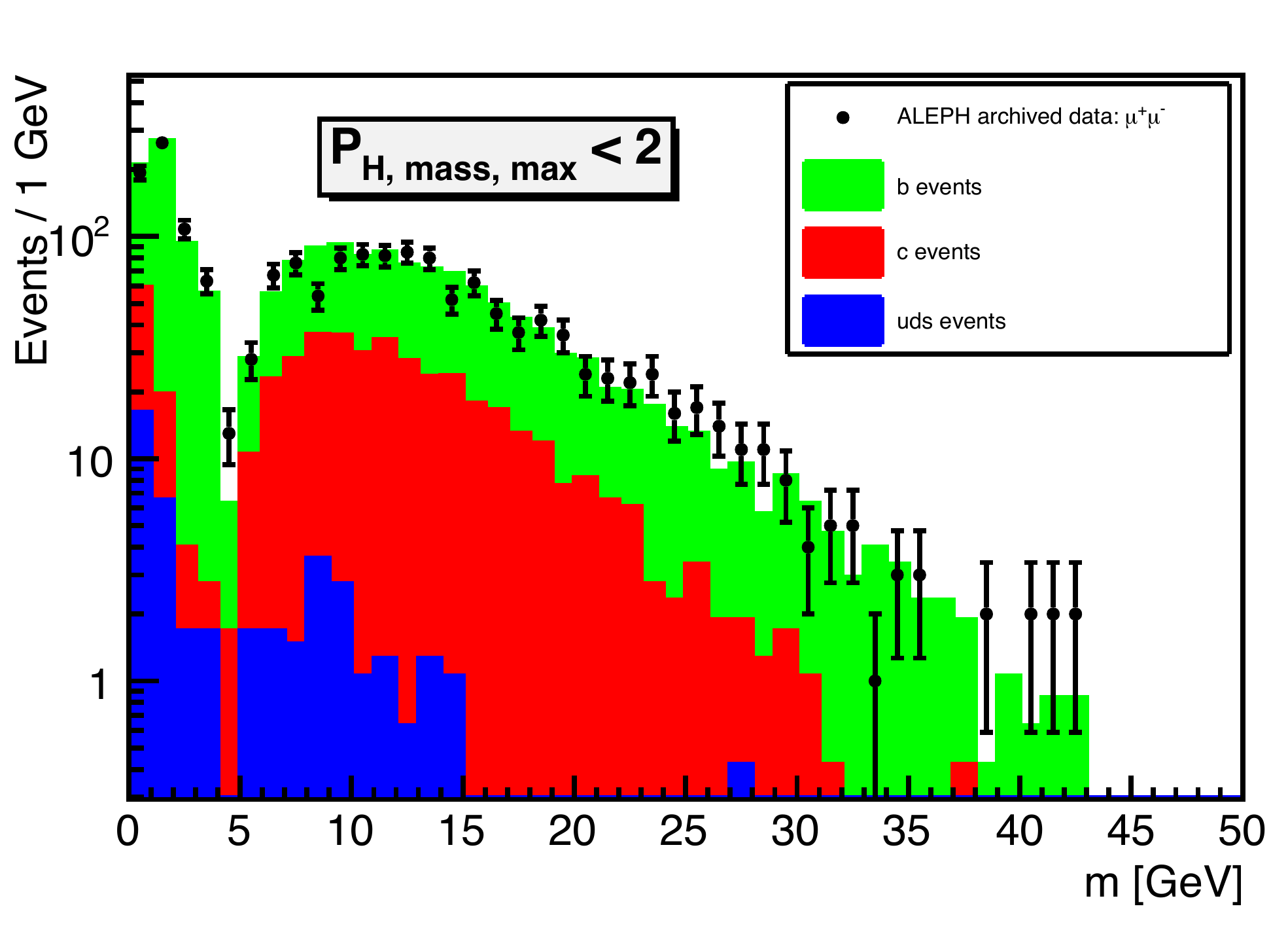}
\caption{The opposite-sign dimuon mass spectrum in $Z \to {\rm hadrons}+\mu^+
  \mu^-$ events in which the $b$-tag has been inverted, indicating no evidence
  for an excess near $30\,\gev$; from Ref.~\cite{Heister:2016stz}.}
  \label{fig:Fig3}
 \end{center}
 \end{figure}

The obvious and simplest explanation of these features of the ALEPH data is
 that the 30~GeV excess is just a statistical fluctuation in semileptonic
 $Z \to \bar bb$ decays. If that possibility is set aside, however, no Monte Carlo of semileptonic $b$-decays
 in $Z \to \bar bb$ at LEP or elsewhere has produced such an excess. It is
 tempting, therefore, to construct a model which can account for the ALEPH
 data and suggest searches by LHC experiments that might confirm -- or refute
 -- the existence of the $30\,\gev$ dimuon in $Z$~decays.

 Two of us proposed such a model~\cite{Lane:2017gry}. It is a two-Higgs
 doublet model (2HDM) in which the heavier CP-even Higgs boson $H$ is the
 $125\,\gev$ Higgs boson discovered in 2012 at the LHC~\cite{Aad:2012tfa,
   Chatrchyan:2012ufa}. The two other neutral Higgs bosons are a CP-even
 one~$h$ and a CP-odd one~$\eta_A$. It is these that will account for the
 ALEPH signal. In this model, one Higgs doublet,
 $\phi_1 = (\phi_1^+,v_1+\rho_1+i\pi_1)/\sqrt{2}$, couples in the usual way
 to all the quark doublets as well as to the $\tau$-lepton doublet. The other
 doublet, $\phi_2$, couples only to the muon and electron doublets. This
 set-up is not one of the commonly studied classes of
 2HDM's~\cite{Branco:2011iw}. Nevertheless, it can be enforced by a softly
 broken $U(1)$ symmetry and it does not give rise to observable charged lepton
 flavor violation.

 By choosing the vacuum expectation values (vevs) so that
 $v \equiv \sqrt{v_1^2 + v_2^2} = 246\,\gev \cong v_1$, the 125~GeV Higgs
 boson $H \cong \rho_1$ and its couplings to the electroweak (EW) gauge
 bosons and all the fermions --- except for muons, electrons and their
 neutrinos --- are very nearly as in the standard model~(SM) with a single
 Higgs doublet. This is effectively the situation referred to as
 ``alignment''; see e.g.~Ref.~\cite{Carena:2013ooa}. The model's additional
 Higgs bosons, $h \cong \rho_2$, $\eta_A \cong \pi_2$ and
 $h^\pm \cong \phi_2^\pm$ couple directly to the~$\mu$ and~$e$ doublets and
 singlets.

 The most natural choice of Higgs self-coupling parameters --- the only one
 with $H \cong \rho_1$ and which can account for the ALEPH dimuon signal ---
 has $M_{\eta_A} \cong M_h = 30\,\gev$.\footnote{This $h$--$\eta$ degeneracy
   in 2HDM's with alignment was noticed earlier, in 
   e.g.~Ref.~\cite{Carena:2013ooa}.} Their main decay modes are
 $h,\eta \to \mu^+ \mu^-$ and, through $\phi_1$--$\phi_2$ mixing,
 $h,\eta_A \to \bar bb$ at the percent level. The decay $Z \to h\eta_A$ can
 then easily account for the $\CO(10^{-4})$ branching ratio in
 Eq.~(\ref{eq:BZbbmm}) by choosing
 $\tan\beta = v_2/v_1 \simeq 1/20$.\footnote{Rates for the ``Higgstrahlung''
   processes such as $Z \to Z^* h$ with $Z^* \to \bar bb$ and
   $h \to \mu^+ \mu^-$ are 5--6 orders of magnitude smaller than
   $Z \to h\eta_A \to \bar bb \mu^+\mu^-$.} This is the origin of the ALEPH
 signal in this model.

 To prevent a very large contribution to the $H$ width from $H \to h^+ h^-$,
 it is simplest to take $M_{h^\pm} > M_H/2$. This can be arranged by taking a
 particular scalar quartic coupling negative, a choice that is consistent
 with vacuum stability. Up to $M_{h^\pm} \simeq 125\,\gev$, its main decay
 mode is $h^\pm \to \mu^\pm \nu_\mu$. No searches for charged Higgses at LEP,
 the Tevatron and the LHC look in this mode, assuming instead that charged
 Higgses decay to the heaviest quark pairs allowed kinematically. The
 strongest limit for a simple $\mu^+ \mu^- + E_T^{\rm miss}$ signal appears
 to come from searches at LEP for pair-production of supersymmetric partners
 of the muon, $\tilde\mu^\pm$.  They limit $M_{h^\pm} > 95\,\gev$; see,
 Ref.~\cite{Abbiendi:2003ji}, e.g. While the $h^+ \bar tb$ coupling in this
 model is suppressed by $\tan\beta$ relative to the large coupling $m_t/v$
 generally assumed in collider searches, it is still large enough to make
 $h^+ \to t\bar b$ an important decay mode if allowed. It is excluded by a
 CMS search for $t(b) h^\pm$ production followed by $h^+ \to t \bar b$ with
 large branching ratio~\cite{Khachatryan:2015qxa}. With the $\tan\beta$
 suppression, we estimate the upper limit implied by this search to be
 $M_{h^\pm} \simle 200\,\gev$.

 Important constraints on the model come from the Higgs couplings to EW
 bosons. The relevant couplings in the unitary gauge are:
\bea\label{eq:phiEW}
\CL_{EW} &=& ie\left[A^\mu + Z^\mu \cot 2\theta_W \right] 
               h^+ \overleftrightarrow{\partial_\mu} h^- \nn\\
&+& \frac{e}{\sin 2\theta_W}\left[\left(h\cos(\beta-\alpha) -
  H\sin(\beta-\alpha)\right)\overleftrightarrow{\partial_\mu}\,\eta_A \right]
Z^\mu \nn\\ 
&+& \frac{e}{2\sin\theta_W}\left[\left(\eta_A \pm i h\cos(\beta-\alpha) \mp
  i H\sin(\beta-\alpha)\right)\overleftrightarrow{\partial_\mu}\,h^\pm \right]
W^{\mp\,\mu} \nn\\
&+&\left[\left(e A_\mu + e\cot 2\theta_W Z_\mu \right)^2 +
  \frac{e^2}{2\sin^2\theta_W} W^{+\,\mu} W_\mu^-\right]h^+ h^- \nn\\
&+& \left[\frac{e^2}{\sin^2 2\theta_W} Z^\mu Z_\mu +
    \frac{e^2}{2\sin^2\theta_W} W^{+\,\mu} W_\mu^-\right]\nn\\
&& \times \left[v(H\cos(\beta-\alpha) + h\sin(\beta-\alpha)) + \thalf(H^2 +
  h^2 + \eta_A^2)\right].
\eea
Here, $\alpha$ is the angle diagonalizing the $H$-$h$ mass matrix.

For small $\alpha$ and $\beta$, the gauge couplings of~$H$ are close to the
SM in all cases that are measurable in the near future. Note the strong
$Z h\eta_A$, $W^\pm h^\mp h$ and $W^\pm h^\mp \eta_A$ couplings, determined
by gauge invariance. This presents an immediate problem for this 2HDM. For
$B(h,\eta_A \to \mu^+\mu^-) \cong 1$,
\be\label{eq:Zheta}
 \Gamma(Z \to h\eta_A \to \mu^+\mu^-\mu^+\mu^-) \cong \Gamma(Z \to h\eta_A) = 
\frac{2\alpha_{EM}\, p^3}{3 M_Z^2\sin^2 2\theta_W} \cos^2(\beta-\alpha),
\ee
where $p$ is the momentum of $h$ in the $Z$~rest frame. For
$M_h = M_{\eta_A} = 30\,\gev$, this gives $B(Z \to 4\mu) \cong 0.0141$, about
3300 times larger that the measured branching ratio of
$4.2 \times 10^{-6}$~\cite{Patrignani:2016xqp}. While it appears impossible
to evade this problem in a 2HDM~\cite{Lane:2017gry}, it may be possible with
the added flexibility of Higgs-fermion couplings in a 3HDM.\footnote{Lane and
  Pritchett are investigating the possibility of this.}

The new Higgs bosons of the model are all in the mass range where they can
easily be produced at the LHC. While no dedicated searches exist,
measurements of final states containing muons have been made and may be
expected to have sensitivity to this model, quite apart from the $Z$ width
issue.  The main focus of this report is to use these measurements to limit
the charged Higgs in the apparently allowed range
$95\,\gev < M_{h^\pm} < 200\,\gev$. The signal processes to be tested are the
Drell-Yan productions
\bea
\label{eq:DYpm}
\bar qq &\to& \gamma^*,Z^* \to h^+ h^- \to \mu^+\mu^- + E_T^{\rm miss}, \\
\label{eq:DYpz}
\bar qq' &\to&W^* \to h^\pm h,\,\, h^\pm \eta_A  \to \mu^\pm \mu^+\mu^-
+ \etmiss.
\eea
In order to investigate the limits from the existing LHC measurements, we use the the Contour analysis tools described in the next section. The analysis itself is in Sec.~\ref{sec2hdm:analysis}.

\section{Analysis Tools}
\label{sec2hdm:tools}
The key tools for our analysis are the Herwig event generator~\cite{Bahr:2008pv,Bellm:2015jjp}, 
the Rivet library of analysis routines~\cite{Buckley:2010ar}, and
the Contur comparison package~\cite{Butterworth:2016sqg}.

Herwig simulates complete LHC events, starting from the matrix element for a hard scatter and including 
leading-logarithmic QCD partons showers, the conversion of partons into hadrons, hadronic decays and a simulation of
the underlying event. An important feature for this analysis is the fact that it provides an interface to read in the 
Universal FeynRules Output (UFO~\cite{Degrande:2011ua}) files produced by the Feynrules~\cite{Christensen:2008py} 
package used to encode the model, 
and incorporates the new matrix elements implied by the model 
into its event generator machinery. It then allows inclusive generation of any or all of the new particles and 
processes, along with SM contributions if desired. In this analysis we use Herwig~7.1.2~\cite{Bellm:2017bvx}.

Rivet contains a library of analysis routines corresponding to published measurements made at colliders.
Many of the analyses are provided by the experiments themselves. The majority of them are particle-level,
differential cross sections made in a fiducial kinematic region. 
This means that the experiments have defined a measurement based on `true' final state particles 
and have corrected for detector effects such as resolution and efficiency, but have not extrapolated beyond 
their acceptance.
These measurements thus have a high degree of model independence. Rivet applies the same analysis as the experiment 
to generated final state particles, in our case from Herwig, and reproduces the measurements as histograms.
Rivet also contains the published data and uncertainties, derived from HEPDATA~\cite{Maguire:2017ypu}, 
which can then be compared to the results from Herwig. In this analysis we use Rivet~2.5.4.

Contur takes the output of Rivet for a range of generated model parameters,
and makes a statistical comparison between the prediction and the data. At
present, the comparison made is between the data alone, and the data plus the
generated BSM contribution. This approach will evaluate the room for new
physics contributions which is left by the uncertainties on the measurement,
under the assumption that the measurement is identical to the SM. Since all
the measurements used have been shown to agree with the SM, this assumption
is not unreasonable, although it neglects the theory uncertainties and will
thus potentially give an over-aggressive exclusion limit when these are
large, or if the data diverge from the SM.~\footnote{Planned future versions
of Contur will allow direct comparison to the SM predictions, with their
uncertainties.} The current aim of Contur is to provide a rapid `health
check' for new physics models, to identify those regions already disfavoured
by existing measurements.

\section{Contur analysis: Limits on the charged Higgs mass}
\label{sec2hdm:analysis}

A key advantage of Herwig is that it is simple to select the inclusive
production of any given particle, including the new particles introduced by
the ALEPH dimuon model of Sec.~\ref{sec2hdm:2hdmtext}. This capability
is particularly well suited to the Contur approach, where all available
measurements can be used simultaneously.  This already led to the unexpected
observation~\cite{Butterworth:2016sqg} that the precise measurements of
vector bosons produced in association with jets have sensitivity to benchmark
Dark Matter models. As we shall see shortly, it also leads to unexpected
results in the current analysis.

As discussed above, events in which a $Z$ boson decays into $h\eta_A$ produce
copious 4-muon final states, inconsistent with the measured branching
ratio~\cite{Patrignani:2016xqp}.  For this reason we exclude those processes
(assuming that a modified model may be constructed to suppress them).  We
focus on the charged sector of the model, requiring all events to have at
least one charged Higgs --- i.e., using Herwig to generate inclusive $h^\pm$
production. The dominant processes are $h^+ h^-$, $h^\pm h$ and
$h^\pm \eta_A$ production. As an additional constraint, for events which
contain a $W$ or a $Z$ boson, only those $W$ and $Z$ decays involving an
electron or a muon are generated. As we shall see, this gives conservative
exclusion limits, as the events containing other decay channels of $W$
bosons also have sensitivity.

We assume the primary data set of interest to be the leptonic ``diboson''
measurements, since the decays $h^\pm \rightarrow \mu^\pm \nu$ and
$h, \eta_A \rightarrow \mu^+\mu^-$ will lead to contributions to
$N \mu + E_T^{\rm miss}$ final states, which will show up in the selections
used in measurements aimed at the fully-leptonic $W^+W^-$ and $WZ$ diboson
processes. The most relevant analysis available in Rivet is the ATLAS 8~TeV
measurement~\cite{Aad:2016wpd}. The equivalent CMS analysis is not available,
although its $H \rightarrow WW$ measurement~\cite{Khachatryan:2016vnn} is,
and was also included. The $7\,\tev$ ATLAS measurements of the low-mass
Drell-Yan process~\cite{Aad:2014qja},
$ZZ \rightarrow \mu^+\mu^- + E_T^{\rm miss}$~\cite{Aad:2012awa} and
fully-leptonic $WW$~\cite{ATLAS:2012mec} are also available, and have some
sensitivity. Since Herwig, Rivet and Contur are designed to generate and
study all processes together with little additional overhead, several other
measurements were also studied for potential sensitivity, most importantly
the ATLAS measurement of the four-lepton line shape~\cite{Aad:2015rka}. Other
potentially useful measurements either have no significant sensitivity, or
were not yet available in Rivet at the time of writing.

\begin{figure}[!t]
 \begin{center}
\includegraphics[width=4.15in, height=3.15in]{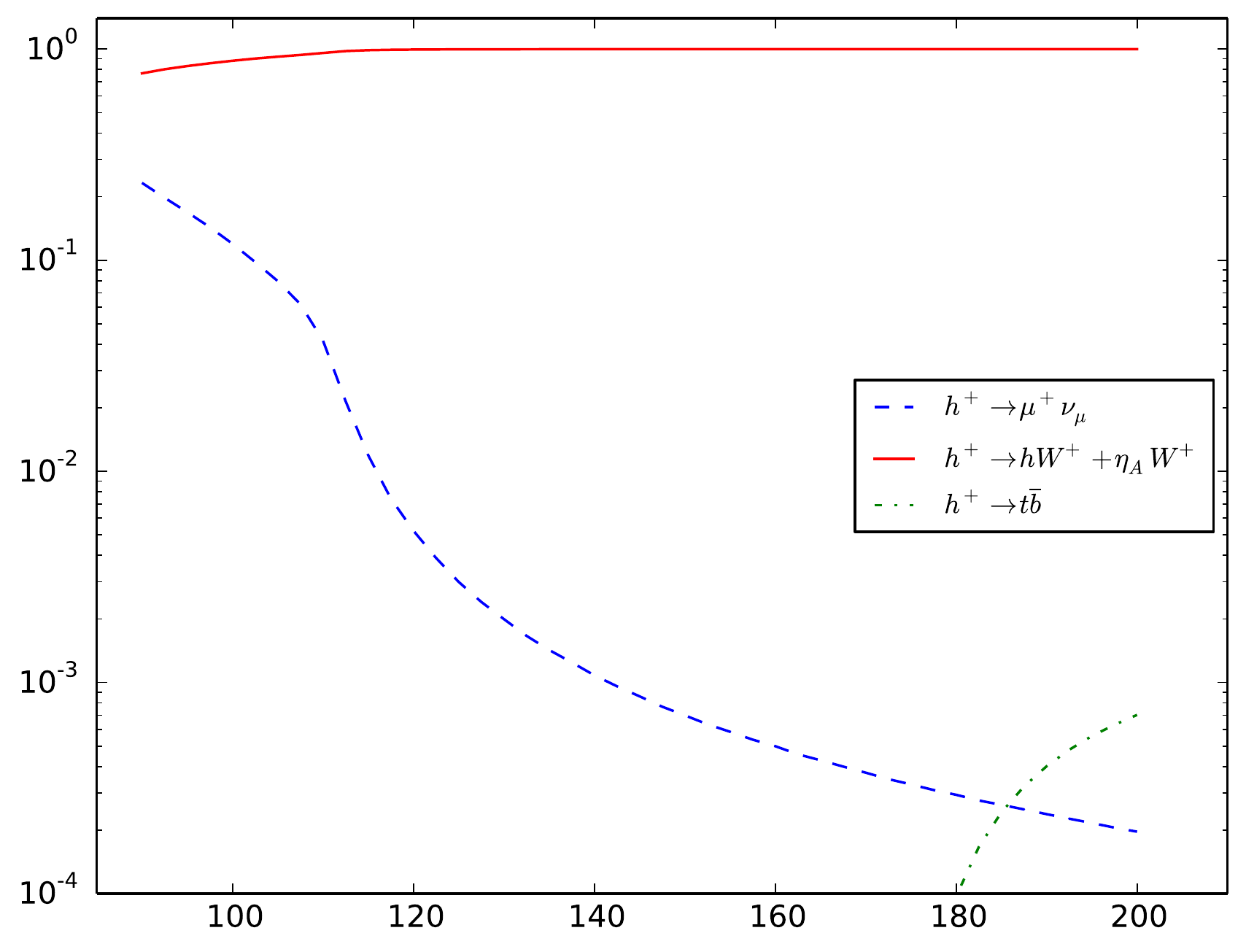}
\caption{Decay branching ratios of the charged Higgs $h^\pm$ in
  Ref.~\cite{Lane:2017gry} to $\mu^\pm \nu$, $W^\pm h + W^\pm \eta_A$ and to
  $t \bar b$. In the model of Ref.~\cite{Lane:2017gry}, the $h^+ \to t \bar
  b$ decay rate is suppressed from its SM value by $\tan^2\beta \simeq
  1/400$.}
  \label{fig:hplusBRs}
 \end{center}
 \end{figure}

 The $W^+W^-$ measurement in the muon channel does indeed have sensitivity,
 excluding the model at the 97\% c.l. for $M_{h^\pm} = 100$~GeV, though this
 sensitivity dies away at higher masses (13\% by $M_{h^\pm} = 200$~GeV).
 But, the model is excluded at a confidence level greater than 99.9\% over
 the otherwise allowed mass range~95 to~$200\,\gev$ discussed in
 Sec.~\ref{sec2hdm:2hdmtext} by the ATLAS four-lepton line shape. This may seem
 surprising, since the ATLAS measurement requires at least one lepton pair
 close to the $Z$ mass ($50 < M_{ll} < 120$~GeV). However, the inclusive
 Herwig calculation reveals that, over the considered mass range, the by far
 dominant decay modes of the charged Higgs are $h^\pm \rightarrow h W^\pm$
 and $h^\pm \rightarrow \eta_A W^\pm$, both at branching fractions close to
 50\%; also see Fig.~\ref{fig:hplusBRs}.  The near 100\% decays of $h$ or
 $\eta_A$ to $\mu^+\mu^-$ give four muons per event. The
 $BR(W^\pm \to \mu^\pm \nu) = 10.6\%$ for decay to a muon or electron and a
 neutrino, mean that $h^+h^-$ events can in fact contain up to six muons, (or
 four muons and an $e^+e^-$ pair, or five muons and an electron); similarly,
 $h^\pm h$ and $h^\pm \eta_A$ events will mostly contain four muons and can
contain up to five. In the absence of vetos on missing energy or if 
additional leptons within the rapidity acceptance do not exceed the 
veto thresholds ($7\,\gev$ for muons, $6\,\gev$ for electrons), 
all these final states can potentially contribute to either a
 $4\mu$ or $2\mu 2e$ final state. In fact, parton luminosities and phase
 space considerations ($M_h = M_{\eta_A} = 30\,\gev \ll M_{h^\pm} =
 100$--$200\,\gev$) imply that the dominant processes being excluded 
are $u \bar 
 d,\, c \bar s \to W^{+\,*} \to h^+h,\, h^+\eta_A \to W^+ hh, \, W^+ h\eta_A
 ,\, W^+ \eta_A \eta_A \to W^+ + 4\mu$. This is borne out by
 Fig.~\ref{fig:sources}.

   We note that the measurement includes the four-electron channels, for which there will be no 
   contribution from our model. This illustrates a feature likely to be common in 
   such studies: while from a SM point of view the combination is of most interest, 
   since the events predominantly involve $Z/\gamma$ propagators; the muon-only measurement, 
   produced without a combination with electrons, would often be even more sensitive to BSM physics.

   While individual pairs of the muons produced in the model primarily come from either low mass $h$
   or $\eta_A$ decays, and other pairings have no mass peak, the high
   multiplicity and high cross section mean that many events pass the
   fiducial selection given in Table~2 of Ref.~\cite{Aad:2015rka} and implemented
   in Rivet.

\begin{figure}
\begin{minipage}[c]{0.5\textwidth}
\begin{center}
\includegraphics[width=0.99\textwidth]{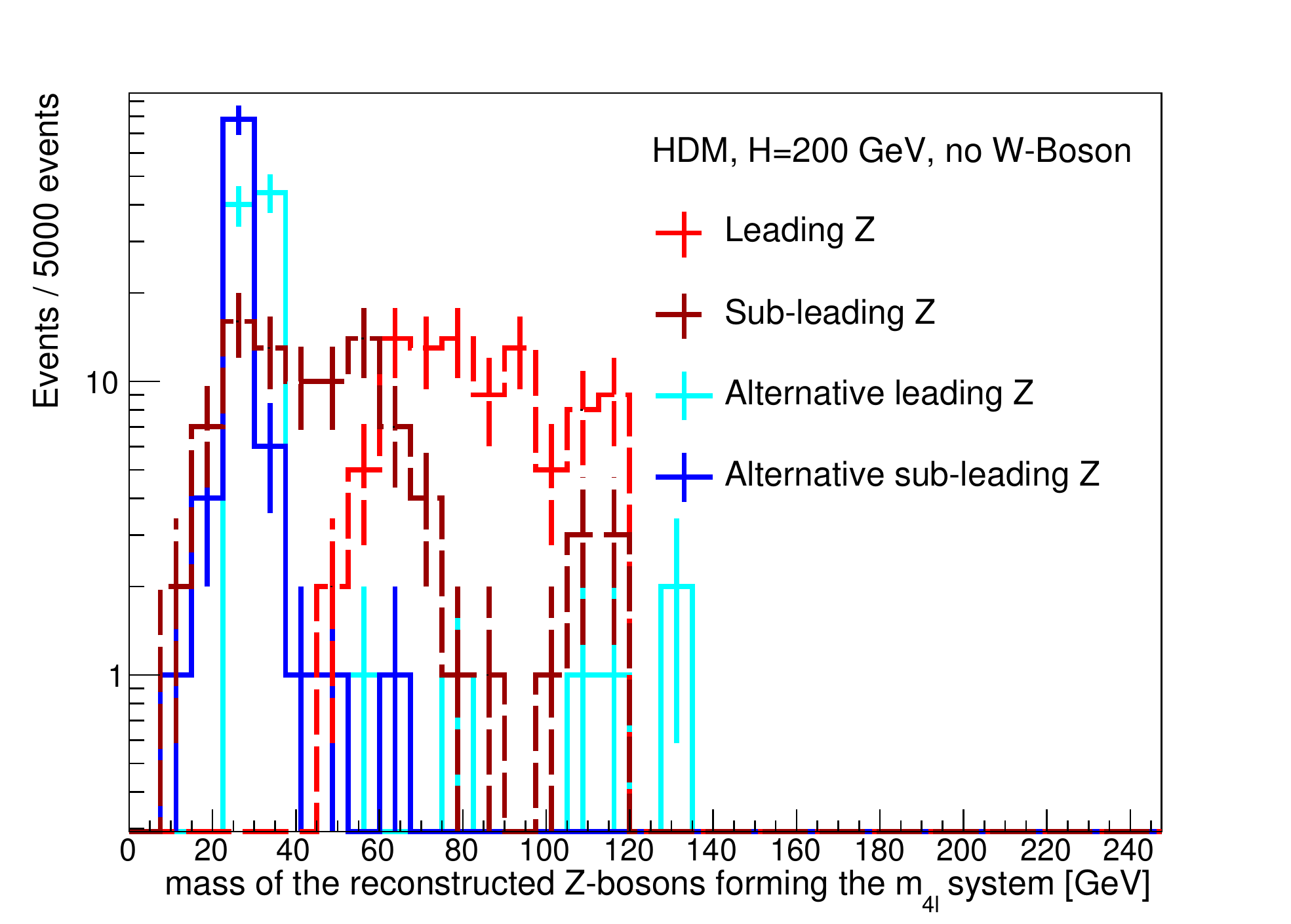}
\end{center}
\end{minipage}
\begin{minipage}[c]{0.5\textwidth}
\begin{center}
\includegraphics[width=0.99\textwidth]{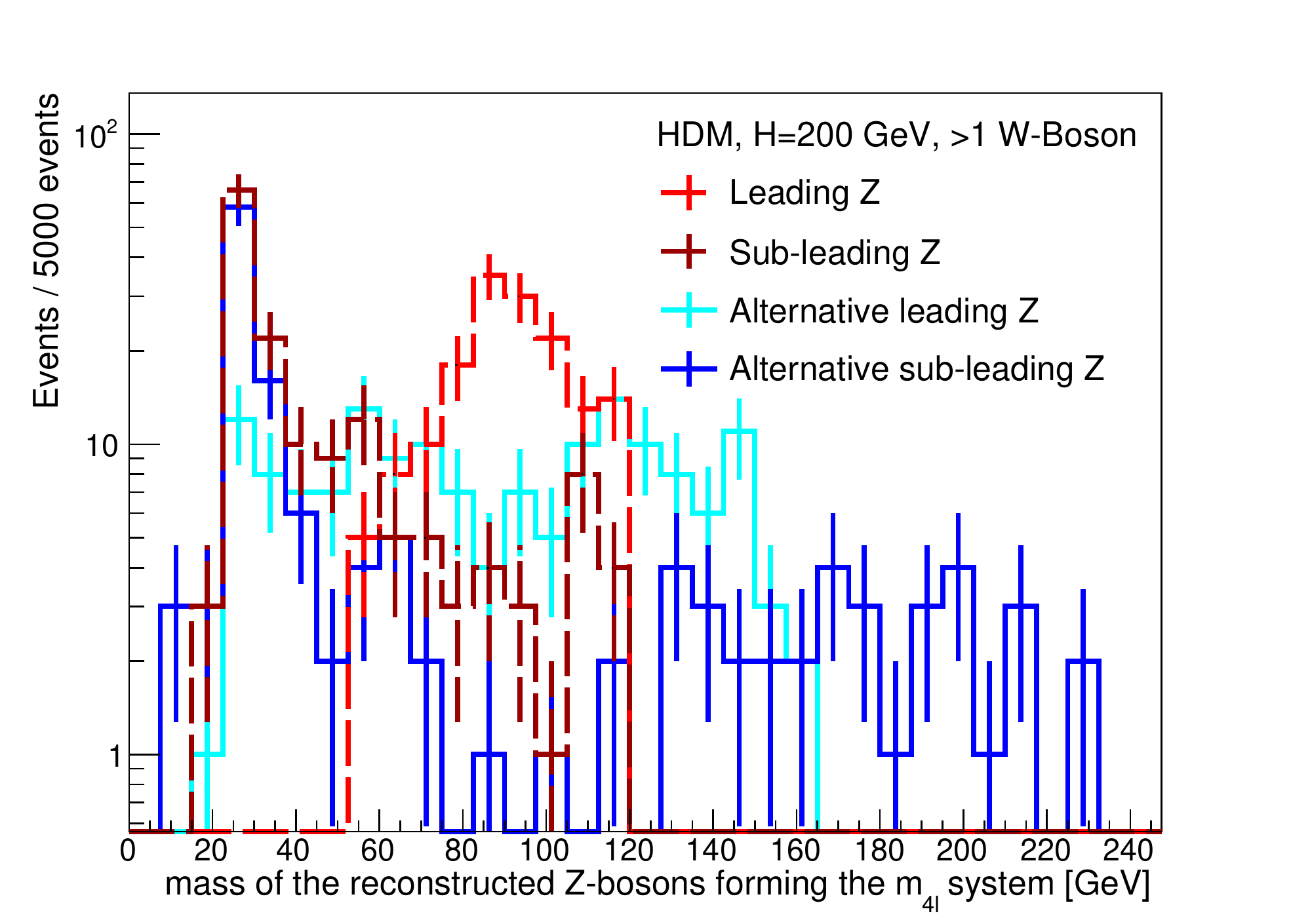}
\end{center}
\end{minipage}
\begin{minipage}[c]{0.5\textwidth}
\begin{center}
\includegraphics[width=0.99\textwidth]{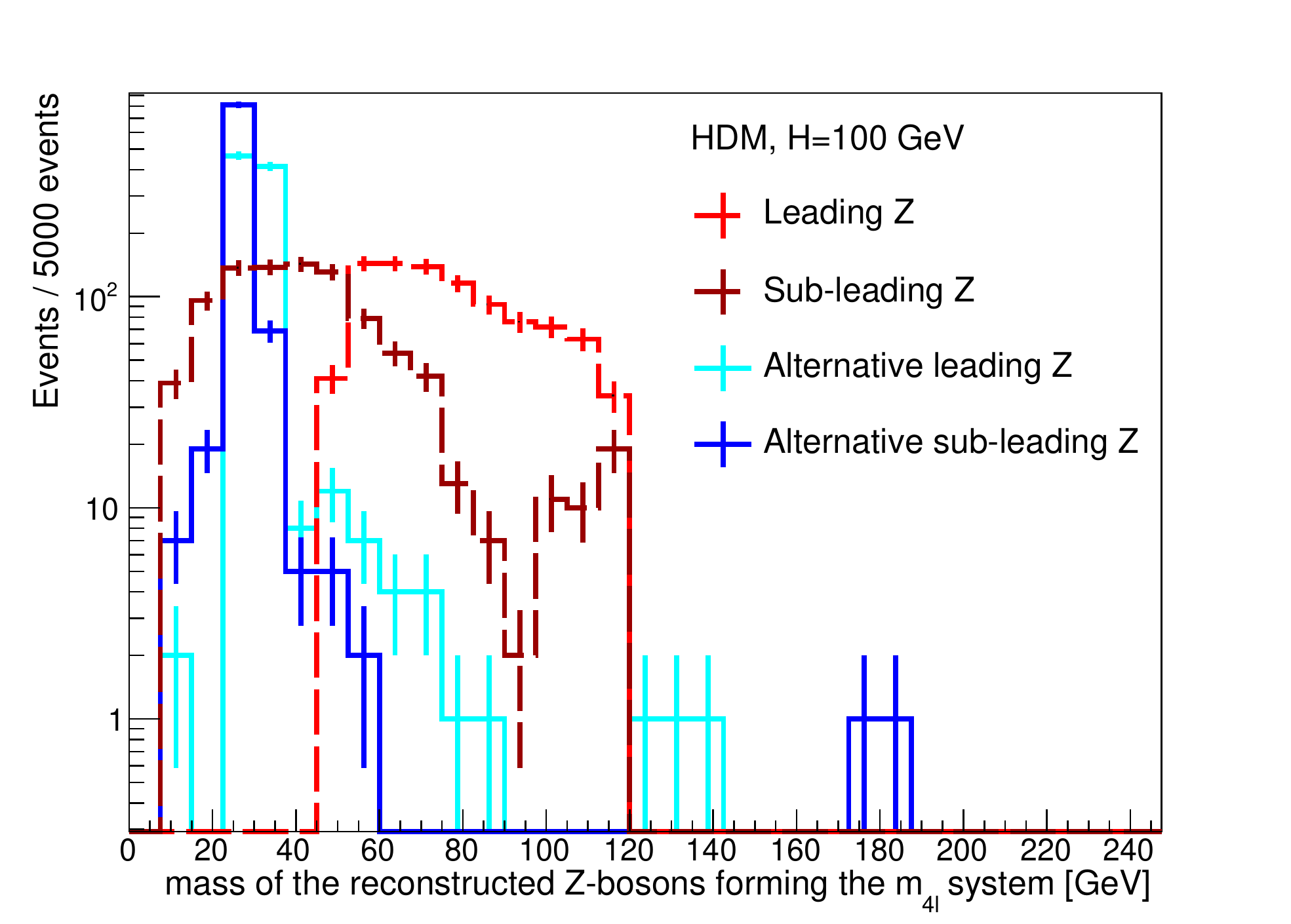}
\end{center}
\end{minipage}
\parbox{0.5\textwidth}{
\center
\begin{minipage}{0.4\textwidth}
\caption[Mass distributions]{
Di-lepton mass distributions for a sample of 5000 generated events passing the four-lepton selection cuts.  Upper left: for $M_{h^\pm} = 100\,\gev$, practically all events have all four muons coming from the $h$ or $\eta_A$ decay; upper right: for $M_{h^\pm} = 200\,\gev$, those events which have all four muons coming from the $h$ or $\eta_A$ decay are shown; lower left: for $M_{h^\pm} = 200\,\gev$, those events which have one muon coming from a $W$ decay are shown. The peaks at $30\,\gev$ are due to the model's $h,\eta_A \to \mu^+\mu^-$.}
\label{fig:sources}
\end{minipage}}
\end{figure}


As noted above, the muons for events passing the analysis cuts
come mainly from the decays of pairs of $h$ and/or $\eta_A$; leptons from $W$
decay make only a small contribution. Therefore, the other decays of $W$
bosons, to jets and $\tau \nu$ --- which were not generated in
this analysis --- are expected to add even further exclusion, since there
is no jet veto in the ATLAS analysis.

The mass distributions in Fig.~\ref{fig:sources} are made after the application of the mass and lepton $p_T$ selections, 
but before any selection on the $p_T$ of the $Z$ bosonss or the $\Delta R$ between leptons. They show the effect of the ATLAS muon selection that
defines the fiducial cross section. In a multi-muon event, the muons are paired according to how 
close the mass of the pair is to the $Z$ mass - the ``leading'' pair is the closest. The ``alternative'' pairings
are those not selected by the ATLAS algorithm, and exhibit the 30~GeV mass peak expected from $h$ and $\eta_A$ decays.
Whenever either of the alternative pairings has a mass differing from 30~GeV, this implies that at least one of the leptons used 
to form the pair does not come directly from a $h$ or $\eta_A$ decay. As demonstrated in Fig.~\ref{fig:sources}, only half the events 
at $M_{h^\pm} = 200\,\gev$ have one lepton not coming from either the $h$ or $\eta_A$ decay and being off-peak.
There are no off-peak events for $M_{h^\pm} = 100\,\gev$.

   The comparisons to data, which give the exclusions, are shown in
   Figs.~\ref{fig:m4l_100} and~\ref{fig:m4l_200} for $M_{h^\pm} = 100\,\gev$
   and $200\,\gev$.

\begin{figure}[htb]
\centering
      \includegraphics[width=0.45\textwidth]{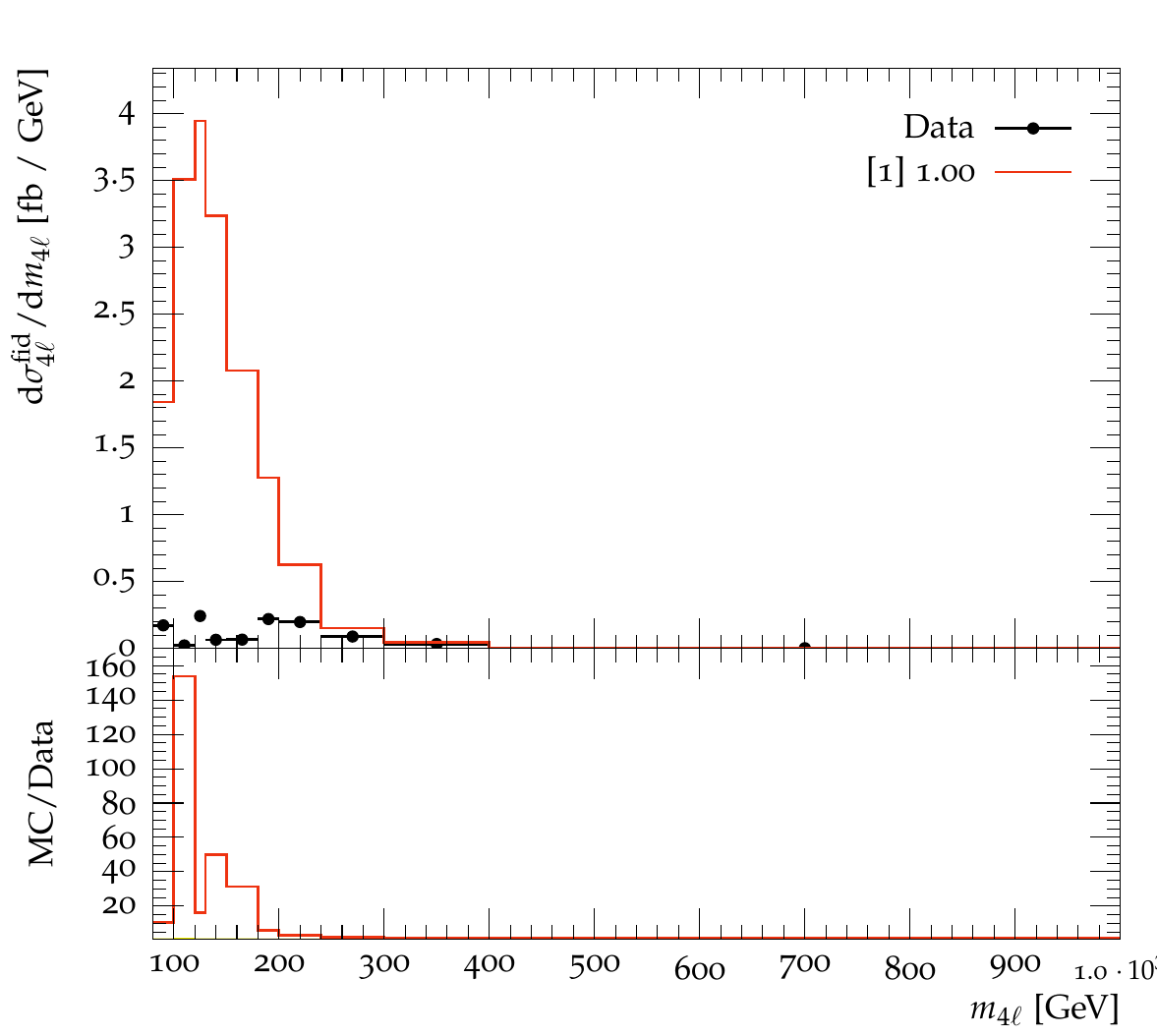}
      \includegraphics[width=0.45\textwidth]{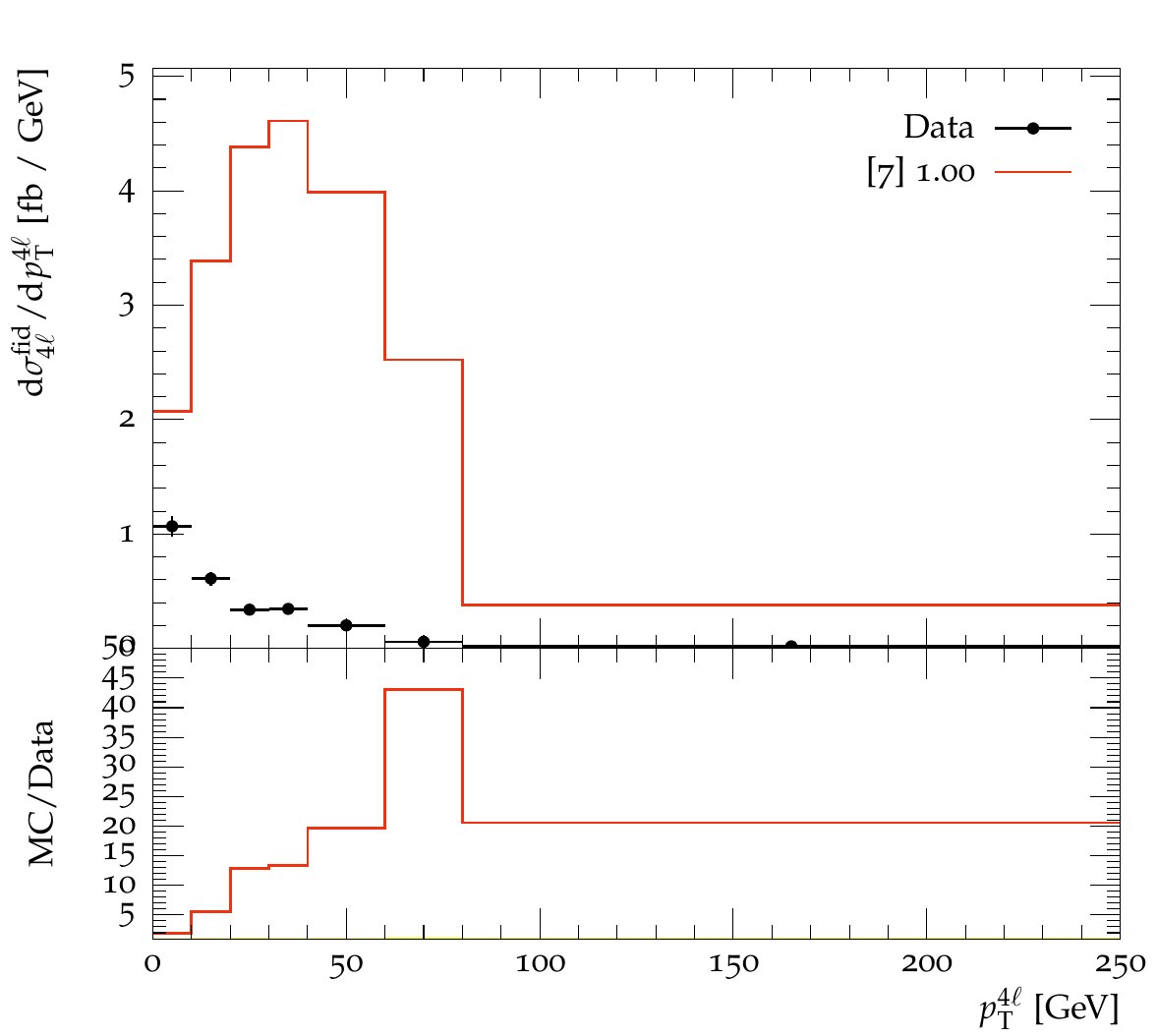}
      \caption{Projection of the contribution of our model, for
        $M_{h^\pm} = 100$~GeV, on to the ATLAS four-lepton differential
        cross-section mass measurement (left) and transverse momentum of the
        four-lepton system (right). Black points indicate the data, the red
        upper histogram is the data+BSM. The lower sections of the plots show
        the ratio of (data+BSM)/data. The uncertainty in the measurement is suppressed by the axis scale. 
        The numbers in the legend show the bin
        number of the most powerful bin, and the exclusion from that bin
        expressed as a probability.}
\label{fig:m4l_100}
\end{figure}

\begin{figure}[htb]
\centering
      \includegraphics[width=0.45\textwidth]{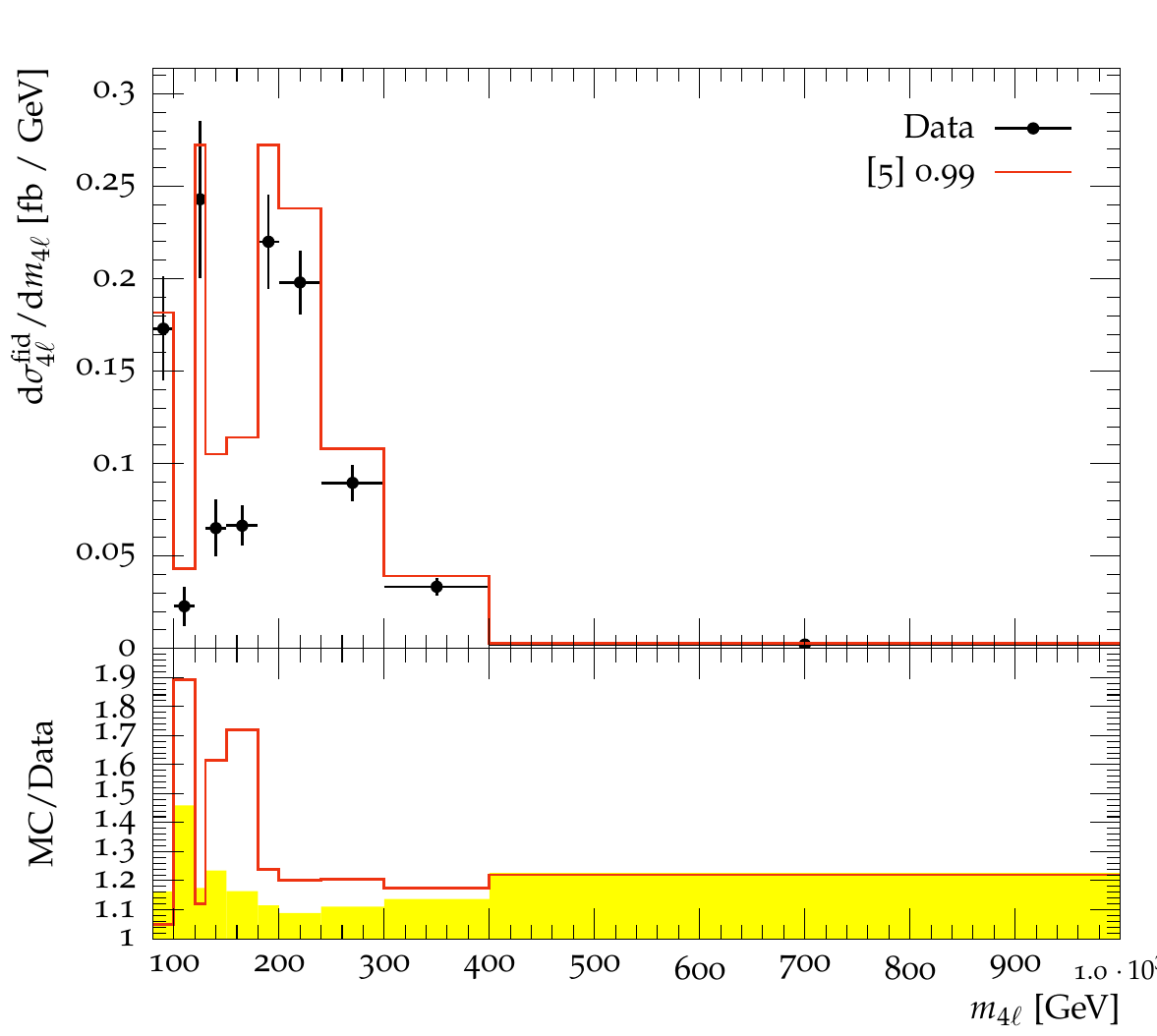}
      \includegraphics[width=0.45\textwidth]{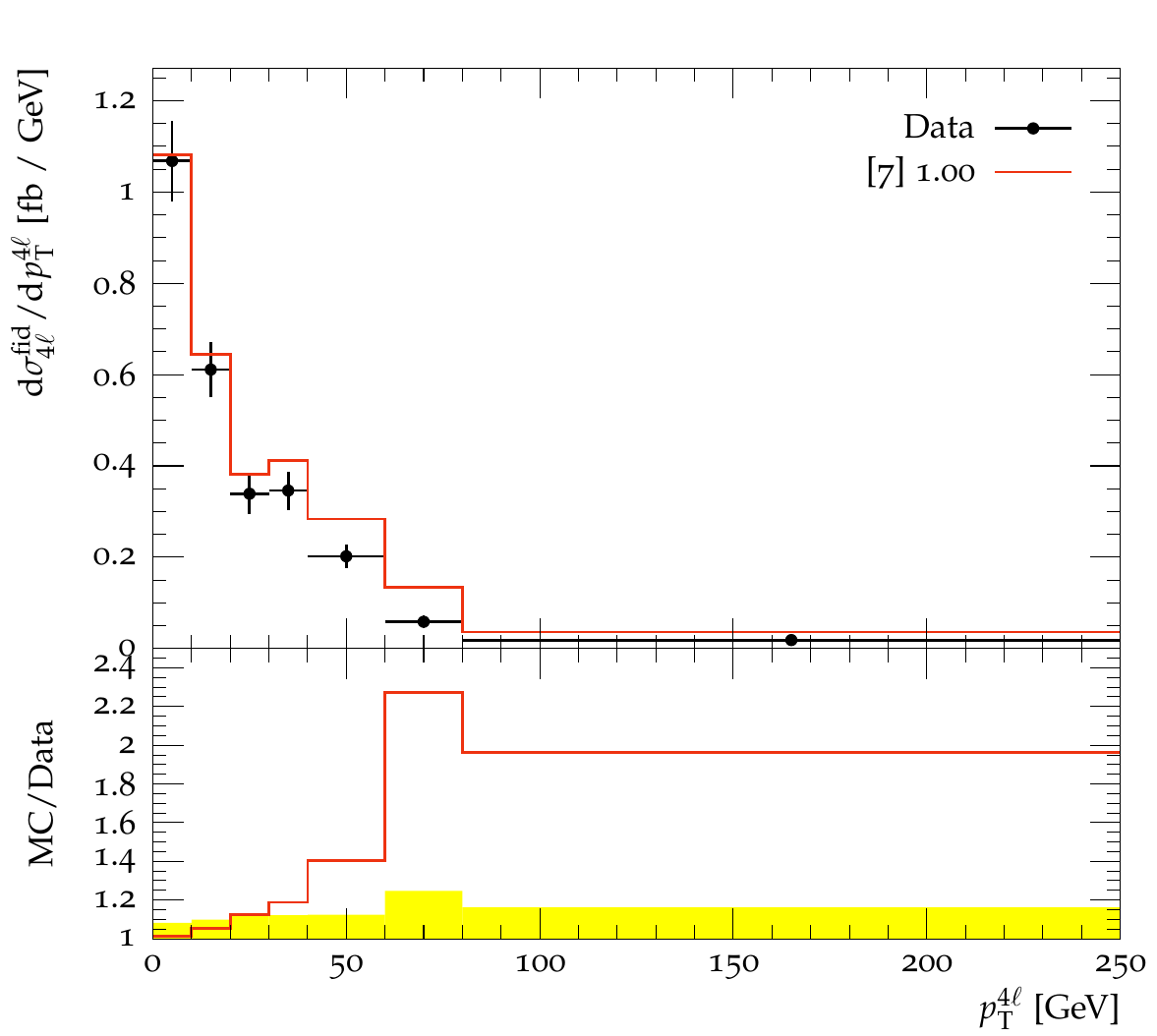}
      \caption{As in Fig.\protect\ref{fig:m4l_100}, but for
        $M_{h^\pm} = 200$~GeV. The yellow error band indicates the
        uncertainty on the measurement.}
\label{fig:m4l_200}
\end{figure}

\section{Conclusions}
\label{sec2hdm:conclusions}

The two Higgs-doublet model considered here leads to large contributions to already-measured differential 
cross sections. Even bearing in mind that the current incarnation of Contur does not fully take into account 
theoretical uncertainties on the SM predictions, the effects of this model would be so large that it can be considered ruled 
out. Other explanations for the ALEPH dimuon excess must be sought. More generally, this study demonstrates the power of
precision cross-section measurements at LHC in terms of constraining BSM physics, when such measurements are made with as
few theoretical assumptions as possible. It hopefully helps motivate a redoubling of efforts to produce such measurements, 
made in fiducial regions and based on final-state particles, as the LHC continues to survey physics above the electroweak 
symmetry-breaking scale. 

\section*{Acknowledgments}

We thank the organizers and conveners of the Les Houches workshop, ``Physics
at TeV Colliders'', for a stimulating meeting.  We also thank T.~Hryn'ova for
a valuable suggestion that helped initiate this study. K.~Lane thanks
A.~Heister for many informative discussions on the ALEPH dimuon data. He
also thanks the Laboratoire d'Annecy-le-Vieux de Physique Th\'eorique (LAPTh)
and the CERN Theory Group for their support and hospitality during the
initial stage of this research. This work has received funding from the
European Union's Horizon 2020 research and innovation programme as part of
the Marie Sklodowska-Curie Innovative Training Network MCnetITN3 (grant
agreement no. 722104) and ERC Grant Agreement no. 715871.



\AddToContent{J.~M.~Butterworth, D.~Grellscheid, K.~Lane, K.~Lohwasser and L.~Pritchett}
\renewcommand{\thesection}{\arabic{section}}

\graphicspath{{hkms_leshouches/}}



\chapter{Collider constraints on light pseudoscalars}

{\it U.~Haisch, J.~F.~Kamenik, A.~Malinauskas, M.~Spira} 



\begin{abstract} 
We investigate   the bounds on light pseudoscalars that arise from a variety of  collider searches. Special attention is thereby devoted to the mass regions $[3, 5] \, {\rm GeV}$ and $[9,11] \, {\rm GeV}$, in which a meaningful theoretical description has to include estimates of non-perturbative effects such as the mixing of the pseudoscalar with QCD bound states. A~compendium of formulas that allows to deal with the relevant corrections  is provided. It  should prove useful for the interpretation of future LHC  searches for light CP-odd spin-0 states.   
\end{abstract}

\section{INTRODUCTION}
\label{sechkms:introduction}

The most significant achievement of the LHC Run-I  physics programme has been the discovery of a new spin-0 resonance~($h$) with a mass of  $125 \, {\rm GeV}$ and with properties consistent with that of the standard model (SM) Higgs boson~\cite{Aad:2012tfa,Chatrchyan:2012xdj,Khachatryan:2016vau}. Besides precision measurements of processes involving a $h$, the LHC Higgs physics programme however also includes a wide spectrum of searches for additional Higgses (a~summary of LHC Run-I results can be found in~\cite{CMS:2016qbe} for instance). Such states  are predicted in many SM extensions such as supersymmetry  or models where the Higgs is realised as a pseudo Nambu-Goldstone boson~(PNGB) of a new approximate global symmetry. 

In fact, if the extended electroweak~(EW) symmetry breaking sector contains a PNGB, this state can be significantly lighter than the other spin-0 particles. A well-known example of a model that includes a light pseudoscalar~($a$) is provided by the next-to-minimal supersymmetric SM~(NMSSM) where this state can arise as a result of an approximate global $U(1)_R$ symmetry~\cite{Dobrescu:2000yn}. Since in this case the amount of symmetry breaking turns out to be proportional to soft breaking trilinear terms,  the  mass of the $a$ can  naturally be less than half of the SM Higgs mass, if the trilinear terms are  dialled to take values in the GeV range. Non-supersymmetric theories that can feature a light pseudoscalar are, to just name a few,   simplified models where  a complex singlet scalar is coupled to the Higgs potential of the SM or the  two-Higgs doublet model~(2HDM), Little Higgs models and hidden valley scenarios (see~\cite{Curtin:2013fra} and references therein for details).

Irrespectively of the precise ultraviolet~(UV) realisation, a light pseudoscalar  can lead to distinctive collider signatures. The most obvious consequence are exotic   decays of the SM Higgs, namely $h \to aa$ for $m_a < m_h/2$~\cite{Dobrescu:2000jt,Dermisek:2005ar} and $h \to aZ$ for $m_a < m_h - m_Z$~\cite{Curtin:2013fra,Christensen:2013dra}.  Another feature that  can have important phenomenological implications is that  in the presence of the heavy-quark transition $a \to b \bar b$ ($a \to c \bar c$) the pseudoscalar $a$ can mix with bottomonium (charmonium) bound states with matching quantum numbers~\cite{Drees:1989du,Domingo:2008rr,Domingo:2010am,Domingo:2011rn,Baumgart:2012pj,Haisch:2016hzu,Domingo:2016yih}.

LHC searches for $h \to aa$ have been performed in the $4 \mu$~\cite{Khachatryan:2015wka,CMS:2016tgd}, $4 \tau$~\cite{Khachatryan:2015nba,Khachatryan:2017mnf}, $2 \mu 2\tau$~\cite{Khachatryan:2017mnf} and $2 \mu 2b$~\cite{Khachatryan:2017mnf} final states.  The obtained results have been used to set upper bounds on the $h \to aa$ branching ratio in 2HDMs with an extra complex singlet~(2HDM+S) for pseudoscalar masses in the range of $[1, 62.5] \, {\rm GeV}$. The analyses~\cite{Khachatryan:2015wka,Khachatryan:2015nba,Khachatryan:2017mnf} however all exclude $m_a$ values in the regions $[3,5] \, {\rm GeV}$ and~$[9,11] \, {\rm GeV}$ for which $a \hspace{0.25mm}$--$\hspace{0.5mm} \eta_c$ and $a \hspace{0.25mm}$--$\hspace{0.5mm} \eta_b$ mixing effects as well as open flavour decays to $D$ and~$B_{(s)}$ meson pairs can be potentially important. 

The main goal of this work is to extend the latter results to the $c \bar c$ and $b \bar b$ threshold regions by including effects that  cannot be properly described in the partonic picture. In order to highlight the complementarity of  different search strategies for a light $a$, we also compare our improved limits  to other  bounds on the~2HDM+S parameter space that derive from the LHC searches for $h \to Z_d Z \to 4 \ell$~\cite{Aad:2015sva}, $h \to Z_d Z \to 2\mu 2 \ell$~\cite{Aaboud:2018fvk}, $pp \to a \to \mu^+ \mu^-$~\cite{ATLAS:2011cea,Chatrchyan:2012am}, $pp \to a b \bar b$ followed by $a \to \tau^+ \tau^-$~\cite{Khachatryan:2015baw} or  $a \to \mu^+ \mu^-$~\cite{Sirunyan:2017uvf}, $pp \to a \to \gamma \gamma$~\cite{CMS-PAS-HIG-17-013,Mariotti:2017vtv}, $pp \to a \to \tau^+ \tau^-$~\cite{CMS-PAS-HIG-16-037}, from the BaBar analyses of  radiative $\Upsilon$ decays~\cite{Lees:2011wb,Lees:2012iw,Lees:2012te} and from the LHCb measurements of  the production of $\Upsilon$ mesons~\cite{Haisch:2016hzu,Aaij:2015awa} as well as the inclusive dimuon cross section~\cite{Ilten:2016tkc,Aaij:2017rft}. 

This article is organised as follows. In Section~\ref{sechkms:generalities} we briefly recall  the structure of the 2HDM+S scenarios. Our recast of the results~\cite{Khachatryan:2015wka,Khachatryan:2015nba,Khachatryan:2017mnf} is presented in Section~\ref{sechkms:results}, where we also derive the constraints on the 2HDM+S parameter space that follow from the measurements and prosposals~\cite{Haisch:2016hzu,Aad:2015sva,Aaboud:2018fvk,ATLAS:2011cea,Chatrchyan:2012am,Khachatryan:2015baw,Sirunyan:2017uvf,CMS-PAS-HIG-17-013,Mariotti:2017vtv,CMS-PAS-HIG-16-037,Lees:2011wb,Lees:2012iw,Lees:2012te,Aaij:2015awa,Ilten:2016tkc,Aaij:2017rft}. We conclude in~Section~\ref{sechkms:conclusions}. The formulas necessary to calculate the partial decay widths of the  pseudoscalar~$a$ are collected  in~Appendix~\ref{app:widths}, while Appendix~\ref{app:mixing} contains a concise discussion of the mixing formalism and of open flavour decays that are relevant in the vicinity of the $b \bar b$ and $c \bar c$ thresholds. 

\section{THEORETICAL FRAMEWORK}
\label{sechkms:generalities}

In the following section we will interpret various searches for light pseudoscalars in the context of 2HDM+S scenarios. In this class of models a complex scalar singlet~$S$ is added to the 2HDM Higgs potential~(see~e.g.~\cite{Gunion:1989we,Branco:2011iw} for 2HDM reviews). The field $S$  couples only to the two Higgs doublets~$H_{1,2}$ but has no direct Yukawa couplings, acquiring all of its couplings to SM fermions through its mixing with the Higgs doublets. A light pseudoscalar $a$ can arise in such a setup from the admixture of the 2HDM pseudoscalar $A$ and the imaginary part of the complex singlet $S$. The corresponding mixing angle will be denoted by $\theta$, and defined such  that for $\theta \to 0$ the mass eigenstate $a$ becomes exactly singlet-like. 

To eliminate phenomenologically dangerous tree-level flavour-changing neutral currents~(FCNCs) the Yukawa interactions that involve the Higgs fields $H_{1,2}$  have to satisfy the  natural flavour conservation hypothesis~\cite{Glashow:1976nt,Paschos:1976ay}.  Depending on which fermions couple to which doublet, one can divide the resulting 2HDMs into four different types. In~all four cases the Yukawa couplings between the pseudoscalar $a$ and the SM fermions take the generic form 
\begin{equation} \label{eq:La}
{\cal L} \supset - \sum_f \frac{y_f}{\sqrt{2}} \, i \hspace{0.25mm} \xi_f^{\rm M} \, \bar f \gamma_5 f \hspace{0.25mm} a \,.
\end{equation}
Here $y_f = \sqrt{2} m_f/v$ denote the SM Yukawa couplings and $v \simeq 246 \, {\rm GeV}$ is the EW vacuum expectation value. The parameters  $\xi_f^{\rm M}$ encode the dependence on the 2HDM Yukawa sector and the factors relevant for the further discussion are given in Table~\ref{tab:xifM}. In this table the shorthand notations $s_\theta = \sin \theta$ and $t_\beta = \tan \beta$ have been used.  Similar abbreviations will also be used in what follows. 

In the presence of~(\ref{eq:La}) the CP-odd scalar $a$ can decay into fermions at tree level and into gluons, photons and EW gauge bosons at loop level. The expressions for the partial decay widths $\Gamma  \hspace{0.25mm}  ( a \to XX )$ that we employ in our study are given in Appendix~\ref{app:widths}. Since in this work we will assume that the  $a$ is lighter than the $W$, $Z$, $h$ and the other 2HDM Higgs mass eigenstates $H$, $A$, $H^\pm$, decays of the $a$ into the latter states are kinematically forbidden. 

If the $a$ is sufficiently light, exotic decays of the SM Higgs into the two final states $aZ$ and $aa$ are however possible.  The partial decay width $\Gamma  \hspace{0.25mm}  ( h \to aZ  )$  is in  2HDM+S scenarios entirely fixed by the 2HDM parameters $\alpha, \beta$ and the mixing angle~$\theta$.  Explicitly, one has at tree level 
\begin{equation}
\Gamma  \hspace{0.25mm}  ( h \to aZ ) = \frac{g_{haZ}^2}{16 \pi} \hspace{0.25mm} \frac{m_h^3}{v^2} \hspace{0.25mm} \lambda^3 \hspace{-0.5mm} \left (m_h^2,m_a^2,m_Z^2 \right ) \,,
\end{equation}
with 
\begin{equation} \label{eq:ghaZ}
g_{haZ} = c_{\beta - \alpha}  \hspace{0.5mm} s_\theta \,,
\end{equation}
and
\begin{equation} \label{eq:lambdaxyz}
\lambda \left (x, y, z \right ) = \sqrt{1  - \frac{2 \left (y + z \right )}{x}  + \frac{(y - z)^2}{x^2} } \,.
\end{equation}
Notice that in the exact alignment/decoupling limit, i.e.~$\alpha =  \beta - \pi/2$, in which the lighter CP-even spin-0 state~$h$  of the 2HDM becomes fully SM-like, the coupling $g_{haZ}$  and thus $\Gamma  \hspace{0.25mm}  ( h \to aZ )$ is precisely zero. However, given that the total decay width of the SM Higgs is only about $4 \, {\rm MeV}$, the process $h \to aZ$ can be  important even if deviations from the alignment/decoupling limit are relatively small.

\begingroup 
\renewcommand{\arraystretch}{1.25}
\setlength\tabcolsep{4pt}
\begin{table}[t!]
\centering
\begin{tabular}{c|c|c|c|c}
type & I & II & III & IV \\
\hline 
up-type quarks& $\phantom{-}s_\theta/t_\beta$ & $\phantom{-} s_\theta/t_\beta$ & $\phantom{-} s_\theta/t_\beta$ &$\phantom{-} s_\theta/t_\beta$\\
down-type quarks & $-s_\theta/t_\beta$ & $\phantom{-} s_\theta \hspace{0.25mm} t_\beta$ & $-s_\theta/t_\beta$ & $\phantom{-} s_\theta \hspace{0.25mm} t_\beta$\\
charged leptons & $-s_\theta/t_\beta$ & $\phantom{-} s_\theta \hspace{0.25mm} t_\beta$ & $\phantom{-} s_\theta \hspace{0.25mm} t_\beta$ & $-s_\theta/t_\beta$\\ 
\end{tabular}
\caption{\label{tab:xifM} Ratios $\xi_f^{\rm M}$ of the Yukawa couplings of the pseudoscalar~$a$ relative to those of the SM Higgs in the four types of 2HDM+S models without  tree-level FCNCs.}
\end{table}
\endgroup

Unlike $g_{haZ}$, the triple Higgs coupling $g_{haa}$ depends  not only on the physical Higgs masses and mixing angles but also on some of the trilinear couplings that appear in the full scalar potential. This feature makes the partial decay width $\Gamma  \hspace{0.25mm}  ( h \to aa )$ model dependent, and in consequence the two exotic branching ratios ${\rm BR} \hspace{0.25mm} ( h \to aZ  )$  and ${\rm BR}  \hspace{0.25mm}  ( h \to aa  )$ can  be adjusted freely by  an appropriate choice of parameters. Following this philosophy  we will  treat  ${\rm BR} \hspace{0.25mm}  ( h \to aZ  )$  and ${\rm BR}  \hspace{0.25mm}  ( h \to aa )$ as free parameters in the remainder of this article. 

\section{NUMERICAL RESULTS}
\label{sechkms:results}

We begin our numerical analysis by interpreting the recent CMS results~\cite{Khachatryan:2015wka,Khachatryan:2015nba,Khachatryan:2017mnf} for the exotic SM Higgs decay $h \to aa$ in the 2HDM+S context. The final states that we consider are $4 \mu$~\cite{Khachatryan:2015wka}, $4 \tau$~\cite{Khachatryan:2015nba,Khachatryan:2017mnf}, $2\mu 2 \tau$~\cite{Khachatryan:2017mnf} and $2 \mu 2b$~\cite{Khachatryan:2017mnf}. These searches probe $m_a$ values in the range~$[0.25, 3.55] \, {\rm GeV}$, $[4, 8] \, {\rm GeV}$, $[5, 15] \, {\rm GeV}$, $[15, 62.5] \, {\rm GeV}$ and $[25, 62.5] \, {\rm GeV}$, respectively. To facilitate a comparison between the results obtained by  the CMS collaboration and by us, we consider like~\cite{Khachatryan:2017mnf} the following four~2HDM+S benchmark scenarios: the type~I model with $t_\beta = 1$,  the type~II model with $t_\beta = 2$, the  type~III model with $t_\beta = 5$ and the type~IV model with $t_\beta = 0.5$. The  fermionic coupling factors~$\xi_f^{\rm M}$  corresponding to each 2HDM+S type are reported in Table~\ref{tab:xifM}. It is important to realise that the $s_\theta$-dependence of~$\xi_f^{\rm M}$ cancels in ${\rm BR}  \hspace{0.25mm}  ( a \to XX )$ and  it is thus possible to translate constraints on signal strengths such as $\sigma  \hspace{0.25mm}  ( pp \to h )  \hspace{0.5mm}  {\rm BR}  \hspace{0.25mm}  (h \to aa  )  \hspace{0.5mm}  {\rm BR}^2  \hspace{0.25mm}  (a \to \mu^+ \mu^- )$ into $s_\theta$-independent bounds on $\mu_h  \hspace{0.5mm} {\rm BR}  \hspace{0.25mm} (h \to aa )$. Here we have defined $\mu_h = \sigma  \hspace{0.25mm}  ( pp \to h  )/\sigma  \hspace{0.25mm}  ( pp \to h )_{\rm SM}$. 

The results of our recast are shown in the  panels of Fig.~\ref{fig:1} and should be compared to the  exclusion plots displayed in Fig.~8 of~\cite{Khachatryan:2017mnf}. The branching ratios ${\rm BR}  \hspace{0.25mm}  ( a \to XX )$ used to interpret the results in the four particular 2HDM+S scenarios are calculated using the formulas given in Appendix~\ref{app:widths} and include the mixing and threshold effects described in Appendix~\ref{app:mixing}. Notice that the inclusion of~$a \hspace{0.25mm}$--$\hspace{0.5mm} \eta_c$ and $a \hspace{0.25mm}$--$\hspace{0.5mm} \eta_b$ mixing is crucial to obtain meaningful predictions in the $m_a$ regions $[3, 5]~{\rm GeV}$ and $[9, 11] \, {\rm GeV}$, which are left unexplored in the CMS analysis~\cite{Khachatryan:2017mnf}. 

\begin{figure}[!t]
\begin{center}
\includegraphics[width=0.95\textwidth]{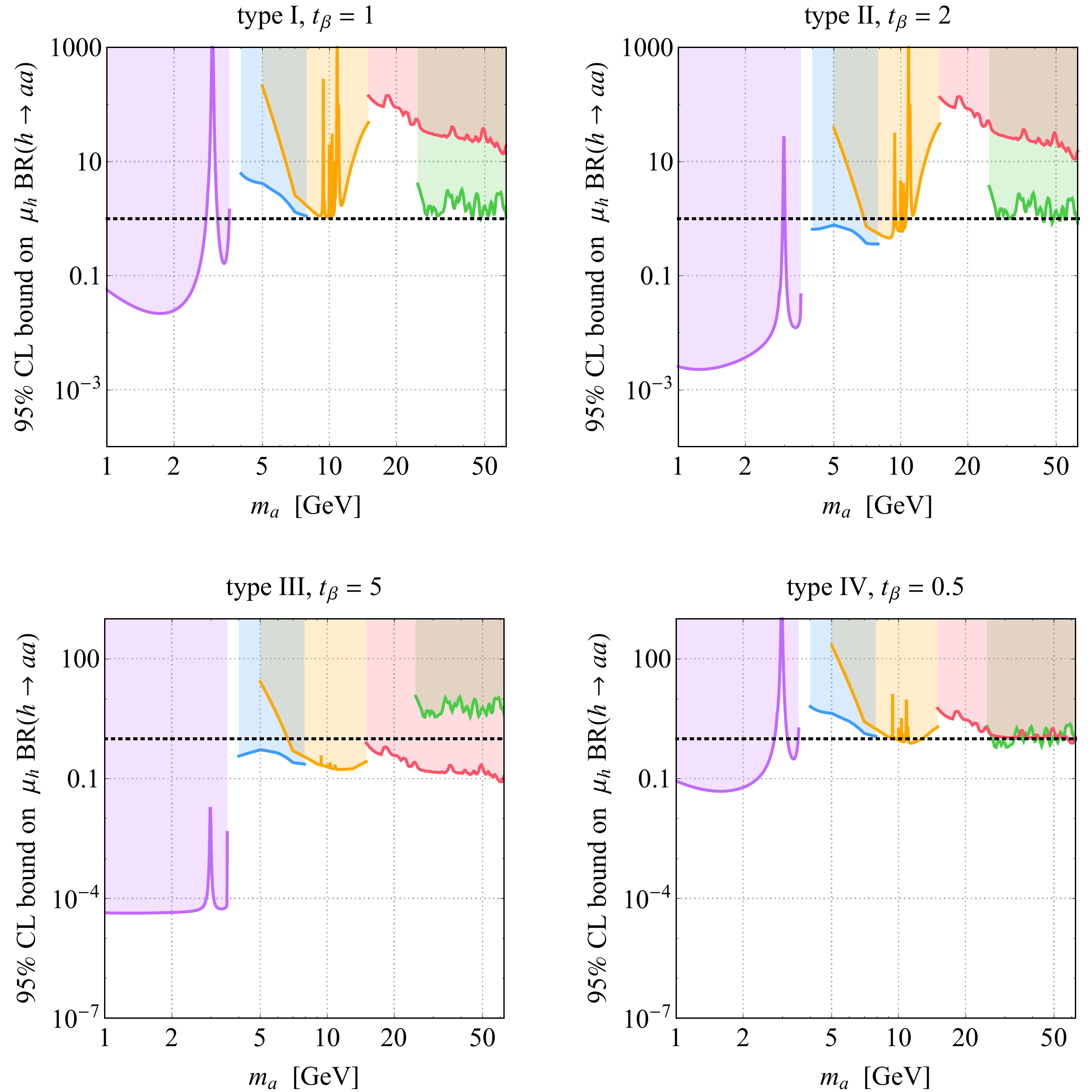}  
\vspace{2mm}
\caption{\label{fig:1} Limits on $\mu_h  \hspace{0.5mm}  {\rm BR}  \hspace{0.25mm} ( h \to aa )$ in the 2HDM+S of type~I with $t_\beta=1$~(top left),  type~II with $t_\beta = 2$~(top right),  type~III with $t_\beta = 5$~(bottom left) and type~IV with $t_\beta = 0.5$ (bottom right). The purple, blue, orange, red and green exclusions correspond to the search for $h \to aa \to 4 \mu$~\cite{Khachatryan:2015wka}, $h \to aa \to 4 \tau$~\cite{Khachatryan:2015nba}, $h \to aa \to 4 \tau$~\cite{Khachatryan:2017mnf}, $h \to aa \to 2\mu 2 \tau$~\cite{Khachatryan:2017mnf} and $h \to aa \to 2 \mu 2b$~\cite{Khachatryan:2017mnf}, respectively. The dashed black lines indicate $\mu_h  \hspace{0.5mm}  {\rm BR}  \hspace{0.25mm} ( h \to aa ) = 1$ and all coloured regions are excluded at 95\%~CL.
}
\end{center}
\end{figure}

While overall we  observe good agreement between the 95\% confidence level~(CL) exclusions set by CMS and by us, some differences in the derived limits are evident. Firstly, our analysis covers the mass region close to the $c \bar c$ ($b \bar b$) threshold, where our limits display a resonance-like behaviour as a result of the mixing of the $a$ with the three~$\eta_c$ (six $\eta_b$) states included in our study. Second, in the $m_a$ range of~$[1, 3] \, {\rm GeV}$ our bounds on $\mu_h  \hspace{0.5mm} {\rm BR}  \hspace{0.25mm}  ( h \to aa )$ tend to be somewhat weaker than those derived in~\cite{Khachatryan:2017mnf}.  The observed difference is  again a consequence of the mixing of the $a$ with QCD bound states. In fact, in the very low mass range the total decay width of the unmixed $a$ is below $10^{-3} \, {\rm MeV}$ in the considered~2HDM+S scenarios, while that of the lightest $\eta_c$ state amounts to around $30 \, {\rm MeV}$~\cite{Patrignani:2016xqp}. Hence even a small $\eta_c$-admixture in the mass eigenstate $a$ can lead to an enhanced total decay width $\Gamma_a$ which in turn results in a suppression of ${\rm BR} \hspace{0.25mm} ( a \to \mu^+ \mu^-)$  and a weakening of the bound on~$\mu_h  \hspace{0.5mm}  {\rm BR}  \hspace{0.25mm} ( h \to aa )$.  

\begin{figure}[!t]
\begin{center}
\includegraphics[width=0.95\textwidth]{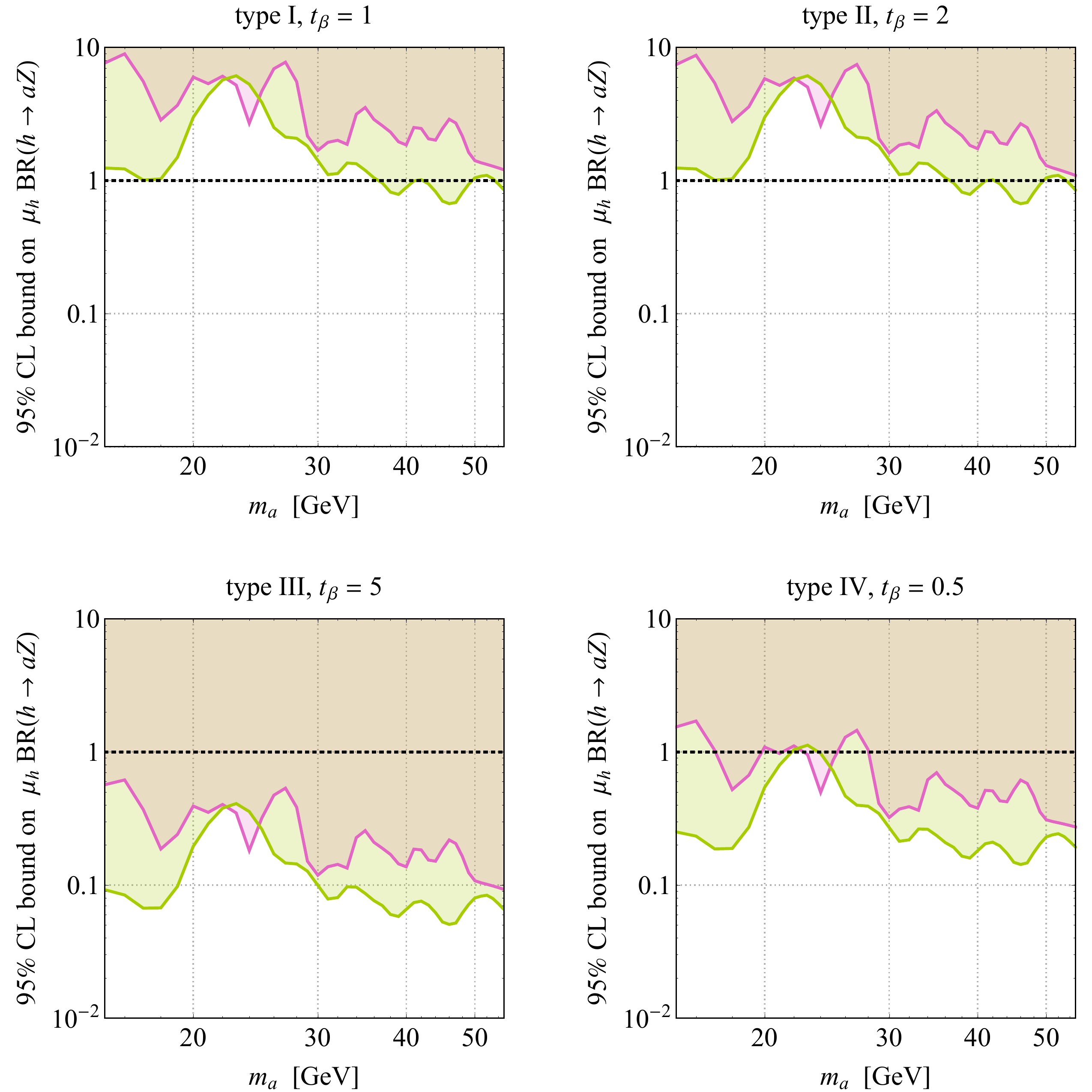}  
\vspace{2mm}
\caption{\label{fig:2} Limits on $\mu_h  \hspace{0.5mm} {\rm BR}  \hspace{0.25mm} ( h \to aZ )$ in the 2HDM+S of type~I with $t_\beta=1$~(top left),  type~II with $t_\beta = 2$~(top right),  type~III with $t_\beta = 5$~(bottom left) and type~IV with $t_\beta = 0.5$ (bottom right). The red and green bounds correspond to the ATLAS search for $pp \to h \to Z_d Z \to 4 \ell$~\cite{Aad:2015sva}  and  $pp \to h \to Z_d Z  \to 2 \mu 2\ell$~\cite{Aaboud:2018fvk}, respectively.   The dashed black lines indicate $\mu_h  \hspace{0.5mm}  {\rm BR}  \hspace{0.25mm} ( h \to aZ ) = 1$ and all coloured regions are excluded at~95\%~CL.}
\end{center}
\end{figure}

A light pseudoscalar $a$ can also be searched for via the decay  $h \to aZ$. The only LHC analyses that presently can be used to set bounds on ${\rm BR} \hspace{0.25mm} ( h \to a Z )$ are the ATLAS searches  for  new dark bosons $Z_d$ produced in $h  \to Z_d Z$~\cite{Aad:2015sva,Aaboud:2018fvk}.  Notice that while the $Z_d$ decays democratically into electrons and muons in the case of the~$a$ one has $\Gamma \hspace{0.25mm} (a \to e^+ e^-)/\Gamma \hspace{0.25mm} (a \to \mu^+ \mu^-) = m_e^2/m_\mu^2 \simeq 2 \cdot 10^{-5}$. As a result $4e$ and $2e2\mu$ events originating from $h \to a Z \to 4e$ and $h \to a Z \to 2e2\mu$ give essentially no contribution to the signal strength in $pp \to h \to aZ \to 4 \ell$. The $8 \, {\rm TeV}$ ATLAS study~\cite{Aad:2015sva} however only provides  exclusion bounds on ${\rm BR} \hspace{0.25mm} (h \to Z_d Z \to 4 \ell)$ from a combination of final states. To correct for this mismatch we have calculated  $r_{{\cal A}\varepsilon} = \sum_{X={4\mu,2\mu2e}} {\cal A}\varepsilon_{X}/\sum_{X={4\mu,4e,2e2\mu,2\mu2e}} {\cal A}\varepsilon_{X}$,  where ${\cal A}\varepsilon_{X}$ denotes the product of acceptance and reconstruction efficiency in the final state $X$ --- the values   for ${\cal A}\varepsilon_{X}$ can be found in the auxiliary material of~\cite{Aad:2015sva}. We find that $r_{{\cal A}\varepsilon}$ has only a mild dependence on $m_a$ and amounts to around $60\%$. The  actual limits are then obtained by equating $r_{{\cal A}\varepsilon}   \,  {\rm BR} \hspace{0.25mm}  (h \to aZ)   \,  {\rm BR} \hspace{0.25mm} (a \to \mu^+ \mu^-) \, {\rm BR} \hspace{0.25mm} (Z \to \ell^+ \ell^-)  = {\rm BR} \hspace{0.25mm} (h \to Z_d Z \to 4 \ell)$ and solving for ${\rm BR} \hspace{0.25mm}  (h \to aZ)$. To~improve upon this naive recast one would need   individual bounds for the different  combinations of final-state lepton flavours. In fact,  the very recent $13 \, {\rm TeV}$ ATLAS analysis~\cite{Aaboud:2018fvk} provides ${\cal A}\varepsilon_{2\mu 2 \ell}$ as well as limits on the relevant fiducial  cross section. Our recast of the latter results thus only has to rely on the assumption that  the product ${\cal A}\varepsilon_{2\mu 2 \ell}$ is roughly the same for the  $Z_d$ model and the 2HDM+S scenario, which we indeed believe to be the case. 

\begin{figure}[!t]
\begin{center}
\includegraphics[width=0.95\textwidth]{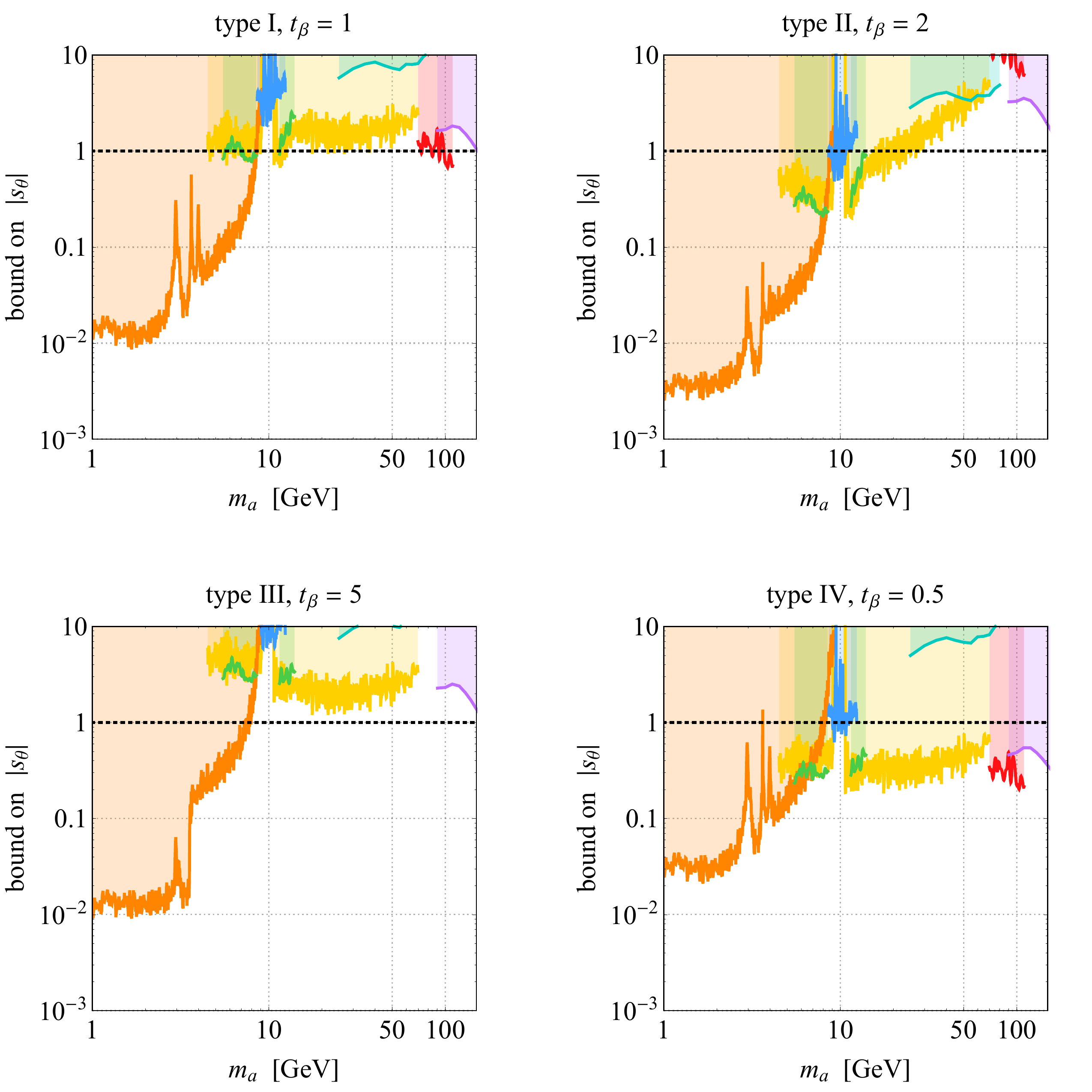}
\vspace{2mm}
\caption{\label{fig:3}   Limits on $|s_\theta|$ in the 2HDM+S of type~I with $t_\beta = 1$~(top left),  type~II with $t_\beta = 2$~(top right),  type~III with $t_\beta = 5$~(bottom left) and type~IV with $t_\beta = 0.5$ (bottom right). The green, turquoise, red, purple, orange, blue and yellow exclusions correspond to the searches for $a \to \mu^+ \mu^-$~\cite{Chatrchyan:2012am}, $pp \to a b \bar b \to\tau^+ \tau^-  b \bar b$~\cite{Khachatryan:2015baw}, $pp \to a \to \gamma \gamma$~\cite{CMS-PAS-HIG-17-013}, $pp \to a \to \tau^+ \tau^-$~\cite{CMS-PAS-HIG-16-037} and $\Upsilon (1S) \to a \gamma \to \mu^+ \mu^- \gamma$~\cite{Lees:2012iw},  the measurements of $\Upsilon$~production~\cite{Haisch:2016hzu,Aaij:2015awa} and the inclusive dimuon cross section~\cite{Aaij:2017rft}, respectively. The dashed black lines indicate $|s_\theta| = 1$ and all coloured regions are excluded at 95\%~CL apart from the orange and yellow contours which only hold at 90\%~CL. 
}
\end{center}
\end{figure}

The exclusion limits  on $\mu_h  \hspace{0.5mm} {\rm BR}  \hspace{0.25mm} ( h \to aZ )$ corresponding to the four  2HDM+S benchmark scenarios discussed earlier are presented in Fig.~\ref{fig:2}. From the panels it is evident that, apart from pseudoscalar masses around $25 \, {\rm GeV}$ where the data~\cite{Aaboud:2018fvk} has a local deficit, the constraints that derive from the $13 \, {\rm TeV}$ analysis~\cite{Aaboud:2018fvk}  are significantly stronger than those that one obtains from the $8 \, {\rm TeV}$ data~\cite{Aad:2015sva}. One also observes that the constraints in the first and second benchmark are weak as they just start to probe the region $\mu_h  \hspace{0.5mm} {\rm BR} \hspace{0.25mm} (h \to aZ) \lesssim 1$, whereas in the third and fourth 2HDM+S scenario already values of $\mu_h  \hspace{0.5mm} {\rm BR} \hspace{0.25mm} (h \to aZ) \lesssim 0.1$ can be probed with the available LHC data sets. Since the  asymmetry between electron and muon  final states from $h \to aZ$ decays is a striking signature of a light pseudoscalar, we strongly encourage  our experimental colleagues to provide as in~\cite{Aaboud:2018fvk} separate bounds for the  $2e2\ell$ and $2\mu2\ell$ final states in future searches for   signatures of the type  $h \to Z_d Z \to 4 \ell$. 

Constraints on the parameter space of the four different types of 2HDM+S scenarios can finally  be derived from the LHC searches for $pp \to a \to \mu^+ \mu^-$~\cite{ATLAS:2011cea,Chatrchyan:2012am}, $pp \to a b \bar b \to\tau^+ \tau^-  b \bar b$~\cite{Khachatryan:2015baw} or $pp \to a b \bar b \to\mu^+ \mu^-  b \bar b$~\cite{Sirunyan:2017uvf}, $pp \to a \to \gamma \gamma$~\cite{CMS-PAS-HIG-17-013}, $pp\to a \to \tau^+ \tau^-$~\cite{CMS-PAS-HIG-16-037}, from the studies of  $\Upsilon  \to a \gamma$ decays performed at BaBar~\cite{Lees:2011wb,Lees:2012iw,Lees:2012te} and from the LHCb measurements of $\Upsilon$  production~\cite{Haisch:2016hzu,Aaij:2015awa} as well as of the inclusive dimuon cross section~\cite{Ilten:2016tkc,Aaij:2017rft}. Since these search strategies all rely on the production of a pseudoscalar $a$ the resulting constraints all scale as $s_\theta^2$. For a given type of 2HDM+S model and a fixed value of $t_\beta$, the measurements~\cite{ATLAS:2011cea,Chatrchyan:2012am,Khachatryan:2015baw,Sirunyan:2017uvf,CMS-PAS-HIG-17-013,CMS-PAS-HIG-16-037,Lees:2011wb,Lees:2012iw,Lees:2012te,Aaij:2017rft} can therefore be used to set limits on $|s_\theta|$ as a function of~the pseudoscalar mass $m_a$. 

For concreteness we study the same four 2HDM+S  scenarios that we have already considered before. The  most stringent limits on $|s_\theta|$ that can be derived at present  are displayed in Fig.~\ref{fig:3}. In order to~recast the results of the CMS searches for $a \to \mu^+ \mu^-$~\cite{Chatrchyan:2012am},  $pp \to a b \bar b \to\tau^+ \tau^-  b \bar b$~\cite{Khachatryan:2015baw}, $pp \to a \to \gamma \gamma$~\cite{CMS-PAS-HIG-17-013}, $pp \to a \to \tau^+ \tau^-$~\cite{CMS-PAS-HIG-16-037}, the LHCb measurements of $\Upsilon$  production~\cite{Haisch:2016hzu,Aaij:2015awa} and the inclusive dimuon cross section~\cite{Aaij:2017rft}, one needs to know the production cross sections of a light $a$ in gluon-fusion and in association with $b \bar b$ pairs.  Our predictions for~$gg \to a$ production are obtained  at next-to-next-to-leading order in QCD using {\tt HIGLU}~\cite{Spira:1995mt}, while the $pp \to a b \bar b$ cross sections are calculated at next-to-leading order~(NLO) in QCD in the four-flavour scheme with {\tt MadGraph5\_aMCNLO}~\cite{Alwall:2014hca} employing an {\tt UFO} implementation~\cite{Degrande:2011ua} of the 2HDM model discussed in the publication~\cite{Bauer:2017ota}. 

Our recast of the results of the LHCb search for dark photons $A^\prime$~\cite{Aaij:2017rft} proceeds as follows. We calculate the inclusive $pp \to A^\prime$ production cross section at NLO in QCD using {\tt MadGraph5\_aMCNLO}~\cite{Alwall:2014hca}, while we extract ${\rm BR} \hspace{0.25mm} (A^\prime \to \mu^+ \mu^-)$ from the well-measured cross section ratio~$R = \sigma \hspace{0.25mm} (e^+ e^- \to {\rm hadrons})/\sigma \hspace{0.25mm} (e^+ e^- \to \mu^+ \mu^-)$~\cite{Patrignani:2016xqp}. Following~\cite{Ilten:2016tkc,Aaij:2017rft}, model-dependent $A^\prime \hspace{0.25mm}$--$\hspace{0.5mm} Z$ mixing effects are included in our calculation employing the formulas given in~\cite{Cline:2014dwa}. We have also  taken into account detector acceptance differences between  $pp \to A^\prime \to \mu^+ \mu^-$  and $pp \to a  \to \mu^+ \mu^-$ by computing the ratio $r_{\cal A} = {\cal A}_a/{\cal A}_{A^\prime}$ of signal acceptances.  We find that $r_{\cal A}$ amounts to around 2.0, 1.3, 1.0 at $m_a = 5 \, {\rm GeV}, 15 \, {\rm GeV}, 70 \, {\rm GeV}$ and scales approximately linear between the quoted $m_a$ values. Concerning the detection efficiencies ${\varepsilon}_{A^\prime}$ and ${\varepsilon}_{a}$ we assume that they are identical for $A^\prime \to \mu^+ \mu^-$ and $a \to \mu^+ \mu^-$, which should be a good approximation when the dimuon signal is prompt~\cite{Aaij:2017rft}. We finally add that in our recast of the LHCb dark photon results, we only consider the mass region $m_a > 4.5 \, {\rm GeV}$  to avoid $a \hspace{0.25mm}$--$\hspace{0.5mm} \eta_c$ mixing contributions to the $pp \to a$ cross section associated to $pp \to \eta_c$ production. The mass region $m_a \in [9.1, 10.6] \, {\rm GeV}$ is also not covered by our recast, because in~\cite{Aaij:2017rft} the LHCb collaboration does not present bounds on the kinetic mixing of the $A^\prime$ close to the $b \bar b$  threshold.

The main conclusion  that can be drawn from the results presented in  Fig.~\ref{fig:3} is that only in the 2HDM+S scenario of  type IV with $t_\beta = 0.5$ it is possible to set physical meaningful bounds on the sine of the mixing angle $\theta$,~i.e.~$|s_\theta| < 1$, over the entire range of studied pseudoscalar masses. One furthermore observes that solely the BaBar search for the radiative decay $\Upsilon (1S) \to a \gamma \to \mu^+ \mu^- \gamma$~\cite{Lees:2012iw}  allows to probe parameter regions with $|s_\theta| < 0.1$. This search is however kinematically limited to $m_a < m_{\Upsilon(1S)} \simeq 9.5 \, {\rm GeV}$.  Improvements in the existing LHC search strategies (and/or new approaches) are needed to reach the same sensitivity on~$|s_\theta|$ for pseudoscalar masses above around $10 \, {\rm GeV}$ in the examined 2HDM+S benchmark models. Measurements of the inclusive dimuon cross section~\cite{Ilten:2016tkc,Aaij:2017rft} seem to be quite promising in this context. 

\section{CONCLUSIONS}
\label{sechkms:conclusions}

Beyond the SM theories with an extended Higgs sector can naturally lead to pseudoscalar resonances with masses significantly below the EW scale if these states serve as PNGBs of  an approximate global  $U(1)$ symmetry. The $R$-symmetry limit in the NMSSM and the case of spontaneously broken $U(1)$ subgroups in Little Higgs models are just two working examples of this general idea. Searches for light CP-odd spin-0 states are thus theoretically well-motivated and in the case of a detection could help to illuminate the structure and dynamics of the underlying UV model. 

The existing  collider searches for pseudoscalars with masses  of approximately $[1, 100]~{\rm GeV}$ fall into two different classes.  Firstly, searches that look for the presence of a light~$a$ in the decay of a SM particle. Searches for $h \to aa$ and $h \to aZ$, but also the radiative decays $\Upsilon \to a \gamma$ belong to this category. In the case of the exotic Higgs decays the resulting signature that the ATLAS and CMS experiments have explored  are four-fermion final states containing at least two opposite-sign leptons~\cite{Khachatryan:2015wka,CMS:2016tgd,Khachatryan:2015nba,Khachatryan:2017mnf,Aad:2015sva,Aaboud:2018fvk}, while what concerns the radiative~$\Upsilon$~decays, BaBar has considered the hadronic, dimuon and ditau decays of pseudoscalars~\cite{Lees:2011wb,Lees:2012iw,Lees:2012te}.  The second type of searches instead relies on the direct production of the $a$ in $pp$ collisions and its subsequent decays to either charged lepton or photon pairs. Both the gluon-fusion channel~\cite{Haisch:2016hzu,ATLAS:2011cea,Chatrchyan:2012am,CMS-PAS-HIG-17-013,Mariotti:2017vtv,CMS-PAS-HIG-16-037,Aaij:2015awa} and $ab \bar b$ production~\cite{Khachatryan:2015baw,Sirunyan:2017uvf} have so far been exploited to look for light pseudoscalars at the LHC in this way. 

In this work, we have performed a global analysis of the present collider constraints on light pseudoscalar states. To facilitate  a comparison with the recent CMS study~\cite{Khachatryan:2017mnf}, we have considered the class of 2HDM+S models, treating the parameters $t_\beta$ and $s_\theta$ as well as the branching ratios ${\rm BR} \hspace{0.25mm} ( h \to aa )$ and ${\rm BR} \hspace{0.25mm} ( h \to aZ )$ as free parameters --- see Section~\ref{sechkms:generalities} for a concise introduction to the  2HDM+S setup. A complication that arises in our analysis is that in the mass regions $[3, 5] \, {\rm GeV}$ and $[9,11] \, {\rm GeV}$, non-perturbative effects such as the mixing of the pseudoscalar with QCD bound states have to be taken into account to allow for a meaningful interpretation of the experimental data. We have worked out the  theoretical formalism necessary to calculate the  most relevant short-distance and long-distance effects and provide a collection of the corresponding formulas in the two Appendices~\ref{app:widths} and~\ref{app:mixing}.  

Our numerical analysis consists of three parts. In the first part, we have derive 95\%~CL exclusion limits  on the signal strength $\mu_h  \hspace{0.5mm} {\rm BR}  \hspace{0.25mm} (h \to aa )$ that follow from the latest CMS searches for the  exotic $h \to aa$ decay~\cite{Khachatryan:2015wka,CMS:2016tgd,Khachatryan:2015nba,Khachatryan:2017mnf}, while in the second part we present the limits on $\mu_h  \hspace{0.5mm}  {\rm BR}  \hspace{0.25mm} (h \to aZ )$ that stem from the ATLAS searches for $h \to Z_d Z \to 4 \ell$~\cite{Aad:2015sva} and $h \to Z_d Z \to 2\mu 2 \ell$~\cite{Aaboud:2018fvk}. The exclusion bounds on~$|s_\theta|$ that arise from the searches~\cite{Haisch:2016hzu,Chatrchyan:2012am,Khachatryan:2015baw,CMS-PAS-HIG-17-013,CMS-PAS-HIG-16-037,Lees:2012iw,Aaij:2015awa,Aaij:2017rft} are finally derived in the third part of our numerical study. In all three cases, we have considered four specific 2HDM+S benchmark scenarios that differ in the choice of Yukawa sector and $t_\beta$.  We have found that the inclusion of $a \hspace{0.25mm}$--$\hspace{0.5mm} \eta_c$ $\big($$a \hspace{0.25mm}$--$\hspace{0.5mm} \eta_b$$\big)$ mixing effects as well as open flavour decays to~$D$~$\big ($$B_{(s)}$$\big )$ meson pairs has a visible impact on the obtained limits only in the mass region of approximately $[1, 4] \, {\rm GeV}$ $\big ($$[10, 15] \, {\rm GeV}$$\big )$, while   perturbative calculations are perfectly adequate  for~$m_a$~values away from the~$c\bar c$ and $b \bar b$ thresholds. 

The main conclusion that can be drawn from the results presented in Figs.~\ref{fig:1},~\ref{fig:2} and~\ref{fig:3} is that existing collider constraints on the parameter space of 2HDM+S models are in general not very strong. Exceptions are the  $[1, 3] \, {\rm GeV}$ region in which $\mu_h \hspace{0.25mm} {\rm BR} \hspace{0.25mm} ( h \to a a )$ is well-constrained by the CMS search for $h \to aa \to 4 \mu$~\cite{Khachatryan:2015wka} and the $[1, 9.5] \, {\rm GeV}$ range where the $\Upsilon(1S) \to a \gamma \to \mu^+ \mu^- \gamma$ search of BaBar~\cite{Lees:2012iw} provides stringent limits on $|s_\theta|$. Much to the opposite, the 2HDM+S parameter space turns out to be least constrained for $m_a $ values in the range of approximately $ [15, 70] \, {\rm GeV}$. The development of improved or new search techniques (such as for instance dedicated searches for $h \to a Z$~\cite{Aaboud:2018fvk}~and inclusive diphoton~\cite{Mariotti:2017vtv} or dimuon~\cite{Ilten:2016tkc,Aaij:2017rft} cross section measurements) that specifically focus  on the latter mass region therefore seems to be a worthwhile scientific goal. 

\section*{ACKNOWLEDGEMENTS}
We are grateful to Kai~Schmidt-Hoberg for providing details on the estimate of $\Gamma  \hspace{0.25mm}  ( a \to KK\pi  )$ as given in~\cite{Dolan:2014ska}. We furthermore thank Ulrich~Ellwanger, Filippo~Sala, Dominik~St{\"o}ckinger and Mike~Williams for their interest in our work, constructive feedback and their useful suggestions. UH appreciates the continued hospitality  and support of the CERN Theoretical Physics Department. JFK acknowledges the financial support from the Slovenian Research Agency (research core funding No.~P1-0035 and J1-8137). MS~would like to thank the organisers of Les~Houches for the great and fruitful atmosphere of the workshop.

\begin{appendix}

\section{Decay width formulas}
\label{app:widths}

In the calculation of the total decay  width $\Gamma_a$ of the unmixed pseudoscalar $a$, we employ the following expressions for the partial decay widths~(see the reviews~\cite{Spira:1997dg,Djouadi:2005gi,Djouadi:2005gj,Spira:2016ztx} for instance)
\begin{align} 
\Gamma\hspace{0.25mm} (a\to \ell^+ \ell^-) & = \frac{\big ( \xi_\ell^{\rm M}  \big )^2  \hspace{0.35mm}  m_\ell^2 \hspace{0.25mm}  m_a}{8\pi v^2} \hspace{0.5mm}   \beta_{\ell/a}  \,, \label{eq:A1} \\[1mm]
\Gamma\hspace{0.25mm} (a\to q\bar q) & = \frac{3 \hspace{0.25mm} \big (\xi_q^{\rm M} \big )^2  \hspace{0.25mm} \overline m_q^2 \hspace{0.25mm}  m_a}{8\pi v^2} \left  (1+\Delta_q + \frac{\xi_t^{\rm M} }{\xi_q^{\rm M}} \hspace{0.25mm} \Delta_t \right )\,, \label{eq:A2} \\[1mm]
\Gamma\hspace{0.25mm} (a\to Q\bar Q) & = \frac{3 \hspace{0.25mm} \big (\xi_Q^{\rm M} \big )^2   \hspace{0.25mm} m_Q^2 \hspace{0.25mm}  m_a}{8\pi v^2} \hspace{0.5mm}  \beta_{Q/a} \left  (1+\Delta_Q  \right )\,, \label{eq:A3} \\[1mm]
\Gamma\hspace{0.25mm} (a\to gg) & = \frac{\alpha_s^2 \hspace{0.25mm} m_a^3}{32\pi^3 v^2} \left| \sum_{q=t,b,c,s} \xi_q^{\rm M}  \hspace{0.5mm}  
\mathcal P (\tau_{q/a}) \right|^2  \hspace{0.25mm} K_g  \,, \label{eq:A4} \\[1mm]
\Gamma\hspace{0.25mm} (a\to \gamma\gamma) & = \frac{\alpha^2 \hspace{0.25mm} m_a^3}{64\pi^3 v^2} \left| \sum_{q=t,b,c,s} 3 \hspace{0.25mm} \xi_q^{\rm M}  Q_q^2 \hspace{0.5mm}  \big ( \mathcal P ( \tau_{q/a})+  \Delta_\gamma \big ) +  \xi_\tau^{\rm M}   \hspace{0.25mm} \mathcal P ( \tau_{\tau/a}) \right|^2 \,, \label{eq:A5} 
\end{align}
where $\overline{\rm MS}$ masses are indicated by a bar while masses without a bar are evaluated in the pole scheme.  We  have furthermore defined $\tau_{f/a} = 4  m_{f} ^2/m_a^2$ and $\beta_{f/a} = \sqrt{1 - \tau_{f/a}}$ and used the symbol~$Q_q$ to denote the electric charge of the quark in question. All $\overline{\rm MS}$ masses  as well as  the coupling constants $\alpha_s$ and $\alpha$ are renormalised at the scale $\mu_R = m_a$.  Table~\ref{tab:xifM} finally contains the coupling assignments~$\xi_f^{\rm M}$ that we consider in our work.  

The  QCD corrections to the partial decay width into light quarks~(\ref{eq:A2}) that are included in our numerical analysis read~\cite{Drees:1989du,Braaten:1980yq,Sakai:1980fa,Inami:1980qp,Gorishnii:1983cu,Drees:1990dq,Gorishnii:1990zu,Gorishnii:1991zr,Kataev:1993be,Surguladze:1994gc,Melnikov:1995yp,Chetyrkin:1996sr}
\begin{equation} \label{eq:A7}
\begin{split}
\Delta_q & = \frac{\alpha_s}{\pi}  \hspace{0.75mm}  5.67  + \left ( \frac{\alpha_s}{\pi}  \right  )^2 \hspace{0.25mm} \big ( 35.94 - 1.36 \hspace{0.25mm}  N_f \big )  +  \left ( \frac{\alpha_s}{\pi}  \right )^3  \Big (164.14 - 25.77 \hspace{0.25mm} N_f + 0.259  \hspace{0.25mm} N_f^2 \Big )   \\[1mm]
& \phantom{xx} + \left ( \frac{\alpha_s}{\pi}  \right )^4 \hspace{0.25mm} \Big ( 39.34 - 220.9  \hspace{0.25mm} N_f  + 9.685 \hspace{0.25mm} N_f^2  - 0.0205 \hspace{0.25mm} N_f^3 \Big )  \,,
\end{split}
\end{equation} 
and~\cite{Chetyrkin:1995pd,Larin:1995sq}
\begin{equation}  \label{eq:A8}
\Delta_t  =  \left ( \frac{\alpha_s}{\pi}   \right )^2 \, \left [ 3.83 + \ln \left ( \frac{m_t^2}{m_a^2} \right ) + \frac{1}{6} \ln^2  \left ( \frac{\overline m_q^2}{m_a^2} \right )  \right ] \,.
\end{equation}
The symbol $N_f$  introduced above denotes the number of light quark flavours that are active at the scale~$m_a$.  For pseudoscalar masses far above the threshold, i.e. $m_a \gg 2 m_q$, the results~(\ref{eq:A7}) and~(\ref{eq:A8}) represent at the moment the most accurate predictions for the QCD corrections to $\Gamma \hspace{0.25mm} ( a \to q \bar q)$. In our numerical analysis, we hence use them to calculate the partonic rate of $a \to s \bar s$.  

In  the case of the partial decay width  into heavy-quark pairs~(\ref{eq:A3}) the  QCD corrections  are given to first order in $\alpha_s$ by ~\cite{Drees:1989du,Braaten:1980yq,Sakai:1980fa,Inami:1980qp,Gorishnii:1983cu,Drees:1990dq} 
\begin{equation} \label{eq:A9}
\Delta_Q  =  \frac{\alpha_s}{\pi} \left ( \frac{4 {\cal Q} (\beta_{Q/a})}{3 \beta_{Q/a}}  - \frac{19+2\beta_{Q/a}^2+3 \beta_{Q/a}^4 }{12 \beta_{Q/a}} \ln x_{\beta_{Q/a}} + \frac{21  - 3 \beta_{Q/a}^2}{6} \right ) \,.
\end{equation}
Here we have introduced the abbreviation $x_{\beta_{Q/a}} = (1 - \beta_{Q/a})/(1 + \beta_{Q/a})$ and  the one-loop function entering~(\ref{eq:A9}) takes the form 
\begin{eqnarray}
{\cal Q} (\beta) = \big (1+\beta^2 \big )\left ( 4 \hspace{0.25mm} {\rm Li}_2 (x_\beta) + 2  \hspace{0.25mm}  {\rm Li}_2 (-x_\beta) + 4 \ln x_\beta \ln \frac{2}{1+\beta} + 2 \ln x_\beta \ln \beta \right ) - 3 \beta \ln \frac{4 \beta^{4/3}}{1-\beta^2} \,, \hspace{0mm}   \label{eq:A10}
\end{eqnarray}
with ${\rm Li}_2 (z)$ denoting the usual dilogarithm. In the threshold region,~i.e.~$m_a \simeq 2 m_Q$, mass effects are important and as a result  the QCD corrections~(\ref{eq:A9}) should be used to describe them.  Following the prescription implemented in {\tt HDECAY}~\cite{Djouadi:1997yw,Djouadi:2018xqq}, the transition between the region close to threshold to that far above threshold is  achieved by a smooth linear interpolation of the results~(\ref{eq:A2}) and~(\ref{eq:A3}).  Because this approach yields an optimised description of $\Gamma \hspace{0.25mm} ( a \to c \bar c )$ $\big(\Gamma \hspace{0.25mm} ( a \to b \bar b )\big)$ for pseudoscalar masses in the vicinity of $m_a \simeq 3.1 \, {\rm GeV}$ ($m_a \simeq 11.5 \, {\rm GeV}$) it is used in our work.  

The one-loop function appearing in~(\ref{eq:A4}) and~(\ref{eq:A5}) is given by 
\begin{equation} \label{eq:A11}
\mathcal P (\tau)  = \tau \arctan^2 \left( \frac{1}{\sqrt{\tau-1}} \right) \,,
\end{equation}
where for analytic continuation it is understood that  $\tau \to \tau - i 0$. 

The multiplicative factor $K_g$ entering (\ref{eq:A4})  takes the following form 
\begin{equation} \label{eq:Kg}
K_g = 1 + 2 \hspace{0.25mm} {\rm Re} \left ( \frac{\sum_{q=t,b,c,s}  \hspace{0.25mm} \xi_q^{\rm M} \hspace{0.5mm}  \Delta_g}{ \sum_{q=t,b,c,s}  \hspace{0.25mm}  \xi_q^{\rm M} \hspace{0.5mm} {\cal P}  (\tau_{q/a}) } \right ) + \frac{\alpha_s}{\pi} \left ( \frac{73}{4} - \frac{7}{6} \hspace{0.25mm} N_f  \right ) \,,
\end{equation}
where the second term encodes the virtual two-loop QCD corrections, while the third term corresponds to the finite part of the real QCD corrections in the heavy-quark limit~\cite{Spira:1997dg,Spira:1995rr}. We have verified that quark mass effects of the real corrections not included in (\ref{eq:Kg}) amount to no more than $5\%$.  The virtual corrections can be written as 
\begin{equation} \label{eq:A12}
\Delta_g  =  \frac{\alpha_s}{\pi}  \left(  {\cal G} (y_{q/a})+  2 \hspace{0.25mm} \tau_{q/a} \hspace{0.25mm}  {\cal P}^\prime (\tau_{q/a}) \, \ln \frac{\mu_q^2}{m_q^2}  \right)\,,
\end{equation}
where $y_{q/a}= -x_{q/a}$ with $\tau_{q/a} \to \tau_{q/a} + i 0$ for analytic continuation and the prime denotes a derivative with respect to $\tau_{q/a}$. To reproduce the position of the $a \to q \bar q$ threshold correctly, we set~$\mu_q = m_a/2$ in our study.  The two-loop function appearing in~(\ref{eq:A12}) reads~\cite{Spira:1995rr,Harlander:2005rq}
\begin{align}
{\cal G} (y) & = \frac{y}{ {\left(1-y\right)}^2}\biggl[ 48 \hspace{0.25mm} {\rm H}(1,0,-1,0;y)+ 4\ln(1-y)\ln^3 y- 24\hspace{0.25mm} \zeta_2 \hspace{0.25mm} {\rm Li}_2( y)- 24 \hspace{0.25mm} \zeta_2 \ln(1- y)\ln y  \nonumber \\[1mm]
& \hspace{2cm} - 72 \hspace{0.25mm}\zeta_3 \ln(1- y) - \frac{220}{3} \hspace{0.25mm} {\rm Li}_3( y) - \frac{128}{3} \hspace{0.25mm} {\rm Li}_3(- y) + 68 \hspace{0.25mm} {\rm Li}_2( y)\ln y \nonumber \\[1mm] 
& \hspace{2cm} + \frac{64}{3} \hspace{0.25mm} {\rm Li}_2(- y)\ln y + \frac{94}{3}\ln(1- y)\ln^2 y  - \frac{16}{3}\hspace{0.25mm}\zeta_2\ln y + \frac{124}{3}\hspace{0.25mm}\zeta_3 + 3\ln^2 y \biggr] \nonumber \\[1mm]
& \phantom{xx} - \frac{ 24  y \left(5+7 { y}^2\right) }{{\left(1- y\right)}^3 \left(1+ y\right)} \hspace{0.25mm} {\rm Li}_4( y)  -\frac{ 24 y \left(5+11 { y}^2\right)}{{\left(1- y\right)}^3 \left(1+ y\right)} \hspace{0.25mm} {\rm Li}_4(-y)  \\[1mm]
&  \phantom{xx} +\frac{ 8 y \left(23+41 { y}^2\right) } {3{\left(1- y\right)}^3 \left(1+ y\right)} \biggl[ {\rm Li}_3( y) +{\rm Li}_3(- y) \biggr] \ln y -\frac{ 4 y \left(5+23 { y}^2\right) }{3{\left(1- y\right)}^3 \left(1+ y\right)}\hspace{0.25mm} {\rm Li}_2( y)\ln^2 y  \nonumber \\[1mm] 
&  \phantom{xx} - \frac{ 32 y \left(1+{ y}^2\right) }{3{\left(1- y\right)}^3 \left(1+ y\right)} \hspace{0.25mm} {\rm Li}_2(- y)\ln^2 y  +\frac{  y \left(5-13 { y}^2\right) }{36 {\left(1- y\right)}^3 \left(1+ y\right)}\ln^4 y +\frac{ 2 y \left(1-17 { y}^2\right) }{3{\left(1- y\right)}^3 \left(1+ y\right)}\hspace{0,25mm} \zeta_2\ln^2 y \hspace{4mm} \nonumber \\[1mm] 
&  \phantom{xx} +\frac{ 4 y \left(11-43 { y}^2\right) }{3{\left(1- y\right)}^3 \left(1+ y\right)} \hspace{0.25mm} \zeta_3\ln y +\frac{ 24 y \left(1-3 { y}^2\right) }{{\left(1- y\right)}^3 \left(1+ y\right)} \hspace{0.25mm} \zeta_4 +\frac{ 2 y \left(2+11 y\right) }{ 3{\left(1- y\right)}^3}\ln^3 y \,. \nonumber 
\end{align}
Here ${\rm{H}}(1,0,-1,0;z)$ is a  harmonic polylogarithm of weight four  with two indices different from zero, which we evaluate numerically with the help of the program {\tt HPL}~\cite{Maitre:2007kp}. The polylogarithm of order three~(four) is denoted by ${\rm Li}_3 (z)$~$\big($${\rm Li}_4 (z)$$\big)$, while  $\zeta_2 = \pi^2/6$, $\zeta_3 \simeq 1.20206$ and $\zeta_4 = \pi^4/90$ are the relevant  Riemann's zeta values. 

In the case of (\ref{eq:A5}) we decompose the relevant QCD corrections as 
\begin{align} \label{eq:Deltagamma}
\Delta_\gamma & =  \frac{\alpha_s}{\pi}  \left(  {\cal A} (y_{q/a})+  2 \hspace{0.25mm} \tau_{q/a} \hspace{0.25mm}  {\cal P}^\prime (\tau_{q/a}) \, \ln \frac{\mu_q^2}{m_q^2}  \right)\,,
\end{align}
with~\cite{Spira:1995rr,Harlander:2005rq,Aglietti:2006tp}
\begin{align}
{\cal A} (y) & = - \frac{ y \left(1+y^2\right) }{{\left(1-y\right)}^3 (1+y)}  \biggl[ 72 \hspace{0.25mm} {\rm Li}_4(y) + 96 \hspace{0.25mm} {\rm Li}_4(-y) - \frac{128}{3} \hspace{0.25mm} \big[  {\rm Li}_3(y) + {\rm Li}_3(-y) \big ] \ln y  \nonumber \\[1mm] 
& \hspace{3.45cm} + \frac{28}{3} \hspace{0.25mm}  {\rm Li}_2(y)\ln^2 y + \frac{16}{3} \hspace{0.25mm}  {\rm Li}_2(-y)\ln^2 y + \frac{1}{18}\ln^4 y \nonumber \\
& \hspace{3.45cm} + \frac{8}{3} \hspace{0.25mm}\zeta_2\ln^2 y + \frac{32}{3}\hspace{0.25mm} \zeta_3\ln y + 12\hspace{0.25mm}\zeta_4 \biggr]   \\[1mm] 
& \phantom{xx}  +\frac{y }{{\left(1-y\right)}^2} \biggl[ -\frac{56}{3} \hspace{0.25mm}  {\rm Li}_3(y) - \frac{64}{3}\hspace{0.25mm}   {\rm Li}_3(-y) + 16 \hspace{0.25mm}  {\rm Li}_2 (y) \ln y + \frac{32}{3} \hspace{0.25mm}  {\rm Li}_2(-y)\ln y  \nonumber \\[1mm] 
& \hspace{2.4cm} +\frac{20}{3} \ln \left ( 1 - y \right ) \ln^2 y -\frac{8}{3} \hspace{0.25mm}\zeta_2\ln y + \frac{8}{3} \hspace{0.25mm} \zeta_3 \biggr] +\frac{2y \left(1+y\right) } {3 {\left(1-y\right)}^3} \ln^3 y \,.\nonumber  
\end{align}

\section{Mixing and threshold effects}
\label{app:mixing}

Even though the decay $a \to b \bar b$ ($a \to c \bar c$) is kinematically forbidden below the open-flavour threshold, the presence of heavy quarks can become relevant through mixing between the pseudoscalar $a$ and bottomonium (charmonium) bound states with the same quantum numbers~\cite{Drees:1989du,Domingo:2008rr,Domingo:2010am,Domingo:2011rn,Baumgart:2012pj,Haisch:2016hzu,Domingo:2016yih}. Such mixings can effectively be described through off-diagonal contributions~$\delta m^2_{a\eta_b(n)}$ to the pseudoscalar mass matrices squared. In the case of $a \hspace{0.25mm}$--$\hspace{0.5mm} \eta_b$ mixing, we employ 
\begin{equation} \label{eq:masssquared}
M_{a \eta_b}^2 = \left( 
\begin{array}{cccc} 
m_a^2 - i m_a \Gamma_a & \delta m^2_{a\eta_b(1)} & \ldots  &  \delta m^2_{a\eta_b(6)} \\
 \delta m^2_{a\eta_b(1)} &  m_{\eta_b(1)}^2 - i m_{\eta_b(1)} \Gamma_{\eta_b(1)} & \ldots & 0 \\
 \vdots & 0 & \ddots & 0 \\
\delta m^2_{a\eta_b(6)} & 0 & 0 & m_{\eta_b(6)}^2 - i m_{\eta_b(6)} \Gamma_{\eta_b(6)} 
  \end{array}  \right) \,,
\end{equation}
with 
\begin{equation} \label{eq:offdiagonal}
\begin{split}
\delta m^2_{a\eta_b(n)}  &= \xi_b^{\rm M} \hspace{0.25mm} \sqrt{\frac{3 }{4\pi v^2}  \hspace{0.25mm} m_{\eta_b(n)}^3 }\hspace{0.25mm}   \big |R_{\eta_b(n)}(0) \big  |\,.
\end{split}
\end{equation}
 The masses  and radial wave functions of the $\eta_b (n)$  states are denoted by $m_{\eta_b(n)}$ and $R_{\eta_b(n)}$, respectively.  The latter quantities can be extracted from the $\Upsilon(n)$ leptonic decay widths (see~\cite{Braaten:2000cm} for instance) which are measured rather precisely~\cite{Patrignani:2016xqp}. In the case of $a \hspace{0.25mm}$--$\hspace{0.5mm} \eta_c$ mixing, we only include the first three states in the pseudoscalar mass matrix squared~(\ref{eq:masssquared}) and rely on the potential model calculations of~\cite{Eichten:1995ch} to  determine the radial wave functions $R_{\eta_c(n)}$. The values of the  $\eta_{b} (n)$ and $\eta_{c} (n)$ masses and radial wave functions that are used in our numerical analysis are collected in Table~\ref{tab:metaReta} for convenience. 
 
 \begingroup 
\renewcommand{\arraystretch}{1.25}
\setlength\tabcolsep{4pt}
\begin{table}[t!]
\centering
\begin{tabular}{c|c|c|c|c}
&  $m_{\eta_b(n)}$ & $\big |R_{\eta_b(n)} (0) \big |$ & $m_{\eta_c(n)}$ & $\big |R_{\eta_c (n)} (0) \big |$ \\[1mm]   
\hline  $n = 1$ & $9.4$ & $2.71$ &$2.98$ & $0.90$ \\
\hline  $n = 2$ & $10.0$ & $1.92$ & $3.64$ & $0.73$ \\
\hline  $n = 3$ & $10.3$ & $1.66$ &$3.99$ & $0.67$ \\
\hline  $n = 4$ & $10.6$ & $1.43$ & --- & --- \\
\hline  $n = 5$ & $10.85$ & $1.41$ & --- & --- \\
\hline  $n = 6$ & $11.0$ & $0.91$ & --- & --- 
\end{tabular}
\caption{Masses of the $\eta_b (n)$ and $\eta_c (n)$ bound states in units of ${\rm GeV}$ and the corresponding values of the radial wave functions in units of ${\rm GeV}^{3/2}$.}
\label{tab:metaReta}
\end{table}
\endgroup

To be able to determine the eigenvalues and eigenvectors of~(\ref{eq:masssquared}) one also needs to know the total decay widths of the $\eta_{b} (n)$ and~$\eta_{c} (n)$ states. The digluon decay widths of  the $\eta_{b} (n)$ states are given to leading order in $\alpha_s$ by (see~\cite{Drees:1989du} for example)
\begin{align} \label{eq:etagg}
\Gamma \hspace{0.25mm} ( \eta_b (n) \to gg ) & = \frac{\alpha_s^2}{3 m_{\eta_b (n)}^2} \, \big |R_{\eta_b (n)} (0) \big  |^2 \,, 
\end{align}
and an analogous formula holds in the case of the charmonium resonances.  

The partial decay widths~(\ref{eq:etagg}) essentially saturate   $\Gamma_{\eta_b (n)}$ with $n \neq 5,6$. For $\eta_b (5)$ and $\eta_b (6)$, however, also decays to final states involving $\pi$ and $B_{(s)}$ mesons are relevant. In the case of  the decays to pion final states, we employ~\cite{Patrignani:2016xqp}
\begin{align}
\Gamma \hspace{0.25mm} (  \eta_b (5)  & \to \pi \; \text{mesons})  = 1.5 \, {\rm MeV} \,, \\[1mm]
\Gamma \hspace{0.25mm} (  \eta_b (6)  & \to \pi \;  \text{mesons})  = 3  \, {\rm MeV}  \,,
\end{align}
while the $B_{(s)}$ decays are incorporated via the approximate relations~\cite{Baumgart:2012pj}
\begin{align}
& \Gamma \hspace{0.25mm} ( \eta_b (5) \to B +  B_{s} \; \text{mesons})   \simeq 0.9 \hspace{0.5mm} \Gamma \hspace{0.25mm} ( \Upsilon(5)  \to B \; \text{mesons}) +  0.65 \hspace{0.5mm} \Gamma \hspace{0.25mm} ( \Upsilon(5)  \to B_{s} \; \text{mesons}) \,, \\[2mm]
& \hspace{6mm} \Gamma \hspace{0.25mm} ( \eta_b (6) \to B + B_{s} \; \text{mesons})   \simeq  \Gamma \hspace{0.25mm} ( \Upsilon(5)  \to B \; \text{mesons}) +  \Gamma \hspace{0.25mm} ( \Upsilon(5)  \to B_{s} \; \text{mesons}) \,, 
\end{align}
in our numerical analysis. Here~\cite{Patrignani:2016xqp}
\begin{align}
\Gamma \hspace{0.25mm} (  \Upsilon(5)  \to B \; \text{mesons} )  & = 42 \, {\rm MeV} \,, \\[1mm]
\Gamma \hspace{0.25mm} (  \Upsilon(5)  \to B_s \; \text{mesons} )  & = 11 \, {\rm MeV}  \,.
\end{align}

In the case of the charmonium bound states, we use directly $\Gamma_{\eta_{c (1)}} = 31.8 \, {\rm MeV}$ and  $\Gamma_{\eta_{c (2)}} = 11.3 \, {\rm MeV}$~\cite{Patrignani:2016xqp}, while for $ \eta_{c } (3)$ we include besides~(\ref{eq:etagg})  an open-charm contribution. Applying the approach of~\cite{Baumgart:2012pj} to relate the $\eta_c (3)$ decays to those of $\psi (3770)$ results in $\Gamma \hspace{0.25mm} ( \eta_{c } (3) \to D  \; \text{mesons}  ) \simeq 30  \hspace{0.5mm} \Gamma \hspace{0.25mm} ( \psi (3770) \to  D \; \text{mesons}  )$. However, the $\psi (3770) $ lies very close to the open-charm threshold and is thus highly susceptible to strong rescattering effects. Using instead the $\psi (4040)$ properties as input, we obtain the approximate result 
\begin{equation} 
\Gamma \hspace{0.25mm} ( \eta_{c } (3) \to D  \; \text{mesons}  ) \simeq 0.5  \hspace{0.5mm} \Gamma \hspace{0.25mm} ( \psi (4040) \to  D \; \text{mesons}  )  \,,
\end{equation}
where  $\Gamma \hspace{0.25mm} ( \psi (4040) \to   D \; \text{mesons} )  \simeq \Gamma_{\psi (4040)}  = 80 \, {\rm MeV}$~\cite{Patrignani:2016xqp}.

We furthermore emphasise that the branching ratios  $\eta_b (n) \to \mu^+ \mu^-$ are all below the $10^{-10}$ level~\cite{Haisch:2016hzu} and therefore can be safely ignored  in the mixing formalism. The effects of the ditau decays of the bottomonium bound states are negligible as well and so are the dilepton decays of the $\eta_c (n)$ mesons. Effects of $a \hspace{0.25mm}$--$\hspace{0.5mm} \eta_b$ mixing in $h \to aa$ such as for instance $h \to 2 \eta_b (n) \to a a$ are part of ${\rm BR} \hspace{0.25mm} (h \to a a )$ and thus effectively included in our numerical analysis. The same is true for contributions of intermediate $\eta_c (n)$ states to the exotic decay $h \to aa$ of the SM Higgs. 

Above the $b\bar b$ ($c \bar c$) threshold a perturbative description of the production and the decay of the pseudoscalar $a$ breaks down. In this region one can however approximate the $b \bar b$ ($c \bar c$) contributions to the total decay width $\Gamma_a$ through a heuristic model that is inspired by QCD sum rules~\cite{Drees:1989du,Baumgart:2012pj,Haisch:2016hzu} and interpolates to the continuum sufficiently above threshold. The interpolations take the form 
\begin{align}  
{\cal N}^{b}_a & = 1 - \exp \left [ - 8.0  \left ( 1 - \frac{(m_B + m_{B^\ast})^2}{m_a^2} \right )^{2.5 \,} \right ] \,, \label{eq:BastB} \\[2mm]
{\cal N}^{c}_a & = 1 - \exp \left [ - 6.5  \left ( 1 - \frac{(m_D + m_{D^\ast})^2}{m_a^2} \right )^{2.5 \,} \right ] \,, \label{eq:DastD} 
\end{align}
with $m_B = 5.28 \, {\rm GeV}$, $m_{B^\ast} = 5.33\, {\rm GeV}$, $m_D = 1.86 \, {\rm GeV}$ and $m_{D^\ast} = 2.01 \, {\rm GeV}$~\cite{Patrignani:2016xqp}.  In our analysis, the interpolation is achieved by simply multiplying the partonic decay width $\Gamma \hspace{0.25mm} (a \to b \bar b)$  and  $\Gamma \hspace{0.25mm} (a \to c \bar c)$  by the factor ${\cal N}^{b}_{a}$ and ${\cal N}^{c}_{a}$, respectively. 

For $m_a > 2 m_K$ decays into kaons become kinematically allowed. The decay $a \to KK$  however violates CP, and as a result $a$ can in practice only decay into three-body final states such as $KK\pi$. Following~\cite{Dolan:2014ska}, we estimate the hadronic width $\Gamma  \hspace{0.25mm} ( a \to s \bar s \to KK \pi )$ by multiplying $\Gamma  \hspace{0.25mm} ( a \to s \bar s )$ by the suppression factor 
\begin{equation}
{\cal N}^{s}_a  = \frac{16 \pi}{m_a^2} \left ( \frac{m_s^\ast}{m_s} \right )^2 \frac{\rho \left ( m_K, m_K, m_\pi, m_a \right ) }{\beta_{s/a}} \,,
\end{equation}
with $m_s^\ast = 450 \, {\rm MeV}$~\cite{McKeen:2008gd}, $m_K = 439 \, {\rm MeV}$ and $m_ \pi = 140 \, {\rm MeV}$~\cite{Patrignani:2016xqp}. Here  $\rho \left ( m_1, m_2, m_3, m_4 \right )$ denotes the phase space for isotropic three-body decays. It can be written as  
\begin{equation}
\begin{split}
\rho \left ( m_1, m_2, m_3, m_4 \right ) & = \frac{1}{(4 \pi)^3} \int_{m_1}^{\frac{m_1^2 + m_4^2 -(m_2+m_3)^2}{2m_4}} \! d E_1 \, 2 \hspace{0.25mm} \sqrt{E_1^2 - m_1^2} \\[2mm] 
& \hspace{2cm} \times \lambda \left ( m_1^2 + m_4^2 - 2 E_1 m_1, m_2^2, m_3^2 \right )  \,,
\end{split}
\end{equation}
with $\lambda \left (x, y, z \right)$ defined in~(\ref{eq:lambdaxyz}). 

\end{appendix}


\AddToContent{U.~Haisch, J.~F.~Kamenik, A.~Malinauskas,~M. Spira}
\renewcommand{\thesection}{\arabic{section}}
\renewcommand{\thesubsection}{\thesection.\arabic{subsection}}
\renewcommand{\thesubsubsection}{\thesubsection.\arabic{subsubsection}}
\renewcommand{\thefigure}{\arabic{figure}}
\renewcommand{\theequation}{\arabic{equation}}
\renewcommand{\thetable}{\arabic{table}}
\renewcommand{\thefootnote}{\arabic{footnote}}

\graphicspath{{LightScalarsZ/}}

\chapter{Light scalar boson searches at the LHC through associated production with a $Z$ boson}

{\it A.~Angelescu, S.~Fichet, L.~Finco, S.~Gascon-Shotkin, G.~Moreau, S.~Zhang}



\begin{abstract}
Light scalar bosons with mass typically below $\sim 65$~GeV, predicted in several scenarios beyond the Standard Model (SM), might have been missed by the present experimental 
searches based on the Large Hadron Collider (LHC) data. Here we show that such generic neutral CP-odd or CP-even scalar fields, thanks to their production 
in association with a detectable Z gauge boson, exhibit specific kinematical distributions which could allow to
distinguish between their signatures at LHC and the SM background. The theoretical framework consists of an effective field theory including various types of interactions, that turns out
to be potentially testable separately at colliders. From the 
experimental side, Monte Carlo simulations for the signal and background are confronted with each other. The production of heavier scalar bosons 
(masses considered here up to $\sim 110$~GeV) in association with a Z boson might 
be useful as well, for measuring their coupling to a Z boson pair. Hence we also derive the exclusion limits on generic heavy scalar production rates for such a process combined
with the vector boson fusion mechanism, by using Large Electron-Positron (LEP) and LHC Run~1 data recorded by the CMS experiment. We demonstrate that the attractive and specific scalar field
example of the radion, arising in the usual SM extensions to a warped extra dimension, is not yet excluded by those data. 

\end{abstract}

\section{INTRODUCTION}

Several extensions of the Standard Model (SM) of elementary particle physics include an electrically 
neutral scalar boson in their field content with a mass below a hundred GeV: light Higgs bosons,
the radion, the dilaton, the axion\dots Although in 2012 the Large Hadron Collider (LHC) discovered a particle compatible with the SM Higgs boson as a $125$~GeV resonance, one should thus consider the 
possibility that some lighter scalar particles might have been missed so far in the LHC data analyses. With scalar boson masses below typically $\sim 65$~GeV, the golden 
scalar decay into a pair of on-shell Z bosons is kinematically closed while its decay 
into a diphoton is extremely difficult to detect experimentally as the LHC diphoton triggers are 
bandwidth-limited for such soft photon production.\footnote{At low diphoton invariant masses, boosted diphoton events with high $p_T$ transverse
momentum can still be triggered, but at the price of a weaker selection
efficiency.}
Therefore, in the present study, instead of considering the dominant gluon-gluon fusion mechanism, 
we consider the promising light scalar boson production in association with a Z gauge boson allowing to trigger on the charged leptons of the Z decay.
We adopt a generic theoretical approach based on an effective field theory for CP-odd and CP-even (under the combined Charge Parity symmetry) scalar fields coupled to 
ElectroWeak (EW) gauge bosons via the main structures of interactions. 
The Monte Carlo simulations of the events for the signal are performed with the {\tt Feynrules} code~\cite{Degrande:2011ua} interfaced with {\tt MadGraph5\_aMC@NLO}~\cite{Alwall:2014hca} while the final-state SM background is generated with {\tt MadGraph5\_aMC@NLO}.

In addition, the production of a scalar boson in association with a Z boson might be interesting also for scalar masses above a hundred GeV, particularly as regards the determination 
of the scalar coupling to the two SM neutral gauge bosons. Motivated by this feature, we also work out generic exclusion limits on heavier scalar production rates for this 
process together with the Vector Boson Fusion (VBF) mechanism (involving also the scalar coupling to EW gauge bosons), 
as deduced from the Large Electron-Positron (LEP)~\cite{Rosca:2002me} and LHC Run~1 ($8$~TeV center-of-mass
energy) CMS~\cite{CMS-PAS-HIG-14-037} data. Finally, we study the precise case of the scalar boson represented by the so-called radion, which arises in any model with at least one
extra spatial dimension, and we focus on the attractive scenario with bulk matter~\cite{Gherghetta:2000qt} in a slice of $AdS_5$ space~\cite{Randall:1999ee} addressing both 
the flavour and gauge hierarchy problems.

\section{THE EFFECTIVE MODEL}  
\label{EffModel}

Diagrammatically, the studied scalar boson production in association with a Z gauge boson occurs through the radiation of the scalar field from a Z boson produced in the s-channel.
This process involves only the scalar coupling to two Z bosons. 
We use an effective theory approach to describe the scalar interaction with SM EW gauge bosons.  
The scalar mass can be smaller than the EW symmetry-breaking scale.
When it is the case, we make the extra assumption that the scalar has large tree-level $SU(2)_L\times U(1)_Y$ couplings, so that the loop-induced EW-breaking contributions 
are subleading. Under this condition, the interactions of a neutral CP-even or CP-odd scalar $\phi$ with the EW gauge bosons are respectively described by the following 
dimension-5 effective gauge invariant Lagrangians 
\begin{equation}
{\cal L}_{\rm eff}\supset\phi \left( \frac{1}{f_G}G^{\mu\nu\,a}G_{\mu\nu}^a+ \frac{1}{f_W}W^{\mu\nu\,b}W_{\mu\nu}^b
+\frac{1}{f_B}B^{\mu\nu}B_{\mu\nu}+\frac{1}{f_H}|D^\mu H|^2
\right)
\label{EFTop1}
\end{equation}
\begin{equation}
{\cal L}_{\rm eff}\supset\phi \left( \frac{1}{\tilde f_G}G^{\mu\nu\,a}\tilde G_{\mu\nu}^a+ \frac{1}{\tilde f_W}W^{\mu\nu\,b}\tilde W_{\mu\nu}^b
+\frac{1}{\tilde f_B}B^{\mu\nu}\tilde B_{\mu\nu}
\right)
\label{EFTop2}
\end{equation}
where $\tilde V^{\mu\nu}=\frac{1}{2}\epsilon^{\mu\nu\rho\sigma}V_{\rho \sigma}$, $H$ represents the SM Higgs doublet, $D^\mu$ the covariant derivative,
the $f$'s denote high-energy scales of new physics, $a,b$ are summed group generator indices whereas 
$\mu$, $\nu$ stand for summed Lorentz indices and the rank-2 tensors are the field strengths for all the SM
gauge bosons before EW symmetry breaking (using standard notations). 
After this breaking, the effective Lagrangian, for example in the CP-even case, contains the $\phi Z Z$ interactions 
\begin{equation}{\cal L}_{\rm eff}\supset \frac{1}{f_Z} \phi  (Z_{\mu\nu})^2 +\frac{m_Z^2}{ 2f_H}\phi (Z_\mu)^2\,,  
\end{equation} where $f_Z^{-1}=s_w^2 f_B^{-1}+c_w^2 f_W^{-1}$ and $s_w^2\equiv{\sin^2(\theta_w)}\approx 0.23$. 
The effective theory is valid as long as the $f$'s (related to Kaluza-Klein mass scales for instance) are larger than the typical energies 
going through the vertices. 
The $\phi$ scalar mixing with the SM Higgs boson is assumed to be small to ensure that the SM Higgs field has SM-like couplings compatible with the 
LHC signal strength measurements. 
The scalar fields entering Eq.~(\ref{EFTop1})-(\ref{EFTop2}) are taken to be the mass eigenstates.

The CP-even couplings might be those of a radion in a model with a warped extra dimension along which matter is propagating.
Notice that if EW brane kinetic terms are negligible in such models, one has $f_W=f_B$ \cite{Fichet:2013ola, Fichet:2013gsa} which implies that the $\phi F^{\mu\nu}Z_{\mu\nu}$ 
coupling vanishes, a property which can be used for model discrimination \cite{Baldenegro:2017aen}.

The CP-odd scalar field can be a pseudo Nambu-Goldstone boson from an approximate global symmetry, just like those appearing in composite Higgs models. The couplings to gauge fields are then induced by the many fermion resonances populating the TeV scale (see e.g Ref.~\cite{Belyaev:2016ftv} or Ref.~\cite{Fichet:2016xvs}).

\section{NUMERICAL ANALYSES}

\subsection{MONTE CARLO SIMULATIONS FOR LIGHT SCALAR FIELDS}

As stated in the introduction, bandwidth considerations at the LHC currently limit the invariant diphoton mass range that can be probed for a new light scalar using pure diphoton triggers to no lower than $\sim 65$~GeV~\cite{CMS-PAS-HIG-14-037,PhysRevLett.113.171801,CMS-PAS-HIG-17-013}. For this reason, it is interesting to envisage a dedicated analysis targeting the associated production of a light scalar with a Z boson, where the Z boson decays to a pair of oppositely-charged electrons or muons. In such a scenario the events could be triggered by dilepton rather than diphoton triggers, thus potentially allowing the lower limit of the light scalar-to-diphoton search range to decrease to $\sim 20$~GeV.  

In order to investigate the capability of such an analysis to be able to distinguish the production of a CP-even from a CP-odd scalar particle as defined in the above effective model, and from SM background processes, we constructed both CP-even and CP-odd instances of the model with the {\tt Feynrules} code in the form of Universal FeynRules Output (UFO) files.  These were then propagated to the {\tt MadGraph5\_aMC@NLO\_v2\_5\_5} program for generation of parton-level events at $\sqrt{s}=8$ TeV of the process $pp\rightarrow\phi +Z$ , $\phi\rightarrow\gamma\gamma$, $Z\rightarrow\mu^+\mu^-$, for each of $m_{\phi}=20$ and $70$ GeV, as well as for generation of events of the process $pp\rightarrow\gamma\gamma +Z$, $Z\rightarrow\mu^+\mu^-$  within the SM. The event generations were performed for the following three choices of parameters:
\begin{itemize}
\item Two different cases of a CP-even scalar boson:
\begin{itemize}
\item $f_B=1$ TeV and $f_{H,W}\to \infty$,  corresponding to the case of a CP-even scalar boson coupling to  two Z bosons via the $(Z_{\mu\nu})^2$ Lorentz structure, called CP-even$_{1/f_Z}$
\item $f_H=1$ TeV and $f_{B,W}\to \infty$,  corresponding to the case of a CP-even scalar boson coupling to  two Z bosons via the $(Z_{\mu})^2$ Lorentz structure, called CP-even$_{1/f_H}$
\end{itemize}
\item  The case of a CP-odd scalar field with $\tilde f_B=1$ TeV and $\tilde f_W\to \infty$, in which the coupling to  two Z bosons always occurs via the $Z_{\mu\nu}\tilde Z^{\mu\nu}$ Lorentz structure.

\end{itemize}
No selection or acceptance criteria were applied.

Figures~\ref{distrib1} and~\ref{distrib2} show kinematical distributions for $m_{\phi}=20$ and $70$ GeV respectively, for each of the above three cases of light scalar:  CP-even$_{1/f_Z}$, CP-even$_{1/f_H}$, and CP-odd, as well as for the SM background.  The areas of all distributions have been normalized to unity. For both $\phi$ mass values, the shape of the distribution of $\Delta R$ (where $\Delta R^2=\sqrt{\Delta\eta^2+\Delta\phi^2}$ and $\eta$ can be approximated by $-ln\tan(\theta /2)$, $\theta$ and $\phi$ here denoting the polar and azimuthal angles, respectively) between the two muons, $\Delta R_{\mu^+\mu^-}$, is clearly different for each of the CP-even$_{1/f_Z}$, CP-odd, and SM cases, whilst the CP-even$_{1/f_H}$ and SM cases can only be distinguished from each other for $m_{\phi}=70$ GeV.  The shapes of the transverse momentum of the dimuon system, $p_{T_{\mu^+\mu^-}}$, of all four cases can be distinguished from each other for both masses (case of  $m_{\phi}=20$ GeV not shown), with the CP-odd light scalar posessing the hardest spectrum. Finally, the invariant mass of the dimuon system, $M_{\mu^+\mu^-}$,  offers some distinction betweeen the CP-even$_{1/f_H}$ case and the other cases, but only for $m_{\phi}=20$ GeV ($m_{\phi}=70$ GeV not shown). However, for both $\phi$ masses, the dimuon invariant mass distributions illustrate the possibility of incorporating a window around the nominal Z boson mass into the trigger.  Where they exist, the nature of the shape differences would seem to favor use in multivariate techniques rather than in the application of sequential selection criteria. Since the distributions shown are at parton level, further studies must be undertaken to determine whether the differences in shape are as marked after parton showering/hadronization, reflecting in particular the influence of the underlying event, and after detector simulation.  
\begin{figure}[h]
\includegraphics[width=7cm]{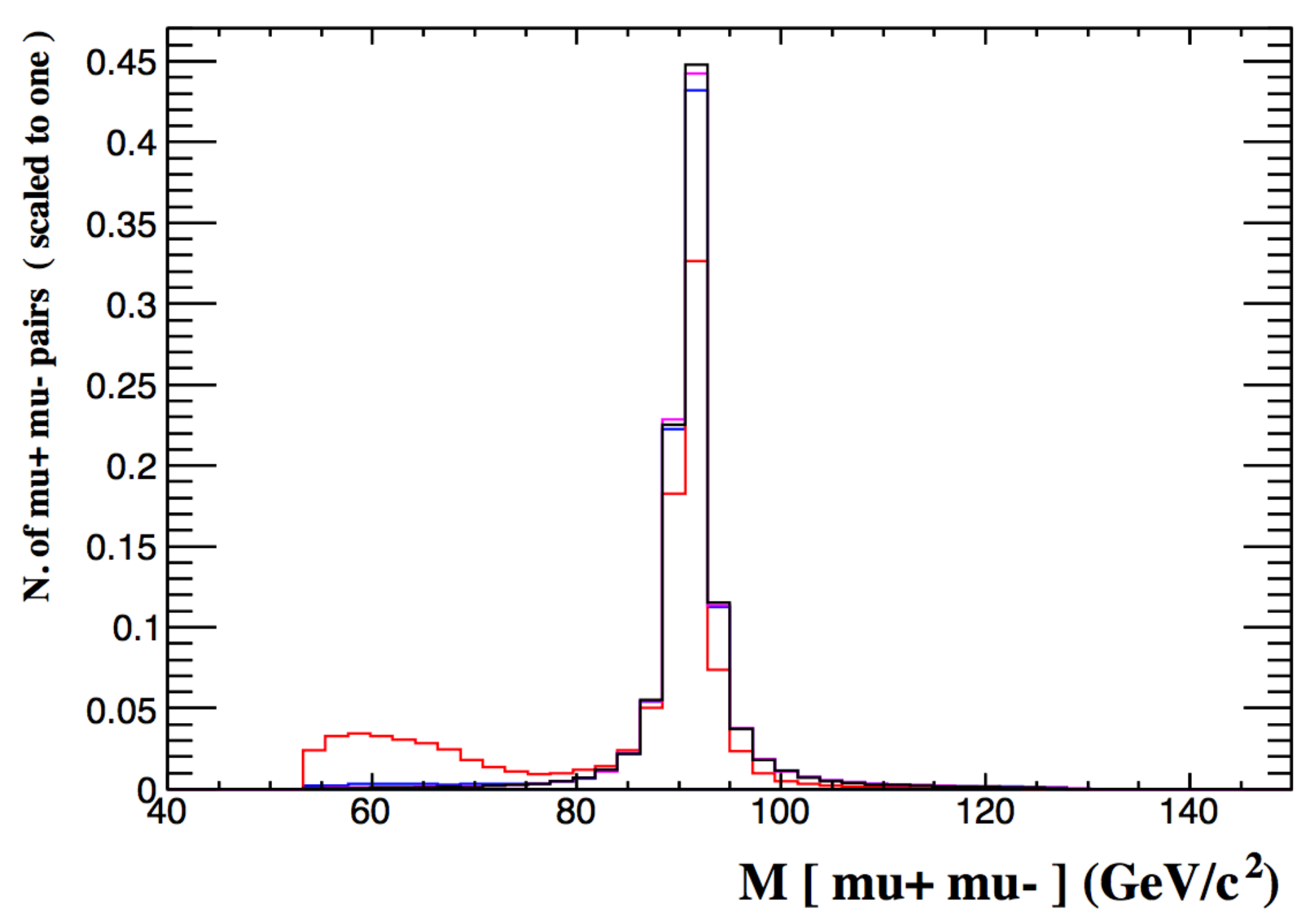}
\hspace{1.5cm}
\includegraphics[width=7cm]{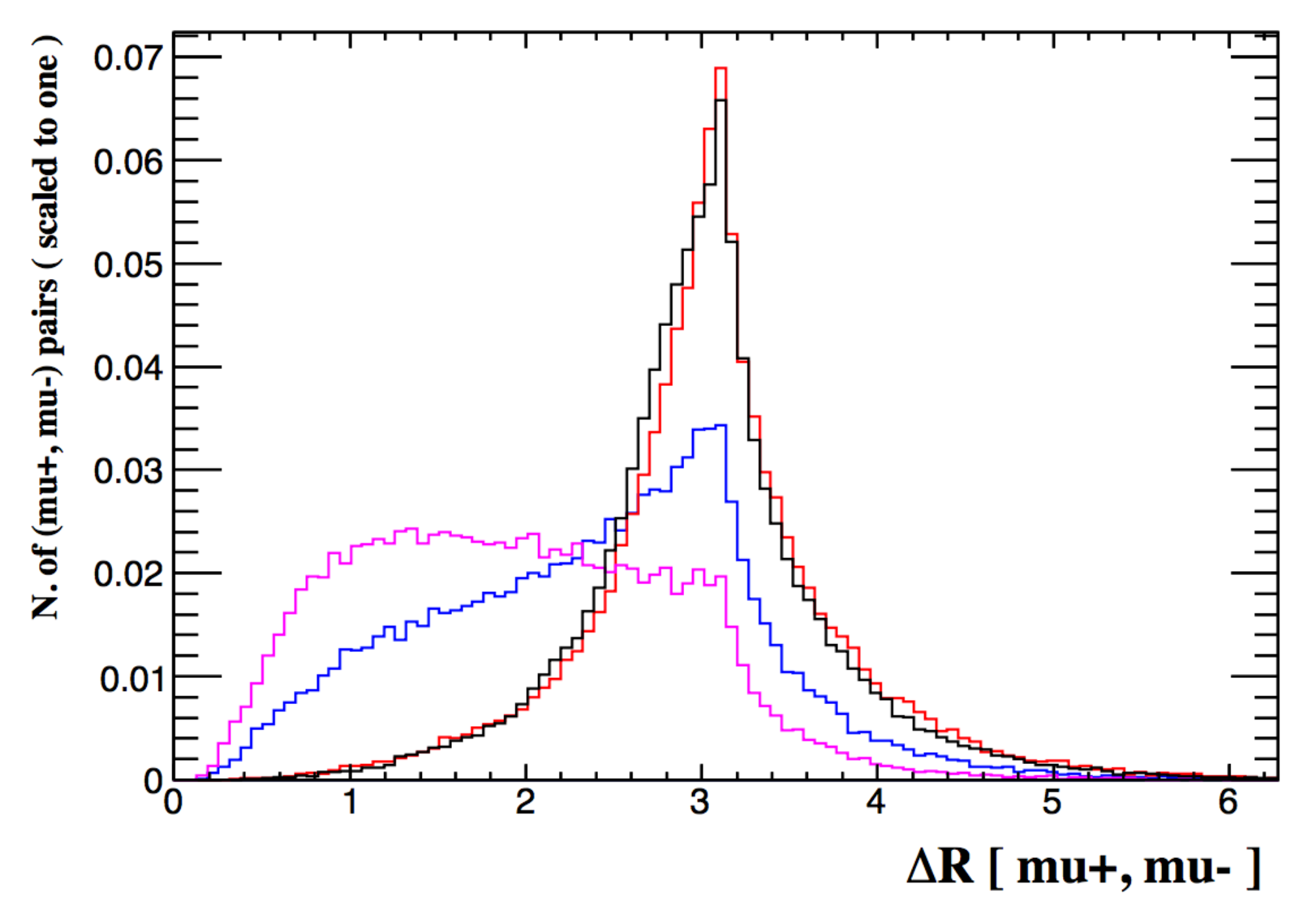}
\caption{Distributions of the dimuon invariant mass, $M_{\mu^+\mu^-}$ [GeV/c$^2$] (left) and $\Delta R$ between the two muons, $\Delta R_{\mu^+\mu^-}$ (right), for a light scalar $\phi$ with $m_{\phi}=20$~GeV for the following cases: CP-even$_{1/f_Z}$ (blue), CP-even$_{1/f_H}$ (red), and CP-odd (purple), superimposed on the same distributions for the SM background (black).}
\label{distrib1}\end{figure}
\begin{figure}
\includegraphics[width=7cm]{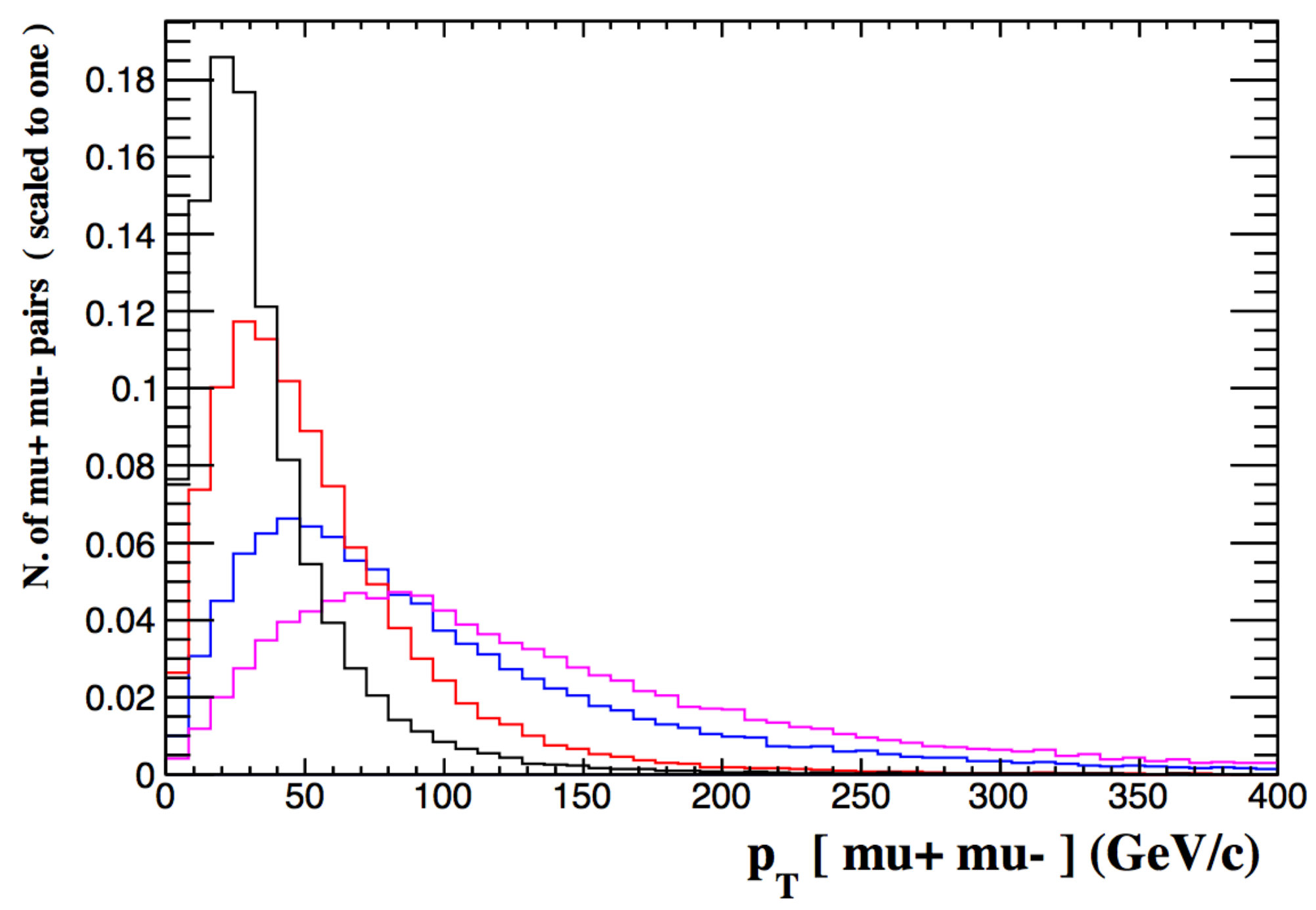}
\hspace{1.5cm}
\includegraphics[width=7cm]{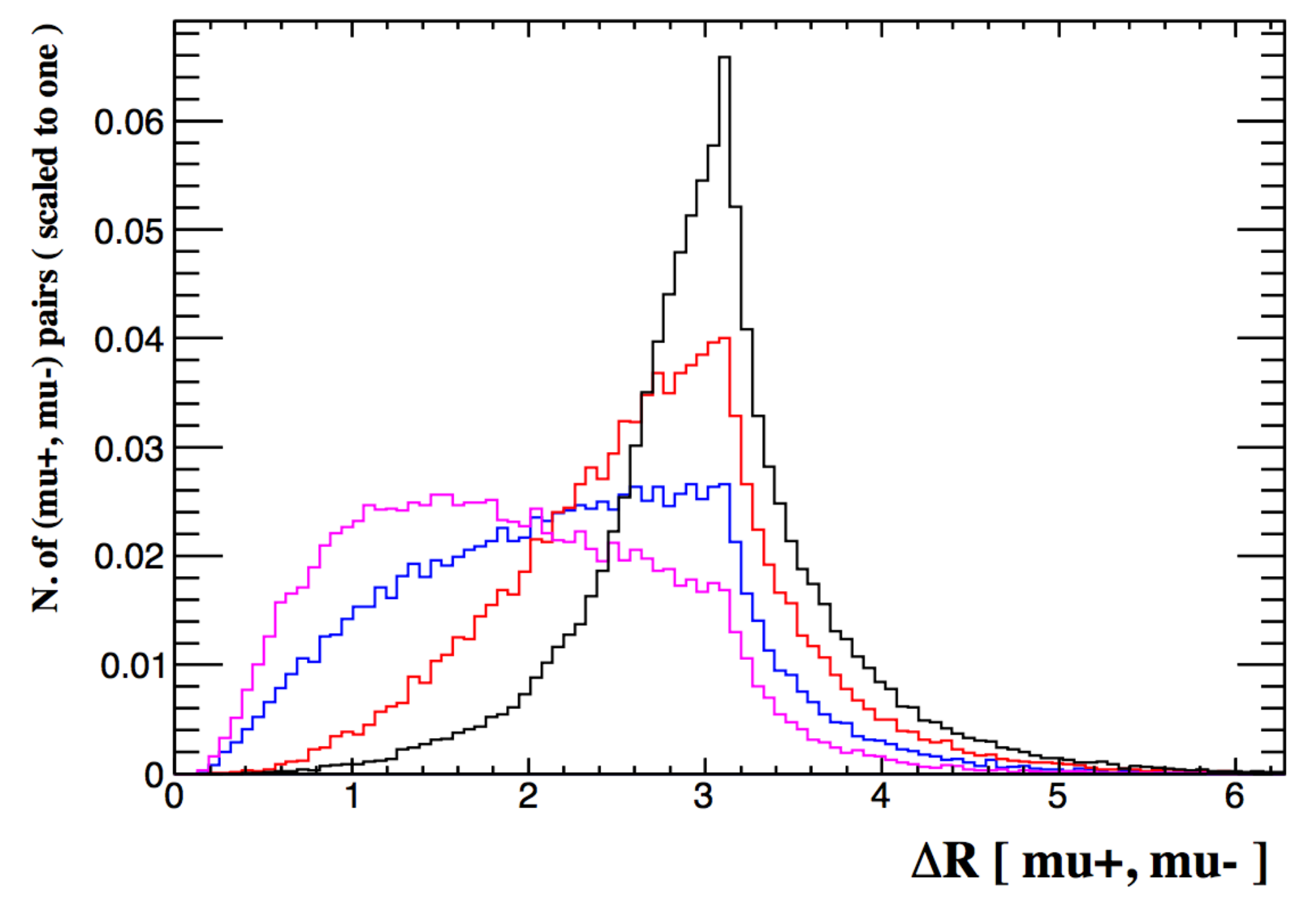}
\caption{Distributions of the dimuon system transverse momentum, $p_{T_{\mu^+\mu^-}}$ [GeV/c] , (left) and $\Delta R$ between the two muons, $\Delta R_{\mu^+\mu^-}$ (right), for a light scalar $\phi$ with $m_{\phi}=70$~GeV for the following cases: CP-even$_{1/f_Z}$ (blue), CP-even$_{1/f_H}$ (red), and CP-odd (purple), superimposed on the same distributions for the SM background (black).}
\label{distrib2}\end{figure}

\subsection{EXCLUSION LIMITS FOR HEAVY SCALAR BOSONS}

Using the LEP~\cite{Rosca:2002me} and LHC Run~1 (CMS Collaboration)~\cite{CMS-PAS-HIG-14-037} data 
combined with the SM background estimates, we have derived conservative upper constraints
on the scalar production rates for scalar masses above $\sim 80$~GeV (as performed in Ref.~\cite{Cacciapaglia:2016tlr} with these LHC data for a lighter Higgs boson 
in the two Higgs doublet model). Both the $pp\to Z\phi$ production and the VBF mechanism, $pp\to\phi qq$ [$q\equiv$~quark], were taken into account. 
In this work we have not turned on the $\phi (W^b_{\mu\nu})^2$ interaction so that $\phi$ does not couple to $W^{\pm}$ (in other words $f_W, \tilde f_W \to \infty$). 
The now experimentally detectable diphoton decay channel $\phi \to \gamma\gamma$ was exclusively used to select the signal, 
but some cuts based on the associated Z boson or quark pair 
should be added to select the two studied production processes and to optimise the signal selection over backgrounds. 
From the obtained rate constraints, we have derived the bounds at $95\%$~C.L. on the combinations of effective parameters and the free diphoton branching ratio
$B_{\phi \to \gamma\gamma}$ entering the scalar rates, as displayed in Fig.~\ref{figuref0B} for the considered $1/\tilde f_Z$ ($1/f_Z$) coupling of the CP-odd (CP-even) 
scalar field and in Fig.~\ref{figuref0H} (left) for the $1/f_H$ coupling of the CP-even scalar boson (for simplification reasons only one coupling is non-vanishing at a time).

\begin{figure}[h]
\includegraphics[width=7.5cm]{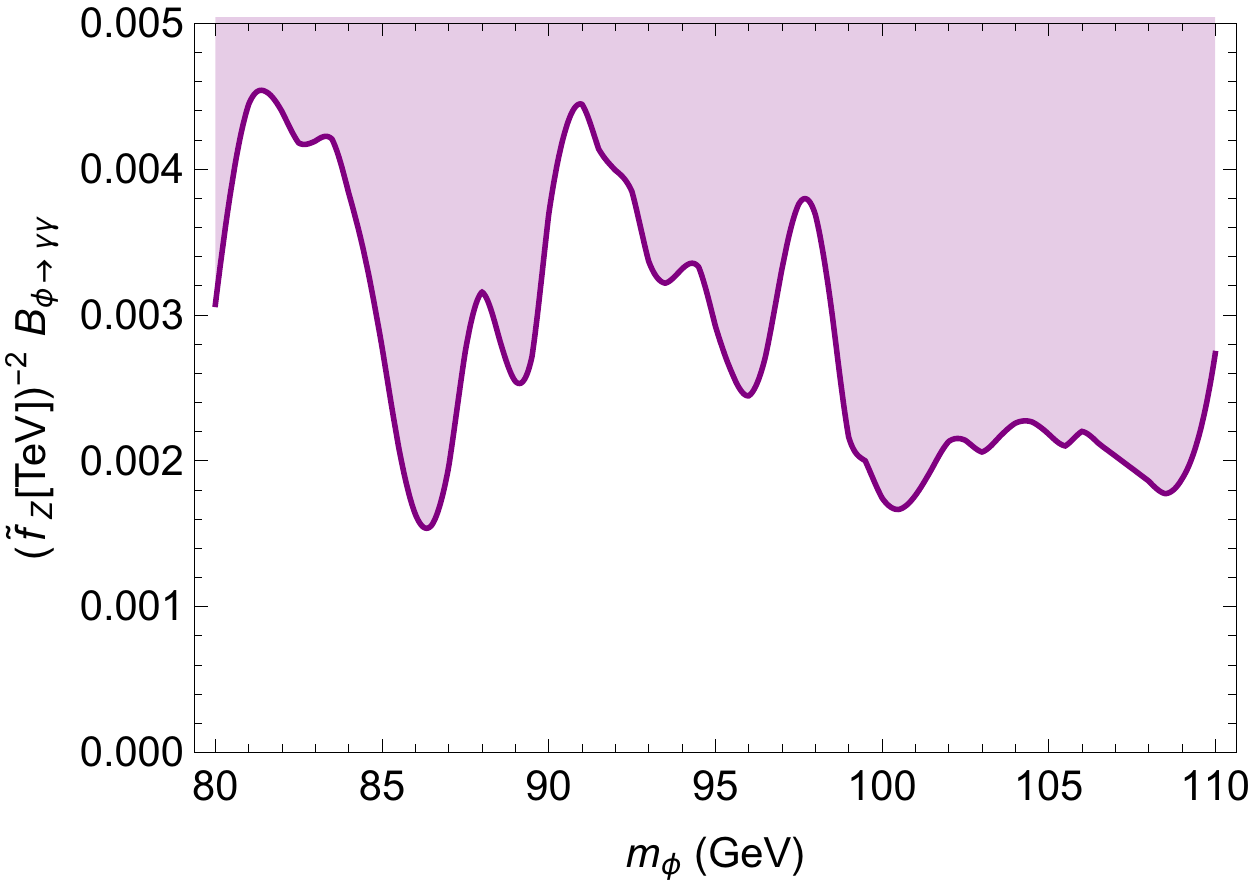}
\hspace{0.5cm}
\includegraphics[width=7.5cm]{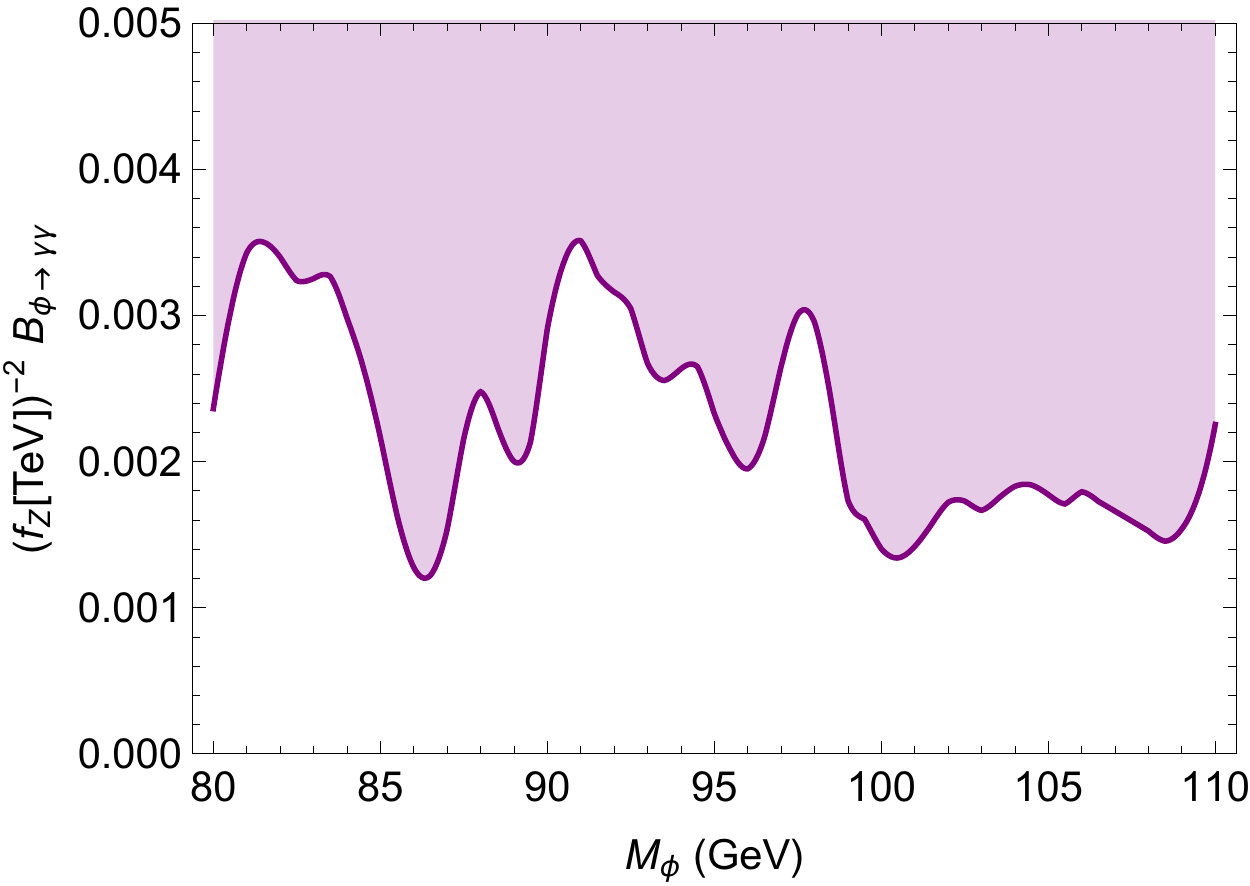}
\caption{Colored regions excluded by the LHC Run~1 (CMS Collaboration) data in the plane $B_{\phi \to \gamma\gamma}/\tilde f_Z^2$ [in TeV$^{-2}$] versus the scalar 
mass $m_\phi$ [in GeV] (left) and $B_{\phi \to \gamma\gamma}/f_Z^2$ [in TeV$^{-2}$] versus $m_\phi$ [in GeV] (right).
}\label{figuref0B}
\end{figure}

\begin{figure}
\includegraphics[width=7.5cm]{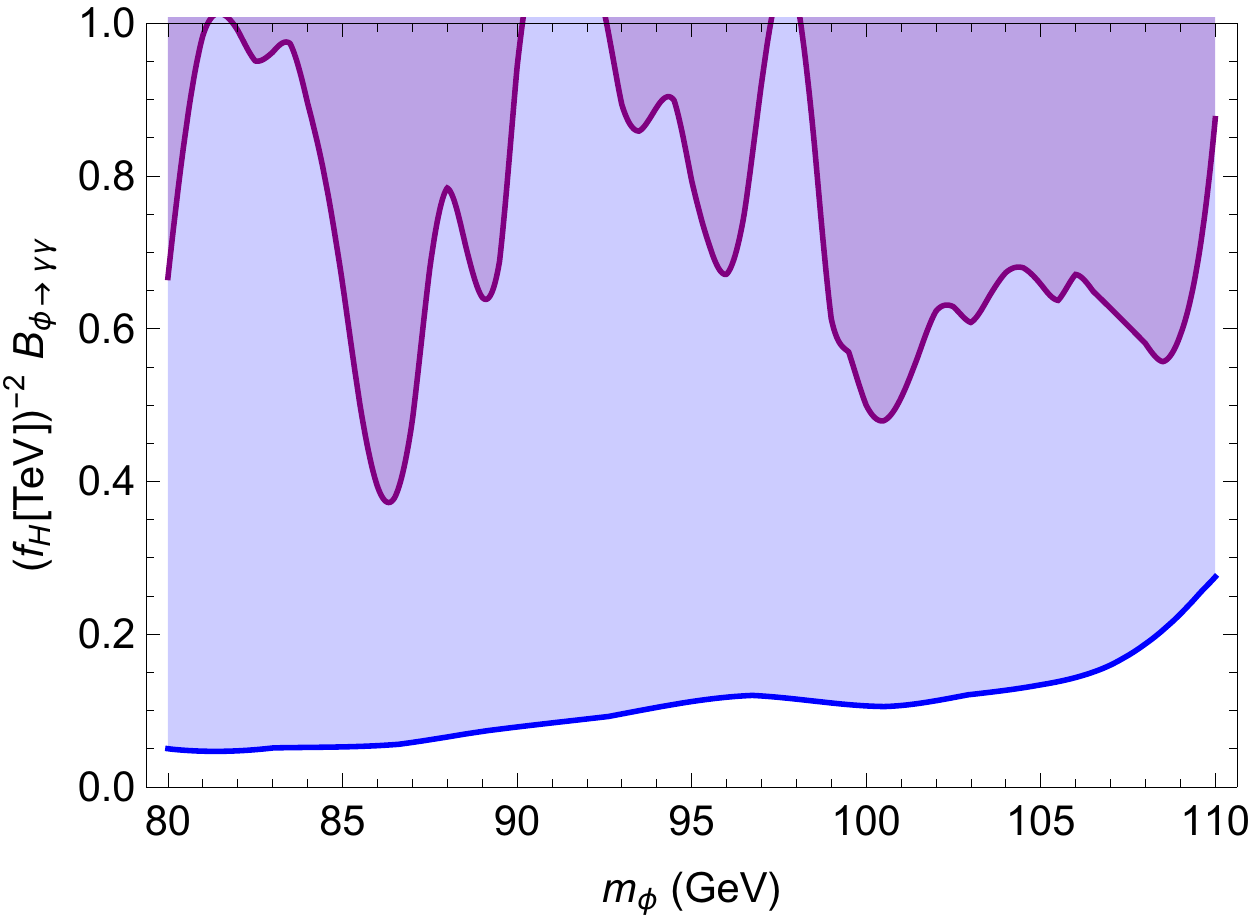}
\hspace{0.5cm}
\includegraphics[width=7.5cm]{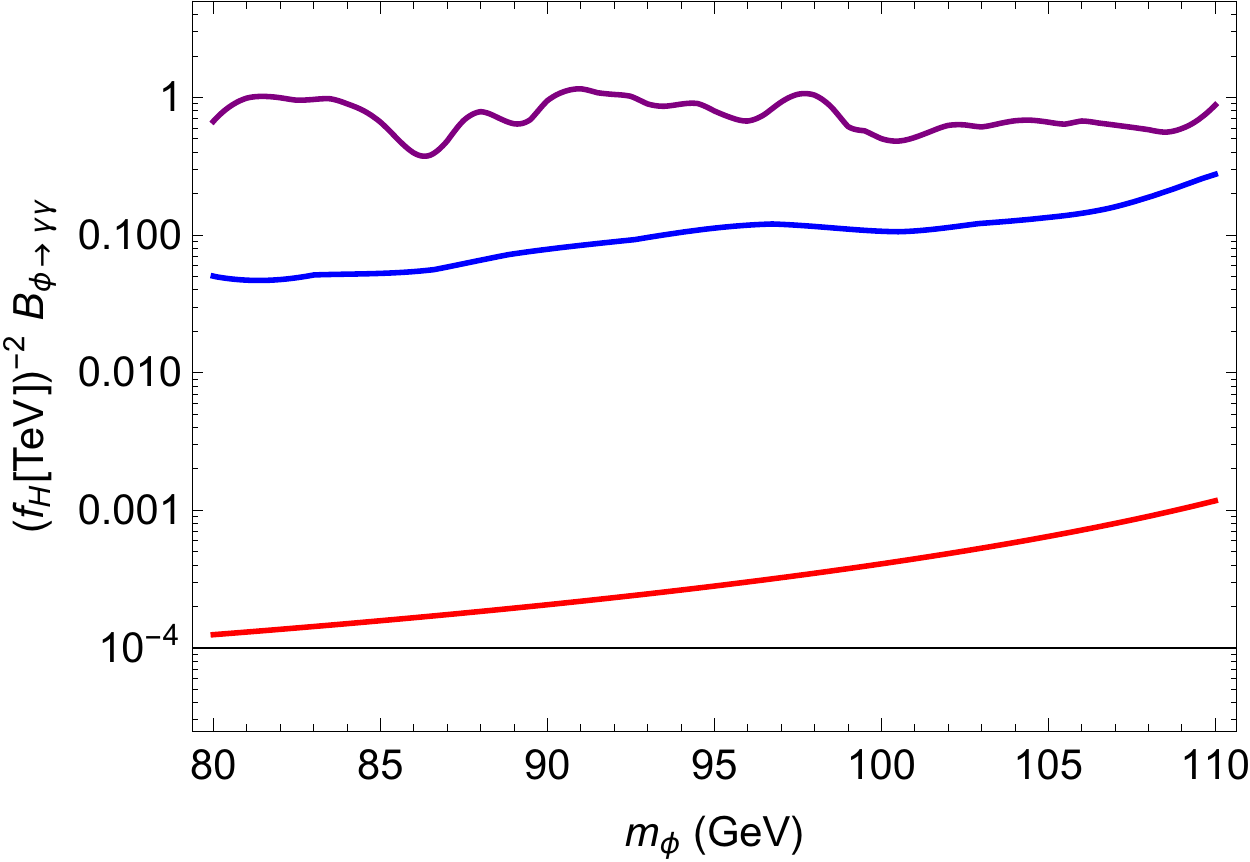}
\caption{Colored domains excluded by the LEP [in blue] and LHC Run~1 (CMS) [in purple] data in the plane $B_{\phi \to \gamma\gamma}/f_H^2$ [in TeV$^{-2}$] 
versus $m_\phi$ [in GeV] (left). The right-hand plot is identical but with a prediction for the radion scalar field superimposed [the plain red line].
}\label{figuref0H}
\end{figure}

We have also studied the precise case of the radion scalar boson arising in the scenario with bulk matter in a slice of $AdS_5$ space~\cite{Angelescu:2017jyj}. 
The radion is in fact associated to the fluctuations of the metric along the extra spatial dimensions (here a unique warped extra dimension). 
As justified in Section~\ref{EffModel}, we take the example of a 
relatively small Higgs-radion mixing generated by a parameter, $\xi =0.3$, a first Kaluza-Klein photon mass quite heavy compared to $m_\phi$, 
$M_{KK}=3$~TeV, and also $\Lambda=10$~TeV (see Ref.~\cite{Angelescu:2017jyj}). We consider 
the limiting case $f_Z\to \infty$ which is realistic as a first approximation since the radion derivative couplings (kinetic-like) to gauge bosons 
are not the dominant ones generally speaking~\footnote{Indeed, the radion derivative couplings to (massive) gauge bosons receive a suppression factor of about $\sim 70$, equal to twice the so-called extra-dimensional volume factor, which is not present in the $f_H^{-1}$ couplings between the radion and gauge bosons. However, for certain values of $\xi$ and $m_{\phi}$, the $f_H^{-1}$ couplings can achieve values close to 0 and thus become 
subdominant compared to the derivative $f_Z^{-1}$ couplings.}.
The prediction for the quantity $1/f_H^2$ for the radion case is then drawn on Fig.~\ref{figuref0H} (right) as a function of the radion mass. For $B_{\phi \to \gamma\gamma}<1$, the prediction lies below the red line, further from the exclusion regions. The conclusion from the comparison between this prediction and the superimposed LEP and LHC limits
is that, with the considered choice of parameters, the radion particle is clearly not excluded by the present combined high-energy collider results.

\section*{CONCLUSIONS}

For two examples of low scalar masses at $20$~GeV and $70$~GeV, we have shown that some selected relevant kinematical distributions for the charged muon pair 
resulting from the Z leptonic decay, subsequent to $Z\phi$ production, have specific shapes which may allow to distinguish them from the displaced SM background (for large
enough production rates / $\phi$ couplings).
The characteristic shapes might even allow to discriminate between the $1/f_Z$ and $1/f_H$ couplings of the CP-even scalar boson and the $1/\tilde f_Z$ coupling 
of the CP-odd field. These results motivate experimental direct and exploratory searches for light neutral scalar particles, using the LHC Run~1 and 
Run~2 data collected at the $8$~TeV and upgraded $13$~TeV center- of-mass energies.

Furthermore, we have provided generic exclusion limits from LEP and LHC Run~1 CMS searches on the (Z associated and VBF) production rates 
(including $B_{\phi \to \gamma\gamma}$) for heavier scalar 
masses (up to $110$~GeV) and translated these limits into constraints on the scalar parameter space, more precisely on combinations of the $f$-scales with 
the free diphoton branching fraction. In particular, the specific radion field, arising from a standard warped model with bulk matter, has been studied quantitatively to 
show that the present collider bounds do not exclude its presence. Those results encourage the experimental LHC collaborations to extend present analyses to the 
(present and future) LHC Run~2 data to determine in particular whether the LHC sensitivity can reach that of LEP for light scalar bosons in the mass range of $80-110$~GeV.

\section*{ACKNOWLEDGEMENTS}
The work of A.A. and G.M. is supported by the ERC advanced grant ``Higgs@LHC''. G.M. would like to thank as well the support from the CNRS LIA 
(Laboratoire International Associ\'e), THEP (Theoretical High Energy Physics) and the INFRE-HEPNET (IndoFrench Network on High Energy Physics) of CEFIPRA/IFCPAR 
(Indo-French Centre for the Promotion of Advanced Research). S.G.-S. and S.Z. would like to acknowledge the support of HiggsTools, an Initial Training Network (ITN) supported by the 7th Framework Programme of the European Commission.
All the authors acknowledge the organisers of the ``Les Houches'' Workshop for the whole organisation and the friendly atmosphere during the workshop.



\AddToContent{A.~Angelescu, S.~Fichet, L.~Finco, S.~Gascon-Shotkin, G.~Moreau, S.~Zhang}
\renewcommand{\thesection}{\arabic{section}}


\renewcommand{\lsim}{\raisebox{-0.13cm}{~\shortstack{$<$ \\[-0.07cm] $\sim$}}~}
\renewcommand{\gsim}{\raisebox{-0.13cm}{~\shortstack{$>$ \\[-0.07cm] $\sim$}}~}
\newcommand{\beq}{\begin{eqnarray}}
\newcommand{\eeq}{\end{eqnarray}}

\chapter{MSSM Higgs Boson Production via Gluon Fusion}

{\it M.~M\"uhlleitner, H.~Rzehak and M.~Spira}


\begin{abstract}
Higgs boson production via gluon fusion is discussed within the minimal
supersymmetric extension of the Standard Model with particular emphasis
on the genuine SUSY--QCD corrections at NLO. When comparing these
corrections with the usual $\Delta_b$ corrections that are absorbed in
effective bottom Yukawa couplings we are left with significant
remainders beyond this approximation.
\end{abstract}

\section{Introduction}
The Standard-Model (SM) Higgs boson production cross section via
gluon-fusion $gg\to H$ is known up to N$^3$LO in QCD
\cite{Djouadi:1991tka, Dawson:1990zj, Graudenz:1992pv, Spira:1995rr,
Harlander:2005rq, Anastasiou:2009kn, Catani:2001ic, Harlander:2001is,
Harlander:2002wh, Anastasiou:2002yz, Ravindran:2003um, Gehrmann:2011aa,
Anastasiou:2013srw, Anastasiou:2013mca, Kilgore:2013gba, Li:2014bfa,
Anastasiou:2014lda, Anastasiou:2015ema, Anastasiou:2015yha,
Anastasiou:2016cez, Mistlberger:2018etf} in the limit of heavy top
quarks and up to NLO QCD \cite{Graudenz:1992pv, Spira:1995rr,
Harlander:2005rq, Anastasiou:2009kn} and NLO electroweak
\cite{Djouadi:1994ge, Chetyrkin:1996wr, Chetyrkin:1996ke,
Aglietti:2004nj, Degrassi:2004mx, Aglietti:2006yd, Actis:2008ug,
Actis:2008ts, Anastasiou:2008tj} including finite quark mass effects
supplemented by soft and collinear gluon resummation up to the N$^3$LL
level \cite{Kramer:1996iq, Catani:2003zt, Moch:2005ky, Ravindran:2005vv,
Ravindran:2006cg, Idilbi:2005ni, Ahrens:2008nc, deFlorian:2009hc,
deFlorian:2012yg, deFlorian:2014vta, Bonvini:2014joa, Bonvini:2014tea,
Catani:2014uta, Schmidt:2015cea}. These results, however, can only
partly be applied to the minimal supersymmetric extension (MSSM), since
for large values of tg$\beta$ the bottom Yukawa couplings are enhanced
leading to the dominance of the bottom-loop contributions where bottom
mass effects are large. In addition scalar squark loops contribute to
the gluon-fusion cross sections of the light and heavy scalar MSSM Higgs
bosons. The QCD corrections to the squark contributions are known up to
NLO in QCD including the full squark mass dependence
\cite{Muhlleitner:2006wx, Anastasiou:2006hc, Aglietti:2006tp,
Bonciani:2007ex}. The pure QCD corrections are large in general. The
full supersymmetric (SUSY--) QCD corrections have been calculated first
in the limit of heavy SUSY particles \cite{Harlander:2003bb,
Harlander:2003kf, Harlander:2004tp, Harlander:2005if, Degrassi:2008zj}
analytically and later involving the full mass dependences numerically
\cite{Anastasiou:2008rm, Muhlleitner:2010nm, Muhlleitner:2010zz}.
However, a rigorous analysis of the results within the MSSM is still
missing. This addresses in particular the well-known $\Delta_b$
approximation of the genuine SUSY--QCD and -electroweak corrections in
comparison to the full results in case of bottom-loop dominance as e.g.
for large values of tg$\beta$.

\section{Effective Bottom Yukawa couplings}
The dominant correction to the bottom Yukawa couplings originates from
the coupling of the 'wrong' doublet $\phi_2$ to the bottom quarks at
one-loop level and can be discussed in terms of the effective Lagrangian
\begin{eqnarray}
{\cal L}_{eff} & = & -\lambda_b \overline{b_R} \left[ \phi_1^0 +
\frac{\Delta_b}{\mbox{tg}\beta} \phi_2^{0*} \right] b_L + h.c. \nonumber
\\
& = & -m_b \bar b \left[1+i\gamma_5 \frac{G^0}{v}\right] b
-\frac{m_b/v}{1+\Delta_b} \bar b \left[ g_b^h \left(
1-\frac{\Delta_b}{\mbox{tg}\alpha~\mbox{tg}\beta}\right) h \right.
\nonumber \\
& & \hspace*{2cm} \left. + g_b^H \left( 1+\Delta_b
\frac{\mbox{tg}\alpha}{\mbox{tg}\beta}\right) H
- g_b^A \left(1-\frac{\Delta_b}{\mbox{tg}^2\beta} \right) i \gamma_5 A
  \right] b
\label{eq:leffdeltab}
\end{eqnarray}
in the low-energy limit where the relations between current ($\phi_1^0,
\phi_2^0$) and mass ($h,H,A,G^0$) eigenstates of the neutral Higgs
components
\begin{eqnarray}
\phi_1^0 & = & \frac{1}{\sqrt{2}}\left[ v_1 + H\cos\alpha - h\sin\alpha
+ iA\sin\beta - iG^0\cos\beta \right] \nonumber \\
\phi_2^0 & = & \frac{1}{\sqrt{2}}\left[ v_2 + H\sin\alpha + h\cos\alpha
+ iA\cos\beta + iG^0\sin\beta \right]
\end{eqnarray} 
have been used with $G^0$ being the neutral would-be Goldstone
component. Here, $\alpha$ and $\beta$ denote the mixing angles in the
CP--even and CP-odd scalar sectors, respectively. The MSSM modifications
of the SM bottom-quark Yukawa coupling can be expressed in terms of the
mixing angles $\alpha,\beta$ as
\begin{equation}
g_b^h = -\frac{\sin\alpha}{\cos\beta}, \quad g_b^H =
\frac{\cos\alpha}{\cos\beta}, \quad g_b^A = \mbox{tg}\beta
\end{equation}
The indices $L,R$ denote the chiralities of the bottom
states, $\lambda_b$ the bottom Yukawa coupling of the MSSM Lagrangian
and $v\approx 246$ GeV the SM vacuum expectation value. The leading
NLO contributions to the correction $\Delta_b$ are given by
\cite{Hall:1993gn, Hempfling:1993kv, Carena:1994bv, Pierce:1996zz,
Guasch:2001wv, DAmbrosio:2002vsn, Buras:2002vd, Barger:2009me,
Christensen:2012ei, Carena:1999py, Guasch:2003cv}
\begin{eqnarray}
\Delta_b & = & \Delta_b^{QCD} + \Delta_b^{elw,t} \nonumber \\
\Delta_b^{QCD} & = & \frac{C_F}{2}~\frac{\alpha_s}{\pi}~m_{\tilde
g}~\mu~\mbox{tg}\beta~ I(m^2_{\tilde{b}_1},m^2_{\tilde{b}_2},m^2_{\tilde
g}) \nonumber \\
\Delta_b^{elw,t} & = &
\frac{\lambda_t^2}{(4\pi)^2}~A_t~\mu~\mbox{tg}\beta~
I(m_{\tilde{t}_1}^2,m_{\tilde{t}_2}^2,\mu^2)
\label{eq:deltab}
\end{eqnarray}
where $C_F=4/3$, $\alpha_s$ denotes the strong coupling constant and
$\lambda_t$ the top Yukawa coupling.  The masses $m_{{\tilde b}_{1,2}}$
and $m_{{\tilde t}_{1,2}}$ are the sbottom and stop masses and $\mu$ the
higgsino mass parameter.  The function $I$ is generically defined as
\begin{equation}
I(a,b,c) = \frac{\displaystyle ab\log\frac{a}{b} + bc\log\frac{b}{c}
+ ca\log\frac{c}{a}}{(a-b)(b-c)(a-c)}
\end{equation}
Two-loop QCD corrections to the leading $\Delta_b^{QCD}$ and
$\Delta_b^{elw,t}$ contributions have been calculated. They modify the
size by a moderate amount of about 10\% and reduce the scale dependence
considerably to the level of a few per-cent \cite{Noth:2008tw,
Noth:2010jy, Ghezzi:2017enb, Mihaila:2010mp, Crivellin:2012zz,
Mihaila:2016lfc} and thus yield a reliable prediction of the effective
bottom Yukawa couplings.

\section{Gluon Fusion}
At leading order the gluon fusion processes $gg\to h/H$ are mediated
by heavy quark and squark triangle loops, {\it cf.}
Fig.\ref{ggh_label3}, the latter contributing significantly for
squark masses $\lsim 400$~GeV. The LO cross section in the
narrow-width approximation can be obtained from the $h/H$ gluonic decay
widths, \cite{Spira:1997dg,Djouadi:2005gi,Djouadi:2005gj,Georgi:1977gs}
\begin{eqnarray}
\sigma_{LO}(pp\to h/H) & = & \sigma^{h/H}_0 \tau_{h/H}\frac{d{\cal
L}^{gg}}{d\tau_{h/H}} \\
\sigma^{h/H}_0 & = & \frac{\pi^2}{8M_{h/H}^3}\Gamma_{LO}(h/H\to gg)
\nonumber \\
\sigma^{h/H}_0 & = & \frac{G_{F}\alpha_{s}^{2}(\mu_R)}{288 \sqrt{2}\pi}
\ \left| \sum_{Q=t,b} g_Q^{h/H} A_Q^{h/H} (\tau_{Q}) +
\sum_{\widetilde{Q}=\widetilde{t}_{1,2},\widetilde{b}_{1,2}}
g_{\widetilde{Q}}^{h/H} A_{\widetilde{Q}}^{h/H} (\tau_{\widetilde{Q}})
\right|^{2}
\end{eqnarray}
\begin{figure}[t]
\begin{center}
\includegraphics[width=0.8\textwidth]{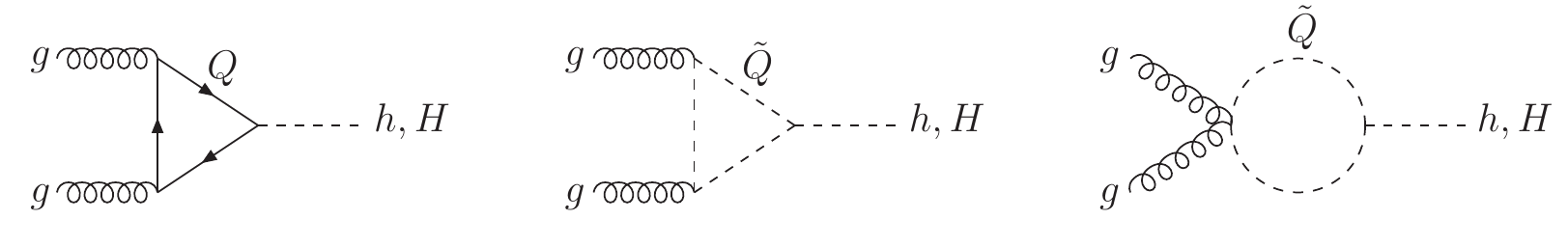}
\caption{Diagrams contributing to $gg\to h,H$ at leading order.}
\label{ggh_label3}
\end{center}
\end{figure}
where $\tau_{h/H} = M_{h/H}^2/s$ with $s$ being the squared hadronic
c.m.\ energy and $\tau_{Q/\tilde{Q}}=4 m_{Q/\tilde{Q}}^2/M_{h/H}^2$.
The MSSM top couplings relative to the SM top-quark Yukawa coupling can
be expressed as
\begin{equation}
g_t^h = \frac{\cos\alpha}{\sin\beta}, \quad g_t^H =
\frac{\sin\alpha}{\sin\beta}, \quad g_b^A = \frac{1}{\mbox{tg}\beta}
\end{equation}
The Higgs couplings to stops and sbottoms are given by
\begin{eqnarray}
m_{\widetilde{Q}_1}^2 g_{\widetilde{Q}_1}^{h/H} & = &
g_{\widetilde{Q},LL}^{h/H} c_Q^2 + g_{\widetilde{Q},RR}^{h/H} s_Q^2 + 2
g_{\widetilde{Q},LR}^{h/H} s_Q c_Q \nonumber \\
m_{\widetilde{Q}_2}^2 g_{\widetilde{Q}_2}^{h/H} & = &
g_{\widetilde{Q},LL}^{h/H} s_Q^2 + g_{\widetilde{Q},RR}^{h/H} c_Q^2 - 2
g_{\widetilde{Q},LR}^{h/H} s_Q c_Q \nonumber \\
g_{\widetilde{Q},LL/RR}^{h} & = & m_Q^2 g_Q^h
\mp M_Z^2 \left( I_{3Q} - e_Q\sin^2\theta_W
\right) \sin (\alpha+\beta) \nonumber \\
g_{\widetilde{Q},LL/RR}^{H} & = & m_Q^2 g_Q^H
\pm M_Z^2 \left( I_{3Q} - e_Q\sin^2\theta_W
\right) \cos (\alpha+\beta) \nonumber \\
g_{\widetilde{t},LR}^{h/H} & = & -\frac{m_t}{2} (\mu g_b^{h/H}
- A_t g_t^{h/H}) \nonumber \\
g_{\widetilde{b},LR}^{h/H} & = & -\frac{m_b}{2} (\mu g_t^{h/H}
- A_b g_b^{h/H})
\end{eqnarray}
The variables $s/c_{t,b} = \sin/\cos \theta_{t,b}$ are related to the
stop/sbottom mixing angles $\theta_{t,b}$, $\theta_W$ is the Weinberg
angle, $I_{3Q}$ denotes the third component of the left-handed isospin
and $e_Q$ the electric charge of the corresponding quark state, while
$A_Q$ are the soft-SUSY-breaking trilinear coupling parameters. The LO
form factors are given by
\beq
A_Q^{h/H}(\tau) & = & \frac{3}{2} \tau [1+(1-\tau)f(\tau)] \nonumber \\
A_{\tilde Q}^{h/H} (\tau) & = & -\frac{3}{4} \tau[1-\tau
f(\tau)] \\
f(\tau) & = & \left\{ \begin{array}{ll}
\displaystyle \arcsin^2 \frac{1}{\sqrt{\tau}} & \tau \ge 1 \\
\displaystyle - \frac{1}{4} \left[ \log \frac{1+\sqrt{1-\tau}}
{1-\sqrt{1-\tau}} - i\pi \right]^2 & \tau < 1
\end{array} \right. \nonumber
\eeq
And the gluon luminosity at the factorization scale $\mu_F$ is defined
as
\begin{displaymath}
\frac{d{\cal L}^{gg}}{d\tau} = \int_\tau^1 \frac{dx}{x}~g(x,\mu_F^2)
g(\tau /x,\mu_F^2)
\end{displaymath}
where $g(x,\mu_F^2)$ denotes the gluon parton density of the proton.
The NLO SUSY-QCD corrections consist of the virtual two-loop
corrections, {\it cf.} Fig.\ref{ggh_label4}, and the real
corrections due to the radiation processes $gg\to gh/H, gq\to qh/H$
and $q\bar{q}\to gh/H$, {\it cf.} Fig.\ref{ggh_label6}.
\begin{figure}[hbtp]
\begin{center}
 \includegraphics[width=0.9\textwidth]{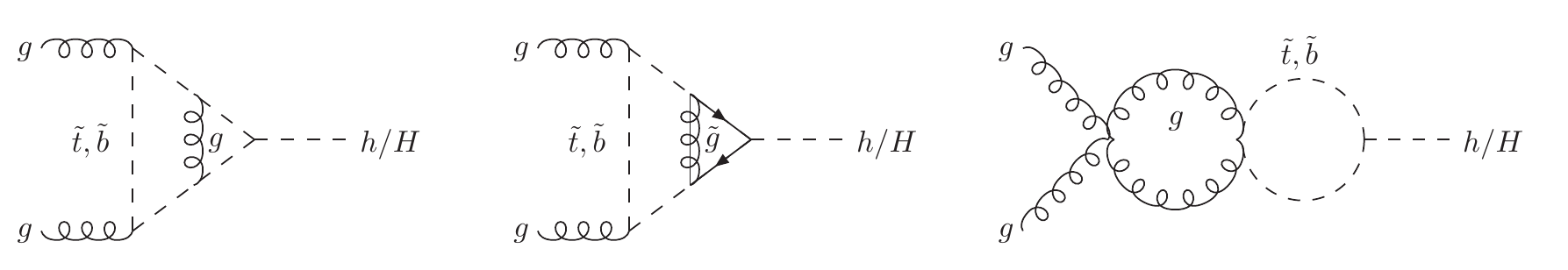}
\caption{Some generic diagrams for the virtual NLO SUSY-QCD corrections
to Higgs boson production via gluon fusion.}
\label{ggh_label4}
\end{center}
\end{figure}
The final result for the total hadronic cross sections can be split
accordingly into five parts,
\beq
\sigma(pp \rightarrow h/H+X) = \sigma^{h/H}_{0} \left[ 1+ C^{h/H}
\frac{\alpha_{s}}{\pi} \right] \tau_{h/H} \frac{d{\cal
L}^{gg}}{d\tau_{h/H}} + \Delta \sigma^{h/H}_{gg} + \Delta
\sigma^{h/H}_{gq} + \Delta \sigma^{h/H}_{q\bar{q}}
\label{ggh_label5}
\eeq
where the second term corresponds to the virtual and the last three
terms to the real corrections.  The strong coupling constant is defined
in the $\overline{\rm MS}$ scheme, with the top quark, gluino and squark
contributions decoupled from the scale dependence. The quark and squark
masses are renormalized on-shell. The parton densities are factorized in
the $\overline{\rm MS}$ scheme with five active flavors, i.e. the top
quark, the gluino and the squarks are not included in the factorization
scale dependence. After renormalization we are left with collinear
divergences in the sum of the virtual and real corrections which are
absorbed in the renormalization of the parton density functions, so that
the result Eq.~(\ref{ggh_label5}) is finite and depends on the
renormalization and factorization scales $\mu_R$ and $\mu_F$,
respectively. The  natural scale choices turn out to be $\mu_R=\mu_F\sim
M_{h/H}$.
\begin{figure}[hbtp]
\begin{center}
 \includegraphics[width=0.8\textwidth]{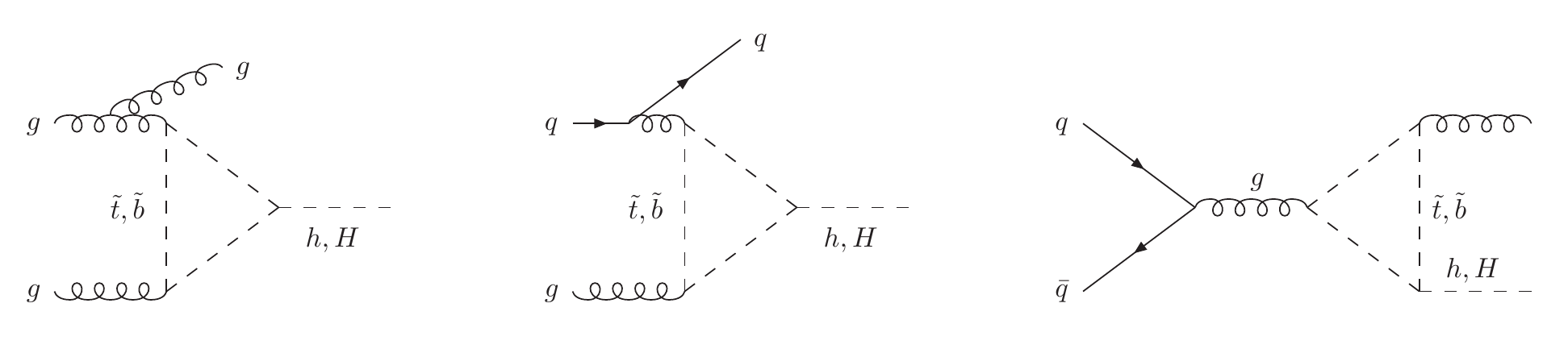}
\caption{Typical diagrams for the real NLO QCD corrections to the squark
contributions to the gluon fusion processes.}
\label{ggh_label6}
\end{center}
\end{figure}

The pure QCD corrections described by the first (generic) diagram of
Fig.~\ref{ggh_label4} for the virtual corrections and the diagrams of
Fig.~\ref{ggh_label6} for the real corrections are known to be large,
i.e.~of similar size as those to the top quark loops
\cite{Muhlleitner:2006wx, Anastasiou:2006hc, Aglietti:2006tp,
Bonciani:2007ex}. The genuine SUSY--QCD corrections are described by the
generic second and third diagrams of Fig.~\ref{ggh_label4} and
contribute only to the virtual corrections at NLO. The renormalization
of the squark sector has been performed along the lines of
Ref.~\cite{Accomando:2011jy} in order to absorb potentially large
contributions of the sbottom sector in the renormalized parameters. In
addition we have absorbed the NLO $\Delta_b$ terms originating from the
effective Lagrangian of Eq.~(\ref{eq:leffdeltab}) in the corresponding
bottom Yukawa couplings. This removes the dominant genuine SUSY--QCD
corrections to the bottom loop contributions and thus also provides a
test of the $\Delta_b$-approximation in terms of the residual
corrections as the remainder.

For the numerical results we have adopted the $\tau$-phobic
scenario \cite{Carena:2013ytb}, defined by the
following choices of MSSM parameters [$m_t=173.2$~GeV],
\beq
\mbox{tg}\beta = 30,\quad M_{\tilde Q} = 1.5~{\rm
TeV},\quad M_{\tilde \ell_3} = 500~{\rm GeV},\quad M_{\tilde{g}} =
1.5~{\rm TeV}, \nonumber \\
M_2 = 200~{\rm GeV}, \quad A_b = A_t = 4.417~{\rm TeV}, \quad A_\tau
= 0, \quad \mu = 2~{\rm TeV}
\eeq
In this scenario the squark masses amount to
\beq
\begin{array}{llllll}
m_{\tilde{t}_1} &=& 1.347\;\mbox{TeV}, & \qquad m_{\tilde{t}_2} &=&
1.739\;\mbox{TeV}, \\
m_{\tilde{b}_1} &=& 1.521\;\mbox{TeV}, & \qquad m_{\tilde{b}_2} &=&
1.583\;\mbox{TeV}
\end{array}
\eeq
Fig. \ref{ggh_label7} displays the genuine SUSY-QCD corrections for the
heavy scalar Higgs boson
\begin{figure}[htp]
\begin{center}
 \includegraphics[width=0.8\textwidth]{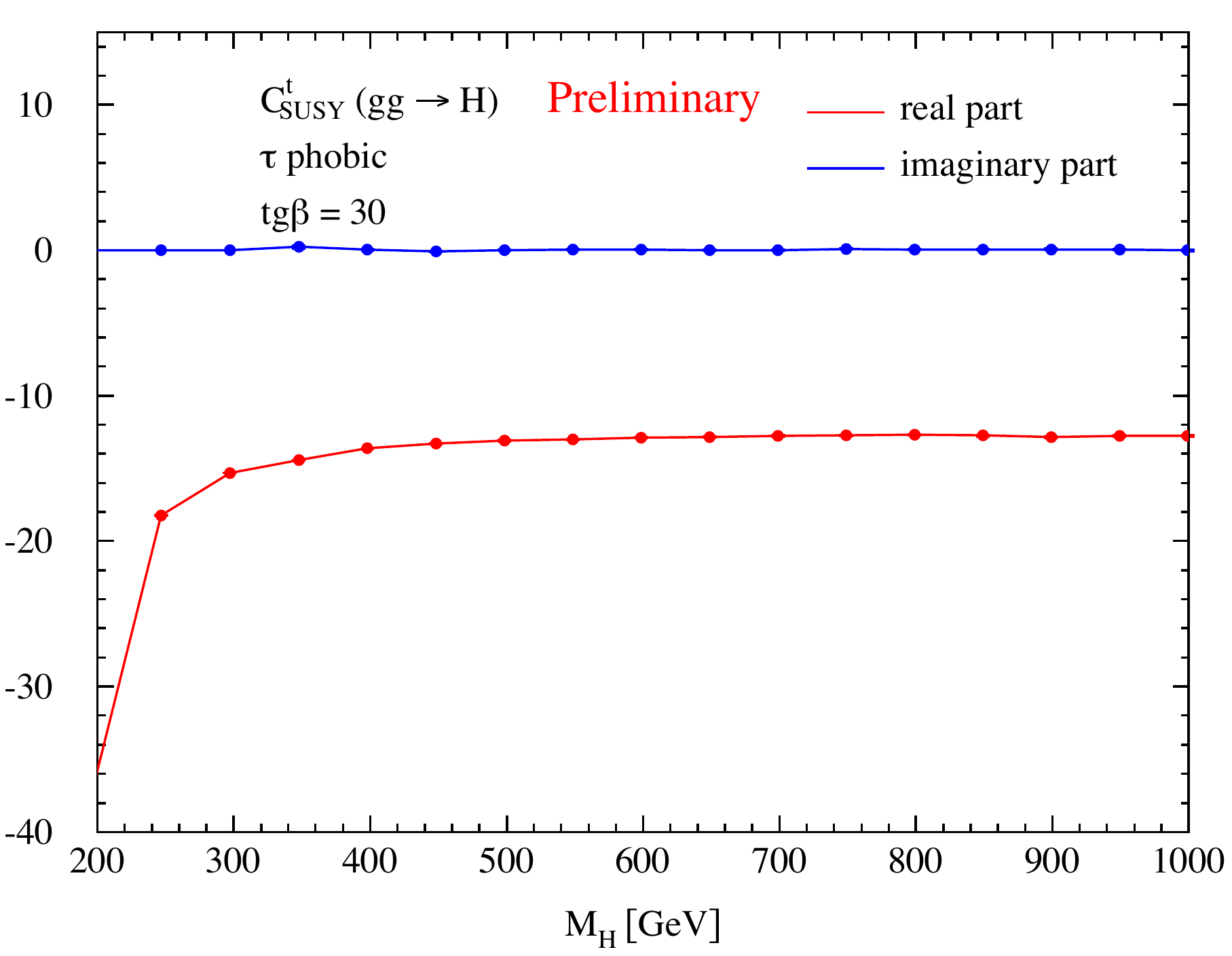}\vspace{0.5cm}
  \includegraphics[width=0.8\textwidth]{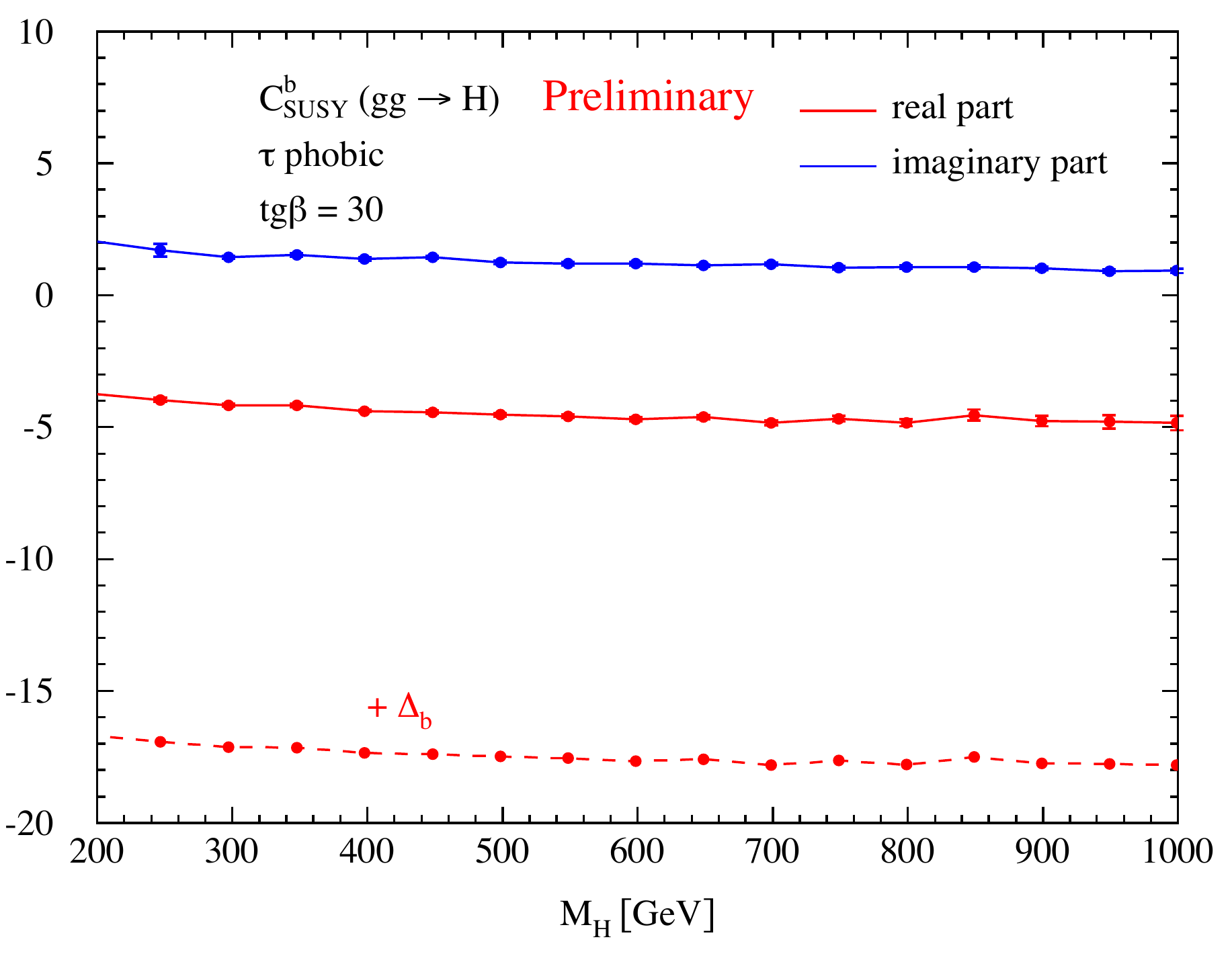}
\caption{The genuine SUSY-QCD corrections to the top (upper) and bottom
(lower) quark form factors normalized to the individual LO expressions.
Real parts: red, imaginary parts: blue, compared to the corrections to
the bottom quark form factor including the fixed-order $\Delta_b$
contribution, i.e.~without absorbing it in the bottom Yukawa coupling
(red dashed line).}
\label{ggh_label7}
\end{center}
\end{figure}
normalized to the  LO quark form factors, {\it i.e.} $A_Q^{h/H} (\tau_Q)
\to A_Q^{h/H} (\tau_Q) \left[1+C^{Q,h/H}_{SUSY}
\frac{\alpha_s}{\pi}\right]~(Q=t,b)$. The corrections can be sizeable.
For the top contributions they amount to about $-15 \frac{\alpha_s}{\pi}
\sim -50\%$. However, the top contribution is strongly suppressed for
these large values of $\mbox{tg}\beta$. The residual corrections to the
bottom contributions range at the level of about $-5
\frac{\alpha_s}{\pi} \sim -15\%$\footnote{It should be noted that these
corrections correspond to the {\it amplitudes} that have to be squared
in order to arrive at the corrections to the cross sections.}. In
addition Fig.~\ref{ggh_label7} shows the corrections to the bottom
contribution without the absorption of the $\Delta_b$ terms into the
bottom Yukawa coupling (red dashed line). In this case the corrections
are much larger thus signalizing that the dominant part of the genuine
SUSY--QCD corrections is indeed originating from these $\Delta_b$ terms.
However, the remainder of the corrections after absorbing these
$\Delta_b$ terms is sizeable and should be taken into account in
reliable analyses.

\section{Conclusions}
We have presented results for the NLO SUSY-QCD corrections to
gluon fusion into CP-even MSSM Higgs bosons, including the full mass
dependence of the loop particles. The genuine SUSY-QCD corrections can
be sizeable and deviate from the results in terms of the effective
bottom Yukawa couplings containing the $\Delta_b$ terms by a significant
amount, i.e.~the remainder should be take into account.




\AddToContent{M.~M\"uhlleitner, H.~Rzehak and M.~Spira}
\renewcommand{\thesection}{\arabic{section}}



\chapter{Extending LHC resonant di-Higgs searches to discover new scalars: \\ $H_1 \to H_2 H_2 \to b \bar{b} b \bar{b}$}

{\it D.~Barducci, K.~Mimasu, J.~M.~No, C.~Vernieri, J.~Zurita}

\begin{abstract}
We extend the coverage of 13 TeV LHC resonant di-Higgs searches in the $b \bar{b} b \bar{b}$ final 
state to the process $H_1 \to H_2 H_2 \to b \bar{b} b \bar{b}$, with both $H_1$ and $H_2$ states beyond the Standard Model. 
The decay $H_1 \to H_2 H_2$ could constitute a joint discovery mode for $H_1$ and $H_2$ within non-minimal Higgs sectors such as two-Higgs-doublet models
or the NMSSM. We present here the first sensitivity study of this channel, using public CMS resonant di-Higgs production data to validate our analysis.  
\end{abstract}

\section{Introduction}

Searches for heavy scalars in non-standard decay channels are needed to fully cover the parameter space of theories beyond the Standard Model (BSM). 
A prime example are resonant di-Higgs searches $p p \to H \to h h$, where the scalar 
$H$ is resonantly produced and decays into a pair of 125 GeV Higgs bosons ($h$).
ATLAS and CMS look for these processes in various final states, 
including $b\bar{b} b\bar{b}$~\cite{Khachatryan:2015yea,Aad:2015uka,Aaboud:2016xco,CMS:2017xxp,CMS:2016tlj}, 
$b\bar{b} W^+ W^-$~\cite{CMS:2016rec,Sirunyan:2017guj,Aad:2015xja}, $b \bar{b} \tau^+ \tau^-$~\cite{CMS:2016knm,Sirunyan:2017djm} 
and $b\bar{b} \gamma \gamma$~\cite{Aad:2014yja,Khachatryan:2016sey,CMS:2016vpz}. 
Searching for such additional Higgs bosons decaying to the SM Higgs at the LHC is a key avenue to probe non-minimal scalar sectors. 
For Higgs sectors with several new states beyond the SM (BSM), such as two-Higgs-doublet models (2HDM) or the 
next-to-minimal supersymmetric Standard Model (NMSSM), 
``{\sl Higgs to Higgs}" decays can occur between several BSM scalars in the presence of sizable mass splittings. 
In this case, such {\sl Higgs to Higgs} decays are potential discovery channels for both 
the decaying particle and its decay products\footnote{This is e.g. the case for $A \to Z H \,/\, H \to Z A$ decays in 2HDM 
scenarios~\cite{Coleppa:2014hxa,Dorsch:2014qja} (see also~\cite{Haber:2015pua}), 
which have been searched for by CMS at 8 TeV~\cite{Khachatryan:2016are} and 13 TeV~\cite{CMS:2016qxc}.}. 
Generalising searches for additional Higgs bosons to probe new scalar decay modes beyond the 
SM Higgs is therefore of great importance for the LHC Higgs physics programme. 

We present here a sensitivity study of the 
channel $H_1 \to H_2 H_2 \to b \bar{b} b \bar{b}$ as a probe for BSM states $H_1$ and $H_2$, where $H_1$ is produced via gluon fusion at the LHC. For $m_{H_1} \gg m_{H_2}$, $H_1 \to H_2 H_2$ can be the dominant decay mode of $H_1$, and moreover 
$H_2 \to b\bar{b}$ generally yields the largest branching fraction (for $m_{H_2} < 2\, m_t$) of $H_2$, making it 
challenging to discover $H_2$ through direct production. A concrete example of such a scenario would be a 2HDM, where $H_1 \equiv H_0$, $H_2 \equiv A$ and $m_{H_0} \gg m_{A}$. The large {\sl Higgs to Higgs} branching fractions induced by significant mass splittings combined with the typically dominant $b\bar{b}$ branching fractions of light scalars make the channel considered here a potentially important, unexplored probe of non-minimal Higgs sectors. 

While no ATLAS/CMS analysis of the $ p p \to H_1 \to H_2 H_2 \to b \bar{b} b \bar{b}$ signature (with both $H_1$ and $H_2$ BSM states) 
exists at present\footnote{We however note there are 
existing LHC analyses for $h \to A A$, with $h$ the 125 GeV Higgs boson (see e.g.~\cite{Khachatryan:2017mnf}).}, we can use its similarity to
resonant di-Higgs searches in the $b \bar{b} b \bar{b}$ final state to validate our analysis for $m_{H_2} = 125$ GeV, 
before extending it to the two-dimensional mass plane ($m_{H_1},\,m_{H_2}$). 
We follow the recent LHC $\sqrt{s} = 13$ TeV CMS search for a narrow di-Higgs resonance in the $b \bar{b} b \bar{b}$ final state 
with $35.9$ fb$^{-1}$ of integrated luminosity~\cite{CMS:2017xxp}, reproducing the reported selection efficiencies with our simulations. 
We then perform an analysis of the expected signal efficiencies for the 
$ p p \to H_1 \to H_2 H_2 \to b \bar{b} b \bar{b}$ process in the mass plane ($m_{H_1},\,m_{H_2}$), and finally provide an estimate of the 
95\% C.L. exclusion sensitivity on the $H_1$ production cross section multiplied by the branching fractions to $H_2$ and $4b$ at 13 TeV LHC with $35.9$ fb$^{-1}$ from this search.

\section{CMS resonant di-Higgs searches for $\sqrt{s} = 13$ TeV in the $b \bar{b} b \bar{b}$ final state}

To validate our analysis, we use the 13 TeV CMS $p p \to X \to h h \to b \bar{b} b \bar{b}$ search~\cite{CMS:2017xxp} (see also~\cite{CMS:2016tlj}).
This considers $X$ to be both a spin-$0$ (``radion") and spin-$2$ (``KK graviton") state, 
and it provides the signal efficiencies at various stages (ranging from the initial event selection to the definition of the final 
signal region) for the spin-$2$ scenario\footnote{The signal efficiencies for spin-$0$ and spin-$2$ are nevertheless found 
to be very similar~\cite{CMS:2017xxp}. We will provide a detailed analysis validation for both spin-$0$ and spin-$2$ 
efficiencies elsewhere.}. 
The search defines two kinematic regions which feature different event selection criteria: 
a low-mass-region (LMR) for masses $m_X \in [250,\, 620]$ GeV, and a medium-mass-region (MMR) for masses $m_X \in [550,\, 1200]$ GeV.
The transition region $m_X \sim 580$ GeV is determined by the respective sensitivities of the LMR and MMR selection strategies~\cite{CMS:2017xxp}.

In this work we concentrate on the MMR selection, for reasons that we explain below. 
As an online trigger selection, Reference ~\cite{CMS:2017xxp} requires either of two conditions: 
\begin{itemize}
    \item[{\it i)}] 4 reconstructed jets of $p_T > 30$ GeV, 
of which two satisfy $p_T > 90$ GeV, and three are $b$-tagged.
    \item[{\it ii)}] 4 reconstructed jets of $p_T > 45$ GeV, of which three are 
$b$-tagged. 
\end{itemize}
The analysis subsequently requires all four selected jets to be $b$-tagged and have $\left|\eta \right| < 2.4$. 
This initial selection stage, labelled {\sl 4$b$}, is common to both LMR and MMR selections.
For the MMR selection, the analysis then identifies two 125 GeV Higgs boson 
candidates ({\sl $HH$ candidate}) by requiring two $b$-jet pairs with $\Delta R_{bb} < 1.5$ for each 
pair\footnote{In case of multiple {\sl $HH$ candidates} in an event, the combination that minimizes $\chi$ as defined in~\eqref{chi_HH} is chosen.}. 
We note that $\Delta R_{bb}$ depends only on the mass ratio $m_X/m_h$~\cite{Gouzevitch:2013qca}, and as such 
the ratio of signal efficiencies at {\sl 4$b$} and {\sl $HH$ candidate} stage can in principle be directly extrapolated to a 
two-dimensional mass plane (this partially justifies our choice of MMR selection). 
Finally, the signal region ({\sl SR}) is defined in the two dimensional space of the reconstructed masses of the Higgs boson
candidates, $m_{h_1}$ and $m_{h_2}$, as the circular region with $\chi < 1$, where $\chi$ is defined as
\begin{equation}
\label{chi_HH}
\chi = \sqrt{\left(\frac{m_{h_1} - C}{R} \right)^2 + \left(\frac{m_{h_2} - C}{R} \right)^2}.
\end{equation}
The values of these parameters are set to $(C,\, R) = (115,\,23)$ GeV in~\cite{CMS:2016tlj} while~\cite{CMS:2017xxp} chooses $(C,\, R) = (125,\,20)$ GeV for the MMR category. 
However, Reference~\cite{CMS:2017xxp} applies a multivariate regression technique to improve the $b\bar{b}$ invariant 
mass resolution\footnote{Incidentally, this is also a reason 
why we focus on the MMR, since for LMR this technique is also applied at the {\sl $HH$ candidate} level.}, which we cannot mimic in the present analysis). We therefore keep the values used in the older analysis which are found to better capture the centre of the Higgs peak post- parton shower and detector simulation.

In order to reproduce the CMS signal efficiencies in~\cite{CMS:2017xxp}, we use the Randall-Sundrum 
model available in {\sc Madgraph5$\_$aMC@NLO}~\cite{Alwall:2014hca} to 
generate $p p \to X \to h h \to b \bar{b} b \bar{b}$ (with $X$ a spin-$2$ state) event samples merged with up to two additional jets using the MLM procedure~\cite{Mangano:2006rw}, with {\sc xqcut}=30 GeV. We simulated $m_X$  
in the range $400$~GeV -- $1200$~GeV with a fixed width of $10$~GeV to match the MMR selection efficiencies provided by~\cite{CMS:2017xxp}.
We then shower/hadronise our events with {\sc Pythia 8.2}~\cite{Sjostrand:2014zea}, matched using the shower-$k_T$ scheme and use {\sc Delphes}~\cite{deFavereau:2013fsa} 
for a simulation of the CMS detector performance which also makes use of {\sc Fastjet}~\cite{Cacciari:2011ma} to cluster anti-$k_T$ jets with radius $0.4$. 
A crucial ingredient in this last step concerns the 13 TeV CMS $b$-tagging efficiencies
(as well as the $c$-jet and light-jet mistag rates) as a function of the jet $p_T$ and $\eta$. We model these using the information from~\cite{Sirunyan:2017ezt},
assuming the performance of the {\it DeepCSV} $b$-tagging algorithm for the same operating point as used in~\cite{CMS:2017xxp}.


%
%

Our simulated signal efficiencies at the 
{\sl $HH$ candidate} and {\sl SR} stages 
are shown in Figure~\ref{Fig_HAA_effs}, together with the corresponding 
CMS efficiencies from~\cite{CMS:2017xxp} for comparison. We find good agreement, with a moderate mismatch at the {\sl $HH$} selection level within the expected accuracy of a fast detector simulation. 
The agreement is particularly good for the 
{\sl SR} selection with $m_X > 600$~GeV, precisely where the MMR category 
is expected to yield the strongest sensitivity. This 
allows for a fairly robust extension of the CMS signal efficiencies to the mass plane ($m_{H_1}$, $m_{H_2}$), which we do in the next section.

\begin{figure}[h!]
\begin{center}
\includegraphics[width=0.7\textwidth]{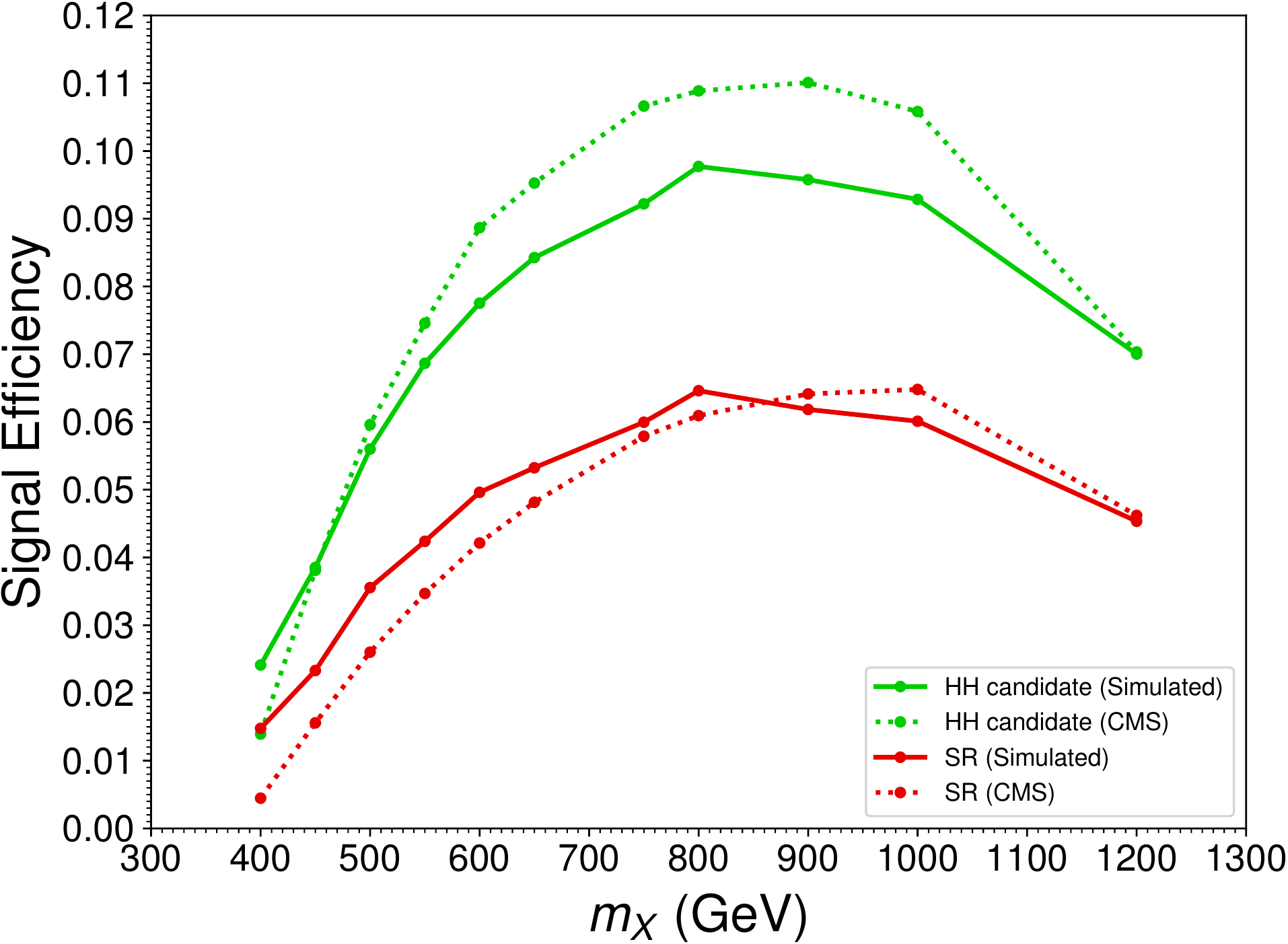}
\caption{\small Signal efficiencies as a function of $m_X$ 
at {\sl $HH$} (green) and {\sl $SR$} (red) selection stages (as defined in the text) for spin-2 $X$ resonant di-Higgs production in 
the $b \bar{b} b \bar{b}$ final state. Solid lines correspond to our simulation, while dotted lines correspond to the efficiencies from the 
13 TeV CMS analysis~\cite{CMS:2017xxp}.}
\label{Fig_HAA_effs}
\end{center}

\vspace{-2mm}

\end{figure}

\section{Extending CMS resonant di-Higgs searches to search for new scalars: $H_1 \to H_2 H_2 \to b \bar{b} b \bar{b}$}

Once we have validated the CMS analysis, we can proceed to extend the search to the two dimensional mass plane: 
we implement the (Type I) 2HDM model in {\sc FeynRules}~\cite{Alloul:2013bka} and generate $p p \to H_1 \to H_2 H_2 \to b \bar{b} b \bar{b}$ event samples 
with {\sc Madgraph5$\_$aMC@NLO} (we follow the procedure discussed in the previous section with the only difference being the generation of samples merged with up to one additional jet to reduce computing time)
in a ($m_{H_1}$, $m_{H_2}$) mass grid, in order to obtain the signal efficiencies at the {\sl $SR$} selection stage
in the ($m_{H_1}$, $m_{H_2}$) plane. The widths of the two scalar states are kept at $1$~GeV to ensure consistency with the narrow width hypothesis of the search.

Our mass grid scan is defined as follows: we vary $m_{H_2}$ in the range $m_{H_2} \in [65,\, 290]$~GeV in steps of $15$~GeV; 
For each $m_{H_2}$, the minimum value of $m_{H_1}$ we consider is given by
$m^{\mathrm{min}}_{H_1} = m_{H_2} \times 400/125 = 3.2\times m_{H_2}$, since the minimum mass ratio 
in the CMS MMR region is $m_X/m_h = 400/125$, and we choose not to extrapolate the CMS efficiencies from~\cite{CMS:2017xxp} outside of regions that we did not explicitly validate for the $125$~GeV case.
Similarly, the maximum value of $m_{H_1}$ we consider is given by $m_{H_2} \times 1200/125$. Since we only perform our search up to 
$m_{H_1} = 1200$~GeV, we define $m^{\mathrm{max}}_{H_1} = \mathrm{Min}\left(1200\,\mathrm{GeV},\, m_{H_2} \times 1200/125\right)$, 
and consider values of $m_{H_1}$ in the 
range $m_{H_1} \in [m^{\mathrm{min}}_{H_1}, m^{\mathrm{max}}_{H_1}]$. In principle, given that we have validated the analysis over a fairly broad mass range, the possibility of extrapolating our analysis outside of the validation region is not unreasonable, however we opt to remain somewhat conservative and postpone this to a future investigation.

Next, we implement the same MMR category event selection described in the previous section for the case of $p p \to X \to h h \to b \bar{b} b \bar{b}$, up to the 
{\sl $HH$} selection stage. In order to perform the {\sl $SR$} selection, we use Equation~\eqref{chi_HH} 
with a varying $C = m_{H_2} - 10$ GeV to account for the shift in the signal di-jet invariant mass peak as $m_{H_2}$ changes 
(since now the invariant mass distributions $m_{h_1}$ and $m_{h_2}$ from Equation~\eqref{chi_HH} are observed to peak around $\sim m_{H_2} - 10$~GeV).
The resulting signal efficiency map for the {\sl $SR$} selection stage in the ($m_{H_1}$, $m_{H_2}$) plane is shown in Figure~\ref{Fig_HAA_2D_Effs}.
We note that, similarly to the case of the previous section, we expect an analysis based on the LMR selection 
from~\cite{CMS:2017xxp} to yield higher signal efficiencies than those shown in Figure~\ref{Fig_HAA_2D_Effs} for the lower mass ratios 
$m_{H_1} < m_{H_2} \times 580/125 \simeq 4.6 \times m_{H_2}$. 
At the same time, for $m_{H_1} > m_{H_2} \times 900/125 = 7.2\times m_{H_2} $ an analysis clustering each $H_2$-candidate into a fat-jet would also
be more sensitive~\cite{CMS:2017xxp} than the present one. Still, the present analysis provides a conservative estimate of the prospective 
sensitivity a search for $p p \to H_1 \to  H_2 H_2\to b \bar{b} b \bar{b}$ could yield.

\begin{figure}[h]
\begin{center}
\includegraphics[width=0.65\textwidth]{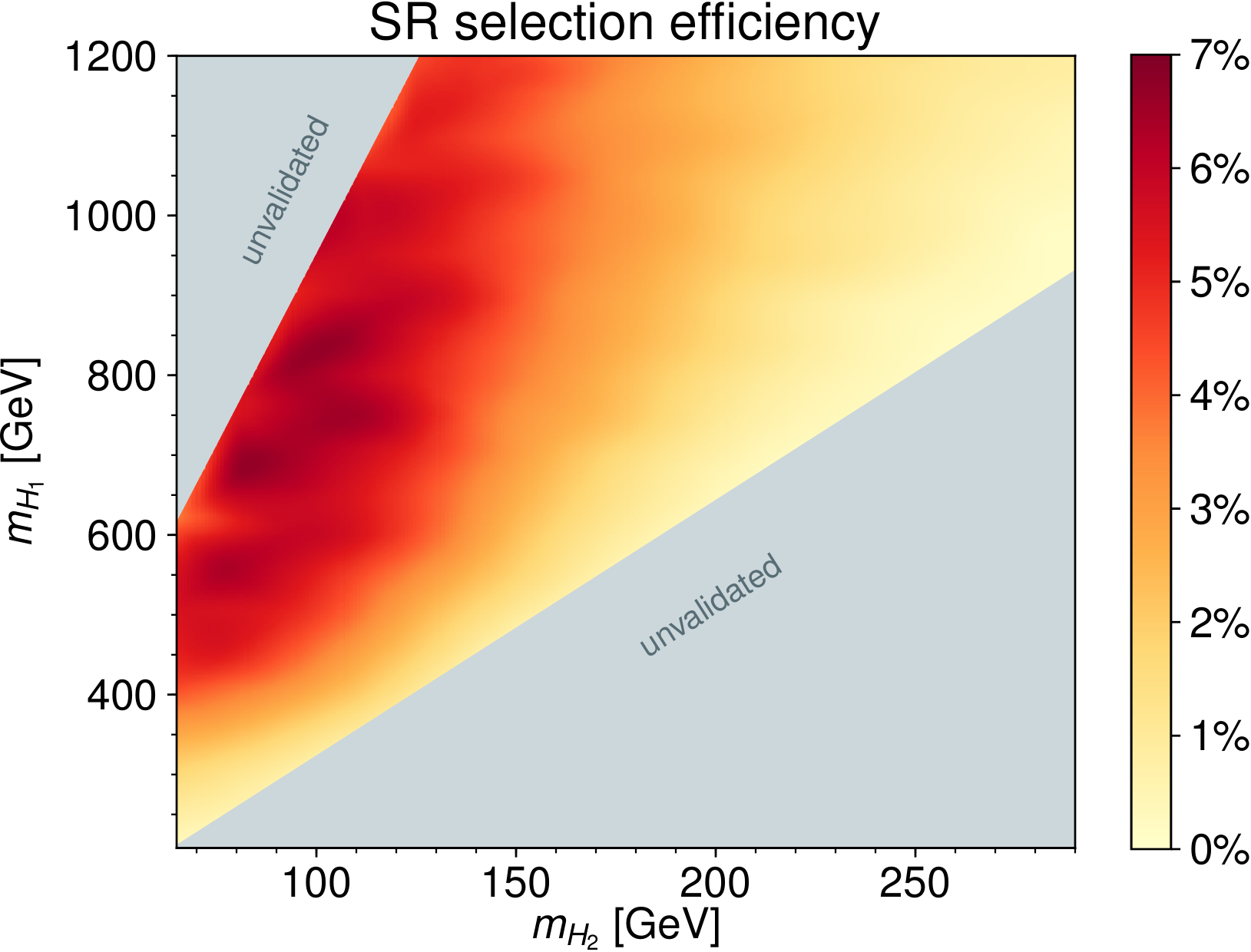}
\caption{\small Signal ($p p \to H_1 \to  H_2 H_2 \to b \bar{b} b \bar{b}$) efficiency at the {\sl $SR$} selection stage for the MMR selection category 
(see text for details), in the ($m_{H_1}$, $m_{H_2}$) mass plane.}
\label{Fig_HAA_2D_Effs}
\end{center}

\vspace{-2mm}

\end{figure}

\vspace{3mm}

The 13 TeV CMS $p p \to X \to h h \to b \bar{b} b \bar{b}$ analysis~\cite{CMS:2017xxp} also 
provides a data-driven estimate of event yields for the dominant SM multi-jet background after {\sl $HH$} 
selection as a function of the invariant mass of the reconstructed $b$-jet pairs, 
$m_{h_1}$ and $m_{h_2}$. This allows us to directly apply our modified 
{\sl $SR$} selection to the multi-jet background. 
Using the similar mass grid to the one discussed above for the signal (varying $m_{H_2}$ in the range $m_{H_2} \in [65,\, 290]$ GeV), 
we compute the expected number of SM background events after {\sl $SR$} selection as a function of $m_{H_2}$. 
With this information, we can estimate the number of signal events after {\sl $SR$} selection that would be excluded at 
$95\%$ C.L. as a function of $m_{H_2}$, based on the number 
of signal and background events after {\sl $SR$} selection and assuming that only the multi-jet yield is observed. We use here an approximate $S/\sqrt{S+B} = 2$ criterion, and leave a more 
precise assessment, including also background systematic uncertainties, for the future. 
We stress that our estimate does not make use of any information 
regarding $m_{H_1}$ for the signal, and as such could be significantly improved by a dedicated search.
Combining the corresponding number of excluded signal events with the signal efficiency map from Figure~\ref{Fig_HAA_2D_Effs}, 
we finally extract the $95\%$ C.L. cross section times branching ratio exclusion sensitivity (in fb) for $p p \to H_1 \to H_2 H_2 \to b \bar{b} b \bar{b}$
in the ($m_{H_1}$, $m_{H_2}$) plane, for LHC 13 TeV with an integrated luminosity of 35.9 fb$^{-1}$, shown in Figure~\ref{Fig_HAA_2D_Limits}.

While we leave model-dependent interpretations of the projection to a future study, it is interesting to see that production cross sections times branching fractions in the region of several tens of fb can be accessed in the optimal high $m_{H_1}$ region above 600 GeV in this albeit simplified approximation of the analysis at hand. The very large multijet background at the lower end of the invariant mass spectra has as significant degrading effect on the overall sensitivity which combines with the reduced selection efficiency in this region, as seen in Figure~\ref{Fig_HAA_2D_Effs}. One can compare the $\sim$pb sensitivity to the typical SM Higgs production cross section by gluon fusion of around 40 pb, suggesting that one may still retain sensitivity to realistic models of new scalars around this mass. In this regard, it will be important to re-examine the validity of the narrow width approximation employed throughout this analysis in future work given the fact that bosonic partial widths tend to grow relatively fast with the resonance mass. 

%
%
%
%
%

\begin{figure}[t]
\begin{center}
\includegraphics[width=0.65\textwidth]{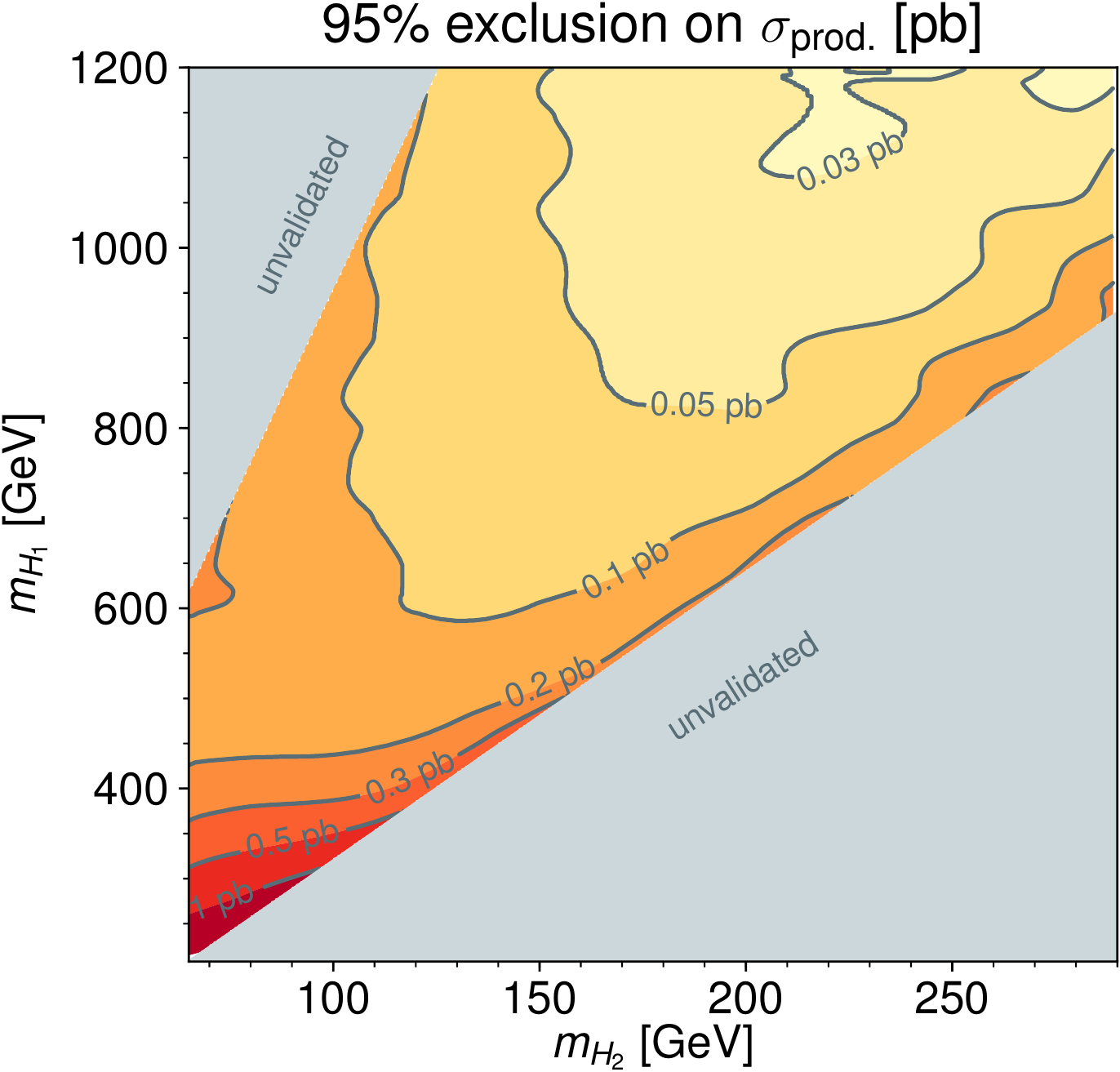}
\caption{\small Estimated $95\%$ C.L. $\sigma \times$ BR exclusion sensitivity for $p p \to H_1 \to H_2 H_2 \to b \bar{b} b \bar{b}$ 
in the ($m_{H_1}$, $m_{H_2}$) plane for LHC 13 TeV with an integrated luminosity of 35.9 fb$^{-1}$.}
\label{Fig_HAA_2D_Limits}
\end{center}

\vspace{-2mm}

\end{figure}

\section{Conclusions}

Searches for additional Higgs bosons at the LHC via new scalar decay modes are a key avenue to explore extensions of the SM Higgs sector.
We have presented here the first study of the $p p \to H_1 \to H_2 H_2 \to b \bar{b} b \bar{b}$
channel, with both $H_1$ and $H_2$ being BSM states.
A relatively precise estimate of the LHC sensitivity of such a search is possible given its similarity with 
CMS and ATLAS resonant di-Higgs searches, which we have benefited from to 
validate our analysis, specifically using for this purpose the latest $\sqrt{s} = 13$ TeV CMS resonant di-Higgs search 
in the $b \bar{b} b \bar{b}$ final state~\cite{CMS:2017xxp}.

Production cross sections time branching fractions for the heavy scalar state  ranging from the picobarn to tens of femtobarns appear accessible at the price of a simple generalisation of an existing LHC search. Further optimisation taking into account the heavy resonance mass is likely to yield appreciable improvements in this sensitivity.  Our study shows promising prospects for this yet unexplored probe of heavy Higgs bosons, 
and shows explicitly how extending the coverage of current LHC searches for new scalars
can yield new avenues to search for non-minimal Higgs sectors.

\section*{Acknowledgements}

K.M. is supported in part by the Belgian Federal Science Policy Office through the Interuniversity Attraction 
Pole P7/37 and by the European Union’s Horizon 2020 research and innovation programme
under the Marie Skłodowska-Curie grant agreement No. 707983.
J.M.N. is supported by the European Research Council under the European Union’s Horizon 2020 program (ERC Grant Agreement no.648680 DARKHORIZONS).
The work of C.V. is supported by Fermi Research Alliance, LLC under Contract 
No. DE-AC02-07CH11359 with the U.S. Department of Energy, Office of Science, Office of High Energy Physics.

\AddToContent{D.~Barducci, K.~Mimasu, J.~M.~No, C.~Vernieri, J.~Zurita}
\renewcommand{\thesection}{\arabic{section}}




\chapter{Interferences in searches for heavy Higgs bosons}

{\it A.~Carvalho, R.~Gr\"ober, S.~Liebler, J.~Quevillon}


\begin{abstract}
We study the relevance of interferences in the search for heavy Higgs bosons
in $hh$ and $Zh$ final states, where $h$ labels the SM-like Higgs boson at $125$\,GeV
and $Z$ is the $Z$ boson. We study generic scenarios in terms
of a few parameters, which we choose in accordance with simple extended Higgs sectors.
Interferences do not only enhance or weaken signal contributions, but through a peak-dip-like
structure can also shift the invariant mass distributions in terms of the final
state in both directions. For their classification we introduce three parameters.
We find the signal-over-background ratio very helpful
in discriminating the relevance of interference effects.
\end{abstract}

\section{INTRODUCTION}
\label{sec:intro;heavyhiggsinter} 

After the discovery of the Standard Model (SM)-like Higgs boson at $125$\,GeV
the quest for additional Higgs bosons,
part of extended Higgs sectors, is ongoing. 
Extensions of the SM Higgs sector can contain several new CP-even $H$
and/or CP-odd Higgs bosons $A$, or in case of CP violation also mixtures of them.
In the following we denote such additional Higgs bosons generically by $\phi$.
Searches for them are performed in
a variety of final states involving SM particles, among them also the
newly discovered SM-like Higgs boson $h$.
In such studies most experimental analysis use a narrow width approximation, which allows
to split production $gg\to \phi$ and decay $\phi\to F$, where $F$ denotes a
generic final state. In that case the cross section for $gg\to \phi\to F$ is obtained 
according to
$\sigma=\sigma(gg\to \phi)$BR$(\phi\to F)$, which allows to take
into account higher order corrections independently for the initial
state production of the particle $\phi$ and it subsequent decay into $F$.
The narrow-width approximation on the other hand misses the inclusion of
interference effects of Feynman diagrams involving the $s$-channel Higgs
boson~$\phi$ and SM background diagrams $gg\to F$, which can also involve
an $s$-channel SM Higgs boson $h$.
Such interferences do not only enhance or lower the signal
contribution $gg\to \phi \to F$, but also distort the peak structure
in the invariant mass distribution $m_F$ of the final state substantially.
Interference effects on the $\phi$ line-shape might be able to provide crucial information on both the real and imaginary
parts of the $gg\to \phi\to F$ amplitude, providing supplementary constraints on the properties of the new state $\phi$.

There is an extensive literature on interference effects on the corresponding signals of the SM-like $125$\,GeV Higgs boson~$h$,
in the $\gamma\gamma$ and $ZZ^{\star}$ final states, which might generate an observable difference between the apparent
masses measured in these final states~\cite{Dicus:1987fk,Dixon:2003yb} and/or provide loose indirect
constraints on the total width of the SM Higgs~\cite{Heinemeyer:2013tqa}.

There have also been pioneering studies of possible interference effects in the decays of a heavy Higgs boson into
$t\bar{t}$ final states, in both the SM~\cite{Gaemers:1984sj,Dicus:1994bm,Moretti:2012mq} and Two-Higgs-Doublet
models \cite{Bernreuther:1997gs,Barger:2006hm,Frederix:2007gi,Barcelo:2010bm,Figy:2011yu,Jung:2015gta,Gori:2016zto}.
Within the last years various theoretical works pointed out the relevance
of interference effects in the search for heavy Higgs bosons.
Such effects are dependent on the initial and final state, and model dependent.
Rather generic studies beyond concrete model realizations were performed for the $t\bar t$ and the $\gamma\gamma$
final states in Refs.~\cite{Djouadi:2016ack,Carena:2016npr}. Interference effects in $gg\to \phi\to t\bar t$ in
a Two-Higgs-Doublet Model were studied including NLO QCD effects in Refs.~\cite{Hespel:2016qaf,BuarqueFranzosi:2017jrj}.
An exhaustive discussion with polarization and spin effects can also be found in Refs.~\cite{Bernreuther:2015fts,Bernreuther:2017yhg}.
A search for heavy pseudoscalar and scalar Higgs bosons decaying into a top quark pair including 
interference effects has been performed by ATLAS with $20.3\,\text{fb}^{-1}$
of collected data at a center-of-mass energy of $\sqrt{s} = 8$\,TeV \cite{Aaboud:2017hnm}.

For rather model independent and generic studies for $pp\to\phi\to gg$ and to different fermionic final states
we refer to Ref.~\cite{Martin:2016bgw} and Ref.~\cite{Jung:2015gta}, respectively.
The final state $VV$, where $V$ denotes a vector boson, is more involved, since
generic models miss to properly account for the right unitarization of the cross
section at high invariant masses. It is thus recommendable to work in a concrete
model setup, which are usually the extension of the SM Higgs sector with
one singlet or one doublet. Note also that in this case $\phi$ is supposed to be CP-even.
For $gg\to \phi\to VV$ in the SM+singlet
we refer to Refs.~\cite{Maina:2015ela,Kauer:2015hia}, for vector boson fusion
in the same model setup to Ref.~\cite{Ballestrero:2015jca}, for a study
with decays into two leptons and two quarks, i.e. $gg\to\phi\to VV\to 2l2q$, to Ref.~\cite{Kauer:2015dma}.
In the context of the Two-Higgs-Doublet Model studies were carried out in
Refs.~\cite{Jung:2015sna,Greiner:2015ixr}.
The final state $hh$ in $gg\to \phi\to hh$
was covered in the context of a Two-Higgs-Doublet in Ref.~\cite{Hespel:2014sla}
and in the SM+singlet in Ref.~\cite{Carena:2018vpt} and in Ref.~\cite{Dawson:2015haa} taking into account NLO corrections.
Lastly the final state involving a $Z$ boson and the SM-like Higgs boson, i.e. the process $gg\to A\to Zh$
shows similar effects. Here $\phi$ is CP-odd in concrete model interpretations.
In the context of the Two-Higgs-Doublet Model we refer to Refs.~\cite{Harlander:2013mla,vhnnlov2}.
Finally another interesting aspect is the interference of Higgs bosons among themselves,
which was for example discussed in Refs.~\cite{Fuchs:2014ola,Fuchs:2017wkq,Patel:2018tdq}.

In this work we focus on the two final states $gg\to hh$ and $gg\to Zh$, which are of relevance for
SM Higgs precision measurements and the search for new physics in various aspects:
The process $gg\to hh$ allows for a measurement of the triple
Higgs coupling \cite{Djouadi:1999rca, Baglio:2012np, Dolan:2012rv}, which can be strongly modified in scalar extensions
of the SM \cite{DiLuzio:2017tfn}, or for an anomalous $t\bar{t}hh$ coupling \cite{Dib:2005re, Grober:2010yv, Contino:2012xk, Carvalho:2017vnu}.
The $gg\to Zh$ process, which is formally part of the next-to-next-to leading order corrections to associated production of $Zh$,
where the leading order process is the Drell-Yan process, allows for the measurement of couplings of the Higgs
boson to vector boson and the bottom-quark or anomalous interactions of the Higgs bosons and of top quarks to SM states
\cite{Harlander:2013mla, Hespel:2015zea, Englert:2016hvy}. In order to reduce background experimental
analysis often consider high transverse momenta of the vector boson, where
the relevance of the gluon-induced component $gg\to Zh$ compared to the Drell-Yan like component is
substantially enhanced~\cite{Harlander:2013mla}.

For the two mentioned processes we want to classify interferences
and provide a recipe that allows to estimate when such interference effects
are of relevance. The relevance of interferences is not only classified by the
ratio of the width $\Gamma_\phi$ and the mass $m_\phi$, $\Gamma_\phi/m_\phi$,
but in particular the strength of
the signal contribution over the background is a strong indication, how large
interference effects are expected to be. 
On the other hand the narrow-width approximation usually only assumes the ratio $\Gamma_\phi/m_\phi$
to be small.
Moreover a distortion of the peak structure
is usually visible already before a significant effect on the total peak-integrated
signal cross section including interferences is observed. For this purpose we define
two asymmetric parameters.

Our contribution is structured as follows: We first describe our calculational setup
including the simplified parametrization of the process that we employ. We continue
with its implementation into our codes used for the numerical analysis. Subsequently
we provide the classification of interferences in terms of three parameters
and then present our results, before we comment on their relevance for the experimental searches.

\section{CALCULATIONAL SETUP}
\label{sec:calc;heavyhiggsinter} 

\subsection{Simplified parametrization of the processes}
We discuss the two processes $gg\to H\to hh$ and $gg\to A\to Zh$ in this manuscript.
Relevant sample Feynman diagrams are depicted in Fig.~\ref{feynman;heavyhiggsinter}.
\begin{figure}
\begin{center}
\includegraphics[width=0.33\textwidth]{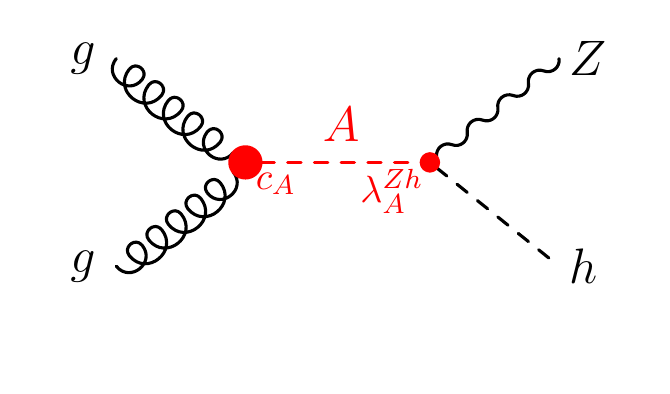}\hfill
\includegraphics[width=0.33\textwidth]{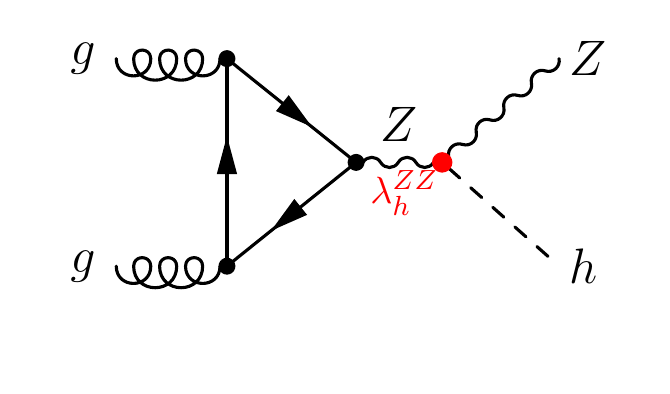}\hfill
\includegraphics[width=0.33\textwidth]{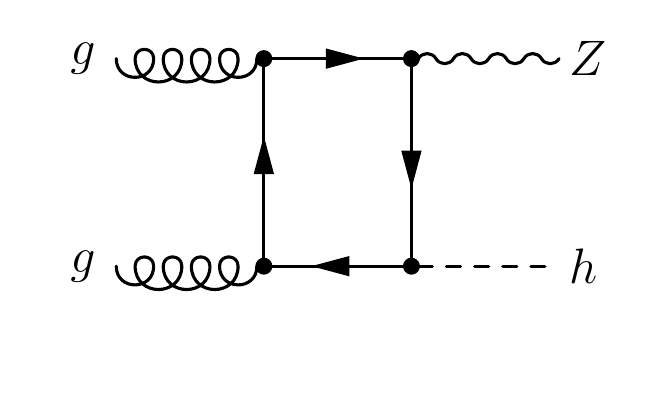}\\
\includegraphics[width=0.33\textwidth]{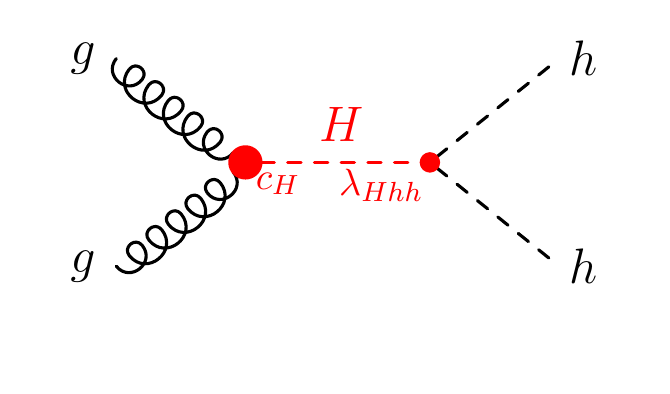}\hfill
\includegraphics[width=0.33\textwidth]{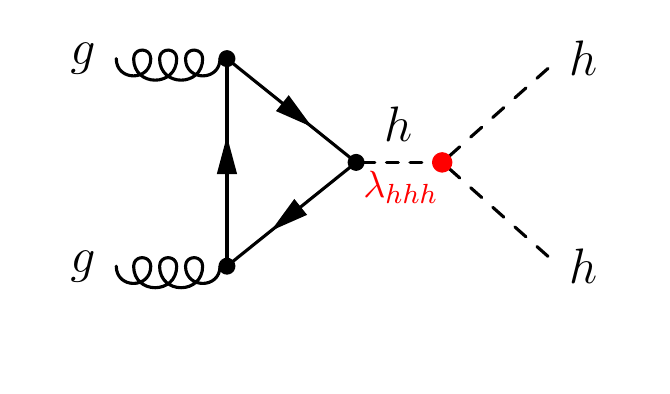}\hfill
\includegraphics[width=0.33\textwidth]{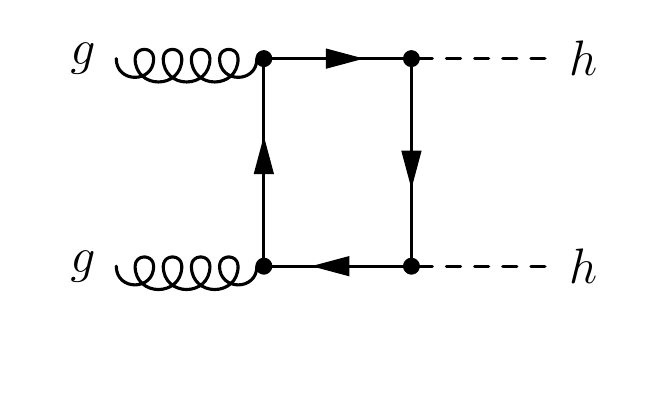}
\caption{Feynman diagrams of the two processes $gg\to Zh$ (upper row) and $gg\to hh$ (lower row).
The two left Feynman diagrams form the signal amplitudes $A_S$, see Sct.~\ref{classification;heavyhiggsinter},
the other diagrams enter the background amplitudes $A_B$.}
\label{feynman;heavyhiggsinter}
\end{center}
\end{figure}
We assume effective couplings of the Higgs bosons $H$ and $A$ to two gluons according to
\begin{equation}
L \supset \frac{\alpha_s}{12\pi v}c_H\,HG^a_{\mu\nu}G^{a,\mu\nu} +
\frac{\alpha_s}{8\pi v}c_A\,AG^a_{\mu\nu}\tilde G^{a,\mu\nu}\,. 
\end{equation}
These formulas include the strong-coupling constant $\alpha_s$, the vacuum expectation
value $v=1/\sqrt{\sqrt{2}G_F}\approx 246$\,GeV and Wilson coefficients $c_H$ and $c_A$, which are normalized
such, that for $c_A=c_H=1$ the coupling of $H$ and $A$ to two gluons resembles the one
obtained through a top-quark loop with an infinitely heavy top-quark mass.
$G^a_{\mu\nu}$ denotes the gluonic field strength tensor with
color index $a$ and Lorentz indices $\mu$ and $\nu$, and $\tilde
G_{\mu\nu}^a\equiv \varepsilon_{\mu\nu\rho\sigma}G^{a,\rho\sigma}$ is
its dual with $\varepsilon^{0123}=+1$.

We allow the two coefficients $c_\phi=|c_\phi|e^{i\theta_\phi}$
to be complex, since particles $Q$, which run in a potential loop and couple $\phi$ to two gluons,
induce a complex contribution to the amplitude for $2m_Q\leq m_\phi$. Formally the description through
an effective operator for such loop contributions is not valid, but since we
are not restricted to a specific model and treat $c_\phi$ as a generic parameter it makes sense
to condense the amplitude in a (complex-valued) Wilson coefficient of the given form.
We assume the masses and total widths of $H$ and $A$ to be free parameters.
In summary we have the following parameters, that are identical for both processes:
\begin{equation}
 |c_\phi|, e^{i\theta_\phi},m_\phi,\Gamma_\phi
\end{equation}
In addition we have the following process-specific parameters:
The process $gg\to hh$ involves the trilinear Higgs self-couplings $\lambda_{hhh}$
and $\lambda_{Hhh}$, which are both normalized with respect to the SM Higgs self-coupling.
We allow both to be free, but any variation of $\lambda_{Hhh}$ can also be condensed
into the Wilson coefficient $c_H$. The process~$gg\to Zh$ includes
the coupling of the light Higgs boson to gauge bosons, $\lambda_{h}^{VV}$, which we
normalize with respect to the SM coupling. In addition the pseudoscalar coupling to the
light Higgs and the $Z$ boson is of relevance, which we name $\lambda_Z^{Ah}$ and which
corresponds to $g_Z^{Ah}$ in the Appendix of Ref.~\cite{Harlander:2013mla}.
Inspired by the Higgs sector of a Two-Higgs-Doublet Model we set 
the relative strength of the coupling $\lambda_Z^{Ah}$ equal to $\lambda_H^{VV}=\sqrt{1-(\lambda_h^{VV})^2}$.
The coupling $\lambda_h^{VV}$ is experimentally restricted to be close to $1$.
Practically again any variation of $\lambda_Z^{Ah}$ can also be shifted into the Wilson coefficient $c_A$,
but due to the assumed correlation of Higgs and gauge boson couplings we change its value.
For both processes
the light Higgs $h$ is assumed to couple with SM strength to all quarks in our simplified parametrization.

\subsection{Employed codes}
For the evaluation of the differential cross sections
we use a modified version of {\tt HPAIR}~\cite{hpair} for di-Higgs production
and {\tt vh@nnlo}~\cite{Brein:2012ne} for $Zh$ production. Both
{\tt vh@nnlo} and {\tt HPAIR} include the $s$-channel propagator $gg\to \phi\to F$, i.e. $gg\to A\to Zh$ and $gg\to H\to hh$ respectively, in the form of a Breit-Wigner propagator
\begin{equation}
\label{eq:bw;heavyhiggsinter}
 \frac{1}{m_F^2-m_\phi^2+i\Gamma_\phi m_\phi}
\end{equation}
with the final state invariant mass $m_F$.
Even though higher order corrections to gluon fusion processes are generically quite high, 
we restrict ourselves to the leading-order result. In the infinite top mass limit the $K$-factors
in beyond-the Standard Model extensions are not expected to vary much with respect to the SM,
even in the presence of a new resonance, \cite{Dawson:2015haa, Grober:2015cwa, deFlorian:2017qfk, Grober:2017gut},
so when showing ratios we can assume them to drop out. 
Our results are based on hadronic cross sections for the LHC integrated over the gluon luminosities.
Still, for simplicity our studies could be performed at the partonic level also, since only the relative importance of
interferences in the vicinity of the internal masses are investigated. This on the other hand
implies that our results are mostly independent of the center-of-mass energy of a hadron collider
and even more of the employed parton distribution functions.

\subsection{Classification of interferences}
\label{classification;heavyhiggsinter}

In order to classify the interferences we split
the cross section as a function of the invariant mass of the final state $d\sigma/dm_F$
in three contributions
\begin{equation}
 \frac{d\sigma}{dm_F}=\frac{d\sigma_S}{dm_F}+\frac{d\sigma_I}{dm_F}+\frac{d\sigma_B}{dm_F}\,.
\end{equation}
Therein, the signal contribution $S$ only includes the $s$-channel Feynman diagram $gg\to\phi\to F$
involving the heavy scalar $\phi$, whereas the background $B$ sums up the square of all other Feynman diagrams,
including the $s$-channel Feynman diagrams involving SM particles, i.e. $h$ and $Z$.
With background we mean the non-resonant di-Higgs contribution or the non-resonant $gg\to Zh$ production.

The interference contribution $I$ is proportional to $2$Re$(A_SA_B^*)$, where $A_S$ and $A_B$
denote the amplitudes of signal and background diagrams, respectively. This split of amplitudes
is gauge-invariant. We define
\begin{equation}
\begin{split}
 &\eta=\left.\int_{m_\phi-10\Gamma_\phi}^{m_\phi+10\Gamma_\phi}dm_F\left(\frac{d\sigma_S}{dm_F}+\frac{d\sigma_I}{dm_F}\right)\right/
 \int_{m_\phi-10\Gamma_\phi}^{m_\phi+10\Gamma_\phi}dm_F\left(\frac{d\sigma_S}{dm_F}\right)\\
 &\eta_-=\left.\int_{m_\phi-10\Gamma_\phi}^{m_F^I}dm_F\left(\frac{d\sigma_S}{dm_F}+\frac{d\sigma_I}{dm_F}\right)\right/
 \int_{m_\phi-10\Gamma_\phi}^{m_F^I}dm_F\left(\frac{d\sigma_S}{dm_F}\right)\\
 & \eta_+=\left.\int_{m_F^I}^{m_\phi+10\Gamma_\phi}dm_F\left(\frac{d\sigma_S}{dm_F}+\frac{d\sigma_I}{dm_F}\right)\right/
 \int_{m_F^I}^{m_\phi+10\Gamma_\phi}dm_F\left(\frac{d\sigma_S}{dm_F}\right)\,.
\end{split}
\end{equation}
The definition includes
the overall factor $\eta$, which is a relative factor that, if multiplied with the signal cross section $\sigma_S$,
yields the overall change of the signal cross section due to interference effects.
Still, as already indicated, interference effects also distort the peak structure substantially.
If the two curves $d(\sigma_S+\sigma_I)/dm_F$ and $d\sigma_S/dm_F$ intersect once, at $m_F=m_F^I$,
we in addition split the integrals into two components and define the corresponding factors $\eta_-$ and $\eta_+$.
Example for both cases are given in Fig.~\ref{etaclass;heavyhiggsinter}.
If the two curves do not intersect, we set $\eta_\pm=0$. If non-zero,
$\eta_\pm$ can be quite large, whereas the overall effect of the interference remains small.
An example is given in Fig.~\ref{etaclass;heavyhiggsinter} (right), where
$\eta=1.34$, $\eta_-=35.02$ and $\eta_+=-30.28$. If the peak structure of the
the heavy scalars $\phi$ can be experimentally resolved, the factors $\eta_\pm$
thus yield a useful classification of interference effects, since they
allow to deduce in which direction the peak shift occurs and in which way a peak-dip structure appears.
The boundaries of the integrals being $m_\phi\pm 10\Gamma_\phi$ capture the majority
of the peak structure, which is suppressed by the form of the Breit-Wigner propagator, see Eq.~\ref{eq:bw;heavyhiggsinter}.

In our subsequent scans over the parameter space we will thus deal with the
three factors $\eta$, $\eta_-$ and $\eta_+$ to classify interference effects. For very large width the boundaries
$\pm 10\Gamma_\phi$ span a large invariant mass range. In case it includes thresholds in the background
Feynman diagrams, like at $m_F\approx 2m_t$ in the triangle loop producing a $Z$ boson or the SM-like Higgs,
we leave out the corresponding width choice, since such
cases need a more thorough study of the interference effects.

We will present the parameters $\eta$, $\eta_\pm$ as a function of the ratio $\Gamma_\phi/m_\phi$
and as a function of the signal-over-background cross sections. For this purpose we define
\begin{equation}
\begin{split}
\sigma_{sig}=\int_{m_{\phi}-10 \Gamma_{\phi}}^{m_{\phi}+10 \Gamma_{\phi}}dm_F \frac{d\sigma_{S}}{dm_F} \qquad\text{and}\qquad
\sigma_{back}=\int_{m_{\phi}-10 \Gamma_{\phi}}^{m_{\phi}+10 \Gamma_{\phi}}dm_F \frac{d\sigma_{B}}{dm_F}\,,
\end{split}
\end{equation}
which are integrated signal and background cross sections within an invariant mass window of
$\pm 10\Gamma_\phi$.\footnote{For a comparison with the experiment, it might instead be useful to define $\sigma_{sig}$ and $\sigma_{back}$ 
in terms of the bin width in the invariant mass of different experimental searches. We leave this for future work.} 
The choice $\pm 10\Gamma_\phi$ captures the majority of the signal contribution. Normalizing each, $\sigma_{sig}$ and $\sigma_{back}$, to $20\Gamma_\phi$
is not needed, since we only consider their ratio $\sigma_{sig}/\sigma_{back}$ subsequently.
Lastly keep in mind that the background in the $Zh$ case includes only the gluon-induced component, but omits
the Drell-Yan like component starting with light quarks in the initial state.

\begin{figure}
\begin{center}
\includegraphics[width=0.48\textwidth]{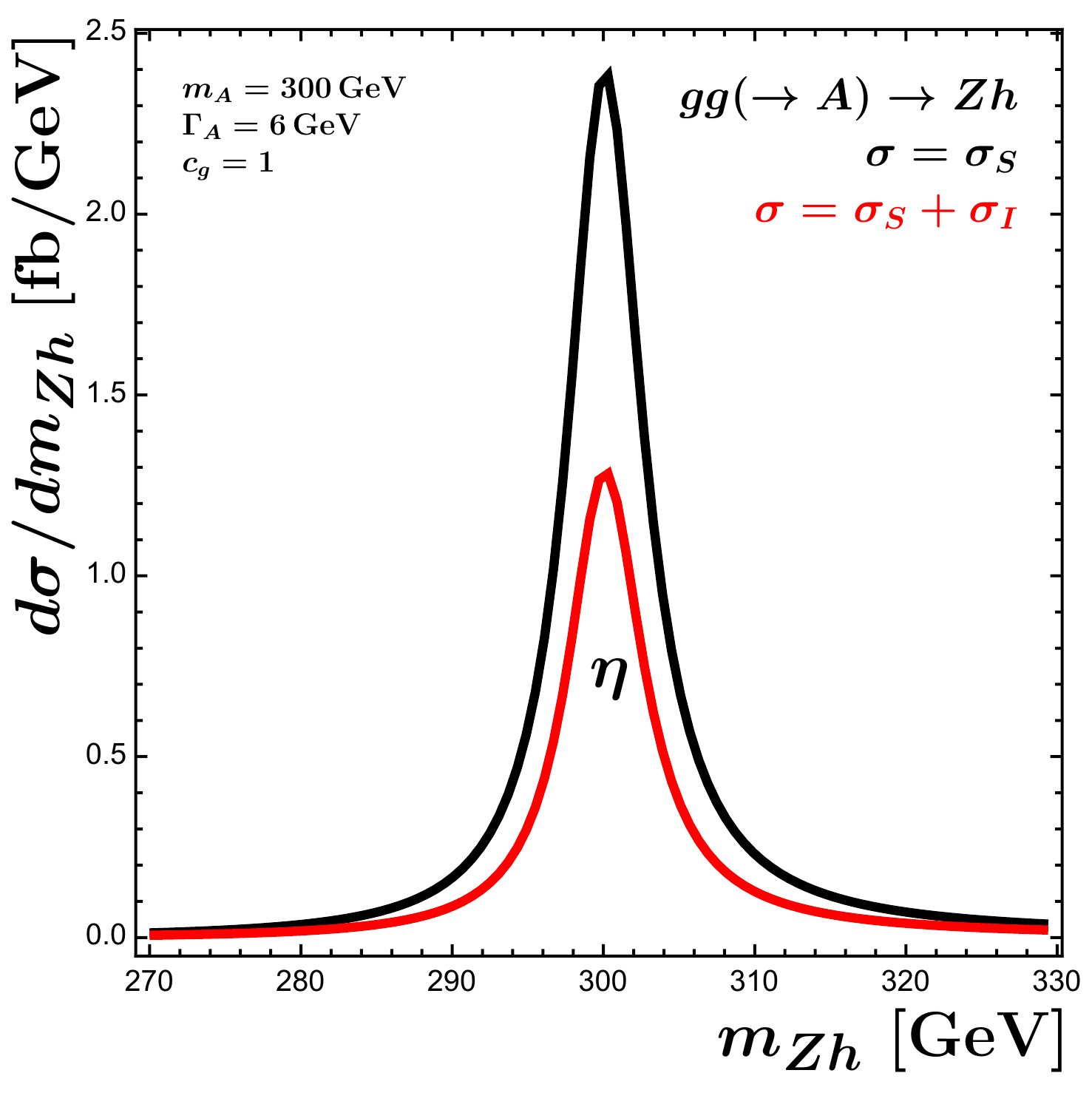}\hfill
\includegraphics[width=0.48\textwidth]{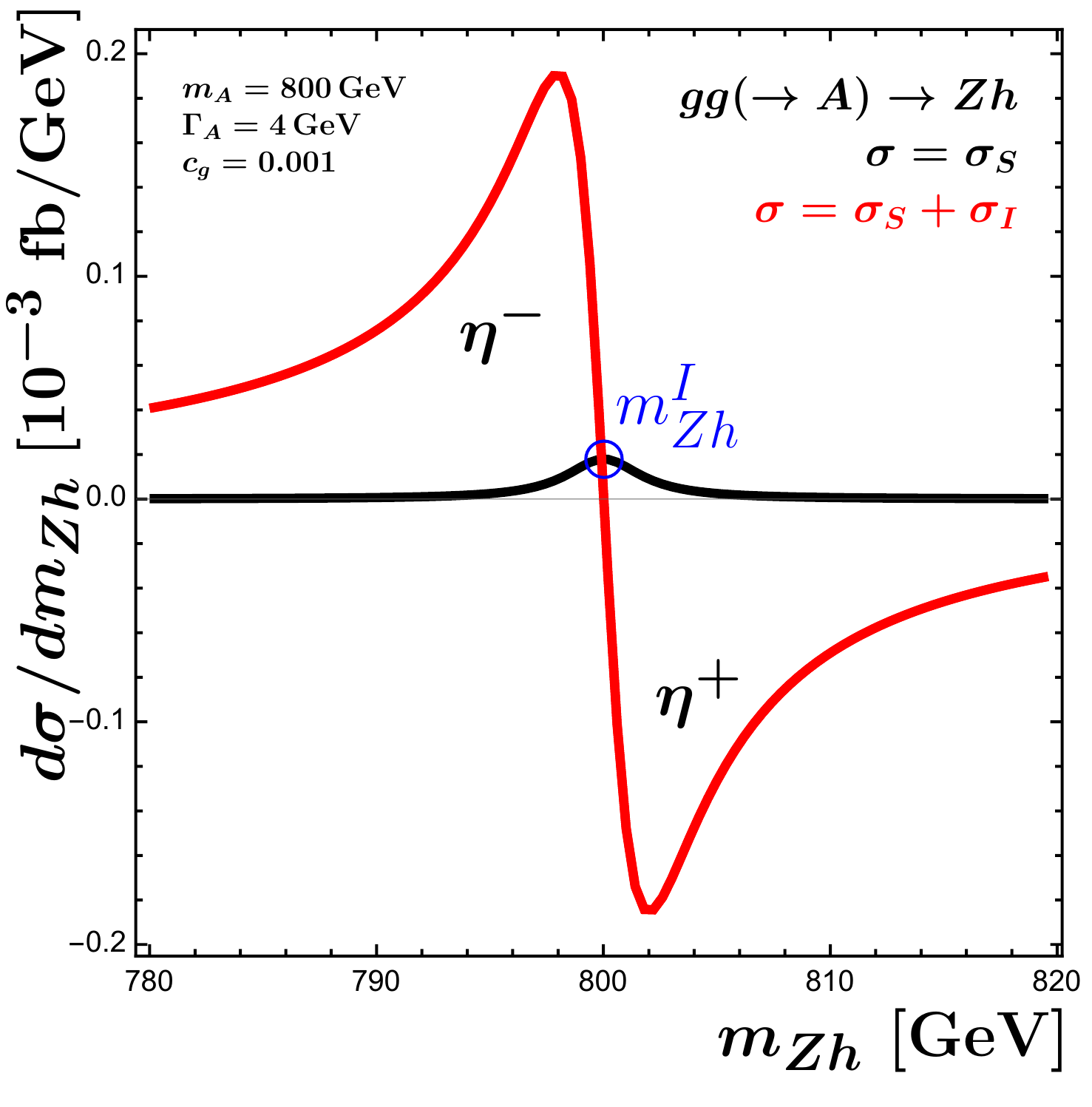}
\caption{Classification of interference effects through $\eta$, $\eta_+$ and $\eta_-$ for two parameter sets of $gg(\to A)\to Zh$. Only a window of $\pm 5\Gamma_A$ in $m_{Zh}$ is shown.}
\label{etaclass;heavyhiggsinter}
\end{center}
\end{figure}

\section{NUMERICAL RESULTS}
In order to show numerical results we vary the parameters involved in both processes within the following ranges
\begin{equation}
\begin{split}
|c_{\phi}|\in [0.001, 5]\,, \hspace*{0.5cm} \theta_{\phi} \in \{0,\frac{\pi}{4},\frac{\pi}{2}\}, \hspace*{0.5cm} m_{\phi}\in [0.3,1.4] \text{ TeV}, \hspace*{0.5cm}\Gamma_{\phi}/m_{\phi}\in [10^{-4},0.2] \,.
\end{split}
\end{equation}
As already mentioned, we abandon very large width choices of $\Gamma_\phi/m_\phi=\{0.1,0.2\}$, if they capture thresholds in background diagrams.
For $\lambda_{hhh}$ we choose the values $\lambda_{hhh}=\{0,1,2\}$, and we leave $\lambda_{Hhh}=1$.
Keep in mind that both are normalized to the SM, i.e. $\lambda_{hhh}=1$ yields the SM Higgs self coupling.
Any variation in $\lambda_{Hhh}$ can be shifted to the Wilson coefficient $c_H$.
For the $Zh$ process we pick two choices of $\lambda_{h}^{VV}$, namely $0.97$ and $0.995$, which yield a SM-like Higgs coupling to gauge bosons $\lambda_{h}^{VV}$
compatible with the experimental results, namely differing by $6\%$ and $1\%$ from the SM expectation, respectively.
It is well possible that some of the choices are non-physical in the sense that large values of the Wilson coefficient
or the involved couplings would also trigger a large decay width, since the heavy intermediate resonance can decay at
least into $gg$ and $hh$ or $Zh$. However, our choices of small widths generally induce a large signal cross section,
for which in turn interference effects are small. We thus leave such points in our scan and emphasize that a
concrete model realization would properly correlate the total width $\Gamma_\phi$ with the other parameters.

\subsection{The process $gg\to hh$}

We show the impact of the interference for the $hh$ final state in Fig.~\ref{result1;heavyhiggsinter}, where in different colors the values of the
overall interference factor $\eta$ as a function of both ratios $\Gamma_H/m_H$ and $\sigma_{sig}/\sigma_{back}$ is presented.
We split the range of $\eta$ into four regions, namely in one region in which $\eta$ differs from $1$ by less than $3\%$, one
with more than $3\%$, one with more than $10\%$ and the fourth one with more than $50\%$. It is apparent that for all values
of $\Gamma_H/m_H$ large interferences can occur. This implies that $\Gamma_H/m_H$ is not ideal for discriminating 
interference effects. On the other hand $\eta$ clearly correlates with the value of the ratio $\sigma_{sig}/\sigma_{back}$.
Even for relatively large ratios of $\sigma_{sig}/\sigma_{back}>1$ interferences of $50\%$ are observed.

In Fig.~\ref{result2;heavyhiggsinter} we show $|\eta|$ (left side) and $|\eta_+|$ (right side) as a function of $\Gamma_H/m_H$.
For any value of $\Gamma_H/m_H$ the interference factors $|\eta_{(\pm)}|$ can vanish. On the other hand their largest values are only reached for
large width $\Gamma_H$.
Taking Fig.~\ref{result1;heavyhiggsinter} and Fig.~\ref{result2;heavyhiggsinter}
together we can see that the interference effects mostly depend on the ratio of signal-over-background rather
than $\Gamma_H/m_H$, however, for lower $\Gamma_H/m_H$ we usually find larger signal-over-background ratios. 
Lastly we show
$|\eta|$ (black points), $|\eta_+|$ (red points) and $|\eta_-|$ (blue points)
as a function of $\sigma_{sig}/\sigma_{back}$ in Fig.~\ref{result3;heavyhiggsinter}, where the right figure is a zoom of the left figure.
It can be inferred, that the interference increases with decreasing 
$\sigma_{sig}/\sigma_{back}$. The figure can hence give indication when the interference needs to be taken into account 
in experimental searches. We see that already for $\sigma_{sig}=10\,\sigma_{back}$ we can have interference effects leading
to a cross section increased by a factor of $1.5$.
The interference factors $\eta_+$ and $\eta_-$ take generally larger values than $\eta$, i.e. the peak structure is
already distorted before an overall effect on the signal cross section gets significant.

\begin{figure}
\begin{center}
\includegraphics[width=0.55\textwidth]{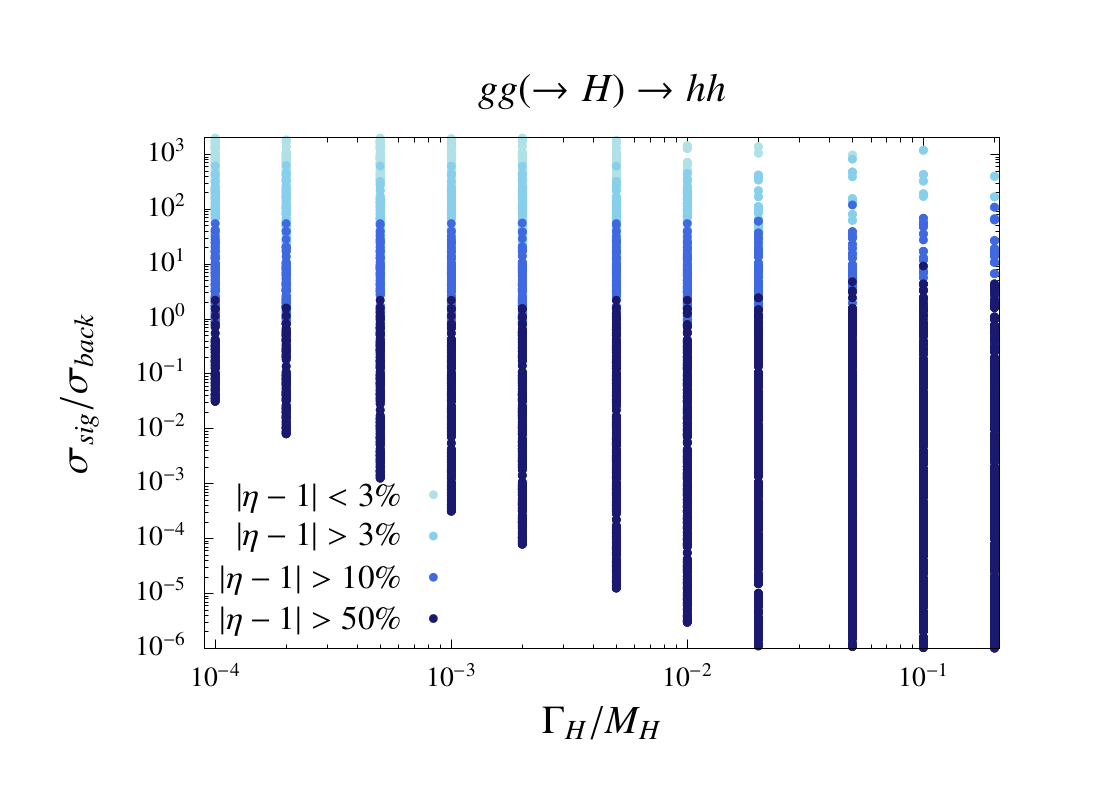}
\caption{Relative difference of the interference factor $\eta$ for $gg\to H\to hh$ from $1$ in percent in the $\Gamma_H/m_H$-$\sigma_{sig}/\sigma_{back}$ plane.
The scan was performed in a simplified model for $gg\to H\to hh$, see text.}
\label{result1;heavyhiggsinter}
\end{center}
\end{figure}
\begin{figure}
\begin{center}
\includegraphics[width=0.52\textwidth]{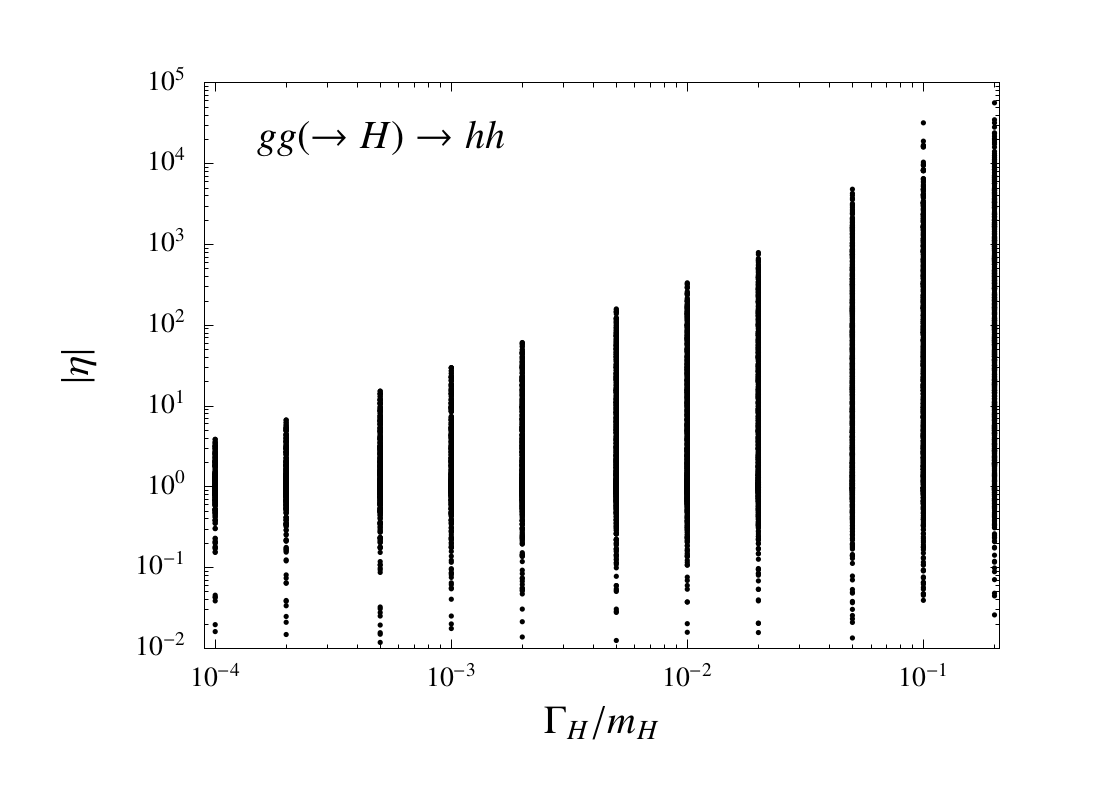}\hspace*{-0.6cm}
\includegraphics[width=0.52\textwidth]{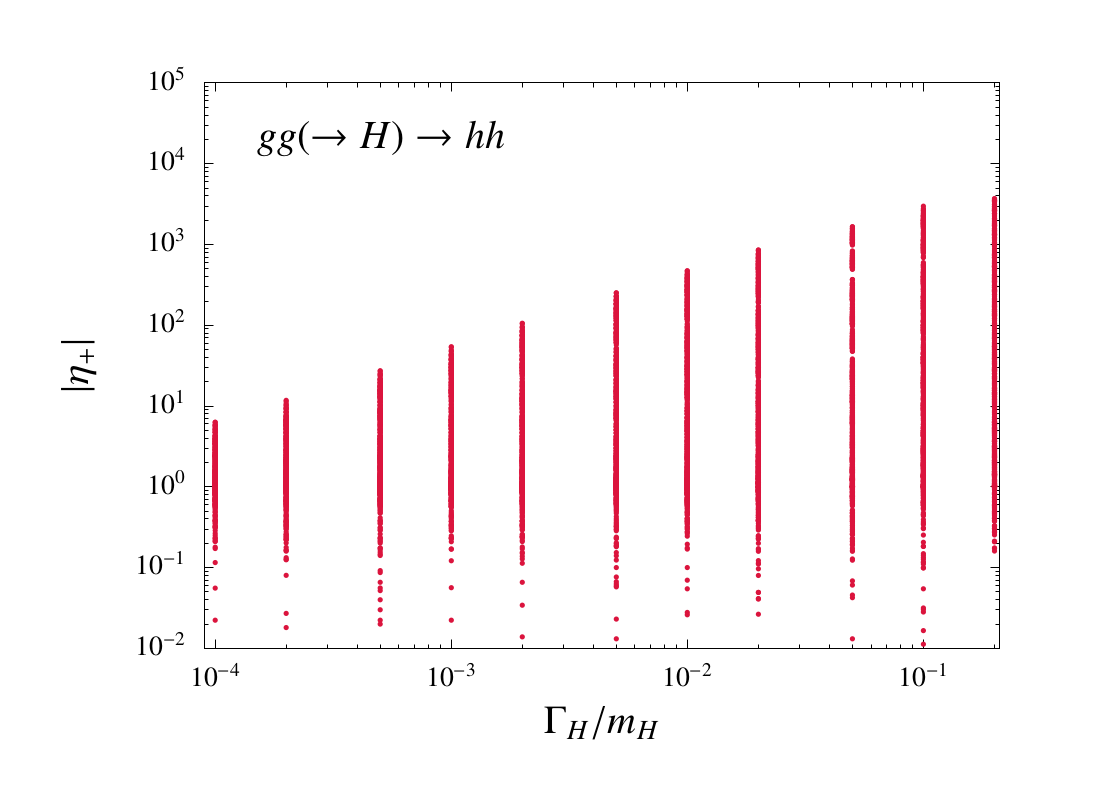}
\caption{Interference factors~$\eta$ (left side) and $\eta_+$ (right side) as a function of $\Gamma_H/m_H$ for $gg\to H\to hh$.}
\label{result2;heavyhiggsinter}
\end{center}
\end{figure}
\begin{figure}
\begin{center}
\includegraphics[width=0.52\textwidth]{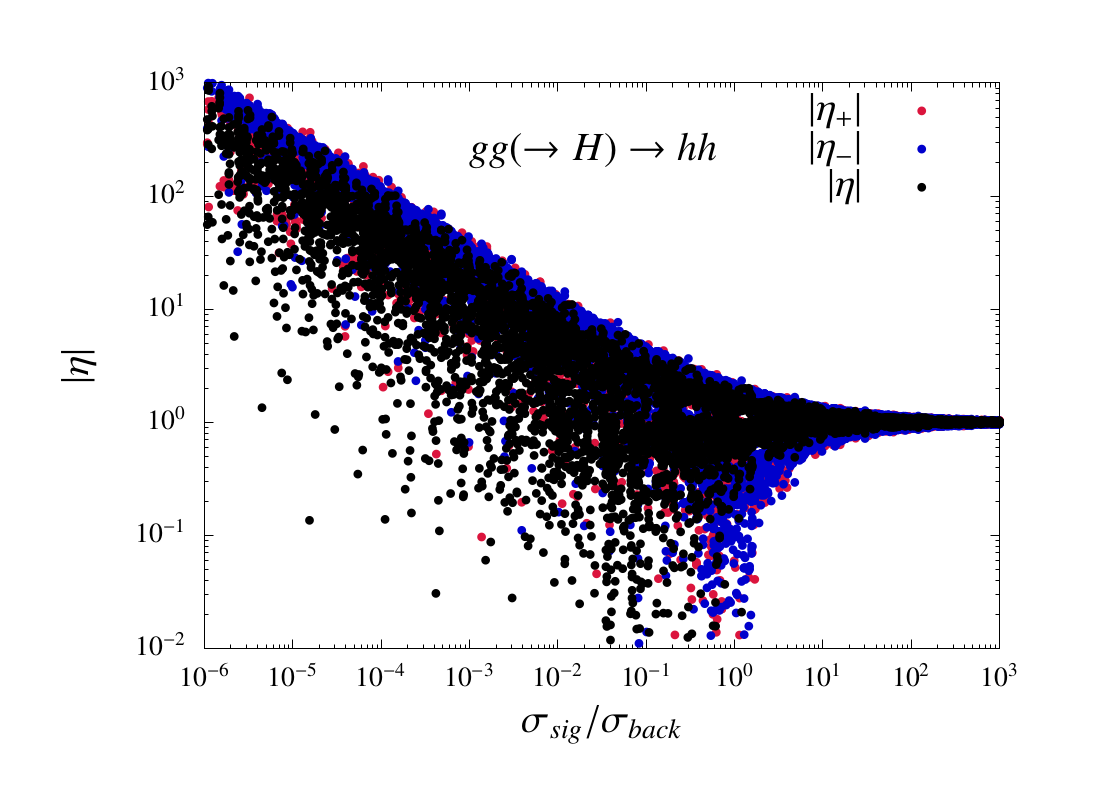}\hspace*{-0.6cm}
\includegraphics[width=0.52\textwidth]{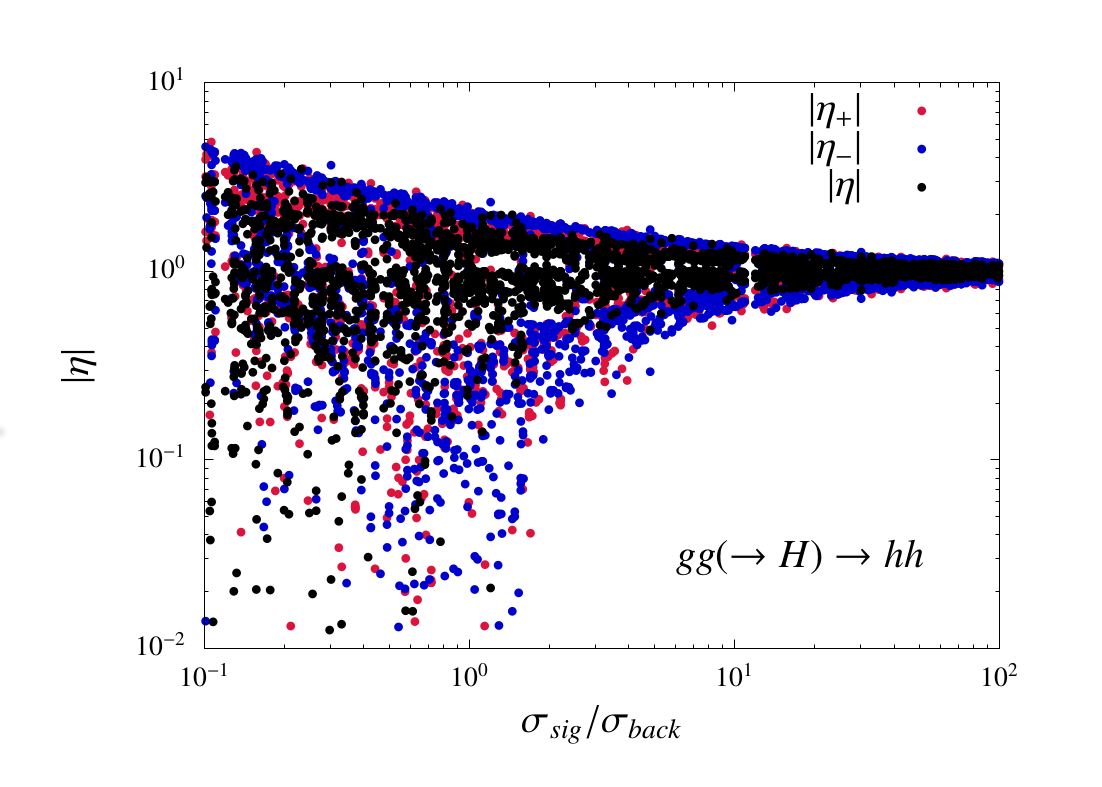}
\caption{Interference factors~$\eta$, $\eta_+$ and $\eta_-$ as a function of $\sigma_{sig}/\sigma_{back}$ for $gg\to H\to hh$. The right figure is a zoomed version of the left figure.}
\label{result3;heavyhiggsinter}
\end{center}
\end{figure}

\subsection{The process $gg\to Zh$}

We continue with a presentation of our results for $gg\to Zh$. Again we present
the relative difference of $\eta$ from $1$ in the  $\Gamma_A/m_A$ and $\sigma_{sig}/\sigma_{back}$ plane
in Fig.~\ref{result4;heavyhiggsinter} on the left side. The right side shows $|\eta_{(\pm)}|$ as a function
of $\Gamma_A/m_A$. Even for small width $\Gamma_A/m_A\sim 10^{-4}$ the factors $\eta_\pm$ are non-zero
and clearly differ from $1$.
Finally in Fig.~\ref{result5;heavyhiggsinter} we depict $|\eta|$ and $|\eta_{\pm}|$
as a function of $\sigma_{sig}/\sigma_{back}$. 
Significant interferences are observed already at large values of $\sigma_{sig}/\sigma_{back}$.
The peak distortion sets in before a significant effect on the overall signal cross section is observed,
i.e. at even larger $\sigma_{sig}/\sigma_{back}$.
In the context of this figure we emphasize that the peak distortion is strongly dependent on
the phase of the Wilson coefficient, for both discussed processes.
In almost all cases either $\eta_+$ or $\eta_-$ turns negative, which is mainly
dependent on the phase and partially the width $\Gamma_A$. A detailed analysis of such effects
in concrete model realizations is left for future work.

\begin{figure}
\begin{center}
\includegraphics[width=0.5\textwidth]{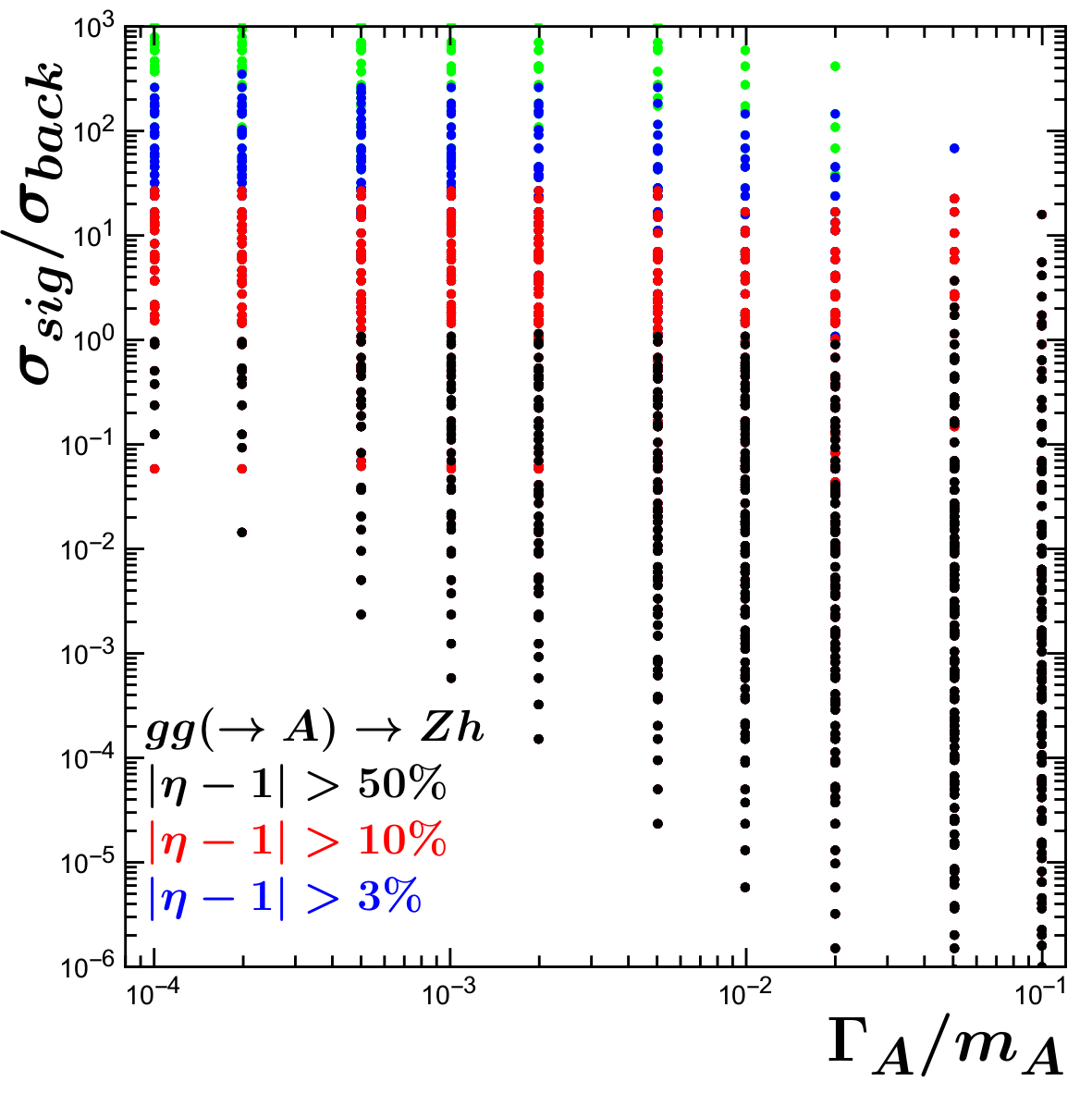}\hfill
\includegraphics[width=0.5\textwidth]{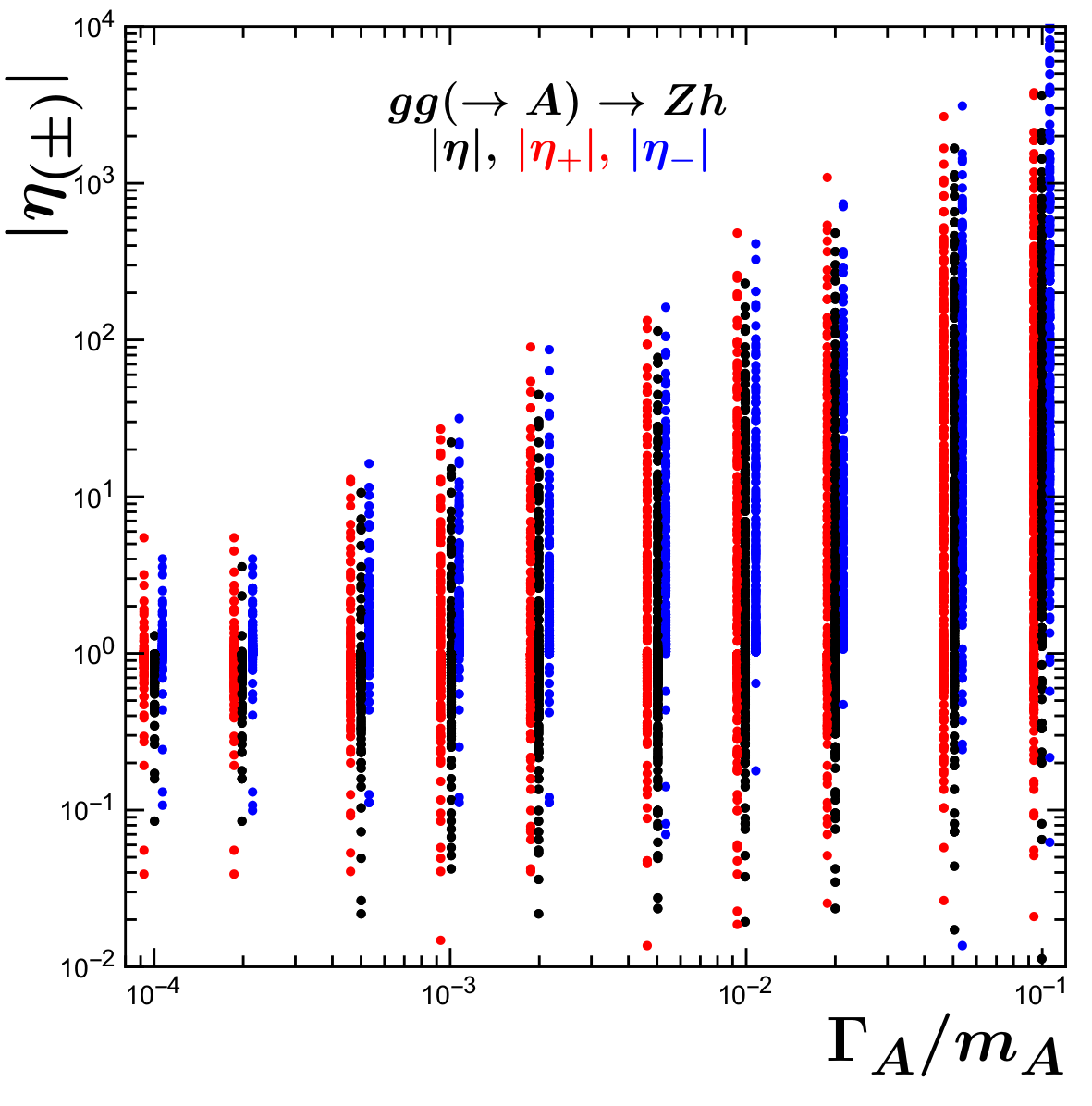}
\caption{Relative difference of the interference factor $\eta$ for $gg\to A\to Zh$ from $1$ in percent in the $\Gamma_A/m_A$-$\sigma_{sig}/\sigma_{back}$ plane (left side).
Interference factors $\eta$, $\eta_+$ and $\eta_-$ for $gg\to A\to Zh$ as a function of $\Gamma_A/m_A$.
The scan was performed in a simplified model for $gg\to A\to Zh$, see text.}
\label{result4;heavyhiggsinter}
\end{center}
\end{figure}
\begin{figure}
\begin{center}
\includegraphics[width=0.5\textwidth]{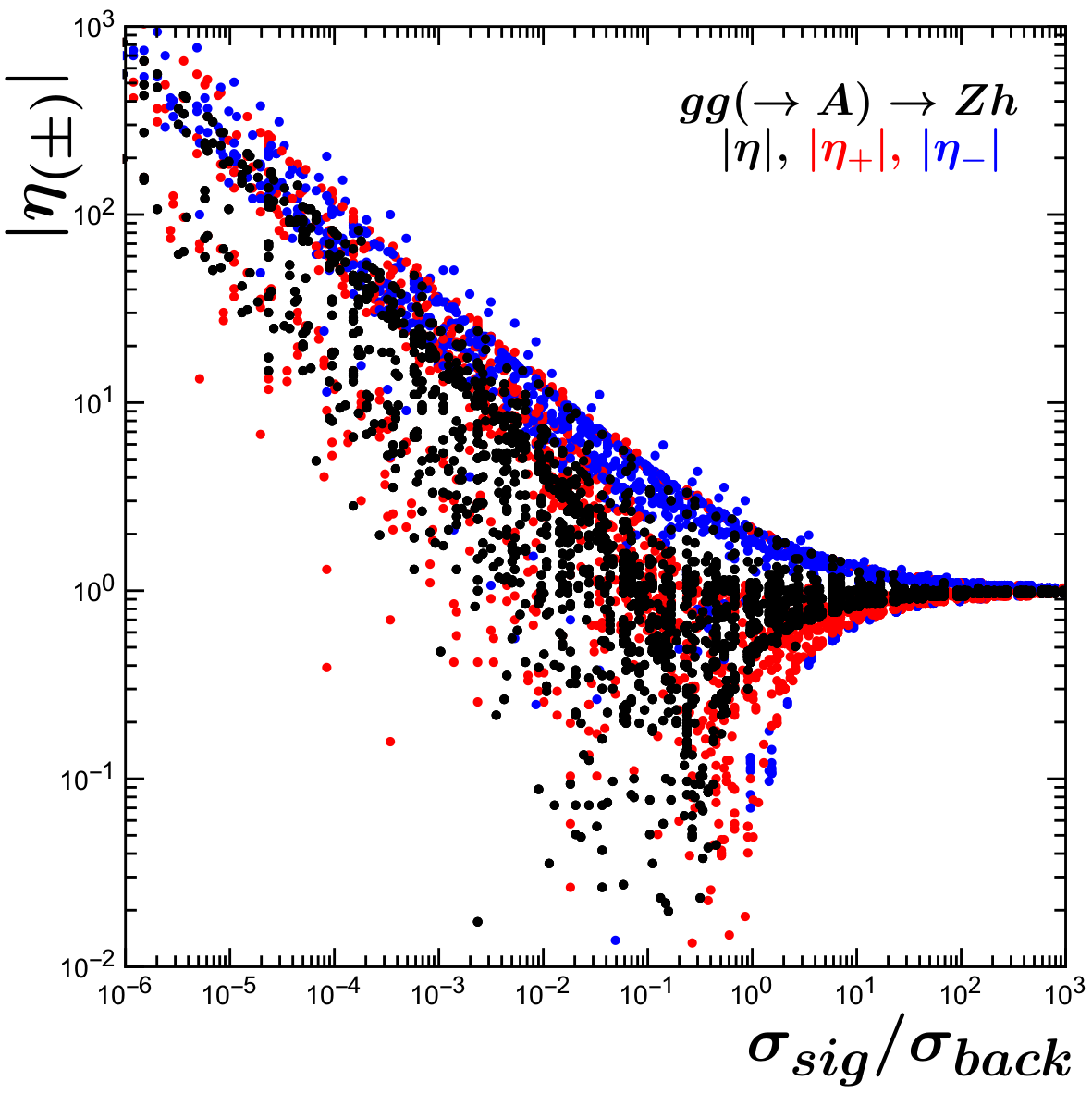}\hfill
\includegraphics[width=0.5\textwidth]{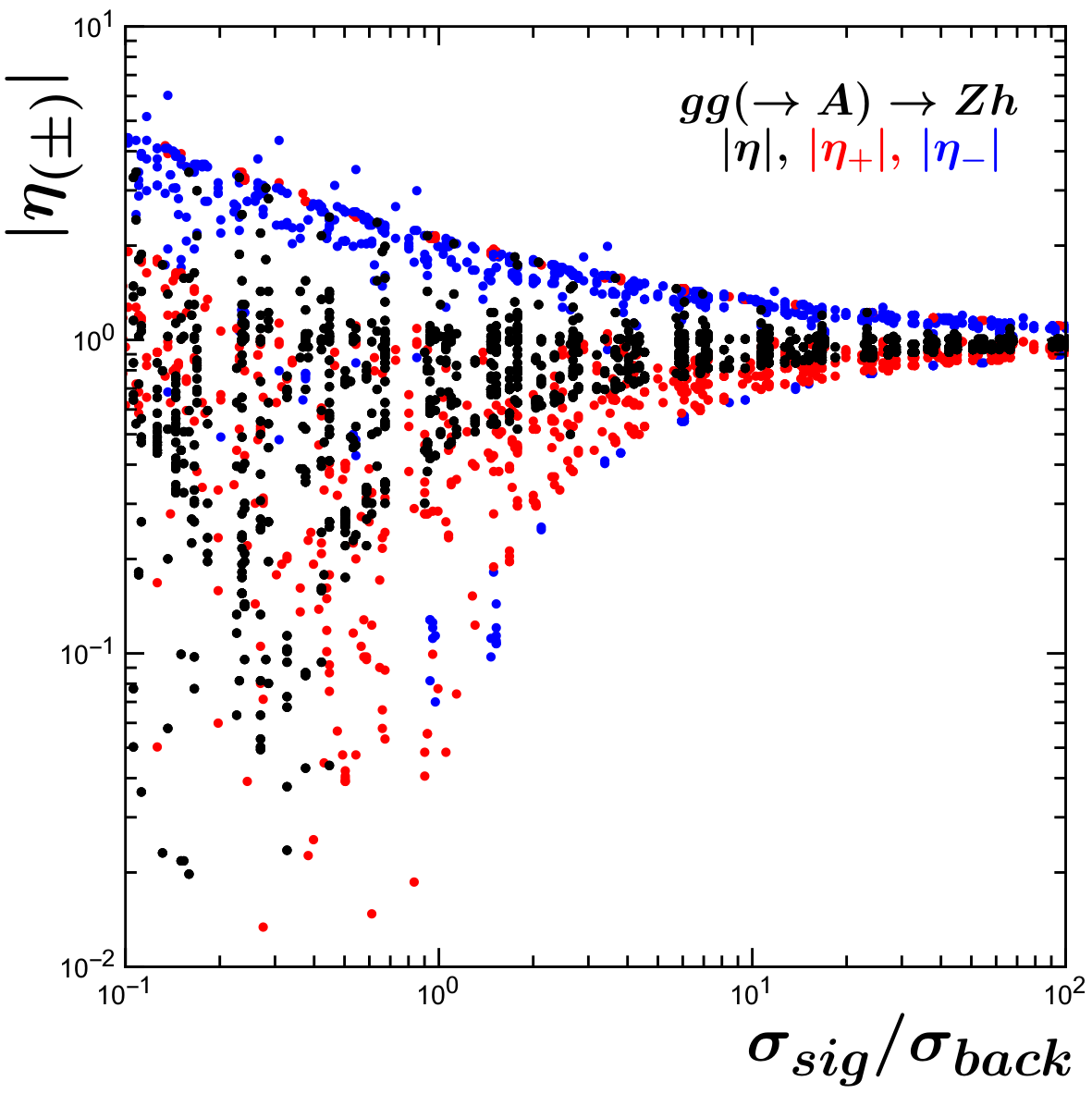}
\caption{Interference factors~$\eta$, $\eta_+$ and $\eta_-$ as a function of $\sigma_{sig}/\sigma_{back}$ for $gg\to A\to Zh$. The right figure is a zoomed version of the left figure.}
\label{result5;heavyhiggsinter}
\end{center}
\end{figure}

\section{APPLICATION AT THE LHC}

In this study we proposed a parametrization that allows to infer the size of the interference
between a heavy intermediate resonance and the SM background in the two processes $gg\to hh$ and $gg\to Zh$.
Our results are based on a generic setup and are thus mostly model independent.
For a bump search the experimental sensitivity depends mainly on
the total width of the resonance and the peak position of the resonance with respect to the
true resonance mass. 
Up to width values of roughly $10\%$ of the mass of the intermediate resonance, the
experimental sensitivity is independent of the width, since the peak structure is
experimentally not resolvable. 
When interpreting the experimental results the factor $\eta$, being a function of all parameters
of the model under consideration, can then be multiplied to the theoretical signal cross section
to yield exclusion bounds including interference effects.

On the other hand, our setup is not sufficient to quantify peak shifts, since $\eta_\pm$
only give an estimate of the peak distortion.
Again assuming an experimental resolution of $10\%$ in the invariant mass spectrum
a shift in the mass peak can only be observed if the peak is shifted by more
than $10\%$ from the true resonance mass.
In such a case only a proper (Monte-Carlo) modeling of the peak-dip like structure allows
to infer the true resonance mass. Such a setup was for example used in a heavy resonance
search decaying to a top-quark pair, see the ATLAS analysis in Ref.~\cite{Aaboud:2017hnm},
where the signal from a scalar resonance interfere with a huge $t\bar{t}$ background continuum.
Therein the changes to the line-shape are drastic
and a fit of the mass line-shape to determine the peak mass
needs to account for the parameter dependence of the
interferences, which is non-trivial. A simplified parametrization of the underlying
relevant parameters in terms of a few free parameters, as done here for $hh$ and $Zh$,
is very helpful in this context.
Still keep in mind, that the values $\eta_{\pm}$ alone do not allow to reconstruct
the true resonance mass.

For what concerns the $hh$ and $Zh$ final state
it is also crucial, which subsequent decays of the final state particles $h$ and $Z$
are considered in the experimental analysis.
The invariant mass of the resonance is not always directly used 
in the signal extraction method of the experimental analysis,
or even in intermediate selections. In most of the channels for $\phi\to hh$ and $\phi\to Zh$
the mass of $\phi$ is not 
fully reconstructed and alternatively transverse variables or multivariate approaches are used.
There are also cases where the invariant mass of the final state is fully reconstructed, but the resolution might not 
be good enough for the signal extraction when compared to a multivariate approach.
One example of the former for both $hh$ and $Zh$ channels is the final state $\tau^+\tau^- b\bar{b}$, where the
most sensitive sub-channel is the one where one of the $\tau$'s decays leptonically~\cite{Sirunyan:2017djm,Aad:2015xja}.
One example for multivariate approaches are $Zh\to l^+l^-b\bar{b}$ from CMS~\cite{Khachatryan:2015lba}, where a Boosted
Decision Tree is used, or $hh\to b\bar{b} b\bar{b}$ from ATLAS where the invariant mass of the reconstructed
$H\to b\bar{b}$ decays is used in a $2$-dimensional selection~\cite{Aad:2015uka}. 
There are indeed examples where the mass of the new resonance is fully reconstructable with
a fair resolution to allow it to be used for signal extraction. Examples for the $hh$ final state 
are $hh\to\gamma\gamma b\bar{b}$~\cite{ATLAS-CONF-2016-004,CMS-PAS-HIG-16-032}.

\section{CONCLUSIONS}

We classified interference effects between heavy resonances and SM background contributions
for the two processes $gg\to hh$ and $gg\to Zh$ through three parameters. We find that
the ratio $\Gamma_\phi/m_\phi$ involving the width and mass of the intermediate resonance $\phi$
is not the best observable to judge if interference effects are large, but at the same time
the ratio of the signal-over-background ratio should be considered. 
In a concrete model realization $\sigma_{sig}/\sigma_{back}$ can be predicted.
It can also be experimentally accessed when providing
experimental exclusions on the signal cross section.
Even for large signal contributions, where $\sigma_{sig}/\sigma_{back}$ is larger than $1$, interference
effects are sizable and should be taken into account in current experimental analysis.
An overall interference factor $\eta$ to be multiplied with the signal cross section
can then be employed while interpreting the bounds in a concrete model realization.
More difficult are peak distortions, which we classified in terms of the two parameters $\eta_\pm$.
Since $\eta_-$ and $\eta_+$ usually mostly cancel, they individually can be much larger than
the overall interference factor $\eta$ and thus point at large interference effects 
for even higher ratios $\sigma_{sig}/\sigma_{back}$.
We expect very similar observations for the process $gg\to VV$, which we leave to future work.
Another aspect for future studies are intermediate spin-$2$ resonances, which are often considered in the di-Higgs final state.
Moreover our generic parameter set might not cover any concrete model realization.
A more thorough discussion of concrete models and their mapping to our generic parameter set is thus desirable.

\section*{ACKNOWLEDGEMENTS}
The authors thank the organizers of the Les Houches Workshop Series
``Physics at TeV Colliders'' 2017 for the fruitful and pleasant atmosphere.
RG is supported by a European Union COFUND/Durham Junior Research Fellowship under the EU grant number 609412.
SL acknowledges support from ”BMBF Verbundforschung Teilchenphysik“ under grant number 05H15VKCCA. 



\AddToContent{A.~Carvalho, R.~Gr\"ober, S.~Liebler, J.~Quevillon}
\renewcommand{\thesection}{\arabic{section}}

\superpart{ Tools and Methods}

\graphicspath{{lightscalar/}}

\chapter{Sensitivity of current (and future?) LHC measurements to a new light scalar particle}

{\it J.~M.~Butterworth, S.~Fichet, L.~Finco, S.~Gascon-Shotkin, D.~Grellscheid,  G.~Moreau, P.~Richardson, D.~Yallup, S.~Zhang}



\begin{abstract}
Additional scalar particles are a generic feature of many well-motivated extensions of Standard Model. 
Here we use a simplified model in which a light scalar particle couples to electroweak gauge bosons via dimension-5 operators. For the masses considered, decays to pairs of weak bosons are suppressed,
and the $\gamma\gamma$ mode dominates. We find that existing measurements from Run I of the LHC already exclude the model over a significant parameter range. 

\end{abstract}

\section{INTRODUCTION}

Additional light scalar particles are a common feature in extensions of the SM, for example appearing in composite 
Higgs scenarios, or as the radion in models with extra dimensions~\cite{Angelescu:2017jyj}. 
Consideration of precision electroweak measurements, collider searches and flavour physics
does not completely exclude the existence of light neutral CP-odd or CP-even scalar particles below the mass of the 
observed Higgs boson\cite{Cacciapaglia:2016tlr}. In this contribution we use a simplified  model 
to examine whether measurements from Run I at the LHC can give information about such possible particles.

\section{THE MODEL}

We use an effective theory (EFT) approach to describe a scalar with mass $M_\phi$ interacting with gauge bosons. The effective theory has $SU(2)\times U(1)_Y$ symmetry. This EFT gives a generic parametrization if $M_\phi\gg v$ \cite{Fichet:2015yia}, where $v$ is the electroweak scale.
 Whenever the scalar is light so that $M_\phi\gg v$ is not true, we make the extra assumption that the scalar has large tree-level $SU(2)\times U(1)_Y$ couplings, so that the loop-induced electroweak-breaking contributions are subleading. 
Under these conditions the interactions of a CP-even and CP-odd scalars with gauge bosons are respectively described by the following dimension-5 effective Lagrangians 
\begin{equation}
{\cal L}_{\rm eff}\supset\phi \left( \frac{1}{f_G}G^{\mu\nu\,a}G_{\mu\nu}^a+ \frac{1}{f_W}W^{\mu\nu\,I}W_{\mu\nu}^I
+\frac{1}{f_B}B^{\mu\nu}B_{\mu\nu}+\frac{1}{f_H}|D^\mu H|^2
\right)
\end{equation}
\begin{equation}
{\cal L}_{\rm eff}\supset\phi \left( \frac{1}{f_G}G^{\mu\nu\,a}\tilde G_{\mu\nu}^a+ \frac{1}{f_W}W^{\mu\nu\,I}\tilde W_{\mu\nu}^I
+\frac{1}{f_B}B^{\mu\nu}\tilde B_{\mu\nu}
\right)
\end{equation}
where $\tilde V^{\mu\nu}=\frac{1}{2}\epsilon^{\mu\nu\rho\sigma}V_{\rho \sigma}$.
The effective theory is valid as long as the $f$'s are larger than the energy going through the vertices. 
 Mixing with the SM Higgs is assumed to be small to ensure that the SM Higgs has SM-like  couplings compatible with  observations.

The CP-even scalar can for instance be identified as the radion mode present in warped extra-dimension models with bulk gauge fields. Interestingly, if EW brane kinetic terms are negligible in such models, one has $f_W=f_B$ \cite{Fichet:2013ola, Fichet:2013gsa}, which implies that the $\phi F^{\mu\nu}Z_{\mu\nu}$ coupling vanishes, a property which can be used for model discrimination \cite{Baldenegro:2017aen}. 
The CP-odd scalar is typically a pseudo Nambu Goldstone boson from an approximate global symmetry, just like those appearing in composite Higgs models. The couplings to gauge fields are induced by the many fermion resonances populating the TeV scale (see e.g \cite{Belyaev:2016ftv} or also \cite{Fichet:2016xvs}).

In the following, as a first exercise, we assume a common scale $\Lambda$ for all couplings,
\begin{equation}
f_G \sim f_B\sim f_W\sim f_H\sim \Lambda \,,
\end{equation}
and similarly for the CP-odd case.

\section{SENSITIVITY OF EXISTING MEASUREMENTS}

\subsection{Herwig Implementation}

The new processes defined by the model described above are exported as UFO file~\cite{Degrande:2011ua} which is 
read by Herwig~7.2.1~\cite{Bellm:2015jjp,Bahr:2008pv}. This requires the four-boson vertices, 
which were added to the Herwig UFO interface as part of this work 
and are now available in this subsequently released version. The five-boson vertices implied by the model 
are not yet implemented
but are assumed not to have a major impact. 
This assumption is supported by a cross-check using MadGraph5\_aMC@NLO\_v2\_5\_5~\cite{Alwall:2014hca} for a selection of the parameter points considered. MadGraph5\_aMC@NLO
includes the full set of vertices, as well as some higher-order QCD contributions; the cross sections predicted by MadGraph5\_aMC@NLO are 
generally higher than the Herwig values, but are consistent with a factor of two. Thus any limits derived using Herwig are likely to
be somewhat conservative.

The scale which suppresses couplings to the 
Higgs and weak bosons is varied across the range $1 < \Lambda < 10$~TeV; all other BSM coupling are heavily 
suppressed ($\Lambda = 1000$~TeV). All
allowed $\phi$-production processes are generated inclusively.

\subsection{Rivet and Contur}

Generated Herwig events are passed to the Rivet library of analysis routines~\cite{Buckley:2010ar}. This contains 
a signficant number of published ATLAS and CMS analyses. Measurements which have been corrected for detector effects
to a particle-level fiducial phase space are rather model-independent. Rivet allows the particle-level analysis as performed
by the experiments to be applied to the BSM events generated by Herwig. Since the measurements considered 
have all been compared to precision SM calculations and shown to agree, there is limited room for additional BSM contributions.
The Contur comparison package~\cite{Butterworth:2016sqg} quantifies the level of contribution which could still be 
consistent with the data. Currently this is done on the assumption that the data are identical to the SM; a more
complete approach would be to use the SM predictions and their uncertainties directly; such a capability is a planned
future development of Contur, but the present implementation is enough to give a reasonable indication of the 
sensitivity of the data to BSM models.

\subsection{Measurements}

All available ATLAS and CMS Rivet analyses are used to study the data. However, since the branching ratio 
$\phi \rightarrow \gamma\gamma$ is $\approx 1$, the measurements of interest are those involving
isolated photons, or pairs of photons, in the final state. These have been measured 
inclusively~\cite{Aad:2012tba,Aad:2013zba,Aad:2016xcr}, and in 
association with jets~\cite{Chatrchyan:2013mwa,Aad:2013gaa,ATLAS:2012ar}, $W$ or $Z$ bosons\cite{Aad:2013izg,Aad:2016sau} (i.e. leptons and/or missing energy). 

The Higgs fiducial diphoton measurements~\cite{Aad:2014lwa} are also of interest. These were studied and in principle 
have some sensitivity -- events generated by the models considered do contribute to the fiducial region. However, since 
the value of $M_\phi$ considered here lie below the SM Higgs mass, the events which will enter the fiducial phase space
of the Higgs measurement will arise from combinatorial backgrounds of pairs of photons, and thus will not exhibit a 
peak at the Higgs mass. Because of this, they are likely to removed as part of the background fitting and subtraction 
process in that analysis. We therefore do not include the Higgs cross sections when calculating the exclusion limits.

\subsection{Results}

For the CP-even scalar, the cross section in 8 TeV $pp$ collisions calculated by Herwig ranges from 
110~pb for $\Lambda = 1$~TeV to 1.3~pb for $\Lambda = 10$~TeV for $M_\phi = 10$~GeV,
and from 
8.2~pb for $\Lambda = 1$~TeV to 0.12~pb for $\Lambda = 10$~TeV for $M_\phi = 90$~GeV.
For the CP-odd scalar, the cross section in 8 TeV $pp$ collisions calculated by Herwig ranges from 
15~pb for $\Lambda = 1$~TeV to 0.26~pb for $\Lambda = 10$~TeV for $M_\phi = 10$~GeV,
and from 
4.3~pb for $\Lambda = 1$~TeV to 0.077~pb for $\Lambda = 10$~TeV for $M_\phi = 90$~GeV. In all cases, the 
associated production of $\phi$ with a $Z$ or $W$ boson makes the biggest contribution to the cross section, 
although the $\phi + \gamma$ process is significant (10-20\%), and the $\phi + g$ process contributes up to
20\% (40\%) for the highest scale and mass values considered for the CP even (odd) scalar.

At low $M_\phi$ and low-ish $\Lambda$, one of the most sensitive measurements is the $\gamma+E_T^{\rm miss}$ measurement 
from \cite{Aad:2013izg}. The differential cross section as a function of the transverse momentum of the photon 
is shown in Fig.~\ref{fig:cpo_lowlow}, and alone is enough to excludes the model at the 97\% cl.
The inclusive photon measurements are also sensitive, with the 7~TeV diphoton measurement extending to
the lowest mass and $p_T$ values, and the 8~TeV measurement (shown) playing a role once $M_\phi \geq 20$ GeV.

\begin{figure}
\begin{center}
\includegraphics[width=0.49\textwidth]{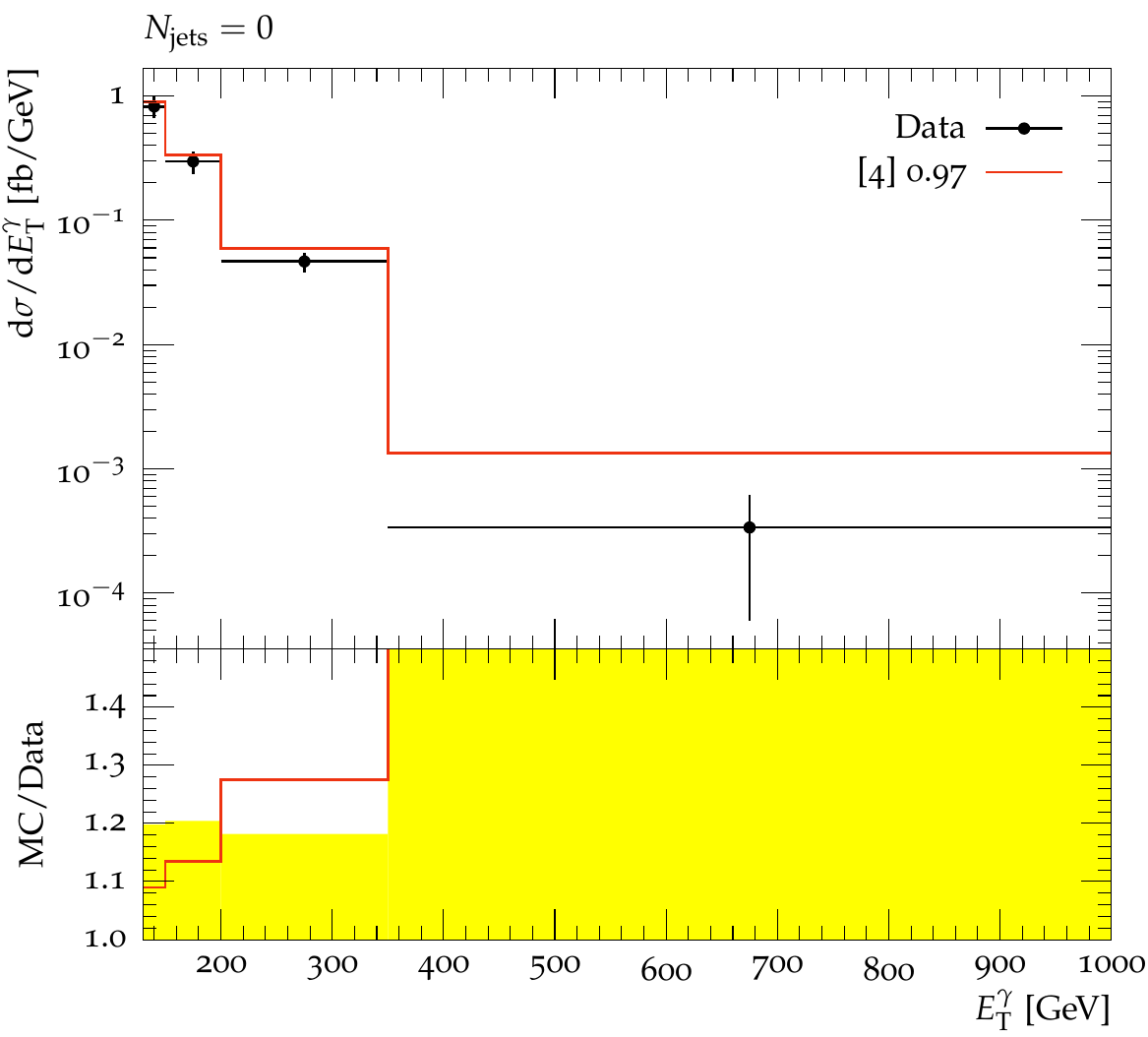}
\includegraphics[width=0.49\textwidth]{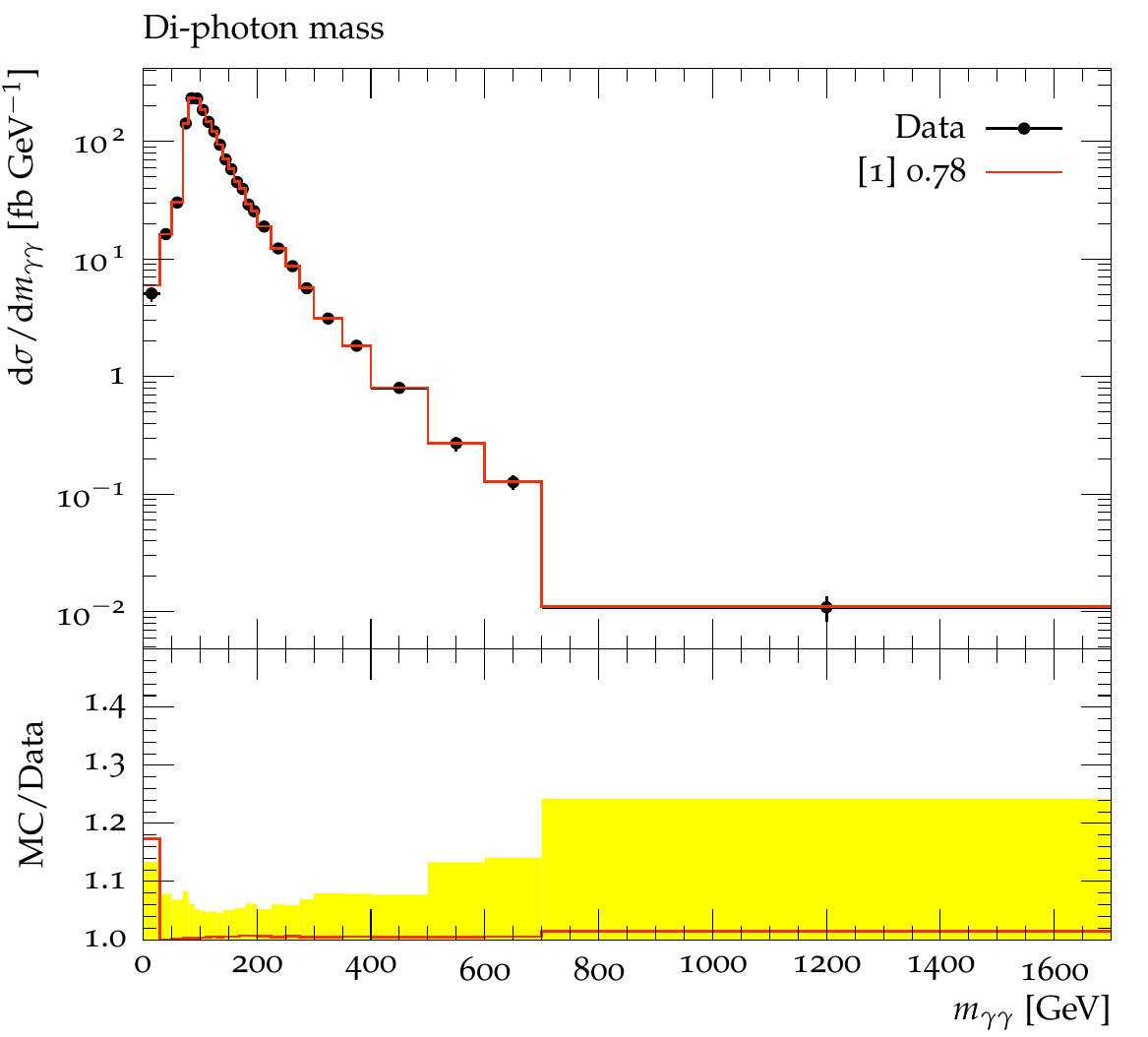}
    \caption{Projection of the contribution of the CP-odd model, (left) for $M_{\phi} = 10$~GeV and $\Lambda = 3500$~TeV, on to the 
8 TeV ATLAS $\gamma+E_T^{\rm miss}$ differential $E_T^\gamma$ cross-section measurement  and (right) on the 
diphoton mass measurement, now with $M_{\phi} = 20$~GeV -- which brings the mass peak from the $\phi$ within
the range of the measurement.
Black points indicate the data, the red upper histogram is the data+BSM. The lower sections of the plots show the ratio of 
(data+BSM)/data, with the yellow band indicating the uncertainty in the measurement. 
The numbers in the legend show the bin number of the most powerful bin, and the exclusion from that bin expressed as a 
probability.}
\label{fig:cpo_lowlow}
\end{center}
\end{figure}

As mentioned in the previous section, events from the model can contribute to the Higgs fiducial two-photon cross section,
and this is seen in the Rivet routine. The major contribution occurs for relatively low $M_\phi$, presumably due to 
$pp \rightarrow \gamma \phi + X \rightarrow \gamma \gamma \gamma + X$ processes in which one pair 
of photons has a mass close to 125~GeV. An example, for $M_\phi = 20$~GeV, $\Lambda = 3.5$~TeV, is shown in Fig.~\ref{fig:higgs_and_cpo}.
The event contribute mainly at low values of $p_T$ for the photon pair. As discussed, this analysis is not used in deriving
the final sensitivity, and is shown only for illustration. {\bf TODO check this. Looks like it is used.}

\begin{figure}
\begin{center}
\includegraphics[width=0.49\textwidth]{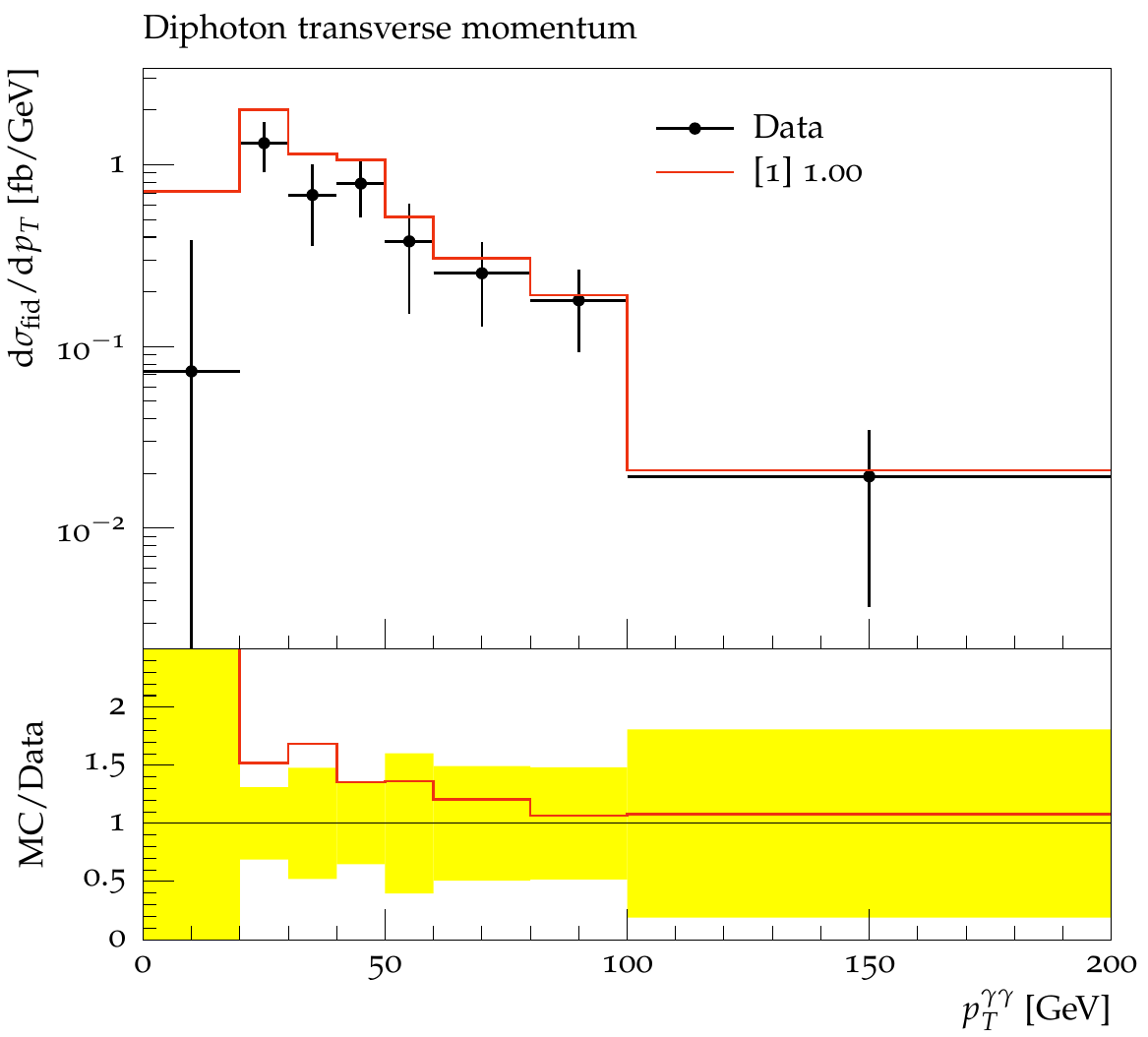}
    \caption{Projection of the contribution of the CP-odd model, for $M_{\phi} = 20$~GeV and $\Lambda = 3500$~TeV, on to the 
8 TeV ATLAS $H \rightarrow \gamma\gamma$ differential $p_T^{\gamma\gamma}$ cross-section measurement.
Legend as Fig.\protect\ref{fig:cpo_lowlow}}
\label{fig:higgs_and_cpo}
\end{center}
\end{figure}

The CP-even model contributes to the same final states, but with a larger cross section for a given coupling. The distributions
for the this model with the same parameter settings as Fig.~\ref{fig:cpo_lowlow} are shown in Fig.\ref{fig:cpe_lowlow}

\begin{figure}
\begin{center}
\includegraphics[width=0.49\textwidth]{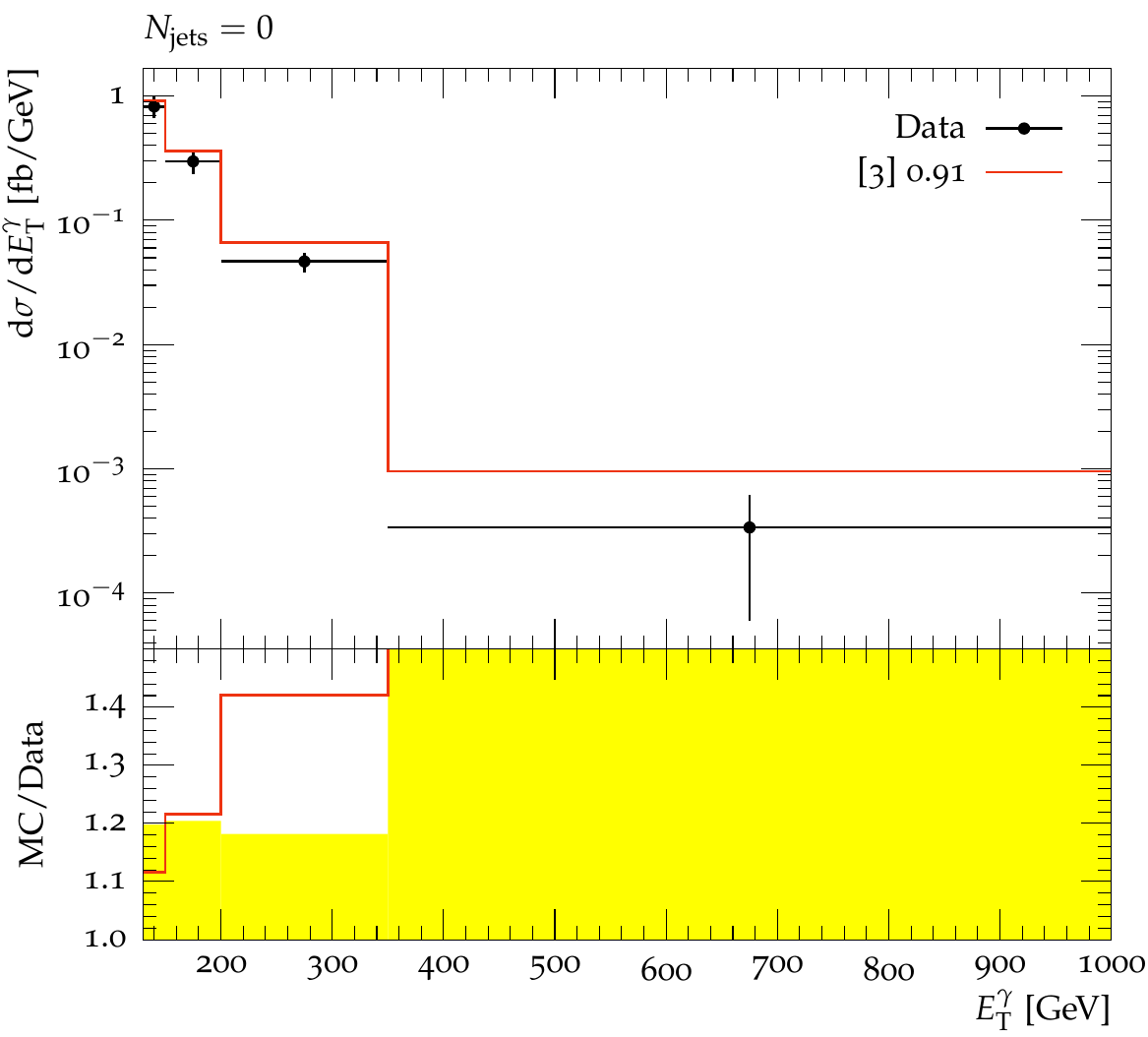}
\includegraphics[width=0.49\textwidth]{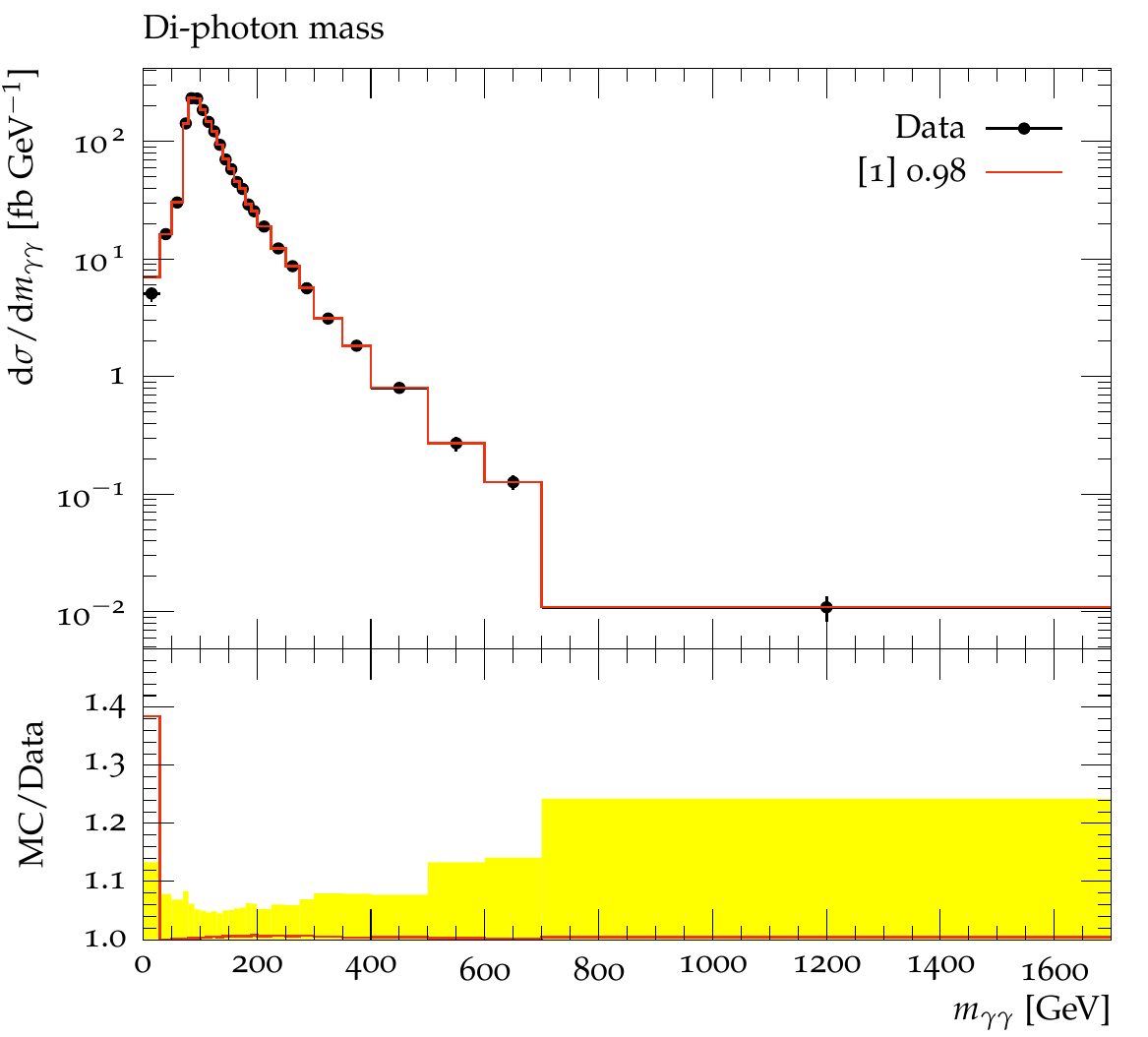}
    \caption{Projection of the contribution of the CP-even model, for $M_{\phi} = 10$~GeV and $\Lambda = 3500$~TeV, on to the 
8 TeV ATLAS $\gamma+E_T^{\rm miss}$ differential $E_T^\gamma$ cross-section measurement (left) and (right) the 
diphoton mass measurement playing a role now with $M_{\phi} = 20$~GeV, which brings the mass peak from the $\phi$ within
the range of the measurement.
Legend as Fig.\protect\ref{fig:cpo_lowlow}}
\label{fig:cpe_lowlow}
\end{center}
\end{figure}

The sensitivity of the combined 7 and 8 TeV data to the CP-odd scalar model is illustrated in Fig.~\ref{fig:cpomaps}.
Dependent on $M_\phi$, the $\Lambda$ values up to 4.5 to 8.5~TeV are excluded, under the assumptions of our procedure. 
Similar sensitivity plots for the CP-even model are shown in Fig.~\ref{fig:cpemaps}. comment on the range.
Precision 13~TeV data can be expected to extend the reach still further; possible dedicated analyses which might extend the
sensitivity still further are discussed in the following section.

\begin{figure}
\begin{center}
\includegraphics[width=0.49\textwidth]{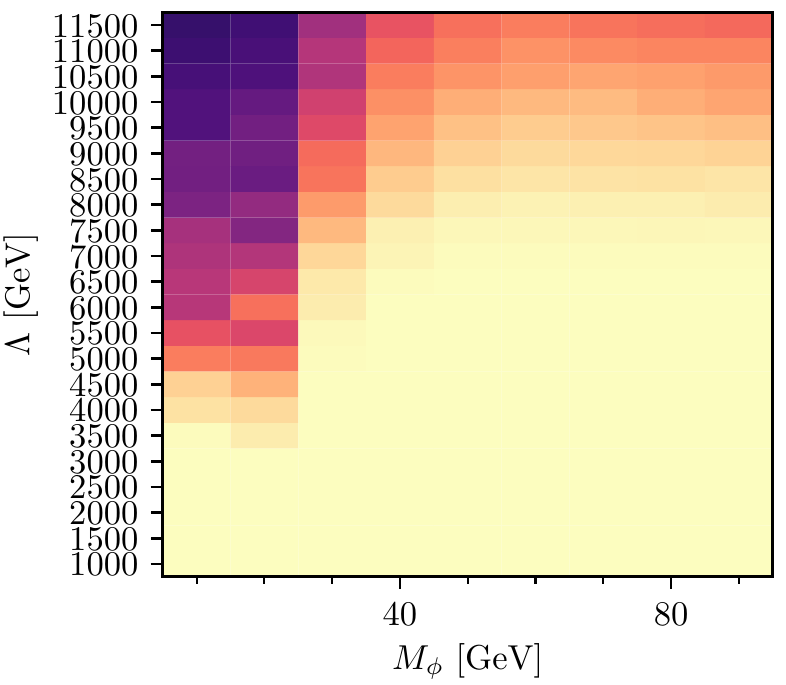}
\includegraphics[width=0.49\textwidth]{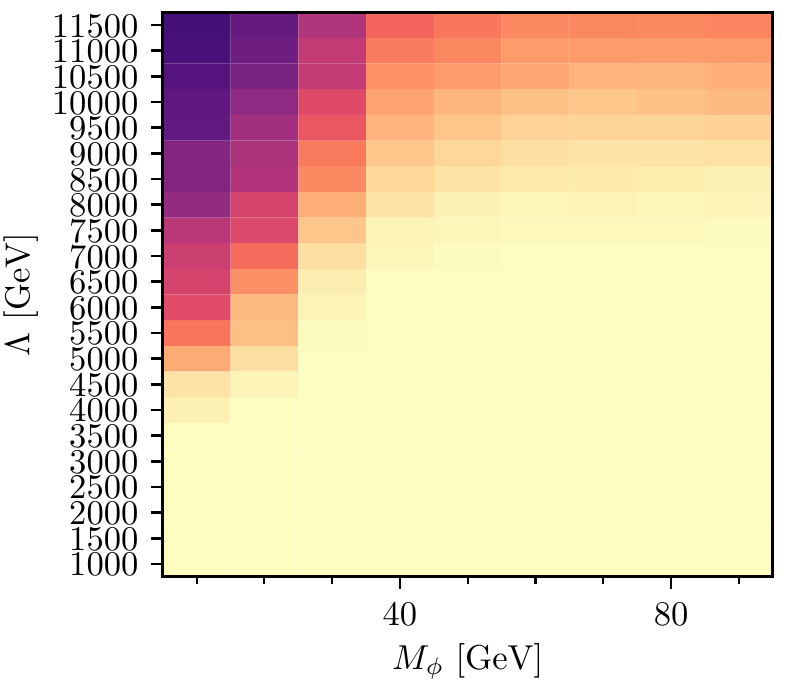}
\includegraphics[width=0.49\textwidth]{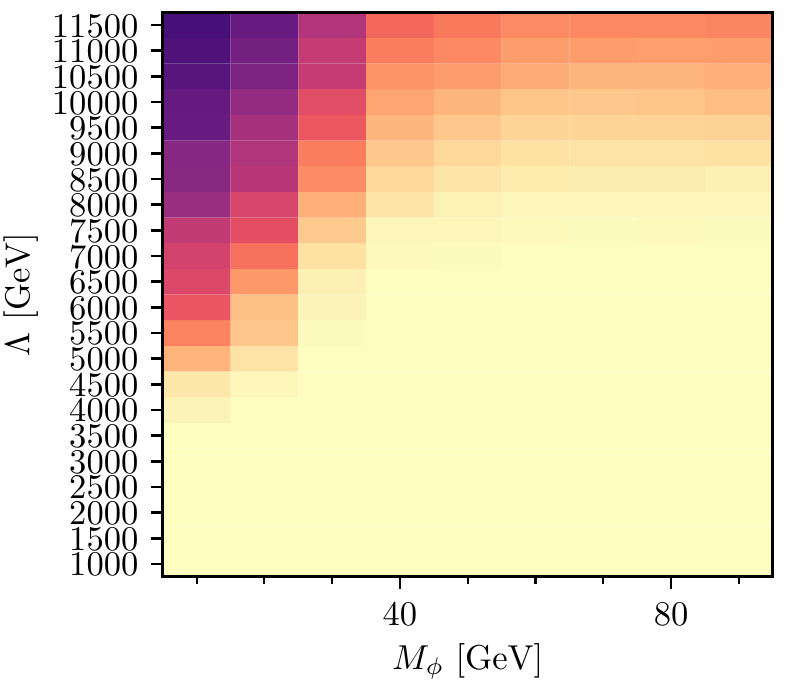}
\includegraphics[width=0.49\textwidth]{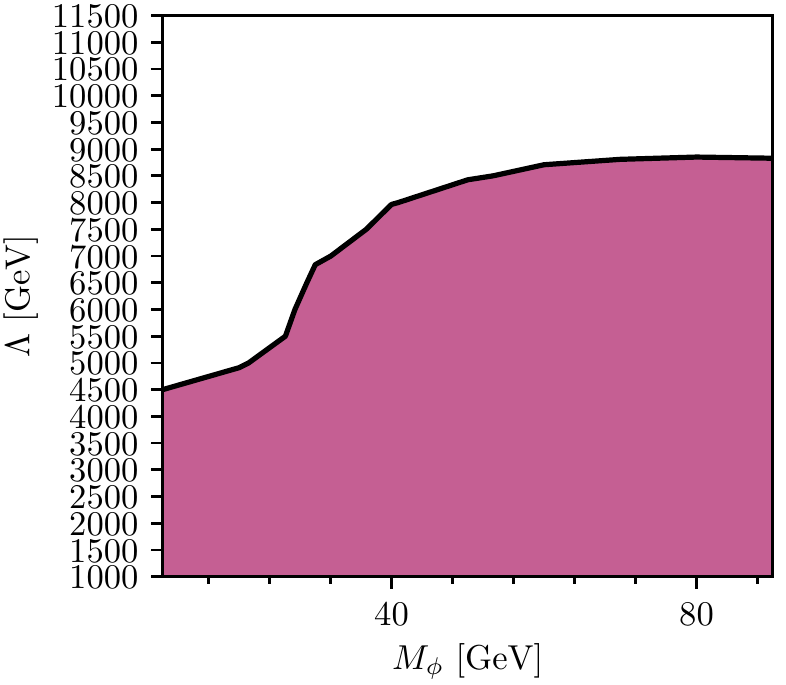}\\
\includegraphics[width=0.99\textwidth]{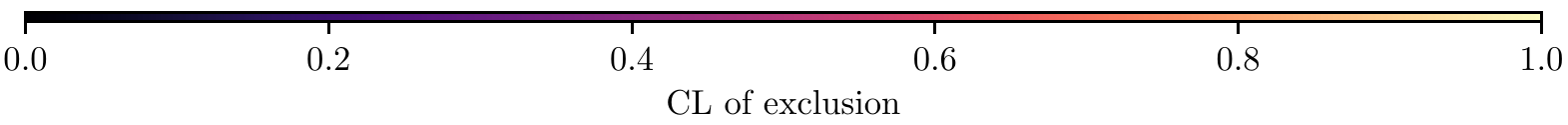}
    \caption{CP-odd scalar model: Top left, exclusion heatmap (with the key below the figure) for 
7 \& 8 TeV diboson measurements (i.e. final states consistent with $WW, ZZ, W+\gamma(\gamma), Z+\gamma(\gamma)$) 
from ATLAS and CMS. Top right, exclusion heatmap for 7 \& 8 TeV photon and diphoton measurements, lower left combined 
exclusion heatmap, bottom right, combined exclusion contour at 95\% c.l.}
\label{fig:cpomaps}
\end{center}
\end{figure}

\begin{figure}
\begin{center}
\includegraphics[width=0.49\textwidth]{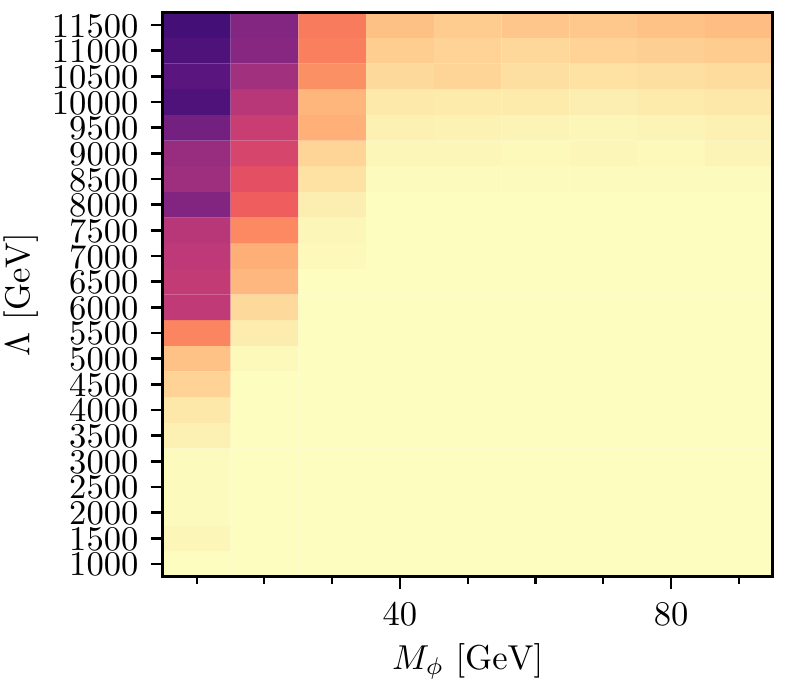}
\includegraphics[width=0.49\textwidth]{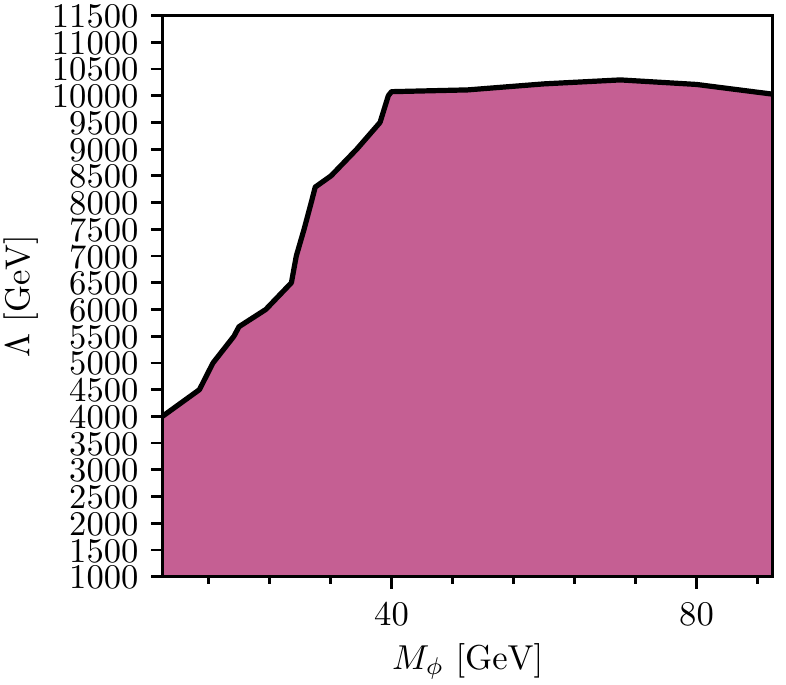}\\
\includegraphics[width=0.99\textwidth]{colorbarkey.pdf}
    \caption{As the lower row of Fig.\protect\ref{fig:cpomaps}, but for the CP-even scalar model.}
\label{fig:cpemaps}
\end{center}
\end{figure}



\section*{CONCLUSIONS}
The generic light scalar models considered here imply significant contributions to differential cross sections involving 
weak bosons and/or isolated photons which have already been measured at the LHC and shown to be consistent with the Standard Model. 
While a rigorous exclusion would require a treatment of the theory uncertainties on the SM photon cross sections, these models
can be considered highly disfavoured for scales below about 4~TeV for $M_\phi = 10$~GeV and up to about 8.5~TeV for 
$M_\phi = 90$~GeV.

\section*{ACKNOWLEDGEMENTS}
We thank the organizers and conveners of the Les Houches workshop, ``Physics
at TeV Colliders'', for a stimulating meeting.
This work has received funding from STFC (UK), and the European Union's Horizon 2020 research and innovation programme 
as part of the Marie Sklodowska-Curie Innovative Training Network MCnetITN3 (grant agreement no. 722104).



\AddToContent{J.~M.~Butterworth, S.~Fichet, L.~Finco, S.~Gascon-Shotkin, D.~Grellscheid,  G.~Moreau, P.~Richardson, D.~Yallup, S.~Zhang}
\renewcommand{\thesection}{\arabic{section}}

\graphicspath{{recasting/}}



\setlength\parindent{0pt}

\chapter{Recasting activities at LH2017}

{A.~Buckley, N.~Desai, B.~Fuks, P.~Gras, D.~Grellscheid, F.~Maltoni, O.~Mattelaer, L.~Perrozzi, P.~Richardson, S.~Sekmen}




\begin{abstract}
We discuss a first benchmark comparison assessing the performance of different public recasting tools in reproducing ATLAS and CMS analysis results.
\end{abstract}

\section{Introduction}

Searches for new physics constitute a primary objective of the LHC physics program.
Their large number and variety pose severe challenges to both the experimental and theory communities. 
In fact, a plethora of searches in different final states are performed by different physics groups in ATLAS and CMS, 
while new ideas to probe new models and non-trivial signatures and to improve the sensitivity of existing searches constantly emerge.
The ultimate goal of this effort is to discover new physics if such
exists within the reach of the LHC, and to test the widest possible range of hypothetical new physics models.

A typical analysis defines quantities to classify events as signal
or background. They include properties of analysis objects such as
jets, electrons, muons, or global event variables such as object multiplicities,
transverse momenta or transverse masses.
An analysis can be very complex and feature many intricate definitions of object and event
variables, some of which cannot be expressed in closed algebraic form and must be defined
algorithmically. This complexity renders the tasks of visualizing, understanding, developing and
interpreting analyses increasingly challenging.

In the paper publications describing the analyses and their results, the experimental collaborations
provide interpretations of the results in terms of one or more theoretical scenarios the analysis has been designed for. 
Often this is done in the context of so-called simplified models, which consider just a subset of physics states and 
production/decay modes out of a full theory.   
There are, however, a multitude of theories beyond the Standard Model and they come in ever increasing variants. 
To fully assess the implications of the LHC searches for new physics requires the interpretation of the experimental results 
in the context of all these models. This is a very active field with close theory-experiment interaction, 
see e.g.~\cite{reinterpretationforum}, and with several public tools being developed for the (re)interpretation of the 
experimental results.\footnote{This includes also dedicated efforts at Les Houches to provide ''Recommendations for Presentation of LHC Results''~\cite{Brooijmans:2012yi,Kraml:2012sg,Brooijmans:2016vro}.} 
In particular,  
CheckMate~\cite{Drees:2013wra,Cacciari:2005hq},
MadAnalysis~\cite{Conte:2012fm,Conte:2014zja,Dumont:2014tja}
and Rivet~\cite{Waugh:2006ip,Buckley:2010ar} 
aim at reproducing experimental analyses in Monte Carlo simulation, including an approximate emulation of detector effects, 
as new physics searches, which have given only null results so far, are typically not unfolded.  
The scope of this contribution is to provide a first benchmark to compare different public tools in reproducing ATLAS and CMS analysis results.\footnote{It is highly  appreciated that many of these results are provided numerically through HEPDATA~\cite{Maguire:2017ypu} or on the collaboration twiki pages.}

\section{Benchmarking tools and comparison strategy}

The idea behind the exercise described in this section is the implementation of LHC analyses of increasing complexity, in different frameworks followed by a comparison of the results. The exercise is performed with three frameworks, CheckMate~\cite{Drees:2013wra,Cacciari:2005hq}, MadAnalysis~\cite{Conte:2012fm,Conte:2014zja,Dumont:2014tja} and Rivet~\cite{Waugh:2006ip,Buckley:2010ar}, followed by a comparison of the results. We choose two analyses for which a detailed cutflow and detector effects were available. In the future it might be beneficial to use dedicated parsers to convert the analysis described in a common format (denoted LHADA in Fig.~\ref{fig:exercise}) into different recasting codes using for instance the technique described in Contribution~\ref{LHADA}.
Once the analysis are available in the needed format, we attempt to reproduce the new physics
interpretations presented in the original experimental research papers,
validating in this way our reimplementations.
A further step consists in the recasting of the analyses within different new
physics contexts and compare the results among the different frameworks.
A sketch of the recasting exercise workflow is presented in Fig.~\ref{fig:exercise}.

Aside the current scope of the exercise, it is interesting to check how the performance of the Delphes simulation behave across different phase spaces, since they are generally referred to as analysis-spedfic.

\begin{figure}
\begin{center}
\includegraphics[width=0.65\textwidth]{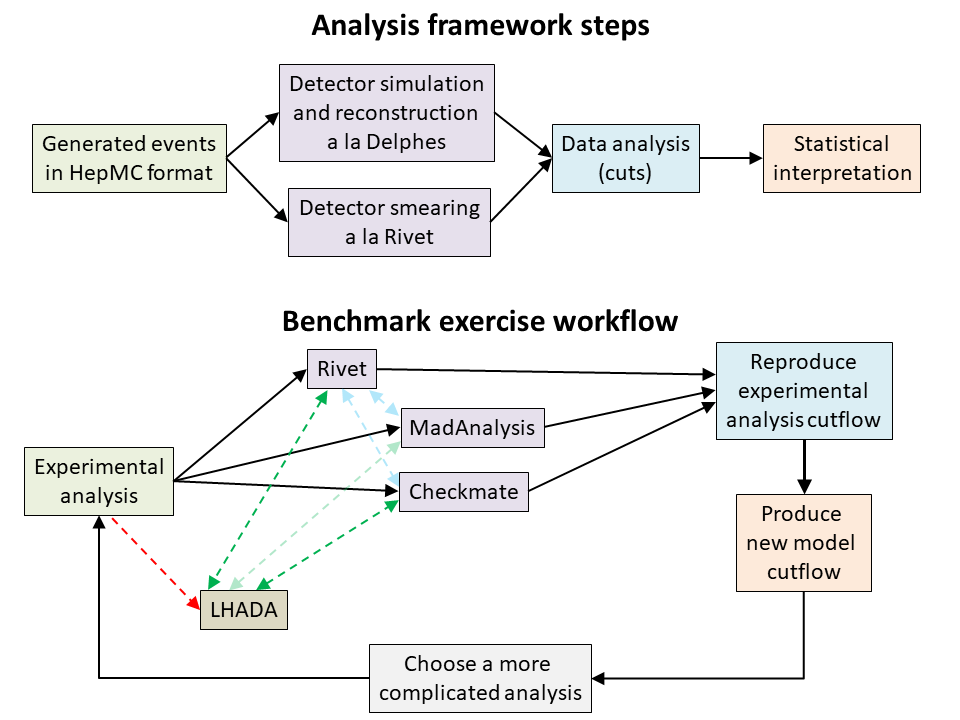}
 \caption{Sketch of the recasting exercise workflow.
}
\label{fig:exercise}
\end{center}
\end{figure}

\subsection{Analysis frameworks and tools}
In this section we describe the analysis frameworks and tools used for the comparison and benchmarking

\subsubsection{CheckMate}
CheckMATE~\cite{Drees:2013wra,Cacciari:2005hq} takes simulated event files in .hep or .hepmc for any model as input and simply returns if the underlying model is `excluded' or `allowed' after
performing a detector simulation and testing various implemented analyses. The embedded AnalysisManager allows for the embedding of additional current and
prospective future LHC results from ATLAS and CMS which have not yet been implemented.
Detector effects are modeled by Delphes with a tune containing efficiency functions for lepton reconstruction and flavour tagging. The soon-to-be published version 2.0 of the code adds the possibility of using Pythia 8~\cite{Sjostrand:2007gs} to
generate supersymmetric events on-the-fly or to shower provided Les Houches
event files for any model. Currently, the collaboration is working on an extension to enable the on-the-fly simulation of events for any model.

\subsubsection{MadAnalysis}
MadAnalysis 5~\cite{Conte:2012fm,Conte:2014zja} is a generic user-friendly
framework for phenomenological investigations at particle colliders, {\it i.e.}
to perform physics analyses of Monte Carlo event files. While prospective
analyses of hard scattering events, parton showered events, hadronized events or
reconstructed events can be designed easily thanks to its Python-based
meta-language, MadAnalysis also allows for the recasting of LHC analyses on new
physics signals provided under the form of .hep and .hepmc event files. The
output here consists in the confidence level at which the model signals are
excluded.
Its Public Analysis Database~\cite{Dumont:2014tja} comprises a growing
collection of LHC analyses which have been implemented in the MadAnalysis 5
framework for the purpose of recasting. Delphes is used for the detector
simulation. For each implemented analysis, a detailed validation note is
provided and the public analysis database follows an open-source policy. Only
contributed codes provided with a detailed validation note are published, and
they are moreover citable via Inspire. The framework being integrated within
MadGraph5\_aMC@NLO~\cite{Alwall:2014hca}, it provides a full recast chain linking a model
and its associated signatures to limit setting.

\subsubsection{Rivet}
Originally developed as a toolkit for the validation of Monte Carlo event
generators, Rivet~\cite{Waugh:2006ip,Buckley:2010ar} (Robust Independent Validation of Experiment and
Theory) has become a standard for documenting (unfolded) Standard Model (SM)
measurements. The top and Higgs physics working groups of all LHC experiments
are increasingly providing Rivet routines for their analyses. Rivet analyses are
written in a user-friendly subset of C++11, and are picked up at runtime as
`plugin libraries'; they can be executed on an event stream either through a Python script interface, or by direct code interfacing to a C++ API.
The original SM-focused requirement of unfolded observables made Rivet
inappropriate for beyond the Standard Model (BSM) searches (other than those
using just jets and missing energy) until the addition of
detector-smearing/efficiency machinery in Rivet~2.5.0. This detector machinery provides equivalent efficiency effects to a Delphes-type simulation, and imitates the less important kinematic smearing of physics objects to within a few percent. A novel feature is that the Rivet detector implementation allows for using
different jet algorithms, lepton and b-tagging operating points, full-detailed object isolation algorithms, and resolutions/efficiencies specific to each
analysis procedure and event selection. This hence allows for a more accurate detector modelling and more robust analysis preservation than `global' detector simulations in addressing some experiment requests for `official fast-sim' tools. The aim is to encourage Rivet code provision directly from BSM data analysers, as is already the case for SM results: additional tools to assist BSM analysis implementation are being added on request.

\section{Analyses benchmarking, comparisons and results}

\subsection{An ATLAS search for supersymmetry in a final
state with jets and missing energy (13 TeV, 3.2~fb$^{-1}$)}

In the analysis of Ref.~\cite{Aaboud:2016zdn}, the ATLAS collaboration targets
the production of the strongly-interacting superpartners of the Standard Model
QCD partons, followed by their decay into jets and missing energy carried by
neutralinos. 3.2~fb$^{-1}$ of proton-proton LHC collisions at a center-of-mass
energy of 13~TeV are analyzed.

The analysis focuses on jets reconstructed by means of the anti-$k_T$
algorithm~\cite{Cacciari:2008gp} with a radius parameter set to $R=0.4$, with
a transverse momentum larger than 20~GeV and a pseudorapidity $|\eta|<2.8$.
Events featuring loosely reconstructed electrons and muons are vetoed.
Event preselection requires a significant amount of missing energy,
$\slashed{E}_T > 200$~GeV and the transverse-momentum of the leading jet is
imposed to be larger than 200~GeV and 300~GeV if two or more than two jets are
reconstructed, respectively.

\begin{table*}[b!]
\footnotesize
 \centering
  \begin{tabular}{ | l || l | l | l || l | l | l || l | }
\hline
    &  \multicolumn{3}{c||}{\bf Rivet} & \multicolumn{3}{c||}{\bf MadAnalysis 5} &   {\bf CheckMATE}   \\ \hline
  Description       & \#evt & tot.eff & rel.eff & \#evt & tot.eff & rel.eff &   tot.eff   \\ \hline \hline
{\bf 2jl cut-flow}                  & 31250 & 1 & - & 31250 & 1 & - & \   \\ \hline
Pre-sel+MET+pT1   & 28592 & 0.91 & 0.91 & 28626 & 0.92 & 0.92 & \   \\ \hline
Njet              & 28592 & 0.91 & 1 & 28625 & 0.92 & 1 & \   \\ \hline
Dphi\_min(j,MET)   & 17297 & 0.55 & 0.6 & 17301 & 0.55 & 0.6 & \   \\ \hline
pT2               & 17067 & 0.55 & 0.99 & 17042 & 0.55 & 0.99 & \   \\ \hline
MET/sqrtHT        & 8900 & 0.28 & 0.52 & 8898 & 0.28 & 0.52 & \   \\ \hline
m\_eff(incl)       & 8896 & 0.28 & 1 & 8897 & 0.28 & 1 & \   \\ \hline
\hline
{\bf 2jm cut-flow} & 31250 & 1 & - & 32150 & 1 & - & 1  \\ \hline
Pre-sel+MET+pT1   & 28472 & 0.91 & 0.91 & 28478 & 0.91 & 0.91 & 0.91  \\ \hline
Njet              & 28472 & 0.91 & 1 & 28477 & 0.91 & 1 & 0.91  \\ \hline
Dphi\_min(j,MET)   & 22950 & 0.73 & 0.81 & 22889 & 0.73 & 0.8 & 0.73  \\ \hline
pT2               & 22950 & 0.73 & 1 & 22889 & 0.73 & 1 & 0.73  \\ \hline
MET/sqrtHT        & 10730 & 0.34 & 0.47 & 10710 & 0.34 & 0.47 & 0.33  \\ \hline
m\_eff(incl)       & 10630 & 0.34 & 0.99 & 10609 & 0.34 & 0.99 & 0.32  \\ \hline
\hline
{\bf 2jt cut-flow} & 31250 & 1 & - & 31250 & 1 & - & \   \\ \hline
Pre-sel+MET+pT1   & 28592 & 0.91 & 0.91 & 28626 & 0.92 & 0.92 & \   \\ \hline
Njet              & 28592 & 0.91 & 1 & 28625 & 0.92 & 1 & \   \\ \hline
Dphi\_min(j,MET)   & 17297 & 0.55 & 0.6 & 17301 & 0.55 & 0.6 & \   \\ \hline
pT2               & 17067 & 0.55 & 0.99 & 17042 & 0.55 & 0.99 & \   \\ \hline
MET/sqrtHT        & 5083 & 0.16 & 0.3 & 5098 & 0.16 & 0.3 & \   \\ \hline
Pass m\_eff(incl)  & 4861 & 0.16 & 0.96 & 4889 & 0.16 & 0.96 & \   \\ \hline
		\end{tabular}
 \caption{Number of events surviving each selection, total and relative
  selection efficiencies as obtained with Rivet and MadAnalysis 5 for the dijet
  signal regions of the multijet+missing energy ATLAS analysis of
  Ref.~\cite{Aaboud:2016zdn}. Partly available Checkmate results for the total
  efficiencies are also indicated.}
	\label{tab:1605.03814-2j}
\end{table*}

\begin{table*}
\begin{tabular}{ | l || l | l | l || l | l | l || l | }
\hline
    &  \multicolumn{3}{c||}{\bf Rivet} & \multicolumn{3}{c||}{\bf MadAnalysis 5} &   {\bf CheckMATE}   \\ \hline
  Description       & \#evt & tot.eff & rel.eff & \#evt & tot.eff & rel.eff &   tot.eff   \\ \hline \hline
\hline
{\bf 4jt cut-flow} & 31250 & 1 & - & 31250 & 1 & - & 1  \\ \hline
Pre-sel+MET+pT1   & 28592 & 0.91 & 0.91 & 28626 & 0.92 & 0.92 & 0.91  \\ \hline
Njet              & 27322 & 0.87 & 0.96 & 27128 & 0.87 & 0.95 & 0.87  \\ \hline
Dphi\_min(j,MET)   & 18929 & 0.61 & 0.69 & 18829 & 0.6 & 0.69 & 0.6  \\ \hline
pT2               & 18715 & 0.6 & 0.99 & 18825 & 0.6 & 1 &     --       \\ \hline
pT4               & 16610 & 0.53 & 0.89 & 16430 & 0.53 & 0.87 & 0.52  \\ \hline
Aplanarity        & 11849 & 0.38 & 0.71 & 11395 & 0.36 & 0.69 & 0.36  \\ \hline
MET/m\_eff(Nj)     & 8334 & 0.27 & 0.7 & 7971 & 0.26 & 0.7 & 0.25  \\ \hline
m\_eff(incl)       & 7201 & 0.23 & 0.86 & 6972 & 0.22 & 0.87 & 0.21  \\ \hline
\hline
{\bf 5j cut-flow} & 31250 & 1 & - & 31250 & 1 & - & 1 \\ \hline
Pre-sel+MET+pT1   & 28592 & 0.91 & 0.91 & 28626 & 0.92 & 0.92 & 0.91 \\ \hline
Njet              & 21234 & 0.68 & 0.74 & 21185 & 0.68 & 0.74 & 0.68 \\ \hline
Dphi\_min(j,MET)   & 14294 & 0.46 & 0.67 & 14292 & 0.46 & 0.67 & 0.45 \\ \hline
pT2               & 14146 & 0.45 & 0.99 & 14289 & 0.46 & 1 &    --       \\ \hline
pT4               & 13229 & 0.42 & 0.94 & 13228 & 0.42 & 0.93 & 0.42 \\ \hline
Aplanarity        & 9836 & 0.31 & 0.74 & 9576 & 0.31 & 0.72 & 0.3 \\ \hline
MET/m\_eff(Nj)     & 4643 & 0.15 & 0.47 & 4506 & 0.14 & 0.47 & 0.13 \\ \hline
m\_eff(incl)       & 4620 & 0.15 & 1 & 4476 & 0.14 & 0.99 & 0.13 \\ \hline
\hline
{\bf 6jm cut-flow} & 31250 & 1 & - & 31250 & 1 & - & 1  \\ \hline
Pre-sel+MET+pT1   & 28592 & 0.91 & 0.91 & 28626 & 0.92 & 0.92 & 0.91  \\ \hline
Njet              & 13235 & 0.42 & 0.46 & 13236 & 0.42 & 0.46 & 0.41  \\ \hline
Dphi\_min(j,MET)   & 8520 & 0.27 & 0.64 & 8553 & 0.27 & 0.65 & 0.26  \\ \hline
pT2               & 8436 & 0.27 & 0.99 & 8551 & 0.27 & 1 &    --        \\ \hline
pT4               & 8135 & 0.26 & 0.96 & 8217 & 0.26 & 0.96 & 0.25  \\ \hline
Aplanarity        & 6365 & 0.2 & 0.78 & 6307 & 0.2 & 0.77 & 0.19  \\ \hline
MET/m\_eff(Nj)     & 2675 & 0.09 & 0.42 & 2665 & 0.09 & 0.42 & 0.08  \\ \hline
m\_eff(incl)       & 2670 & 0.09 & 1 & 2656 & 0.08 & 1 & 0.08  \\ \hline
\hline
{\bf 6jt cut-flow} & 31250 & 1 & - & 31250 & 1 & - & \   \\ \hline
Pre-sel+MET+pT1   & 28592 & 0.91 & 0.91 & 28626 & 0.92 & 0.92 & \   \\ \hline
Njet              & 13235 & 0.42 & 0.46 & 13236 & 0.42 & 0.46 & \   \\ \hline
Dphi\_min(j,MET)   & 8520 & 0.27 & 0.64 & 8553 & 0.27 & 0.65 & \   \\ \hline
pT2               & 8436 & 0.27 & 0.99 & 8551 & 0.27 & 1 & \   \\ \hline
pT4               & 8135 & 0.26 & 0.96 & 8217 & 0.26 & 0.96 & \   \\ \hline
Aplanarity        & 6365 & 0.2 & 0.78 & 6307 & 0.2 & 0.77 & \   \\ \hline
MET/m\_eff(Nj)     & 3900 & 0.12 & 0.61 & 3839 & 0.12 & 0.61 & \   \\ \hline
m\_eff(incl)       & 3715 & 0.12 & 0.95 & 3672 & 0.12 & 0.96 & \   \\ \hline
 \end{tabular}
 \caption{Same as in Table~\ref{tab:1605.03814-2j} but for the signal regions
  targeting final states containing four, five and six jets.}
	\label{tab:1605.03814-nj}
\end{table*}

The analysis is then divided into seven signal regions focusing on different jet
multiplicities (from 2 to 6) with different transverse-momentum thresholds. The
missing transverse momentum is then enforced to be well separated from the
leading reconstructed jets, and its significance is constrained for events
featuring only two jets. For cases where at least four jets are reconstructed,
additional selections on the aplanarity variable
and the effective mass, {\it i.e} the scalar sum of the transverse momenta of the
reconstructed and the missing transverse energy.

Implementations of this analysis are available in Checkmate, MadAnalysis\,5 (recast code \cite{ma5-multijet}) 
and Rivet (ATLAS\_2016\_I1458270~\cite{rivet-multijet}). 

We generated signal events for a gluino pair production in the simplified model considered in Ref~\cite{Aaboud:2016zdn} with a direct decay of the gluino into SM particles and the lighest supersymmetric particle (LSP). The gluino mass is set to 1.6$\,$TeV mass and the LSP is assumed to be massless. The pseudo-data samples have been generated by using MadGraph5\_aMC@NLO~\cite{Alwall:2014hca} and Pythia8~\cite{Sjostrand:2014zea}.

The comparison of
predictions for the cutflows as obtained with MadAnalysis\,5  and Rivet are
reported in Table~\ref{tab:1605.03814-2j} and Table~\ref{tab:1605.03814-nj} for
all seven signal regions.  The tables include the total number of events
surviving each selection, the associated cut efficiency and the total efficiency
evaluated with respect to the initial number of events. Partially available
Checkmate results are also indicated for what concern the total efficiencies and
for a few signal regions.
An excellent agreement between the three codes has been obtained.

\subsection{An ATLAS search for dark matter in the monophoton final state
  (13~TeV,  36.1~fb$^{-1}$)}

In the analysis of Ref.~\cite{Aaboud:2017dor}, the ATLAS collaboration has
searched for dark matter when it is produced in association with a very
energetic photon. The search results have been reinterpreted in dark matter
simplified scenarios in which a pair of dark matter particles is produced in
association with a photon arising from initial state radiation. 36.1~fb$^{-1}$
of proton-proton LHC collisions at a center-of-mass energy have been analyzed.

The analysis requires the presence of at least one tightly-isolated photon with
a transverse energy $E_T > 150$~GeV and with a pseudorapidity satisfying
$|\eta| < 2.37$, the pseudorapidity region $1.37 < |\eta| < 2.37$ being
excluded. Events featuring loose eletrons and muons and more than one jets with
a transverse momentum larger than 30~GeV and a pseudorapidity $|\eta|<4.5$ are
vetoed. As in the previous analysis, jets are reconstructed by means of the
anti-$k_T$ algorithm~\cite{Cacciari:2008gp}  and a radius parameter set to
$R=0.4$. In addition, event selection requires a missing transverse energy
significance larger than 8.5~GeV$^{1/2}$, and the missing transverse momentum
has top be well separated from the photon and the jet (for events featuring one
reconstructed jet).

Five signal regions are defined according to different requirements on the
amount of missing transverse energy, namely three inclusive regions and two
non-overlapping exclusive regions.

We generated events using the simplified model of dark matter (DM) production involving an axial-vector operator, Dirac DM and couplings g$_{\mathrm{q}}= 0.25$ and g$_{\chi} = 1$ with m$_{\chi} = 10\,$GeV and m$_{\mathrm{med}}=800\,$GeV described in Ref.~\cite{Aaboud:2017dor}.

In Table~\ref{tab:1704.03848}, we compare the total number of events surviving each selection, the associated cut
efficiency and the total efficiency evaluated with respect to the initial number
of events as obtained with MadAnalysis5 (recast code~\cite{ma5-monophoton}) 
and Rivet. 
Whilst a fair agreement is
obtained between two codes, differences of 5\%--10\% are observed for a few 
cuts. This can be traced back to the missing energy modelling that is complicated
to reproduce. 
The final acceptances of about 40\% (Rivet) and 38\% (MadAnalysis) are however 
in good agreement.

\begin{table*}
 \centering
  \begin{tabular}{ | l || l | l | l || l | l | l | }
\hline
                  &  \multicolumn{3}{c||}{\bf Rivet} & \multicolumn{3}{c||}{\bf MadAnalysis 5}    \\ \hline

Description       & \#evt & tot.eff & rel.eff & \#evt & tot.eff & rel.eff    \\ \hline \hline

Initial                    &  	1198	& 1		  & -     & 1198	& 1	 &    -      \\ \hline
ETmiss $>$ 150 GeV           &   	798.3	& 0.67	& 0.67	& 736	& 0.61 &  0.61     \\ \hline
Photon w/ ET $>$ 150 GeV     &   	703.5	& 0.59	& 0.88	& 700	& 0.58 &  0.95     \\ \hline
Pass Tight photon          &   	598.1	& 0.50	& 0.85	& 658	& 0.55 & 	0.94     \\ \hline
Pass Isolated photon       &   	598.1	& 0.50	& 1.00	& 620	& 0.52 & 	0.94     \\ \hline
Pass $\delta\phi$(gamma,MET) $>$ 0.4 &   	597.5	& 0.50	& 1.00	& 596	& 0.50 & 	0.96     \\ \hline
Pass MET/sqrt(SET) $>$ 8.5   &   	538.2	& 0.45	& 0.90	& -	  &  -   &     	     \\ \hline
Pass Jet veto              &   	476.8	& 0.40	& 0.89	& 461	& 0.38 & 	0.77     \\ \hline
Pass Lepton veto           &   	475.5	& 0.40	& 1.00	& 460	& 0.38 & 	1.00     \\ \hline
  \end{tabular}
 \caption{Number of events surviving each selection, total and relative
  selection efficiencies as obtained with Rivet and MadAnalysis 5 for the SRI1
  signal region of the monophoton ATLAS analysis of Ref.~\cite{Aaboud:2017dor}.}
 \label{tab:1704.03848}
\end{table*}


\section*{CONCLUSIONS}
We presented a first benchmark comparison of the performance of different recasting tools 
which reproduce LHC analyses in Monte Carlo simulation.
For the two cases treated here, good agreement is found between the different frameworks and detector simulation techniques. 
The comparison is ongoing with several more analyses which are currently being validated. 
It will also be interesting to compare performances for different signal scenarios, to assess the reliability of the 
recasting methods in, e.g.\  extreme regions of phase space and/or for very different signal hypotheses the the 
one the analyses have been designed for.

\section*{ACKNOWLEDGEMENTS}
The authors thank the organizers of the Les Houches Workshop Series ``Physics at TeV Colliders 2017'' 
for the fruitful and pleasant atmosphere, and Sabine~Kraml for the numerous discussions and the help in finalising the contribution.



\AddToContent{A.~Buckley, N.~Desai, B.~Fuks, P.~Gras, D.~Grellscheid, F.~Maltoni, O.~Mattelaer, L.~Perrozzi, P.~Richardson, S.~Sekmen}
\renewcommand{\thesection}{\arabic{section}}

\graphicspath{{LLP_recast/}}
\newcommand{\com}[1]{\emph{\color{red}[#1]}}  	
\newcommand{\prop}[1]{{\color{red}#1}}			
\newcommand{\drop}[1]{{\color{red}\sout{#1}}}


\chapter{Recasting Long-Lived Particles Searches}

{\it G.~Cottin, N.~Desai, J.~Heisig, A.~Lessa}


\begin{abstract}
Long-lived particles (LLPs) arise in several
beyond the Standard Model (BSM) theories, and they provide striking (non-standard) signatures
at colliders. Several LHC searches look for LLP models in a broad range
of final states, and limits have been presented for specific BSM models.
However, extrapolating such limits to other scenarios often proves to be
a difficult task outside the experimental collaborations.
This note discusses the recasting of three types of LLP signatures: displaced
vertices, displaced leptons and heavy stable charged particles.
Several conclusions are obtained from these recasting attempts and recommendations
to the experimental collaborations are made.  
\end{abstract}

\section{INTRODUCTION}

The key scientific goal for the second run of the LHC is to explore
physics beyond the Standard Model (BSM). Motivated by a variety of BSM theories,
there has been a growing interest in non-standard signatures, such as long-lived
particles (LLP). A large variety of LLP signatures have been explored by the
experimental searches and the re-interpretation of these results
within the context of new BSM theories is extremely relevant
to exploit the full potential of the LHC experiment.
However, in most cases it is not possible to reinterpret LLP searches 
using fast detector simulation and a cut-and-count based analysis, as it is
usually done for prompt searches.
In particular, efficiencies for object reconstruction (such as displaced
vertices) and event selection are much more analysis dependent in LLP searches
and difficult to reproduce using fast simulation.
Furthermore, trigger and pile-up vetoes included in prompt searches
are difficult to reproduce and can invalidate
the extrapolation of prompt search results and limits to long-lived (displaced)
scenarios.

In this note we discuss the difficulties of recasting LLP searches.
In order to make the discussion concrete, we present
results for the recasting of three distinct LLP signatures: displaced vertices,
displaced leptons and heavy stable charged particles.
As a way to discuss the typical issues encountered  when recasting LLP
searches, we try to reproduce the official exclusion curves presented in the
13~TeV ATLAS displaced vertex plus missing energy~\cite{ATLAS:2017bvh}, 
the 8~TeV CMS displaced lepton~\cite{Khachatryan:2014mea} and 
the 13~TeV CMS heavy stable charged particle~\cite{CMS-PAS-EXO-16-036} searches.
Although the issue of recasting prompt searches within the context of LLP
models is a very relevant one, we do not discuss it here.

The first difficulty related to recasting LLP searches
concerns the detector simulation of such signatures.
While the relevant signatures for stable particles (in detector scales)
are charged tracks and missing energy,
for a particle decaying within the detector several
signatures are possible, depending on the LLP nature and its decay.
The lifetime of the particle and its boost are also 
essential features, since only certain parts of the detector
are capable of observing specific decay products.
Furthermore, the Standard Model background typically decreases as
the LLP decay moves further away from the primary vertex.
In general the LLP signatures can be classified as follows:
\begin{itemize}
    \item {\it charged track} (stable charged particle) 
    \item {\it disappearing track} (charged LLP decaying to a neutral/soft final
    state)
    \item {\it displaced vertex} (charged or neutral LLP decaying to charged
    final states)
    \item {\it track kink} (charged LLP decaying to neutral and charged final
    states)
    \item {\it trackless jets, displaced leptons} (neutral LLP decaying to
    charged final states)
    \item {\it missing energy}\footnote{The missing energy signature is usually
    not classified as a LLP search, since it is covered by several prompt searches.
    Nonetheless, we include it here for completeness.} (stable neutral particle) 
\end{itemize} 
Typical fast detector simulators do not yet include the
information required for dealing with the above signatures.
As a result, dedicated recasting tools must be developed to deal with LLP searches.
In this note we mainly make use of {\sc MadGraph5}~\cite{Alwall:2014hca} and
\textsc{Pythia}~\cite{Sjostrand:2006za,Sjostrand:2014zea} to simulate hadron level
events and no fast detector simulation is employed.
Several approaches are then discussed in order to emulate the experimental
selection and reconstruction efficiencies and reproduce
the official exclusion curves presented in the corresponding analyses.

The recasting of the ATLAS displaced vertex search is presented
in Section~\ref{sec:DV}, where two methods are employed in order to
reproduce the exclusion curves for a simplified long-lived gluino scenario.
While the first method makes use of the limited information provided by the
corresponding conference note, the second uses the full efficiencies provided
by the ATLAS auxiliary material. The goal of comparing these two approaches is
to illustrate how lack of experimental information drastically decreases the recasting
performance.
Sections~\ref{sec:Dlep} and~\ref{sec:HCP} discuss searches
based on isolated tracks, which, in principle, are  much
simpler to recast.
Although CMS has provided detailed efficiencies for the 8~TeV searches,
we will show that these can not be easily extrapolated to the 13~TeV results.
We illustrate this by recasting two CMS searches --- displaced lepton search (Section~\ref{sec:Dlep}) and the long-lived charged
particle search (Section~\ref{sec:HCP}).
Finally, in Section~\ref{sec:llpconclusions} we present the overall conclusions and
recommendations drawn from the recasting of these particular searches.

\section{DISPLACED VERTEX SEARCH}
\label{sec:DV}

The ATLAS displaced vertex $+$ missing energy analysis presented
in Ref.~\cite{ATLAS:2017bvh} investigate an important BSM
scenario: long-lived gluinos decaying to jets and missing energy.
The analysis searches for displaced vertices (DV) in association with large missing transverse
energy, a signature present in several BSM models (long-lived stops,
hidden valley scenarios and others).
Hence it is relevant to investigate how well it is possible
to recast this analysis and extend its constraints to other LLP models.
One important feature of the ATLAS search is that it does not rely on
the gluino (or $R$-hadron) charge, so it can be directly applied to both charged
and neutral LLPs.

In Ref.~\cite{ATLAS:2017bvh} the results are interpreted in the long-lived gluino scenario with
R-Parity conservation, the lightest neutralino being the LSP and the
gluino the NLSP. Hence, after being pair produced, the gluinos hadronize and
then decay to jets and the LSP with a $100\%$ branching ratio:
\begin{equation}
p p \to \tilde{g} \tilde{g} \to (j j \tilde{\chi}_1^0) + (j j
\tilde{\chi}_1^0)
\end{equation}
Limits are presented for gluino lifetimes between 0.02~ns and 20~ns, gluino
masses of 1.4 and 2 TeV and several values of $m_{\tilde \chi_1^0}$.

The search imposes the following criteria for
selecting displaced vertices:

\begin{enumerate}
  \item {Missing energy selection in the event: $E_\text{T}^{\text{miss}} >
  250$~GeV. }
  \item {{\it{Base vertex selection}} (at least one DV in the event): }
  \vspace{0.2cm}
  \begin{itemize}
    \item The DV coordinates must satisfy: $R_\text{DV} = \sqrt{x^2 + y^2} <
    300$~mm and  $|z_\text{DV}| < 300$~mm.
    \item The DV must not fall into a material rich area. This criterion
    corresponds to discarding approximately 42\% of the fiducial volume.
    \item The vertex must be separated by more than 4\,mm from all primary vertices.
   \end{itemize}
   \item Signal region selection:
   \vspace{0.2cm}
   \begin{itemize}
     \item the invariant mass of the DV must be $m_\text{DV} > 10$~GeV, assuming all its tracks have the pion mass.
     \item $n_\text{tracks} \geq 5$, where $n_\text{tracks}$ corresponds to the number of tracks
			originating from the vertex and satisfying:	$p_\text{T} > 1$~GeV  and $|d_0| > 2$~mm.
   \end{itemize}
\end{enumerate}

After applying the above selections, no displaced vertices were observed 
in $32.1$~fb$^{-1}$ of data at $\sqrt{s} = 13$~TeV.
The number of expected background displaced vertices is:
\begin{equation}
N_\text{DV}^\text{BG} = 0.02 \pm 0.02
\end{equation}

This analysis is particularly interesting because it provides a large set of information
useful for recasting.\footnote{We point out that the conference note ATLAS-CONF-2017-026,
which has been superseded by Ref.~\cite{ATLAS:2017bvh}, included only a subset of the information
present in the publication.}
Specially useful are the efficiency grids provided for DV reconstruction and event selection
as a function of the relevant (truth level) variables: number of tracks, DV mass, DV position
and missing energy.
We point out that these detailed efficiencies are usually not provided for most LLP searches.
The corresponding 8 TeV search in Ref.~\cite{Aad:2015rba}, for instance, only provided reconstruction
and event-level efficiencies as a function of a single parameter.
Therefore, we will discuss below how recasting performs when distinct levels of information
are available. In particular, we will discuss two approaches:
\begin{itemize}
    \item {\it Method 1}: recasting using correction functions for the vertex reconstruction
        efficiency and the track efficiency.
    \item {\it Method 2}: recasting using the efficiency grids provided by ATLAS in Ref.~\cite{ATLAS:2017bvh}.
\end{itemize}

\subsection{Method 1: Recasting using correction functions}
\label{method1}

The first approach only makes use of the displaced vertex reconstruction efficiency
provided by ATLAS and shown as red points in Fig.\ref{fig:RDVeff}.
We point out that this data is provided only for $m_{\tilde g} = 1.2$~TeV
and $\tau = 1$~ns.
The method discussed below tries to construct functions which aim to approximate
the experimental track and vertex reconstruction efficiencies. These will then
be applied to the event selection and used to compute 
upper limits for the same gluino scenarios considered by ATLAS.

In order to recast the ATLAS displaced vertex $+$ missing energy search, we
use \textsc{Pythia}~8~\cite{Sjostrand:2014zea} for the simulation of gluino
production, hadronization and decays. No smearing is applied and the truth-level $E_\text{T}^{\text{miss}}$
is considered.  All charged particles generated by the $R$-hadron decay
are considered as potential charged tracks and only
these are included when computing $m_\text{DV}$.
Furthermore, $d_0$ is calculated with the assumption
of a zero magnetic field. The veto of decays in material rich area
is implemented simply as an overall fiducial volume cut.
Since a dedicated algorithm was used by the collaboration to identify and
select DV candidates, it can not be easily reproduced.
However the vertex reconstruction efficiency as a function of $R_\text{DV}$ has
been provided in Ref.~\cite{ATLAS:2017bvh} for  
$\big( m_{\tilde{g}}, m_{\tilde{\chi}_1^0}, \tau  \big) =  \left( 1200 \mbox{
GeV}, 100 \mbox{ GeV}, 1 \mbox{ ns} \right)$
and
$\big( m_{\tilde{g}}, m_{\tilde{\chi}_1^0}, \tau  \big)  =  \left( 1200 \mbox{
GeV}, 1170 \mbox{ GeV}, 1 \mbox{ ns} \right)$.

As a first step in the recasting procedure, we test how well the vertex
reconstruction efficiency ($\epsilon_\text{DV}$) can be reproduced under the
assumption that all displaced vertices satisfying the {\it base
vertex selection} cuts are reconstructed.
The efficiencies provided by ATLAS and the result obtained from recasting
are shown in Fig.\ref{fig:RDVeff}.
As we can see, $\epsilon_\text{DV}$ decreases with $R_\text{DV}$, except for
the first bin (0 mm $< R_\text{DV} < 5$ mm).
The slope seen in the recasting curve is purely due to
the {\it base vertex selection} cuts applied.
The same behavior is seen in the official data, although the
efficiency falls much faster, likely due to detector effects and the
reconstruction algorithm.
For the low bins (except for the first one) the two curves differ by
$\simeq 20\%$, while for the largest bins the difference is almost an order of
magnitude.
The high level of agreement in the first bin ($R_\text{DV} < 5$~mm) is an
artificial effect due to the requirement that the displaced and primary vertices
must be separated by at least 4 mm.

\begin{figure}[t]
\begin{center}
\includegraphics[scale=0.6]{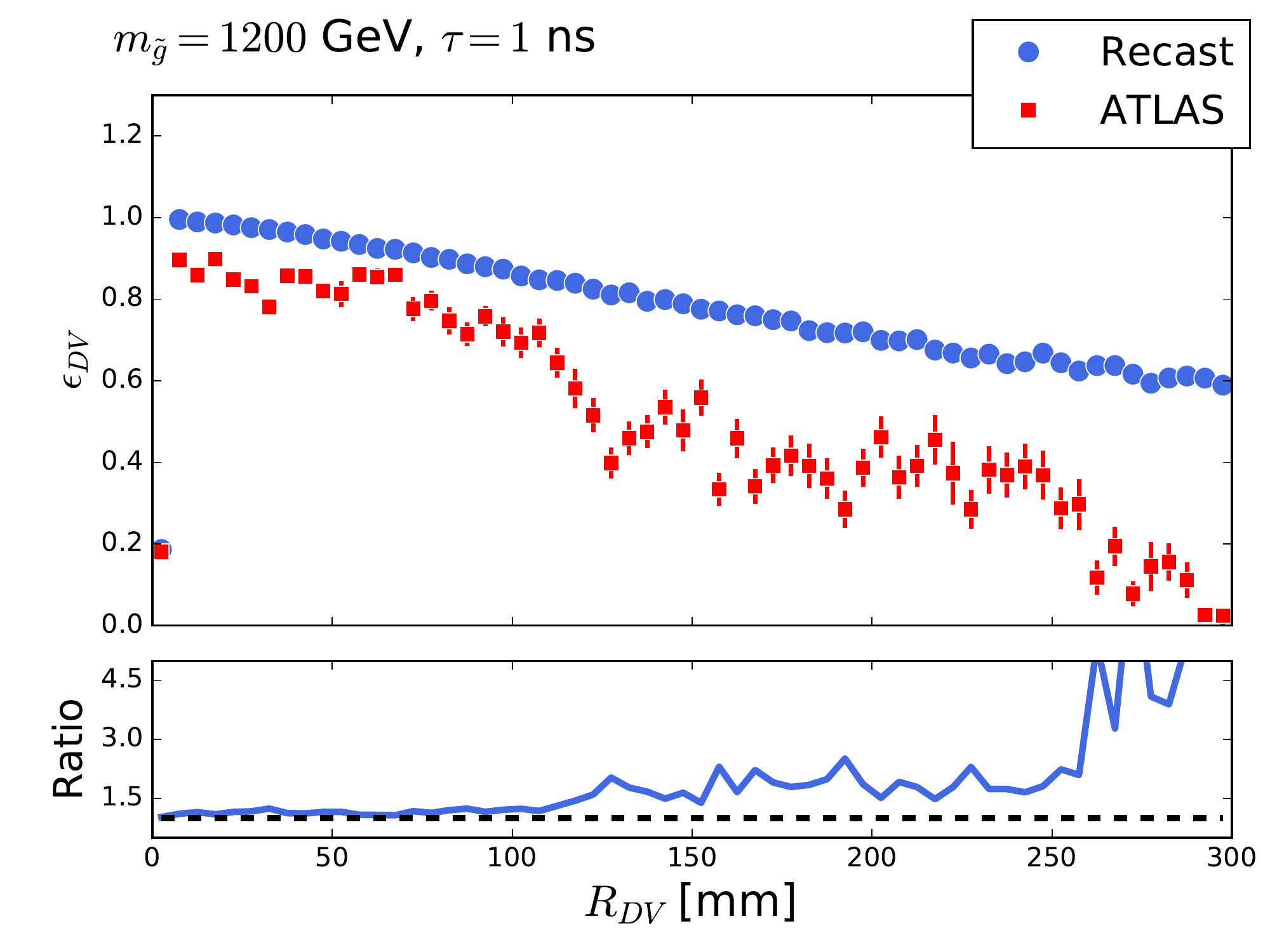}
\caption{Vertex reconstruction efficiency as a function of the
displaced vertex transverse position, $R_\text{DV}=\sqrt{x^2 + y^2}$.
The red points correspond to the results obtained by
ATLAS in Ref.~\cite{ATLAS:2017bvh}, while the blue points correspond to
the results obtained through recasting after the {\it base vertex selection}
cuts have been applied. The results refer to the benchmark
point $\big( m_{\tilde{g}}, m_{\tilde{\chi}_1^0}, \tau  \big)  =  \left( 1200 \mbox{
GeV}, 100 \mbox{ GeV}, 1 \mbox{ ns} \right)$.}
\label{fig:RDVeff}
\end{center}
\end{figure}

As shown in Fig.\ref{fig:RDVeff}, the reconstruction efficiency obtained
by recasting can be overestimated by more than an order of magnitude for high
values of $R_\text{DV}$. One is then tempted to directly apply the vertex
reconstruction efficiencies provided by ATLAS (red points in Fig.\ref{fig:RDVeff})
to the hadron level events generated in \textsc{Pythia}.
However these have been derived for the benchmark point
$m_{\tilde{g}} = 1.2$~TeV, $m_{\tilde{\chi}_1^0} = 100$~GeV, $\tau = 1$~ns
and are not necessarily valid for other values of the gluino and LSP masses or
the lifetime.
In fact, as shown in Ref.~\cite{ATLAS:2017bvh}, the efficiencies can be
affected by the mass difference $m_{\tilde{g}}-m_{\tilde{\chi}_1^0}$.
We have also computed the reconstruction efficiency using distinct values of
the gluino lifetime. The results are shown in Fig.\ref{fig:RDVeff2}.
Despite the fluctuations at high $R_\text{DV}$ (due to limited statistics),
we can see that the efficiency also has a strong dependence on the gluino
lifetime. 
Therefore we conclude that the ATLAS efficiencies shown in
Fig.\ref{fig:RDVeff} can not be directly applied to any input model.
Also we can not neglect the large impact of detector response and the
reconstruction algorithm, as illustrated by the difference between the blue
and red points in Fig.\ref{fig:RDVeff}.
In order to proceed with the recasting (using only the data from Fig.\ref{fig:RDVeff}) we will
adopt the following {\it strategy}:

\begin{enumerate}[label=\roman*.]
  \item Apply the {\it base vertex selection} cuts.
  \item Rescale the DV reconstruction efficiencies obtained from the vertex
  selection cuts above by a factor $r\left(R_{\text{DV}}\right)$. This
  ``correction factor'' ($r$) is defined by the ratio of the red and blue points in
  Fig.\ref{fig:RDVeff}. This factor aims to encapsulate the experimental
  features which are not captured by the base vertex selection cuts and
  is assumed to be model independent.
  \item Apply a constant track efficiency\footnote{The track efficiency clearly depends on the
   track $p_\text{T}$ and production position, hence a constant efficiency is
   an oversimplification. We have tried distinct functional forms
   for $\epsilon_\text{track}$ following a procedure similar to the one
   described in Refs.~\cite{Allanach:2016pam,Liu:2015bma}. None of these, however, perform much better
   (for all values of $\tau$) than the results presented here assuming a constant efficiency.},
   which aims to approximate how many of the 
   truth level charged tracks are actually reconstructed at detector level. This efficiency
   impacts the event selection efficiency, since events are required to have $n_\text{tracks} \geq 5$.
   \item Apply the missing energy and signal region cuts to the surviving
   events.
\end{enumerate} 

    With the above procedure we aim to capture the impact of detector effects
and the reconstruction algorithm in a model independent way. 
The validity of this approach clearly relies on strong assumptions about the
detector performance and the relation between truth level and detector level
observables. However, without further information we believe it is not possible
to significantly improve the recasting.
Once we apply all the cuts and the correction factor to the events, it is
possible to compute signal efficiencies for any input model.
These efficiencies can then be
used to extract 95\% CL upper limits on the total visible 
cross section using the number of observed events ($N_\text{obs} = 0$) and the
expected background ($N_\text{DV}^\text{BG} = 0.02 \pm 0.02$).
For reference, a $100\%$ efficiency corresponds to the upper limit
$\sigma_{\tilde g \tilde g} < 0.091$ fb.

\begin{figure}[t]
\begin{center}
\includegraphics[scale=0.6]{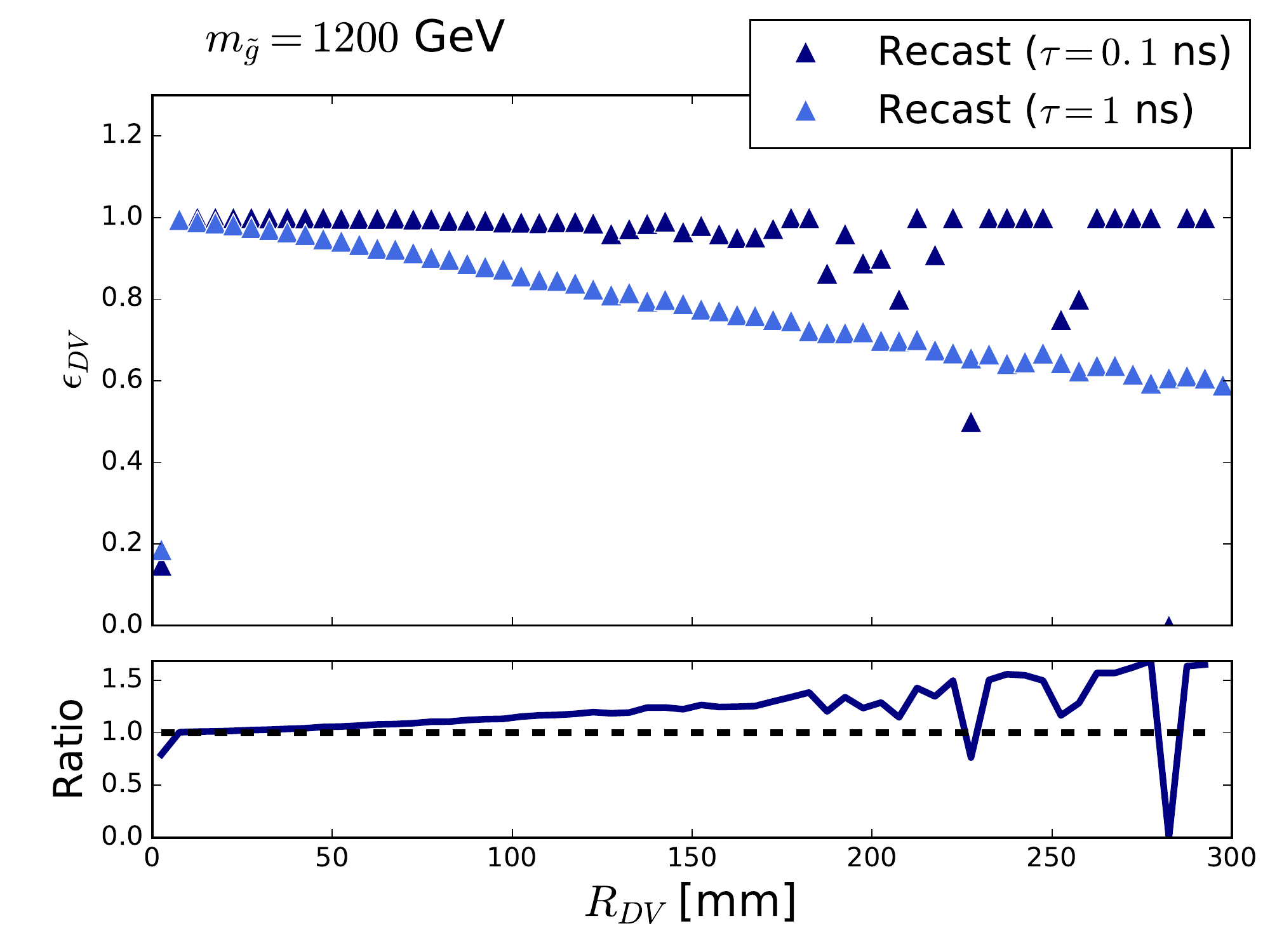}
\caption{Vertex reconstruction efficiency as a function of the
displaced vertex transverse position ($R_\text{DV}$).
The results were obtained through recasting after the {\it base vertex
selection} cuts have been applied.
The plot compares the recasting results for a fixed
gluino and LSP mass ($m_{\tilde{g}} = 1200$~GeV, $m_{\tilde \chi_1^0} =
100$~GeV) and two lifetime values, $\tau = 1$~ns and $\tau = 0.1$~ns.}
\label{fig:RDVeff2}
\end{center}
\end{figure}

The results obtained by the above approach for the benchmark points 
$m_{\tilde{g}} = 1.4$~TeV and $m_{\tilde{g}} = 2$~TeV are shown in
Fig.~\ref{fig:excCurve}.
We present recasting curves for three distinct values of the (constant)
track efficiency: $\epsilon_\text{track} = 15\%,\,25\%,\,100\%$.
As we can see, a 25\% efficiency can reproduce the official exclusion curve
(solid black line) within $\sim 50\%$ for most of the lifetime values.
However, in the regions where the efficiency drops significantly ($\tau < 10^{-2}$~ns),
the recasting curve is wrong by more than an order of magnitude.
Therefore, we conclude that this procedure is not satisfactory for a general purpose recasting
of the search presented in Ref.~\cite{ATLAS:2017bvh}.
In the next section we discuss how the recasting improves once we include the
detailed efficiencies provided by ATLAS.

\begin{figure}
\begin{center}
\includegraphics[width=0.49\textwidth]{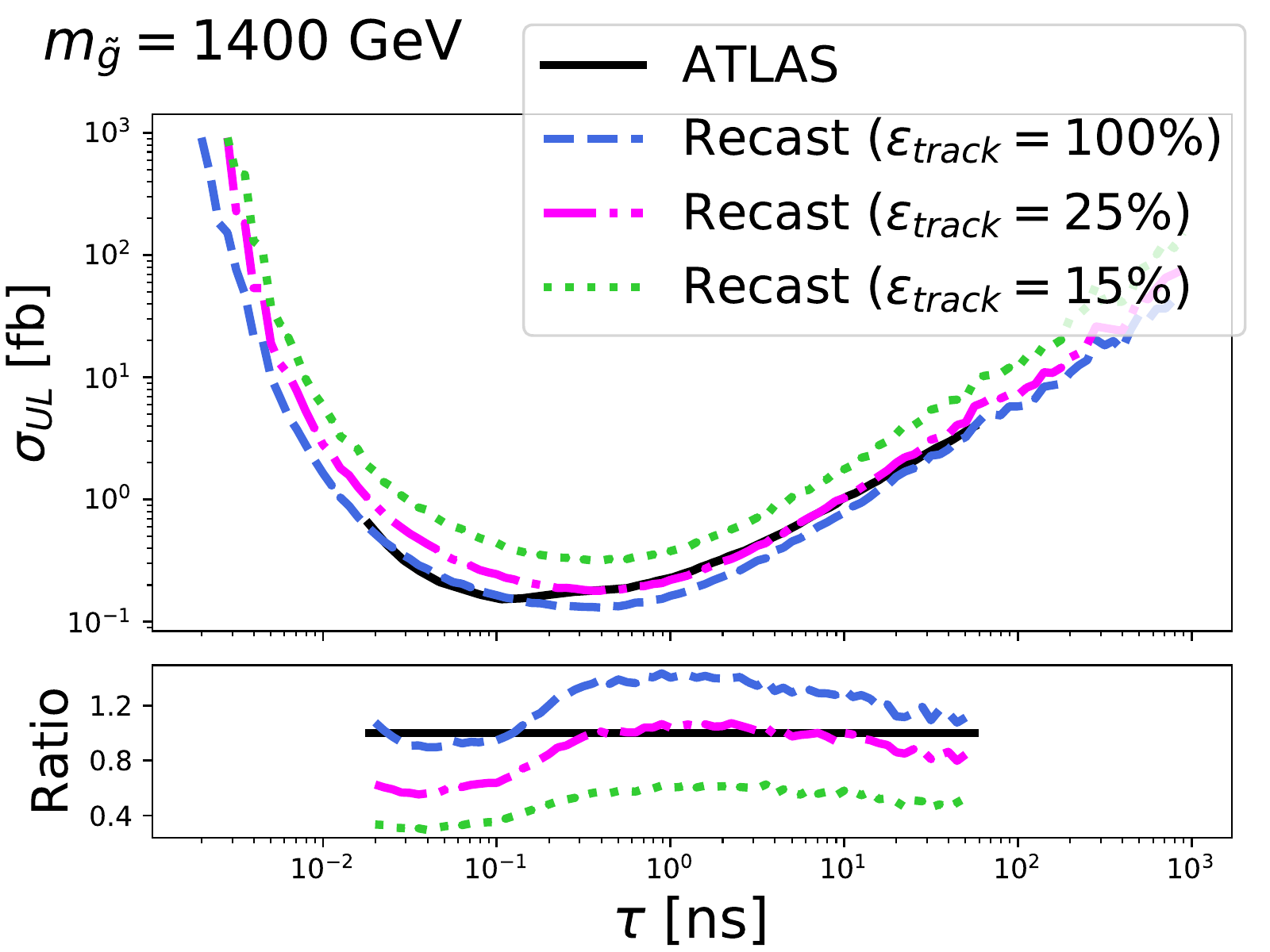}
\includegraphics[width=0.49\textwidth]{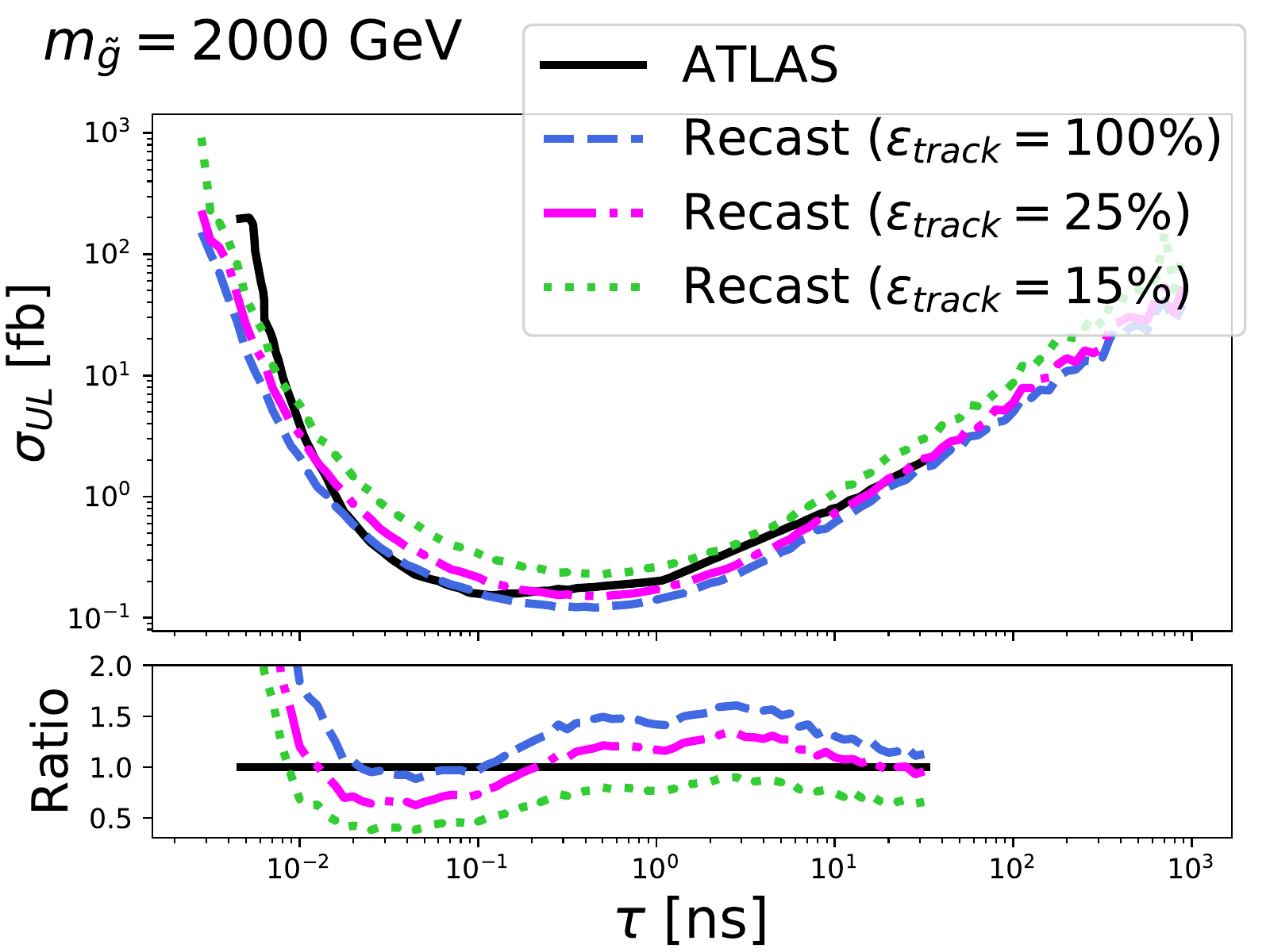}
\caption{Comparison between the official exclusion curve and the one obtained
by recasting the analyses with Method 1 for two values of gluino masses.}
\label{fig:excCurve}
\end{center}
\end{figure}

\subsection{Method 2: Recasting using ATLAS efficiency grids}
\label{method2}

This approach makes use of the full information provided in the auxiliary
material\footnote{This material can be directly access from: \\
\texttt{https://atlas.web.cern.ch/Atlas/GROUPS/PHYSICS/PAPERS/SUSY-2016-08/hepdata\_info.pdf}} of Ref.~\cite{ATLAS:2017bvh}: efficiency grids for the event selection (as a function of $R_\text{DV}$ and $E_\text{T}^{\text{miss}}$) and for the vertex
reconstruction efficiency (as a function of $R_\text{DV}$, $m_\text{DV}$ and
$n_\text{track}$). These parametrized efficiencies are also given for different
regions in the detector, encapsulating the effect of the material veto cut.

According to the note provided by ATLAS, these efficiencies can be applied at
truth level once some fiducial cuts have been applied. We use truth level
missing energy and identify the truth $R-$hadron decay position and decay products.

The selection of events requires:

\begin{itemize}
\item{truth level missing energy $E_\text{T}^{\text{miss}}>200$ GeV.}
\item{one trackless jet with $p_{T}>70$ GeV, or two trackless 
jets with $p_{T}>25$ GeV.  A trackless jet is defined as a jet
 for which the scalar sum of the $p_{T}$ of all charged particles 
 inside the jet does not exceed $5$ GeV. These jet requirements are applied to
 75\% of the data. The remaining 25\% do not need to satisfy any jet cuts.}
\end{itemize}

In addition, events must have at least one displaced vertex with:

\begin{itemize}
\item{distance between the interaction point and the decay position $>4$ mm.}
\item{the decay position must lie in the fiducial region $R_\text{DV}<300$ mm and $|z_\text{DV}|<300$ mm.}
\item{the number of selected decay products must be at least 5, where selected decay products are charged and stable, with $p_{T}>1$ GeV and $|d_{0}|>2$ mm.}
\item{the invariant mass of the truth vertex must be larger than 10 GeV, and is constructed assuming all decay products have the mass of the pion.}
\end{itemize}

After imposing the above fiducial cuts, the vertex reconstruction and event
selection efficiencies provided by ATLAS can then be applied to compute the
final signal efficiencies.
The results for the two benchmark points are shown in Fig.~\ref{fig:effVsTauRecasted}.
Finally, using these efficiencies,
we can extract 95\% CL upper limits on the total visible 
cross section using the same procedure described in the previous Section.
The results for the exclusion curves are shown in Fig.~\ref{fig:sigmaLimits},
where we can see that the recasting reproduces
the official exclusion curves fairly
well for most of the lifetime values. The largest discrepancies are
within $\sim40\%$, which corresponds to a major improvement with respect
to the results obtained using only the limited information provided
by the ATLAS conference note, as discussed in Section \ref{method1}.

\begin{figure}[t]
\begin{center}
\includegraphics[width=0.45\textwidth]{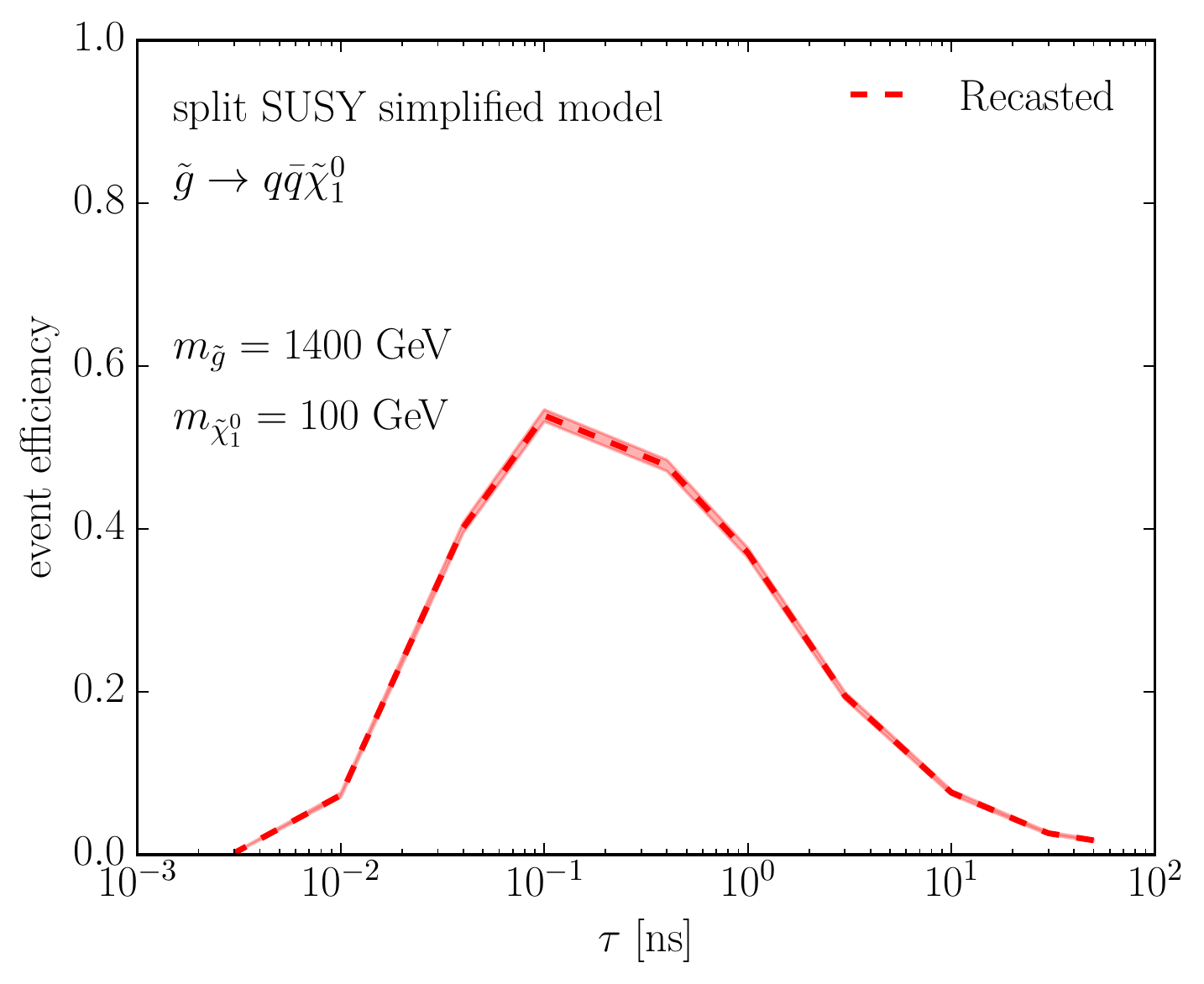}
\includegraphics[width=0.45\textwidth]{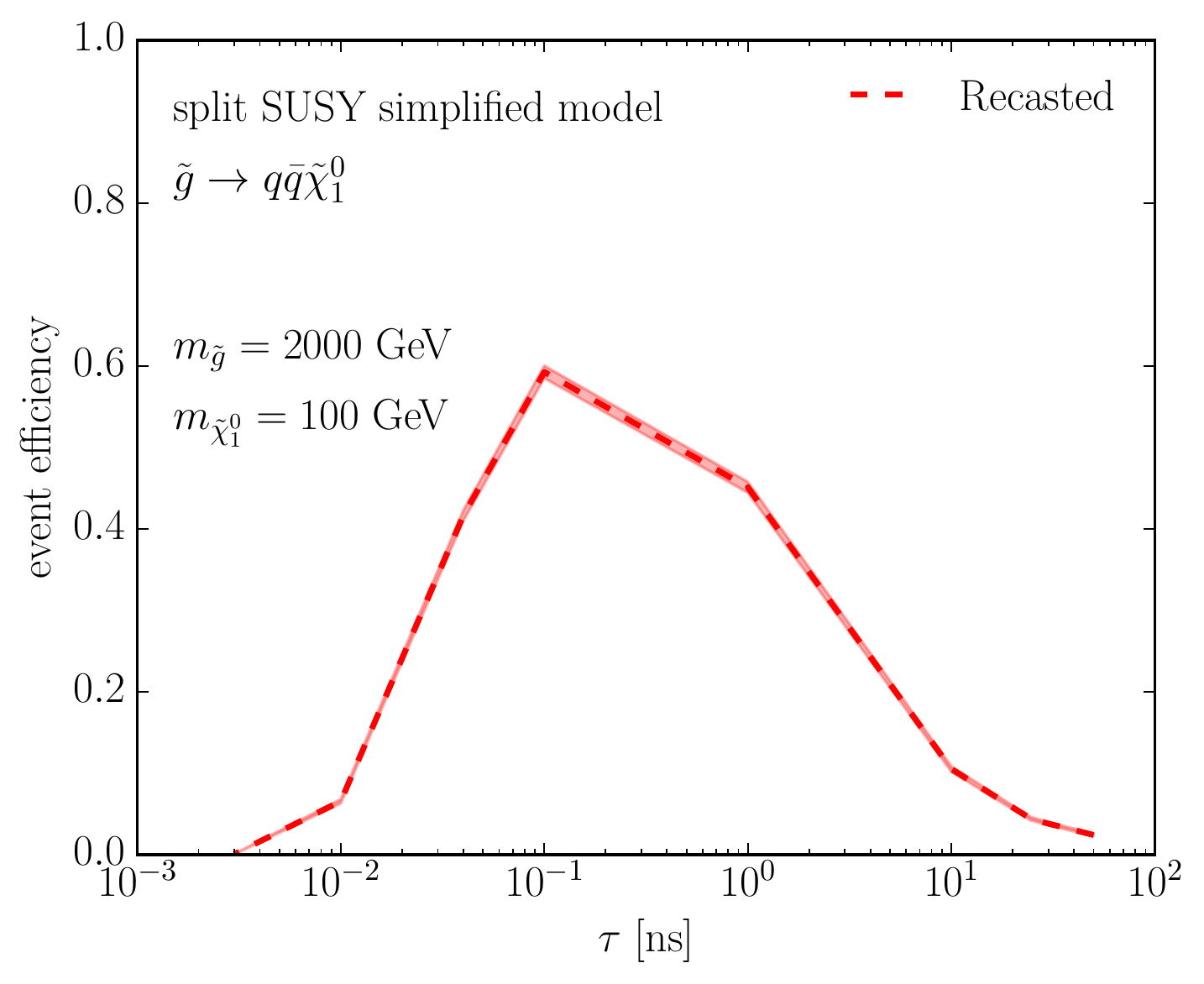}
\caption{ Recasted event-level efficiencies against gluino proper decay lifetime for two values of gluino masses. }
\label{fig:effVsTauRecasted}
\end{center}
\end{figure}

\begin{figure}[t]
\begin{center}
\includegraphics[width=0.45\textwidth]{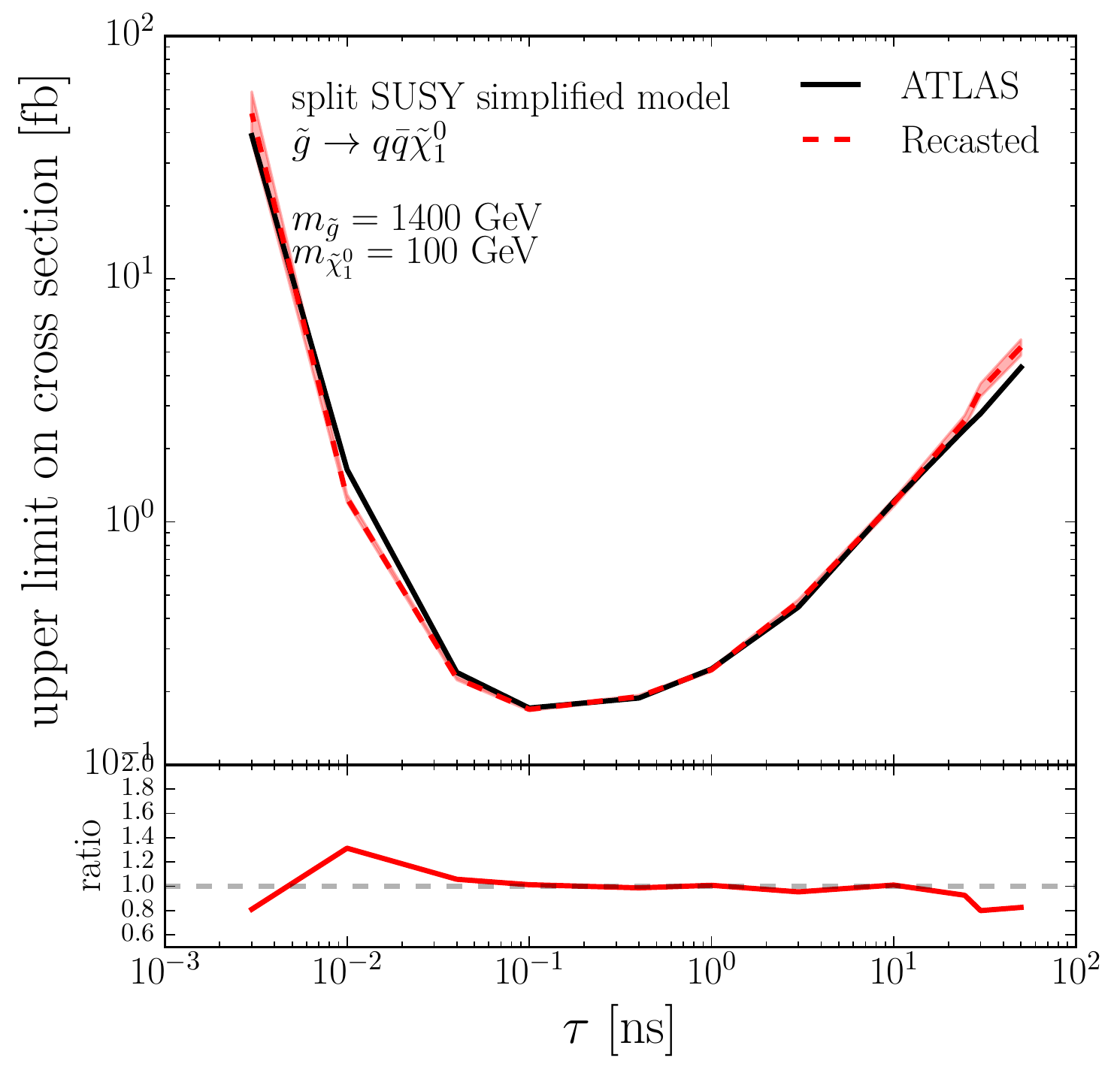}
\includegraphics[width=0.45\textwidth]{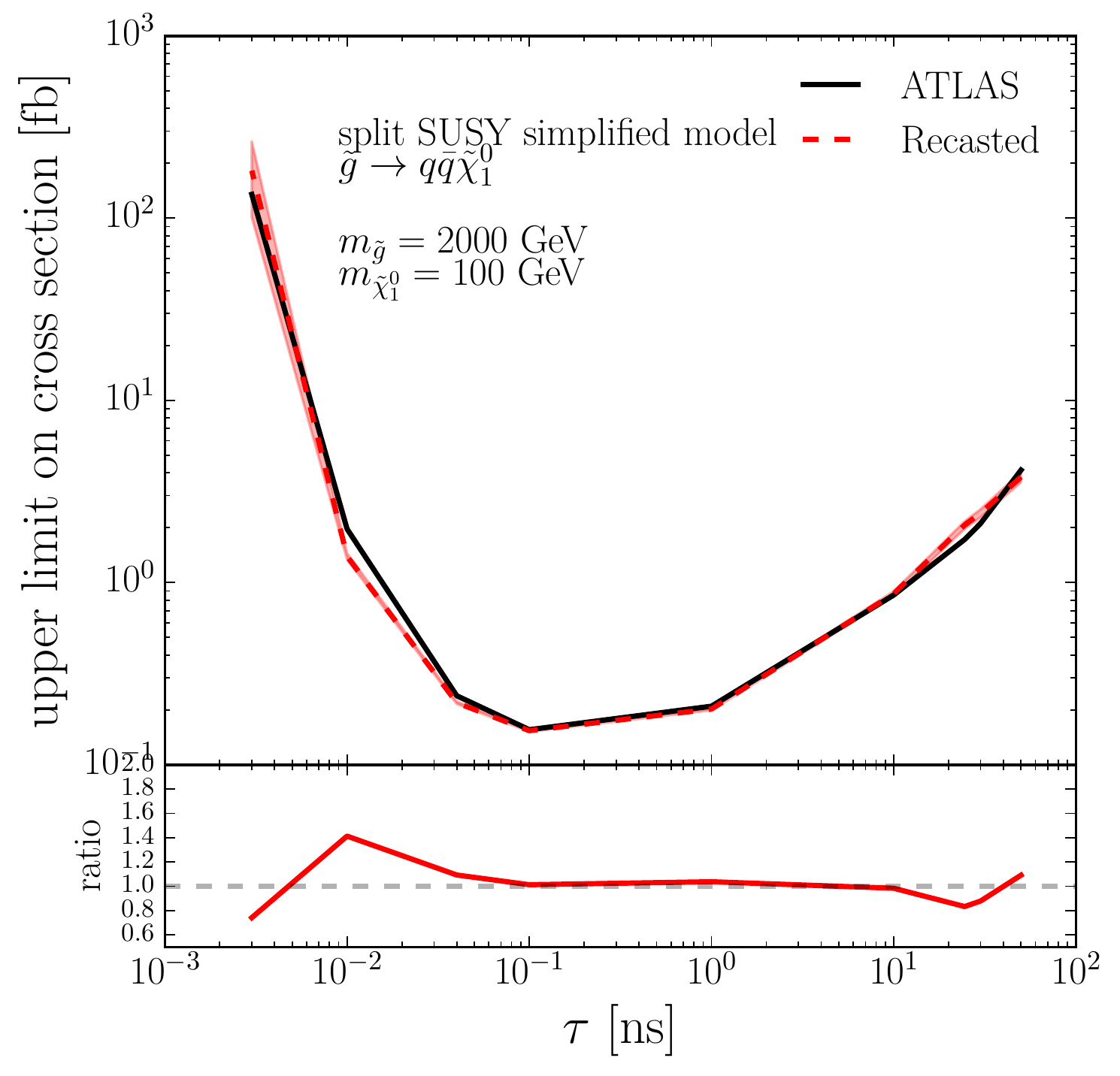}
\caption{Comparison between the official exclusion curve (solid black line) and the one obtained
by recasting the analyses with Method 2 (dotted red line) for two values of gluino masses.}
\label{fig:sigmaLimits}
\end{center}
\end{figure}

\subsection{Comments on the recasting procedure} 

The results presented in Fig.\ref{fig:excCurve}
show that it is not possible to recast the ATLAS displaced vertex $+$ missing energy search 
making use only of the simple vertex reconstruction efficiencies provided in 
the conference note ATLAS-CONF-2017-026.
Although different correction functions can be applied at the vertex and track
level to try to capture the impact of detector
effects and the vertex reconstruction algorithm,
we find that the recasting of the exclusion curves is still off by an order
of magnitude at small lifetimes.
On the other hand, using the ATLAS efficiency grids highly improves the level of
agreement, limiting the discrepancies to be under $\sim 40\%$, as shown by
Fig.\ref{fig:sigmaLimits}.
We therefore find that the parametrization of the efficiency grids
in terms of truth level variables is 
extremely useful for recasting, being also straightforward to implement.

Finally, we note that, even though the parametrized selection
efficiencies can be in principle used for any model 
(and are said to be model independent according to the ATLAS note), we encourage
the experimental collaborations to present exclusion curves (or signal
efficiencies) for a second model with a distinct event topology. This
information is essential to validate the recasting procedure 
and accurately assess the level of model independence provided by the
parametrized efficiencies.

\section{DISPLACED LEPTON SEARCHES}
\label{sec:Dlep} 

One of the cleanest search strategies for long-lived particles decaying into leptons is to look for leptons with a non-zero impact parameter with the primary vertex.  Such tracks are normally vetoed to remove contamination from underlying event, cosmics etc., and therefore may not be attributed to the right collision event in standard prompt searches.  The CMS displaced lepton search ($\sqrt s = 8$ TeV, $\mathcal{L} = 19.7 \pm 0.5 \mathrm{fb}^{-1}$) \cite{Khachatryan:2014mea} follows a simple strategy of requiring two isolated, oppositely charged, lepton tracks that have a significant impact parameter with respect to the primary vertex.  The accompanying material includes efficiency for identification of electrons and muons based on the transverse impact parameter ($d_0$) and the transverse momentum ($p_T$) of the lepton.  The benchmark scenario for the analysis is stop-pair production $ p p \rightarrow \tilde t \tilde t^*$ ($m_{\tilde t} = 500$~GeV) followed by an R-parity violating decay via $\lambda'$-type coupling $\tilde t \rightarrow b \ell$, with equal probabilities for $\ell = e, \mu~\text{and}~\tau$.  

The recommended recasting procedure is to apply the cuts below on generator-level leptons and reweight the event with the four identification efficiencies and an overall trigger efficiency of 0.95.
\begin{enumerate}
\item Select events with one $e$ and one $\mu$, oppositely charged, both coming from a stop
\item Require both leptons to have $|\eta| < 2.5$
\item Require decay vertex of stop to have transverse position $v_0 < 40~\text{mm}$, and z-position $v_z < 300~\text{mm}$.
\item Require $p_T^{(e,\mu)} > 25$ GeV and $\Delta R_{e\mu} > 0.5 $
\item Jet isolation: for each jet (anti-kt, $R=0.5$, $p_\text{T}^{\mathrm{min}} = 10$ GeV), require $\Delta R_{\ell j} > 0.5$
\item Transverse impact parameter $0.1~\text{mm} < d_0  < 20~\text{mm} $.
\item Signal regions are further defined as:
\begin{description}
\item [\textsc{SR3:}] Both leptons satisfy  $1.0~\text{mm} < d_0 < 20~\text{mm}$.
\item [\textsc{SR2:}] One or both leptons fail \textsc{SR3} but satisfy $ d_0 > 0.5 ~\text{mm}$
\item [\textsc{SR1:}] One or both leptons fail \textsc{SR2} but satisfy $ d_0 > 0.2 ~\text{mm}$
\end{description}
\end{enumerate}

\begin{figure}[t]
\begin{center}
\includegraphics[scale=0.4]{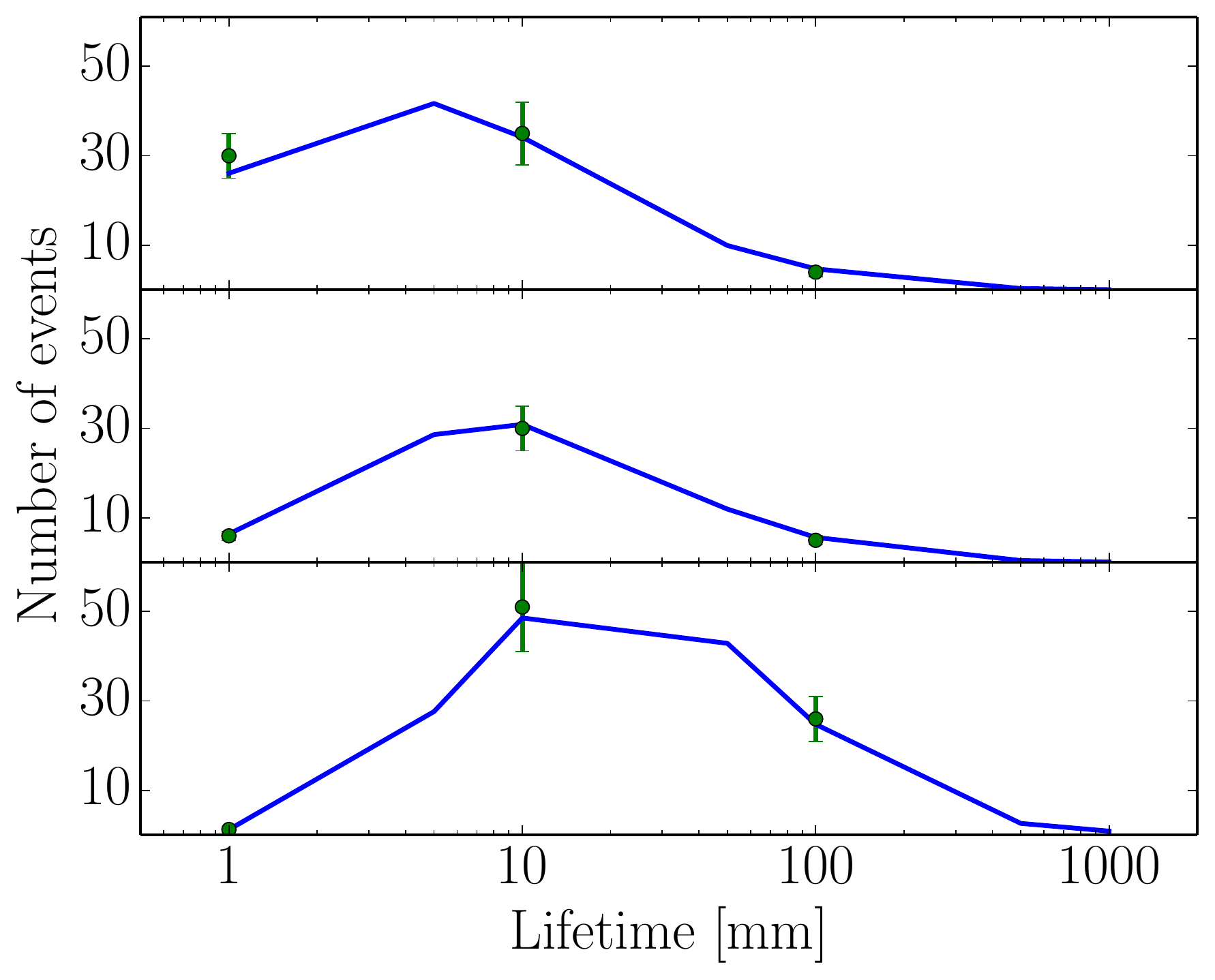}
\includegraphics[scale=0.4]{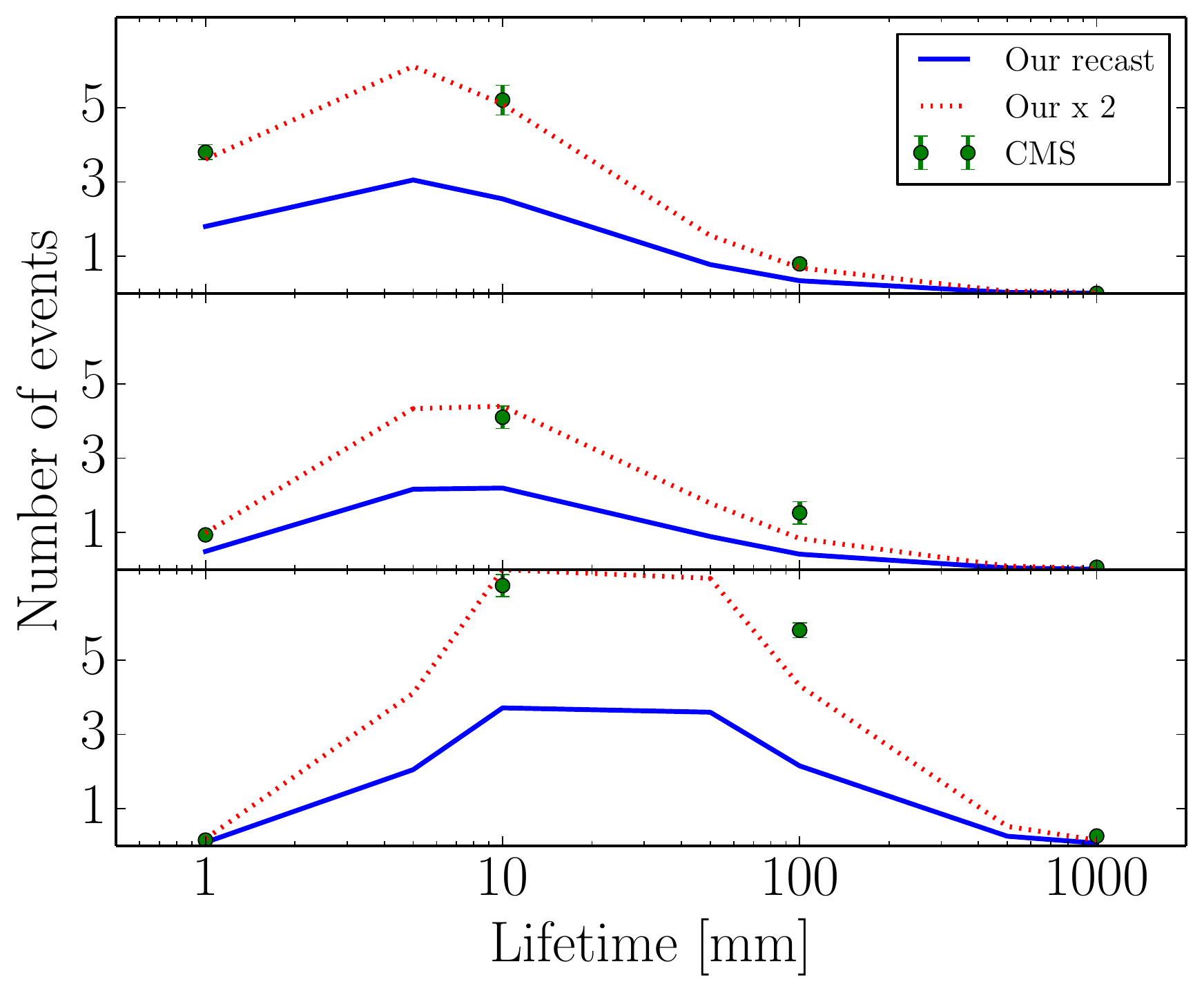}
\caption{Result from recasting the 8 TeV (left) and 13 TeV (right) displaced lepton analysis comparing number of expected signal events from our simulation (blue) with the published CMS values (green dots). The panels refer to SR1 (top), SR2 (middle) and SR3 (bottom). }
\label{fig:DispLep}
\end{center}
\end{figure}

\vspace{0.1in}

\noindent The updated analysis at 13 TeV ($\sqrt s = 13$ TeV, $\mathcal{L} = 2.6 \mathrm{fb}^{-1}$, conference note \textsc{CMS-PAS-EXO-16-022}) \cite{CMS-PAS-EXO-16-036}, applies stronger $p_\text{T}$ cuts by requiring $p_T^{e(\mu)} > 42(45) $ GeV. It also improves isolation cuts on the leptons by requiring that the sum of $p_\text{T}$ (of all particles) in a cone of  0.3 should be less than 3.5\% (6.5\%) in the barrel (encdap) for electrons. For muons, cone size is taken to be 0.4 and the sum of $p_T$ is required to be less than 15\%. Separation between the two leptons is $\Delta R_{e\mu} > 0.5 $, same as before.  The mass of the stop in the benchmark point is increased to 700~GeV, however, the decay branching fractions remain unchanged.

A comparison of our validation of the analysis can be seen in Fig.~\ref{fig:DispLep}. At 8~TeV, the agreement is well within the quoted errors. Using the same efficiency parametrization as 8 TeV for the 13~TeV analysis, we are unable to reproduce the benchmark efficiencies and we find that in general, we have an overall mismatch of a factor of about 2 with respect to the published expected number of signal events.  Moreover, we find that the mismatch is higher for longer lifetimes, prompting the inference that the $d_0$ dependence of the efficiency may have changed between the two runs.

\section{CHARGED HEAVY PARTICLE SEARCH}
\label{sec:HCP}

Another important example of LLPs which have been
searched for at the LHC are heavy stable charged particles (HSCPs).
Here we will consider the searches performed by CMS using 
8~TeV~\cite{Khachatryan:2015lla} and 13~TeV~\cite{CMS-PAS-EXO-16-036} data.
These searches are targeted at detector-stable HSCPs and are based on
the signature of highly ionizing tracks and anomalous time-of-flight.
Below we will discuss the recasting of both of these searches.
We point out, however, that while the 8~TeV analysis provided detailed
efficiencies for the HSCP reconstruction, the 13~TeV results do not include this information.
Therefore, for the 13~TeV search the recasting will make use of
an extrapolation of the 8~TeV efficiencies.

\subsection{Recasting the 8~TeV search}
\label{sec:hscp8}

As mentioned above, the 8~TeV CMS analysis~\cite{Khachatryan:2015lla} provides
efficiency grids for HSCP reconstruction. In particular, probabilities for events to pass the
on- and off-line selection criteria are given
as a function of the truth-level HSCP kinematics.
Due to the
inclusive nature of the search, this provides a powerful way to reinterpret the
search for arbitrary models containing
detector-stable~\cite{Heisig:2015yla,Evans:2016zau,Bagnaschi:2016afc,Hessler:2016kwm,Heisig:2017lik}
or metastable~\cite{Garny:2017rxs} LLPs.

The recasting follows the procedure described in
Ref.~\cite{Khachatryan:2015lla}. We simulate events for HSCP production
using {\sc MadGraph5}~\cite{Alwall:2014hca} for the
parton level process and \textsc{Pythia}~6~\cite{Sjostrand:2006za} for
showering and hadronization. For each event we first identify
isolated HSCP candidates imposing the following isolation criteria:
\begin{equation}
\left( \sum_{i}^{\overset{\text{charged}}{\Delta R<0.3}} \!\!\!\!\! p_\text{T}^{i} 
\right)  < 50\,\text{GeV}
\;,\quad
\left(
\sum_{i}^{\overset{\text{visible}}{\Delta R<0.3}} \!\! \frac{E^i}{
|\vec{p}|} \right)  < 0.3\,,
\label{eq:GenTkIso1}
\end{equation}
on the truth level events. 
The sums include all charged and visible particles, respectively,
in a cone of $\Delta R=\sqrt{\Delta\eta^2+\Delta \phi^2}<0.3$ around the
direction of the HSCP candidate, $p_T^{i}$ denotes their transverse
momenta and $E^i$ their energy. 
Muons are not con\-si\-dered as visible particles and the HSCP candidate itself is not included in either sum.
Once the HSCP candidates are identified, the final signal efficiency ($\epsilon$) is then given by:
\begin{equation}
\label{eq:Technique}
\epsilon = \frac{1}{N} \sum_{i}^{N}
P_{\text{on}}(\vec{k}) \times
P_{\text{off}}(\vec{k})\,,
\end{equation}
where the sum runs over all generated events and $P_{\text{on}}$
($P_{\text{off}}$) is the on-line (off-line) selection efficiency provided by the 8~TeV CMS analysis
as a function of the HSCP truth level kinematics ($\vec{k}$).
For events containing two HSCP candidates, the above probabilities must be
replaced by~\cite{Khachatryan:2015lla}
\begin{equation}
\label{eq:EventAcceptance}
P^{(2)}_{\text{on}/\text{off}}(\vec{k}^1, \vec{k}^2) 
= P_{\text{on}/\text{off}}(\vec{k}^1)  + P_{\text{on}/\text{off}}(\vec{k}^2) 
- P_{\text{on}/\text{off}}(\vec{k}^1)  P_{\text{on}/\text{off}}(\vec{k}^2)  \,,
\end{equation}
where $\vec{k}^{1,2}$ are the kinematical vectors of the HSCPs.

In order to validate the above procedure we compute exclusion curves for the
same benchmark models considered by the 8~TeV CMS search~\cite{Heisig:2015yla}.
The results are shown in Figure~\ref{fig:gmsbComp} for the the gauge-mediated supersymmetry 
breaking (GMSB) model containing long-lived staus.
As we can see both the signal efficiencies (left panel) and the 95\% CL upper
limits on the inclusive production cross section (right panel) 
agree with the official CMS results within 5\% or less, thus providing an
excellent approximation.

\begin{figure}[!h]
\centering
\setlength{\unitlength}{1\textwidth}
\begin{picture}(1,0.45)
 \put(0.003,-0.015){\includegraphics[width=0.99\textwidth]{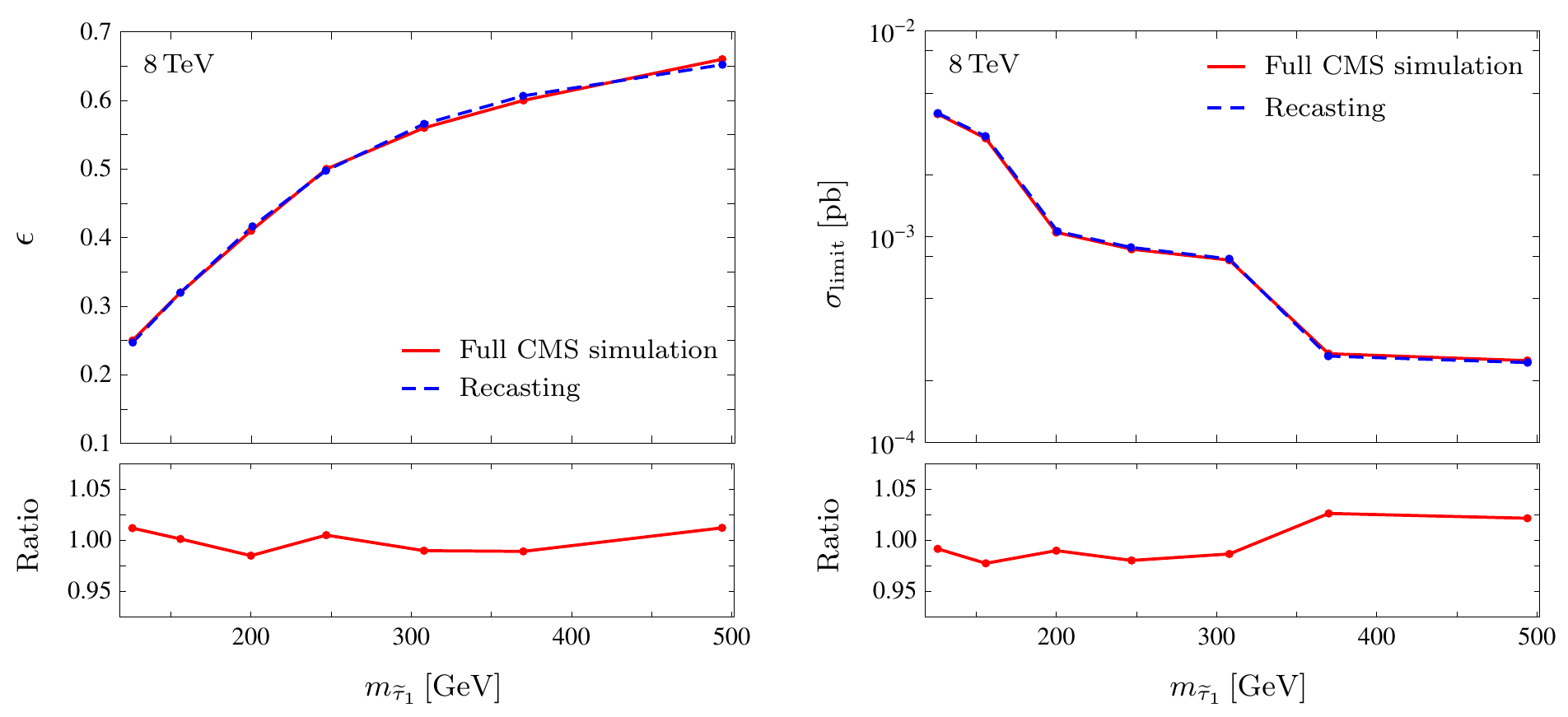}}
\end{picture}
\caption{
Signal efficiency $\epsilon$ (left panel) and 95\% CL cross section upper limit (right panel)
for the GMSB model as the function of the stau mass. 
We compare the CMS analysis~\cite{Khachatryan:2015lla}
from the full detector simulation (red solid lines) with the 
recasting from~\cite{Khachatryan:2015lla} (blue dashed lines).
In the lower frames we show the respective ratios
$\epsilon^\text{Full}/\epsilon^\text{Recast}$,
$\sigma^\text{Full}_\text{limit}/\sigma_\text{limit}^\text{Recast}$.
Taken from~\cite{Heisig:2015yla}.
}
\label{fig:gmsbComp}
\end{figure}

\subsection{Recasting the 13~TeV search}
\label{sec:hscp13}

Given the excellent performance of the recasting method discussed above for the
8\,TeV LHC run it would be appealing to be able to use a similar method for the 13\,TeV 
analysis~\cite{CMS-PAS-EXO-16-036}. However, the provision of object efficiencies
for the 13\,TeV search is so far not pursued. 
In this section we discuss an attempt to recast the 13~TeV search
based on an extrapolation of the object efficiencies from the 8\,TeV to the
13\,TeV run.

Our aim is to use the 8\,TeV efficiencies
($P^{8\,\text{TeV}}_{\text{on}/\text{off}}$) for the 13\,TeV recasting through
the introduction of a correction function that accounts for the differences
between both runs. Since the selection criteria for the 8~TeV and 13~TeV
searches are very similar, we require the same isolation cuts listed in
Eq.~\ref{eq:GenTkIso1}.
We then assume the following ansatz for the 13~TeV efficiencies
($P^{13\,\text{TeV}}_{\text{on}/\text{off}}$):
\begin{equation}
P^{13\,\text{TeV}}_{\text{on}} (\vec{k}) = P^{8\,\text{TeV}}_{\text{on}}
(\vec{k}) \;,\; P^{13\,\text{TeV}}_{\text{off}} (\vec{k}) = F(\beta) \times
P^{8\,\text{TeV}}_{\text{off}} (\vec{k})\,,
\label{eq:introF}
\end{equation}
where $F$ is the correction function.
The above relations assume that the on-line probability does not change
drastically and the main differences between the runs happen in the off-line
(trigger) selection. This is a reasonable assumption, since the on-line cuts
of both analyses are similar. Furthermore, we expect our
treatment to leave enough freedom to account for small corrections 
in the on-line probability as well. 
At least for events with only one HSCP candidate there is no distinction, as the
two probabilities are just multiplied.
Finally, we assume that $F$ only depends on the HSCP velocity ($\beta$) and we
parametrize this correction function by eight parameters ($C_{\beta_n}$),
which corresponds to the value of the function for $\beta =
0,0.47,0.6,0.7,0.77,0.83,0.89,1.0$. For other values of $\beta$ 
we interpolate linearly.

In order to compute the correction function $F$ introduced above
we will make use of the official signal efficiencies
reported in 13\,TeV LHC analysis~\cite{CMS-PAS-EXO-16-036} for specific
benchmark points.
We determine the parameters $C_{\beta_n}$ of the correction function in a 
global fit to these efficiencies using the $\chi^2$ defined as:
\begin{equation}
\label{eq:chi2}
\chi^2(C_{\beta_n}) = \sum_m \frac{(\epsilon_m(C_{\beta_n}) - \epsilon_m^\text{CMS})^2}{\sigma_\epsilon^2}\,,
\end{equation}
where $\epsilon_m(C_{\beta_n})$ is the efficiency for the benchmark point $m$ 
using the correction function parameters $C_{\beta_n}$, $\epsilon_m^\text{CMS}$ is the 
respective efficiency reported in~\cite{CMS-PAS-EXO-16-036} and $\sigma_\epsilon$ is
the characteristic size of the uncertainty which we (arbitrarily) set to $0.02$.
We include in the fit the 6 direct stau benchmark points used in the CMS
analysis and minimize the $\chi^2$
using \textsc{Multinest}~\cite{Feroz:2008xx,Feroz:2013hea} for an efficient exploration of the parameter space.
The best-fit correction function, $F_\text{best-fit}$, and its $1\sigma$ uncertainty\footnote{%
As stated above the $1\sigma$ uncertainty corresponds $\sigma_\epsilon=0.02$ which 
is roughly the level of accuracy we aim at in the fit. Note, however, that intrinsic systematic
uncertainties in the determination of the efficiencies might be larger, up to around 10\%.}
are shown in figure~\ref{fig:F}. The deviation of $F$ from 1 implies a decrease or increase of the 
respective detector and signal efficiencies between the 8 and 13\,TeV analysis.
We find a slight decrease of efficiencies for large velocities $\beta\gtrsim0.85$,
which is, however, not significant. More surprisingly, the efficiencies at low
velocities ($\beta\simeq0.5$) contain a large and significant increase.
From these results it appears that the CMS detector in Run 2 performs
significantly better at low velocities.
We also point out that the uncertainties for low values of $\beta$ are quite 
large, which illustrate the fact that the signal efficiencies provided by the
13~TeV CMS results are not sufficient to fully constraint the correction
function $F$. However, since no additional information is provided by the
13~TeV CMS analysis, we will use the best fit for $F$ in the following.

\begin{figure}[t]
\begin{center}
\includegraphics[width=0.6\textwidth]{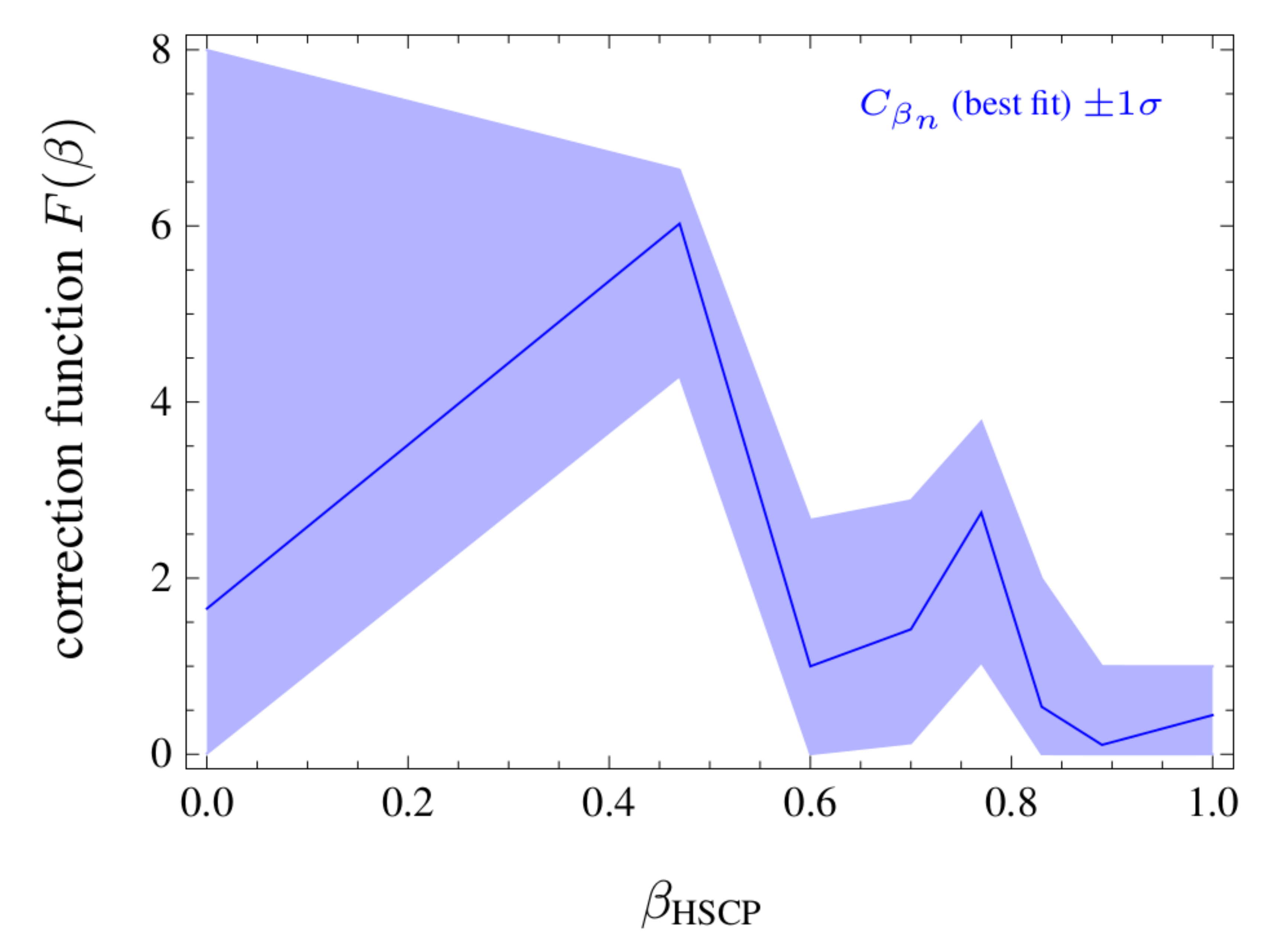}
\caption{Best-fit correction function $F(\beta)$ and its $ \pm 1\sigma$ band.
}
\label{fig:F}
\end{center}
\end{figure}

Using the off- and on-line probabilities defined in
Eq.~\ref{eq:introF} and the best fit for the correction function
($F_\text{best-fit}$) shown in Fig.~\ref{fig:F} we can then
compute the 13~TeV signal efficiencies for any input model.
We first compute efficiencies for the same 6 benchmark
points used to obtain $F$. These benchmarks 
correspond to direct production of long-lived status with distinct values of the
stau mass. The results for the final signal efficiencies for these six
benchmark points are shown in Fig.~\ref{fig:eff} as a function of the HSCP
(stau) mass. The solid black line shows the efficiencies
reported by CMS in Ref.~\cite{CMS-PAS-EXO-16-036}, while the dashed blue
curve shows the efficiencies obtained by recasting using the extrapolation of
the 8~TeV probabilities and the best fit for $F$. We also show
the recasting efficiencies obtained without the inclusion of the correction
function (dashed magenta curve).
As expected, $F_\text{best-fit}$ reproduces the 
efficiencies for the six mass points 
well within the expected fit uncertainties (2\%). In particular, 
it significantly improves the agreement with respect to the
naive extrapolation without a correction, \emph{i.e.}~for $F=1$.
Formally, in our fit this is reflected in a decrease in the
$\chi^2$ from around 130 for $F=1$ to 0.1 for $F_\text{best-fit}$.

\begin{figure}[t]
\begin{center}
\includegraphics[scale=0.6]{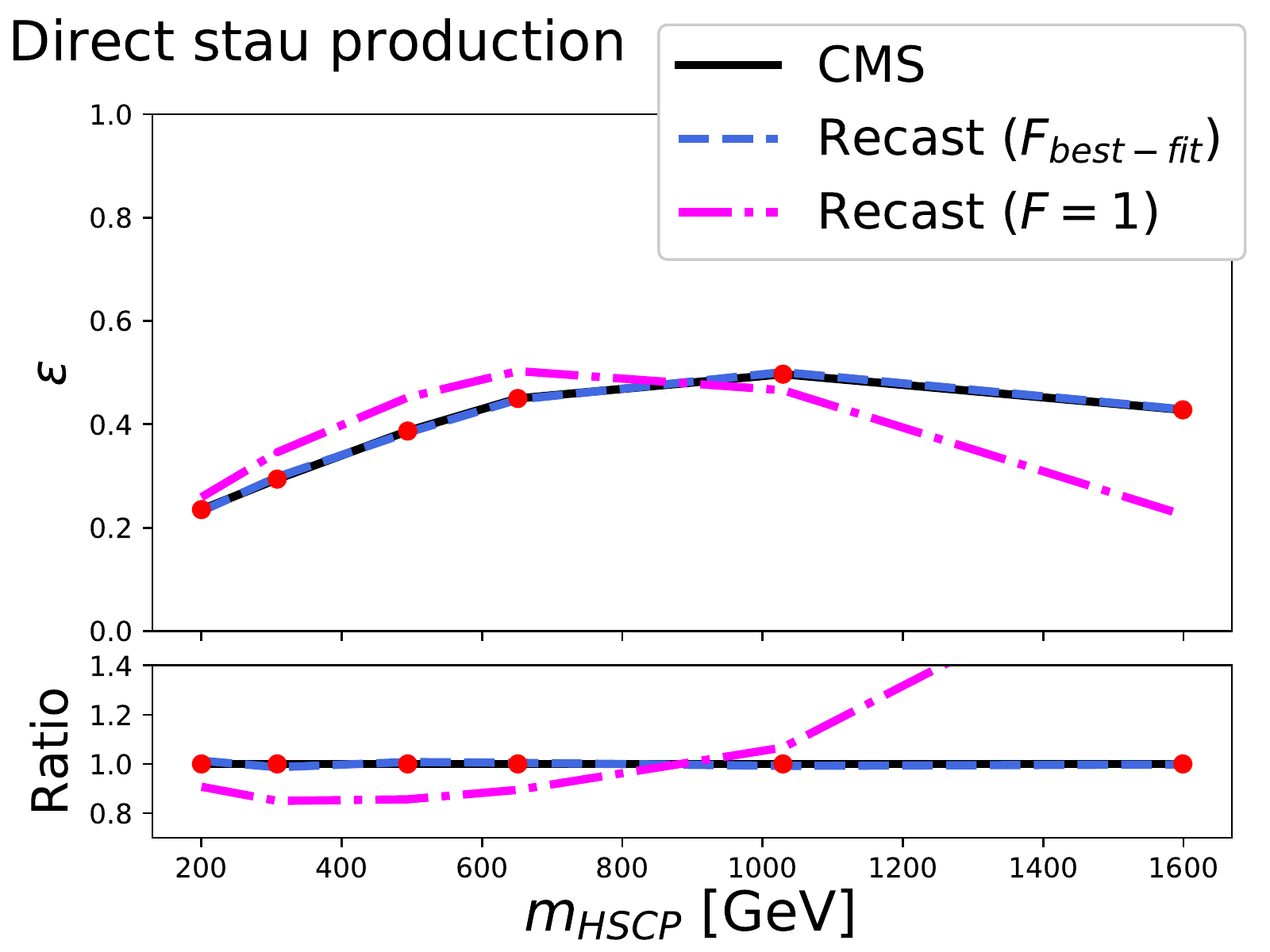}
\caption{Efficiencies for the six benchmark masses in the direct stau production
scenario. The solid black curve shows the efficiencies
from~\cite{CMS-PAS-EXO-16-036}, while the dashed curves show the efficiencies
obtained through recasting. The blue curve corresponds to the best-fit for the
correction function, while the magenta shows the
efficiencies for $F=1$.}
\label{fig:eff}
\end{center}
\end{figure}

%
%
 
The good agreement shown in Fig.~\ref{fig:eff} between the official CMS results
and $F_\text{best-fit}$ is expected, since the same benchmark points were used
to fit the correction function $F$. Therefore, 
a crucial test of the validity of the recasting
is its application to an independent set of models. Fortunately, in
Ref.~\cite{CMS-PAS-EXO-16-036} CMS has also reported the signal efficiencies for
six GMSB points with long-lived staus.
These points include both direct production of staus and production through
cascade decays of heavier sparticles, resulting in a broader spectrum
of event topologies.
Using again the extrapolation of the 8~TeV efficiencies and the best fit for the
correction function from Fig.\ref{fig:F}, we compute the final signal
efficiencies for these six GMSB points.
The results are shown in Fig.~\ref{fig:effGMSB}. 
As we can see, the efficiencies using $F_\text{best-fit}$ now deviate by up to
20\% for large HSCP masses, where our recasting undershoots the efficiencies reported
by CMS.
Although the overall agreement is improved by the correction function, 
$\chi^2(F_\text{best-fit})\simeq 88$ versus 
$\chi^2(F=1)\simeq 240$, the result is much worse than the ones obtained at
8~TeV, where the uncertainties were below 5\%.

\begin{figure}[t]
\begin{center}
\includegraphics[scale=0.6]{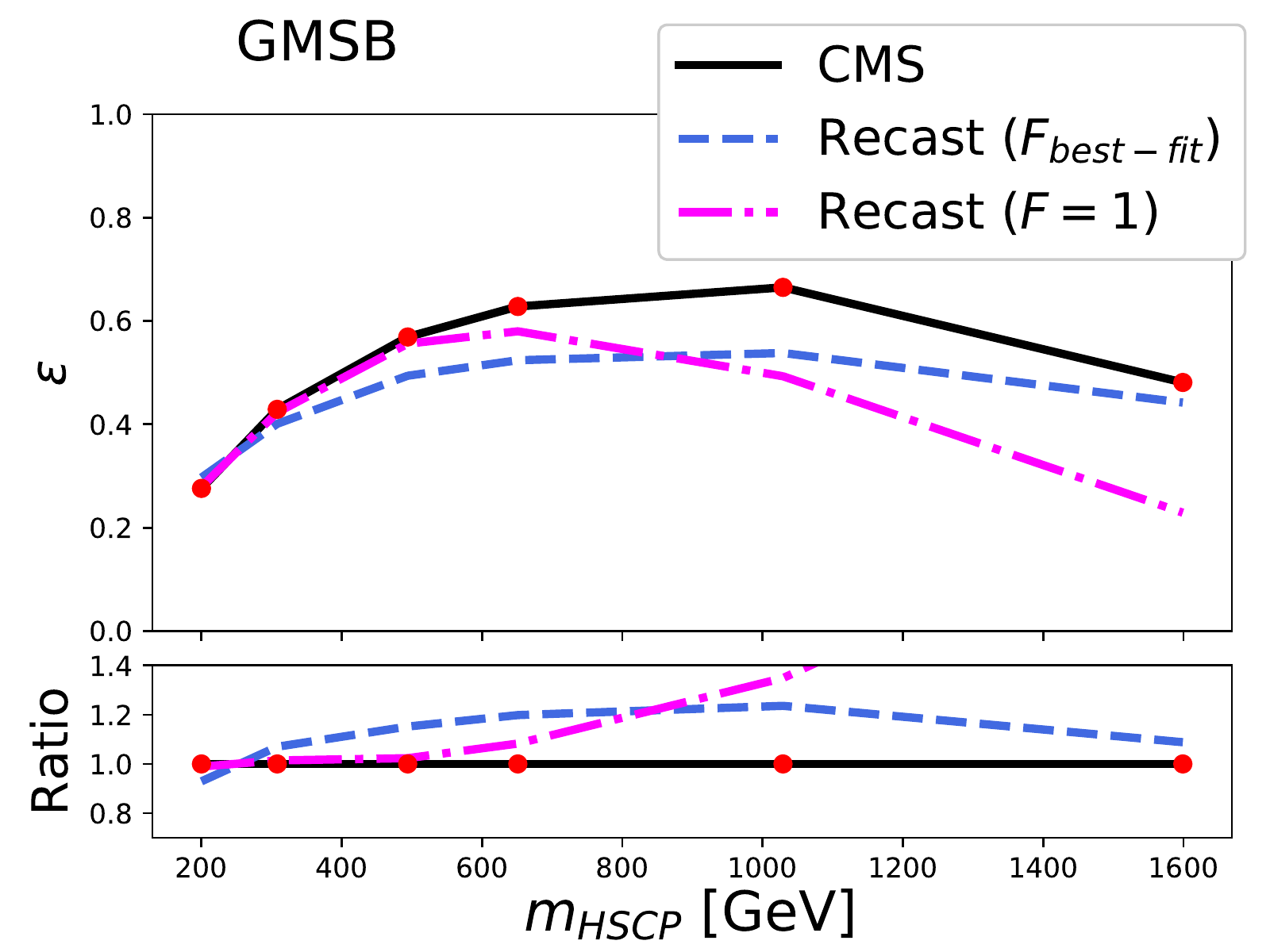}
\caption{Efficiencies for the six benchmark masses in the GMSB 
scenario. The solid black curve shows the efficiencies
from~\cite{CMS-PAS-EXO-16-036}, while the dashed curves show the efficiencies
obtained through recasting. The blue curve corresponds to the best-fit for the
correction function, while the magenta shows the
efficiencies for $F=1$.}\label{fig:effGMSB}
\end{center}
\end{figure}


The differences between the recasting efficiencies and the ones reported by CMS
for the GMSB model might arise from several shortcomings in our
recasting procedure.
First, we assume $F$ to only dependent on $\beta$ whereas the full probability maps
are parametrized as a function of three kinematic variables: $\beta,
p_\text{T}$ and $|\eta|$.
A parametrization in terms of all three variables, however, is
clearly not feasible given the limited amount of information provided in
Ref.~\cite{CMS-PAS-EXO-16-036}. As illustrated by Fig.~\ref{fig:F}, even the
parametrization in terms of a single variable can not be fully constrained using
only the signal efficiencies reported by CMS.
Second, assigning the correction function to the off-line probabilities only
might be an over-simplification. For events containing two HSCP candidates, a single
correction function $F$ may not be sufficient to parametrize the differences
between the two runs.
Again, only a better understanding of the underlying changes (\emph{e.g}~in the trigger settings)
can resolve these ambiguities.

\subsection{Comments on the recasting procedure}

As illustrated by the results in Section~\ref{sec:hscp8}, the uncertainties in
the recasting of HSCP searches can be reduced to the few percent level if
efficiencies for the reconstruction and selection of HSCP tracks are provided.
This is the case for the 8~TeV CMS analysis, where we were able to successfully
recast the HSCP results with a high degree of accuracy~\cite{Heisig:2015yla}.
The situation is drastically distinct if these officiencies are not provided by
the experimental collaboration, as illustrated by the 13~TeV results in
Section~\ref{sec:hscp13}. Even though the CMS analyses for both runs are very
similar, the 8~TeV object efficiencies cannot be easily 
extrapolated to the run 2. Therefore,
in order to establish robust reinterpretations of HSCP searches, further information on detector
efficiencies are required.

We also point out that the recasting uncertainties for the 
13~TeV search could only be properly assessed using the  official CMS
signal efficiencies for two distinct scenarios (the direct stau and GMSB
models).
This is an important point, since several LLP searches present results
for a single model or type of signal topology (see 
Section~\ref{sec:DV}, for instance).
The second model is essential for estimating the validity of the
recasting procedure and if it can indeed be extrapolated to other models
containing distinct event topologies. As shown by the results in
Figs.~\ref{fig:eff} and \ref{fig:effGMSB}, the recasting uncertainties would
have been highly underestimated if the CMS values for the GMSB scenario were
not available.

\section{CONCLUSIONS}
\label{sec:llpconclusions}

In this note we have investigate the feasibility of recasting LLP
searches within the context of three distinct signatures: displaced vertices,
displaced leptons and charged tracks.
Although each signature presents its own challenges for recasting,
we showed that without detailed object reconstruction and selection efficiencies
a satisfactory recasting can not be performed outside the experimental
collaborations. The final signal efficiencies or limits in this case can be inaccurate
by almost an order of magnitude.
On the other hand, for the cases where the object efficiencies were available,
such as the ATLAS displaced vertex and the 8~TeV CMS displaced lepton and HSCP searches, we were able
to reproduce the official results within 5\% to 40\%.

In summary, our overall conclusions regarding the relevant information 
required for a proper recasting of LLP searches are as follows:
\begin{enumerate}
\item {\em Object efficiencies} for reconstruction of the LLP signature must be provided in terms of the relevant truth level observables.
As discussed above, without these efficiencies the recasting uncertainties can be over 100\%.
\item {\em Cut-flow tables for expected signal events} would greatly improve the ability to recast these searches.  In the event of a mismatch in the calculation of the final expected signal yield, it is at the moment impossible to pinpoint the source of the problem.
\item {\em Limits for at least two models (or signal topologies)} should be published so a sanity check can be made before using object efficiencies on a different model.  Current standard practice  has been to provide object efficiencies with respect to Monte-Carlo truth-level objects for the benchmark model.  It is therefore still difficult to understand how well these can be trusted when applied to models or topologies other than the benchmark (e.g.\,see Figs.~\ref{fig:eff} and \ref{fig:effGMSB}).  
\end{enumerate}

\section*{ACKNOWLEDGEMENTS}

This research was supported in part by the S\~ao Paulo Research Foundation (FAPESP), projects
2015/20570-1 and 2016/50338-6. GC acknowledges support by the Ministry of Science and Technology of Taiwan under grant No. MOST-106-2811-M-002-035. JH acknowledges support by the German Research Foundation DFG through the 
research unit ``New physics at the LHC''.  ND acknowledges the support of the OCEVU Labex (ANR-11-LABX-0060) and the A*MIDEX project (ANR-11-IDEX-0001-02) funded by the "Investissements d'Avenir" French government program managed by the ANR.



\AddToContent{G.~Cottin, N.~Desai, J.~Heisig, A.~Lessa}
\renewcommand{\thesection}{\arabic{section}}


\chapter{Analysis description for LHC result reinterpretations}
\label{LHADA}

{\it P.~Gras, H.~B.~Prosper, S.~Sekmen}



\begin{abstract}
{\sc Lhada} is a language to describe LHC analysis which has a wide range of applications. In this work, this language is investigated for its usage in the context of LHC result reinterpretation. It would be employed to describe in an unambiguous and concise manner a data analysis including all the details needed for a reinterpretation of the result in the context of a physics theory not considered in the original analysis. A specialisation of the language dedicated to reinterpretation is introduced. The specialisation defines extra syntax rules and constitutes a subset of the language. Three different analyses used as benchmarks are described with this language. Automatic generation of code reproducing the analysis on Monte-Carlo samples for the purpose of result reinterpretation is investigated. We demonstrate that programs that generates code to be used in a result reinterpretation tool can be easily developed and a prototype is presented. In addition, the generated code can be used to validate the accuracy of the analysis description.
\end{abstract}

\section{Introduction}

The need for a standard to describe analyses of LHC data in an unambiguous way together with the definition of its requirements has been studied at the 2015 session of Les Houches PhysTeV workshop~\cite{Brooijmans:2016vro}. The study includes a proposal for this standard ({\sc Lhada}). In this work, we investigate this proposal in the context of analysis reinterpretation. Three questions are addressed: the coverage of the language, that is its ability to implement a large spectrum of analyses, the completeness of the analysis description, and the capacity to validate this description. The first question is addressed by implementing the description of example analyses with different levels of complexity. The second and third questions were addressed by developing two machine interpreters. The interpreters generate c$++$ code that reproduces the analysis on an input samples. There is an alternative approach, taken by the {\tt CutLang}~\cite{Sekmen:2018ehb} interpreter, which does a direct runtime interpretation of its internal text based analysis description language.  One of the two interpreters detailed in the following study produces a module, so-called ``Rivet analysis'', for the {\sc Rivet}~\cite{Buckley:2010ar} framework, while the second produces a standalone code based on the ROOT~\cite{Brun:1997pa} framework. The {\sc Rivet} based code is meant to be used for result reinterpretation. It can also be used to validate an analysis description by reproducing reference numbers provided by the analysis authors. In particular the cut flow, that is the acceptances of the subsequent selections (the ``cuts'') of the analysis, is well suited for such validation~\cite{Kraml:2012sg}. The completeness of the description is validated at the same time. The second interpreter, called {\sc Lhada2TNM} is aimed towards running a {\sc Lhada} analysis on any given type of input ROOT ntuple, and targets a more generic use, including design and implementation of analyses with the experimental data.

\section{Describing analyses in {\sc Lhada}}

In order to test the suitability of the {\sc Lhada} language to describe LHC analyses, three different new physics searches have been considered. The first analysis is the {\em Search for new physics in the all-hadronic final state with the MT2 variable} from Ref.~\cite{CMS:2016xva}. The two other analyses are the {\em Search for squarks and gluinos in final states with jets and missing transverse momentum at $\sqrt{s} = 13\,$TeV  with the ATLAS detector}~\cite{Aaboud:2016zdn} and {\em Search for dark matter at $\sqrt{s}=13\,$TeV in final states containing an energetic photon and large missing transverse momentum with the ATLAS detector}~\cite{Aaboud:2017dor}, which are used for the comparison of the reinterpretation tools performed in Contribution~\verb|\ref{sec:recast}|.

We have described the three analyses with the {\sc Lhada} language and the descriptions can be found in the analysis descriptions database~\cite{bib:lhada_git} respectively under {\tt lhada/analyses/CMS-PAS-SUS-16-015}, {\tt lhada/analyses /ATLASSUSY1605.03814}, and {\tt lhada/analyses/ATLASEXOT1704. 0384}. No particular difficulty has been encountered. The {\sc Lhada17} language subset has been used and a cut flow table has been included [editor's note: to be added] in the description to allow the validation of the descriptions using code generated with the interpreter.

\section{Generating a Rivet analysis from {\sc Lhada}}

{\sc Lhada} is a multipurpose and flexible language. In the case of {\sc Lhada2rivet}, the interpreter that generates a Rivet analysis, we have chosen to limit ourselves to the analysis reinterpretation use case and to specify accurately the analysis description language understood by the program. For this purpose, we have derived a subset of the {\sc Lhada} language, called {\sc Lhada17}. In this section, we will first draw the specifications of this sublanguage. We will then present the automatic generation of Rivet analyses.

\subsection{Describing the description language}\label{sec:desc}


The Extended Backus-Naur Form~\cite{bib:ebnf,bib:ebnf-wiki} (EBNF) notation has been used to specify the syntax and grammar of {\sc Lhada}17. The syntax is given in Appendix~\ref{app:ebnf}. The following rules which have not been included in the EBNF syntax to simplify the notation apply:
\begin{itemize}
\item A hash sign (\#) can be used to include comments in the {\sc Lhada} files: all characters of a line starting from a hash sign are comments and ignored for the interpretation based on the EBNF description.
\item If the last non-space character of a line is a backslash ($\backslash$), then the line is merged with the following line before being interpreted according to the EBNF description.
\item An entity ({\tt function}, {\tt object}  or {\tt cut}) should be declared before being used. For instance if a function is used in a ``cut'' definition, the corresponding {\tt function} block should appear before the {\tt cut} line. This rule is meant to simplify the parsing and also to avoid circular definitions. 
\end{itemize}

A specificity of {\sc Lhada} is the usage of programming languages to describe algorithms, via the {\sc Lhada} {\tt functions} while the main structure of the analysis is described with the dedicated language. A reference implementation of the algorithm is provided in a ``commonly used'' programming language. The implementation is given in a code source file, which can group implementations of different {\sc Lhada} {\tt functions} and can be shipped along with the {\sc Lhada} description file or provided as an http link.  In order to ease machine interpretation {\sc Lhada}17 includes the following restrictions for the reference {\tt function} implementations.

\begin{itemize}
\item the implementation is written in c$++$11~\cite{bib:c++11};
\item the implementation must depend only on the code provided in the file and libraries from the restricted set defined below; the file should be compilable with c$++$11 compliant compilers.
\item the allowed types for the function parameters are: {\tt int}, {\tt float}, {\tt double}, {\tt std::string}, {\tt LHADAParticle}, {\tt LHADAJet}, {\tt FourMomentum} and {\tt std::std::vector} of the last three types. Parameters are passed by copy (no modifier) or by constant reference (with {\tt const \&} modifier, like {\tt const LHADAParticle\&}); this rules exclude the use of a templated function; function templates are allowed for auxiliary functions; 
\item {\tt \#include} statements can be used to include header files from the allowed libraries;
\item the file, where the functions is defined should be compilable with c$++$11 compliant compilers;
\item self-contained function are encouraged but not mandatory; by self contained we mean that the file is compilable with c$++$11 compliant compilers after having removed all code except the function and the {\tt \#include} that precedes it;
\item the function can use the random object to draw pseudo-random numbers, whose scope is global to all functions and which provides the methods described in Table~\ref{tab:rand}; 
\end{itemize}

The {\tt LHADAParticle} and {\tt FourMomentum} types are two classes storing the properties listed in Tables~\ref{tab:part} and \ref{tab:mom}. The {\tt LHADAJet} type is identical to {\tt LHADAParticle}, but without the {\tt pdgid} property; it is introduced to distinguish jets from particles. In the {\tt select} and {\tt reject} statements, the property is referred to with its name, while in the c++ code, a method sharing the same name is used, e.g. {\tt name$()$} for the property {\tt name}. The complete class definition can be found in the {\sc Lhada} github repository under {\tt code\_lib/include}.

The restricted set of libraries includes the libraries that comes with the c$++$ standard ({\tt std} libraries) and a common library provided in the {\sc Lhada} github repository in the directory {\tt code\_lib}.  The set of libraries can evolve without requiring a revision of the {\sc Lhada17} language standard and will be defined by a list stored in the {\sc Lhada} repository.

\begin{table}
  \center
  \caption{Definition of the FourMomentum type: list of properties.\label{tab:mom}}
  \begin{tabular}{l|l}
    {\sc Lhada} & Description \\
    \hline
    mass 	& Mass\\
    e 	        & Energy\\
    px 	        & momentum x-component\\
    py 	        & momentum y-component\\
    pz 	        & momentum z-component\\
    pt 	        & absolute transverse momentum\\
    eta         & pseudorapidity\\
    rapidity    & rapidity\\
  \end{tabular}
\end{table}

\begin{table}
  \center
  \caption{Definition of the LHADAParticle type: list of properties coming in addition to the ones of the FourMomentum type.\label{tab:part}}
  \begin{tabular}{l|l}
    Property    & Description \\
    \hline
    pdgid       & PDG particle id\\
    charge      & charge\\
    x           & particle production vertex x-coordinate\\
    y           & particle production vertex y-coordinate\\
    z           & particle production vertex z-coordinate  \\
  \end{tabular}
\end{table}

\begin{table}
  \caption{Definition of the random object interface: list of provided methods. \label{tab:rand}}
  \begin{tabular}{l|p{20em}}
    c$++$ method & Description \\
    \hline
    uniform(double x)  & Returns a pseudorandom number following a uniform distribution over the $[0, \text{x}]$ interval.\\
    gauss(double mean, double sigma)    & Returns a pseudorandom number following a Gaussian distribution.\\
    poisson(int mean) & Returns a pseudorandom number following a Poisson distribution.\\
    breitWigner(double mean, double gamma) & Returns a pseudorandom number following a Breit-Wigner distribution. \\
    exp(double tau) & Return a pseudorandom number following the $\exp(-t/\text{tau})$ distribution. \\
    landau(double mean, double sigma) & Returns a pseudorandom number following a Landau distribution. \\
    binomial(int ntot, double prob) & Returns a pseudorandom number in the $[0, \text{ntot}]$ interval following a Binomial distribution.
  \end{tabular}
\end{table}


The {\sc Lhada} language does not explicitly specify how the arguments listed in a {\tt function} block are matched to the arguments of the reference implementation of the function. To prevent confusion, {\sc Lhada17} requires that the arguments appear in the {\tt function} block in the order of the c$++$ function argument list of its reference implementation and with the same name. If the names differ, the arguments should be matched according to their order, though in such case the file can simply be considered as invalid.

An {\tt object} block defines an entity, typically a collection of particles, starting from the input defined by the {\tt take} statement that is transformed by a sequence of {\tt apply}, {\tt select}, and {\tt reject}  statements. The {\tt apply} statement specifies a function that transforms the entity. In {\sc Lhada17}, the function must take as first argument the entity to transform. This argument is specified in the function definition, but not on the {\tt apply} statement, where it is implicit. Collections are filtered with {\tt select} and {\tt reject} statements. A condition to respectively keep or reject a collection element is provided in the form of a boolean expression. In addition to arithmetic and boolean operations, the expression can contain calls to functions. In the examples given in Ref.~\cite{Brooijmans:2016vro}, the functions take the collection element as an implicit argument, but this might not always be the case. In {\sc Lhada17} the following rule applies: if the number of passed arguments is less by one with respect to the expected one, then the collection element to filter is assumed to be implicitly passed as first argument. A mismatch between the expected argument type and the collection element type is considered as ill-formed. While the {\tt apply} statement is valid for an entity which is a single object (e.g. missing transverse momentum), the {\tt select} and {\tt reject} statements are restricted to collections. Blocks without a {\tt take} statement are also allowed. In this case there is no implicit argument in the {\tt apply} statement. It is strongly discourage to use this form when the one with a {\tt take} statement can be used.

The event selection is defined with {\tt cut} blocks. In order to simplify the description of the selection flow, the reference in a {\tt cut} block to another {\tt cut} block, introduced in {\tt Lhada} to allow branching, is allowed in the first statement of the block only.

Three extensions to {\sc Lhada} are introduced in {\sc Lhada17}. The first is the backslash line continuation marker described above. The second extension concerns the {\tt take external} statement of the {\tt object} block. In {\sc Lhada17} is it followed by a label identifying the object. A record of possible objects with their definition and properties (e.g. reconstruction efficiency and resolution in case of reconstructed particles) is kept in the {\tt Lhada} repository. The repository is updated when new objects are needed. Finally, we have introduced two aliases for the keyword {\tt object}: {\tt variable} and {\tt collection}. The {\tt object} block can represent several types of entities. Providing the possibility to use a name reflecting the type of entities, {\tt collection} when dealing with a collection of physics object, {\tt variable} when dealing with a single observable, like an event shape variable, should help in writing more intelligible analysis descriptions. The choice of the name is left to the discretion of the analysis description author.

\subsection{Automated generation of Rivet analysis code}

The {\sc Lhada} language will play a role for LHC result reinterpretations only if it is interfaced to commonly used reinterpretation frameworks. The interface can be done in two different ways. The first approach is to interpret the analysis description at run time. The second one, which is adopted here, is to generate code from the description.

An application, called {\sc Lhada2Rivet}, that produces a Rivet analysis from its description in {\sc Lhada17} is being developed. The analysis produces a cut flow table. Special care has been taken to produce code in the {\sc Rivet} style using facilities provided by the framework, like the projections or the {\tt CutFlow} class.

The detector response effect can be included in two different ways. The interpreter supports a list of reconstructed objects, defined in the {\sc Lhada} repository, using the {\sc Rivet} built-in feature. The {\sc Lhada} description will take reconstructed objects as {\tt external}. Alternatively, the detector effects (efficiency and resolution) can be defined in the {\sc Lhada} description: it then takes generator-level object (typically {\sc HepMC}~\cite{Dobbs:2001ck} particles) as {\tt external} and the detector effect is included using {\tt apply} statements and c$++$ functions. The random object was introduced in {\sc Lhada17} for this purpose and the common library includes a help function that can be called to apply efficiency and resolution effects.

The code produced by the {\sc Lhada2Rivet} interpreter for the first analysis considered in the previous section can be found in appendix~\ref{app:code}. The code was produced with an early version of the interpreter that did not include the detector effect simulation. The code is shown for illustration only and is not fully valid. Each {\sc Lhada} cut block is mapped to a c$++$ method. {\sc Rivet} built-in tools, as Projections and Cutflow are used, leading to a clean code that is well integrated in the framework. The calculation of the event count passing each cut using the Cutflow object is not correct and a proper support of cut flow is under development.  The {\sc Rivet} interface to the fastjet~\cite{Cacciari:2011ma} library is used to cluster the jets.

With this prototype we have investigated the different aspects that a result reinterpretation code generator based on a {\sc Lhada} analysis description should cover. We can conclude from this exercise that the development of such a generator that takes as input a description compliant with the {\sc Lhada17} specifications is possible with a reasonable effort. The code produced by such a generator can be used to validate the analysis description using analysis cut flow, which needs to be provided by the analysis authors.

\section{The {\sc Lhada2TNM} interpreter}
%

Two key design features of {\sc Lhada} are human readability and analysis framework independence. As noted above, framework independence can be tested by attempting to implement tools that automatically translate analyses described using {\sc Lhada} into analyses that can be executed in different analysis frameworks. Human readability is enhanced by limiting the number of rules and syntactical elements in {\sc Lhada}. But, since we also demand that {\sc Lhada} be sufficiently expressive to capture the details of LHC analyses, it pays to follow Einstein's advice: ``Everything should be made as simple as possible, but not simpler".    In the prototype of the  {\sc Lhada2TNM} translator, we have tried to place the burden where it properly belongs, namely, on the translator. For example, it is expected that physicists will write {\sc Lhada} files so that blocks appear in a natural order. However, the {\sc Lhada2TNM} translator does not rely on the order of blocks within a {\sc Lhada} file. The appropriate ordering of blocks is handled by the translator. Given a {\tt cut} block called {\tt signal}, which makes use of another called {\tt preselection}, {\sc Lhada2TNM} places  the code for {\tt preselection} before the code for {\tt signal} in the resulting C++ file.

Another example of placing the burden on the translator rather than on the author of a {\sc Lhada} file, concerns statements that span multiple lines. Many computer languages have syntactical elements to identify such statements. However, since {\sc Lhada} is a keyword-value language, continuation marks are not needed because it is possible to identify when the value associated with a statement ends. In order to determine where a statement ends, {\sc Lhada2TNM} looks ahead one record in the {\sc Lhada} file during translation.

The {\sc Lhada2TNM} translator is a {\tt Python} program that translates a {\sc Lhada} file to a C++ program that can be executed within the {\sc TNM} $n$-tuple analysis framework. This framework is the analysis component of a tool developed as a generic mapper from CMS analysis data objects to ntuples comprising integers and floats and arrays thereof. Note, however, that the framework depends on {\sc ROOT} only and not on any CMS data structures. {\sc TNM} therefore serves as a generic ntuple-based analysis framework. The {\sc Lhada2TNM} translator extracts all the blocks from a {\sc Lhada} file and places them within a data structure that groups the blocks according to type. The {\tt object} and {\tt cut} blocks are ordered according to their dependencies on other object or cut blocks. It is assumed that a standard, extensible, type is available to model all analysis objects and that an adapter exists to translate input types, e.g., {\sc Delphes}, {\sc ATLAS}, {\sc CMS}, etc., types  to the standard type. This assumption is not an imposition on the {\sc Lhada} language, but rather is an aid to the writing of translators and, or, interpreters for {\sc Lhada}. One benefit is that the C++ implementations provide a clear separation between the analysis code, viewed as an algorithm applied to instances of a standard type, and the input types.   

In the current version of {\sc Lhada2TNM}, instances of the standard, extensible, type as well as functions are placed in the global namespace of the C++ program so that the object and cut code blocks
that need them can access them without the need to pass objects between code blocks. The name of a function defined in {\sc Lhada} is assumed  to be identical with that of a function, which, ultimately, will be accessed from an online code repository. However, this assumption can be relaxed if warranted in a later iteration of {\sc Lhada}; for example, the appropriate function can be specified by its {\sc DOI}.  While the technical details of the automatic access of
codes from an online repository need to be worked out, we see no insurmountable hurdles. 

One of the purposes of the standard, extensible, type is to accomodate the reality that different input types can, and do, have different attributes and sometimes identical attributes with different names. For example, the transverse momentum of a particle may be called {\tt PT}, in {\sc Delphes}, while the same attribute may be called {\tt Pt} in other input types. It can  be argued that we should try to agree on naming conventions. But, in the real world of particle physics, we cannot even agree on whether the signal strength is to be defined as measured over predicted cross section or the inverse. Trying to enforce naming conventions, at least until such time as {\sc Lhada} has become mainstream, would be decidedly counter-productive. Therefore, the extensible type used by {\sc Lhada2TNM} uses the attribute names of the input types. The attributes are modeled as map between a name (as a string) and a floating point value.

\section{Conclusion}

The sustainability of the {\sc Lhada} language to be used in the context of analysis reinterpretation has been studied. Three questions have been addressed: the analysis range the language can cover, the completeness of the analysis description, and the capacity to validate the analysis description. The analysis coverage question has been tackled by taking three different LHC searches and implementing their description. The language turns out to be very flexible and no difficulty has been encountered in this exercise giving confidence that the proposed language covers the need. Nevertheless, it will be wise to extend the exercise with a larger number of and more sophisticated analyses. The two other questions have been addressed by developing the prototypes of two applications that interpret analysis descriptions written in {\sc Lhada}. The development of an application generating reinterpretation code out of a {\sc Lhada} analysis description can easily be done provided that the syntax used by the description is well-defined and not too flexible. A specialisation, {\sc Lhada17}, of the {\sc Lhada} language that fulfils this requirement has been set up. The restrictions introduced by {\sc Lhada17} have not been a limitation for the description of the analysis considered in the coverage test. 

\section*{Acknowledgements}

Since the first discussions in Les Houches 2015, many people contributed to developing the {\sc Lhada} concept.  
We would especially like to thank Daniel Dercks, Nishita Desai, Sabine Kraml, Suchita Kulkarni, Gokhan Unel, Federico Ambrogi, Wolfgang Waltenberger, Lukas Heinrich, Roberto Leonardi, Luca Perrozzi, Jim Pivarski, Kati Lassila-Perini, Tibor Simko and the CERN Analysis Preservation Support Group for their valuable input.  We also would like to thank the organizers of Les Houches PhysTeV 2017 workshop for a stimulating and fruitful workshop.


\section*{Appendices}

\appendix

\section{{\sc Lhada17} language syntax}\label{app:ebnf}

\fvset{fontsize=\footnotesize}
\begin{Verbatim}
(* Definition of the LHADA file structure *)

lhada file = info block, {function block}, {object block}, {cut block}, {table block}


(* Syntax of an info block *)

info block = "info analysis", EOL,
             {space, analysis info}

analysis info = analysis info key, text

analysis info key = "id" | "doc" | "experiment" | "publication" | "sqrtS" | "lumi" |
                     "arXiv" | "hepdata"


(* Syntax of a function block *)

function block = "function", space, function name, EOL,
                 {space}, "#", {char}, EOL,
		 {arg statement}
		 indent, "code", space, extented file path, EOL

function name = identifier
arg statement = indent, "arg", space, arg name, {space, arg name}, EOL
arg name = identifier

(* Syntax of an object block *)

object block = internal objecct | external object

internal object block = object, space, object name, EOL,
                        indent, "take", space, internal input, EOL,
	                {object optional statement}

object = "object" | "collection" | "variable"

object optional statement = apply statement | cut statement

internal input = "Particles" |  defined object

defined object = identifier (* the object must be defined in an object block in
                               the preceding text *)

external object name = identifier

external object block = object, space, object name, EOL,
                        indent, "take", space, "external", external object name, EOL,

apply statement = indent, "apply", space, function call

cut statement  = indent, "cut", expression

(* Syntax of a cut block *)
cut block = "cut",  space, cut name, EOL
             {"select", space, expression, EOL}

cut name = identifier

(* Syntax of a table block*)

table block = "table", space, table name, EOL,
               indent, "type", space, table type, EOL,
               indent, "columns", {space, column name}, EOL (* header *)
	       {indent, "entry", {space, cell content}, EOL} (* lines *)
	       (* The number of fields in table lines must match
	          with the number of columns defined in the header*)
               [indent, "hepdata", space, extended file path]

table name = identifier
table type = "events" | "limits" | "cutflow" | "corr" | "bkg"
cell content = (char - space}  

(* Syntax for a function call *)

function call =  defined function, "(", arg list, ")"
defined function = function name (* The function must be defined
                                    with a function block before in
				    the file *)
arg list = "" | arg, { ", ", arg}
arg = arg name, {space}, "=", {space}, expression, { ",", expression }


(* Miscellaneous *)

extended file path = ? http, https, or ftp url as defined in RFC 2386 ? |
                     ? unix-like path name ?


(* Syntax of a mathematical expression *)
(* Note: we have not expressed the operator precedence rules within the grammar below *)
(* The same precedence as the one of c/c++ language applied *)

expression = primary expression, {primary expression}

primary expression = real number | variable | unary operation | binary operation |
                     function call | "(", expression, ")"

variable = identifier

operator = "+" | "-" | "*" | "/" | "**" | "^" | "and" | "or" | "&&" | "||" | "<" | ">" |
           "<=" | ">=" | "!="
unary operator = "-" | "not" | "!"
unary operation = unary operator, ( real number | variable | unary operation |
                                    function call | "(" expression ")" )
binary operation = expression, binary operator, expression

(* Syntax of identifiers*)
identifier = ( letter | "_" ), {letter | "_" | digit }

(* Definition of character sets *)

letter = "A" | "B" | "C" | "D" | "E" | "F" | "G" | "H" | "I" | "J" | "K" | "L" | "M" |
         "N" | "O" | "P" | "Q" | "R" | "S" | "T" | "U" | "V" | "W" | "X" | "Y" | "Z" |
	 "a" | "b" | "c" | "d" | "e" | "f" | "g" | "h" | "i" | "j" | "k" | "l" | "m" |
	 "n" | "o" | "p" | "q" | "r" | "s" | "t" | "u" | "v" | "w" | "x" | "y" | "z"

digit = "0" | "1" | "2" | "3" | "4" | "5" | "6" | "7" | "8" | "9"

(* symbols include all 7-bit ASCII characters other than letters, digits and spaces *)
symbol =  "!" | """ | "#" | "$" | "%" | "&" | "'" | "(" | ")" | "*" | "+" | "," | "-" |
          "." | "/" | ":" | ";" | "<" | "=" | ">" | "?" | "@" | "[" | "\" | "]" | "^" |
	  "_" | "`" | "{" | "|" | "}" | "~"

space char = " " | (? ISO 6429 character Horizontal tab ?)

char = letter | digit | symbol | space char

sign = "+" | "-"

exp10 sign = "e" | "E"

EOL = ( ?  ISO 6429 character Line Feed ?)

(* Definition of numbers *)

natural number = digit, {digit}

integer = [sign], ( digit - "0" ) , { digit }

real number = decimal, { exp10 sign, integer }

decimal = ( ( integer, [ "." ] ) | ( { integer }, ".", natural number) )

(* Definition of space *)

space = {space char}
indent = space (* Alias used for a space at beginning of a line *)
\end{Verbatim}

\section{Example of code produced by the {\sc Ladha2rivet} interpreter}\label{app:code}

\begin{Verbatim}
// -*- C++ -*-
#include <iostream>
#include "Rivet/Analysis.hh"
#include "Rivet/Tools/Cutflow.hh"

#include <math.h>
#include "Rivet/Projections/FastJets.hh"
#include "Rivet/Projections/FastJets.hh"
#include "Rivet/Projections/MissingMomentum.hh"
namespace Rivet {


  class CMS_PAS_SUS_16_015: public Analysis {
  public:

    ///Cut ids
    enum {kmt2_cut, kdeltaphi_etmiss_jet, kveto, kpreselection,
	  kone_jet_selection, kat_least_two_jet_selection,
	  k1j_loose_1, k1j_loose_2, k1j_medium_1, k1j_medium_2,
	  k1j_medium_3, k1j_medium_4, k1j_medium_5, k1j_medium_6,
	  k1j_medium_7, k1j_medium_8} CutIds;
    /// Constructor
    CMS_PAS_SUS_16_015():      Analysis("CMS_PAS_SUS_16_015"),
      cutflow("CutFlow", {"kmt2_cut", "kdeltaphi_etmiss_jet",
	    "kveto", "kpreselection", "kone_jet_selection",
	    "kat_least_two_jet_selection", "k1j_loose_1",
	    "k1j_loose_2", "k1j_medium_1", "k1j_medium_2",
	    "k1j_medium_3", "k1j_medium_4", "k1j_medium_5",
	    "k1j_medium_6", "k1j_medium_7", "k1j_medium_8"})
    {  }


    /// Book histograms and initialise projections before the run
    void init() {

      FinalState fs;
      VisibleFinalState visfs(fs);

      addProjection(FastJets(fs, FastJets::ANTIKT, 0.4), "jets_eta47");
      addProjection(FastJets(fs, FastJets::ANTIKT, 0.4), "bjets");
      addProjection(MissingMomentum(fs), "met");
    }

    template<typename P>
    double  scalar_pt_sum(const std::vector<P>& momenta){
        double ht  = 0;
        for(const auto& p: momenta){
          ht += p.pt();
        }
        return ht;
    }


    double  btag_eff(const Particles& bjets){
        return 1.;
    }


    template<typename P>
    double  mt2(const P& particle1, const P& particle2, const P& met){
        return P();
    }


    template<typename P1, typename P2>
    double  dphi(const P1& p1, const P2& p2){
        double r = acos(p1.px()*p2.px() + p1.px()+p2.py() )
          / sqrt((p1.px()*p1.px() + p1.py()*p1.py())
		 * (p1.px()*p1.px() + p1.py()*p1.py()));
        return isnan(r) ? 0 : r;
    }


    bool cut_mt2_cut(){
        bool r = true;
        return cutflow.fill(kmt2_cut, r);
    }

    bool cut_deltaphi_etmiss_jet(){
        bool r = true;
        r &= (dphi(met, jets[1 - 1]) > 0.3)  && (dphi(met, jets[2 - 1]) > 0.3)
	  && (dphi(met, jets[3 - 1]) > 0.3)  && (dphi(met, jets[4 - 1]) > 0.3);
        return cutflow.fill(kdeltaphi_etmiss_jet, r);
    }

    bool cut_veto(){
        bool r = true;
        return cutflow.fill(kveto, r);
    }

    bool cut_preselection(){
        bool r = true;
        r &= cut_mt2_cut();
        r &= cut_deltaphi_etmiss_jet();
        r &= cut_veto();
        return cutflow.fill(kpreselection, r);
    }

    bool cut_one_jet_selection(){
        bool r = true;
        r &= cut_preselection();
        r &= jets.size() == 1;
        r &= jets[1 - 1].pt() > 200;
        return cutflow.fill(kone_jet_selection, r);
    }

    bool cut_at_least_two_jet_selection(){
        bool r = true;
        r &= cut_preselection();
        r &= jets.size()> 1;
        r &= cut_mt2_cut()> 200;
        return cutflow.fill(kat_least_two_jet_selection, r);
    }

    bool cut_1j_loose_1(){
        bool r = true;
        r &= cut_one_jet_selection();
        r &= ht> 575;
        return cutflow.fill(k1j_loose_1, r);
    }

    bool cut_1j_loose_2(){
        bool r = true;
        r &= cut_at_least_two_jet_selection();
        r &= bjets.size() < 3;
        r &= ht> 575;
        r &= ht < 1000;
        r &= cut_mt2_cut()> 200;
        return cutflow.fill(k1j_loose_2, r);
    }

    bool cut_1j_medium_1(){
        bool r = true;
        r &= cut_one_jet_selection();
        r &= bjets.size() < 1;
        r &= ht> 1000;
        return cutflow.fill(k1j_medium_1, r);
    }

    bool cut_1j_medium_2(){
        bool r = true;
        r &= cut_one_jet_selection();
        r &= bjets.size()> 0;
        r &= ht> 575;
        return cutflow.fill(k1j_medium_2, r);
    }

    bool cut_1j_medium_3(){
        bool r = true;
        r &= cut_at_least_two_jet_selection();
        r &= jets.size() < 4;
        r &= bjets.size() < 1;
        r &= ht> 575;
        r &= ht < 1000;
        r &= cut_mt2_cut()> 800;
        return cutflow.fill(k1j_medium_3, r);
    }

    bool cut_1j_medium_4(){
        bool r = true;
        r &= cut_at_least_two_jet_selection();
        r &= jets.size() < 4;
        r &= bjets.size()> 0;
        r &= bjets.size() < 3;
        r &= ht> 575;
        r &= ht < 1000;
        r &= cut_mt2_cut()> 600;
        return cutflow.fill(k1j_medium_4, r);
    }

    bool cut_1j_medium_5(){
        bool r = true;
        r &= cut_at_least_two_jet_selection();
        r &= jets.size() < 4;
        r &= bjets.size() < 2;
        r &= ht> 1000;
        r &= ht < 1500;
        r &= cut_mt2_cut()> 800;
        return cutflow.fill(k1j_medium_5, r);
    }

    bool cut_1j_medium_6(){
        bool r = true;
        r &= cut_at_least_two_jet_selection();
        r &= jets.size() < 4;
        r &= bjets.size() == 2;
        r &= ht> 1000;
        r &= ht < 1500;
        r &= cut_mt2_cut()> 400;
        return cutflow.fill(k1j_medium_6, r);
    }

    bool cut_1j_medium_7(){
        bool r = true;
        r &= cut_at_least_two_jet_selection();
        r &= jets.size() < 4;
        r &= bjets.size() < 2;
        r &= ht> 1500;
        r &= cut_mt2_cut()> 400;
        return cutflow.fill(k1j_medium_7, r);
    }

    bool cut_1j_medium_8(){
        bool r = true;
        r &= cut_at_least_two_jet_selection();
        r &= jets.size() < 4;
        r &= bjets.size() == 2;
        r &= ht> 1500;
        r &= cut_mt2_cut()> 200;
        return cutflow.fill(k1j_medium_8, r);
    }
    /// Perform the per-event analysis
    void analyze(const Event& event) {

       const FastJets& jets_eta47Proj = applyProjection<FastJets>(event, "jets_eta47");
       Jets jets_eta47 = jets_eta47Proj.jetsByPt();
       Jets jets;
       for(const auto& p: jets_eta47){
         if((p.eta() < 2.4)){
           jets.push_back(p);
         }
       }

       const FastJets& bjetsProj = applyProjection<FastJets>(event, "bjets");
       Jets bjets = bjetsProj.jetsByPt();
       const MissingMomentum metProj = applyProjection<MissingMomentum>(event, "met");
       met = metProj.missingMomentum();
       ht = scalar_pt_sum(jets);

       cut_1j_loose_1();
       cut_1j_loose_2();
       cut_1j_medium_1();
       cut_1j_medium_2();
       cut_1j_medium_3();
       cut_1j_medium_4();
       cut_1j_medium_5();
       cut_1j_medium_6();
       cut_1j_medium_7();
    }


    /// Normalise histograms etc., after the run
    void finalize() {
       std::cout << "Analsyis cut flow:\n"
                 << "-----------------\n\n"
                 << cutflow << "\n";
    }


  protected:

    //@{
    /** Collections and variables
     */
    //@}
     /** Analysis objects
      * @{
      */
     Jets jets_eta47;

     Jets jets;

     Jets bjets;

     FourMomentum met;

     double  ht;
     /** @}
     */    //@{
    /** Histograms
     */
    //@}

    ///Tracks the event counts after each cut
    Cutflow cutflow;

  };
  DECLARE_RIVET_PLUGIN(CMS_PAS_SUS_16_015);
}



\end{Verbatim}


\AddToContent{P.~Gras, H.~B.~Prosper, S.~Sekmen}
\renewcommand{\thesection}{\arabic{section}}
\renewcommand{\thesubsection}{\thesection.\arabic{subsection}}
\renewcommand{\thesubsubsection}{\thesubsection.\arabic{subsubsection}}
\renewcommand{\thefigure}{\arabic{figure}}
\renewcommand{\theequation}{\arabic{equation}}
\renewcommand{\thetable}{\arabic{table}}
\renewcommand{\thefootnote}{\arabic{footnote}}

\clearpage


\bibliography{LH_NewPhysics_Biblio}

\end{document}